\documentclass[a4paper,openany,dvips,bibliography=totoc]{scrbook}

\usepackage{tocloft}
\usepackage{setspace}
\usepackage{ifthen}
\usepackage[]{datetime} \settimeformat{ampmtime} \usdate

\newboolean{doDraft} \setboolean{doDraft}{false}  

\usepackage{scrpage2}
\lehead{\sffamily\headmark} 
\rohead{\sffamily\headmark} 
\lefoot{\sffamily\pagemark} 
\rofoot{\sffamily\pagemark} 
\ifthenelse{\boolean{doDraft}} { 
  \lofoot{\normalfont\sffamily\bfseries\textcolor{Mahogany}{DRAFT}} 
  \refoot{\normalfont\sffamily\bfseries\textcolor{Mahogany}{DRAFT}} 
}{}
\changefontsizes{11.5pt}

\usepackage{rotating}
\usepackage{packages/a4p}

\usepackage{epsfig}
\usepackage{xspace}
\usepackage{graphicx}
\usepackage{packages/physicsDef}

\usepackage{mathptmx}

\usepackage{amssymb}
\usepackage{amsmath}
\usepackage{multirow}
\usepackage{wrapfig}
\usepackage{maybemath}
\usepackage{capt-of}

\usepackage{bm}
\usepackage[percent]{overpic}

\usepackage{hhline}

\usepackage[format=plain,labelsep=space]{caption}
\setkomafont{captionlabel}{\sffamily\bfseries}

\usepackage{subfig}
\newcommand{\Subref}[1]{\protect\subref{#1}}

\usepackage[dvips,dvipsnames,usenames]{color}
\usepackage{varioref}
\usepackage{colortbl}

\usepackage[dvips, breaklinks={true}, bookmarks, colorlinks={true}, pdfpagemode={NON}, pdffitwindow=true, pdfstartview={FitB}, linkcolor={BlueViolet}, menucolor={BlueViolet}, citecolor={Mahogany}, filecolor={OliveGreen}, pagecolor={Mahogany}, urlcolor={MidnightBlue}, pdftitle={Double parton scattering in four-jet events in pp collisions at 7 TeV with the ATLAS experiment at the LHC}, pdfauthor={Iftach Sadeh}, pdfsubject={Double parton scattering in four-jet events in pp collisions at 7 TeV with the ATLAS experiment at the LHC}, pdfkeywords={ATLAS,Jets,MPI,DPS}]{hyperref}
\usepackage{packages/breakurl}
\usepackage[all]{hypcap}

\providecommand{\boldsymbol}[1]{\mbox{\boldmath $#1$}}
\labelformat{equation}{\textup{(#1)}}
\labelformat{enumi}{\textup{(#1)}}

\let\orgautoref\autoref
\providecommand{\Autoref}
        {\def\equationautorefname{Equation}%
         \def\figureautorefname{Figure}%
         \def\subfigureautorefname{Figure}%
         \def\chapterautorefname{Chapter}%
         \def\sectionautorefname{Section}%
         \def\subsectionautorefname{Section}%
         \def\subsubsectionautorefname{Section}%
         \def\Itemautorefname{Item}%
         \def\tableautorefname{Table}%
         \def\appendixautorefname{Appendix}%
         \orgautoref}

\providecommand{\Autorefs}
        {\def\equationautorefname{Equations}%
         \def\figureautorefname{Figures}%
         \def\subfigureautorefname{Figures}%
         \def\chapterautorefname{Chapters}%
         \def\sectionautorefname{Sections}%
         \def\subsectionautorefname{Sections}%
         \def\subsubsectionautorefname{Sections}%
         \def\Itemautorefname{Items}%
         \def\tableautorefname{Tables}%
         \def\appendixautorefname{Appendices}%
         \orgautoref}

\renewcommand{\autoref}
        {\def\equationautorefname{equation}%
         \def\figureautorefname{figure}%
         \def\subfigureautorefname{figure}%
         \def\chapterautorefname{chapter}%
         \def\sectionautorefname{section}%
         \def\subsectionautorefname{section}%
         \def\subsubsectionautorefname{section}%
         \def\Itemautorefname{item}%
         \def\tableautorefname{table}%
         \def\appendixautorefname{appendix}%
         \orgautoref}

\providecommand{\autorefs}
        {\def\equationautorefname{equations}%
         \def\figureautorefname{figures}%
         \def\subfigureautorefname{figures}%
         \def\chapterautorefname{chapters}%
         \def\sectionautorefname{sections}%
         \def\subsectionautorefname{sections}%
         \def\subsubsectionautorefname{sections}%
         \def\Itemautorefname{items}%
         \def\tableautorefname{tables}%
         \def\appendixautorefname{appendices}%
         \orgautoref}

\usepackage{packages/tautitle}
\usepackage{cite,packages/mcite}

\newcommand{\ul}{\kern -0.1em\_}

\newcommand{\ATLAS}{\text{ATLAS}\xspace}  \newcommand{\atlas}{\ATLAS}
\newcommand{\CMS}{\text{CMS}\xspace}

\newcommand{\cms}{\text{center-of-mass}\xspace}
\newcommand{\fastjet}   {\textsc{FastJet}\xspace}

\newcommand{\SISCONE}   {SISCone\xspace}
\newcommand{\CAalg}     {Cambridge/Aachen\xspace}
\def\AKT{\hbox{anti-${k_{\perp}}$}\xspace}
\def\KT{\hbox{$k_{\perp}$}\xspace}

\newcommand{\geant}     {\textsc{Geant}\xspace}          
\newcommand{\pythia}    {\textsc{Pythia}\xspace}        \newcommand{\PYTHIA}    {\pythia}
\newcommand{\sherpa}    {\textsc{Sherpa}\xspace}        
\newcommand{\alpgen}    {\textsc{Alpgen}\xspace}        
\newcommand{\jimmy}     {\textsc{Jimmy}\xspace}         
\newcommand{\herwig}    {{\textsc{Herwig}}\xspace}      
\newcommand{\herwigpp}  {\textsc{Herwig++}\xspace}      
\newcommand{\herwigJimmy}{\textsc{Herwig+Jimmy}\xspace}
\newcommand{\nlojet}    {\textsc{NLOJET++}\xspace}
\newcommand{\applgrid}  {\textsc{APPLGR}\xspace}
\newcommand{\RooUnfold} {\text{RooUnfold}\xspace}
\newcommand{\ahj}       {\textsc{AHJ}\xspace}        \newcommand{\AHJ}    {\ahj}

\newcommand{\ie}        {i.e.\/,\xspace}
\newcommand{\eg}        {e.g.\/,\xspace}
\newcommand{\insitu}    {\textit{in-situ}\xspace}
\newcommand{\Insitu}    {\textit{In-situ}\xspace}

\newcommand{\Ajet}{\ensuremath{\mathcal{A}_{\mrm{jet}}}}
 
\def\sqs{\ensuremath{\sqrt{s}=7~\mathrm{TeV}}\xspace}

\def\etaphi{\hbox{$\eta$,$\phi$}\xspace}

\def\lumiUnits{\hbox{$\mathrm{cm}^{-2}\mathrm{s}^{-1}$}\xspace}

\newcommand{\mrm}[1]{\mathrm{#1}}

\def\ystr{\hbox{$y^{\ast}$}\xspace}

\newcommand{\headFont}[1]{\textbf{\sffamily\bfseries#1}}

\newcommand{\ue}    {underlying event\xspace}

\newcommand{\pu}    {pile-up\xspace}
\newcommand{\Pu}    {Pile-up\xspace}
\newcommand{\itpu}  {in-time pile-up\xspace}

\newcommand{\otpu}  {out-of-time pile-up\xspace}

\newcommand{\pp}    {\ensuremath{pp}\xspace}
\def\Npv{\ensuremath{N_{\mathrm{PV}}}\xspace}
\def\NpvL{\ensuremath{N_{\mathrm{PV}}^{\mathrm{loose}}}\xspace}
\def\Mu{\ensuremath{\mu}\xspace}
\def\Rho{\ensuremath{\rho}\xspace}
\def\Eta{\ensuremath{\eta}\xspace}

\newcommand{\topos}    {topo-clusters\xspace}
\newcommand{\Topos}    {Topo-clusters\xspace}
\newcommand{\topo}    {topo-cluster\xspace}

\renewcommand{\TeV}{~\text{Te\kern -0.1em V}\xspace}
\renewcommand{\GeV}{~\text{Ge\kern -0.1em V}\xspace}
\renewcommand{\MeV}{~\text{Me\kern -0.1em V}\xspace}
\renewcommand{\keV}{~\text{ke\kern -0.1em V}\xspace}
\renewcommand{\eV}{~\text{e\kern -0.1em V}\xspace}

\newcommand{\fb}{~\text{fb}\xspace}
\newcommand{\mb}{~\text{mb}\xspace}
\newcommand{\ns}{~\text{ns}\xspace}
\newcommand{\wrt}{\text{with regard to}\xspace}

\newcommand{\inv}[1]{\ensuremath{{{#1}^{-1}}}}

\newcommand{\pow}[2]{\ensuremath{{{#1}\cdot10^{{#2}}}}}
\newcommand{\etaLower}[1]{\ensuremath{{\left|\eta\right|<{#1}}}}
\newcommand{\etaHigher}[1]{\ensuremath{{\left|\eta\right|>{#1}}}}
\newcommand{\etaRange}[2]{\ensuremath{{{#1}<\left|\eta\right|<{#2}}}}
\newcommand{\muRange}[2]{\ensuremath{{{#1}<\mu<{#2}}}}
\newcommand{\yLower}[1]{\ensuremath{{\left|y\right|<{#1}}}}

\newcommand{\ptRange}[2]{\ensuremath{{{#1}<\pt<{#2}}}\GeV}
\newcommand{\ptRangeT}[2]{\ensuremath{{{#1}<\pt<{#2}}}\TeV}
\newcommand{\ptLower}[1]{\ensuremath{{\pt<{#1}}}\GeV}
\newcommand{\ptHigher}[1]{\ensuremath{{\pt>{#1}}}\GeV}

\newcommand{\Scale}    {\texttt{S}\xspace}
\newcommand{\EM}    {\texttt{EM}\xspace}
\newcommand{\EMJES}    {\texttt{EM\kern -0.04em +\kern -0.18em JES}\xspace}
\newcommand{\JES}    {\texttt{JES}\xspace}
\newcommand{\LC}    {\texttt{LCW}\xspace}
\newcommand{\LCJES}    {\texttt{LCW\kern -0.06em +\kern -0.18em JES}\xspace}

\newcommand{\figQaud}    {\qquad \qquad \qquad \qquad \qquad \quad \,}

\newcommand{\ttt}[1]{\texttt{#1}}
\newcommand{\vect}[1]{\boldsymbol{#1}}

\newenvironment{Tabular}[2][1]
  {\def\arraystretch{#1}\tabular{#2}}
  {\endtabular}
\newcolumntype{C}[1]{>{\centering\let\newline\\\arraybackslash\hspace{0pt}}m{#1}}

\newcommand\scalemath[2]{\scalebox{#1}{\mbox{\ensuremath{\displaystyle #2}}}}

\makeatletter \newcommand\mynobreakpar{\par\nobreak\@afterheading} \makeatother 

\DeclareMathAlphabet{\mathcal}{OMS}{cmsy}{m}{n}

\newcommand{\fDPS}{\ensuremath{f_\mathrm{DPS}}\xspace}
\newcommand{\sigeff}{\ensuremath{\sigma_\mathrm{eff}}\xspace}

\newcommand{\twofour}{\ensuremath{(2\kern -0.25em \to \kern -0.2em 4)}\xspace}
\newcommand{\threefour}{\ensuremath{(3\kern -0.25em \to \kern -0.2em 4)}\xspace}
\newcommand{\twotwo}{\ensuremath{(2\kern -0.25em \to \kern -0.2em 2)}\xspace}
\newcommand{\dtwot}{\ensuremath{(2\kern -0.25em \to \kern -0.2em 2)^{\kern -0.12em\times\kern -0.05em 2}}\xspace}

\newcommand{\twothree}{\ensuremath{(2\kern -0.25em \to \kern -0.2em 3)}\xspace}
\newcommand{\onetwo}{\ensuremath{(1\kern -0.25em \to \kern -0.2em 2)}\xspace}

\newcommand{\Xsec}{Cross section\xspace}
\newcommand{\xsec}{cross section\xspace}
\newcommand{\Xsecs}{Cross sections\xspace}
\newcommand{\xsecs}{cross sections\xspace}
\newcommand{\com}{center-of-mass\xspace}

\begin{document}


\maketitle 

\chapter*{\begin{center}Abstract\end{center}}
%
%
The dijet double-differential \xsec is measured as a function of the dijet invariant mass, using
data taken during 2010 and during 2011 with the \atlas experiment at the LHC, with a \com energy, \sqs.
The measurements are sensitive to invariant masses between~$70\GeV$ and~4.27\TeV with \com jet rapidities up to~3.5.
A novel technique to correct jets for \pu (additional proton-proton collisions) in the 2011 data is developed and
subsequently used in the measurement.
The data are found to be consistent over~12 orders of magnitude
with fixed-order NLO pQCD predictions provided by \nlojet.
The results constitute a stringent test of pQCD, in an energy regime previously unexplored.

The dijet analysis is a confidence building step for the extraction of the signal of
hard double parton scattering in four-jet events, and subsequent extraction of the
effective overlap area between the interacting protons, expressed in terms of the variable, \sigeff.
The measurement of double parton scattering is performed using the 2010 \atlas data.
The rate of double parton scattering events is estimated using a neural network. 
A clear signal is observed, under the assumption
that the double parton scattering signal can be represented by a random combination of exclusive dijet production.
The fraction of double parton scattering candidate events is determined to be
$\fDPS = 0.081~\pm~0.004~\mrm{(stat.)}~^{+\;0.025}_{-0.014}~\mrm{(syst.)}$ in the analyzed phase-space of
four-jet topologies.
Combined with the measurement of the dijet and four-jet \xsecs in the appropriate phase-space regions,
the effective \xsec is found to be
${\sigeff = 16.0~^{+\;0.5}_{-0.8}~\mrm{(stat.)}~^{+\;1.9}_{-3.5}~\mrm{(syst.)}~\mrm{mb}}$.
This result is consistent within the quoted uncertainties with previous measurements of \sigeff
at \com energies between~63\GeV and~7\TeV, using several final states.

\chapter*{\begin{center}Acknowledgements\end{center}}
%
%
It is with immense gratitude that I acknowledge
my supervisors, Prof.\/~Halina Abramowicz and Prof.\/~Aharon Levy, not only for
teaching me physics, but for making me feel like part of the family. This thesis is a tribute
to their steadfast encouragement, guidance and support, making the course of my studies
a pleasure which is difficult to leave behind. I could not have hoped for a better pair
of physicists, or for a more caring pair of people, to be a student of.

I consider it an honor to have worked on the double parton scattering analysis with my colleague and friend, Orel Gueta.
Defeats and victories, late night insights and countless discussions, have all culminated in a measurement to be proud of.
Great thanks is also owed to Dr.\/~Arthur Moraes, for inspiring us with his ideas about
multiple interactions, and for keeping faith when all seemed lost. In addition,
the shared experience of Dr.\/~Frank Krauss and of Dr.\/~Eleanor Dobson has been instrumental
in the success of the analysis, and is much appreciated.
On the theoretical side of things, the fruitful discussions
with Prof.\/~Yuri Dokshitzer, Prof.\/~Mark Strikman, Prof.\/~Leonid Frankfurt and Prof.\/~Evgeny Levin
were of great help in formulating the strategy of the analysis.

I would like to show special appreciation to Dr.\/~Sven Menke and Dr.\/~Teresa Barillari
for the stimulating discussions on all things calorimeter (and then some),
and to Prof.\/~Allen Caldwell in addition. They have made my visits to 
the Max-Planck-Institut f\"{u}r Physik in Munich an experience to remember, intellectually and otherwise.

Several people have contributed greatly to my understanding of the workings of the \atlas experiment,
be it in terms of performance or of physics. The debates with Dr.\/~Ariel Schwartzman
and Dr.\/~Peter Loch  have been
of great help in my development of a \pu correction using the jet area/median method.
To start me off on the invariant mass measurement, the guidance of Dr.\/~Christopher Meyer via many
hours of discussions and a multitude of emails, has been invaluable.
I would like to thank Dr.\/~Pavol Strizenec for starting me off on \atlas software. I would also like
to Acknowledge Dr.\/~Eric Feng and Dr.\/~Serguei Yanush, for developing the software framework for
calculation of the non-perturbative corrections
to the invariant mass theoretical predictions.

I am grateful to my colleagues and the members of the group at Tel Aviv University,
Prof.\/~Gideon Alexander, Prof.\/~Erez Etzion, Prof.\/~Abner Soffer, Dr.\/~Gideon Bella,
Dr.\/~Sergey Kananov, Dr.\/~Zhenya Gurvich, 
Dr.\/~Ronen Ingbir, Amir Stern, Rina Schwartz and Itamar Levy, for allowing me to bounce ideas off of,
for the stimulating discussions, and for the pleasure of their company.

Finally, completing this thesis would not have been possible without the love and support of my family. As with
everything else which is good in my life, it all starts and ends with them. Thank you for being you.

\chapter*{\begin{center}Preface\end{center}}
%
%
The goal of this thesis is the study of double-parton scattering
(DPS) in four-jet events with the \atlas experiment.
In order to extract DPS in this channel, a good understanding of
the reconstruction and calibration of jets is needed. A comprehensive framework exists in \atlas
for this purpose, featuring two main calibration schemes, referred to as the electromagnetic (\EM)
and the local-hadron (\LC) calibrations.
These rely on extensive test-beam and simulation campaigns, which are
the result of the efforts of a large number of individual researchers and analysis subgroups.

Test-beam data taken during~2004 served to test the detector performance and to validate the description of the data by simulations.
Due to changing software models, these data became incompatible with current \atlas reconstruction tools. In order to maintain future access,
the information had to be made \textit{persistent}, \ie compatible with all future software.
The first project which had been undertaken by the author of this thesis in \atlas, was persistification of the test-beam data.
This was followed by continued support and maintenance of the \atlas calorimeter reconstruction software,
as part of the operational contribution of the author to the experiment.

One of the major challenges in the calibration of jets in \atlas, is the existence of \textit{\pu}, additional proton-proton ($pp$) collisions,
which coincide with the hard scattering of interest. The effects of \pu on final states which involve
jets are complicated. \Pu tends to both bias the energy of jets which originate from the hard interaction, and
to introduce additional jets which originate from the extraneous $pp$ collisions.
The current \pu subtraction method in \atlas involves a simulation-based scheme; it affects an average
correction for jet energies, based on the instantaneous luminosity and on the number of reconstructed vertices in an event.
An alternate, event-by-event-based correction, has been developed by the author for the \LC calibration scheme.
In the new correction, referred to as the \textit{jet area/median method},
the area of a given jet and the ``local'' energy-density, are used in order to subtract \pu energy from the jet.
The method takes advantage both of the average response of the calorimeter to \pu energy, and of
the observed energy in the vicinity of the jet of interest. The median method is completely data-driven.
Consequently, compared to the nominal \pu correction, the uncertainties on the energy correction
associated with the simulation of \pu were reduced.

The new \pu correction was developed by the author, initially by using the \atlas detector simulation.
It was subsequently validated by the author with the 2011~\atlas dataset, in which the rate of \pu is high.
The validation included \insitu measurements of several observables, one of which was the invariant
mass of dijets,
the system of the two jets with the highest transverse momentum in an event. 
The author also performed measurements of the dijet double-differential \xsec for different \com jet rapidities,
as a function of the invariant mass of dijets.
The invariant mass spectra had previously been measured in \atlas using the 2010~data. The new measurements, performed for
the first time on the 2011~data, were found to be compatible with the previous observations,
showing that the \pu corrections were under control.
In addition, the measurement using the larger dataset recorded during 2011,
served to extend the experimental reach to higher values of the invariant mass.

Using the improved jet calibration, events with four-jets in the final state were investigated
as part of the DPS analysis.
It was shown by the author that the deterioration of the energy resolution due to \pu
distorted greatly the observables of the measurement. The scope of the measurement had, therefore, to be limited.
As this is a first measurement of double parton scattering in this channel in \atlas, it was decided
to choose a conservative
approach and to limit the systematic uncertainties as much as possible.
The analysis, therefore, used only single-vertex events from the 2010~dataset,
for which \pu corrections are small.

Several strategies of the DPS analysis were explored by the author. For instance, the author attempted to
measure the fraction of DPS events, by exploiting
the sum of the pair-wise transverse momentum balance between different jet-pair combinations in a four-jet event, as
previously done by the CDF collaboration.
The problem with this type of approach was the absence of an appropriate simulation sample,
in which DPS events within the required phase-space were generated.
In order to avoid dependence on simulated DPS events, and in addition, to increase the robustness of
the analysis, the author decided to utilize a neural network.
Two input samples were prepared by the author for the neural network. The first consisted
of simulated events in which multiple-interactions were switched off. The second, which
stood for the DPS signal, was comprised of overlaid simulated dijet events.
Utilizing the neural network, the fraction of double parton scattering events was
extracted from the data, and the effective \xsec for DPS was subsequently  measured.
The result was found to be compatible with
previous measurements
which had used several final states,
at \com energies between 63\GeV and 7\TeV.

\newpage

\begin{spacing}{0.97}
\renewcommand*\contentsname{Table of Contents} \phantomsection \addcontentsline{toc}{chapter}{Table of Contents}{}
\tableofcontents
\end{spacing}
\newpage

\newboolean{do:introduction}      \setboolean{do:introduction}{false}       \setboolean{do:introduction}{true}   
\newboolean{do:atlasDetector}     \setboolean{do:atlasDetector}{false}      \setboolean{do:atlasDetector}{true}
\newboolean{do:MCsimulation}      \setboolean{do:MCsimulation}{false}       \setboolean{do:MCsimulation}{true}
\newboolean{do:jetReconstruction} \setboolean{do:jetReconstruction}{false}  \setboolean{do:jetReconstruction}{true}
\newboolean{do:dataSelection}     \setboolean{do:dataSelection}{false}      \setboolean{do:dataSelection}{true}
\newboolean{do:jetMedian}         \setboolean{do:jetMedian}{false}          \setboolean{do:jetMedian}{true}
\newboolean{do:jetMass}           \setboolean{do:jetMass}{false}            \setboolean{do:jetMass}{true}
\newboolean{do:fourJetDPS}        \setboolean{do:fourJetDPS}{false}         \setboolean{do:fourJetDPS}{true}
\newboolean{do:summary}           \setboolean{do:summary}{false}            \setboolean{do:summary}{true}
\newboolean{do:plotAppendix}      \setboolean{do:plotAppendix}{false}       \setboolean{do:plotAppendix}{true}
%
\newboolean{do:includeAllGraphics}      \setboolean{do:includeAllGraphics}{false}     \setboolean{do:includeAllGraphics}{true}
%
%

\ifthenelse{\boolean{do:introduction}}      { 
\chapter{Introduction\label{chapIntroduction}}
%
The Standard Model (SM)~\cite{nachtmann1990elementary} is one
of the major intellectual achievements of the twentieth century. In the late 1960s
and early 1970s, decades of path breaking experiments culminated in the emergence of 
a comprehensive theory of particle physics. This theory identifies the fundamental constituents of
matter and combines the theory of electromagnetic, weak and strong interactions.

Numerous measurements at energy scales from a few\eV to several\TeV
are reproduced by the SM, and many of its predictions, \eg the existence of the $W$ and $Z$
bosons, have been found to be realised in nature. By now, only the source of electroweak
symmetry breaking, which in the SM is attributed to the Higgs mechanism, has not been
verified experimentally. A particle with properties consistent with those of
the Higgs boson has recently been discovered at the LHC by the
\atlas~\cite{:2012gk} and \CMS~\cite{:2012gu} experiments; the final piece
of the puzzle is therefore within reach.

The Standard Model falls short of being a complete theory of fundamental interactions because
it does not incorporate the full theory of gravitation, as described by general relativity;
nor does it predict the accelerating expansion of the universe, as possibly described by dark energy.
The theory does not contain any viable dark matter particle that possesses all of the required
properties deduced from observational cosmology. It also does not correctly account for
neutrino oscillations or for the non-zero neutrino masses.
Although the SM is believed to be theoretically self-consistent,
it has several apparently unnatural properties, giving rise to puzzles like the
\textit{strong CP problem} and the \textit{hierarchy problem}~\cite{ramond2004journeys}.
The SM is therefore viewed as an effective
field theory that is valid at the lower energy scales where measurements have been
performed, but which arises from a more fundamental theory at higher scales.
It is therefore expected that a more fundamental theory exists beyond the\TeV energy scale that explains
the missing features.

Verification of the Higgs mechanism and the search for new physics beyond the SM are two of the primary enterprises in
particle physics today.
Any experimental search for the Higgs boson or for new interactions or particles,
requires a detailed understanding of the strong interactions, described by Quantum Chromodynamics (QCD).
It is common to discuss high-energy phenomena involving QCD
in terms of partons (quarks and gluons), yet partons are never visible in their own right.
Almost immediately after being produced, a quark or gluon fragments and hadronizes,
leading to a collimated spray of energetic hadrons, which may be characterized by a \textit{jet}~\cite{Salam:2009jx}.
Since partons interact strongly, jet production is the
dominant hard scattering process in the SM. \Autoref{FIGxsLHC} shows the production
cross section for various processes as a function of the \cms energy of
an accelerator.
\begin{figure}[ht]
\begin{center}
\ifthenelse{\boolean{do:includeAllGraphics}} {
  \includegraphics[trim=0mm 0mm 0mm 0mm,clip,width=.6\textwidth]{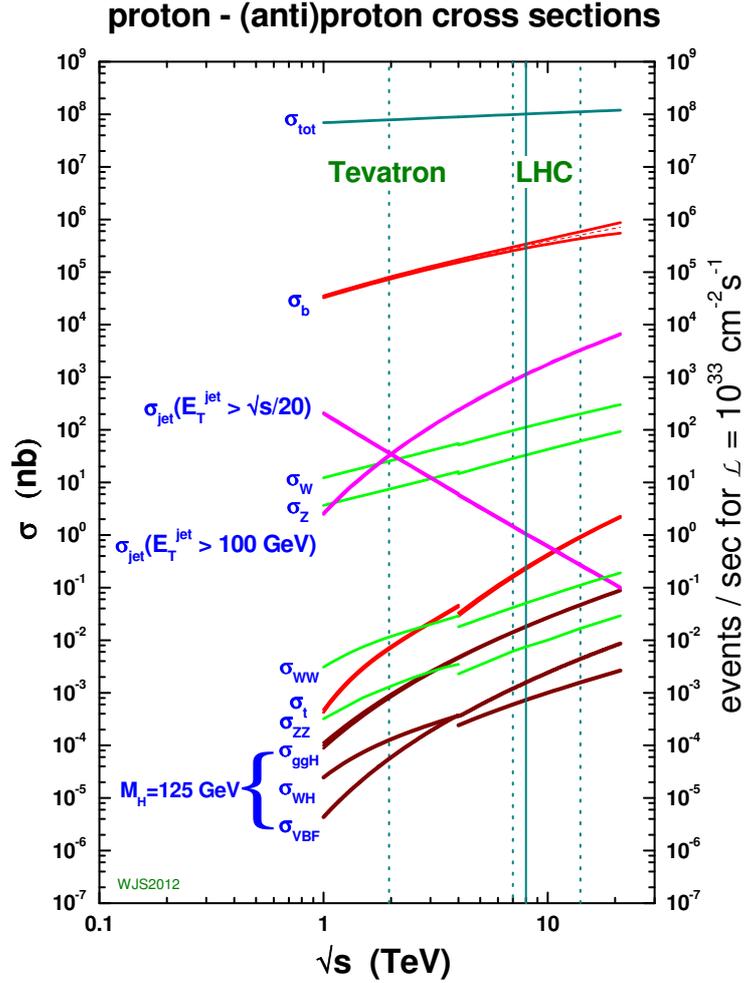}
}{
  \includegraphics[width=0.4\textwidth]{figures/plotFiller}
  }
\caption{\label{FIGxsLHC}
\Xsecs for different physics processes as a function of \cms energy, $\sqrt{s}$, assuming the existance of
a Higgs boson with a mass of~125~GeV.
(Figure by {J.~Stirling}, taken from~\protect\url{http://projects.hepforge.org/mstwpdf/}.)
}
\end{center}
\end{figure} 
One may compare the production cross section for energetic jets with transverse energy above~$100\GeV$,
with those of other processes ($W$ and $Z$) involving leptons.
At the \cms energy of the LHC, the former is much larger than the latter.

Because of the large cross sections, jet production provides an ideal avenue to probe
QCD and parton distribution functions~\cite{Stump:2003yu,PDF-MRST,PDF-CTEQ,Pumplin:2002vw},
which describe the distribution of the momenta of quarks and gluons within a proton.
Processes involving jets also serve as large backgrounds in many searches
for new physics.
Many models predict the production of new heavy and coloured particles at
the LHC. These particles are expected to decay into quarks and gluons, which are detected 
as particle jets. 
Such models may therefore be tested by measuring the rate of
jet production and comparing it to the expectations from QCD~\cite{Aad:2010ae}.
Other models, such as quark \textit{compositeness} (the hypothesis that quarks are composed of more fundamental particles),
may be realised through \textit{contact interactions} between quarks. Such new phenomena may be
discovered by measuring the kinematic distributions of jets~\cite{Collaboration:2010eza}.

In the most common final state involving jets at the LHC,
two jets with high transverse momentum, \pt, emerge from the interaction, as illustrated in \autoref{FIGfourJetEventDisplay1}. These \textit{dijet} events are
particularly useful for measuring quantities associated with the initial interaction, such as the polar scattering
angle in the two-parton \cms frame, and the dijet invariant mass.
Precise tests of perturbative QCD (pQCD) at high energies, may therefore be carried out by comparing the
theoretical predictions to the experimental distributions.
In addition, new physics may manifest itself with \eg the production of a new massive
particle, which subsequently decays into a high-mass dijet system~\cite{Aad:2011aj}.
Final states composed of low-\pt or forward jets are also interesting.
These topologies are sensitive to QCD effects which can not be calculated using perturbation theory;
as such, their measurement may help to improve the phenomenological models which are currently in use
in this regime.


%
%
\begin{figure}[ht]
\begin{center}
\ifthenelse{\boolean{do:includeAllGraphics}} {
  \subfloat[]{\label{FIGfourJetEventDisplay1}\includegraphics[trim=0mm 121mm 0mm 00mm,clip,width=.49\textwidth]{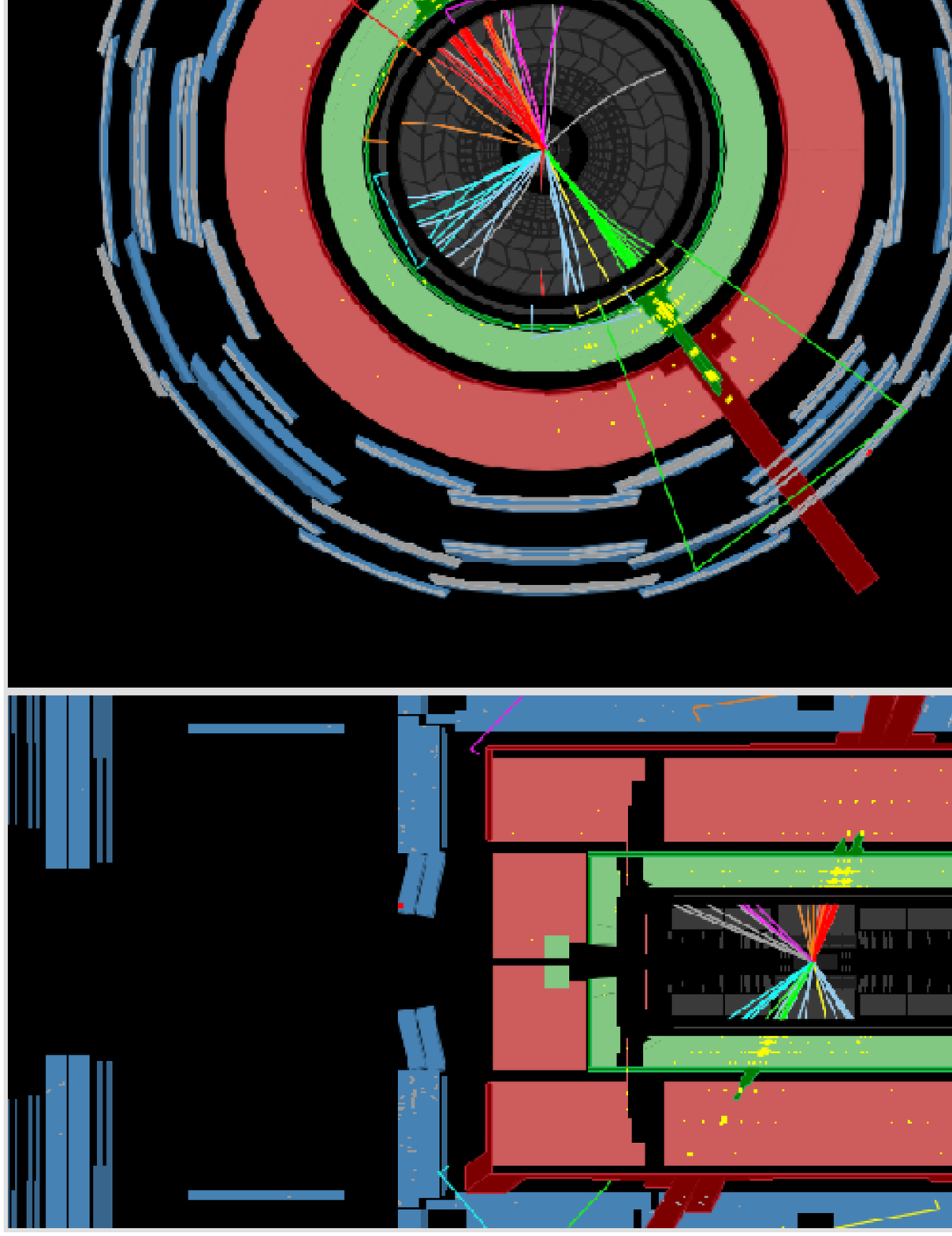}} \;
  \subfloat[]{\label{FIGfourJetEventDisplay2}\includegraphics[trim=0mm   2mm 0mm 00mm,clip,width=.49\textwidth]{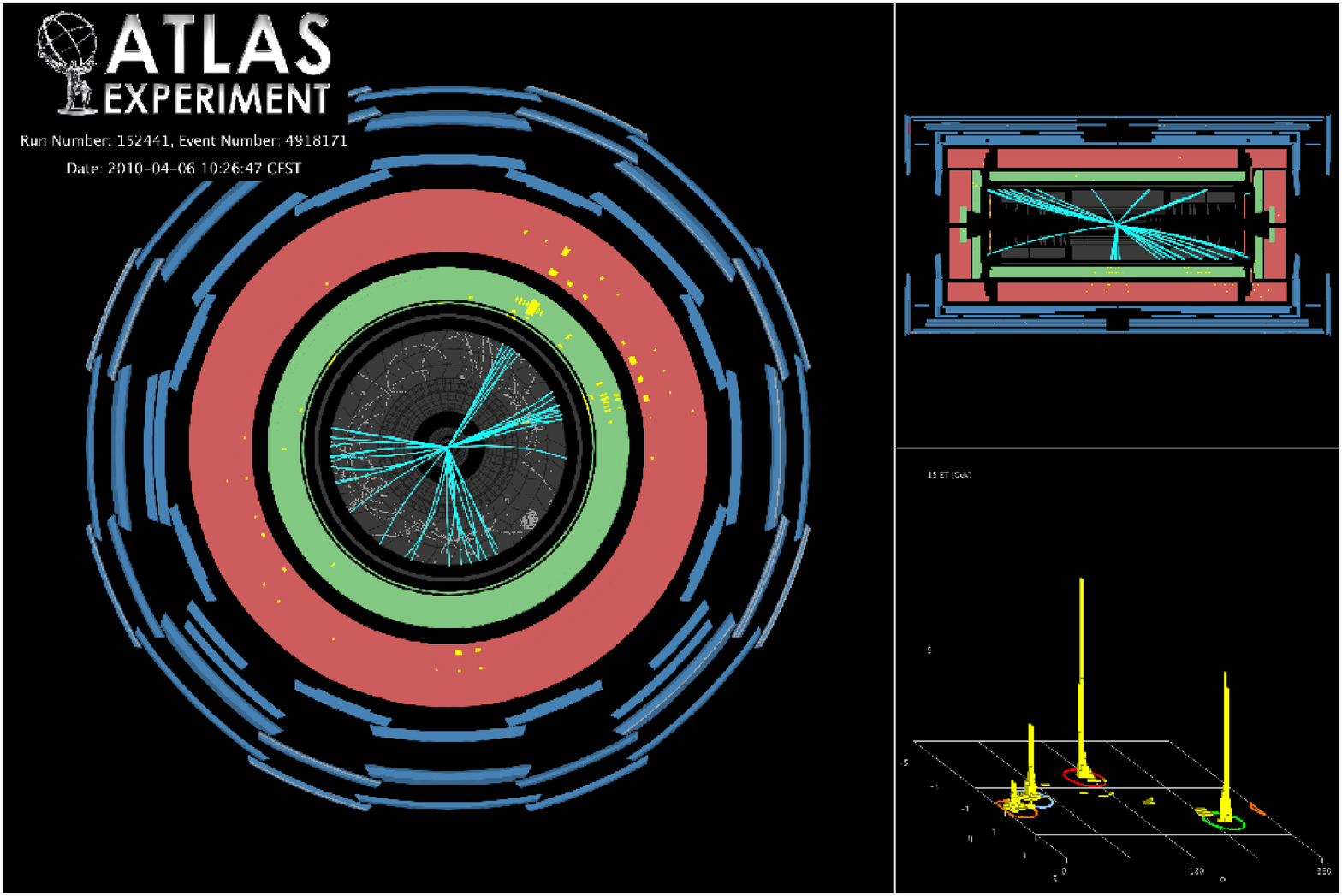}} 
}{
  \subfloat[]{\label{FIGfourJetEventDisplay1}\includegraphics[width=0.4\textwidth]{figures/plotFiller}}
  \subfloat[]{\label{FIGfourJetEventDisplay2}\includegraphics[width=0.4\textwidth]{figures/plotFiller}}
  }
  \caption{\label{FIGfourJetEventDisplay}\Subref{FIGfourJetEventDisplay1} Event display of a 
  high-mass (4.04\TeV) dijet event taken in April~2011. The two jets which comprise the dijet system are marked by the colours red and green;
  these respectively have transverse momenta, $\pt=1.85\;\mrm{and}\;1.84\TeV$,
  pseudo-rapidities, $\eta=0.32\;\mrm{and}\;-0.53$ and azimuthal angles, $\phi=2.2\;\mrm{and}\;-0.92$.
  Shown are a view along the beam axis (left), and the angular distribution of transverse energy in~$\eta$ and~$\phi$ (bottom right).\\
  \Subref{FIGfourJetEventDisplay2} Event display of an event with four reconstructed jets, taken in April~2010.
  The four jets all have \ptHigher{50}, where the highest-\pt jet has transverse momentum of~108\GeV.
  Depicted points-of-view are along the beam axis (left), parallel to the beam axis (top right), and the angular
  distribution of transverse energy in~$\eta$ and~$\phi$ (bottom right).
  }
\end{center}
\end{figure} 

One of the important sources of background for physics searches at the LHC
are multiple-parton interactions (MPI).
In a generic hadron-hadron collision, several partons
in one hadron scatter on counterparts from the other hadron. Most of these interactions are soft,
and so generally do not result in high-\pt jets. However, the kinematic reach of the LHC, which extends to
high energies and low fractions of proton momenta, enhances the probability of hard MPI
relative to past experiments~\cite{Berger:2009cm}.
Hard MPI constitute a source of background for \eg Higgs production~\cite{DelFabbro:1999tf,Levin:2008jf},
and can possibly influence observed rates of other final states, involving the decay of heavy objects.
The existing phenomenology of MPI is based on several simplifying assumptions.
Recent interest has produced some
advancements~\cite{Calucci:1997uw,Gaunt:2009re,Calucci:1999yz,Gaunt:2010vy,Gaunt:2012wv,Humpert:1983fy,Ametller:1985tp},
however, a systematic treatment within QCD remains to be developed.

The simplest case of MPI is that of double parton scattering (DPS).
Measurements of DPS at energies between 63\GeV and 1.96\TeV
in the four-jet and $\gamma$~+~three-jets channels have previously been
performed~\cite{Åkesson:173908,Alitti1991145,PhysRevD.47.4857,PhysRevD.56.3811,Abazov:2009gc}. In addition,
a measurement using 7\TeV \atlas data with $W$~+~two-jet production in the final state
has recently been released~\cite{Sadeh:1498427}.
These have proven helpful in constraining the phenomenological models which describe DPS, and in
the development of double parton distribution functions, as \eg discussed in~\cite{Gaunt:2009re}.
An additional channel in which DPS may be measured at the LHC is four-jet production.
An \atlas event display of a four-jet event is presented in \autoref{FIGfourJetEventDisplay2} for illustration.
A four-jet final state may
arise due to a single parton-parton collision, accompanied by additional radiation. Alternatively,
it can also originate from two separate parton-parton collisions, each producing a pair of jets.
The latter case has distinguishing kinematic characteristics, and so
the rate of DPS may be estimated on average. The DPS-rate is related to a so-called
effective \xsec. The latter holds information about the transverse momentum distributions of partons in the
proton, and about the correlations between partons.
A measurement of DPS in four-jet events has been performed using a sub-sample of
the 2010 \atlas dataset, and is presented in this thesis.

\pagebreak
\minisec{Outline of the thesis}
%
%
This thesis presents a measurement of hard double parton scattering in four-jet events. The analysis
involves measurement of the dijet and four-jet \xsecs, and the extraction of the rate of double
parton scattering from the four-jet sample.

In order to measure the individual \xsecs, a measurement of the inclusive dijet invariant
mass distribution is first performed.
Part of the experimental challenge of measurements involving jets in \atlas, is
the calibration of the energy of jets, as these suffer from detector background due to
multiple simultaneous $pp$ collisions, referred to as \pu. A novel method
to reduce the \pu background using jet areas is developed and subsequently utilized to correct the energy of jets. The re-calibrated
jets serve as input for the dijet mass distribution analysis of the 2011 \atlas dataset, for which the \pu
background is severe.
The invariant mass measurement serves as a confidence building process for understanding
the reconstruction of jets, handling of the trigger and the luminosity.
The measurement is also performed on the 2010 dataset, and found to be compatible
with the previously published results of \atlas, and with the present measurement, using the 2011 data.

The 2010 dijet data-sample is also an essential element in the analysis of double parton scattering.
In spite of the good understanding of how to handle the 2011 data, the measurement of double parton scattering
is limited to the 2010 data, and even to a sub-sample of these data, which include single-vertex events. This 
choice is dictated by the inherent difficulty in extracting the double parton scattering signal,
and by the shortcomings of the available simulation.

Following the short introduction given here, a theoretical overview of the Standard Model and of QCD
is presented in \autoref{chapTheoreticalBackground}. The \atlas experiment is described in \autoref{chapAtlasexperimentAtTheLHC},
followed by a summary of the physics event generators and detector simulation which are used in the analysis in \autoref{chapMCsimulation}.
In \autoref{chapJetReconstruction} the concept of a jet is rigorously defined, followed by a discussion of calorimeter
jets in the context of the \atlas detector; this includes a description of the energy calibration of jets and of the systematic
uncertainties associated with the calibration.
\Autoref{chapDataSelection} details the data sample which is used in the analysis, with an emphasis on the
trigger selection procedure and on calculation of the luminosity.
In \autoref{chapJetAreaMethod} the jet areas method to subtract \pu background is introduced, and the performance of the \pu correction is
estimated in simulation. Several \insitu measurements are also used for validation of the performance in data. The new jet energy calibration which
is thus developed is used to perform a measurement of the differential dijet invariant mass \xsec in \autoref{chapMeasurementOfTheDijetMass}.
In \autoref{chapDoublePartonScattering}, the rate of double parton scattering events is extracted from the data using a
neural network, and the effective \xsec, \sigeff, is measured.
A summary of the results is finally given in \autoref{chapSummary}.

\chapter{Theoretical background\label{chapTheoreticalBackground}}
%
\section{The Standard Model of particle physics\label{chap}}
%
The Standard Model (SM)~\cite{Griffiths:1987tj} is the most successful theory describing the properties and
interactions (electromagnetic, weak and strong) of the elementary particles. 
The SM is a gauge quantum field theory based in the symmetry group
$\mathrm{SU}(3)_{\mathrm C} \times \mathrm{SU}(2)_{\mathrm L} \times \mathrm U(1)_{\mathrm Y}$, 
where the electroweak sector is based in the $\mathrm{SU}(2)_{\mathrm L} \times \mathrm U(1)_{\mathrm Y}$ group, 
and the strong sector is based in the $\mathrm{SU}(3)_{\mathrm C}$ group.

\begin{wrapfigure}{r}{0.4\textwidth}
\vspace{-10pt}
\centering
\includegraphics[trim=20mm 0mm 15mm 10mm,clip,width=.4\textwidth]{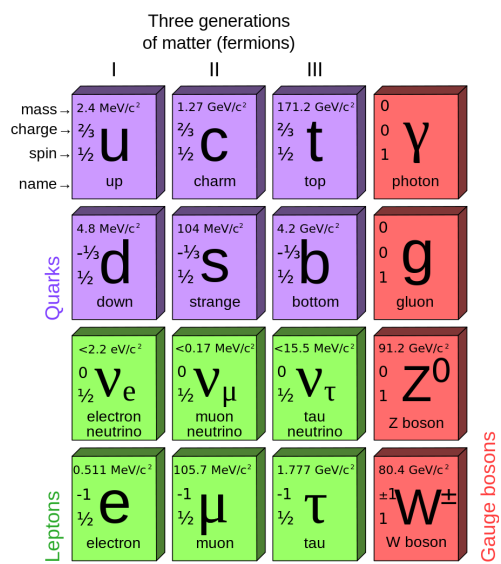}
\caption{\label{FIGstandardModelElementaryParticles}The fundamental particles of the Standard Model,
  sorted according to family, generation and mass.
}
\noindent \hrulefill
\label{}
\end{wrapfigure}
Interactions in the SM occur via the exchange of integer spin bosons. The mediators of the electromagnetic and strong interactions, the photon
and eight gluons respectively, are massless. The weak force acts via the exchange of three massive
bosons, the $W^{\pm}$ and the $Z$.

The other elementary particles in the SM are half-integer spin fermions, six quarks and six leptons. While both groups
interact via the electroweak force, only quarks feel the strong interaction.
Electrons ($e$), muons ($\mu$) and taus ($\tau$) are
massive leptons and have electrical charge $Q = -1$. Their associated neutrinos, respectively $\nu_{e}$, $\nu_{\mu}$ and $\nu_{\tau}$,
do not have electrical charge. Quarks can be classified
depending on their electrical charge; quarks $u$, $c$ and $t$ have $Q = 2/3$ and quarks $d$, $s$ and $b$ have $Q = -1/3$.
For each particle in the SM, there is an anti-particle with opposite quantum numbers.
The fundamental particles of the Standard Model, sorted according to family, generation and mass,
are listed in \autoref{FIGstandardModelElementaryParticles}.

The SM formalism is written for massless particles. The Higgs mechanism 
of spontaneous symmetry breaking is proposed for generating non-zero boson and fermion 
masses. The symmetry breaking requires the introduction of a new field that leads to 
the existence of a new massive boson, the Higgs boson. A particle with properties consistent with those of
the Higgs boson has recently been discovered at the LHC by the \atlas~\cite{:2012gk} and \CMS~\cite{:2012gu} experiments.

\section{Quantum Chromodynamics\label{chapTheoryQCD}}
%
%
Quantum chromodynamics (QCD)~\cite{Ellis:1991qj} is the theory of strong interactions. Its fundamental constituents are
quarks and gluons, which are \textit{confined} in the nucleon but act as free at sufficiently small scales
(and high energies). The latter behaviour is called \textit{asymptotic freedom}. The direct consequence
of confinement is that free quarks and gluons are never observed experimentally, and their final
state is a collimated shower of hadrons.

The development of QCD was posterior to that of quantum electrodynamics (QED); while the latter was highly successful
in the mid-Sixties, no information about the components of the nucleus was available.
Strong interactions were commonly described using general principles and the exchange of
mesons~\cite{PTPS.1.1}, although the basis for theories that could eventually accommodate QCD had also been
developed~\cite{PhysRev.96.191}. A framework called the \textit{Eightfold Way}~\cite{gelnee_1964} had been developed to organize subatomic
baryons and mesons into octets. Its connection to an underlying point-like structure of hadrons
came after the so-called \textit{heroic age} of deep inelastic scattering (DIS) measurements, interpreted using
the \textit{parton model}~\cite{Feynman:1969:BHC}. These experiments and subsequent interpretations showed that the probes
scattered against point-like, spin~$1/2$ constituents of the nucleons that are the quarks.
The presence of spin~$1$ gluons was also inferred
using kinematic considerations in terms of the total momentum shared by the quarks. The
QCD equivalent of the electromagnetic charge is the colour charge; (anti-)quarks can
take three (anti-)colours (red, green and blue, and their counterparts); the eight interacting
gluons exist in a superposition of colour and anti-colour states.
$\mathrm{SU}(3)$ QCD was established as a theoretical framework for strong interactions only following
the discovery of asymptotic freedom as a consequence of the renormalisability of the theory~\cite{Gross28062005}.
A short overview of these concepts follows.

\minisec{The lagrangian of QCD, confinement and asymptotic freedom}
%
%
QCD is the renormalizable gauge field theory that describes 
the strong interaction between colored particles in the SM,  based in the $\mathrm{SU}(3)$ symmetric group.
The lagrangian of QCD is
\begin{equation}
\mathcal{L}_{\mathrm{QCD}}  = \sum_{q}\bar{\psi}_q (i\gamma^{\mu}D_{\mu}-m_q)\;\psi_q
                            - \frac{1}{4}G_{\alpha\beta}^{A}G_{A}^{\alpha\beta} \;,
\end{equation}
where quarks and anti-quark fields are respectively denoted by $\psi_q$ and $\bar{\psi}_q$, quark mass is denoted by $m_q$,
$\gamma^{\mu}$ are the Dirac matrices, and $D_{\mu}$ stands for a covariant derivative.
The sum runs over the six different flavors of quarks. 
The gauge invariant gluonic field strength tensor,
\begin{equation}
G_{\alpha\beta}^{A} = [\partial_{\alpha}\mathcal{G}_{\beta}^{A}
                    - \partial_{\beta}\mathcal{G}_{\alpha}^{A}
                    - g_s f^{ABC}\mathcal{G}_{\alpha}^{B}\mathcal{G}_{\beta}^{C}] \;,
\end{equation}
is derived from the gluon fields, $\mathcal{G}_{\alpha}^{A}$,
where $g_s$ is a coupling constant, $f^{ABC}$ are the structure constants of $\mrm{SU}(3)$,
and the indices $A$, $B$ and $C$ run over the eight color 
degrees of freedom of the gluon field. The third term originates from the non-abelian character of the $\mrm{SU}(3)$ group.
It is responsible for the gluon self-interaction, giving rise to triple and quadruple gluon vertexes. This leads to 
a strong coupling, $\alpha_{s} = g_s^{2}/4\pi$, that is large at low energies and small at
high energies (discussed further in the next section).
Two consequences follow:
\begin{list}{-}{}
\item \headFont{confinement -} the color field potential increases linearly with distance. Quarks and gluons
can never be observed as free particles, and are always inside hadrons, either as
mesons (quark-antiquark) or as baryons (three quarks, each with a different color). 
If two quarks separate far enough, the field energy increases and new quarks are created, forming colorless hadrons;
\item \headFont{asymptotic freedom -} at small distances the strength of the strong coupling
is low, such that quark and gluons behave as if they were free.
This allows the perturbative approach to be used in the regime where $\alpha_{s} \ll 1$.
\end{list}

Confinement and asymptotic freedom have relevant experimental consequences; quarks and
gluons require interactions with high energy probes to be ejected from the nucleon, and they
cannot be observed directly. What one detects instead of quarks and gluons are \textit{jets}, which are collimated
showers of particles. These particles are the product of a series of steps~\cite{Sjostrand:2004pf},

\begin{list}{-}{}
\item two hadrons collide with a large momentum transfer;
\item two \textit{incoming} partons from the hadrons collide and produce the hard
process; at leading order this is a $2 \rightarrow 2$ process, for which the \xsec is calculable in
pQCD. It is assumed that the probabilities of finding partons in the proton, given by the parton density functions (PDFs),
are know;
\item additional semi-hard or hard interactions may occur between the remaining partons.
These arise naturally in that the integrated \xsec for
hard scattering diverges for low transverse momentum; the \xsec becomes
bigger than the total \xsec, unless multi-parton interactions
(MPI) are invoked. The remaining partons are referred to as \textit{outgoing} partons;
\item the incoming and outgoing partons may radiate, due to their having electromagnetic and color charges,
as described using pQCD;
\item when partons are sufficiently ``far'' from each other, \textit{hadronisation} takes place.
That is, confinement comes into play and additional color charges (quark/anti-quark pairs) are created
such that all free partons combine into hadrons.
Many of these hadrons are unstable and decay further at various timescales; they or their decay products
may be observed in a detector.
\end{list}
The products of the collision that are not directly identified with the hard scattering
(hadron remnants, products of soft multiple parton interactions and radiation) are conventionally
defined as the \textit{underlying event} (UE). 

Each of the steps described here is subject to modeling, and is accompanied by uncertainties that are hard to quantify.
Various Monte Carlo (MC) generators~\cite{Sjostrand:2006za,Bopp:1998rc,Corcella:2002jc,Gaunt:2010pi,Paige:2003mg}
adopt different approaches, in particular for the treatment of MPI and the UE.
A detailed discussion is given in \autoref{chapIntroductionSimulatioOfQCD}.

\minisec{Renormalisation}
%
%
Gluons in QCD are massless, the theory therefore results in 
divergences of theoretical \xsec calculations. A \textit{renormalisation} procedure is
necessary in order to allow the theory to give meaningful (non infinite) results that can be
compared to experimental measurements. This is achieved by effectively subtracting these
infinities through counter-terms embedded in so-called bare parameters that are not measurable.
The renormalisation procedure introduces a correction to the renormalised parameter, depending
on the renormalisation scale, $\mu_{\mrm{R}}$, (interpreted as the scale at which the subtraction
is made), and on the physical scale at which the measurement is made; the latter is taken as the squared
momentum transfer, $Q^2$, in the following. Imposing the independence of the final result in all
orders of perturbation theory from the renormalisation scale,
allows one to derive an explicit form for the renormalised parameter.
As an example of a renormalised parameter, the strong coupling constant in the one loop approximation (first
order in perturbation theory) is
\begin{equation}
\alpha_{s}\left(Q^2\right) = \frac{ \alpha_{s}\left(\mu_{\mrm{R}}^2\right)   }
                                  { 1 + (\beta_{1}/4\pi) \;\alpha_{s}\left(\mu_{\mrm{R}}^2\right)\; \ln\left( Q^2/\mu_{\mrm{R}}^2\right)  } 
                           = \frac{4\pi}
                                  { \beta_{1} \; \ln\left( Q^2/\Lambda_{\rm QCD}^2 \right) }
                           \;,
\label{eqAlphSrenormalisation} \end{equation}
where
\begin{equation}
\Lambda_{\rm QCD} = \mu_{\mrm{R}} \exp{\left(-\frac{2\pi}{\beta_{1} \alpha_{s}\left(\mu_{\mrm{R}}^2\right) }\right)}
\label{eqLambdaQCD} \end{equation}
sets the scale of the coupling.
The parameter $\beta_{1}$ is a constant, computed by Gross Wilczek and Politzer~\cite{PhysRevD.8.3633,PhysRevD.9.980,PhysRevLett.30.1346};
it is the first term in the \textit{beta function} of the strong coupling constant,
\begin{equation}
\beta \left(\alpha_{s}\right) = - \sqrt{4\pi \alpha_{s}(\mu^2)} 
                                  \left( \frac{\alpha_{s}}{4\pi} \beta_{1} + \left(\frac{\alpha_{s}}{4\pi}\right)^2\beta_{2}+ ... \right) \;,
\label{eqBetaFunctionAlphaS} \end{equation}
which encodes the dependence of $\alpha_{s}$ on the energy scale.

The coupling constant, initially scale-invariant, becomes a function of the scale of the process,
commonly referred to as a \textit{running coupling constant}. The current theoretical
and experimental results for the running $\alpha_{s}$~\cite{Beringer:1900zz} are shown in \autoref{FIGalphaS}. 
\begin{figure}[htp]
\begin{center}
\includegraphics[width=.6\textwidth]{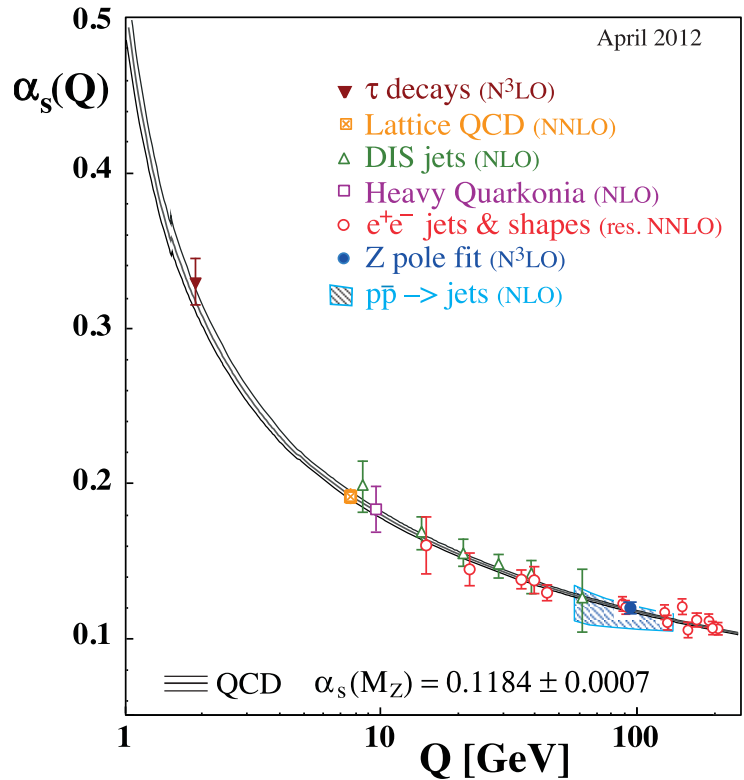}
  \caption{\label{FIGalphaS}Compilation of measurements of the coupling constant, $\alpha_{s}$, as a function of the energy
  scale, $Q$. The degree of QCD perturbation theory used in the extraction
  of $\alpha_{s}$ is indicated in parenthesis in the figure.
  (Figure taken from~\protect\cite{Beringer:1900zz}.)
  }
\end{center}
\end{figure} 
Contrary to QED, where the coupling constant increases with the scale of the process,
gluon self-interactions result in negative values of $\beta\left(\alpha_{s}\right)$. The coupling constant
therefore is sizeable at low values of $Q^2$, leading to confined partons, and decreases as $Q^2$ increases,
leading to asymptotic freedom~\cite{Prosperi:2006hx}.

\minisec{The parton model, parton distribution functions and evolution}
%
%
Asymptotic freedom allows QCD to be described using point-like constituents at sufficiently
large energies. The first evidence of this behaviour was given by the SLAC experiments~\cite{citeulike:6775206}
and interpreted by Feynman using the \textit{parton model}~\cite{Feynman:1969:BHC}. Later on, it was realised
that the momentum scale introduced by renormalisation needed to be accommodated, and the \textit{improved parton model}
was developed~\cite{Collins:1989gx}. Starting from these ideas, the perturbative evolution of
quarks and gluons can be predicted independently of the soft, non-perturbative physics,
allowing for theoretical calculation of QCD processes.

The differential \xsec for lepton-hadron ($lh$) inelastic scattering
can be parameterised, starting from that of elastic scattering of fundamental
particles,
\begin{equation}
\frac{d^2\sigma^{lh}}{dxdQ^2} = \frac{1}{q^4} \left( f(y) x F_{1} \left(x,Q^2\right)  
                                                     + g(y) F_{2} \left(x,Q^2\right) \right) \;,
\label{eqDISxs} \end{equation}
where $F_{1}$ and $F_{2}$ are the \textit{structure functions}, which reflect the structure
of the nucleon. They are parametrized in
terms of the momentum transfer, $Q^2$, and of $x$, which represents (at leading order) 
the fraction of hadron momentum carried by the massless struck quark. Similarly, $f(y)$ and $g(y)$
are functions which depend on the kinematics of the scattering, where the parameter $y$
measures the ratio of the energy transferred to the hadronic system, to the total leptonic energy
available in the target rest frame.

The structure function $F_{2}$ can be written as
\begin{equation}
F_{2} = \sum_{i}^{N_q}{e^2_i x f_i (x)} \;,
\label{eqStructureFuncF2} \end{equation}
where $e^2_i$ and $f_i(x)$ are respectively the squared charge and momentum distribution of the $i^{\mrm{th}}$ quark,
and the sum goes over all quarks in the hadron (in total $N_q$ quarks).
The $f_i(x)$ functions can be interpreted (at leading order of perturbation theory) as the
probability densities of finding a quark with flavour $i$, carrying a fraction $x$ of the hadron
momentum. The momentum distribution for a given quark or gluon is also called the \textit{parton distribution function} (PDF).
The two structure functions are not independent, due to the fact that quarks have spin $1/2$. This is expressed
by the Callan-Gross relation~\cite{Manohar:1992tz},
\begin{equation}
F_{2} = 2 x F_{1} \;.
\label{eqCallanGross} \end{equation}

Probability conservation requires that
\begin{equation}
\int_{0}^{1}x\sum_{i}^{N_q}{f_{i}(x)dx} = 1 \;,
\label{eqMomentumSumRule} \end{equation}
also called the \textit{momentum sum rule}. However, calculation of the integral comes up to about~$0.5$.
This calls for the presence of gluons.
Conventionally, partons composing a hadron are divided between gluons, \textit{valence quarks} and
\textit{sea quarks}. Valence quarks are responsible for the quantum numbers of the hadron, while sea quarks
are quark/anti-quark pairs that are generated due to quantum fluctuations.

The independence of the structure function from momentum transfer, $Q^2$, at fixed values of
$x$, is known as \textit{Bjorken scaling}.
Interactions among the partons lead to deviation from the naive parton model in terms of
\textit{scaling violations}. The latter have been observed experimentally, as seen in \autoref{FIGscalingViolationsDIS},
where measurements at a given value of $x$ are shown to depend on $Q^2$.
\begin{figure}[ht]
\begin{center}
\includegraphics[width=.8\textwidth]{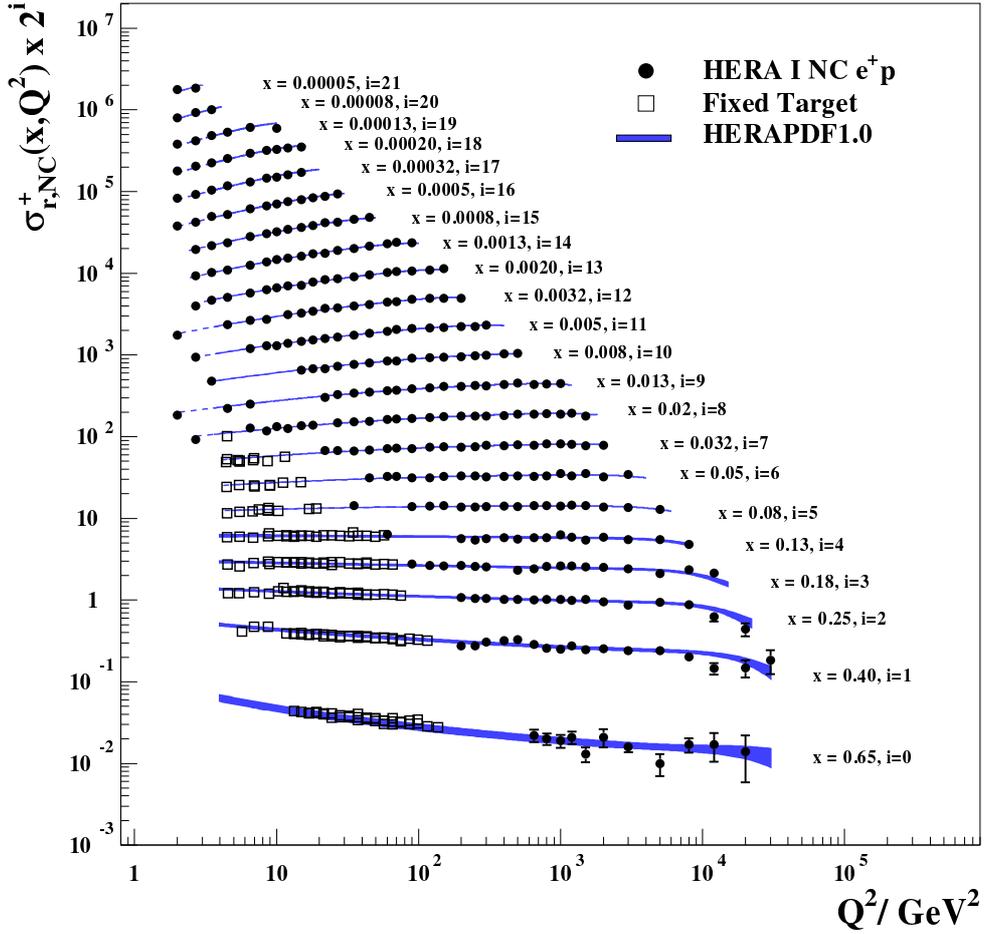}
  \caption{\label{FIGscalingViolationsDIS}HERA combined neutral current (NC) $e^{+}p$ reduced \xsec
  and fixed-target measurements as a function
  of momentum transfer, $Q^2$, at fixed values of Bjorken-$x$.
  The error bars indicate the total experimental uncertainty. A PDF set, called HERAPDF1.0~\protect\cite{Aaron:2009aa},
  is superimposed. The bands represent the total uncertainty of the PDF fit. Dashed lines are shown for
  $Q^2$ values not included in the QCD analysis. (See~\protect\cite{Aaron:2009aa} for further details.)
  }
\end{center}
\end{figure} 
Intuitively, an increase in $Q^2$ can be seen as an increase in the resolving power
of the probe; if the internal structure of the hadron can be probed at smaller distances, then
the number of partons ``seen'' increases, as gluons can produce quark/anti-quark pairs and
quarks or anti-quarks can radiate gluons.
The DGLAP (Dokshitzer, Gribov, Lipatov, Altarelli and Parisi) formalism~\cite{Dokshitzer:1977sg,Gribov:1972ri,Altarelli:1977zs} models these interactions
through splitting functions, and uses them to \textit{evolve} perturbatively the renormalised parton
densities that contain the $Q^2$ dependence.
The implication of PDF evolution is that measuring parton distributions for one scale, $\mu_0$,
allows their prediction for any other scale, $\mu_1$,
as long as both $\mu_0$ and $\mu_1$ are large enough
for both $\alpha_{s}(\mu_0)$ and $\alpha_{s}(\mu_1)$ to be small.

The DGLAP formalism gives information on the evolution of the PDFs, though not on their
shape. The latter is derived using a combination of experimental data on the structure functions at
a given scale, $Q^2=Q_0^2$, and an initial analytical form.
Seven functions should be determined, one for the gluon and the others for each one of the light quarks and 
anti-quarks.
Typically, specific functional forms are postulated for the PDFs with a set of free parameters.
The functional form assumed for several PDF sets (such as CTEQ~\cite{PDF-CTEQ}),
motivated by counting rules~\cite{Brodsky:1974vy} and Regge theory~\cite{Abarbanel:1969eh}, is
\begin{equation}
f_{i}(x,Q_{0}^{2}) = x^{\alpha_{i}}(1-x)^{\beta_{i}}g_{i}(x) \;,
\end{equation}
where $\alpha_{i}$ and $\beta_{i}$ are fit parameters and $g_{i}(x)$ is a function that 
asymptotically (${x \rightarrow 0 \;,\; x \rightarrow 1}$) tends to a constant.

\minisec{QCD Factorisation}
%
%
One of the reasons for the success of QCD as a predictive theory, is that the short-distance
component of the scattering process described by pQCD can be separated from the
non-perturbative long-distance component; this result is known as
the \textit{factorisation theorem}~\cite{Collins:1989gx}.
Factorisation implies that perturbation theory can be used to calculate the hard scattering cross
section, while universal functions such as the PDFs\footnote { It
has been shown experimentally that due to factorisation, PDFs are universal~\cite{Theorems:of:Perturbative:QCD}; that is,
they can be derived from different physics processes and then used to provide full
theoretical predictions independently from the calculation of the hard scattering \xsec.
} can be included a posteriori to obtain the full theoretical prediction. This takes the form,
\begin{equation}
d\sigma_{\rm full} \left( p_A,p_B,Q^2 \right) = \sum_{ab}{\int{ dx_a dx_b f_{a/A}\left(x_a,\mu_{\mrm{F}}^2 \right)
                                                     f_{b/B}\left(x_b,\mu_{\mrm{F}}^2 \right)
                                                     \times d\sigma_{ab \rightarrow cd} \left(
                                                         \alpha_s \left(\mu_{\mrm{R}}^2\right) , Q^2/\mu_{\mrm{R}}^2 \right) }}
\;,
\label{eqFactorization} \end{equation}
where the full and hard scattering \xsecs are denoted by $\sigma_{\rm full}$ and $\sigma_{ab \rightarrow cd}$ respectively.
Two partons, $a$ and $b$, respectively originating from hadrons $A$ and $B$ with momenta $p_A$ and $p_B$, are the constituents of the hard
process. The integral is performed over the respective parton momenta fractions $x_a$ and $x_b$, weighted by $f_{a/A}$ and $f_{b/B}$, 
which denote the parton momentum densities for the two interacting partons. The sum is over parton flavours in the hadrons.

Factorisation is a byproduct of a procedure that absorbs
singularities into physical quantities in the same fashion as renormalisation. A new scale, $\mu_{\mrm{F}}^2$,
called the \textit{factorisation scale}, is therefore introduced in addition to the renormalisation scale $\mu_{\mrm{R}}^2$ and
the momentum transfer, $Q^2$.
Both $\mu_{\mrm{R}}^2$ and $\mu_{\mrm{F}}^2$ are generally chosen to be of the order of $Q^2$.
When truncating calculations at a given order, the uncertainties due to the choice
of scale need to be calculated in order to account for higher order terms.
The primary sources of uncertainty on QCD calculations are discussed next.

\minisec{Uncertainties on QCD calculations}
%
%
There are three main sources of uncertainties in the calculation of pQCD observables, summarized in the following:
\begin{list}{-}{}
\item \headFont{higher order terms and scale dependence -}
the lack of knowledge of higher order terms, neglected in the calculation,
is estimated by varying the renormalization scale, $\mu_{\rm R}$, usually by a 
factor of two with respect to the default choice. The factorization scale, $\mu_{\rm F}$,
is independently varied in order to evaluate the sensitivity to the choice of scale, where the PDF
evolution is separated from the partonic \xsec. The envelope 
of the variation that these changes introduce in an observable is taken as a systematic uncertainty;
\item \headFont{knowledge of the parameters of the theory -}
uncertainties on parameters of QCD, such as the coupling constant, $\alpha_{s}$,
and the masses of heavy quarks, are propagated into a measured observable;  
\item \headFont{PDF uncertainties -}
PDF uncertainties on an observable are evaluated differently for different PDF sets. They account
for several factors; uncertainties on the data used to evaluate the PDFs; tension between input data sets;
parametrisation uncertainties; and various theoretical uncertainties.
\end{list}

For the LHC, a major source of uncertainty is linked to the fact that
$pp$ interactions at \cms energies \sqs,
probe very low momenta of the partons in the proton, as illustrated in \autoref{LHCKinRegimeFIG}.
\begin{figure}[htp]
\begin{center}
\includegraphics[trim=0mm 0mm 0mm 10mm,clip,width=.8\textwidth]{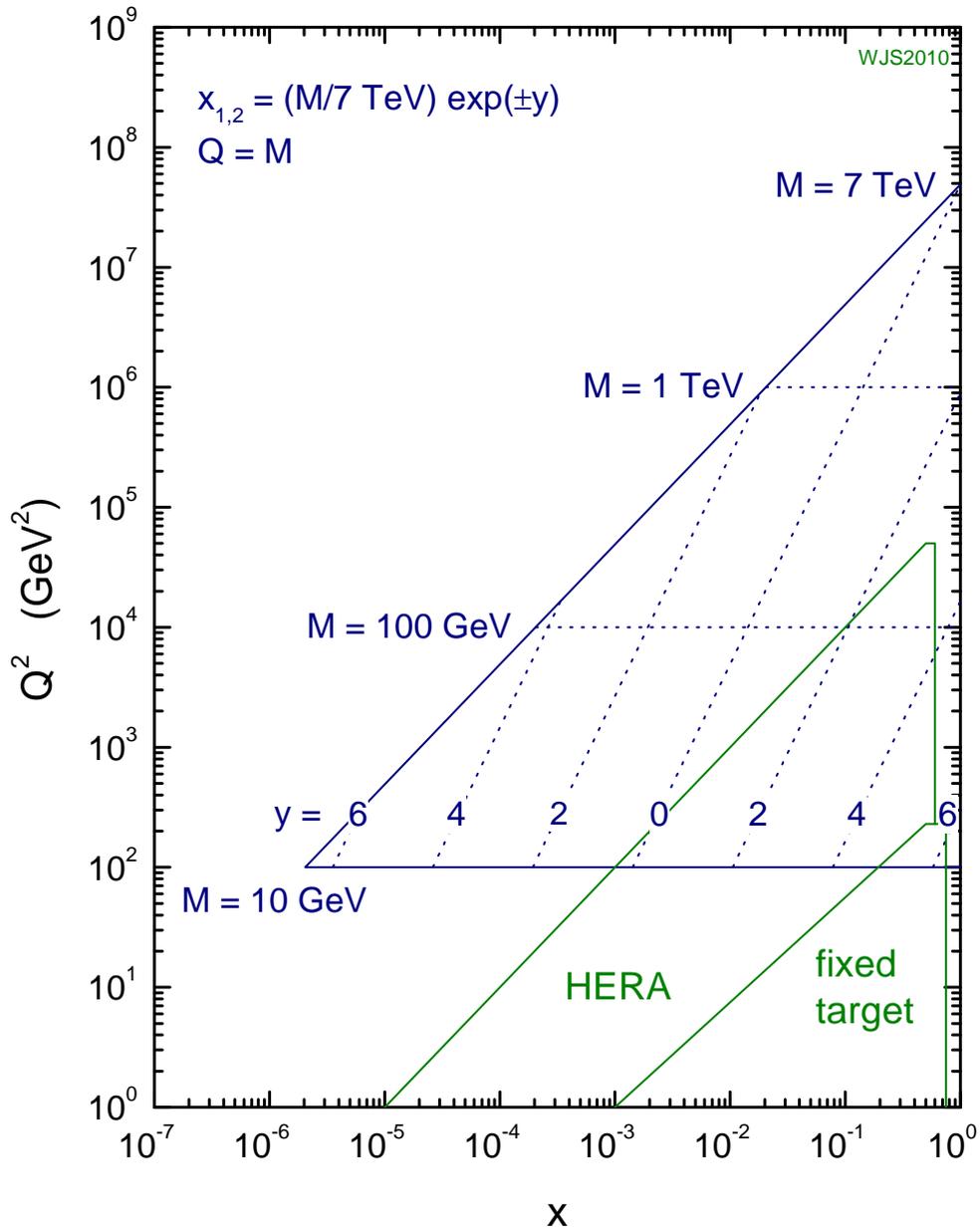}
\caption{\label{LHCKinRegimeFIG}The $(x,Q^2)$ kinematic plane for LHC at \sqs, where $x$ denotes the
fraction of the momentum of the proton carried by an interacting parton, and $Q^2$ stands
for the momentum transfer of the interaction. Shown also are the ranges covered by HERA and fixed-target experiments.
The $x$ and $Q^2$ ranges probed by the production of mass, $M$, or
a hard final state at fixed rapidity, $y$, are indicated by the dashed lines.
(Figure taken from~\protect\url{http://projects.hepforge.org/mstwpdf/}, courtesy of {J.~Stirling}.)
}
\end{center}
\end{figure} 
This low-$x$ region is dominated by the gluon density~\cite{Aktas:2007bv},
which is less well constrained by measurements than the quark densities.

Example are shown in \autoref{gluonDensityToXSFIG}, where the PDFs being
compared differ in the data samples from which they were derived, and in the assumed value of $\alphas$.
\begin{figure}[htp]
\begin{center}
\subfloat[]{\label{gluonDensityToXSFIG1}\includegraphics[width=.49\textwidth]{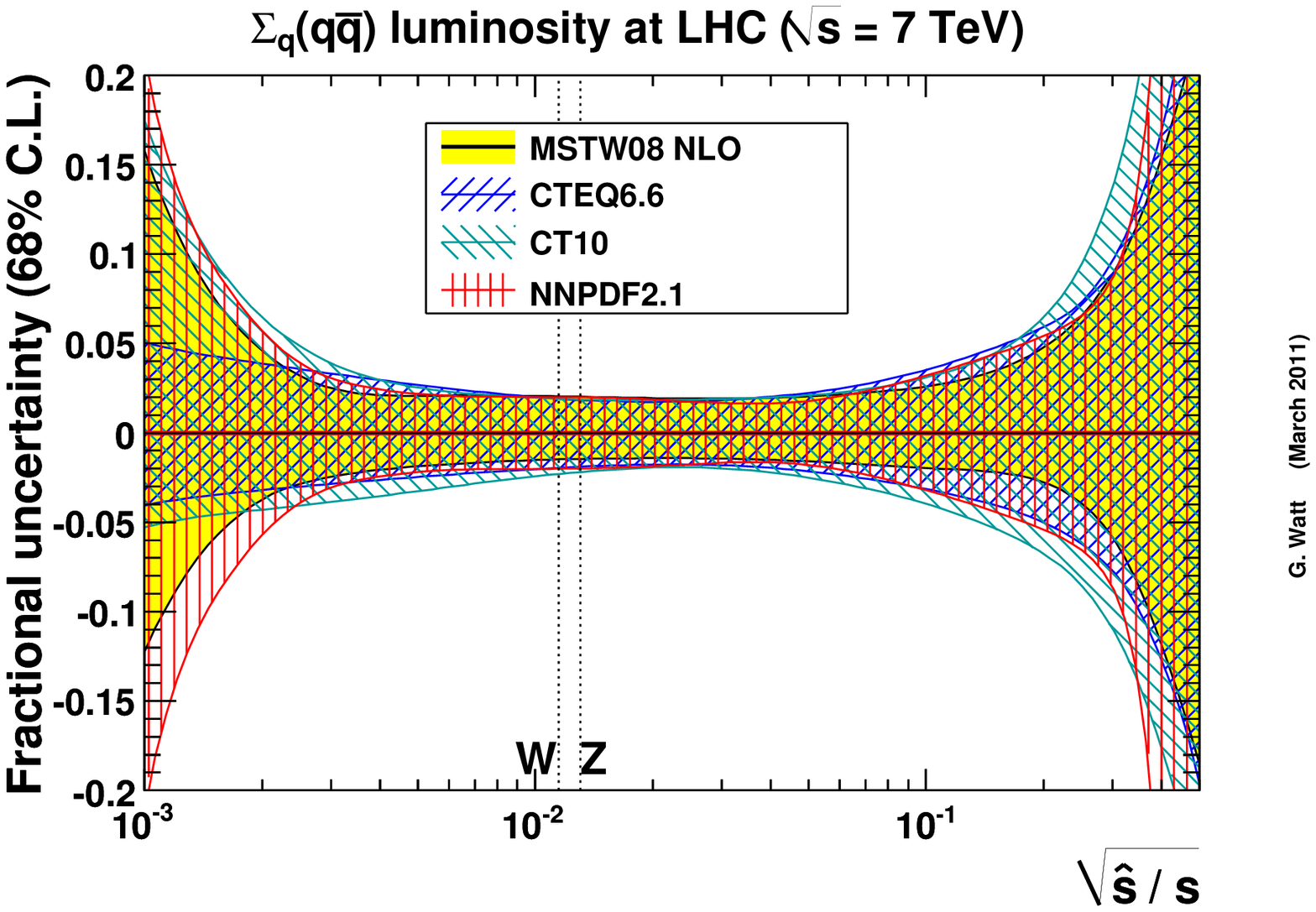}}
\subfloat[]{\label{gluonDensityToXSFIG2}\includegraphics[width=.49\textwidth]{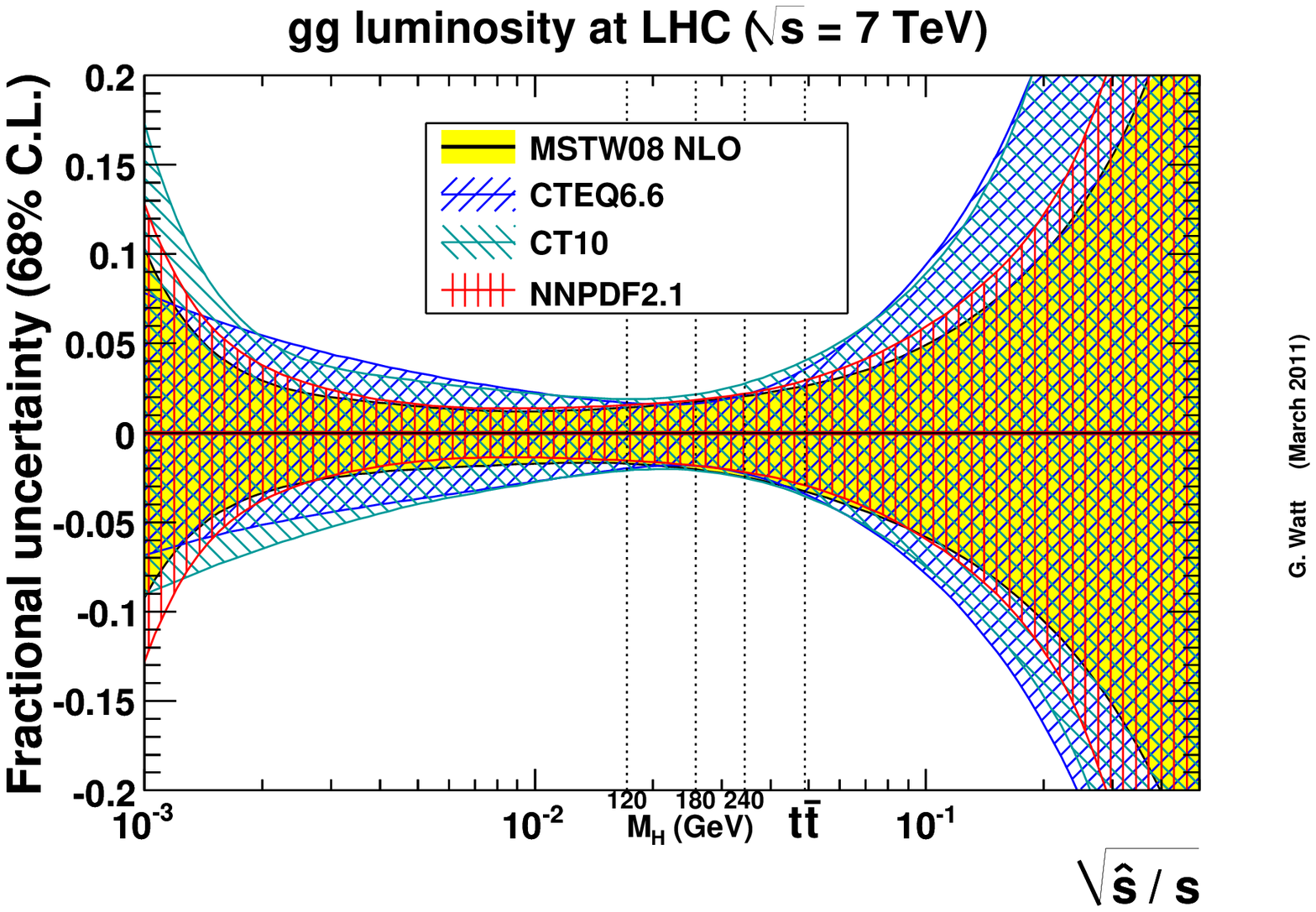}} \\
\subfloat[]{\label{gluonDensityToXSFIG3}\includegraphics[width=.49\textwidth]{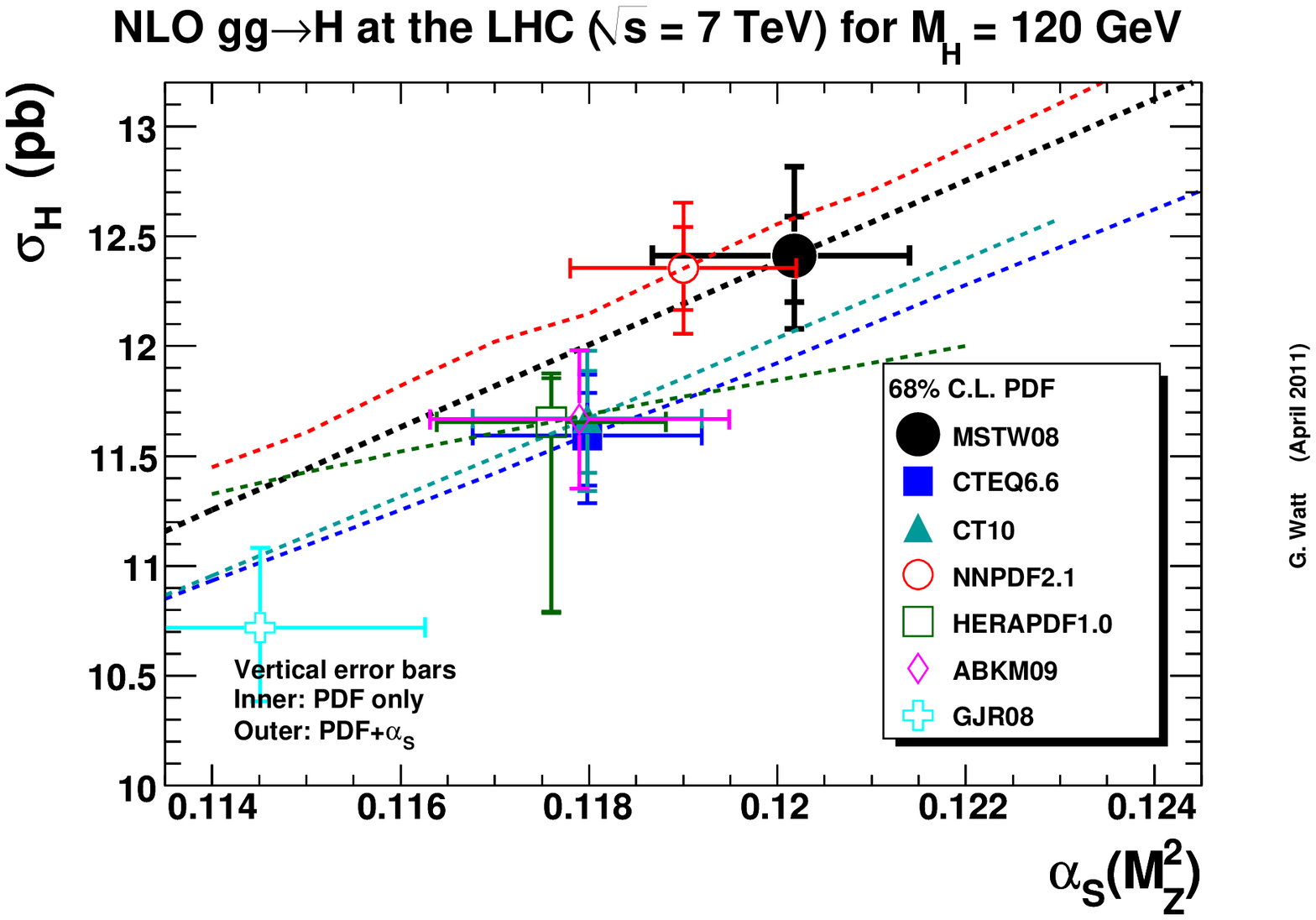}}
\subfloat[]{\label{gluonDensityToXSFIG4}\includegraphics[width=.49\textwidth]{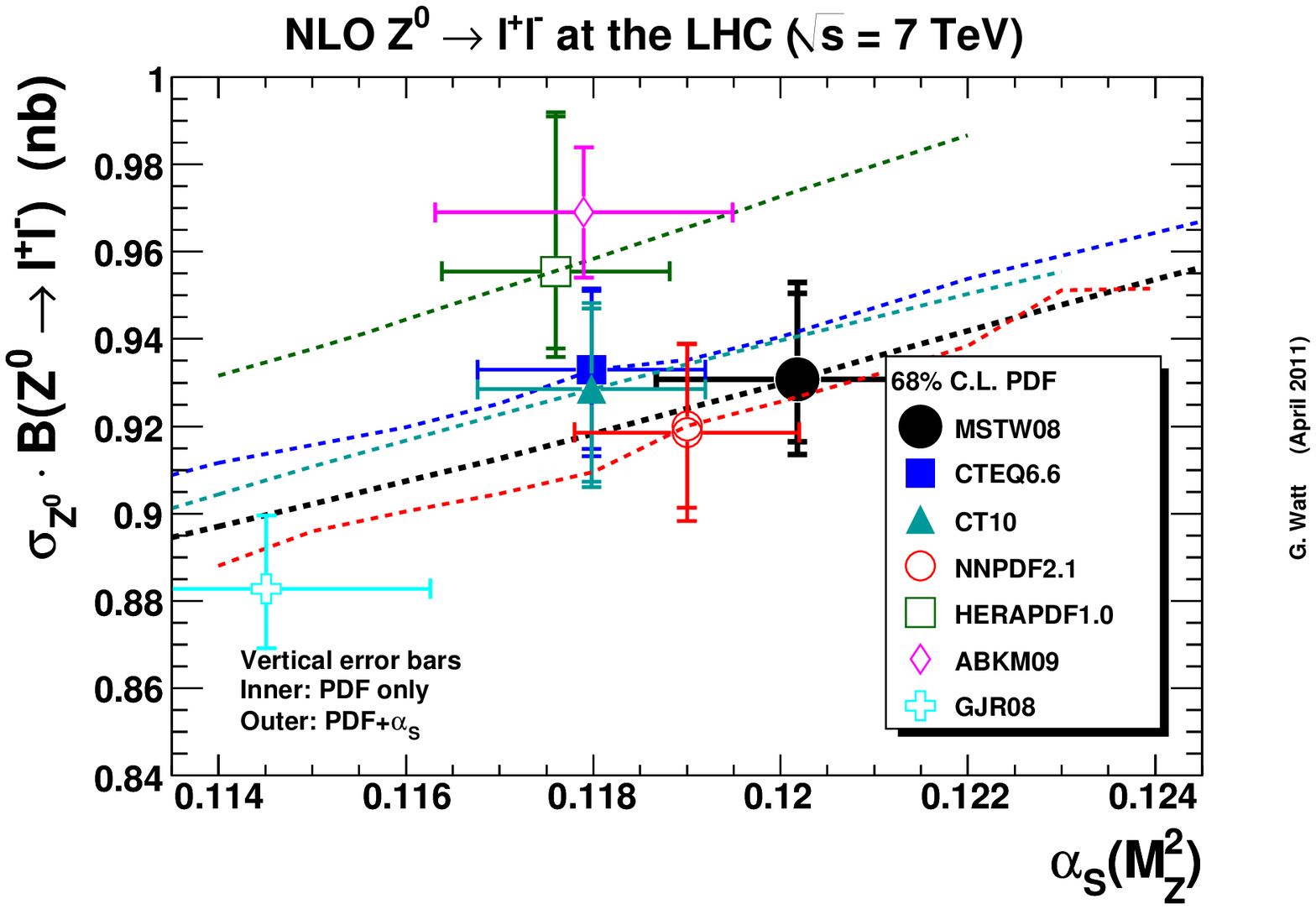}}
\caption{\label{gluonDensityToXSFIG}\Subref{gluonDensityToXSFIG1}-\Subref{gluonDensityToXSFIG2}
  Fractional uncertainties on the NLO parton-parton luminosities for quarks
  and for gluons, derived using different PDF sets, as indicated in the figures.
  The uncertainties are shown as a function of the fractional \com energy of the
  partonic system responsible for the scattering, denoted by $\sqrt{\hat{s}/s}$. \\
  \Subref{gluonDensityToXSFIG3}-\Subref{gluonDensityToXSFIG4}
  \Xsecs for Higgs and $Z_{0}$ production at \sqs for different PDF sets, as indicated in the figures.
  The \xsecs are shown as a function of the strong coupling constant (at the scale of the $Z$), $\alphas(M_Z^2)$,
  which is used in the derivation of the PDFs.
  The error bars represent the uncertainties on the respective
  PDFs excluding (inner) or including (outer) the uncertainty on $\alphas(M_Z^2)$.
  The dashed lines interpolate the \xsec predictions calculated with
  each PDF set for different values of $\alphas(M_Z^2)$.  
  \\
  (Figures are taken from~\protect\cite{Watt:2011kp}.)  
}
\end{center}
\end{figure} 
In \autorefs{gluonDensityToXSFIG1}~-~\ref{gluonDensityToXSFIG2}
the fractional uncertainties of the luminosities of quarks and gluons may be compared
for several PDF sets. The uncertainty on the gluon luminosities is larger
than on those of the quarks, especially for low and high values of~$x$.
An alternative way to illustrate this point is presented in \autorefs{gluonDensityToXSFIG3}~-~\ref{gluonDensityToXSFIG4},
where the effect of the uncertainties on observable \xsec predictions may be deduced;
the $gg$ \xsec for Higgs production is compared to that
of $Z$ production, the latter originating from $q\qbar$ interactions.
The relative spread of \xsec perdictions depending on the gluonic PDFs is
roughly twice that of the quarks'.

HERA measurements at low-$x$, and in particular forward jet production~\cite{Chekanov:2007pa,Aktas:2005up}, indicate
that the DGLAP dynamic used in deriving the PDFs may not be sufficient to describe
the interactions in this region~\cite{Abramowicz:2009pk}.
This may indicative the need to include higher order pQCD corrections,
or use of the formalisim
of Balitsky, Fadin, Kuraev and Lipatov (BFKL)~\cite{Lipatov:1976zz,Kuraev:1977fs,Balitsky:1978ic}.

\section{Full description of a hard $pp$ collision\label{chapIntroductionSimulatioOfQCD}}
%
A full description of the final state of a $pp$ collision incorporates two elements.
The first is the hard scattering, involving a large transfer of transverse momentum and 
calculable in pQCD. The second part pertains to non perturbative effects,
taking into account low-\pt interactions and hadronization.
The two aspects of the computation are discussed in the following sections.

\subsection{The hard scattering}
%
The hard scattering is computed at a fixed order (in the strong coupling constant) in perturbation theory.
The dominant contributions to jet \xsecs arise from Feynman
diagrams that contribute to jet production at leading order (LO), known as \twotwo diagrams.
Some examples of LO diagrams are shown for quark/quark $t$-channel scattering, quark/anti-quark $s$-channel annihilation,
and gluon/gluon $t$-channel scattering in \autorefs{FIGJetFeynmanDiagrams1}~-~\ref{FIGJetFeynmanDiagrams3}.
A calculation at next-to-leading order (NLO) may include either a \twotwo process with
one virtual loop, as shown in \autoref{FIGJetFeynmanDiagrams4}, or a \twothree interaction, as in \autoref{FIGJetFeynmanDiagrams5},
where one of the incoming or outgoing partons radiates a third parton.
\begin{figure}[htp]
\begin{center}
\subfloat[]{\label{FIGJetFeynmanDiagrams1}\includegraphics[trim=0mm -1mm 0mm 0mm,clip,width=.3\textwidth]{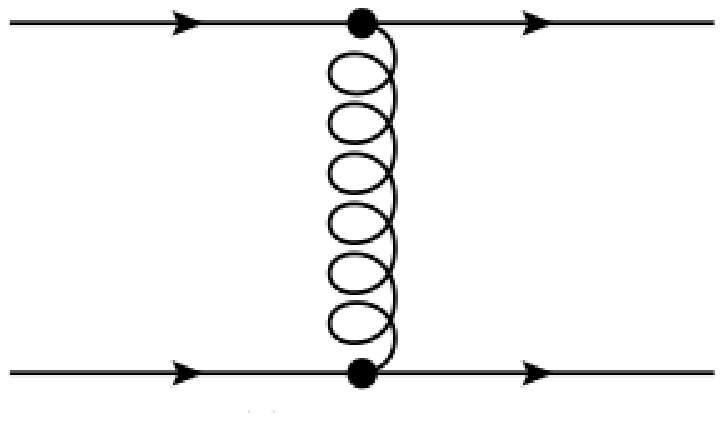}} \quad
\subfloat[]{\label{FIGJetFeynmanDiagrams2}\includegraphics[trim=0mm -1mm 0mm 0mm,clip,width=.3\textwidth]{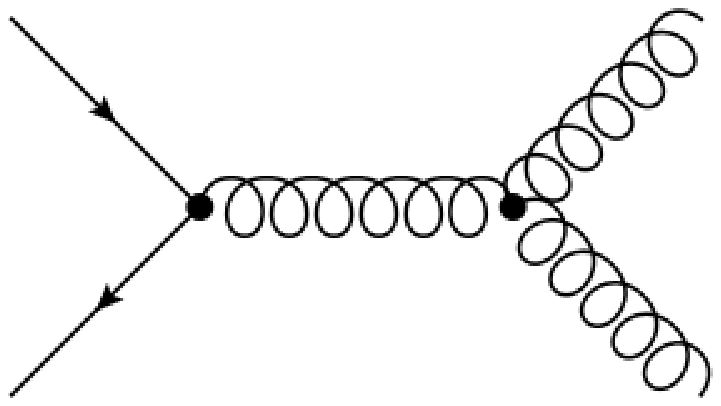}} \quad
\subfloat[]{\label{FIGJetFeynmanDiagrams3}\includegraphics[width=.3\textwidth]{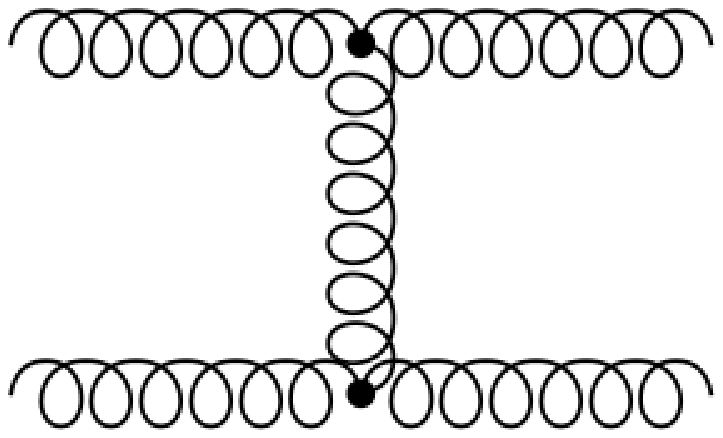}} \\
\subfloat[]{\label{FIGJetFeynmanDiagrams4}\includegraphics[width=.3\textwidth]{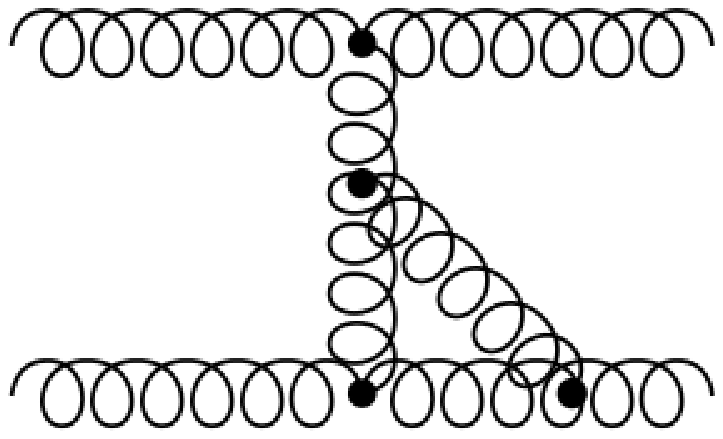}} \quad
\subfloat[]{\label{FIGJetFeynmanDiagrams5}\includegraphics[width=.3\textwidth]{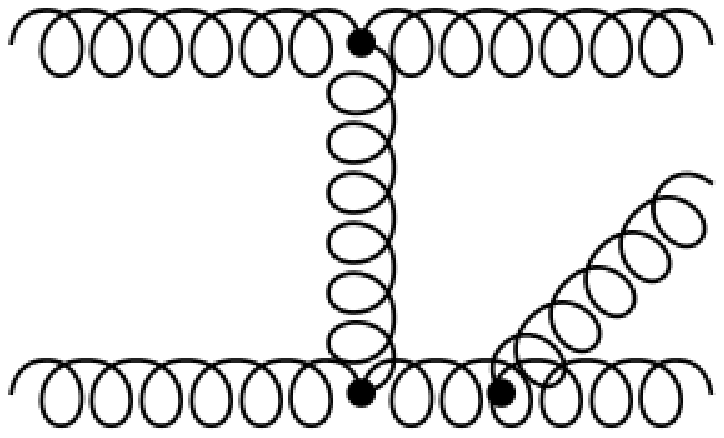}} 
\caption{\label{FIGJetFeynmanDiagrams}Examples for leading-order (\Subref{FIGJetFeynmanDiagrams1}-\Subref{FIGJetFeynmanDiagrams3})
  and next-to-leading order (\Subref{FIGJetFeynmanDiagrams4}-\Subref{FIGJetFeynmanDiagrams5})
  Feynman diagrams for jet production in proton-proton collisions at the LHC;
  $t$-channel scattering \Subref{FIGJetFeynmanDiagrams1},
  quark/anti-quark $s$-channel annihilation \Subref{FIGJetFeynmanDiagrams2},
  gluon/gluon $t$-channel scattering \Subref{FIGJetFeynmanDiagrams3},
  a \twotwo diagram with a virtual loop \Subref{FIGJetFeynmanDiagrams4},
  a \twothree diagram where the third outgoing parton is produced via real emission from another parton \Subref{FIGJetFeynmanDiagrams5}.
}
\end{center}
\end{figure} 

The results of a full NLO calculation of the differential dijet mass \xsec,
using the \nlojet package~\cite{Nagy:2003tz}, are compared to \atlas data in \autoref{chapMeasurementOfTheDijetMass}.

\subsection{Non-perturbative effects}
%
As mentioned above, complete pQCD calculations are performed up to a fixed order in $\alpha_{s}$.
However, the enhanced soft-gluon radiation and collinear configurations at higher orders can not be neglected. 
These are taken into account in the parton shower (PS) approximation, that sums the leading contributions of 
such topologies to all orders. MC generator programs include the PS approximation, 
as well as models to reproduce non-perturbative effects, such as hadronization and
the underlying event.

Since non-perturbative physics models are by necessity deeply phenomenological, they
usually account for the majority of parameters incorporated into event generators.
For instance, typical hadronization models
require parameters to describe \eg the kinematic distribution of transverse momentum
in hadron fragmentation, baryon-to-meson ratios, strangeness and suppression
of $\eta$ and $\eta^{\prime}$ mesons, and the assignment of orbital angular momentum to final state particles.
Event generators are therefore tuned to data in specific regions of phase-space.
At times it is not possible to match all features of the data, \eg simultaneous description
of both the transverse momentum and the multiplicity distributions of charged particles
in hadron-hadron collisions~\cite{Leyton:2012em}.
Different tunes are usually compared in order to asses the theoretical uncertainty associated with
non-perturbative processes.

The different elements of event generators, apart from the hard scattering itself, are discussed next.

\minisec{Parton showers}
%
%
The PS approximation describes successive parton emission from the partons taking part in the hard interaction.
In principle, the showers represent higher-order corrections to the hard subprocess.
However, it is not feasible to calculate these corrections exactly. Instead, an approximation
scheme is used, in which the dominant contributions are included in each order.
These dominant contributions are associated with collinear parton splitting or soft gluon emission,
illustrated in \autoref{FIGpartonSplittingDiagram}.
\begin{figure}[htp]
\begin{center}
\subfloat[]{\label{FIGpartonSplittingDiagram1}\includegraphics[width=.3\textwidth]{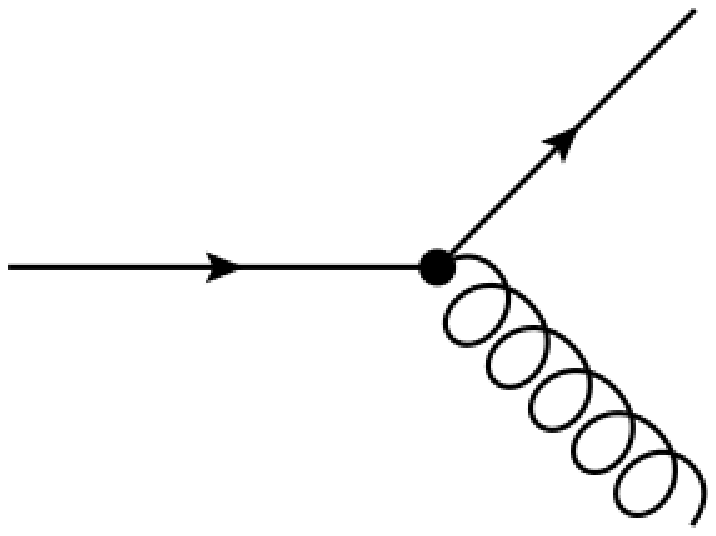}} \quad
\subfloat[]{\label{FIGpartonSplittingDiagram2}\includegraphics[width=.3\textwidth]{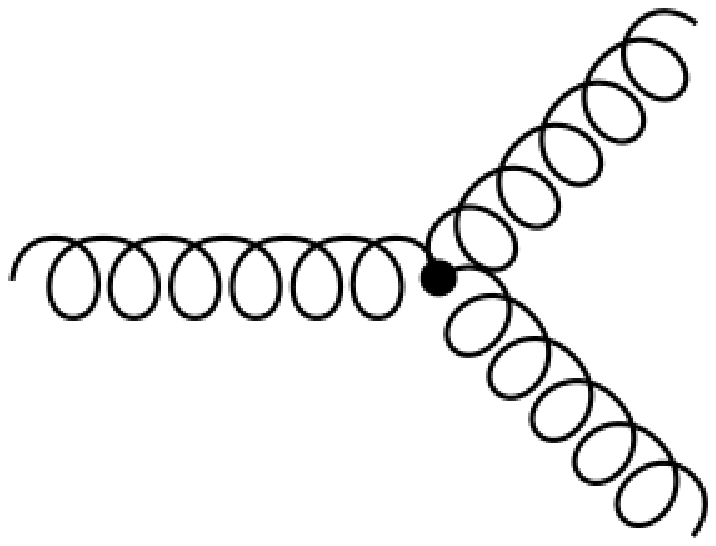}} \quad
\subfloat[]{\label{FIGpartonSplittingDiagram3}\includegraphics[width=.3\textwidth]{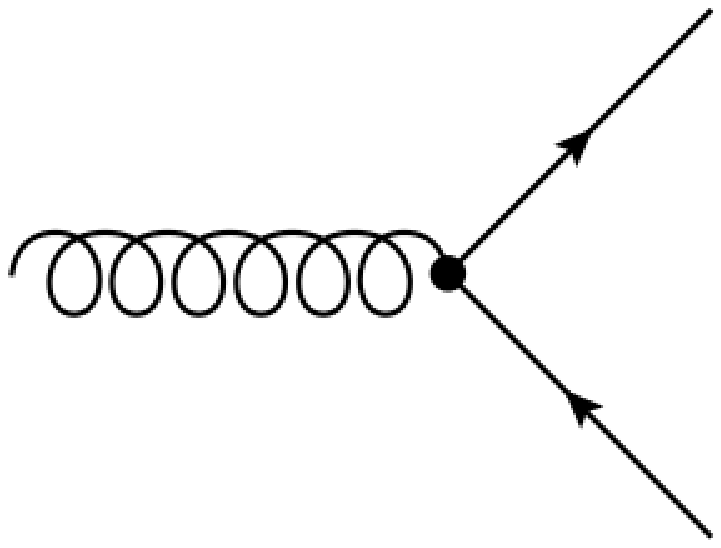}} 
\caption{\label{FIGpartonSplittingDiagram}Examples of leading-order \onetwo Feynman diagrams for the splitting of quarks and gluons;
  a quark emitting a gluon \Subref{FIGpartonSplittingDiagram1},
  a gluon emitting another gluon \Subref{FIGpartonSplittingDiagram2},
  a gluon splitting into a quark/anti-quark pair \Subref{FIGpartonSplittingDiagram3}.
}
\end{center}
\end{figure} 
%
%

The evolution of the shower is usually governed by the DGLAP equations, though
at low-$x$, as mentioned above, BFKL dynamics may be required.
Numerical implementation of the parton shower is achieved using
the \textit{Sudakov form factors}~\cite{Sudakov:1954sw}; these represent the probability that a 
parton does not branch between some initial scale and another, lower scale.
In each step, as a branching $a \rightarrow bc$ occurs from scale $t_{a}$,
subsequent branchings are derived from the scales $t_{b}$ and $t_{c}$ of the products of the initial state.
Branchings can be angle- or transverse momentum-ordered.
For the former, the opening angles between each successive branchings become smaller; for the
latter, emissions are produced in decreasing order of intrinsic $\pt$.
Successive branching stops at a cutoff scale of the order of $\Lambda_{\rm QCD}$, 
after producing a high-multiplicity partonic state.  

Since quarks and gluons can not exist isolated, MC programs contain models for the hadronization of the 
partons into colorless hadrons, discussed in the following.

\minisec{Hadronization}
%
%
The hypothesis of local \textit{parton-hadron duality} states that the momentum and quantum numbers of hadrons follow those 
of their constituent partons~\cite{dokshitzer1991basics}. This makes up the general guideline of all hadronization models.
There exist two main models of hadron production in the popular physics generators, the string model
and the cluster model.

\headFont{The string model}~\cite{lundstring}
for hadronization, depicted schematically in \autoref{FIGhadronizationModels1},
is based on an observation from lattice simulations of QCD;
at large distances, the potential energy of colour sources, such as heavy quark/anti-quark ($q\bar{q}$) pairs,
increases linearly with their separation.
This indicates a distance-independent force of attraction, thought to be due to the self-attraction
of the gluonic field.

In the model, the field between each $q\bar{q}$ pair is represented by a string with uniform energy per unit length.
As the $q$ and the $\bar{q}$ move apart from each other, the energy of the color field increases and
the string connecting the two is tightened. Eventually the string breaks, and its two ends form a new
quark/anti-quark pair.
If the invariant mass of either of these string pieces is large enough, further breaks may occur in addition. 
The string break-up process is assumed to proceed until only on-mass-shell hadrons remain.
In the simplest approach of baryon production, a diquark ($D$) is treated 
just like an ordinary anti-quark; a string can break either into a quark/anti-quark or into
a diquark/anti-diquark pair, leading to three-quark states.

\headFont{The cluster model}~\cite{herwigclustermodel}
for hadronization is based on the so-called \textit{preconfinement} property of QCD,
discovered by Amati and Veneziano~\cite{Amati:1979fg}.
They showed that at evolution scales, $q$, much smaller than the scale of the
hard subprocess, the partons in a shower are clustered in colourless groups. These groups have an
invariant mass distribution that is independent of the nature and scale of the hard subprocess,
depending only on $q$ and on $\Lambda_{\rm QCD}$. It is then natural to identify these clusters
at the hadronization scale, $Q_0$, as \textit{proto-hadrons} that later decay into the observed final-state hadrons.

In practical terms, at a scale around $Q_0$, gluons from the PS are 
split into light quark/anti-quark or diquark/anti-diquark pairs, as illustrated in \autoref{FIGhadronizationModels2}. 
Color-singlet clusters are formed from the different pair combinations; mesonic ($q\bar{q}$ and $D\bar{D}$),
barionic ($qD$) and anti-barionic ($\bar{q}\bar{D}$). 
The clusters thus formed are fragmented into two hadrons. If a cluster is too light 
to decay into two hadrons, it is taken to represent the lightest single hadron of its
flavor; its mass is therefore shifted to the appropriate value by an exchange of momenta with a neighboring cluster. 
If the cluster is too heavy, it decays into two clusters, which are further fragmented into hadrons.
\begin{figure}[htp]
\begin{center}
\subfloat[]{\label{FIGhadronizationModels1}\includegraphics[trim=60mm 0mm 0mm 0mm,clip,angle=90,width=.4\textwidth]{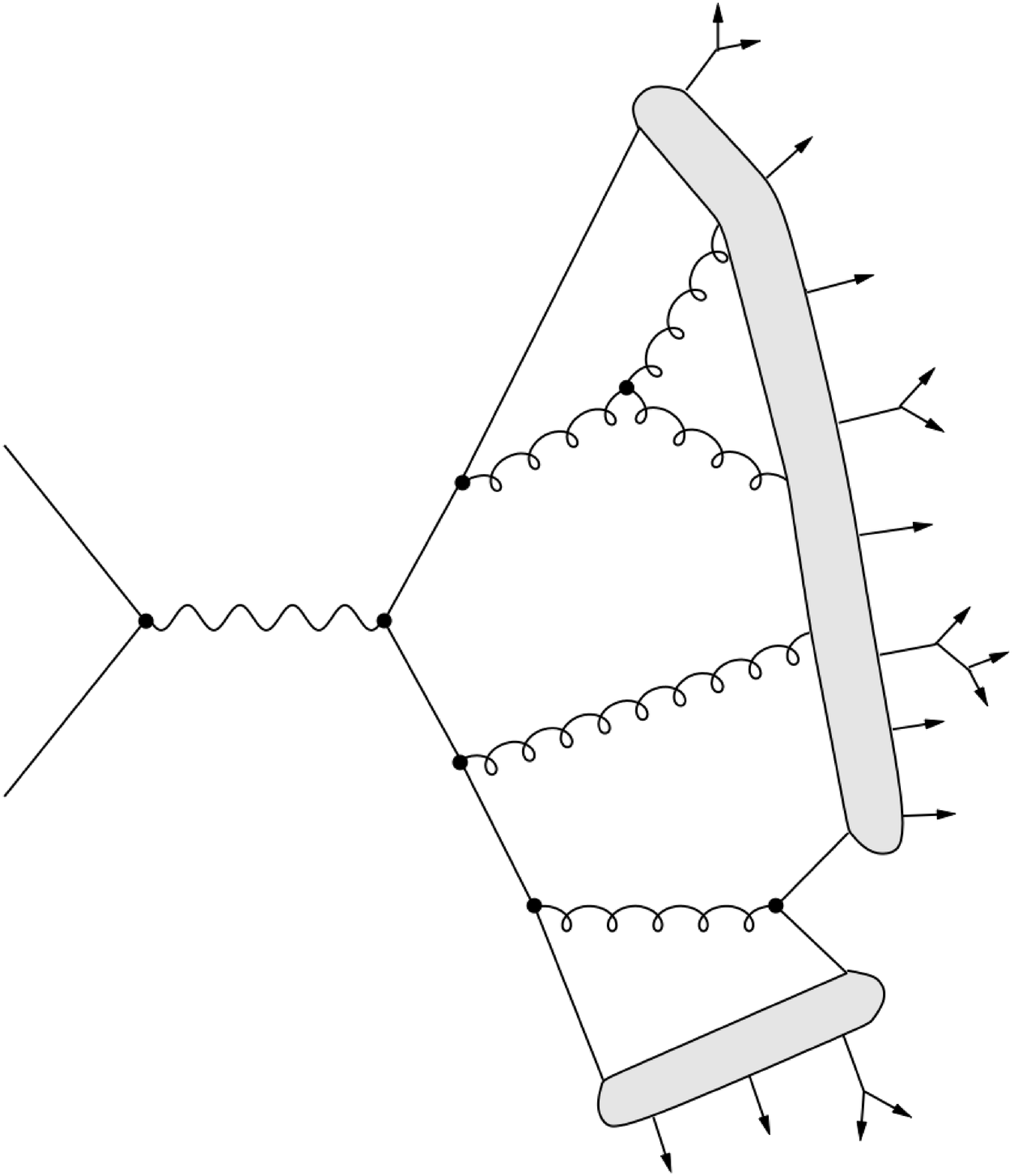}} \qquad\qquad
\subfloat[]{\label{FIGhadronizationModels2}\includegraphics[trim=55mm 0mm 0mm 0mm,clip,angle=90,width=.4\textwidth]{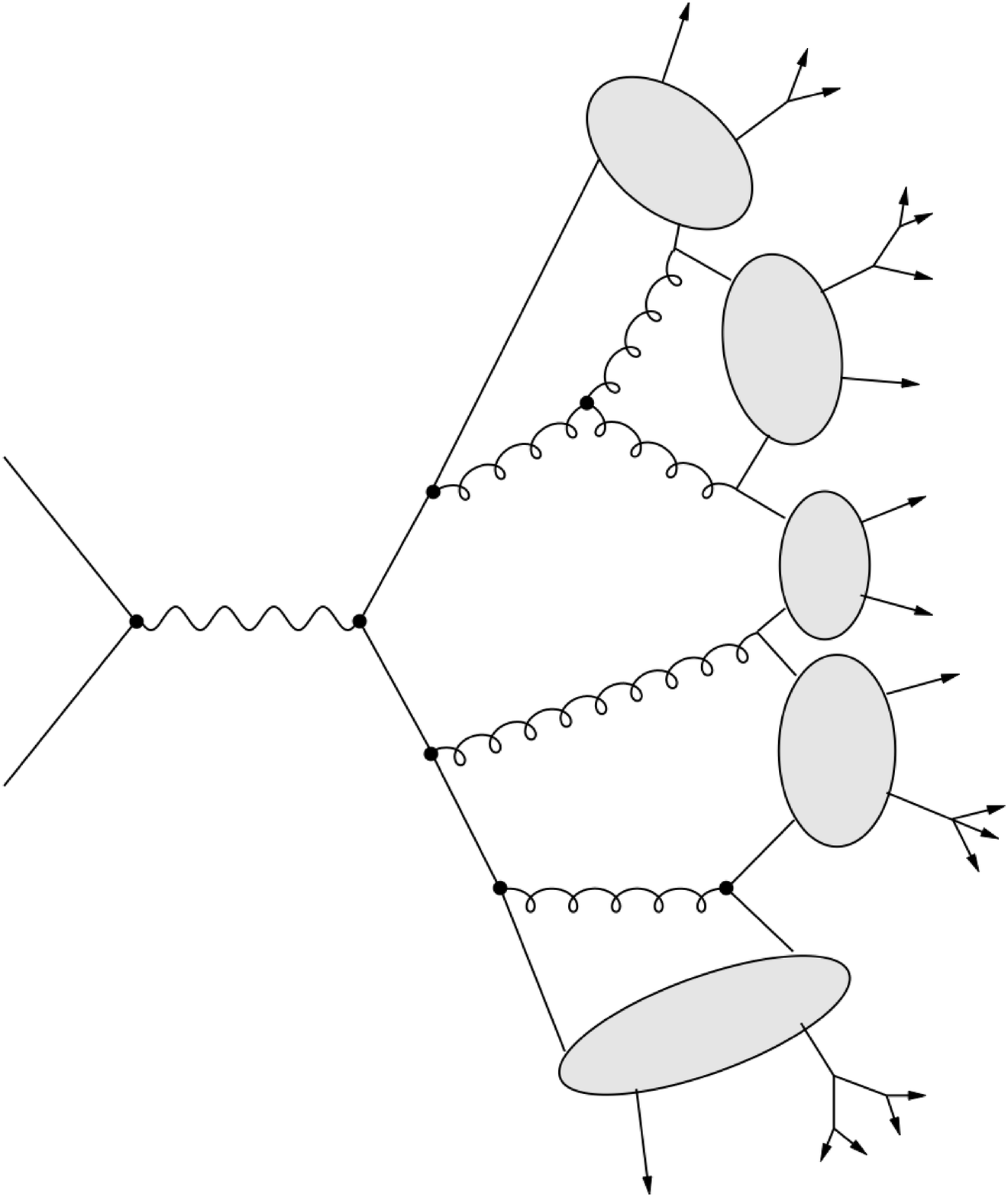}}
  \caption{\label{FIGhadronizationModels}Schematic illustration of hadronization in the string \Subref{FIGhadronizationModels1} and in the
  cluster \Subref{FIGhadronizationModels2} models.
  }
\end{center}
\end{figure} 
%
%

\minisec{The underlying event}
%
%
In events that contain a hard subprocess, there is extra hadron production that
cannot be ascribed to showering from the coloured partons participating in the subprocess,
known as the underlying event (UE).
Furthermore, this extra activity is greater than that in
so-called minimum-bias events (collisions that do not yield an identifiable hard subprocess).
The UE is believed to arise from collisions between those partons in the incoming
hadrons that do not directly participate in the hard subprocess.

The most common hard subprocess at the LHC
is elastic gluon-gluon scattering, $gg \rightarrow gg$.
The leading-order differential \xsec for this subprocess diverges at zero momentum transfer,
due to the exchange of a massless virtual gluon. This divergence is presumably regularized below some momentum
transfer, $t_{\rm min}$, by higher-order and non-perturbative effects. Nevertheless, for reasonable values of $t_{\rm min}$,
the integrated \xsec for gluon-gluon scattering is very large, larger even than the total proton-proton (\pp)
scattering \xsec. This result indicates that the average number of gluon-gluon collision per \pp collisions
is greater than one.
Including the \xsecs for elastic scattering of quarks, anti-quarks and gluons in all possible combinations
(all of which diverge in leading order), multiple parton interactions (MPI) are found to be highly probable

This is the basis on which modern event generators model both minimum-bias collisions and the UE.
To account for the extra hadron production when a hard subprocess is present, an event generator
must model the impact parameter structure of hadron-hadron collisions. The partons in each
incoming hadron are distributed over a transverse area of the order of $1~\mrm{fm}^2$.
The impact parameter of a collision is the transverse distance between the centroids of
these areas before the collision.
When the impact parameter is large, the areas overlap little
and the collision is peripheral; this configuration is associated with a low probability of a hard parton-parton interaction and
few MPI.
On the other hand, at small impact parameter values, the collision is central and has
a large overlap of areas; several multiple interactions and a higher probability of a hard interaction are therefore expected.

To summarize, the presence of a hard subprocess is correlated with more MPI
and a higher level of UE activity. Most of the multiple-interactions are soft, though
hard MPI is also possible.
As mentioned in \autoref{chapIntroduction}, hard MPI are an important background for
\eg new physics signals involving milti-jet final states.
As such, hard MPI are given special attention beyond the general discussion in
the context of the UE. The most simple (and often most prominent) case, that of double parton scattering,
is reviewed in the next section.

\section{Double parton scattering\label{chapMultiPartonInteractionsInGenerators}}
%
%
The formalism to deal with (semi-)hard double parton scattering in hadronic interactions at \cms
energy $\sqrt{s}$ ~\cite{Humpert:1983fy,Ametller:1985tp} may be summarized by 
\begin{eqnarray}
\sigma_{\text{(A,B)}}^{\text{DPS}}(s) & = &
\frac{m}{2}\sum\limits_{i,j,k,l}  \int
                                    \Gamma_{ij}(x_{1},x_{2},\vect{d};Q_{\text{A}},Q_{\text{B}})
                                    \; \hat{\sigma}^{\text{(A)}}_{ik}(x_{1},x_{1}^{\prime},s)
                                    \; \Gamma_{kl}(x_{1}^{\prime},x_{2}^{\prime},\vect{d};Q_{\text{A}},Q_{\text{B}})  \nonumber\\[-10pt]
                                   & & \qquad\qquad\qquad \times \; \hat{\sigma}^{\text{(B)}}_{jl}(x_{2},x_{2}^{\prime},s)
                                    \; dx_{1} \; dx_{2} \; dx_{1}^{\prime} \; dx_{2}^{\prime} \; d^{2}\vect{d}
                                    \;,
\label{EQgeneralSigmaMPI}\end{eqnarray}
where $\sigma_{\text{(A,B)}}^{\text{DPS}}$ is the differential double parton scattering \xsec for the inclusive production of
a combined system $A+B$ at a given $\sqrt{s}$;
the terms $\hat{\sigma}^{\text{(A)}}_{ik}$
denote the differential partonic \xsecs for the production of a system $A$ in the
collision of partons $i$ and $k$;
the terms ${\Gamma_{ij}(x_{1},x_{2},\vect{d};Q_{\text{A}},Q_{\text{B}})}$ represent
double parton distribution functions (DPDFs);
and the parameter $m$ is a symmetry factor
such that ${m = 1}$ if ${\text{A} = \text{B}}$ and ${m = 2}$ otherwise.
The integration over the momentum fractions $x_1$ and $x_2$ is constrained by 
energy conservation, such that ${(x_1 + x_2 \le 1)}$. Summation over all possible parton combinations is performed.
A sketch of double parton scattering is shown in \autoref{FIGschematicDPS} for illustration.
\begin{figure}[htp]
\begin{center}
\includegraphics[width=.4\textwidth]{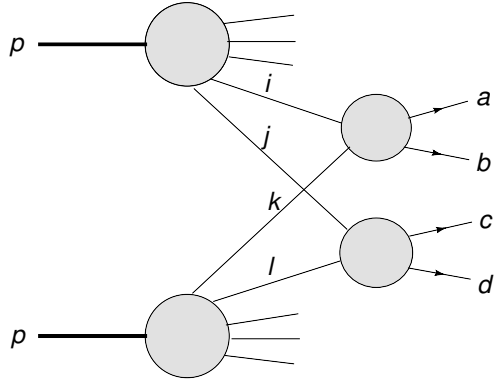}
\caption{\label{FIGschematicDPS}Sketch of a double parton scattering process, in which the active partons
originating from one proton are $i$ and $j$ and from the other are $k$ and $l$.
The two hard scattering subprocess are $A(i~k \rightarrow a~b)$ and $B(j~l \rightarrow c~d)$.}
\end{center}
\end{figure}

The DPDFs, $\Gamma_{ij}(x_{1},x_{2},\vect{d};Q_{\text{A}},Q_{\text{B}})$, may be loosely interpreted as the inclusive probability
distribution to find a parton $i\;(j)$ with longitudinal momentum fraction $x_{1}\;(x_{2})$ at
scale $Q_{\text{A}}\;(Q_{\text{B}})$ in the proton, with the two partons separated by a
transverse distance $\vect{d}$. The scale $Q_{\text{A}}\;(Q_{\text{B}})$ is given by the characteristic
scale of subprocess $\text{A}\;(\text{B})$.  
It is assumed that the DPDFs may be decomposed into longitudinal and transverse components,
\begin{equation}{
\Gamma_{ij}(x_{1},x_{2},\vect{d};Q_{\text{A}},Q_{\text{B}}) \simeq 
            D_{ij}(x_{1},x_{2};Q_{\text{A}},Q_{\text{B}}) \; F(\vect{d}) \;.
\label{DecomposeTransversePart}} \end{equation}

The longitudinal component, ${D_{ij}(x_{1},x_{2};Q_{\text{A}},Q_{\text{B}})}$, has a rigorous interpretation in leading order
pQCD, as the inclusive probability of finding a parton $i$ with momentum fraction
$x_1$ at scale $Q_{\text{A}}$, as well as and a parton $j$ with momentum fraction $x_2$ at scale $Q_{\text{B}}$ in a proton. Accurate
prediction of double parton scattering \xsecs and of event signatures requires good
modelling of $D_{ij}(x_{1},x_{2};Q_{\text{A}},Q_{\text{B}})$ and of the transverse component, $F(\vect{d})$.
In particular, one must correctly take account of the effects of
correlations in both longitudinal momenta and transverse positions in these functions.

Correlations between the partons in transverse space are highly significant; at the very
least, they must tie the two partons together within the same hadron. 
However, their precise calculation is not possible using perturbation theory. Existing models typically
use Gaussian or exponential forms (or their combination) to describe $F(\vect{d})$~\cite{Sjostrand:2004pf,HerwigppUI}.
The transverse component is usually expressed simply as
\begin{equation}{
\sigeff(s) = \left[\int d^{2}\vect{d}\left(F(\vect{d})\right)^{2}\right]^{-1} \;.
\label{EQsigmaEff}} \end{equation}
The quantity $\sigma_{\rm{eff}}(s)$ is defined at the parton-level, and has the units of a \xsec. In the 
formalism outlined here, it is independent of the process and of the phase-space under consideration.
Naively, it may be related to the geometrical size of the proton. That is,
given that one hard scattering occurs, the probability of the other hard scattering
is proportional to the flux of accompanying partons; these are confined to the colliding
protons, and therefore their flux should be inversely proportional to the area (\xsec)
of a proton.
This leads to an estimate of ${\sigeff\approx \pi R_p^2 \approx 50}$~mb, where $R_p$ is the proton 
radius.
%
Alternatively, $\sigeff$ may be connected to the inelastic \xsec,
which would lead to ${\sigeff\approx\sigma_{\rm{inel}}\approx 70}$~mb at \sqs~\cite{Tompkins:2011bv,Antchev:2011vs}.  
For hard interactions, assuming uncorrelated scatterings, \sigeff can be estimated from the gluon form factor
of the proton~\cite{Frankfurt:2003td} and comes out to be~$\sim30$~mb.

A number of measurements of $\sigeff$ have been performed in $pp$ or ${p\bar p}$ collisions
at different \com energies, as specified  in \autoref{sigmaEffTable}.
\begin{table}[htp]
\vspace{10pt}
\begin{center}
\begin{Tabular}[1.45]{|cc|c|c|c|} \hline
\multicolumn{2}{|c|}{Experiment}                  & $\sqrt{s}$~[GeV]  & Final state        & $\sigma_{\text{eff}}$~[mb]      \\ \hline \hline
\cite{Åkesson:173908}		& AFS ($pp$),        1986 & \qquad63\quad\quad\qquad 	& \qquad4 jets\qquad\quad  & $\sim$5           \\  \hline
\cite{Alitti1991145}	  & UA2 ($p\bar{p}$),  1991 & 630 			        & 4 jets 		         & \qquad$>$8.3 (95\% C.L)\qquad\quad \\  \hline
\cite{PhysRevD.47.4857}	& CDF ($p\bar{p}$),  1993 & 1800 			        & 4 jets 		         & 12.1 $^{+10.7}_{-5.4}$ 		       \\  \hline
\cite{PhysRevD.56.3811}	& CDF ($p\bar{p}$),  1997 & 1800 			        & $\gamma$~+~3-jets	 & 14.5 $\pm$ $1.7\;^{+\;1.7}_{-2.3}$  \\  \hline
\cite{Abazov:2009gc}	  & D\O\ ($p\bar{p}$), 2010 & 1960 			        & $\gamma$~+~3-jets  & 16.4 $\pm$ 0.3 $\pm$ 2.3 		   \\  \hline
\cite{Sadeh:1498427}    & \atlas ($pp$), 2012     & 7000 			        & $W$~+~2-jets       & 15   $\pm$ $3\;^{+\;5}_{-3}$ 		     \\  \hline
\end{Tabular} 
\caption{Summary of published measurements of $\sigma_{\text{eff}}$.}
\label{sigmaEffTable}
\end{center}
\end{table}
The energy dependence of the measured values of \sigeff yielded an increase
from about 5~mb at the lowest energy~(${63\GeV}$) to about 15~mb at LHC energies~(${7\TeV}$).
Attempts to explain the differences between these values have used the
Constituent Quark Model~\cite{Levin:1965mi,Lipkin:1965fu}, such as
in~\cite{Bondarenko:2002pp}, or have introduced non-trivial correlations between the two
scattering systems, as in~\cite{Calucci:1997uw,Gaunt:2009re,Calucci:1999yz,Gaunt:2010vy,Gaunt:2012wv}.
A complete explanation, however, is still elusive.
A recent argument, suggested in~\cite{Ryskin:2011kk,PhysRevD.86.014018,Blok:2012mw} may resolve the discrepancy; it goes as follows.
In DPS, two partons from one proton collide with two partons from the other proton.
The two partons from a given proton can originate from the non-perturbative hadron wave function or, alternatively,
emerge from perturbative splitting of a single parton from the hadron.
The former represents a double-\twotwo interaction, and the latter a \threefour interaction.
The \threefour process could explain the difference between the experimental value of \sigeff of~$\sim15$~mb,
and the expected value of~$\sim30$~mb.

In the absence of a rigorous formalism, measurements of \sigeff typically assume a simple factorization ansatz
for the DPDFs,
%
%
\begin{equation}{
D_{ij}(x_{1},x_{2};Q_{\text{A}},Q_{\text{B}}) \simeq D_{i}(x_{1};Q_{\text{A}})D_{j}(x_{2};Q_{\text{B}}) \;.
\label{EQfactorizationDPDF}} \end{equation}
The differential double parton scattering \xsec defined in \autoref{EQgeneralSigmaMPI} therefore reduces to
\begin{equation}
\sigma_{\text{(A,B)}}^{\text{DPS}} = 
\frac{m}{2}\frac{\sigma_{\text{A}}\sigma_{\text{B}}}{\sigma_{\text{eff}}} \;.
\label{EQsimpleSigmaMPI}
\end{equation}

The assumption of factorization is problematic.
For one, this naive representation does not obey the relevant momentum and number sum rules.
In addition, while previous experiments suggest that approximate factorisation holds
at moderately low~$x$~\cite{PhysRevD.56.3811}, this can not be true for all values of~$x$. 
Namely, if ${D_{i}(x_{1};Q_{\text{A}})}$ and ${D_{j}(x_{2};Q_{\text{B}})}$ each satisfy DGLAP evolution,
then the naive product, ${D_{i}(x_{1};Q_{\text{A}})D_{j}(x_{2};Q_{\text{B}})}$, can not
be a solution of the \textit{double-DGLAP equations} (dDGLAP), suggested \eg in~\cite{Gaunt:2009re}.
In order to use the factorization ansatz, the unknown correlations are absorbed into \sigeff, which then possibly becomes
dependant on the process (and respective phase-space) under consideration.

A measurement of the rate of double parton scattering in four jet events in \atlas is presented in this thesis, 
using the simplified form of the effective \xsec given in \autoref{EQfactorizationDPDF}.
Only the double-\twotwo topology is considered in the analysis, as discussed in \autoref{chapDoublePartonScattering}.


           }{}
\ifthenelse{\boolean{do:atlasDetector}}     { 
\chapter{The \ATLAS experiment at the LHC\label{chapAtlasexperimentAtTheLHC}}
%
\section{The Large Hadron Collider}
%
%
The Large Hadron Collider (LHC) is the world's largest and highest-energy particle accelerator.
LHC is a proton-proton ($pp$) collider, located at the Franco-Swiss border near Geneva, Switzerland. 
It lies in a tunnel 27~km in circumference at an average depth of 100~meters. 
The tunnel houses 1232~superconducting bending dipole magnets, cooled using liquid helium to an operating
temperature of 1.9~K, producing a magnetic field of about 8~T. 
The use of dipole magnets allows to keep protons traveling clockwise and counter-clockwise on orbit at the same time. 
In total, 392~quadrupole magnets are used to keep the beams focused and to collide them at the four interaction points (IPs) of the LHC experiments.
The design center-of-mass energy of the LHC is ${\sqrt{s} = 14\TeV}$.

Protons are produced by ionizing hydrogen atoms in an electric field. They are injected into RF cavities and accelerated to 750\keV.
The beam is then transmitted to the LINAC~2, a linear accelerator, which increases the energy to 50\MeV.
The protons are accelerated to 1.4\GeV by the Proton Synchrotron Booster and then further to 26\GeV by the Proton Synchrotron. Next, the Super
Proton Synchrotron accelerates the protons to 450\GeV,
the minimum energy required to maintain a stable beam in the LHC. Finally, the LHC accelerates them to the operating energy.

During the first two years of operation (2010-2011), the LHC operated at a center-of-mass energy of $\sqrt{s} = 7$\TeV,
with 3.5\TeV per proton. This center-of-mass energy has been chosen to ensure a safe operating margin for the magnets in the accelerator.
So far, up to the end of the 2011 run, the LHC delivered 5.61~\inv{\fb} total integrated
luminosity with a peak luminosity of $3.65\times10^{33}$~\lumiUnits~\cite{LumiTwiki}. The bunch separation was 50\ns for most of the running period.
The full 2010 and most of the 2011 datasets are used in this analysis.

\section{The \ATLAS detector\label{chapTheATLASdetectorSect}}
%
%
ATLAS (A Toroidal Lhc ApparatuS)~\cite{Aad:2008zzm,Aad:2009wy} is a general-purpose detector surrounding IP~1 of the LHC.
ATLAS consists of three main sub-systems, the inner detector (ID),
the calorimeters and the muon spectrometer (MS).
\Autoref{atlasDetectorCalorimeter1} shows a schematic view of the \ATLAS detector and its sub-systems.

\subsubsection{Detector sub-systems}
%
%
The ID is used to measure the tracks of charged particles. It covers the pseudo-rapidity range, $|\eta| < 2.5$~\footnote{
The coordinate system used by \ATLAS is a right-handed Cartesian coordinate system.
The positive $z$-direction is defined as the direction of the anti-clockwise beam.
Pseudo-rapidity is defined as $\eta = \ln\tan(\theta/2)$, where $\theta$ is the angle with respect to the $z$-axis.
The azimuthal angle in the transverse plane $\phi$ is defined to be zero along the $x$-axis, which points
toward the center of the LHC ring.},
and has full coverage in azimuth. It is made of three main components, arranged in concentric layers,
all of which are immersed in a 2~T field provided by
the inner solenoid magnet. Three layers of silicon pixel detectors
provide a two-dimensional hit position very close to the interaction point.
Silicon microstrip detectors are then used in the next four layers, providing excellent position resolution for charged tracks.
A transition-radiation detector is the final component of the tracker, with poorer position
resolution with respect to the silicon, but providing many measurement points and a large
lever-arm for track reconstruction in addition to particle identification capabilities.

The \ATLAS calorimeter, shown schematically in \autoref{atlasDetectorCalorimeter2}, is the principal tool used in the analysis.
\begin{figure}[htp]
\begin{center}
\ifthenelse{\boolean{do:includeAllGraphics}} {
  \subfloat[]{\label{atlasDetectorCalorimeter1}\includegraphics[width=1\textwidth]{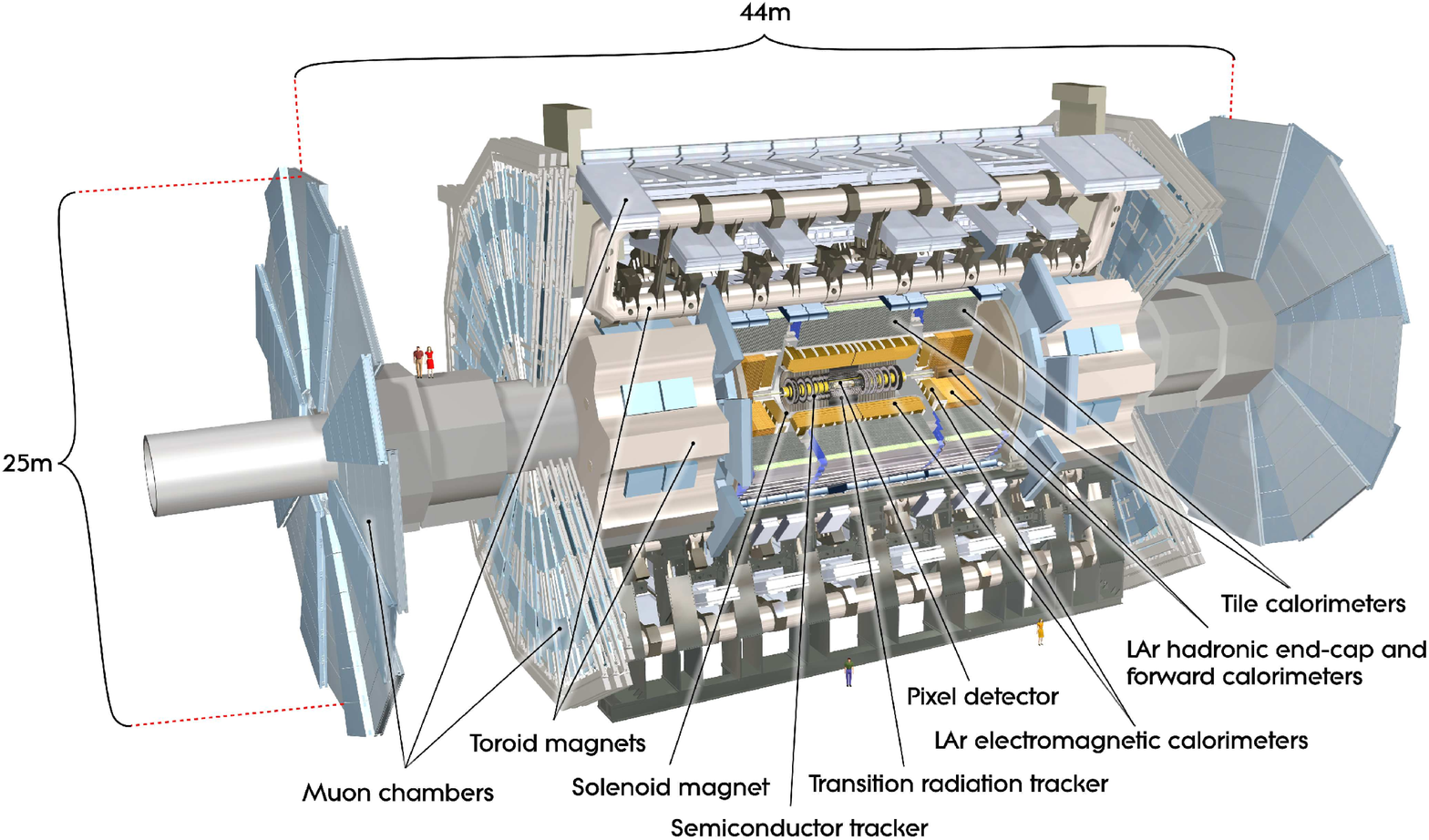}} \newline\newline
  \subfloat[]{\label{atlasDetectorCalorimeter2}\includegraphics[width=1\textwidth]{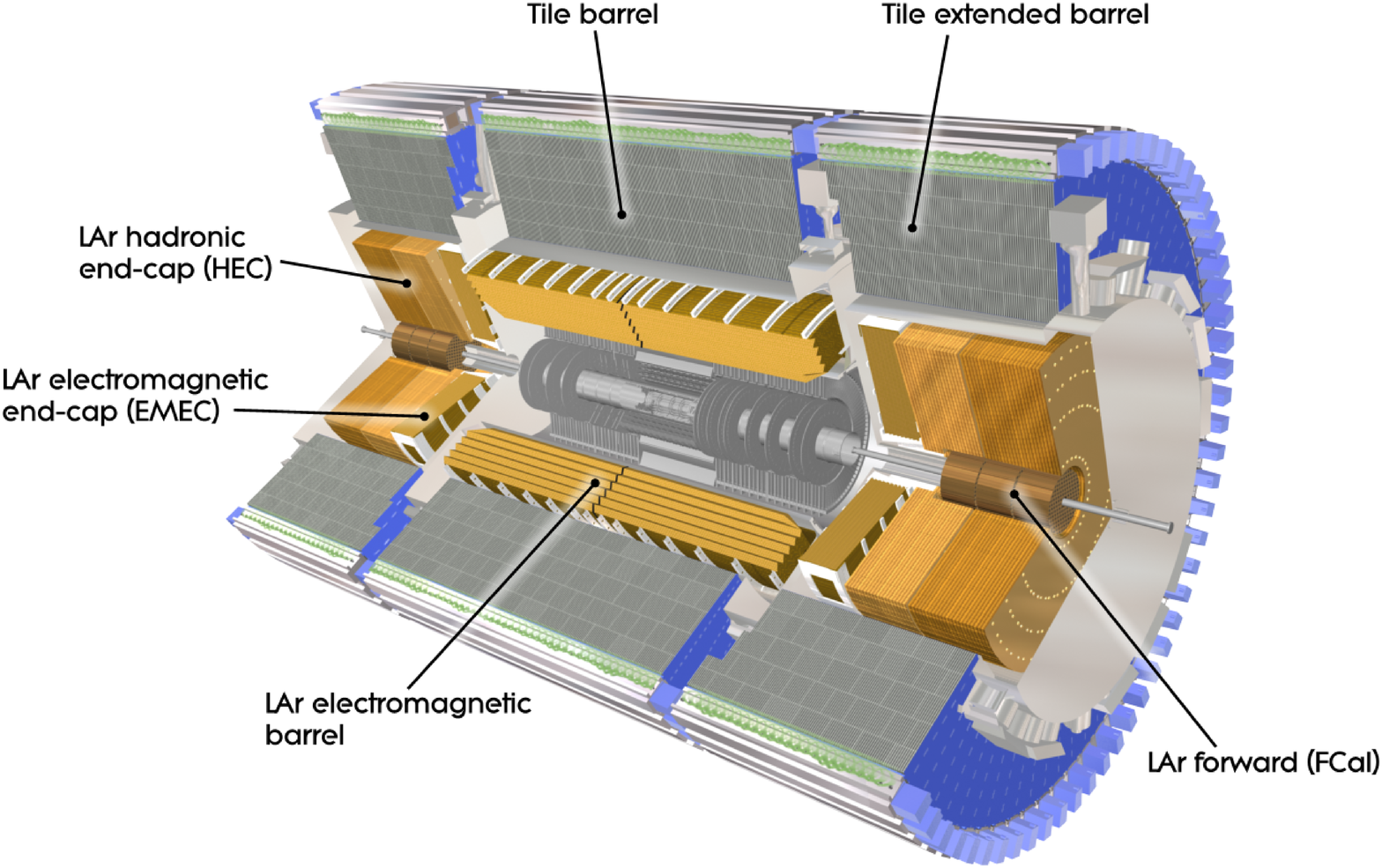}} 
}{
  \includegraphics[width=.4\textwidth]{figures/plotFiller} \\  \includegraphics[width=.4\textwidth]{figures/plotFiller} 
  }
\caption{\label{atlasDetectorCalorimeter}Schematic view of the \ATLAS detector \Subref{atlasDetectorCalorimeter1} and of
  the \atlas calorimeter system \Subref{atlasDetectorCalorimeter2}. }
\end{center}
\end{figure}
The calorimeter is composed of several sub-detectors.
The liquid argon (LAr) electromagnetic (EM) calorimeter is divided into one barrel
($|\eta| < 1.475$) and two end-cap components ($1.375 < |\eta| < 3.2$). 
It uses an accordion geometry to ensure fast and uniform response and fine segmentation
for optimum reconstruction and identification of electrons and photons. 
The hadronic scintillator tile calorimeter consists of a barrel covering the
region, $|\eta| < 1.0$, and two extended barrels in the range $0.8 < |\eta| < 1.7$. 
The LAr hadronic end-cap calorimeter ($1.5 < |\eta| < 3.2$) is located behind the end-cap electromagnetic calorimeter. 
The forward calorimeter covers the range $3.2 < |\eta| < 4.9$ and also uses LAr as the active material.

The MS forms the outer part of the \ATLAS detector and is designed to detect
muons exiting the barrel and end-cap calorimeters and to measure their momentum in
the pseudo-rapidity range $|\eta| < 2.7$. It is also designed to trigger on muons in the
region $|\eta| < 2.4$. The MS operates inside an air-core toroid magnet system with a peak field in the coil windings of 4~T.
The precision momentum measurement is performed by the Monitored Drift Tube chambers covering
the pseudo-rapidity range $|\eta| < 2.7$ (except in the innermost end-cap layer where their coverage
is limited to $|\eta| < 2.0$). In the forward region ($2.0 < |\eta| < 2.7$), Cathode-Strip Chambers are used in the innermost tracking layer.
The capability to trigger on muon tracks is achieved by a system of fast trigger chambers capable
of delivering track information within a few tens of nanoseconds after the passage of the particle.
In the barrel region ($|\eta| < 1.05$), Resistive Plate Chambers were selected
for this purpose, while in the end-cap ($1.05 < |\eta| < 2.4$) Thin Gap Chambers were chosen.

\subsubsection{The Trigger System}
%
%
The \ATLAS detector has a three-level trigger system consisting of Level~1
(\ttt{L1}), Level~2 (\ttt{L2}) and Event Filter (\ttt{EF}). 
The \ttt{L1} trigger rate at design luminosity is approximately 75~kHz. The \ttt{L2} and \ttt{EF}
triggers reduce the event rate to approximately 200~Hz.
Another trigger system used in \ATLAS relies on the minimum-bias trigger scintillators (\ttt{MBTS}).
The \ttt{MBTS} trigger requires one hit above threshold from either one of the sides of the
detector. The different triggers used in the analysis are described in further detail
in \autoref{chapDataSelection}, \autoref{chapTriggerAndLuminosity}.

\subsubsection{Luminosity Measurement}
%
%
Accurate measurement of the delivered luminosity is a key component of the \ATLAS physics program. 
For cross section measurements of SM processes, the uncertainty on the delivered luminosity
is often one of the dominant systematic uncertainties.

The instantaneous luminosity of proton-proton collisions can be calculated as
\begin{equation}{
\mathcal{L} = \frac{R_{\text{inel}}}{\sigma_{\text{inel}}} \;,
}\label{eqInelasticLumi} \end{equation}
where $R_{\text{inel}}$ is the rate of \pp interactions and $\sigma_{\text{inel}}$ 
is the inelastic cross section. Any detector sensitive to inelastic \pp
interactions can be used as a source for relative luminosity measurement.
However, these detectors must be calibrated using an absolute measurement of the luminosity. 

The recorded luminosity can be written as
\begin{equation}{
\mathcal{L} = \frac{\mu_{\text{vis}}n_{b}f_{\text{rev}}}{\sigma_{\text{vis}}} \;,
}\label{eqRecordedLumi} \end{equation}
where $\mu_{\text{vis}}$ is the visible number of interactions per bunch crossing in the
detector, $n_{b}$ is the number of bunch pairs colliding in \ATLAS, ${f_{\text{rev}} = 11245.5~\mrm{Hz}}$
is the LHC revolution frequency and $\sigma_{\text{vis}}$
is the visible cross section, to be determined via calibration for each detector.

The calibration is done using dedicated beam separation scans, also known as \textit{van~der~Meer} scans~\cite{ATLAS-CONF-2011-116},
where the two beams are stepped through each other in the horizontal and vertical planes to measure
their overlap function. The delivered luminosity is measured using beam parameters,
\begin{equation}{
\mathcal{L} = \frac{n_{b}f_{\text{rev}}n_{1}n_{2}}{2\pi\Sigma_{x}\Sigma_{y}} \;,
}\label{eqAbsoluteLumi} \end{equation}
where $n_{1}$ and $n_{2}$ are respectively the bunch populations (protons per bunch) in beam~1 and in beam~2
(together forming the bunch charge product), and $\Sigma_{x}$ and $\Sigma_{y}$ respectively characterize the
horizontal and vertical profiles of the colliding beams.

By comparing the delivered luminosity to the peak interaction rate, $\mu_{\text{vis}}^{\text{Max}}$,
observed by a given detector during the van~der~Meer scan, it is possible to determine the visible cross section,
\begin{equation}{
\sigma_{\text{vis}} = \mu_{\text{vis}}^{\text{Max}}\frac{2\pi\Sigma_{x}\Sigma_{y}}{n_{1}n_{2}} \;.
}\label{eqSigmaVis} \end{equation}

Two detectors are used to make bunch-by-bunch luminosity measurements in 2010 and in 2011, LUCID and BCM.
LUCID is a Cerenkov detector specifically designed for measuring the luminosity in \ATLAS. It is made up of sixteen mechanically
polished aluminum tubes filled with $\mrm{C}_{4}\mrm{F}_{10}$
gas surrounding the beampipe on each side of the IP at a distance of 17~m,
covering the pseudo-rapidity range, $5.6 < |\eta| < 6.0$.
The beam conditions monitor (BCM) consists of four small diamond
sensors on each side of the \ATLAS IP arranged around the beam-pipe in a cross pattern. The BCM is a fast device, primarily designed
to monitor background levels. It is also capable of issuing a beam-abort request in cases of possible damage to \ATLAS detectors, due to beam losses.

The relative uncertainty on the luminosity scale for \pp collisions during 2010~\cite{Aad:2011dr}
and 2011~\cite{ATLAS-CONF-2011-116} was found to be, respectively~$\pm$3.4 and~$\pm$3.9\%\footnote{ The uncertainty on the 
luminosity in 2011 quoted here is larger than the value given in~\cite{ATLAS-CONF-2011-116}.
The increased value is a consequence of recent studies by the Luminosity group in \ATLAS, which show
significant non-linear correlations between the van~der~Meer scans in the horizontal and vertical directions~\cite{lumiUncertaintyBug}.
These correlations were not previously known and are not yet fully understood.}.

          }{}
\ifthenelse{\boolean{do:MCsimulation}}      { 
\chapter{Monte Carlo simulation\label{chapMCsimulation}}
%
Any analysis involving a complex detector such as \atlas, requires a detailed detector Monte Carlo (MC) simulation. The MC
is used in order to compare distributions of observables, as simulated by physics generators, with the data. In addition,
it is used to study the performance of the detector by estimating reconstruction efficiencies, geometrical coverage, the
performance of triggers etc.\

A description of the MC generators and of the \atlas detector simulation
used in the analysis is presented in the following.

\section{Event generators\label{chapEventGenerators}}
%
%
The four-momenta of particles produced in \pp collisions at the LHC
are simulated using various event generators. An overview of MC
event generators for LHC physics can be found in~\cite{mcforlhc}.
The different MC samples used for comparison with data taken during
2010 and during 2011 are denoted respectively as MC10 and MC11. 
The following event generators, using different theoretical models,
are utilized in the analysis.

\minisec{\pythia}
%
\pythia~\cite{pythia} simulates non-diffractive \pp collisions using a $2 \to 2$ matrix element
at leading order in the strong coupling to model the hard subprocess. It
uses \pt-ordered parton showers to model additional radiation in the leading-logarithmic approximation~\cite{pythiapartonshower}. 
Multiple parton interactions~\cite{Sjostrand:2004pf,Sjostrand:2004ef}, as well as fragmentation and hadronization,
based on the Lund string model~\cite{lundstring}, are also simulated. 

Several \pythia~6 tunes utilizing different parton distribution function (PDF) sets are used.
The nominal version used in this analysis employs the modified leading-order PDF set, MRST~LO*~\cite{PDF-MRST}.
The parameters used for tuning the underlying event include  
charged particle spectra measured by \ATLAS in minimum bias collisions~\cite{mc10chargedparticles}.
The samples used to compare with the 2010 and with the 2011 data respectively use the AMBT1~\cite{MC10}
and AMBT2B~\cite{ATL-PHYS-PUB-2011-009} tunes.

Several additional combinations of \pythia versions, PDF sets and underlying event tunes are used in \autoref{chapMeasurementOfTheDijetMass}
in order to evaluate non-perturbative corrections to the dijet mass \xsec.
These include the following; \pythia~6.425 with the MRST~LO* PDF set and the AUET2B~\cite{ATL-PHYS-PUB-2011-009} tune;
\pythia~6.425 with the CTEQ6L1~\cite{PDF-CTEQ} PDF set and the AMBT2B and AUET2B tunes;
\pythia~8 (v150)~\cite{Sjostrand:2007gs} with the MRST~LO** PDF set and 4C~\cite{Corke:2010yf} tune.

\minisec{\herwig , \herwigpp}
%
\herwig uses a leading order \twotwo matrix element supplemented 
with angular-ordered parton showers in the leading-logarithm approximation~\cite{herwig3,herwig2,herwig}. 
The so called cluster model~\cite{herwigclustermodel} is used for the hadronization.
Multiple parton interactions are modelled using \jimmy~\cite{jimmy}.
The model parameters of \herwigJimmy have been tuned to \ATLAS data (AUET1 tune)~\cite{AUET1}.
The MRST~LO* PDF set is used.

\herwigpp~\cite{Herwigpp} (v2.5.1) is based on \herwig,       
but redesigned in the {\textit C++} programming language.
The generator contains a few modelling improvements.                    
It also uses angular-ordered parton showers, but with an updated 
evolution variable and a better phase-space treatment.
Hadronization is performed using the cluster model, as in \herwig. 
The underlying event and soft inclusive interactions are described using a 
hard and soft multiple partonic interactions model~\cite{HerwigppUI}.
The MRST~LO* and the CTEQ6L1 PDF sets are used with the UE7000-3~\cite{ATL-PHYS-PUB-2011-009}
underlying event tune, in order to evaluate non-perturbative corrections
to the dijet mass \xsec in \autoref{chapMeasurementOfTheDijetMass}.

\minisec{\alpgen}
%
\alpgen is a tree level matrix-element generator for hard multi-parton processes ($2 \to n$) 
in hadronic collisions~\cite{alpgen}. 
It is interfaced to \herwig to produce parton showers in the leading-logarithmic 
approximation. Parton showers are matched to the matrix element with the MLM matching scheme~\cite{MLM}. 
For the hadronization, \herwig is interfaced to \jimmy in order to model soft multiple parton interactions.
The PDF set used is CTEQ6L1.
\alpgen is used in \autoref{chapMeasurementOfTheDijetMass} in order to evaluate the systematic uncertainty associated
with the unfolding procedure of the dijet mass \xsec measurement.

\minisec{\sherpa}
%
\sherpa is a general-purpose tool for the simulation of 
particle collisions at high-energy colliders, using a 
tree-level matrix-element generator for the calculation of hard scattering 
processes~\cite{Gleisberg:2008ta}.
The emission of additional QCD partons off the initial and final states is 
described through a parton-shower model. To consistently combine 
multi-parton matrix elements with the QCD parton cascades, the approach of 
Catani, Krauss, Kuhn and Webber~\cite{Krauss:2005re} is employed. 
A simple model of multiple 
interactions is used to account for underlying events in hadron--hadron 
collisions. The fragmentation of partons into primary hadrons is described 
using a phenomenological cluster-hadronization model~\cite{Winter:2003tt}. A comprehensive 
library for simulating tau-lepton and hadron decays is provided. Where 
available form-factor models and matrix elements are used, allowing for 
the inclusion of spin correlations; effects of virtual and real 
QED corrections are included using the approach of Yennie, Frautschi 
and Suura~\cite{1961AnPhy..13..379Y}.
The CTEQ6L1 PDF sets together with the default underlying event tune are used. 
The CKKW matching scale is set at 15\GeV, where the latter refers to the energy scale in which matching of matrix elements
to parton showers begins. The implication of this choice is that partons with transverse momentum above
15\GeV in the final state, necessarily originate from matrix elements, and not from the parton shower.

\sherpa is employed in the analysis as part of the double parton scattering measurement in \autoref{chapDoublePartonScattering}.
events are generated without multiple interactions, by setting the internal flag, \ttt{MI\ul HANDLER=None}. The
generated events serve as input to a neural network.

\section{Simulation of the \ATLAS detector\label{chapSimulationOfTheATLASdetector}}
%
%
Event samples produced using the different event generators described above are 
passed through the full \ATLAS detector simulation and are reconstructed as the data.
The \geant software toolkit~\cite{Geant4} within the \ATLAS simulation framework~\cite{simulation} 
propagates the generated particles through the \ATLAS detector and simulates their interactions with 
the detector material. 
The energy deposited by particles in the active detector material is 
converted into detector signals with the same format as the \ATLAS detector read-out.
The simulated detector signals are in turn reconstructed with the same reconstruction software
as used for the data.

In \geant the model for the interaction of hadrons with the detector material
can be specified for various particle types and for various energy ranges.
For the simulation of hadronic interactions in the detector, the \geant set of 
processes called \texttt{QGSP\ul BERT}~\cite{geanthadronic} is chosen. 
In this set of processes, the \textit{Quark Gluon String model}~\cite{QGS,QGSP2,QGSP3,QGSP4,QGSP5} is used for the 
fragmentation of the nucleus, 
and the \textit{Bertini cascade} model~\cite{Bertini,Bertini1,Bertini2,Bertini3} for the description 
of the interactions of  hadrons in the nuclear medium. 

The \geant simulation, and in particular the hadronic interaction model
for pions and protons, has been validated with test-beam measurements for the
barrel~\cite{Tile2002, Tile2002pionproton, CTB2004topology, CTB04pion, CTB2004vlepion} 
and endcap~\cite{EndcapTBelectronPion2002,Pinfold:2008zzb, Kiryunin:2006cm} calorimeters. 
Agreement within a few percent is found between simulation and data
for pion momenta between~2 and~350\GeV.
Further tests have been carried out \insitu comparing the single hadron response, measured using isolated tracks
and identified single particles~\cite{singleparticle900,atlassingleparticle2011}.
Good agreement between simulation and data is found for particle momenta from a few hundred\MeV to 6\GeV.

Studies of the material of the inner detector in front of the calorimeters have been performed using
secondary hadronic interactions~\cite{Aad:2011cxa}.
Additional information is obtained from studying photon conversions \cite{PhotonConversions900} and
the energy flow in minimum bias events \cite{MaterialBudgetMinBias7}.
The \ATLAS detector geometry used in the 2010 and in the 2011 simulation campaigns reflects the best current knowledge of the detector
at the time the simulations were made. Subsequently, compared to MC10, a more detailed description
of the geometry of the LAr calorimeter was used for MC11. The improvement in understanding of the geometry introduced an increased
calorimeter response to pions below 10\GeV of about 2\%.

The MC11 simulation is made up of different \textit{MC-periods}.
Each MC-period represents different data-taking conditions, such as malfunctioning hardware or changes in the
amount of \textit{\itpu}, additional $pp$ collisions, coinciding with the hard interaction.
Events in the different MC-periods are given relative weights according to the integrated luminosity of the
respective data-taking periods they represent.

A note regarding the simulation in \autoref{chapJetAreaMethod};
before the start of data-taking in 2011, a transitional MC simulation was created, denoted as MC10b.
MC10b was generated as part of the MC10 campaign. It used the 2010 simulation settings as described above, with 
the exception of highly increased
\pu conditions matching the expected characteristics of the 2011 data. The study described in \autoref{chapJetAreaMethod}
was originally performed using an MC10b \pythia sample. When MC11 became available,
most of the results were reproduced; those which were not, are presented here using the previous version of the simulation.
Unless otherwise indicated, MC11 is used in conjecture with the 2011 data.

\section{Bunch train structure and overlapping events}
%
%
The LHC bunch train structure of the 2010 and the 2011 data is modelled in MC.
In MC10 the simulated collisions are organised in double trains with 225\ns separation.
Each train contains eight filled proton bunches with a bunch separation of 150\ns.
In MC11 the simulation features four trains with 36~bunches per train and 50\ns spacing between the bunches.

The MC samples are generated with additional minimum bias interactions, using \pythia~6
with the AMBT1 underlying event tune and the MRST~LO* PDF set for MC10, and \pythia~6
with the AUET2B tune and the CTEQ6L1 PDF set for MC11.
These minimum bias events are used to form \pu events, which are overlaid onto the hard scattering
events. The number of overlaid events follows a Poisson distribution around the average number of additional \pp collisions per bunch
crossing, as measured in the experiment. The average number of additional interactions depends on the instantaneous luminosity
(see \autoref{eqDefinitionMu} in \autoref{chapInAndOutOfTimePU}); it is thus greater in MC11 compared to MC10, following the trend in the data.

The small separation between bunches in the 2011 data (and respectively in MC11)
requires inclusion of the effect of \textit{\otpu}\footnote{ A feature of the 2011 data
is that the signal in the calorimeter is sensitive to collisions from several consecutive
bunch crossings; \otpu refers to this fact. See \autoref{chappuInAtlas} for a detailed discussion.}
in the simulation.
The properties of this effect depend on the distance of the
hard scatter events from the beginning of the bunch train. The first ten (approximately) bunch crossings
are characterized by increasing \otpu contributions from the previous collisions. 
For the remaining~26 bunch crossings in a train, the \otpu is stable within the bunch-to-bunch
fluctuations in proton intensity at the LHC.

           }{}
\ifthenelse{\boolean{do:jetReconstruction}} { 
\chapter{Jet reconstruction and calibration\label{chapJetReconstruction}}
%
\section{Jet reconstruction algorithms }
%
Jets originate from the fragmentation of partons.
Due to color flow at the parton-level, there is no one-to-one correspondence between the hadrons inside a
jet and the partons which initiated the jet.
Because the measurements are performed at the hadron-level and the theoretical expectations at the
parton-level, a precise definition of jets is required.
The algorithm that relates a jet of hadrons to partons must satisfy,
\begin{list}{-}{}
\item 
  \headFont{infrared safety - }
  the presence or absence of additional soft particles between two particles belonging to the same
  jet should not affect the recombination of these two particles into a jet.
  Generally, any soft particles not coming from the fragmentation of a hard scattered parton
  should not affect the number of produced jets;
  \item 
  \headFont{collinear safety -}
  a jet should be reconstructed independently whether the transverse
  momentum is carried by one particle, or if that particle is split into two collinear particles;
\item 
  \headFont{order independence -}
  the algorithm should be applicable at parton- hadron- or detector-level, and lead
  to the same origin of the jet.
\end{list}
An illustration of infrared and collinear sensitivity in jet-finding is given in \autoref{FIGjetSafetyIllustrationFIG}. 
\begin{figure}[htp]
\begin{center}
\subfloat[]{\label{FIGjetSafetyIllustrationFIG1}\includegraphics[width=.49\textwidth]{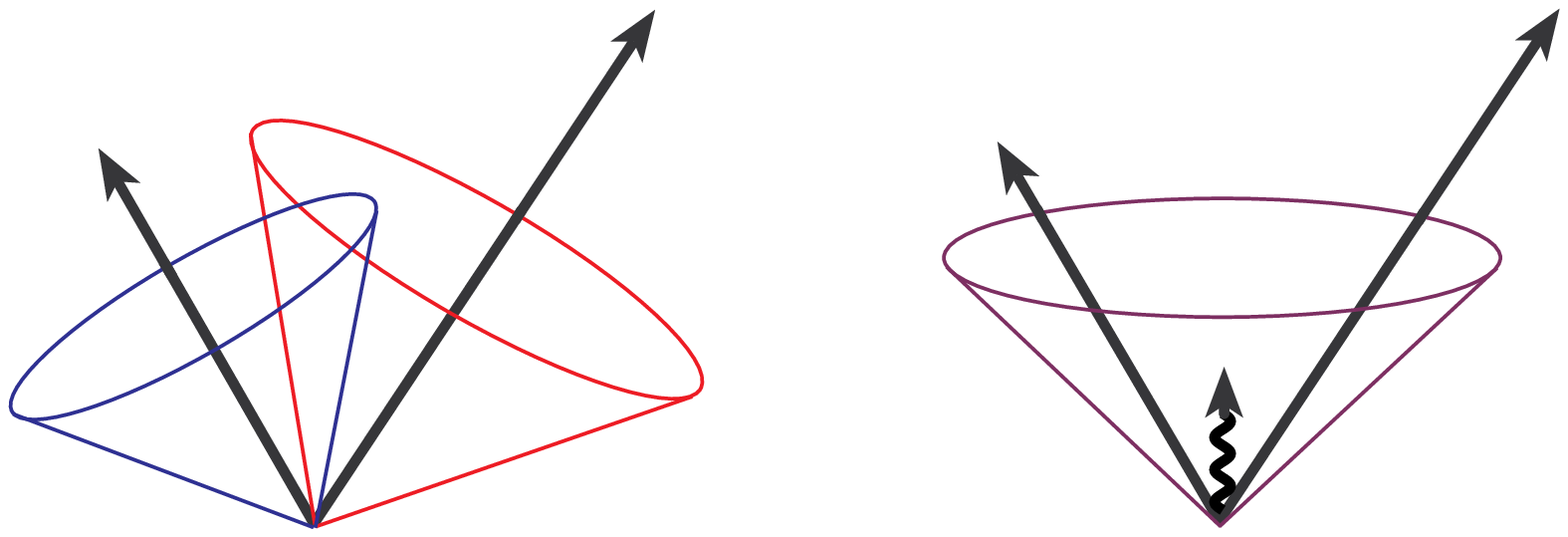}} \hspace{20pt} 
\subfloat[]{\label{FIGjetSafetyIllustrationFIG2}\includegraphics[trim=0mm 0mm 0mm 0mm,clip,width=.45\textwidth]{figures/jetReconstruction/collinearJetSafety.epsi}}
\caption{\label{FIGjetSafetyIllustrationFIG}An illustration of infrared \Subref{FIGjetSafetyIllustrationFIG1} 
and collinear \Subref{FIGjetSafetyIllustrationFIG2} sensitivity in jet finding.}
\end{center}
\end{figure} 

There are many jet algorithms which have been proposed over the years.
These include fixed sized cone algorithms as well as sequential recombination algorithms (cluster algorithms), 
which are based on event shape analysis~\cite{Buttar:2008jx}.
The term cone algorithm is applied to a wide range of jet algorithms which broadly aim at maximizing
\et inside a geometric circle in the \etaphi coordinates space.
The radius of the circle, $R$, is a key parameter of the algorithm.
On the other hand, clustering algorithms are based upon pair-wise clustering of the input constituents.
In general, an algorithm defines a distance measure between constituents, as well as
some condition upon which clustering into a jet should be terminated.

The \AKT~\cite{Cacciari:2008gp}, \KT~\cite{Ellis:1993tq} and \CAalg~\cite{Dokshitzer:1997in,Wobisch:1998wt}
clustering algorithms are used in this analysis.
The clustering algorithms begin by computing for each jet constituent, 
the quantity $d_{iB}$ and for each pair of constituents the quantity $d_{ij}$, defined as
\begin{eqnarray}
d_{iB} &=& k_{\mathrm{T}i}^{2p} \; , \\
d_{ij} &=& \min(k_{\mathrm{T}i}^{2p},k_{\mathrm{T}j}^{2p})\frac{(\Delta R)^2_{ij}}{R^2} \; , 
\label{eq:dist}
\end{eqnarray}
where
\begin{equation*}
(\Delta R)^2_{ij} = (y_i - y_j)^2 + (\phi_i - \phi_j)^2  \; .
\end{equation*}
The variables $R$ and $p$ are constants of the algorithm,
and $k_{\mathrm{T}i}$, $y_i$ and $\phi_i$ are respectively the transverse momentum, the rapidity and the
azimuth of the $i^{\mrm{th}}$~constituent.
The $d_{ij}$ parameter, represents the distance between
a pair of jet constituents, while $d_{iB}$ is the distance between a given constituent
and the beam. 

The algorithm proceeds by comparing for each constituent the different $d$ values. In the case that
the smallest entry is a $d_{ij}$, constituents $i$ and $j$ are
combined and the list is remade. If the smallest entry is $d_{iB}$, this constituent is considered a complete
jet and is removed from the list.

The variable $R$ sets the resolution at which jets are resolved from each other as compared to the beam.
For large values of $R$, the $d_{ij}$ are smaller, and thus more merging takes place before jets are complete.
The variable $p$ can also take different values; for the \KT algorithm $p = 1$,
for the \CAalg algorithm $p = 0$, and for the \AKT algorithm $p = -1$.

The nominal jet collection used for physics in the analysis uses the \AKT algorithm with size parameter, $R=0.6$,
reconstructed with the \fastjet package~\cite{fastjet}. Jets built with the
\KT algorithm are also used, but only as part of the jet energy calibration procedure, as discussed
in \autoref{chapJetAreaMethod}. The \CAalg algorithm is used in order to evaluate the systematic uncertainty of said calibration.

\section{Inputs to jet reconstruction\label{chapInputJetReconstruction}}
%
\subsection{Calorimeter jets}
%
The input to {\it calorimeter jets} are topological calorimeter clusters ({\it \topos})~\cite{EndcapTBelectronPion2002,TopoClusters}.
\Topos are groups of calorimeter cells that are designed to follow shower
development in the calorimeter.
A \topo is defined as having an energy equal to the energy sum of all the included calorimeter cells, zero mass and a reconstructed
direction calculated from the weighted averages of the pseudo-rapidities and azimuthal angles of its constituent cells.
The weight used in the averaging is the absolute cell energy. The positions of the cells
are relative to the nominal \ATLAS coordinate system.

\minisec{Clustering algorithm}
%
The \topo algorithm is designed to suppress calorimeter noise. The algorithm starts from a
\textit{seed} cell, whose signal-to-noise ($S/N$) ratio is above a threshold, $S/N = 4$.
Cells neighbouring the seed (or the cluster being formed) that have a signal-to-noise ratio, $S/N \geq 2$,
are included iteratively. Finally, all calorimeter cells neighbouring the formed \topo are added.

The noise is estimated as the absolute value of the energy deposited in the cell, divided by
the RMS of the energy distribution of the cell, measured in events triggered at random bunch crossings.
For data taken during 2010, cell-noise was dominated by electronic noise.
Due to the shortened bunch crossing interval and increased instantaneous luminosity during 2011 data-taking,
the noise increased. An additional component was added, originating from 
energy deposited in a given cell from previous collisions, but
inside the window of sensitivity of the calorimeters. This added energy is referred to as
\textit{\pu} (see \autoref{chapJetAreaMethod} for a detailed discussion).
Subsequently, different nominal noise thresholds, $\sigma_{\mathrm{noise}}$, are used
to reconstruct \topos in 2010 and in 2011,
\begin{equation}
        \renewcommand{\arraystretch}{1.75}
        \sigma_{\mathrm{noise}}= \left\{\begin{array}{ll}
                                 \sigma_{\mathrm{noise}}^{\mathrm{elc}} & \quad \mathrm{(2010)} \\
                                 \sqrt{\left(\sigma_{\mathrm{noise}}^{\mathrm{elc}}\right)^{2} + \left(\sigma_{\mathrm{noise}}^{\mathrm{pu}}\right)^{2}} &
                                 \quad \mathrm{(2011)}
                                 \end{array}\right. .
\label{eqTrigJetDistanceDef} \end{equation}
Here $\sigma_{\mathrm{noise}}^{\mathrm{elc}}$ is the electronic noise, and $\sigma_{\mathrm{noise}}^{\mathrm{pu}}$ is the noise from pile-up.
On an event-by-event basis, the magnitude of the latter term may vary, as the \pu is sensitive to
the instantaneous luminosity, characterized by the average number of interactions per
bunch crossing, \Mu, (see \autoref{chapInAndOutOfTimePU}).
A constant baseline is chosen for cluster reconstruction throughout 2011. This baseline corresponds to
the noise which results from eight additional \pu interactions,
$\mu = \mu^{\mrm{ref}} = 8$, and reflects the conditions under which the average effects of \pu in the 2011 data are minimal.

The change in time of the total nominal noise and its dependence on calorimeter pseudo-rapidity is observed 
by comparing \autorefs{FIGcaloNoise1} and~\ref{FIGcaloNoise2}.
\begin{figure}[htp]
\begin{center}
\subfloat[]{\label{FIGcaloNoise1}\includegraphics[trim=5mm 5mm 0mm 0mm,clip,width=.52\textwidth]{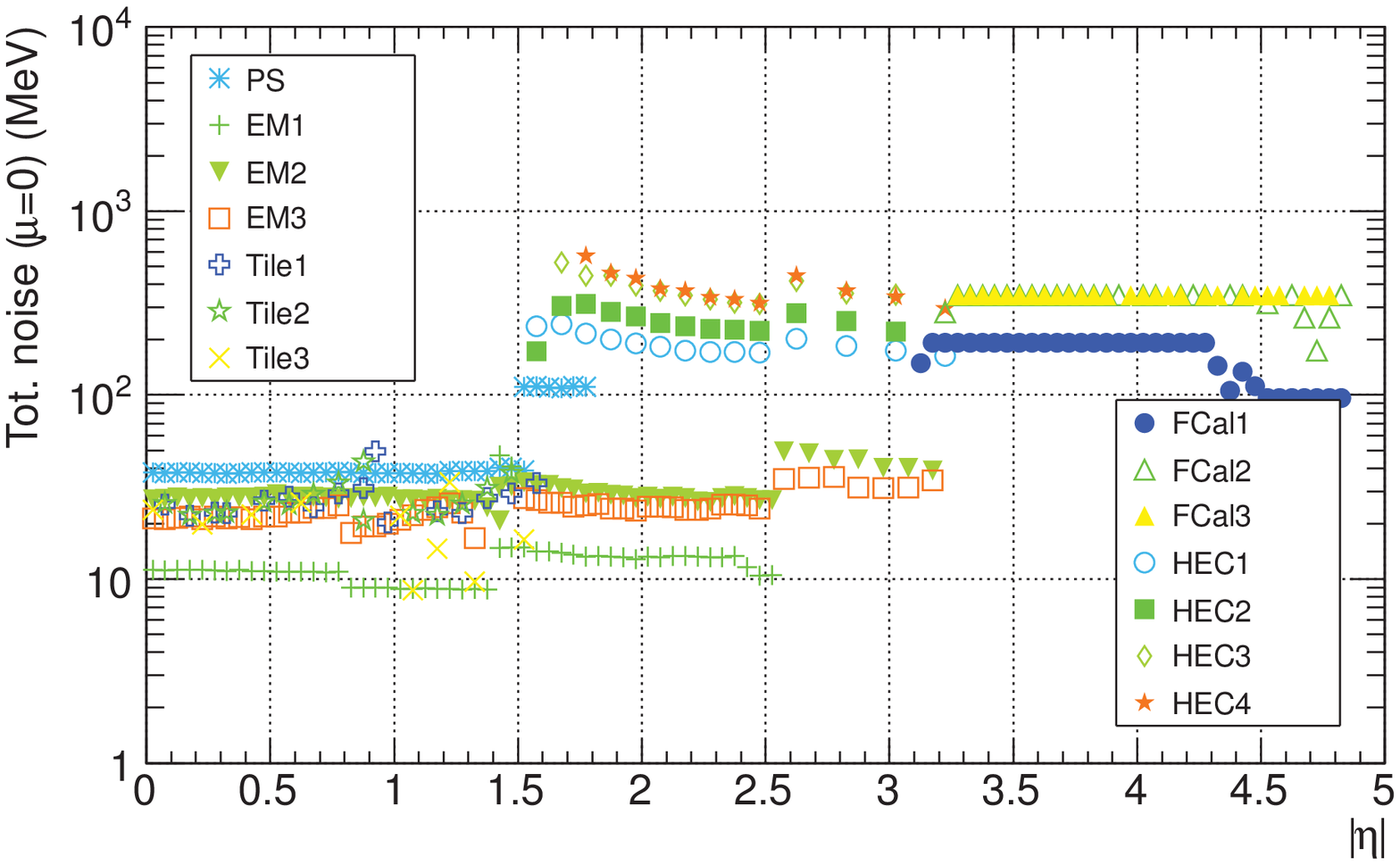}}
\subfloat[]{\label{FIGcaloNoise2}\includegraphics[trim=5mm 5mm 0mm 0mm,clip,width=.52\textwidth]{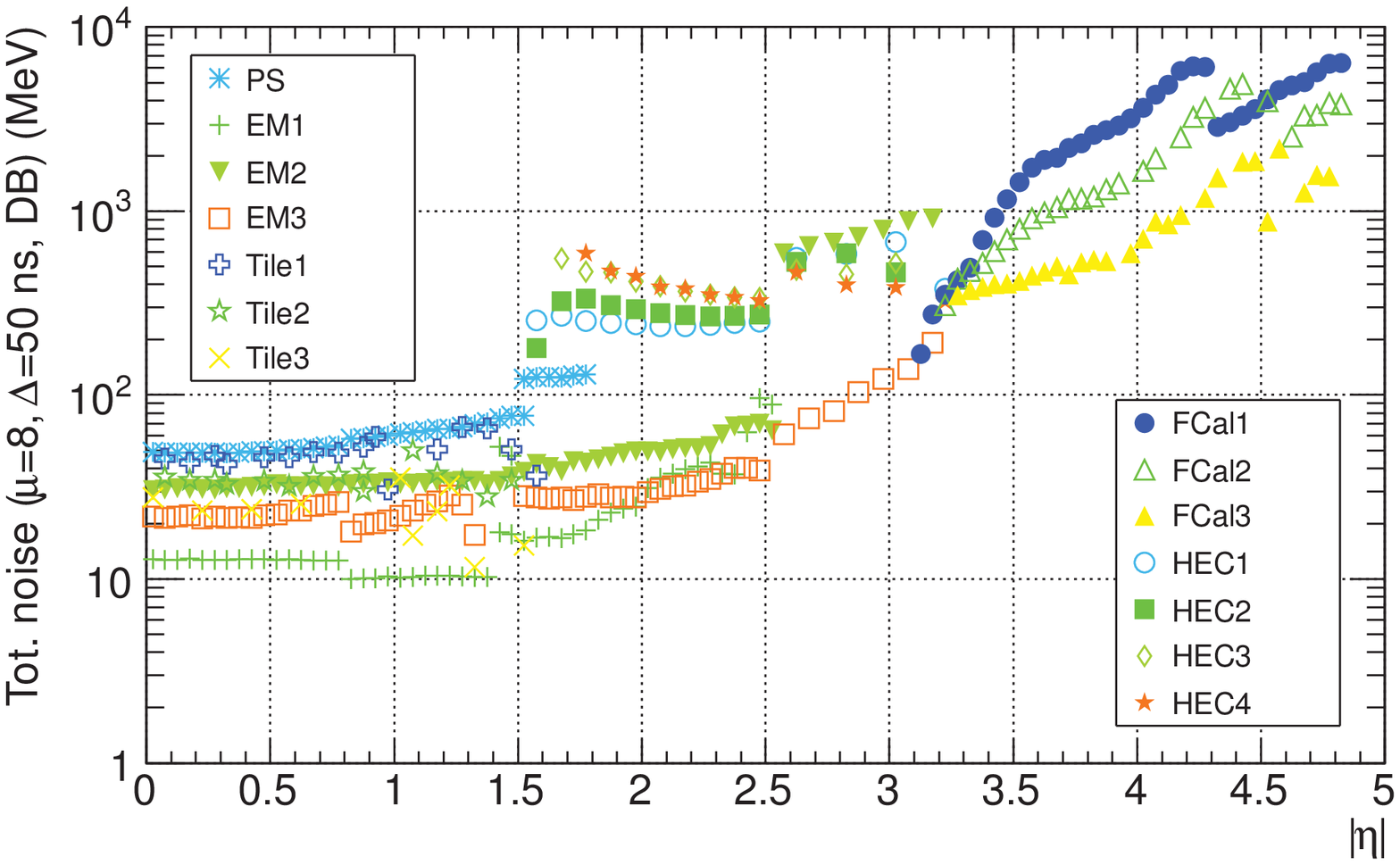}}
  \caption{\label{FIGcaloNoise}The energy-equivalent cell-noise in the \ATLAS calorimeters at the electromagnetic (\EM) scale, 
  as a function of pseudo-rapidity, \Eta. The magnitude of the noise represents the
  configurations used for cluster reconstruction in 2010 \Subref{FIGcaloNoise1} and in
  2011 \Subref{FIGcaloNoise2}. These respectively use $\mu = 0$ and $\mu = 8$ as baseline values for the average number of interactions.
  The various data points indicate the noise in different calorimeter elements;
  the pre-sampler (PS); the first three layers of the \EM calorimeter (EM1, EM2 and EM3);
  the first three layers of the Tile calorimeter (Tile1, Tile2 and Tile3);
  the first three layers of the forward FCal calorimeter (FCal1, FCal2 and FCal3);
  the first four  layers of the hadronic end-cap calorimeter (HEC1, HEC2 and HEC3). (Figures are taken from~\protect\cite{ATLAS-CONF-2012-064}.)
  }
\end{center}
\end{figure} 
The former shows the energy-equivalent cell-noise in different calorimeter regions in 2010, representing
the baseline electronic noise; the latter shows
the respective cell-noise in 2011, for which \pu contributes as well.
In most calorimeter regions, the \pu induced noise is smaller or of the same magnitude as the electronic noise.
The exception is the forward calorimeters, where $\sigma_{\mathrm{noise}}^{\mathrm{pu}} \gg \sigma_{\mathrm{noise}}^{\mathrm{elec}}$.

The \topo algorithm also includes a splitting step in order to
optimise the separation of showers from different close-by particles:
All cells in a \topo are searched for local maxima in terms of energy content with
a threshold of 500\MeV. This means that the selected calorimeter cell has to be more energetic than any of its neighbours. 
The local maxima are then used as seeds for a new iteration of topological clustering, which splits
the original cluster into more \topos.

\minisec{Energy scale of clusters}
%
\Topos are initially constructed from energy at the electromagnetic (\EM) scale.
The \EM scale correctly reconstructs the energy deposited by particles
in an electromagnetic shower in the calorimeter.
This energy scale is established using test-beam measurements for
electrons in the barrel~\cite{ctb2004electronseoverp,ctb2004electrons, LArTB02uniformity, LArTB02linearity,Tile2002}
and the endcap calorimeters~\cite{Pinfold:2008zzb,EndcapTBelectronPion2002}.
The absolute calorimeter response to energy deposited via electromagnetic processes
was validated in the hadronic calorimeters using muons, both from 
test-beams~\cite{Tile2002,LArTB02muons} and produced \insitu by cosmic rays~\cite{TileReadiness}.
The energy scale of the electromagnetic calorimeters
is corrected using the invariant mass of $Z$ bosons produced \insitu 
in proton-proton collisions ($Z \rightarrow e^+ e^-$ events)~\cite{Atlaselectronpaper}.

An additional calibration step may also be used, where so called \textit{Local Calibration Weights} (\LC)
are applied to \topos, bringing the clusters to the \LC calibration scale~\cite{citeulike:4403592}.
Starting at the \EM scale, \topos are classified as either \textit{\EM-like} or \textit{hadron-like},
using cluster shape moments.
Clusters classified as \EM-like are kept at the original energy scale. On the other hand,
hadron-like \topos are subject to recalibration. In the first step this entails a cell weighting procedure,
which aims to compensate for the lower response of the calorimeter to hadronic deposits.
Next, out-of-cluster corrections are applied. These try to account for lost energy deposited in calorimeter cells
outside of reconstructed clusters. Finally, dead material
corrections are applied, accounting for energy deposits outside of active
calorimeter volumes, \eg in the cryostat, the magnetic coil and calorimeter inter-modular cracks.

Calorimeter jets are built using either \EM or \LC scale \topos. In either case, further corrections need
to be applied to calibrate jets to the particle-level. This final step is referred to
as the \textit{jet energy scale} (\JES) calibration.

\subsection{Other types of jets}
%

\minisec{Track jets}
%
Jets can also be built using inner detector tracks as inputs to the jet finding
algorithm. Track-based jets, called \textit{track jets}, are reconstructed using the \AKT algorithm with size parameter, $R=0.6$.
Track jets are required to be composed of at least two tracks; tracks are required to have
transverse momentum above~500\MeV, at least one pixel detector hit, at least six hits in the silicon-strip detector,
and impact parameters in the transverse plane and in the $z$-direction, both smaller than~1.5~mm. In order to be fully contained
in the tracking region, track jets are required to have \etaLower{1.9}. Track jets are also required to
have transverse momentum larger than~4\GeV.

\minisec{Truth jets}
%
Monte Carlo simulated particle jets are referred to as \textit{truth jets}. They
are defined as those built from stable interacting particles with a lifetime longer 
than 10~ps in the Monte Carlo event record (excluding muons and neutrinos) that have not been passed through 
the simulation of the \ATLAS detector. Truth jets are built using the \AKT algorithm with size parameter $R=0.6$,
and are required to have transverse momentum larger than 10\GeV.

\section{Jet energy calibration\label{chapJetEnergyCalibration}}
%
The jet energy calibration relates the jet energy measured with the 
ATLAS calorimeter to the true energy of the corresponding jet
of stable particles entering the detector.
The jet calibration corrects for the following detector effects that affect the jet energy measurement;
\begin{list}{-}{}
\item
  \headFont{calorimeter non-compensation -} partial measurement of the energy deposited by hadrons;
\item
  \headFont{dead material -} energy losses in inactive regions of the detector;
\item
  \headFont{leakage -} energy of particles reaching outside the calorimeters;
\item
  \headFont{out of calorimeter jet cone -}  energy deposits of particles which are not included in the reconstructed jet,
  but were part of the corresponding truth jet and entered the detector;
\item
  \headFont{noise thresholds and particle reconstruction efficiency -} 
       signal losses in calorimeter clustering and jet reconstruction.
\end{list}

Jets reconstructed in the calorimeter system are formed from calorimeter energy depositions
reconstructed at either the \EM or the \LC scale, as described above.
The correction for the lower response to hadrons is solely based on the topology of the energy 
depositions observed in the calorimeter.
The measured jet energy is corrected on average, using:
\begin{equation}
  E^{\rm calib} = E^{\rm det} \times \mathcal{F}^{{\rm calib}}_{\mrm{\Scale}}\left( E^{\rm det}_{\mrm{\Scale}},\Eta\right)
  \;, \quad {\rm with} \;\;\; E^{\rm det}_{\mrm{\Scale}} = E_{\mrm{\Scale}} - \mathcal{O}_{\mrm{fst}}^{\Scale}
  \;, \;\; {\rm for} \;\;\; \mrm{\Scale} = \EM \;,\; \LC
  \;.
\label{eqCalibJetEnergy} \end{equation}
Here $E_{\mrm{\Scale}}$ is the calorimeter energy measured at the \EM or \LC scales, denoted collectively by \Scale. 
The variable $E_{\rm calib}$ is the hadron-level calibrated jet energy,
and $\mathcal{F}^{{\rm calib}}_{\mrm{\Scale}}$ is a calibration function.
The latter depends on the measured jet energy, and is evaluated in small jet-pseudo-rapidity, \Eta, regions. 
The baseline \PYTHIA MC samples described in \autoref{chapMCsimulation} are used to derive $\mathcal{F}^{{\rm calib}}_{\mrm{\Scale}}$;
the procedure is explained in the following.

The variable $E^{\rm det}_{\mrm{\Scale}}$ denotes the \EM- or \LC-level energy after the contribution of \pu
(additional multiple proton-proton interactions) has been subtracted. The \pu energy is expressed
through the function $\mathcal{O}_{\mrm{fst}}^{\Scale}$, called the \textit{jet offset correction}. 
The offset correction in 2010 and in 2011 depends on the number of \pp collisions which occur within the same bunch
crossing as the collision of interest. This is referred to as \textit{\itpu} and is estimated by the number of primary vertices, \Npv,
in a given event.
In addition, $\mathcal{O}_{\mrm{fst}}^{\Scale}$ exhibits a strong rapidity dependence,
due to the varying calorimeter response in \Eta, and to the rapidity distribution of the \pu interactions.
The offset correction is parametrized separately for \EM and for \LC jets, based in each case on jet constituents
which are calibrated in the respective energy scale.

The offset correction performs well for data taken in 2010~\cite{Aad:2011he}.
For the case of the 2011 data, the nature of \pu becomes more complicated, as
interactions in bunch crossings proceeding and following the event of interest affect
the signal in the calorimeter; this effect is called \textit{\otpu}. 
The offset correction is modified in 2011 in order to account for
this effect, but the performance is nonetheless degraded \wrt 2010.
A detailed discussion of \otpu and the exact functional form of $\mathcal{O}_{\mrm{fst}}^{\Scale}$
are given in \autoref{chappuInAtlas}.
An alternative to the offset correction using a data-driven method, referred to as the \textit{jet area/median method},
is introduced in \autoref{chapJetAreaMethod}. 
The offset correction is used for the calibration of jets in the 2010 data in this analysis;
the jet median approach is used as the nominal \pu correction for the 2011 data.

The two calibrations schemes, which respectively use jets constructed from
either \EM- or \LC-calibrated \topos, are referred to as \EMJES and \LCJES.
For data taken during 2010, only the simpler \EMJES calibration is available for this analysis. For the 2011 data,
the nominal jet collection is calibrated using \LCJES.
The calibration schemes include several steps, described in the following:
\begin{list}{-}{}
\mynobreakpar
\item jet origin correction;
\item jet energy correction;
\item jet pseudo-rapidity correction;
\item residual \insitu calibration (2011 data only).
\end{list}

\minisec{\label{ENUMjetCalibSteps1} Jet origin correction}
%
Calorimeter jets are reconstructed using the geometrical center of the \ATLAS detector as reference to calculate 
the direction of jets and their constituents.
The jet four-momentum is corrected for each event such that
the direction of each \topo points back to the primary (highest-\pt) reconstructed vertex.
The kinematic observables of each \topo are recalculated using 
the vector from the primary vertex to the \topo centroid
as its direction. The raw jet four-momentum is thereafter redefined as the
vector sum of the \topo four-momenta.
This correction improves the angular resolution and results in a small
improvement ($< 1\%$) in the jet \pt response. The jet energy is unaffected.

\minisec{\label{ENUMjetCalibSteps2} Jet energy correction}
%
The principle step of the \EMJES and of the \LCJES jet calibrations restores the reconstructed jet energy
to the energy of the corresponding MC truth jet. Since \pu effects have already been
accounted for, the MC samples used to derive the calibration do not include multiple \pp interactions.

The calibration is derived using isolated calorimeter jets that have a matching isolated truth jet within $\Delta R = 0.3$.
Distance is defined in $\eta$,$\phi$ space as
\begin{equation}
  \Delta R = \sqrt{\left(\phi^{\mrm{rec}}-\phi^{\mrm{truth}}\right)^{2}+\left(\eta^{\mrm{rec}}-\eta^{\mrm{truth}}\right)^{2}} \;,
\label{eqTwoJetDistanceDef} \end{equation}
where $\phi^{\mrm{truth}}$ ($\eta^{\mrm{truth}}$) and $\phi^{\mrm{rec}}$ ($\eta^{\mrm{rec}}$) are respectively
the azimuthal angles (pseudo-rapidities) of truth and reconstructed jets.
An isolated jet is defined as one having no other jet with
${\pt > 7\GeV}$ within ${\Delta R = 2.5\cdot R}$, 
where $R$ is the size parameter of the jet algorithm.

The final jet energy scale calibration is parametrised as a function of
detector-level pseudo-rapidity, $\eta_{\mrm{det}}$,\footnote{ Here, 
pseudo-rapidity refers to that of the original reconstructed jet before the origin correction.}
and of jet energy at the detector-level energy scales,
$E_{\mrm{\Scale}}$.
In the following, as above, \Scale stands for either the \EM or the \LC energy scales.
The detector-level jet energy-response,
\begin{equation}
\mathcal{R}_{\mrm{\Scale}} = E_{\mrm{\Scale}} / E_{\rm  truth} \;,
\end{equation}
is measured for each pair of calorimeter and truth jets in bins of truth jet energy,
$E_{\rm truth}$, and of $\eta_{\mrm{det}}$.
For each $\left(E_{\rm truth},\eta_{\mrm{det}}\right)$-bin, the averaged jet-response, 
$\left< \mathcal{R}_{\mrm{\Scale}} \right>$, is defined as the peak position of a Gaussian fit to the 
$E_{\mrm{\Scale}}/E_{\rm  truth}$ distribution. 
In addition, in the same $\left(E_{\rm truth},\eta_{\mrm{det}}\right)$-bin, the average jet
energy-response, $\left< E_{\mrm{\Scale}} \right>$, is derived from the mean of the distribution of $E_{\mrm{\Scale}}$. 

For a given $\eta_{\mrm{det}}$-bin, $k$, 
the jet energy-response calibration function, $\mathcal{F}^{(k)}_{{\rm calib}}(E_{\mrm{\Scale}})$,
is obtained using a fit of the
$\left(\left< E_{\mrm{\Scale}} \right>_j,\left<\mathcal{R}_{\mrm{\Scale}}\right>_j\right)$ values for each $E_{\rm truth}$-bin $j$.
The fitting function is parameterised as
\begin{equation}
  \mathcal{F}^{(k)}_{{\rm calib}}( E_{\mrm{\Scale}}) =
  \sum_{i=0}^{N_{\rm max}} a_i
  \left(\ln E_{\mrm{\Scale}}\right)^{(i)},
  \label{eq:Rjet}
\end{equation}
where $a_i$ are free parameters, and $N_{\rm max}$ is chosen
between $1$ and $6$, depending on the goodness of the fit.
\index{jet calibration, numerical inversion}

The final jet energy scale correction that relates the measured calorimeter jet energy 
to the true energy, is then defined as $\mathcal{F}^{{\rm calib}}_{\mrm{\Scale}}$ in the following:
\begin{equation}
  E_{\mrm{\ttt{S+JES}}} = \frac{E_{\mrm{\Scale}}}{\mathcal{F}_{\rm calib}^{\mrm{\Scale}}( E_{\mrm{\Scale}} ) |^{k}_{\eta_{\mrm{det}}}}
  =  E_{\mrm{\Scale}} \times \mathcal{F}^{{\rm calib}}_{\mrm{\Scale}}\;,
  \label{eq:JES}
\end{equation}
where $\mathcal{F}_{\rm calib}(E_{\mrm{\Scale}} )|_{\eta_{\mrm{det}}}^{k}$ is
the jet-response calibration function for the relevant $\eta_{\mrm{det}}$-bin~$k$.

As an example, the average jet energy scale correction for \EM jets in 2010, $\left<\mathcal{F}^{{\rm calib}}_{\mrm{\EM}}\right>$, 
is shown as a function of calibrated jet \pt 
for three $\eta$-intervals in \autoref{FIGjesCorrection1}. 
The response of 2010 \EMJES jets, $\mathcal{R}_{\mrm{\EM}}$, is shown for various
energy and $\eta_{\mrm{det}}$-bins in \autoref{FIGjesCorrection2}.
\begin{figure}[htp]
\begin{center}
\subfloat[]{\label{FIGjesCorrection1}\includegraphics[trim=5mm 5mm 0mm 0mm,clip,width=.52\textwidth]{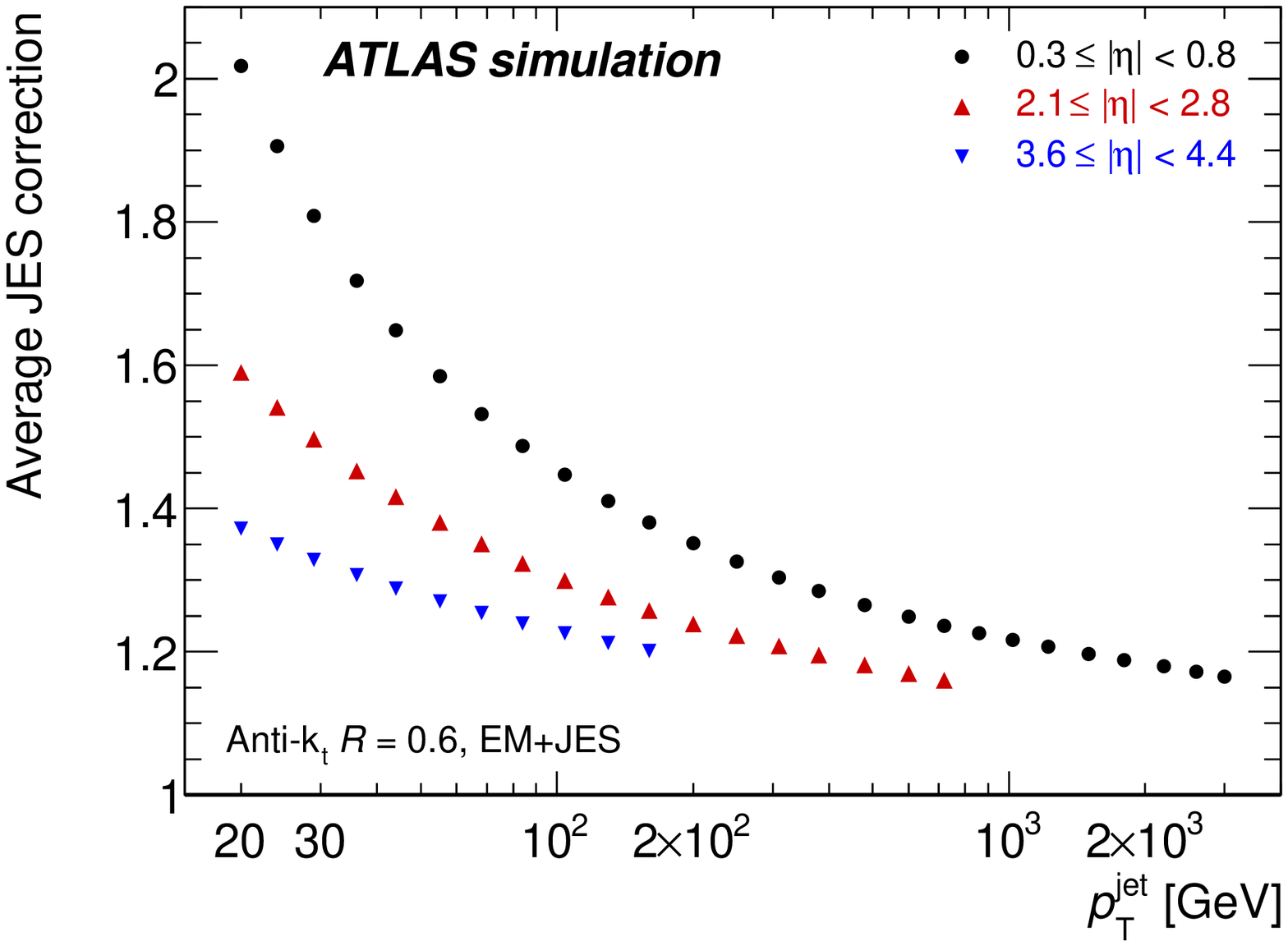}}
\subfloat[]{\label{FIGjesCorrection2}\includegraphics[trim=5mm 5mm 0mm 0mm,clip,width=.52\textwidth]{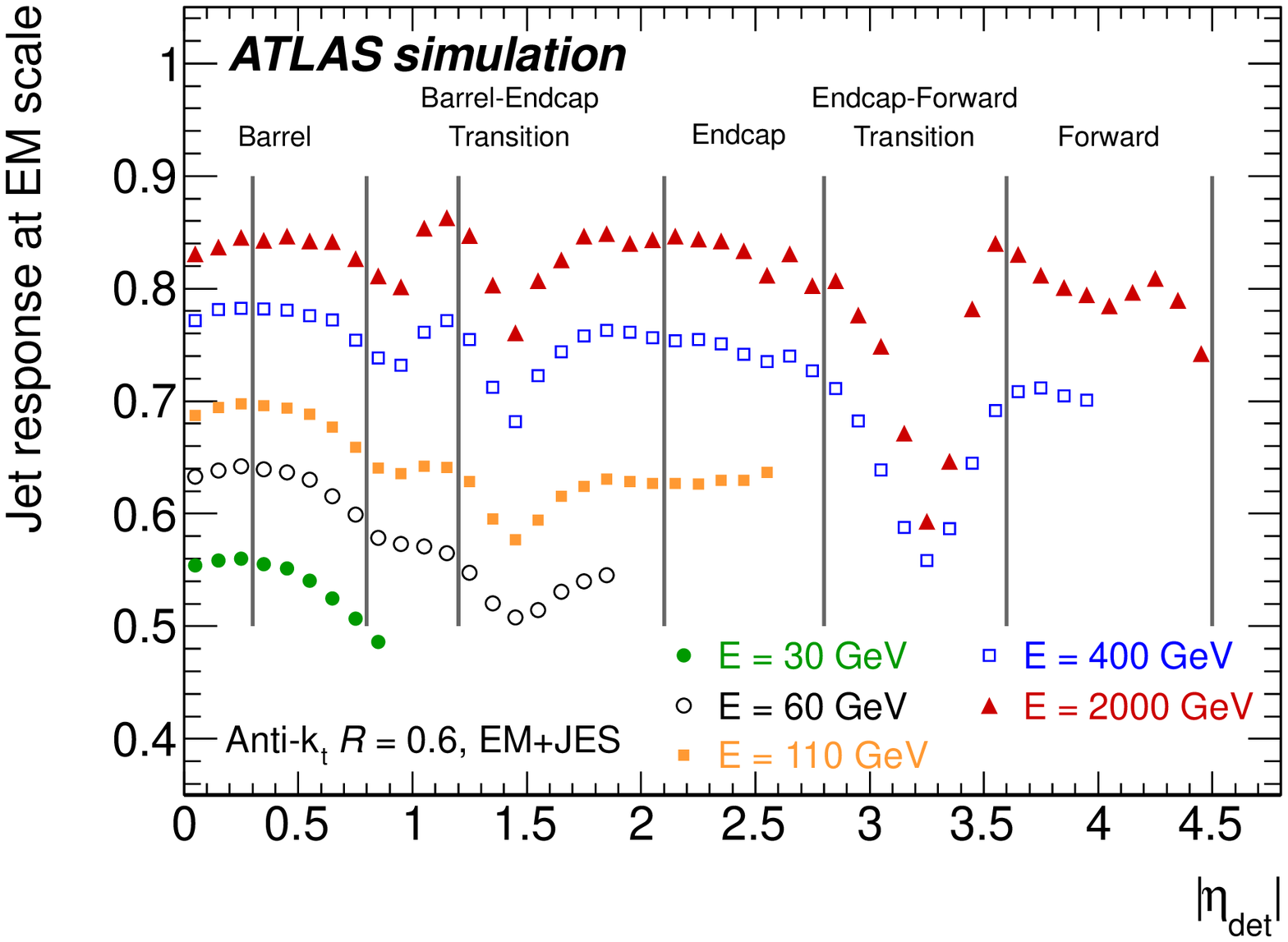}}
  \caption{\label{FIGjesCorrection}\Subref{FIGjesCorrection1} Average jet energy scale correction for \EMJES jets
    as a function of the calibrated jet transverse momentum, $p_{\mrm{T}}^{\mrm{jet}}$, for three pseudo-rapidity, \Eta, regions,
    as indicated in the figure.\\
    \Subref{FIGjesCorrection2} Average simulated jet-response at the \EM scale in bins of detector
    pseudo-rapidity, $\eta_{\mrm{det}}$. Different jet energies, $E$, are represented (using \EMJES jets) as indicated in the figure.
    The parallel lines mark the regions over which the \JES correction is evaluated.
    The inverse of the response shown in each bin is equal to the average jet energy scale correction.\\
    (Figures are taken from~\protect\cite{Aad:2011he}.)
  }
\end{center}
\end{figure} 
As the transverse momentum of jets increases, the response and subsequent \JES correction factor decrease.
The corrections is also highly rapidity-dependant, due in part to the changing structure of calorimeter elements; in transition
regions of the calorimeter (around $|\eta| = 1.5,3.3$) the response is significantly higher, due to energy loss
in dead material or through leakage. Larger correction factors are therefore required in these regions.
Corresponding figures for 2011 \LCJES jets may be found in~\cite{Adomeit:1475233}.

\minisec{\label{ENUMjetCalibSteps3} Jet pseudo-rapidity correction}
%
After the jet origin and energy corrections,
the origin-corrected pseudo-rapidity of jets is further corrected for a bias due to
poorly instrumented regions of the calorimeter.
In these regions \topos are reconstructed with a lower energy 
with respect to better instrumented regions (see \autoref{FIGjesCorrection2}). 
This causes the jet direction to be biased towards the better instrumented calorimeter regions.

The $\eta$-correction is derived as the average difference between the pseudo-rapidity
of reconstructed jets and their truth jet counterparts. It is parametrized according
to the energy and detector-level pseudo-rapidity of reconstructed jets,
$\Delta\eta=\eta_{\rm truth}-\eta_{\rm origin}$.
The correction is very small $(< 0.01)$ for most regions of the calorimeter, and up to five times
larger in the transition regions. 

\minisec{\label{ENUMjetCalibSteps4} Residual \insitu calibration (2011 data only)}
%
Following the pseudo-rapidity correction, the transverse momentum of jets in the 2011 data is compared to that in MC
using \insitu techniques. The latter exploit the balance between the transverse momentum of a jet, \pt, and that of
a reference object, $p_{\mrm{t}}^{\mrm{ref}}$~\cite{Adomeit:1475233}. The ratio,
\begin{equation}
  f_{\mrm{\insitu}} = \frac{  \left<p_{\mrm{t}}/p_{\mrm{t}}^{\mrm{ref}}\right>_{\mrm{data}}  }{ \left<p_{\mrm{t}}/p_{\mrm{t}}^{\mrm{ref}}\right>_{\mrm{MC}} } \;,
\end{equation}
called the \textit{residual \insitu \JES correction}, is used on jets measured in data. It is derived using the following methods:

\begin{list}{-}{}
\mynobreakpar
\item
  the \pt of jets within $0.8 < \left|\eta\right| < 4.5$ is equalized to the \pt of jets within $\left|\eta\right| < 0.8$,
  exploiting the \pt balance between central and forward jets in events with only two high \pt jets;
\item 
  an \insitu \JES correction for jets within $\left|\eta\right| < 1.2$ is derived using the \pt of a photon
  or a $Z$~boson (decaying to $e^{+}e^{-}$ or $\mu^{+}\mu^{-}$) as reference;
\item 
  events where a system of low-\pt jets recoils against a high-\pt jet are used to calibrate jets in
  the\TeV regime. The low- and high-\pt jets are required to respectively be within $\left|\eta\right| < 2.8$ and
  $\left|\eta\right| < 1.2$. 
\end{list}

\section{Jet quality selection\label{chapJetQualitySelection}}
%
%
Jets at high transverse momenta produced in \pp collisions must be distinguished 
from background jets not originating from hard scattering events.
The main sources of background are listed in the following:
\begin{list}{-}{}
\item
  large calorimeter noise;
\item 
  beam-gas events, where one proton of the beam collided with the residual gas within the beam pipe;
\item
  beam-halo events, \eg caused by interactions in the tertiary collimators
  in the beam-line far away from the \ATLAS detector;
\item
  cosmic ray muons overlapping in-time with collision events.
\end{list}
These backgrounds are divided into two categories, calorimeter noise and non-collision interactions.

\headFont{Calorimeter noise -}
two types of calorimeter noise are addressed, sporadic noise bursts and coherent noise.
Sporadic noise bursts in the hadronic endcap calorimeter (HEC) commonly result in a single noisy calorimeter 
cell, which contributes almost all of the energy of a jet.
Such jets are therefore rejected if they have high
HEC energy fractions, denoted by $f_{\mrm{HEC}}$.
The signal shape quality, $\mathcal{S}_{\mrm{HEC}}$, may also be used for rejection, the latter being
a measure of the pulse shape of a calorimeter cell compared to nominal conditions.
Due to the capacitive coupling between channels,
neighbouring calorimeter cells around the noise burst have an apparent negative energy, denoted by $E_{\rm neg}$.
A hight value of $E_{\rm neg}$ is therefore used to distinguish jets which originate in noise bursts.
On rare occasions, coherent noise in the electromagnetic calorimeter develops. 
Fake jets arising from this background are characterised 
by a large \EM energy fraction, $f_{\rm \EM}$, which is
the ratio of the energy deposited in the \EM calorimeter to the total energy. Similar to the case of 
noise bursts in the HEC, a large fraction of calorimeter cells exhibit
poor signal shape quality, $\mathcal{S}_{\mrm{EM}}$, in such cases.

\headFont{Non-collision backgrounds -}
cosmic rays or non-collision interactions are likely to induce events
where jet candidates are not in-time with the beam collision. 
A cut on the jet-time, $t_{\rm jet}$, may therefore be applied to reject such jets.
Jet-time is reconstructed from the energy deposition in the calorimeter 
by weighting the reconstructed time of calorimeter cells forming the jet
with the square of the cell energy. 
The calorimeter time is defined with respect to the event time (recorded by the trigger). 
A cut on $f_{\rm \EM}$ is applied to make sure 
that jets have some energy deposited in the calorimeter layer closest to the interaction region, 
as expected for a jet originating from the nominal interaction point. 
Since real jets are expected to have tracks, 
the $f_{\mrm{EM}}$ cut may be applied together with a cut on the jet charged fraction, $f_{\rm ch}$, 
defined as the ratio of the scalar sum of the \pt of the tracks associated to the jet, divided by the jet \pt.
The jet charged fraction cut is, naturally, only applicable for jets within the acceptance of the tracker. 
A cut on the maximum energy fraction in any single calorimeter layer, $f_{\rm max}$, 
is applied to further reject non-collision background.

Two sets of quality criteria are defined, denoted as \ttt{Loose} and \ttt{Medium} selection.
These incorporate requirements on the rejection variables defined above, and are specified in detail
in \autoref{chapJetReconstructionApp}, \autoref{TBLjetSelectionCriteriaApp}.
For the 2010 data, the tighter, \ttt{Medium} selection, is required. Since these criteria are not fully efficient, an efficiency correction
is applied to jets. The selection efficiency for jets in two pseudo-rapidity regions is shown in \autoref{FIGjetQualityCriteriaEff}
for illustration.
\begin{figure}[htp]
\begin{center}
\subfloat[]{\label{FIGjetQualityCriteriaEff1}\includegraphics[trim=5mm 5mm 0mm 0mm,clip,width=.52\textwidth]{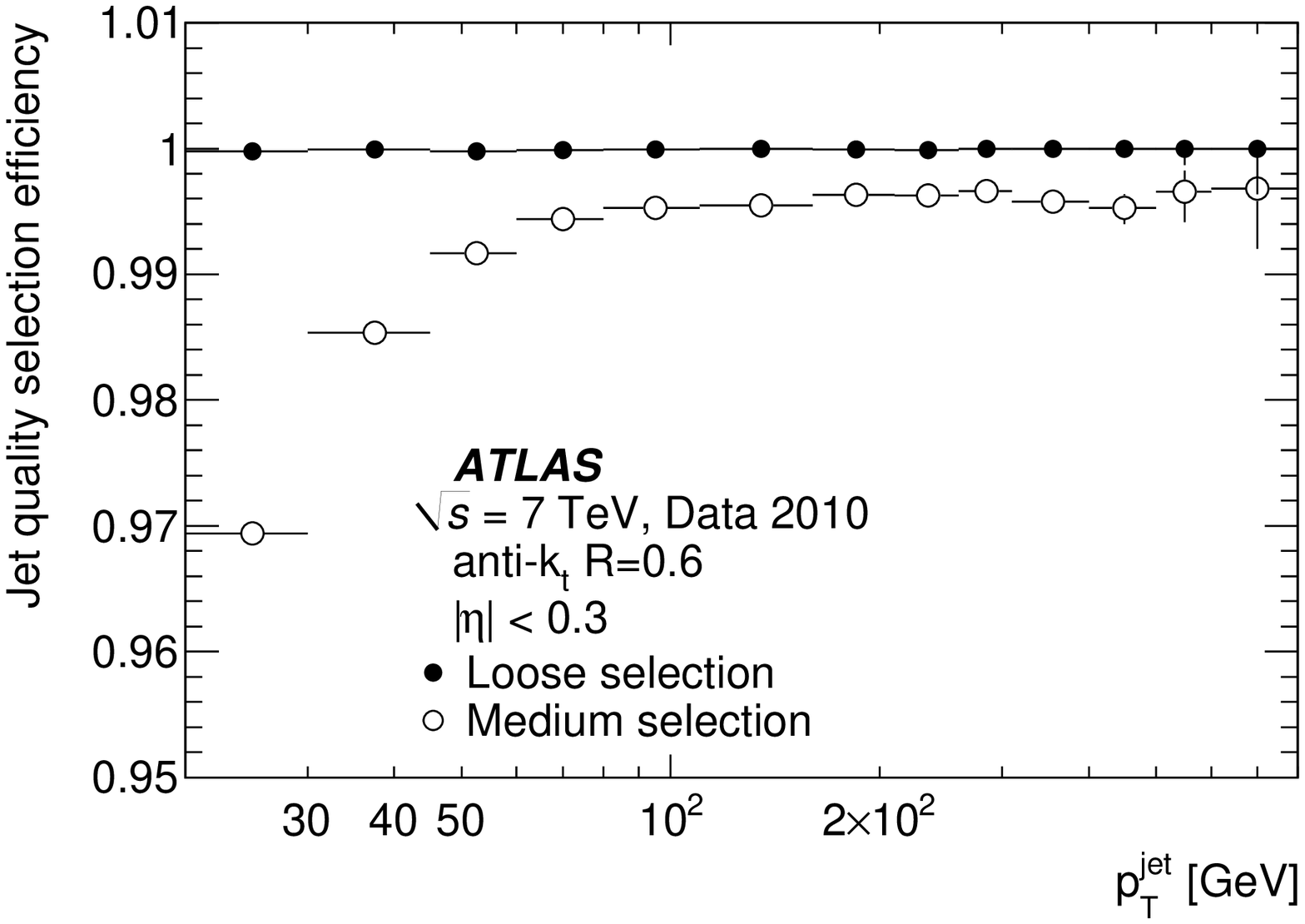}}
\subfloat[]{\label{FIGjetQualityCriteriaEff2}\includegraphics[trim=5mm 5mm 0mm 0mm,clip,width=.52\textwidth]{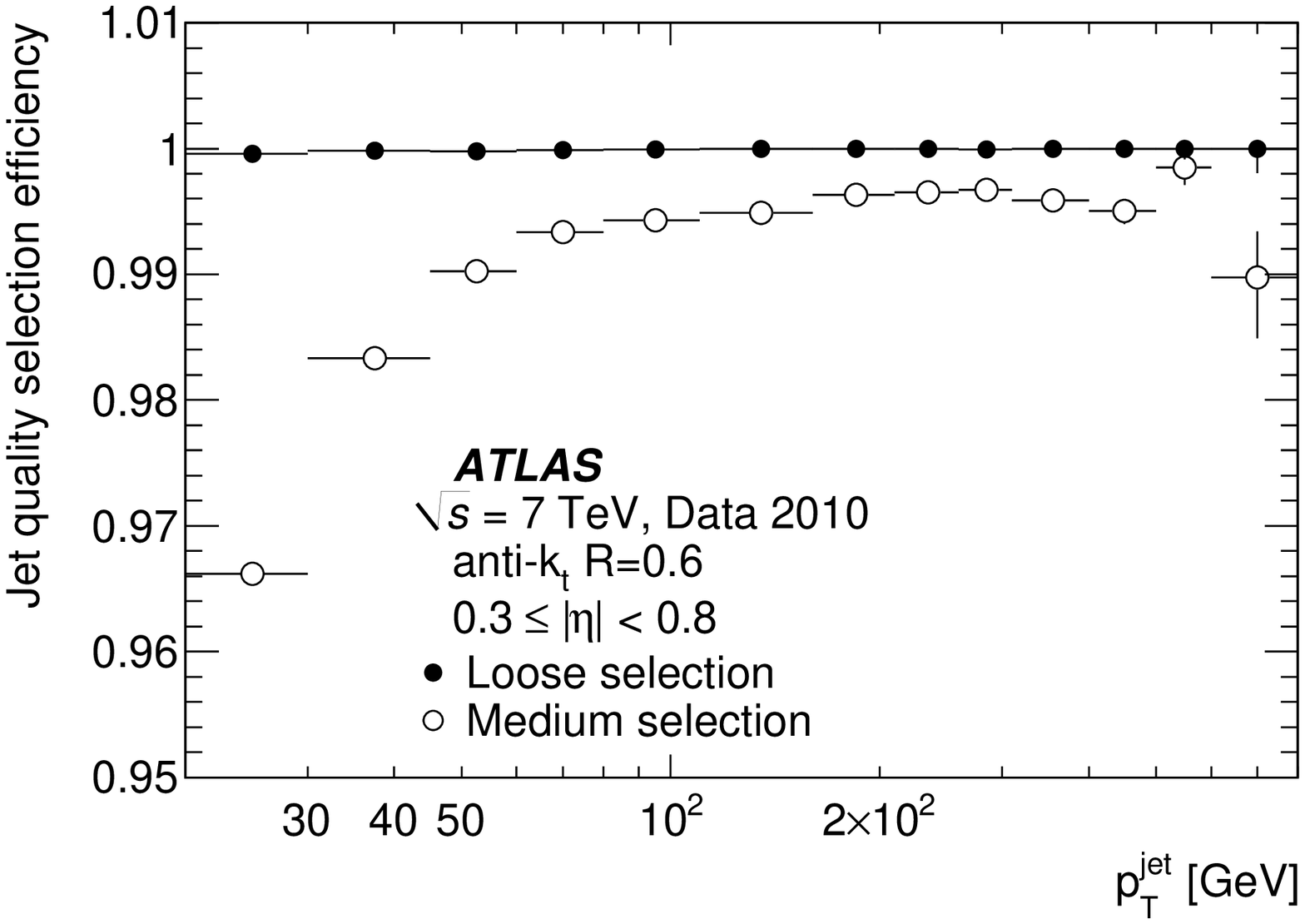}}
  \captionof{figure}{\label{FIGjetQualityCriteriaEff} Jet quality selection efficiency for \AKT jets with size
  parameter, $R = 0.6$, measured 
  as a function of the transverse momentum of jets, $p_{\mrm{T}}^{\mrm{jet}}$, in two pseudo-rapidity, \Eta, regions.
  The \ttt{Loose} and \ttt{Medium} selection criteria used for data taken during 2010 are represented, as indicated in the figures.
  (Figures taken from~\protect\cite{Aad:2011he}.)
  }
\end{center}
\end{figure} 
In general, for transverse momenta larger than $\sim50\GeV$, the efficiency is above 99\% across all rapidities.
For lower jet \pt, efficiencies range between 96-98\% within \etaLower{2.1}, and are above 99\% otherwise.
Further details are available in~\protect\cite{Aad:2011he}.
For the 2011 data, due to improved understanding of the calorimeter,
a reduced version of the 2010 \ttt{Loose} selection is used. The efficiency is above 99.8\%
for jets with \ptHigher{30} across all rapidities.

\section{Systematic uncertainties on the kinematic properties of jets\label{chapSystematicUncertaintiesJets}}
%
\subsection{Jet energy scale uncertainties\label{chapSystematicUncertaintiesJES}}
%
The systematic uncertainty associated with the jet energy scale (\JES) is the primary source of uncertainty in the analysis.
It has been determined for the 2010 and 2011 datasets
in~\cite{Begel:1333972} and in~\cite{Adomeit:1475233}, using integrated luminosities of
$38~\mrm{pb}^{-1}$ and $4.7~\mrm{fb}^{-1}$, respectively.
In total seven sources of uncertainty on the jet energy scale are considered in~2010 and thirteen in~2011.

The sources of uncertainty on the~2010 measurement are the following:
\begin{list}{-}{}
\mynobreakpar
    \item \headFont{single hadron response - }
      uncertainty associated with the response of a single particle entering the calorimeter. Discrepancies
      may arise due to the limited knowledge of the exact detector geometry;
      due to the presence of additional dead material; and due to
      the modelling of the exact way particles interact in the detector;
    \item \headFont{cluster thresholds - }
      uncertainty associated with the thresholds for reconstructing \topos.
      The clustering algorithm is based on the signal-to-noise ratio of calorimeter cells.
      Discrepancies between the simulated noise and the real noise, or changes in time of the noise in data,
      can lead to differences in the cluster shapes and to the presence of fake \topos;
    \item \headFont{Perugia 2010 and Alpgen+Herwig+Jimmy - }
      uncertainty associated with the modelling of fragmentation and the underlying event, or with
      other choices in the event modelling of the MC generator. The response predicted by 
      the nominal \pythia generator are compared to the \pythia Perugia 2010 tune and to
      \alpgen, coupled to \herwig and \jimmy;
    \item \headFont{intercalibration - }
      uncertainty associated with the rapidity-intercalibration method, in which dijet events are used \insitu
      to measure the response in two \Eta regions in the calorimeter. The measurement is done in different rapidity
      intervals simultaneously, by minimizing a response matrix.
      The uncertainty is estimated by comparing the response with that measured \insitu in events
      in which one of the jets is constrained to be central;
    \item \headFont{relative non-closure - }
      uncertainty associated with the non-closure of the energy of jets in MC following the \JES calibration;
    \item \headFont{\itpu\;- }
      uncertainty associated with the simulation and subtraction of \itpu.
\end{list}

For the \LCJES calibration scheme, which is used in this analysis for the 2011 dataset,
the baseline \JES uncertainty is estimated using a combination of \insitu techniques.
In total, the different sources of uncertainty coming from the \insitu techniques
in 2011 data amount to~60 components.
These are combined, assuming they are fully correlated in \pt and \Eta, into six
\textit{effective nuisance parameters} (ENP). The combination procedure involves diagonalizing
the covariance matrix of the \JES correction factors and selecting
the five eigenvectors that have the largest corresponding eigenvalues. An additional sixth
effective parameter represents all residual sources of uncertainty.

The sources of uncertainty on the~2011 measurement are the following:
\begin{list}{-}{}
\mynobreakpar
    \item \headFont{ENP 1-6 - }
      uncertainty associated with the six effective nuisance parameters, which combine the~60 components of the \insitu methods
      used to estimate the \JES uncertainty in 2011;
    \item \headFont{intercalibration, single hadron response and relative non-closure - }
      same as for 2010;
    \item \headFont{close-by jets - } 
      uncertainty associated with event topologies in which two jets are reconstructed in close proximity;
    \item \headFont{in-time PU, out-of-time PU and PU pt - } 
      uncertainty associated with the simulation and subtraction of in- and \otpu.
\end{list}

The final fractional \JES uncertainties for the 2010 and the 2011~data are compared in \autoref{FIGfractionalJESComp} as
a function of the \pt of jets in the central region.
\begin{figure}[htp]
\begin{center}
\subfloat[]{\label{FIGfractionalJESComp1}\includegraphics[trim=0mm 0mm 0mm 0mm,clip,width=.52\textwidth]{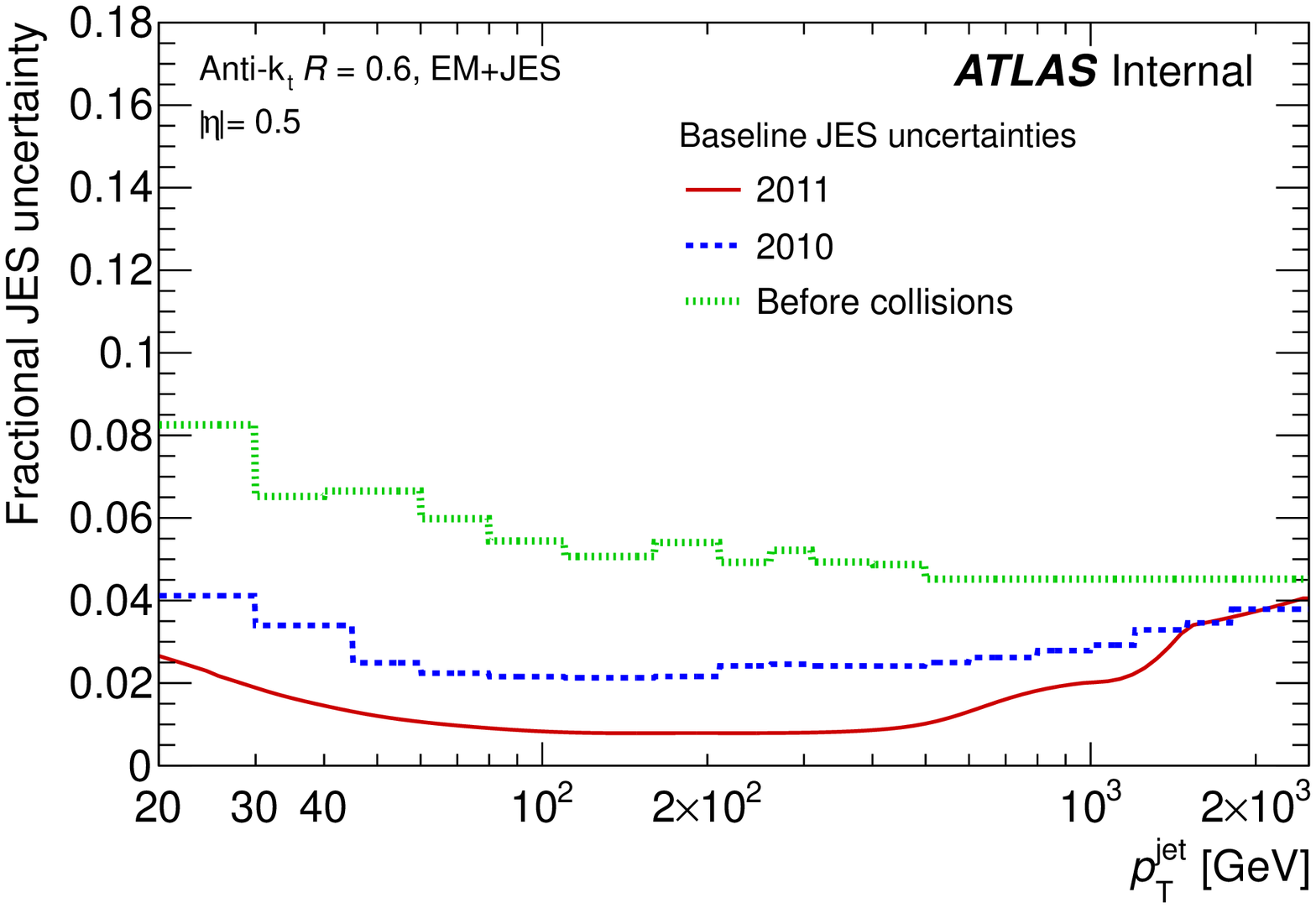}}
\subfloat[]{\label{FIGfractionalJESComp2}\includegraphics[trim=0mm 0mm 0mm 0mm,clip,width=.52\textwidth]{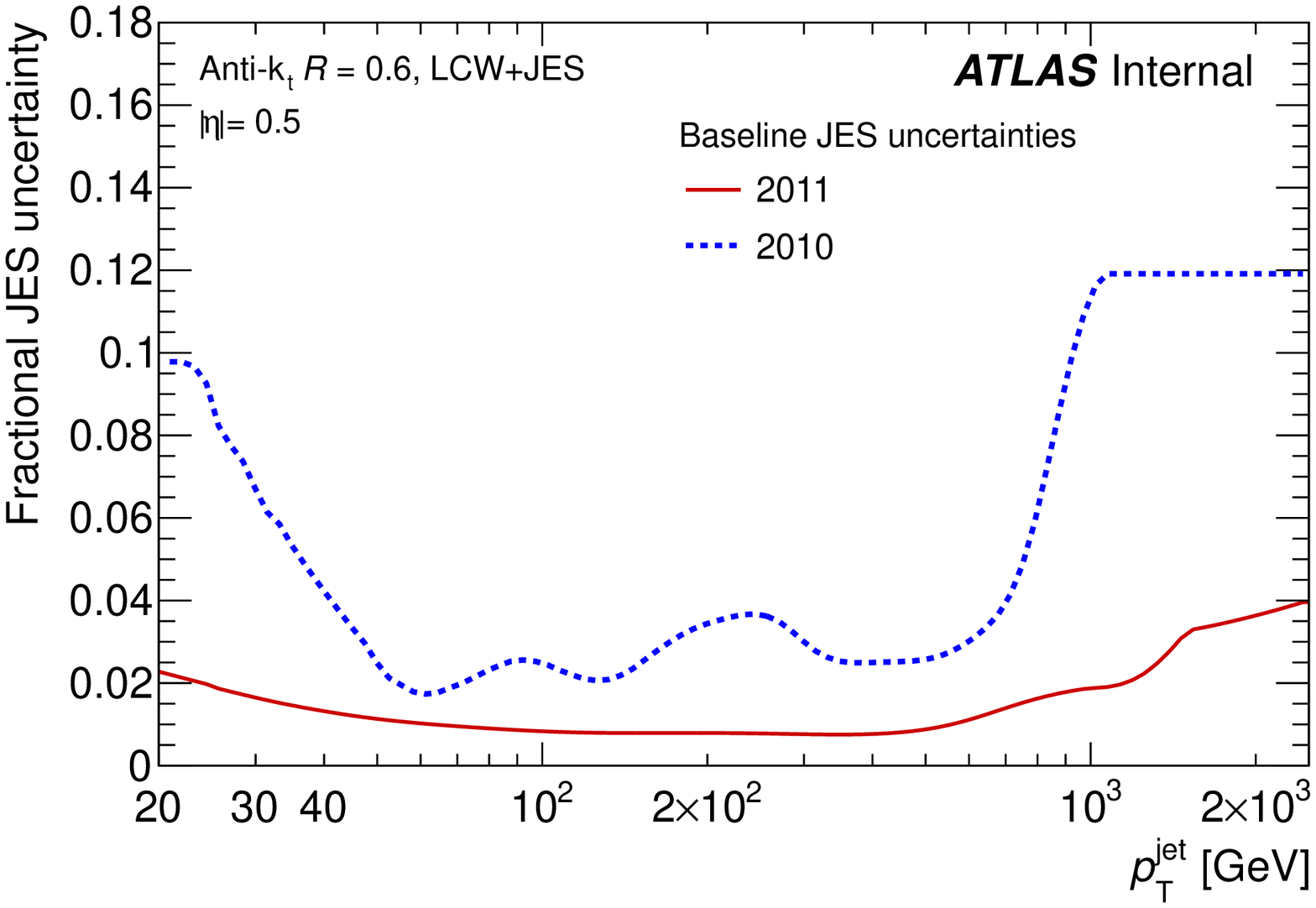}}
  \caption{\label{FIGfractionalJESComp}Fractional jet energy scale systematic uncertainty as a function of
  the transverse momentum of jets, $p_{\mrm{T}}^{\mrm{jet}}$, for jets with pseudo-rapidity, $\left|\eta\right| = 0.5$,
  calibrated under \EMJES \Subref{FIGfractionalJESComp1} and \LCJES \Subref{FIGfractionalJESComp2}. 
  As indicated in the figures, estimates of the uncertainty before beginning of collisions in the LHC,
  for data taken during 2010 and for data taken during 2011, are compared.
  The 2011 uncertainty presented here includes only the \insitu components of the systematic uncertainty.
  (Figure are taken from~\protect\cite{Adomeit:1475233}.)
  }
\end{center}
\end{figure} 
The uncertainty on the 2011 data is much improved compared to 2010. This is due both to the increase in 2011 
in the statistics which are used to derive uncertainty, and to the use of sophisticated \insitu techniques.
For \EMJES jet in 2010 the total uncertainty is roughly~2\% for jets with \ptRange{40}{2\cdot10^3}, increasing up to roughly twice
that for lower or higher transverse momenta. For \LCJES jets in 2011, the uncertainty is roughly~1\%
in the respective mid-\pt range, increasing up to~2\% between~20 and~40\GeV and up to~4\% between~0.5 and~2\TeV.

\subsection{Jet energy and angular resolution\label{chapJetPtAngularResolution}}
%
In addition to the jet energy scale, which is the primary source of uncertainty for most measurements involving jets, the
jet energy and angular resolution must also be taken into account. These are addressed in the following.

\minisec{Jet energy resolution\label{chapJetPtResolution}}
%
The baseline parameterization of the jet energy resolution in MC is derived using 
the nominal \pythia MC simulation for jets with transverse momentum,
\ptRange{30}{500}~\cite{:2012ag}. \Insitu measurements are also made for jets with rapidities, \yLower{2.8}
and \ptHigher{20}.
In this kinematic range, the comparison of the 
resolution measured with the \insitu techniques between the
data and the MC shows agreement better than~10\%.
The uncertainty on the jet energy resolution for each rapidity region is assigned from the
weighted average of the systematic errors on the relative difference between the data and the MC, and is flat as a
function of \pt. Outside the kinematic range of \insitu measurements, the MC parameterization
is kept but the uncertainty is conservatively increased.

In order to illustrate the behaviour of the transverse momentum resolution in MC,
the \textit{relative transverse momentum offset} between the \pt of a
truth jet, $p_{\mrm{t}}^{\mrm{truth}}$, and that of the corresponding matched reconstructed jet,
$p_{\mrm{t}}^{\mrm{rec}}$, is defined as
\begin{equation}
  O_{p_{\mrm{t}}} = \frac{p_{\mrm{t}}^{\mrm{rec}}-p_{\mrm{t}}^{\mrm{truth}}}{p_{\mrm{t}}^{\mrm{truth}}} \,.
\label{eqJetPtOffsetDef} \end{equation}
Matching is performed by selecting a reconstructed jet within distance
${\Delta R_{\mrm{match}} = 0.5 \cdot R}$ of the truth jet, where $R = 0.6$ is the size parameter
of the jet.
In order to keep the sample pure, isolation of the matched truth
jets from their counterparts within ${\Delta R_{\mrm{iso}} = 2 \cdot R}$ is also imposed.

The average relative transverse momentum offset represents the fractional bias in \pt of jets.
The relative \pt resolution is parametrized by the width of the distribution of $O_{p_{\mrm{t}}}$,
denoted by $\sigma\left(O_{p_{\mrm{t}}}\right)$.
The dependence of the average $O_{p_{\mrm{t}}}$ and of $\sigma\left(O_{p_{\mrm{t}}}\right)$
on truth jet \pt is shown in \autoref{FIGjetResolutionPt} for jets in different
pseudo-rapidity regions in MC10 and in MC11.
\begin{figure}[htp]
\begin{center}
  \subfloat[]{\label{FIGjetResolutionPt1}\includegraphics[trim=5mm 14mm 0mm 10mm,clip,width=.52\textwidth]{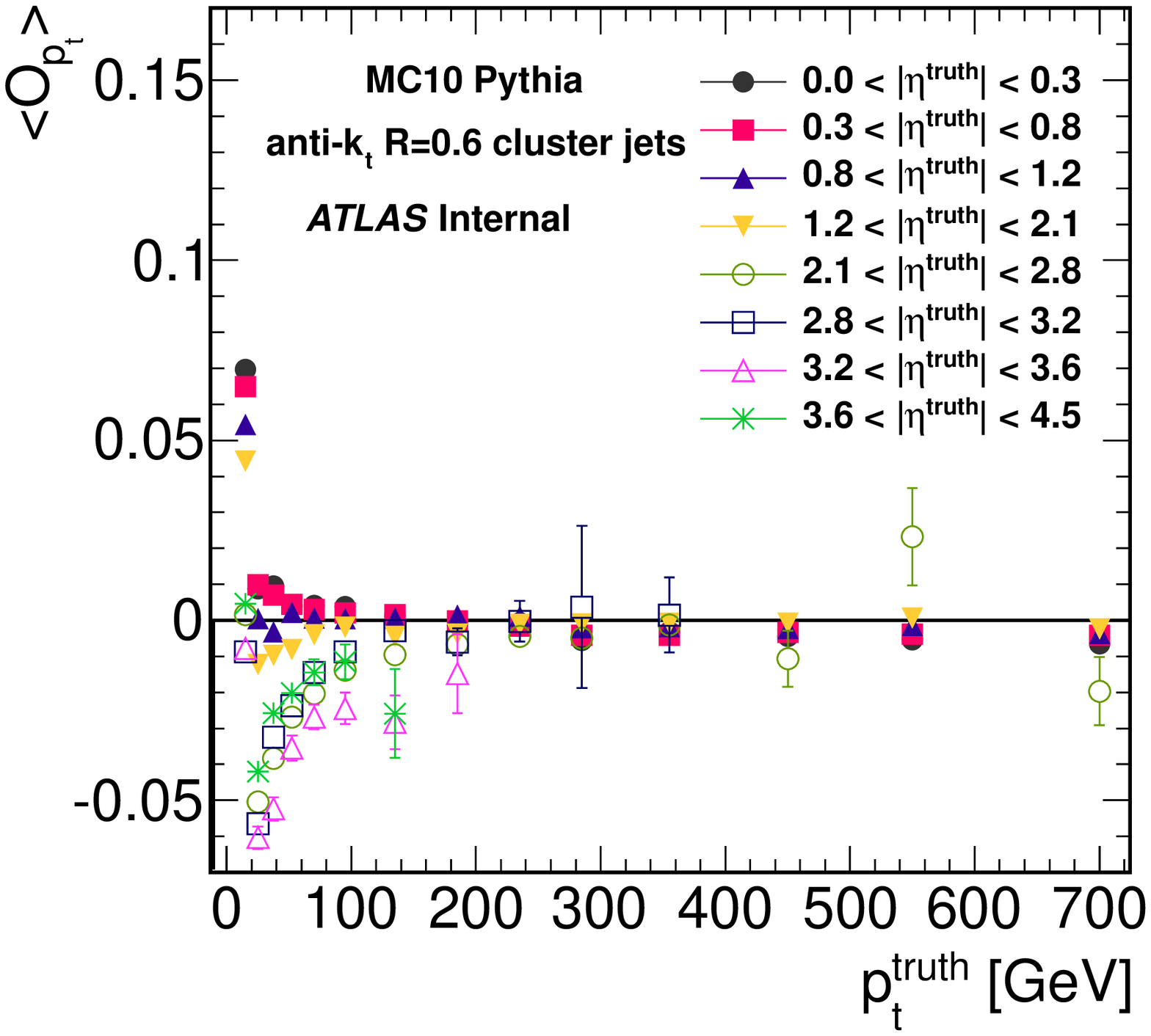}}
  \subfloat[]{\label{FIGjetResolutionPt2}\includegraphics[trim=5mm 14mm 0mm 10mm,clip,width=.52\textwidth]{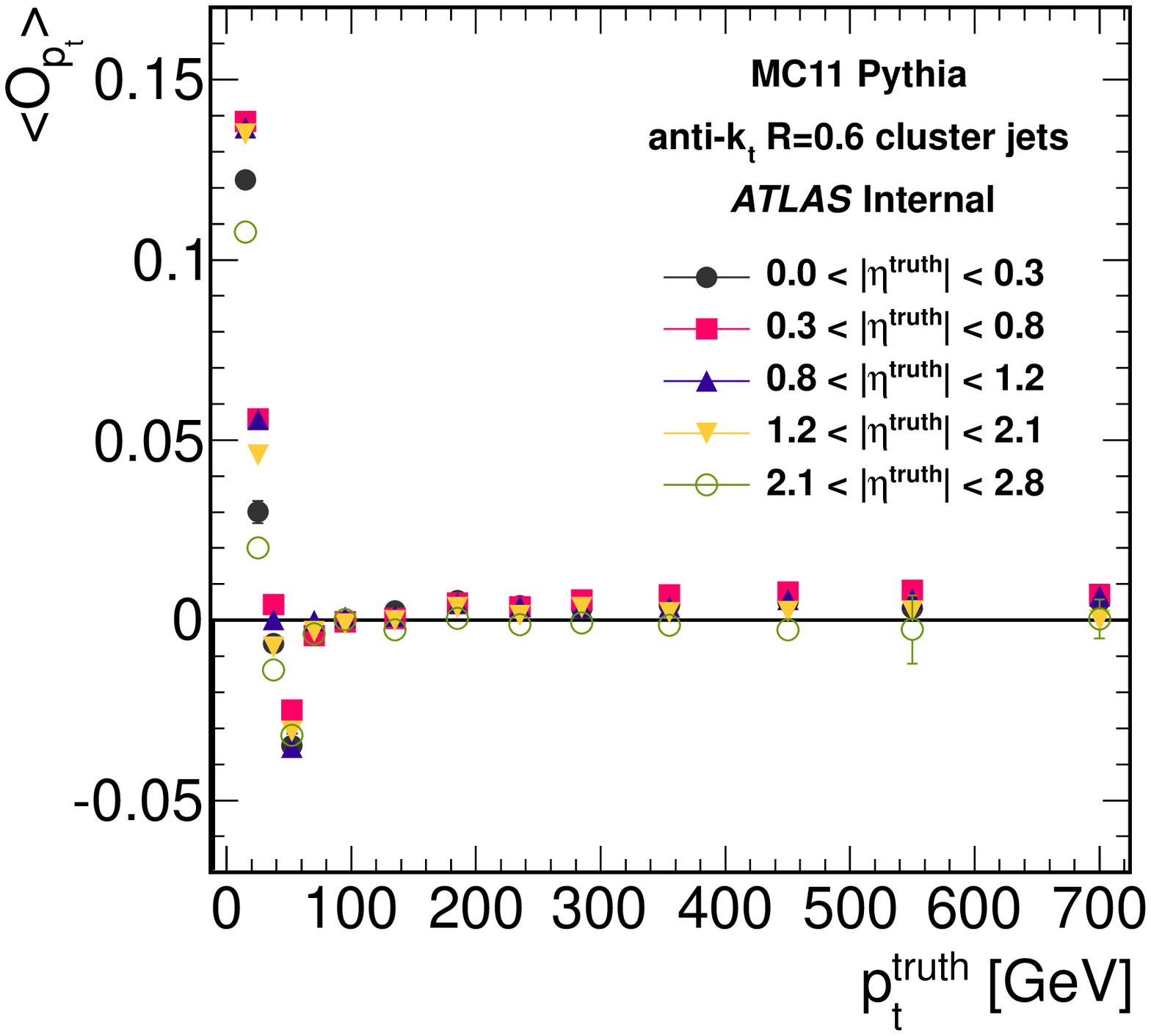}} \\
  \vspace{-10pt}
  \subfloat[]{\label{FIGjetResolutionPt3}\includegraphics[trim=5mm 14mm 0mm 10mm,clip,width=.52\textwidth]{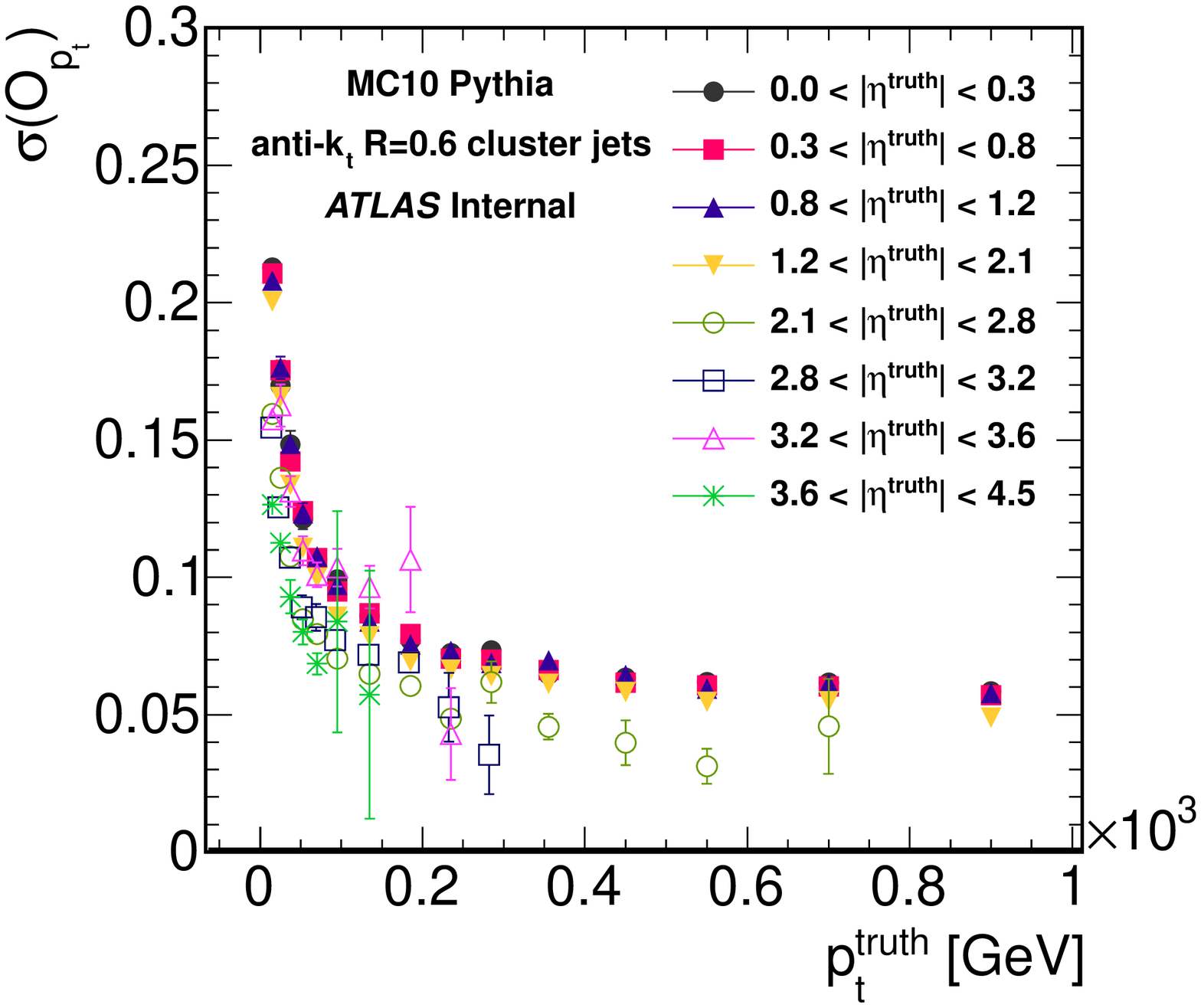}}
  \subfloat[]{\label{FIGjetResolutionPt4}\includegraphics[trim=5mm 14mm 0mm 10mm,clip,width=.52\textwidth]{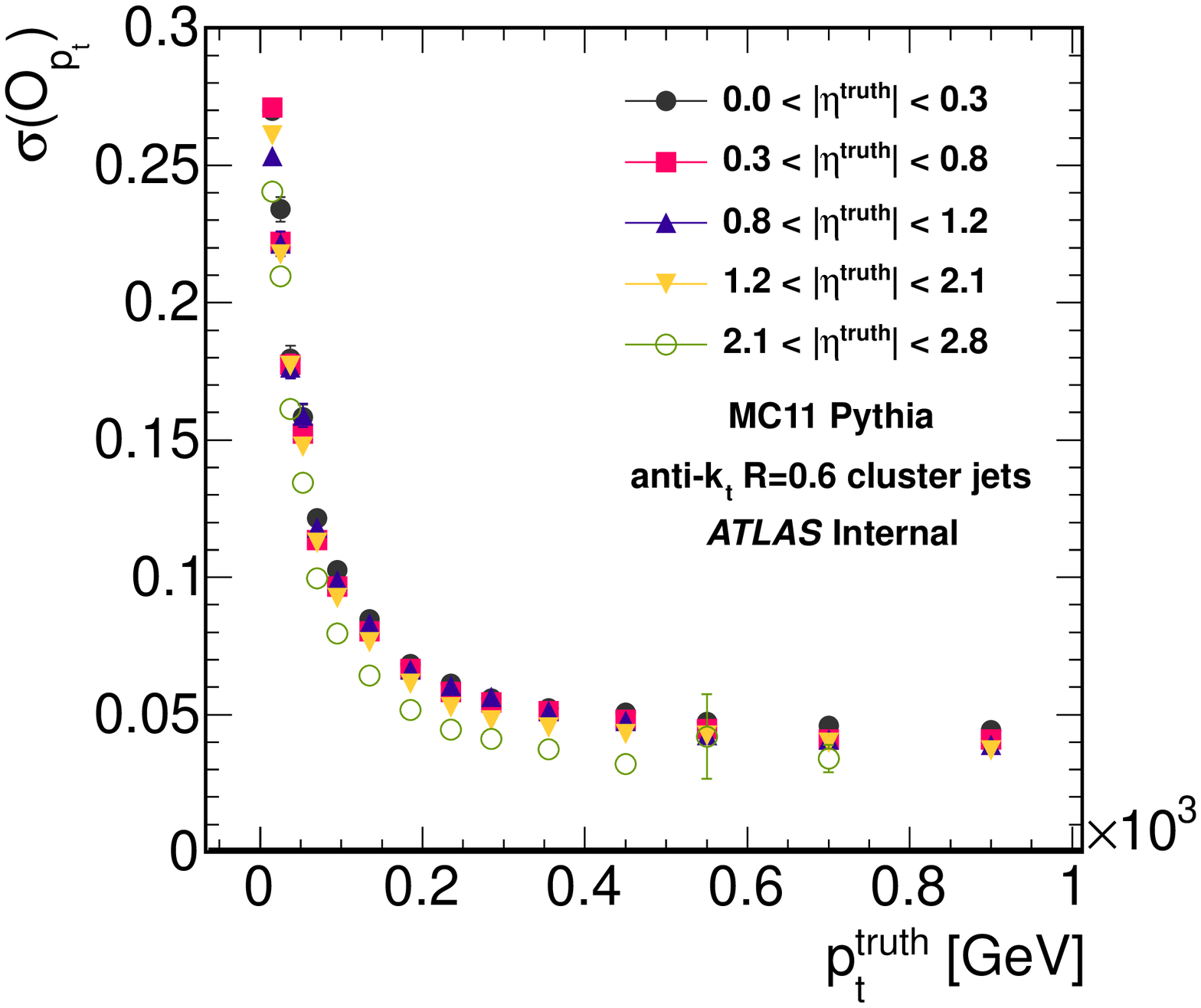}}
  \caption{\label{FIGjetResolutionPt}Dependence of the average relative
    transverse momentum offset, $<O_{p_{\mrm{t}}}>$, (\Subref{FIGjetResolutionPt1} and \Subref{FIGjetResolutionPt2})
    and of the width of the distribution of the relative offset, $\sigma(O_{p_{\mrm{t}}})$,
    (\Subref{FIGjetResolutionPt3} and \Subref{FIGjetResolutionPt4}) on the
    transverse momentum of truth jets, $p_{\mrm{t}}^{\mrm{truth}}$.
    Several regions of truth jet pseudo-rapidity, $\eta^{\mrm{truth}}$, are presented for
    jets in MC10 and in MC11, as indicated in the figures.
  }
\end{center}
\end{figure} 
At \ptLower{20} in 2010 and at \ptLower{30} in 2011, $O_{p_{\mrm{t}}}$ changes sign.
These transverse momentum values mark the
corresponding thresholds in either year for which the energy calibration of jets becomes valid.
The relative bias in transverse momentum is lower in absolute value than~1\%~(6\%) for jets with
\ptLower{100} within \etaLower{2.1} (\etaLower{4.5}) in 2010. In 2011, the bias is lower than~2\% for central
jets within \etaLower{2.8}.
For higher values of transverse momentum the relative bias decreases, becoming negligible for non-central jets in 2010
above~200\GeV and for central jets in 2011 above~100\GeV.
The relative resolution, $\sigma\left(O_{p_{\mrm{t}}}\right)$, is also rapidity- and momentum-dependent. For the various
\Eta-regions the relative resolution decreases from~21\% in 2010 (and~27\% in 2011) for jets with $\pt=20\GeV$, down
to~4-6\% for jets with \ptHigher{300}.
Both $O_{p_{\mrm{t}}}$ and $\sigma\left(O_{p_{\mrm{t}}}\right)$ are generally higher for low-\pt 
jets in MC11 compared to MC10, due to the increase in \pu in 2011.

\minisec{Jet angular resolution\label{chapJetAngularResolution}}
%
In addition to the \pt resolution, the rapidity and azimuthal resolutions are used in \autoref{chapDijetMassSystematicUncertainties}
to estimated a component of the systematic uncertainty associated with the dijet mass measurement.
A parametrization of the angular resolutions for different jet transverse momenta and rapidities is derived
from the following.

The \textit{pseudo-rapidity offset} and the \textit{azimuthal offset} of jets are respectively defined in MC as
\begin{equation}
  O_{\eta} = \eta^{\mrm{rec}}-\eta^{\mrm{truth}} \quad \mrm{and} \quad
  O_{\phi} = \phi^{\mrm{rec}}-\phi^{\mrm{truth}}  \;,
\label{eqJetAngularOffsetDef} \end{equation}
by matching truth and reconstructed jets, as discussed \wrt \autoref{eqJetPtOffsetDef}.
Here, as before, $\eta^{\mrm{truth}}$ ($\phi^{\mrm{truth}}$) and $\eta^{\mrm{rec}}$ ($\phi^{\mrm{rec}}$) are respectively
the pseudo-rapidities (azimuthal angles) of truth and reconstructed jets.
The parameters $O_{\eta}$ and $O_{\phi}$ represents the angular bias of jets.
The width of the distributions of $O_{\eta}$ and $O_{\phi}$ represent the angular
resolution of jets, denoted respectively as $\sigma\left(O_{\eta}\right)$ and $\sigma\left(O_{\phi}\right)$.

The dependence of the average angular offset parameters and of the angular resolution parameters
on truth jet \pt is shown in \autorefs{FIGjetResolutionEta}~-~\ref{FIGjetResolutionPhi} for jets in different
pseudo-rapidity regions in MC10 and in MC11.
\begin{figure}[htp]
\begin{center}
  \subfloat[]{\label{FIGjetResolutionEta1}\includegraphics[trim=5mm 14mm 0mm 10mm,clip,width=.52\textwidth]{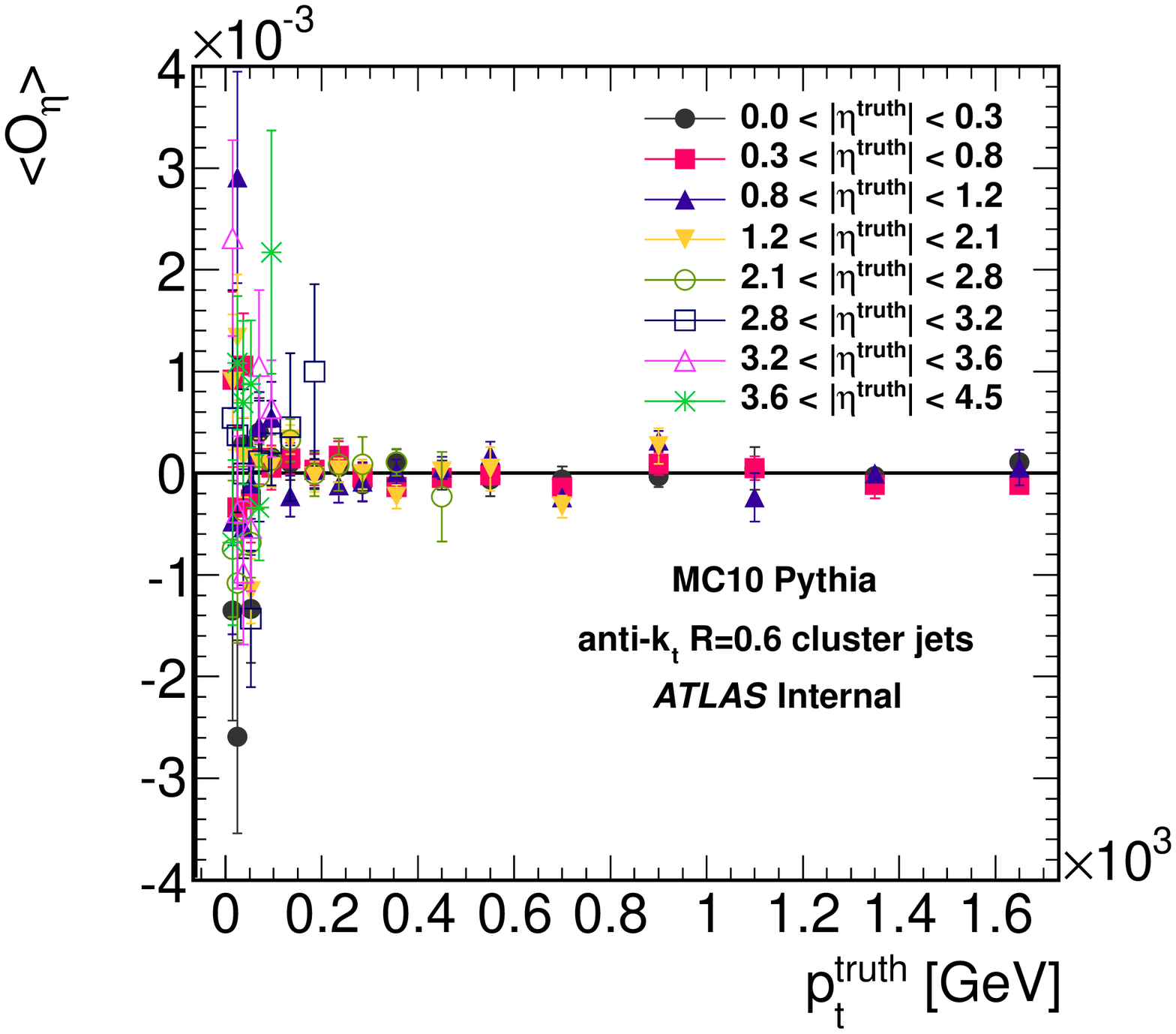}}
  \subfloat[]{\label{FIGjetResolutionEta2}\includegraphics[trim=5mm 14mm 0mm 10mm,clip,width=.52\textwidth]{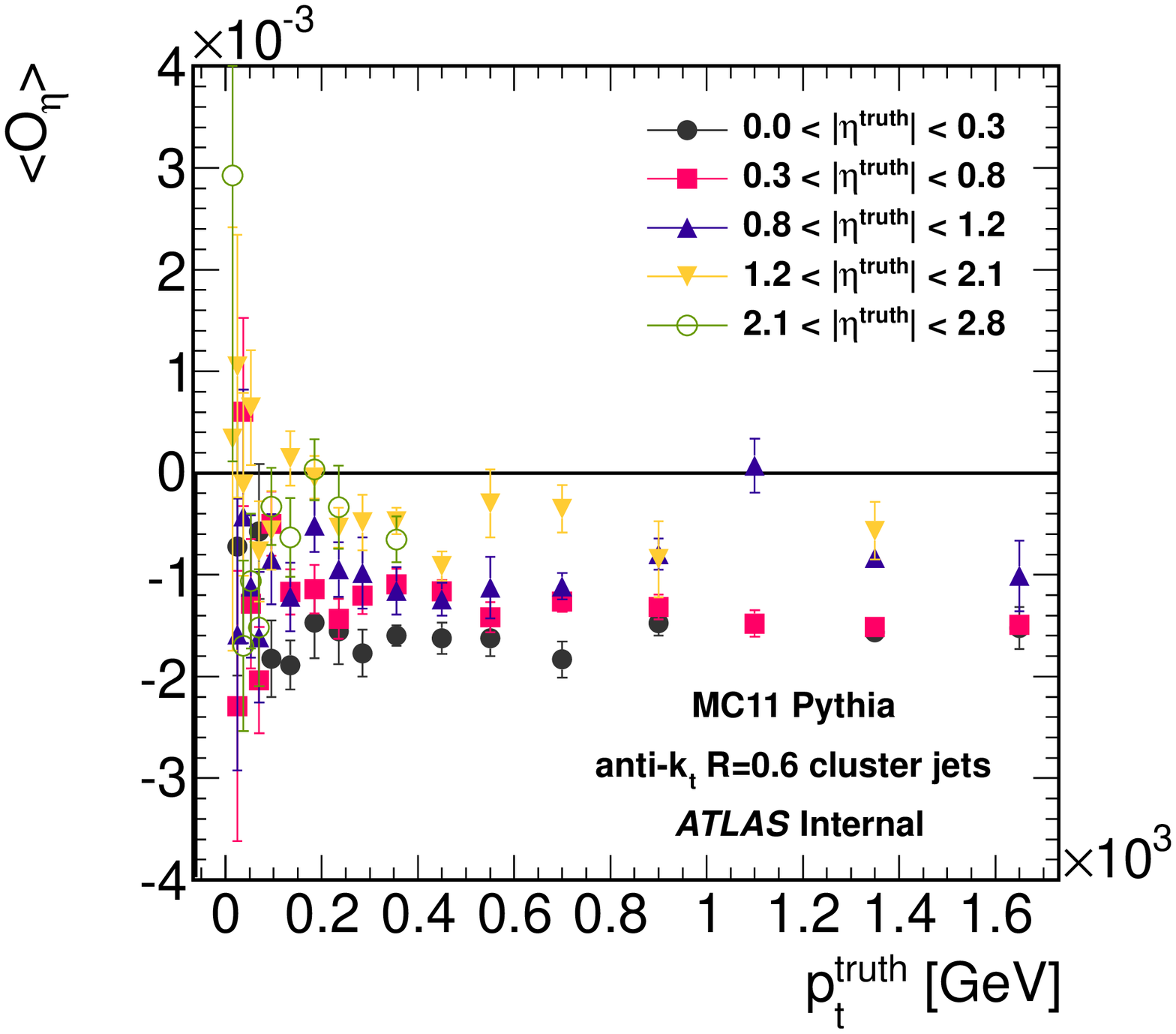}} \\
  \vspace{-10pt}
  \subfloat[]{\label{FIGjetResolutionEta3}\includegraphics[trim=5mm 14mm 0mm 10mm,clip,width=.52\textwidth]{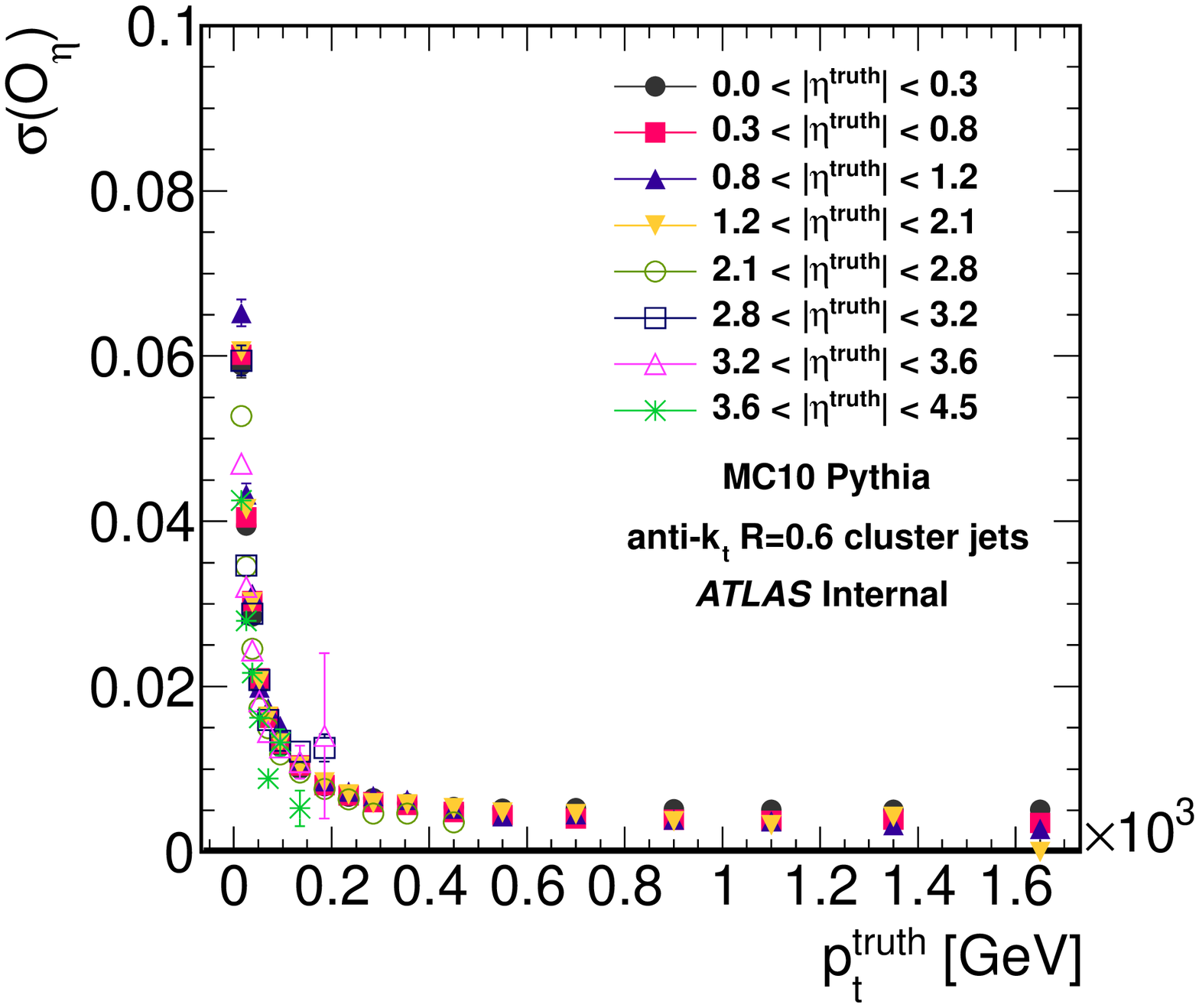}}
  \subfloat[]{\label{FIGjetResolutionEta4}\includegraphics[trim=5mm 14mm 0mm 10mm,clip,width=.52\textwidth]{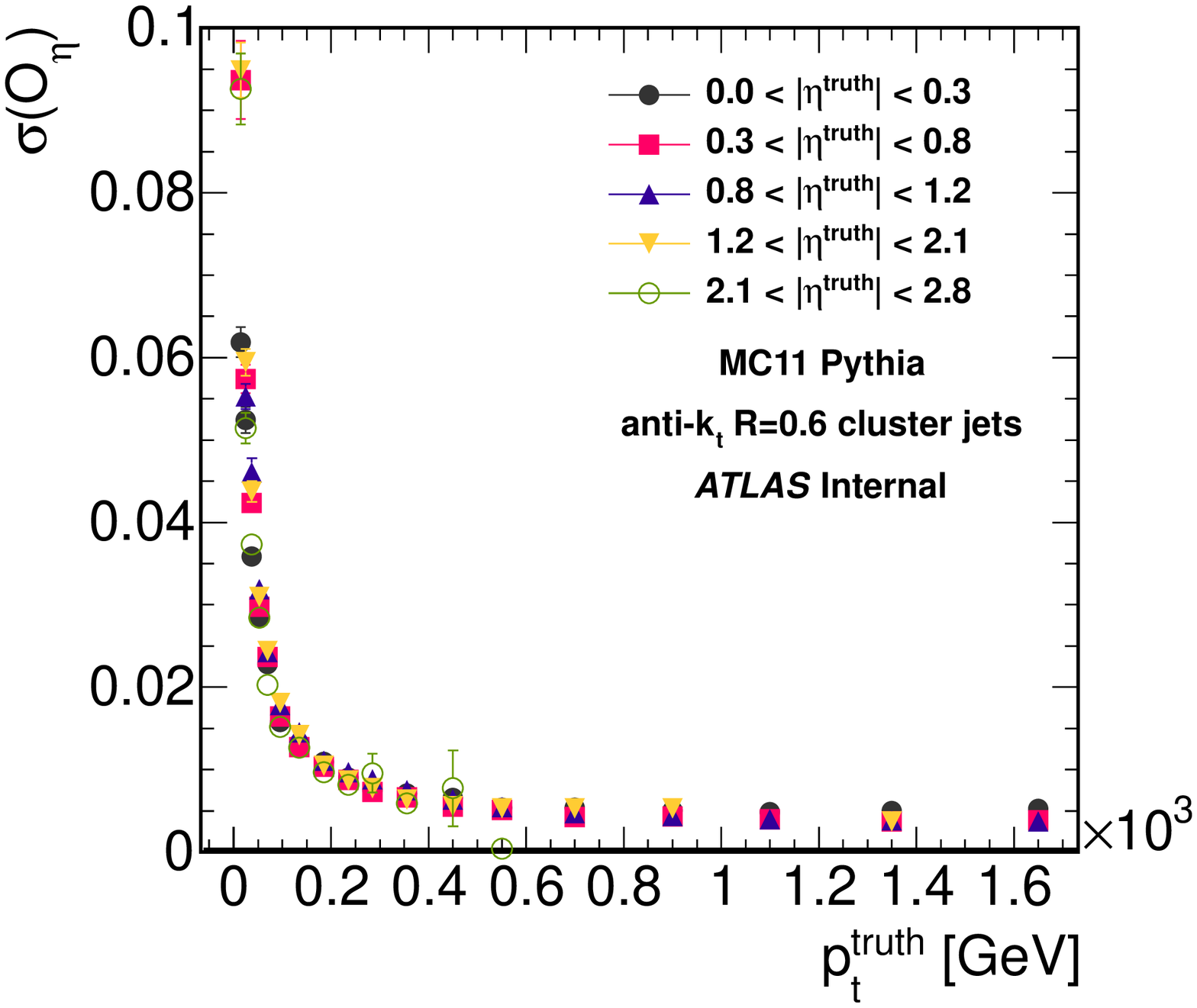}}
  \caption{\label{FIGjetResolutionEta}Dependence of the average
    pseudo-rapidity offset, $<O_{\eta}>$, (\Subref{FIGjetResolutionEta1} and \Subref{FIGjetResolutionEta2})
    and of the width of the distribution of the offset, $\sigma(O_{\eta})$, (\Subref{FIGjetResolutionEta3} and \Subref{FIGjetResolutionEta4}) on the
    transverse momentum of truth jets, $p_{\mrm{t}}^{\mrm{truth}}$.
    Several regions of truth jet pseudo-rapidity, $\eta^{\mrm{truth}}$, are presented for
    jets in MC10 and in MC11, as indicated in the figures.
  }
\end{center}
\end{figure} 
\begin{figure}[htp]
\begin{center}
  \subfloat[]{\label{FIGjetResolutionPhi1}\includegraphics[trim=5mm 14mm 0mm 10mm,clip,width=.52\textwidth]{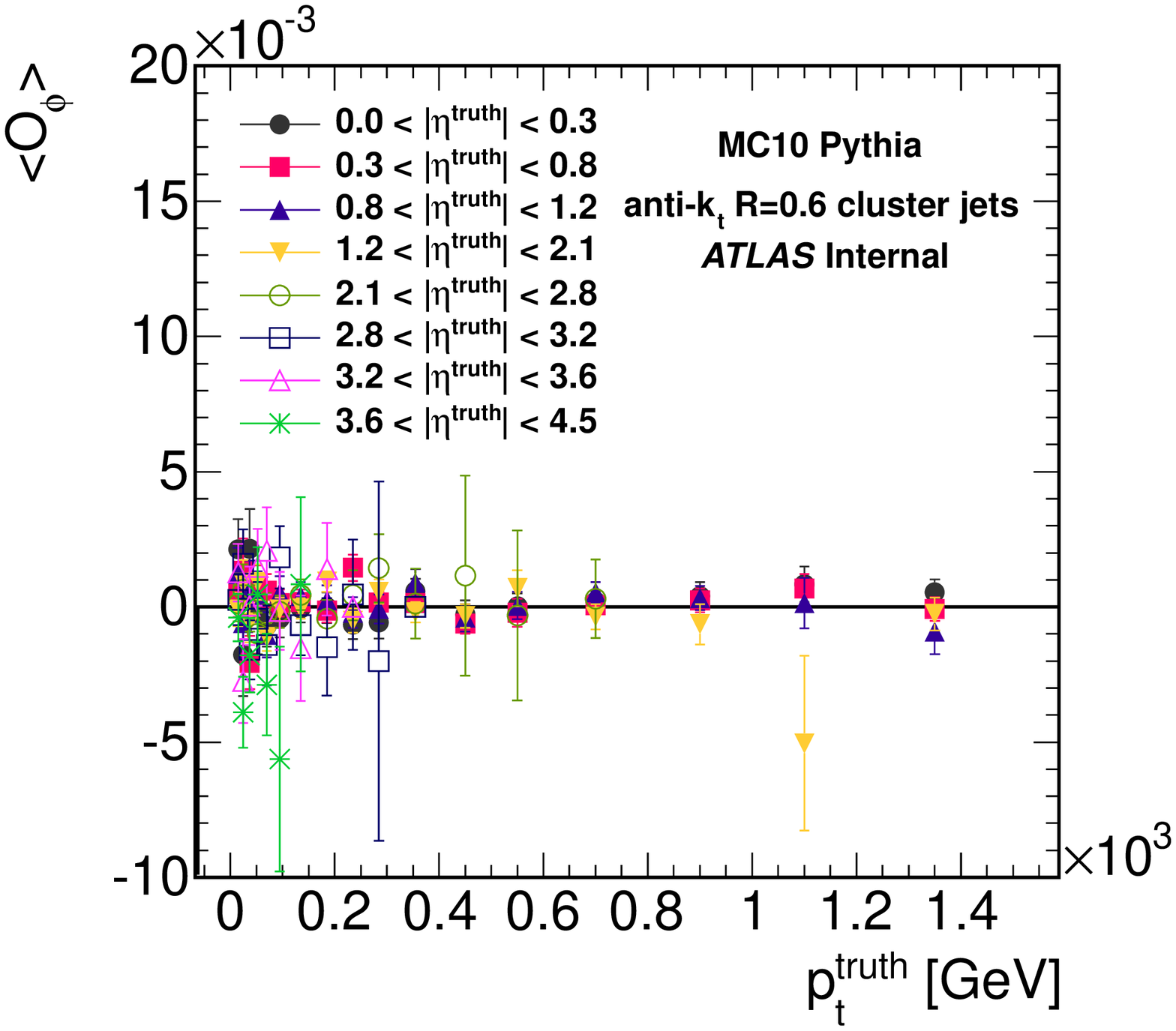}}
  \subfloat[]{\label{FIGjetResolutionPhi2}\includegraphics[trim=5mm 14mm 0mm 10mm,clip,width=.52\textwidth]{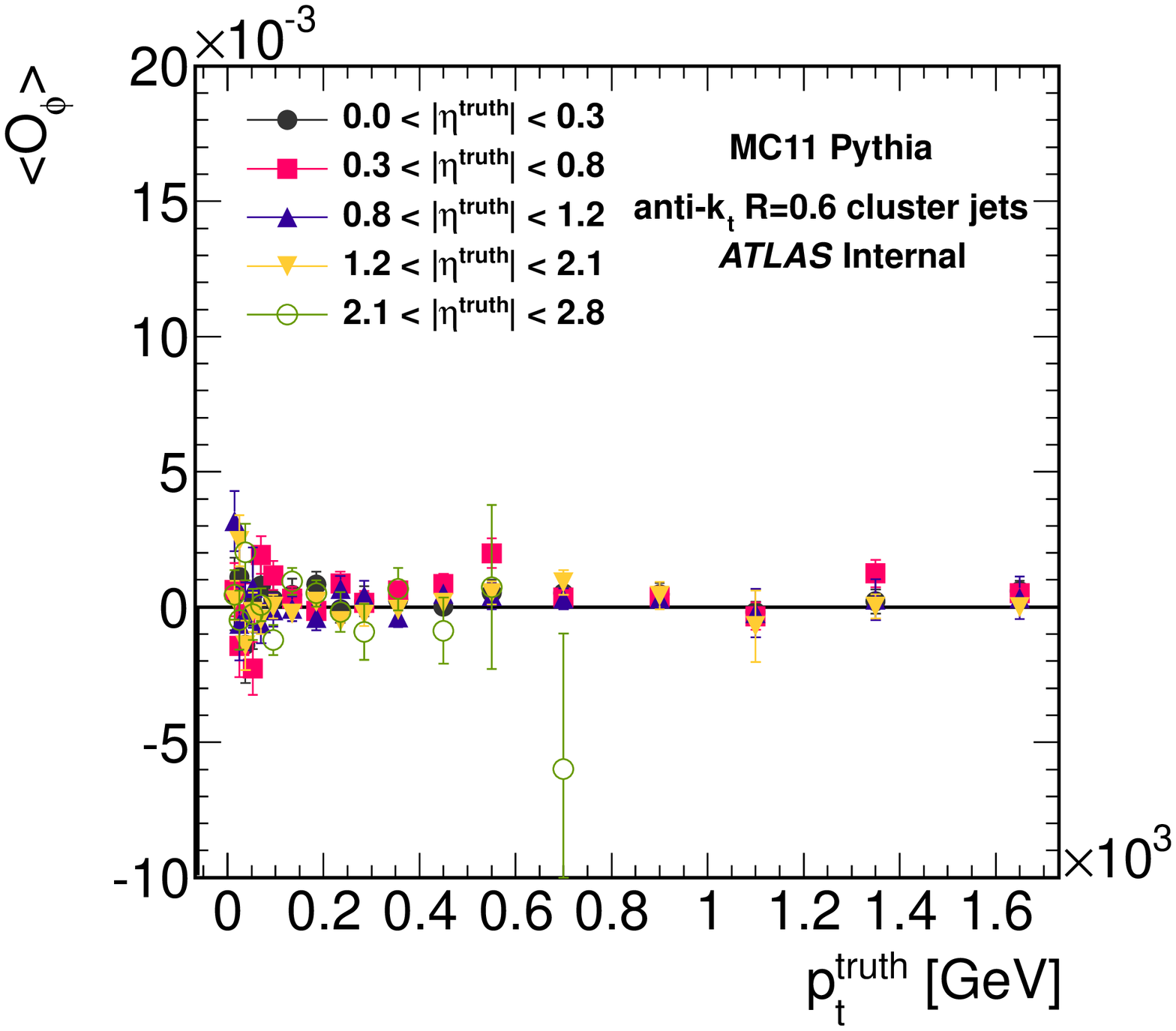}} \\
  \vspace{-10pt}
  \subfloat[]{\label{FIGjetResolutionPhi3}\includegraphics[trim=5mm 14mm 0mm 10mm,clip,width=.52\textwidth]{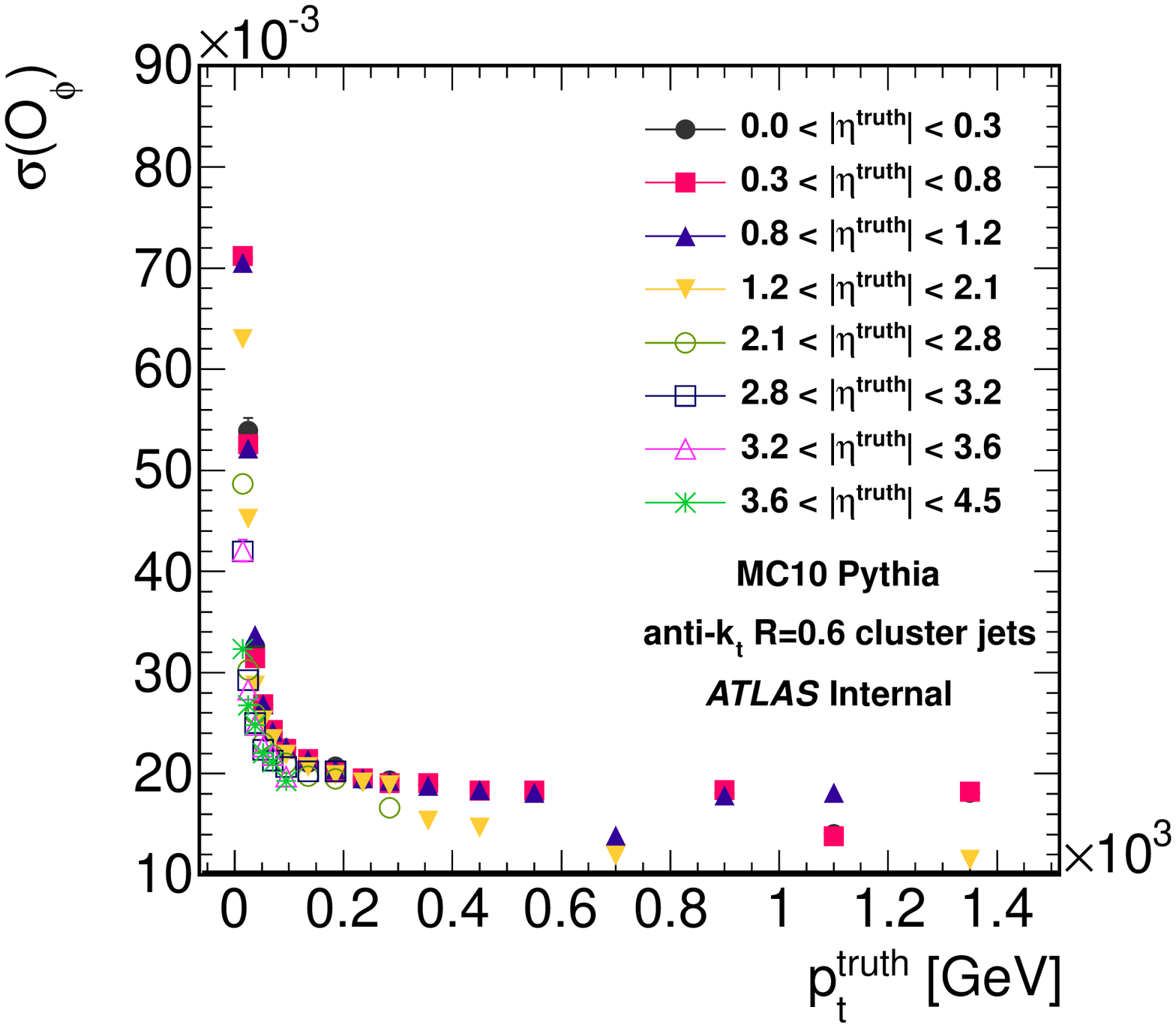}}
  \subfloat[]{\label{FIGjetResolutionPhi4}\includegraphics[trim=5mm 14mm 0mm 10mm,clip,width=.52\textwidth]{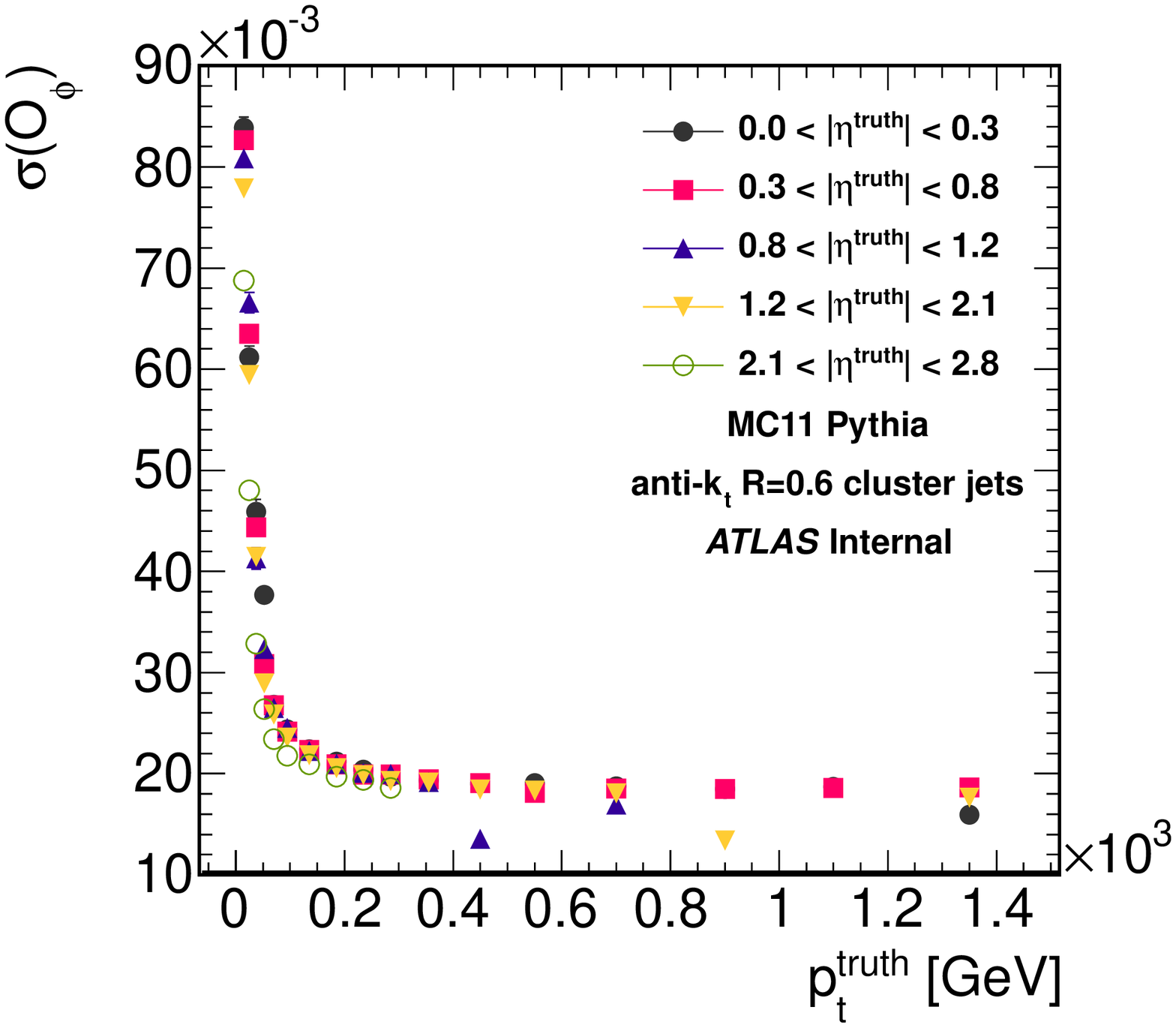}}
  \caption{\label{FIGjetResolutionPhi}Dependence of the average
    azimuthal offset, $<O_{\phi}>$, (\Subref{FIGjetResolutionPhi1} and \Subref{FIGjetResolutionPhi2})
    and of the width of the distribution of the offset, $\sigma(O_{\phi})$, (\Subref{FIGjetResolutionPhi3} and \Subref{FIGjetResolutionPhi4}) on the
    transverse momentum of truth jets, $p_{\mrm{t}}^{\mrm{truth}}$.
    Several regions of truth jet pseudo-rapidity, $\eta^{\mrm{truth}}$, are presented for
    jets in MC10 and in MC11, as indicated in the figures.
  }
\end{center}
\end{figure} 
The pseudo-rapidity offset is smaller than $3\cdot10^{-3}$ in all \Eta regions for jets with low \pt
in 2010, becoming insignificant for transverse momentum values above $200\GeV$.
For jets in 2011 a small bias of roughly $2\cdot10^{-3}$ is observed for all \Eta regions.
The \Eta resolution is roughly 0.04 for jets with $\pt=20\GeV$ in MC10 and for jets with $\pt=30\GeV$ in MC11.
For jets with higher transverse momenta, the resolution decreases, reaching values smaller than 
$3\cdot10^{-3}$ above $200\GeV$.
The slightly degraded performance for jets in 2011 compared to 2010 is due the increase in \pu.

The azimuthal offset and resolution follow similar trends as do the pseudo-rapidity offset and resolution parameters.
The magnitude of the offset is smaller than $3\cdot10^{-3}$ for low \pt jets, becoming insignificant
(consistent with zero within errors) above $\sim100\GeV$.
The azimuthal resolution also improves with growing jet \pt, decreasing from 0.07 for jets with $\pt=20\GeV$ ($\pt=30\GeV$) to roughly 0.02
above $200\GeV$ for jets in MC10 (MC11).

      }{}
\ifthenelse{\boolean{do:dataSelection}}     { 
\chapter{Data selection\label{chapDataSelection}}
%
\section{Dataset}
%
The measurements presented in this thesis are performed on data of $pp$ collisions at \sqs, collected 
by the \ATLAS experiment during 2010 and during 2011.
The former include all data taken during 2010 with two exceptions, for the low-\pt and the forward regions.
For the low-\pt region (events where the highest-\pt jet has $20 < \pt < 60$\GeV),
only data runs taken up to the beginning of June are considered. For these data 
the instantaneous luminosity of the accelerator was low enough that \pu contributions were
negligible, and the majority of the bandwidth was allocated to the minimum bias trigger.
The forward jet trigger (covering rapidities $\left|y\right| > 2.8$) was not active in the first data-taking period.

The 2011 dataset includes data taken during April, and between July and October, corresponding
respectively to 2011~periods~D and~I--M.
Before period~D, the LHC was not operating at the nominal data-taking conditions 
used in this analysis; the solenoid and/or toroid fields were off, the bunch train frequency was varied and the
\cms energy was changed. Data taken during these periods are therefore not included in this analysis.
In data periods E--H, corresponding to~20\% of the luminosity recorded in 2011,
six LAr Front End Boards (FEBs) were lost. LAr FEBs
contain the electronics for amplifying, shaping, sampling, pipelining, and digitizing the liquid argon calorimeter signals.
As a result, a region, $-0.1<\eta<1.5$ in pseudo-rapidity and $-0.9<\phi<-0.5$ in azimuth were affected.
Jets which fell in this area of the calorimeter suffered a~30\% loss in energy and a~$\sim50\%$ increase
in energy resolution at high \pt. Data periods E--H are therefore excluded from the analysis.

All data events considered in the following require good
detector status for the L1 central trigger processor, solenoid magnet, inner detectors (Pixel, SCT, and
TRT), calorimeters (barrel, endcap, and forward), luminosity, as well as tracking, jet, and missing energy
reconstruction performance. In addition, good data quality is required for the high-level trigger during
the periods when this device is used for rejection.
Events have at least one reconstructed vertex; vertices are reconstructed with constraints on
their associated ID tracks, and are consistent with the beam-spot.

In total $37\pm3.4\%~\mrm{pb}^{-1}$ of data taken during 2010,
and~$3776\pm3.9\%~\mrm{pb}^{-1}$ of data taken during 2011, are used in the analysis.


\section{Trigger and luminosity\label{chapTriggerAndLuminosity}}
%
\subsection{Description of the trigger}
%
Four different triggers are used in this analysis, a random trigger,
the central jet trigger ($\left|\eta\right| < 3.2$),
the forward jet trigger ($3.1 < \left|\eta\right| < 4.9$), and the minimum bias trigger (\ttt{MBTS}).

The central and forward jet triggers are composed of three consecutive levels; Level~1
(\ttt{L1}), Level~2 (\ttt{L2}) and Event Filter (\ttt{EF}).
The central and forward jet triggers independently select data using several jet transverse energy thresholds that
each require the presence of a jet with sufficient transverse
energy at the electromagnetic (\EM) scale, $E_{\mrm{t}}^{\mrm{EM}}$.
For each \ttt{L1} threshold, there is
a corresponding \ttt{L2} threshold that is generally placed 15\GeV above the \ttt{L1}
value.
The following convention is used for jet trigger names; names begin with the trigger-level and end with a number. The number stands for the trigger threshold, such
that \eg \ttt{L1\ul J5} is a Level~1 trigger with a threshold, $E_{\mrm{t}}^{\mrm{EM}} > 5$\GeV.
Jet trigger names also contain a ``regional'' identifier, the letter \ttt{J} for central triggers, and the combination \ttt{FJ} for forward triggers.

The \ttt{MBTS} trigger (denoted by \texttt{L1\ul MBTS} at Level~1 and by \ttt{EF\ul MBTS} for the Event Filter)
requires at least one hit in the minimum bias scintillators located in
front of the endcap cryostats, covering $2.09< \left|\eta\right|< 3.84$, and is the primary trigger used to select
minimum-bias events in \ATLAS. It has been demonstrated to have negligible inefficiency for the events
of interest for this analysis~\cite{Aad:2010rd} and is generally used to select events with low-\pt jets.
It is used in this study only for data taken during 2010. Events with low-\pt jets in 2011 data
are selected using a random trigger.

The complete list of triggers used in the analysis in the different data-taking periods is given in
\Autorefs{TBLlumiInTrigBinsApp0}~-~\ref{TBLlumiInTrigBinsApp2} in \autoref{chapDataSelectionAndCalibrationApp}.

\subsection{Trigger efficiency\label{chapTriggerEfficiency}}
%

The per-jet efficiency is determined from the probability that a single
jet passes a given trigger threshold, regardless of what the other jets in the event do.
In order to determine the efficiency, the off-line reconstructed jets are matched to central or forward
trigger objects by the shortest distance,
\begin{equation*}
  \Delta R = \sqrt{\left(\phi_{\mrm{trig}}-\phi_{\mrm{jet}}\right)^{2}+\left(\eta_{\mrm{trig}}-\eta_{\mrm{jet}}\right)^{2}} \;,
\label{eqTrigJetDistanceDef} \end{equation*}
using the pseudo-rapidity and azimuth of the trigger object, $\eta_{\mrm{trig}}$ and $\phi_{\mrm{trig}}$, and of
the selected jet, $\eta_{\mrm{jet}}$ and $\phi_{\mrm{jet}}$. 

There is an ambiguity whether to associate a jet in the transition region
${2.8 < \left|\eta\right| < 3.6}$ to a central or to a forward trigger object. The rapidity resolution
between off-line jets and \ttt{L1} trigger objects, which seed the higher trigger chain, can
therefore lead to central trigger objects that are reconstructed as forward jets, and vice versa.
The problem is aggravated by the fact that the forward \ttt{L1} jet trigger has no \Eta information
in the FCal.
%
In order to resolve the ambiguity, two candidate trigger objects are assigned to an off-line jet,
one from the central and one from the forward trigger systems.
This is done using all available information, that is, central trigger
objects are matched by the smallest $\Delta R$ with respect to the selected jet,
and forward trigger objects by the smallest $\Delta\phi$.
In the next step, the candidate which is closest to the off-line jet between the two is chosen.
A detailed description of the matching procedure may be found in~\cite{Baker:1360174}.

The efficiencies for various central jet triggers are shown in \autoref{FIGTrigEffCentral} for the 2010 and for the 2011 data.
\begin{figure}[htp]
\begin{center}
\subfloat[]{\label{FIGTrigEffCentral0}\includegraphics[trim=5mm 14mm 0mm 10mm,clip,width=.52\textwidth]{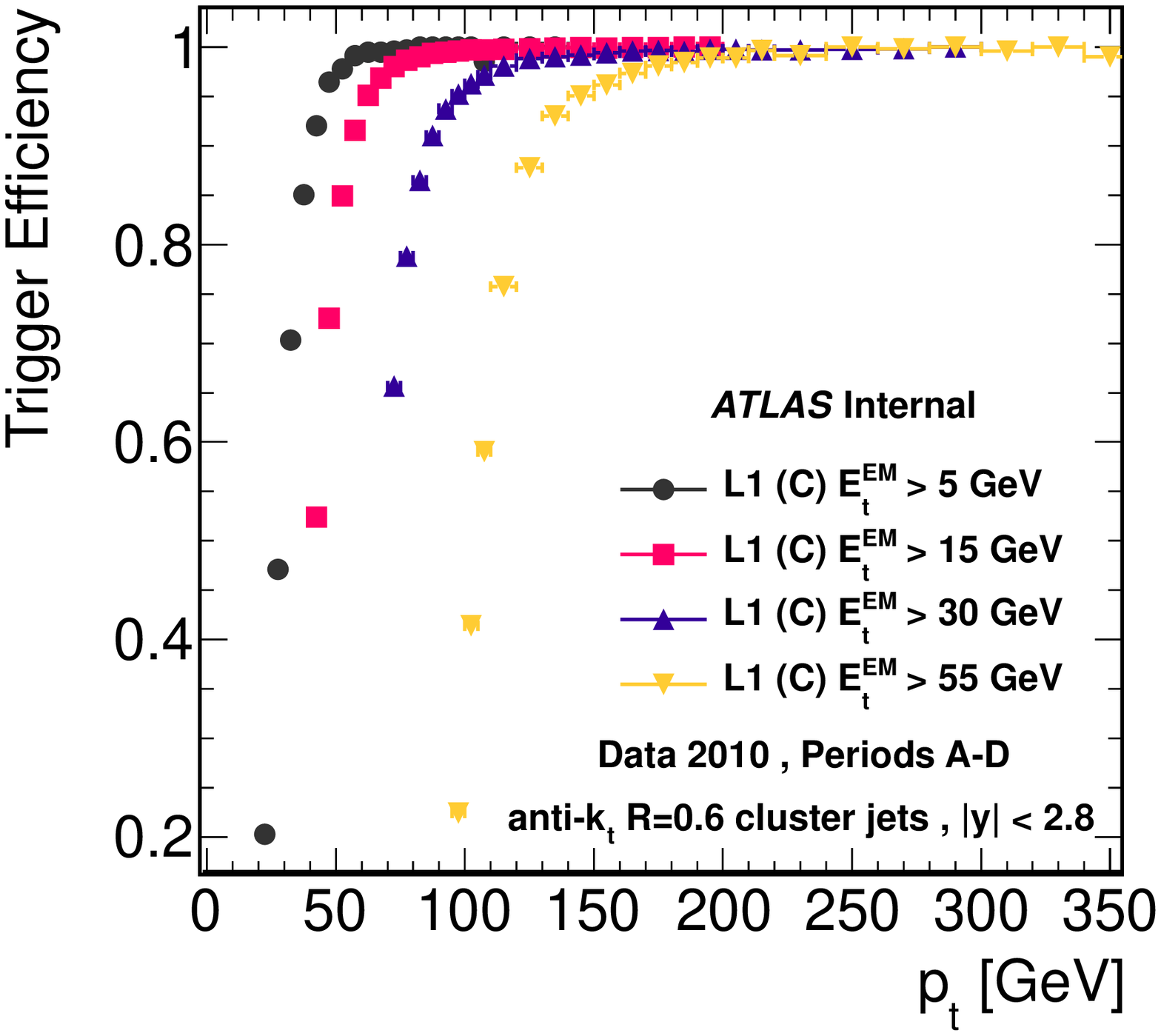}}
\subfloat[]{\label{FIGTrigEffCentral1}\includegraphics[trim=5mm 14mm 0mm 10mm,clip,width=.52\textwidth]{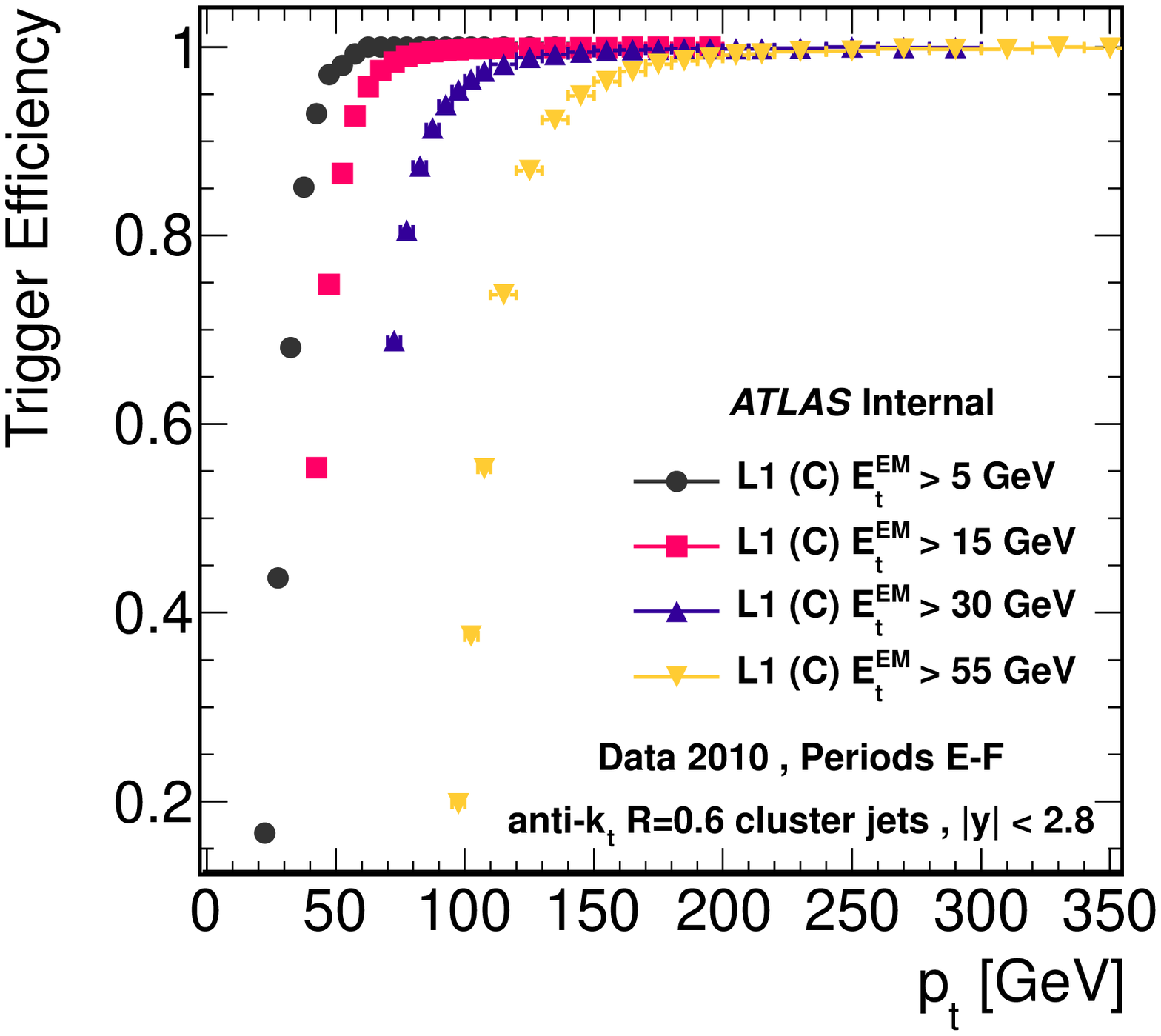}} \\
\subfloat[]{\label{FIGTrigEffCentral2}\includegraphics[trim=5mm 14mm 0mm 10mm,clip,width=.52\textwidth]{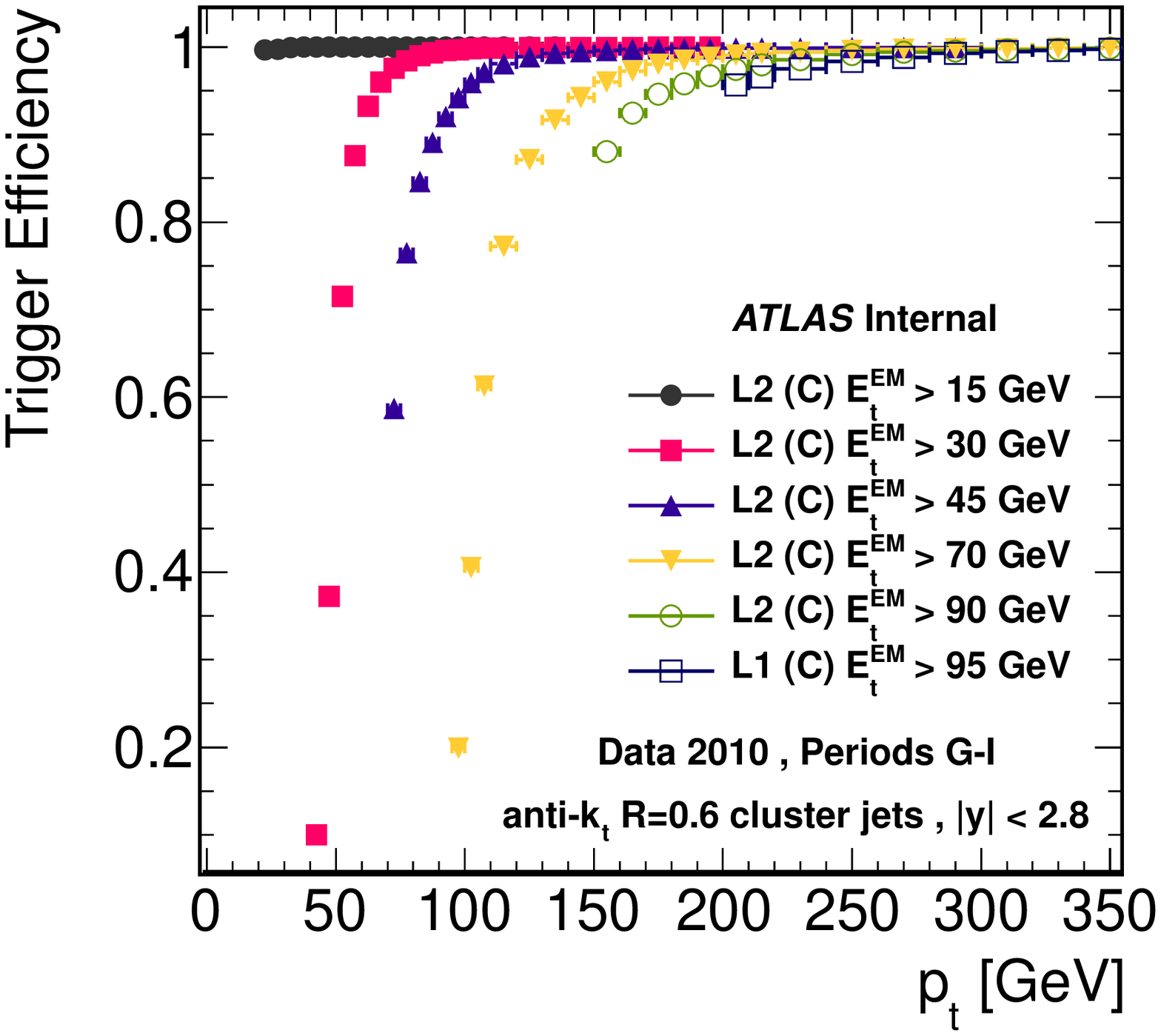}}
\subfloat[]{\label{FIGTrigEffCentral3}\includegraphics[trim=5mm 14mm 0mm 10mm,clip,width=.52\textwidth]{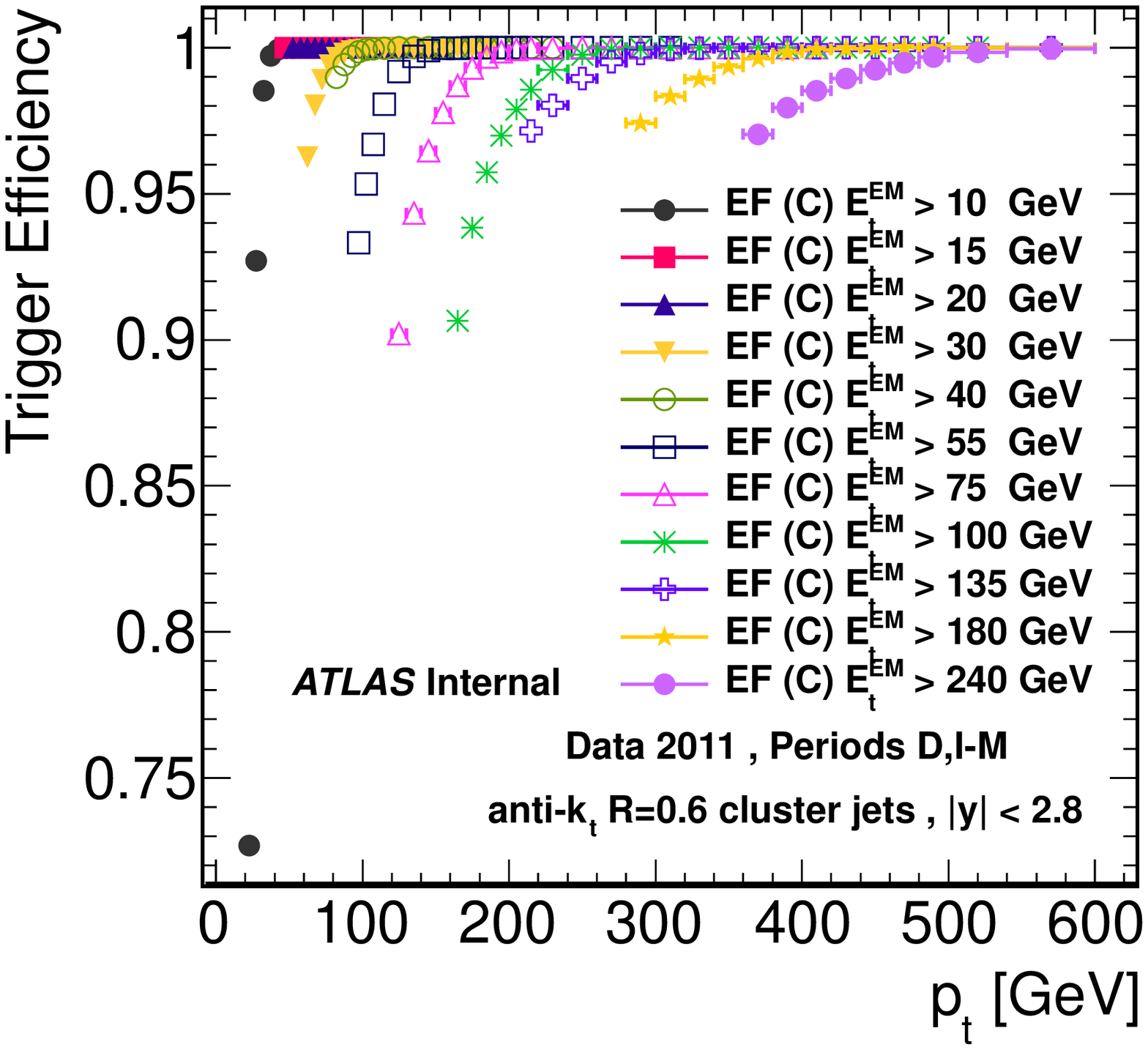}}
\caption{\label{FIGTrigEffCentral}Trigger efficiency curves for various Level~1
  (\ttt{L1}), Level~2 (\ttt{L2}) and Event Filter (\ttt{EF}) triggers. Different 2010 (\Subref{FIGTrigEffCentral0},
  \Subref{FIGTrigEffCentral1} and \Subref{FIGTrigEffCentral2}) and 2011 \Subref{FIGTrigEffCentral3} data
  taking periods are represented, as indicated in the figures.
  The triggers are associated with jets within rapidity $\left|y\right| < 2.8$, belonging to the
  central jet trigger system, denoted by (C).
  Trigger thresholds are denoted by $E_{\mrm{t}}^{\mrm{EM}}$,
  signifying the minimal transverse energy at the electromagnetic scale, of a jet which is required to fire the trigger.
  Additional rapidity bins are presented
  in \autoref{chapDataSelectionAndCalibrationApp}, \autorefs{FIGTrigEffCentralApp}~-~\ref{FIGTrigEffForwardApp}.
}
\end{center}
\end{figure} 
Additional rapidity bins are presented
in \autoref{chapDataSelectionAndCalibrationApp}, \autorefs{FIGTrigEffCentralApp}~-~\ref{FIGTrigEffForwardApp}.
These include three regions:
The first is the crack region between the calorimeter barrel and endcap, ${1.3 < \left|y\right| < 1.6}$; there, the 
per-jet \ttt{L1} and \ttt{L2} trigger efficiencies never become fully efficient due to calorimeter inhomogeneities.
The second is the transition region between the central and the forward trigger systems, ${2.8 < \left|y\right| < 3.6}$; 
in this region, trigger selection is performed by matching jets to either the central or the forward triggers, as described above.
The third is the forward region, ${3.6 < \left|y\right| < 4.4}$, where the forward jet trigger is used;
due to a dead FCal trigger tower that spans a width of ${\Delta\phi = \pi/4}$
in the rapidity region ${\left|y\right| > 3.1}$, the forward jet triggers are not fully efficient in this region. 

As the instantaneous luminosity increased throughout data-taking, it was necessary to prescale triggers with
lower \et thresholds. For each bin
of jet-\pt considered in this analysis, a dedicated trigger threshold is chosen such that the efficiency
is as close as possible to~100\% and the prescale is as small as possible.
\Autoref{TBLcentralTrigNames2010} presents the various triggers used in the analysis as a function of jet
transverse momenta and rapidity, for data taken throughout 2010. The per-jet trigger efficiencies
are shown in \autoref{chapDataSelectionAndCalibrationApp}, \autoref{TBLcentralTrigEfficiency2010App}.
\begin{table}[htp] \begin{center}
\begin{Tabular}[1.3]{ |ccc|c|c|c|c| } \hline \hline
  \multicolumn{3}{|c|}{ \multirow{3}{*}{ \pt~[\GeV] } } & \multicolumn{4}{c|}{Trigger name} \\ \cline{4-7}
                              &&& Period~A     & Periods~A-D          & Periods~E-F   & \multirow {2}{*}{ Periods~G-I }\\
                              &&& Run $<$ 152777 & (Run $\ge$ 152777) & (excl.~E1-E4) & \\
\hline \hline \hline
\multicolumn{7}{c}{ } \\ [-5pt]\hline
\multicolumn{7}{|c|}{ $\left|y\right| < 3.6$ } \\
\hline \hline
\multicolumn{1}{|p{24pt}}{\raggedleft20}    &$<\pt\le$& \multicolumn{1}{p{26pt}|}{\raggedright42.5}& \ttt{L1\ul MBTS} & \ttt{L1\ul MBTS} & \ttt{L1\ul MBTS} & \ttt{EF\ul MBTS} \\ \hline
\multicolumn{1}{|p{24pt}}{\raggedleft42.5}  &$<\pt\le$& \multicolumn{1}{p{26pt}|}{\raggedright70}& \ttt{L1\ul MBTS} & \ttt{L1\ul J5} & \ttt{L1\ul J5} & \ttt{L2\ul J15} \\ \hline
\multicolumn{1}{|p{24pt}}{\raggedleft70}    &$<\pt\le$& \multicolumn{1}{p{26pt}|}{\raggedright97.5}& \ttt{L1\ul MBTS} & \ttt{L1\ul J15} & \ttt{L1\ul J15} & \ttt{L2\ul J30} \\ \hline
\multicolumn{1}{|p{24pt}}{\raggedleft97.5}  &$<\pt\le$& \multicolumn{1}{p{26pt}|}{\raggedright152.5}& \ttt{L1\ul MBTS} & \ttt{L1\ul J30} & \ttt{L1\ul J30} & \ttt{L2\ul J45} \\ \hline
\multicolumn{1}{|p{24pt}}{\raggedleft152.5} &$<\pt\le$& \multicolumn{1}{p{26pt}|}{\raggedright197.5}& \ttt{L1\ul MBTS} & \ttt{L1\ul J55} & \ttt{L1\ul J55} & \ttt{L2\ul J70} \\ \hline
\multicolumn{1}{|p{24pt}}{\raggedleft197.5} &$<\pt\le$& \multicolumn{1}{p{26pt}|}{\raggedright217.5}& \ttt{L1\ul MBTS} & \ttt{L1\ul J55} & \ttt{L1\ul J55} & \ttt{L2\ul J90} \\ \hline
                                     &$\;\;\;\;\pt\ge$& \multicolumn{1}{p{26pt}|}{\raggedright217.5}& \ttt{L1\ul MBTS} & \ttt{L1\ul J55} & \ttt{L1\ul J55} & \ttt{L1\ul j95} \\ \hline

\hline
\multicolumn{7}{c}{ } \\ [-5pt]\hline
\multicolumn{7}{|c|}{ $2.8 < \left|y\right| < 3.6$ } \\
\hline \hline

\multicolumn{1}{|p{24pt}}{\raggedleft20}    &$<\pt\le$& \multicolumn{1}{p{24pt}|}{\raggedright42.5} & \ttt{L1\ul MBTS} & \ttt{L1\ul MBTS} & \ttt{L1\ul MBTS} & \ttt{EF\ul MBTS} \\ \hline
\multicolumn{1}{|p{24pt}}{\raggedleft42.5}  &$<\pt\le$& \multicolumn{1}{p{24pt}|}{\raggedright62.5}& \ttt{L1\ul MBTS} & \ttt{L1\ul MBTS} & \ttt{L1\ul FJ10} & \ttt{EF\ul MBTS} \\ \hline
\multicolumn{1}{|p{24pt}}{\raggedleft62.5}  &$<\pt\le$& \multicolumn{1}{p{24pt}|}{\raggedright72.5}& \ttt{L1\ul MBTS} & \ttt{L1\ul MBTS} & \ttt{L1\ul FJ10} & \ttt{L2\ul FJ25} \\ \hline
\multicolumn{1}{|p{24pt}}{\raggedleft72.5}  &$<\pt\le$& \multicolumn{1}{p{24pt}|}{\raggedright95}& \ttt{L1\ul MBTS} & \ttt{L1\ul MBTS} & \ttt{L1\ul FJ30} & \ttt{L2\ul FJ25} \\ \hline
\multicolumn{1}{|p{24pt}}{\raggedleft95}    &$<\pt\le$& \multicolumn{1}{p{24pt}|}{\raggedright160}& \ttt{L1\ul MBTS} & \ttt{L1\ul MBTS} & \ttt{L1\ul FJ30} & \ttt{L2\ul FJ45} \\ \hline
                                     &$\;\;\;\;\pt\ge$& \multicolumn{1}{p{24pt}|}{\raggedright160}& \ttt{L1\ul MBTS} & \ttt{L1\ul MBTS} & \ttt{L1\ul FJ30} & \ttt{L2\ul FJ70} \\ \hline

\hline
\multicolumn{7}{c}{ } \\ [-5pt]\hline
\multicolumn{7}{|c|}{ $3.6 < \left|y\right| < 4.4$ } \\
\hline \hline

\multicolumn{1}{|p{24pt}}{\raggedleft20}    &$<\pt\le$& \multicolumn{1}{p{24pt}|}{\raggedright42.5}& \ttt{L1\ul MBTS} & \ttt{L1\ul MBTS} & \ttt{L1\ul FJ10} & \ttt{EF\ul MBTS} \\ \hline
\multicolumn{1}{|p{24pt}}{\raggedleft42.5}  &$<\pt\le$& \multicolumn{1}{p{24pt}|}{\raggedright50}& \ttt{L1\ul MBTS} & \ttt{L1\ul MBTS} & \ttt{L1\ul FJ10} & \ttt{L2\ul FJ25} \\ \hline
\multicolumn{1}{|p{24pt}}{\raggedleft50}    &$<\pt\le$& \multicolumn{1}{p{24pt}|}{\raggedright67.5}& \ttt{L1\ul MBTS} & \ttt{L1\ul MBTS} & \ttt{L1\ul FJ30} & \ttt{L2\ul FJ25} \\ \hline
\multicolumn{1}{|p{24pt}}{\raggedleft67.5}  &$<\pt\le$& \multicolumn{1}{p{24pt}|}{\raggedright100}& \ttt{L1\ul MBTS} & \ttt{L1\ul MBTS} & \ttt{L1\ul FJ30} & \ttt{L2\ul FJ45} \\ \hline
                                     &$\;\;\;\;\pt\ge$& \multicolumn{1}{p{24pt}|}{\raggedright100}& \ttt{L1\ul MBTS} & \ttt{L1\ul MBTS} & \ttt{L1\ul FJ30} & \ttt{L2\ul FJ70} \\ \hline
\hline
\end{Tabular} \end{center}
\caption{\label{TBLcentralTrigNames2010}The trigger chains used in the analysis for the 2010 data as a function of jet transverse momentum, \pt, in various
  data-taking periods and rapidity regions, $y$, as indicated in the table. Level~1 (\ttt{L1}) and Level~2 (\ttt{L2}) jet trigger names end
  with a number. This number stands for the trigger threshold in transverse energy at the \EM~scale, $E_{\mrm{t}}^{\mrm{EM}}$; \eg \ttt{L1\ul J5} is
  a Level~1 trigger with a threshold, $E_{\mrm{t}}^{\mrm{EM}} > 5$\GeV. The \ttt{L1} and Event Filter (\ttt{EF}) minimum bias triggers
  are respectively named \ttt{L1\ul MBTS} and \ttt{EF\ul MBTS}.
}
\end{table}
\begin{table}[htp] \begin{center}
\begin{Tabular}[1.3]{ |ccc|c| } \hline \hline
  \multicolumn{3}{|c|}{ \multirow{2}{*}{ \pt~[\GeV] } } & \hspace{40pt} Trigger name \hspace{40pt} \\
                                                     &&& Periods D,I-M \\ [2pt]
\hline \hline
\multicolumn{1}{|p{24pt}}{\raggedleft20}  &$<\pt\le$& \multicolumn{1}{p{26pt}|}{\raggedright47}&    \ttt{EF\ul rd0}    \\ \hline
\multicolumn{1}{|p{24pt}}{\raggedleft47}  &$<\pt\le$& \multicolumn{1}{p{26pt}|}{\raggedright52}&    \ttt{EF\ul J10}    \\ \hline
\multicolumn{1}{|p{24pt}}{\raggedleft52}  &$<\pt\le$& \multicolumn{1}{p{26pt}|}{\raggedright60}&    \ttt{EF\ul J15}    \\ \hline
\multicolumn{1}{|p{24pt}}{\raggedleft60}  &$<\pt\le$& \multicolumn{1}{p{26pt}|}{\raggedright82}&    \ttt{EF\ul J20}    \\ \hline
\multicolumn{1}{|p{24pt}}{\raggedleft82}  &$<\pt\le$& \multicolumn{1}{p{26pt}|}{\raggedright97}&    \ttt{EF\ul J30}    \\ \hline
\multicolumn{1}{|p{24pt}}{\raggedleft97}  &$<\pt\le$& \multicolumn{1}{p{26pt}|}{\raggedright122}&   \ttt{EF\ul J40}    \\ \hline
\multicolumn{1}{|p{24pt}}{\raggedleft122} &$<\pt\le$& \multicolumn{1}{p{26pt}|}{\raggedright160}&   \ttt{EF\ul J55}    \\ \hline
\multicolumn{1}{|p{24pt}}{\raggedleft160} &$<\pt\le$& \multicolumn{1}{p{26pt}|}{\raggedright207}&   \ttt{EF\ul J75}    \\ \hline
\multicolumn{1}{|p{24pt}}{\raggedleft207} &$<\pt\le$& \multicolumn{1}{p{26pt}|}{\raggedright275}&   \ttt{EF\ul J100}    \\ \hline
\multicolumn{1}{|p{24pt}}{\raggedleft275} &$<\pt\le$& \multicolumn{1}{p{26pt}|}{\raggedright365}&   \ttt{EF\ul J135}    \\ \hline
\multicolumn{1}{|p{24pt}}{\raggedleft365} &$<\pt\le$& \multicolumn{1}{p{26pt}|}{\raggedright472}&   \ttt{EF\ul J180}    \\ \hline
                                   &$\;\;\;\;\pt\ge$& \multicolumn{1}{p{26pt}|}{\raggedright472}&   \ttt{EF\ul J240}  \\ \hline
\hline
\end{Tabular} \end{center}
\caption{\label{TBLcentralTrigNames2011}The trigger chains used in the analysis of the 2011 data as a function of jet transverse momentum, \pt,
  as indicated in the table. Event Filter (\ttt{EF}) jet trigger names end
  with a number. This number stands for the trigger threshold in transverse energy at the
  \EM~scale, $E_{\mrm{t}}^{\mrm{EM}}$; \eg \ttt{EF\ul J10} is
  an Event Filter trigger with a threshold, $E_{\mrm{t}}^{\mrm{EM}} > 10$\GeV. The \ttt{EF} random trigger is named \ttt{EF\ul rd0}.
}
\end{table}
The triggers used for the 2011 data are listed in \autoref{TBLcentralTrigNames2011}. 
The \ttt{EF} triggers used in 2011 have efficiency $\gtrsim 99\%$ within the rapidity region
of interest ($\left|y\right| < 2.8$).

\subsection{Luminosity calculation using a two-trigger selection scheme\label{chapTwoTriggerLumiCalcScheme}}
%
A \textit{two-trigger} strategy is used in the analysis, following the prescription employed
in~\cite{Aad:2011fc}. An event may be accepted if at least one of the two leading (highest-\pt)
jets passes a dedicated trigger, determined by the transverse momentum and pseudo-rapidity of the jet.
The event is then assigned to one of two luminosity classes, $\mathcal{L}^{\mrm{jet}}_{\mrm{single}}$ or $\mathcal{L}^{\mrm{jet}}_{\mrm{double}}$. 
The former, a \textit{single-trigger} luminosity class, is
for events where the two jets are assigned to the same trigger. The latter, a
\textit{double-trigger} luminosity class, is for events in which each jet
is assigned to a different trigger.

In order to account for different prescale combinations of the two jets, 
the \textit{Inclusive method for fully efficient combinations}~\cite{Lendermann:2009ah} is used.
This method entails calculation of the integrated luminosity of each trigger over the entire
event sample, and combination of the per-trigger luminosities on an event-by-event basis.
Let $\mathcal{L}_{\mrm{LB}}$ denote the integrated luminosity in a luminosity-block (LB), and
let $\ttt{Ps}_{\mrm{LB}}$ denote the prescale of a given jet trigger within this LB.
The effective luminosity of a single trigger for the entire dataset can then be written as
\begin{equation}
  \mathcal{L}^{\mrm{jet}}_{\mrm{single}}
  = \sum_{   \mrm{LB}}{  \frac{ \mathcal{L}_{\mrm{LB}} } { \ttt{Ps}_{\mrm{LB}} }  }  \;,
\label{eqtrigJetLumi0} \end{equation}
where the summation is over all luminosity-blocks.
The effective luminosity for a pair of triggers takes into account the probability that two
triggers with prescales, ${\ttt{Ps}}^{0}_{\mrm{LB}}$ and $\ttt{Ps}^{1}_{\mrm{LB}}$,
fired simultaneously in a given event; it is defined by
\begin{equation}
  \mathcal{L}^{\mrm{jet}}_{\mrm{double}}
  = \sum_{   \mrm{LB}}{  \frac{ \mathcal{L}_{\mrm{LB}} } {
                                                            \ttt{Ps}^{0}_{\mrm{LB}}  \ttt{Ps}^{1}_{\mrm{LB}}
                                                            / \left(  \ttt{Ps}^{0}_{\mrm{LB}} 
                                                                      + \ttt{Ps}^{1}_{\mrm{LB}} -1 \right)
                                                          }  }  \;.
\label{eqtrigJetLumi0} \end{equation}
\Autorefs{TBLlumiInTrigBinsApp0}~-~\ref{TBLlumiInTrigBinsApp2} in \autoref{chapDataSelectionAndCalibrationApp} show
the integrated single- and double-trigger luminosities for the different trigger-bins used in the analysis.

The combined trigger efficiency of the two leading jets is 
computed in a similar manner as in \autoref{eqtrigJetLumi0}, using the corresponding per-jet trigger
efficiencies (given in \autoref{TBLcentralTrigEfficiency2010App} for the 2010 data and taken as unity for the 2011 data)
and the single-jet reconstruction efficiency (see \autoref{chapJetQualitySelection}).
Together these yield the probability for an event to have fired a given trigger (or trigger combination),
and for a jet (or pair of jets) to have been reconstructed and matched to the trigger(s).
The inverse of the combined trigger and reconstruction efficiency factor is multiplied by the luminosity of the
selected luminosity class; this produces the final luminosity weight for each event.
          }{}
\ifthenelse{\boolean{do:jetMedian}}         { 
\chapter{The jet area/median method for \pu subtraction\label{chapJetAreaMethod}}
%
%
As discussed in \autoref{chapJetEnergyCalibration}, an important component of the energy calibration of jets
is the subtraction of \pu, that is the extra energy which originates from additional $pp$ interactions
that occurred in the same bunch-crossing.
The current strategy in \ATLAS for dealing with \pu involves the subtraction of an average
correction derived from MC simulation, called the offset correction.
The offset correction is parametrized in variables which reflect the magnitude of the different components of the
\pu in a given event. As the parametrization is derived from MC, it suffers from uncertainties due
to miss-modelling of the \pu and requires adjustments based on \insitu measurements.

In the following, an alternative data-driven method is introduced, called the jet area/median method.
The median method combines parametrization of the average
contributions of \pu in data, as well as an event-by-event estimation of the local \pu energy around the jets of interest.
The median correction is used to calibrate the energy of jets in data taken during 2011, and the performance is
compared to the default offset correction.

As mentioned in \autoref{chapSimulationOfTheATLASdetector}, a few of the results shown in the following
use the transitional simulation, MC10b. Unless otherwise indicated, MC11 is used.

\section{\Pu in \ATLAS\label{chappuInAtlas}}
%
\subsection{In- and out-of-time \pu \label{chapInAndOutOfTimePU}}
%
\Pu in \ATLAS is usually separated into \textit{\itpu} and \textit{\otpu}. The former refers to the number of \pp collisions which occur within the same
bunch crossing as the collision of interest, and the latter to interactions in previous or subsequent bunch crossings. The exact nature of \otpu
depends on the bunch-structure of the LHC, and on the specifics of the reconstruction of calorimeter signals.

%
%

%
%
For the data used in this study, the spacing between adjacent bunches was 50\ns.
This number should be compared with the typical signal shaping time of
the liquid Argon calorimeters, which is roughly 400\ns~\cite{ATLASLArTDR}, as illustrated in \autoref{FIGLArPulseShape}.
\begin{figure}[htp]
\begin{center}
\includegraphics[width=.5\textwidth]{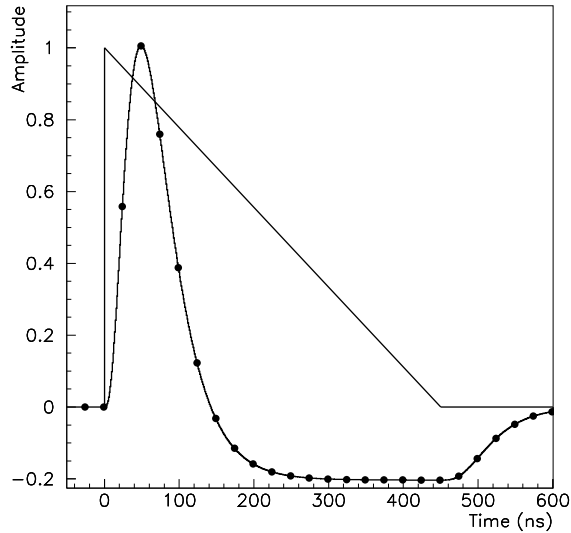}
  \caption{\label{FIGLArPulseShape}Signal shape as produced in the detector (triangle), and after shaping (curve with dots).
  The dots represent the position of the successive bunch crossings for a bunch
  spacing configuration of 25\ns. (Figure taken from~\protect\cite{ATLASLArTDR}.)
  }
\end{center}
\end{figure} 
The figure shows the shaped signal in a typical calorimeter element (cell). The shaping integrates the electronic signal
registered in the cell in two phases. In the first phase, which takes $\sim100\ns$, the signal is integrated with a positive
amplitude. In the second phase, which takes $\sim400\ns$, the signal is integrated with a negative amplitude; the latter
is smaller in magnitude compared to the positive amplitude in the first phase. The shaping is such that the total integral of the positive- and the
negative-amplitude phases is zero, such that for the case of random noise, the integrated signal is minimal.

A consequence of the long integration time (compared to the short bunch crossing rate) is that
the energy deposits from several bunch crossings are typically overlaid with any event of interest.
The situation is further complicated by the mixture of positive and negative amplitudes of the shaping.
The resulting \pu signal can therefore be composed of
both positive- and negative-energy components within a given 50\ns readout window.
The shaping is designed such that \otpu should have an overall negative effect
that cancels the positive effect from \itpu on average.
However, this only works if the bunch intensity and subsequent occupancy of cells
is constant over the integration time of the calorimeter.
By design of the shaping amplifier, the most efficient suppression is achieved for 25\ns bunch spacing in the
LHC beams. The 2011 beam conditions, with 50\ns bunch spacing, do not allow for full \pu suppression by signal shaping.
In addition, a given cell would not necessarily have contributions from \pu from every bunch crossing, and so while there 
is approximate cancellation of positive and negative \pu on average, there are large fluctuations on an even-by-event basis.
Special attention must also be given to the position of the bunch crossing within the bunch train.
For example, at the beginning of a bunch train there is insufficient \otpu to cancel the \itpu;
the calorimeter response in such events is therefore systematically higher.

During reconstruction, calorimeter cells are combined into \topos
(see \autoref{chapInputJetReconstruction}). Due to the presence of the
negative-energy signals, two effects come into play;
the positive energy of positive \topos is reduced, and purely negative-energy
\topos are formed. In the current jet reconstruction scheme of \ATLAS, there is no calibration for
negative-energy jet constituents, therefore these are discarded and not used in the reconstructing of jets.
The energy of a jet which includes positive- and negative-energy cells is subsequently not a simple linear combination
of cell energies. Rather, it is the sum of energies of those positive-energy clusters, which had enough positive-energy
cells to pass the reconstruction threshold.

For the 2010 data, the relatively large spacing between adjacent bunches and the low instantaneous luminosity mean that
the magnitude of the \otpu is negligible. Consequently, \otpu is only discussed here in the context of
the 2011 data and MC.

On average, one can parametrize the amount of energy originating from \pu by the number of interactions, which is measured by
counting the number of reconstructed vertices in a given event, \Npv.

Vertices are reconstructed from ID tracks, with a requirement
of at least five or three tracks per vertex for data taken respectively during 2010 or 2011.
Tracks are selected in 2010 and in 2011 using the following criteria:
\mynobreakpar
\begin{center}
\begin{Tabular}[1.6]{ccc} 
\multirow{2}{*}{2010:} &        & $p_{\mrm{t}}^{\mrm{trk}} >150\MeV \;, \ N_{\rm pixel}+\ N_{\rm SCT} \ge 7 $ \\
                       &        & $|d_0| < 1\,{\rm mm} \;, \ |z_0| < 1.5\,{\rm mm} $ \\ \hline
\multirow{2}{*}{2011:} &        & $p_{\mrm{t}}^{\mrm{trk}} >500\MeV \;, \ N_{\rm pixel} \ge 1 \;, \ N_{\rm SCT} \ge 6 $ \\
                       &        & $|d_0| < 1\,{\rm mm} \;, \ |z_0\sin\theta| < 1\,{\rm mm} \;, \left( \chi^{2}/\mrm{NDF} \right)_{\mrm{trk}} < 3.5 $
\end{Tabular}
\label{eqTrackSelectionCriteria}\end{center}
Here $p_{\mrm{t}}^{\mrm{trk}}$ is the transverse momentum of a track and
$N_{\rm pixel}$ and $N_{\rm SCT}$ are respectively the number of hits from the
pixel and SCT detectors that are associated with the track. The parameters
$d_0$ and $z_0$ are the transverse and longitudinal impact parameters 
measured with respect to the vertex to which the tracks are extrapolated. 
The ratio $\left( \chi^{2}/\mrm{NDF} \right)_{\mrm{trk}}$ represents the quality
of the fit of the track parameters divided by the number of degrees of freedom of the fit.
Vertices are also required to be consistent with the beam-spot.

The constraints on vertex reconstruction used here for the 2011 data are tighter than what is usually used in analyses in \ATLAS,
in that the common 2011 selection criteria do not impose the track selection requirements described above.
Vertices selected with these looser constraints will be referred to as \NpvL in the following. The relation
between the average of \Npv and \NpvL is shown in \autoref{FIGnpvMuAvg1} for reference. The relation between the two
definitions is linear, suggesting that on average the two factors provide equivalent characterization of the amount of \itpu in
an event. The MC describes the data well.

The number of in-time interactions, \Npv, contributes purely-positive \pu energy to a given event.
Depending on the position of a given collision
within the bunch train, \otpu has on average different negative and positive energy components. In order to estimate the amount of
\otpu, one would in principle need to measure \Npv during collisions before and after the event of interest, within a time
window of~$\sim400$\ns. Due to the trigger strategy of \ATLAS, these events
may have been discarded. Subsequently, the only handle on \otpu is the average number of \pp collisions
in a given time-frame, which depends on the instantaneous luminosity. The number of collisions is described by
a Poissonian distribution with some mean value, \Mu. The mean is computed over
a rather large window in time, $\Delta t$, which safely encompasses
the time interval during which the \ATLAS calorimeter signal is sensitive to the activity in the collision
history ($\Delta t \gg 400$\ns for the liquid argon calorimeters). The average number of interactions can
be reconstructed from the average luminosity over this time window, $\mathcal{L}_{\mrm{avg}}$, the total inelastic $pp$ \xsec,
$\sigma_{\mrm{inel}} = 71.5$~\mb~\cite{Aad:2011eu}, the number of colliding bunches in LHC, $N_{\mrm{bunch}}$,
and the LHC revolution frequency, $f_{\mrm{rev}}$,~\cite{Aad:2011dr}
\begin{equation}
  \mu = \frac{\mathcal{L}_{\mrm{avg}} \times \sigma_{\mrm{inel}}}{N_{\mrm{bunch}} \times f_{\mrm{rev}}} \;.
\label{eqDefinitionMu} \end{equation}

The correlation between \Mu and the average \Npv is shown in \autoref{FIGnpvMuAvg2}
for data taken during 2011 and for the nominal MC11 \pythia simulation.
%
%
\begin{figure}[htp]
\begin{center}
\subfloat[]{\label{FIGnpvMuAvg1}\includegraphics[trim=5mm 14mm 0mm 10mm,clip,width=.52\textwidth]{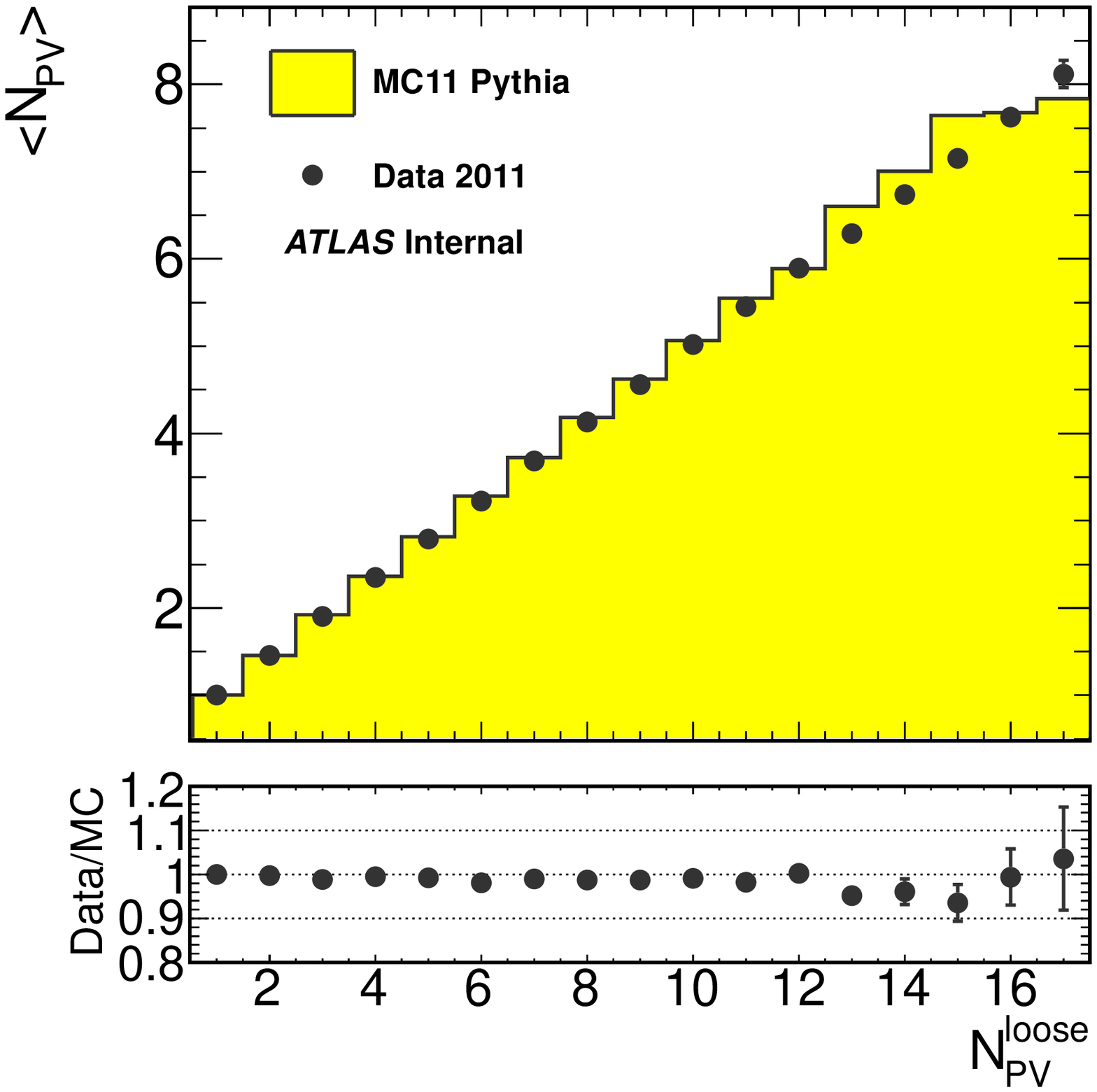}}
\subfloat[]{\label{FIGnpvMuAvg2}\includegraphics[trim=5mm 14mm 0mm 10mm,clip,width=.52\textwidth]{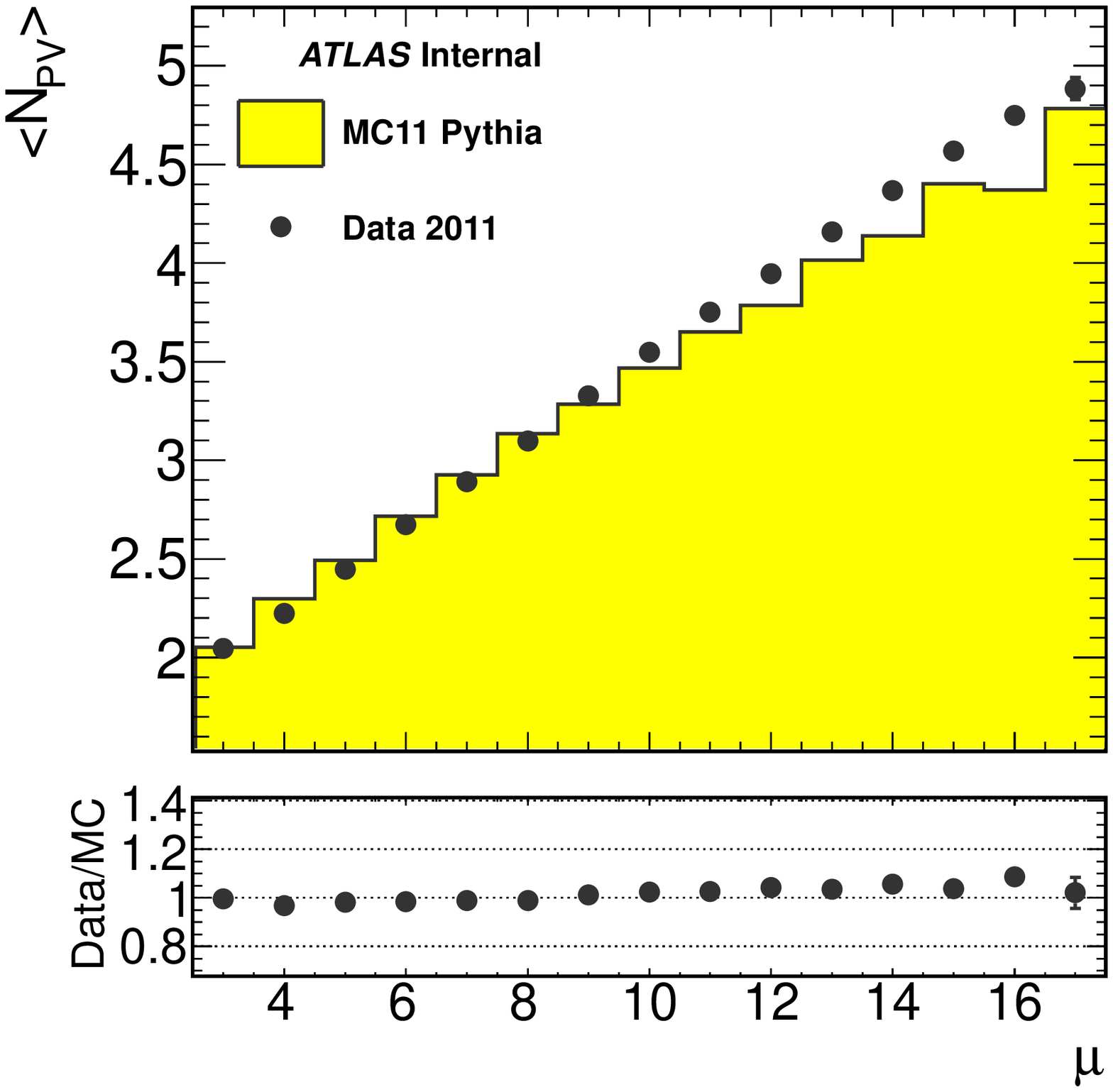}}
\caption{\label{FIGnpvMuAvg}\Subref{FIGnpvMuAvg1} Dependence of the average number of reconstructed vertices, 
  $<\Npv>$, on the number of vertices which are reconstructed with the loose quality criteria, \NpvL. \\
  \Subref{FIGnpvMuAvg2} The average, $<\Npv>$, as a function of the average number of interactions, \Mu. \\
  As indicated in the figures, data taken during 2011 are compared to MC11 in the top panels, and
  the ratio of data to MC is shown in the bottom panels.
}
\end{center}
\end{figure} 
There is strong correlation between the two factors, as expected. Good agreement is observed between the data and the MC.

In order to properly describe \pu in data,
MC samples are weighted such that the distributions of the 
number of reconstructed vertices in MC10 and MC11 match the respective distributions in
the 2010 and in the 2011 data.
The reweighted MC \Npv distributions are compared to the ones in the data in \autoref{FIGNpvDist2011DataMc}.
Full agreement is achieved.
%
%
\begin{figure}[htp]
\begin{center}
\subfloat[]{\label{FIGNpvDist2011DataMc1}\includegraphics[trim=5mm 14mm 0mm 10mm,clip,width=.52\textwidth]{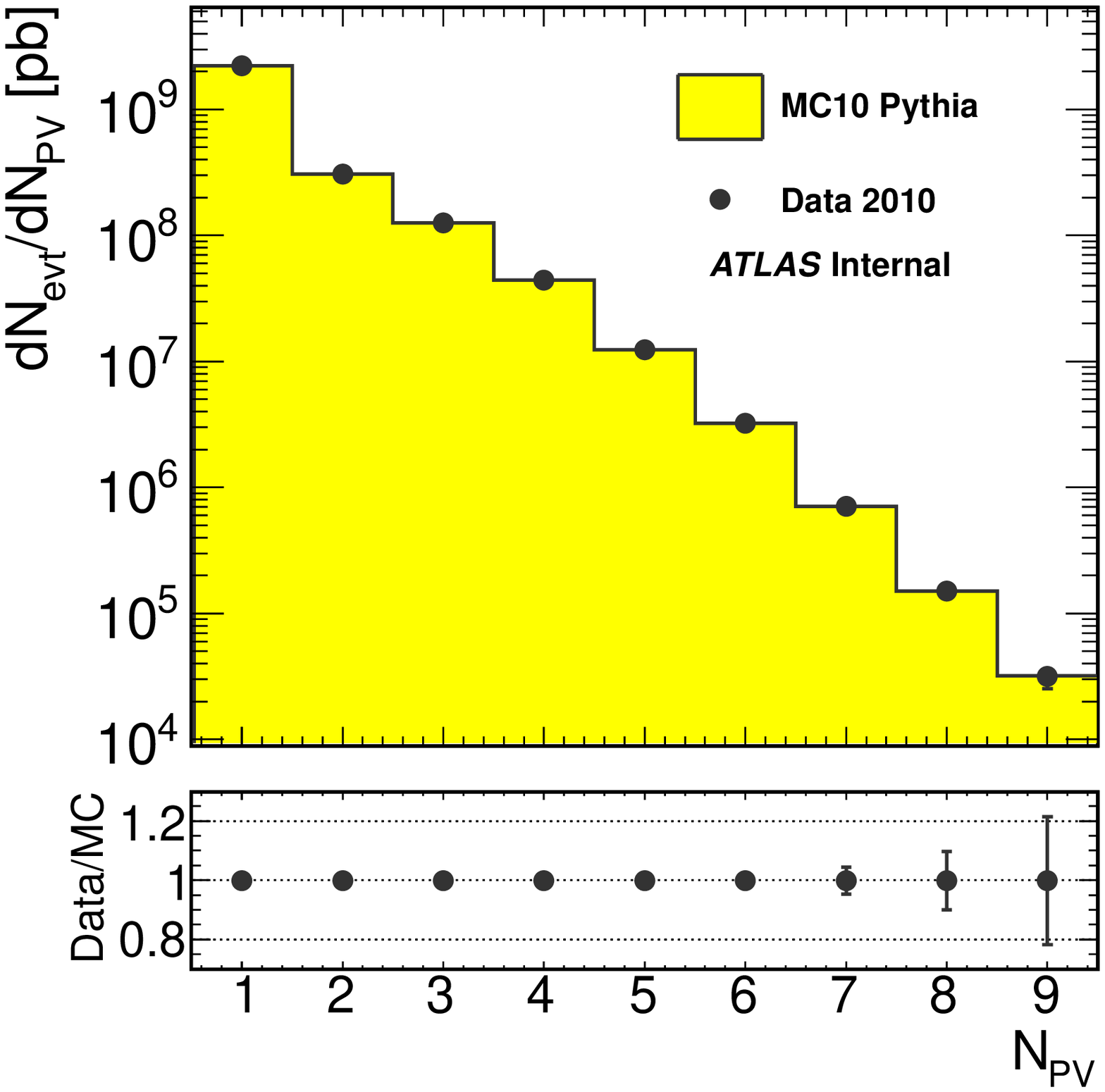}}
\subfloat[]{\label{FIGNpvDist2011DataMc2}\includegraphics[trim=5mm 14mm 0mm 10mm,clip,width=.52\textwidth]{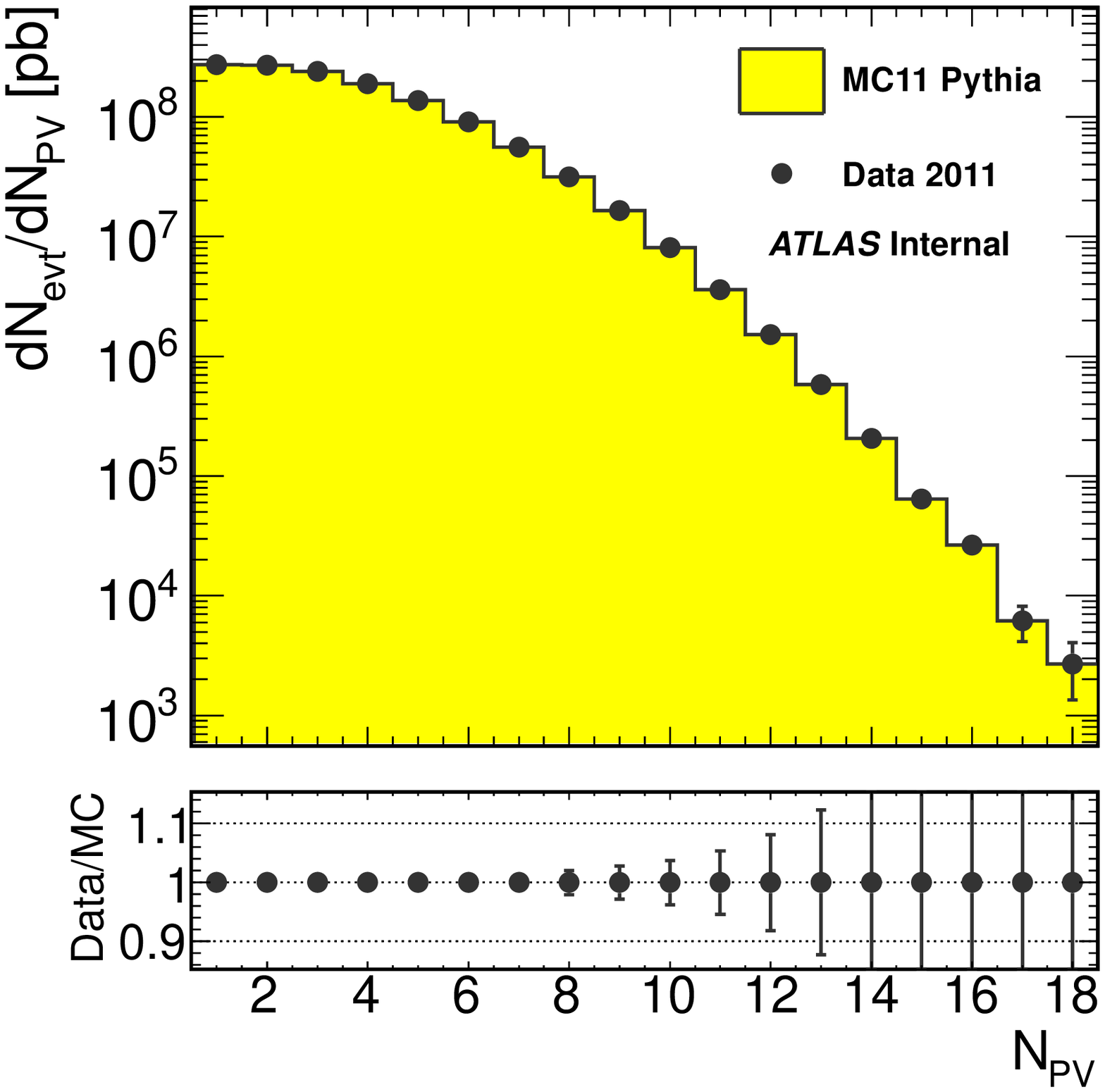}}
  \caption{\label{FIGNpvDist2011DataMc}Distribution of the reconstructed number of vertices, \Npv,
  for data taken in 2010 \Subref{FIGNpvDist2011DataMc1}
  and in 2011 \Subref{FIGNpvDist2011DataMc2}, and the respective distributions in MC10 and MC11. The top panels
  show the \Npv distributions and the bottom panels show the respective ratio of data to MC.
  }
\end{center}
\end{figure} 

Comparison of the \Npv distributions between 2010 and 2011 also
helps to emphasize the increase in \pu in 2011 compared to the previous year.
In 2010, the percentage of single- double- and triple-vertex events out of the full event-sample
are~81.8\%,~11.2\% and~4.6\%. In 2011,
the corresponding values are~20.7\%,~20.5\% and~18.1\%. Thus, more than~79\% of events in 2011 are multi-vertex
events, and require dedicated calibration. The standard method to remove \pu from jets in \atlas is called
the jet offset correction, and is the subject of the following section.

\subsection{The jet offset \pu correction\label{sectDefaultOffsetCorrection}}
%
%
As mentioned in \autoref{chapJetEnergyCalibration}, \pu changes the energy of jets.
An example is shown in \autoref{FIGjesBeforeOffsetCor}.
\begin{figure}[htp]
\begin{center}
\includegraphics[trim=5mm 5mm 0mm 0mm,clip,width=.44\textwidth]{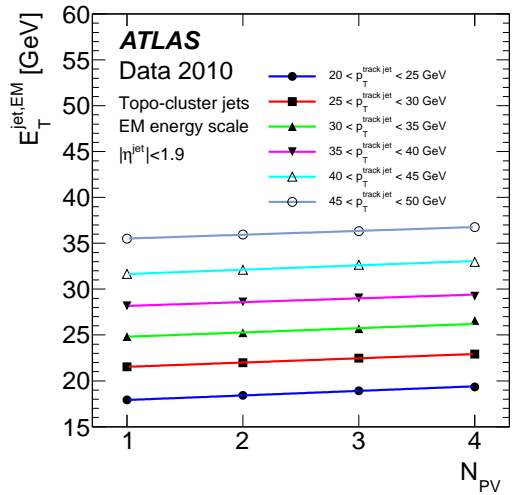}
  \caption{\label{FIGjesBeforeOffsetCor}The most probable value of the 
  transverse jet energy, $E_{\mrm{T}}^{\mrm{jet,EM}}$, of
  calorimeter jets with pseudo-rapidity, $|\eta^{\mrm{jet}}|<1.9$, measured in 2010 at the \EM scale,
  as a function of the number of reconstructed vertices, \Npv.
  The different symbols represent calorimeter jets which are associated to track jets
  with different transverse momenta, $p_{\mrm{T}}^{\mrm{track\;jet}}$, as indicated.
  (Figure taken from~\protect\cite{Aad:2011he}.)
  }
\end{center}
\end{figure} 
The figure shows the transverse energy of calorimeter jets which
are matched with track jets. Track jets are insensitive to \pu effects,
and therefore represent the unbiased energy scale of the respective calorimeter jets. This scale is
represented by the transverse momentum of the track jets, $p_{\mrm{T}}^{\mrm{track\;jet}}$.
Within a given bin of $p_{\mrm{T}}^{\mrm{track\;jet}}$, one would expect to measure a stable (within energy resolution)
value of calorimeter jet \pt. However, as \pu increases (\Npv grows), so does
the energy of calorimeter jets in each $p_{\mrm{T}}^{\mrm{track\;jet}}$ bin.

\Pu must therefore be subtracted as part of the calibration procedure of jets. The current strategy in \ATLAS analyses involves linear
parametrization of the energy content of jets due to \pu, called the \textit{offset} of a jet.
Different methods are used to estimate to offset in 2010 and in 2011, due to the variant nature of \pu discussed above.
The parametrizations are
rapidity dependent due to the differences in calorimeter granularity and occupancy after \topo reconstruction;
the differences in the signal shaping in the calorimeter; and the differences in the energy flow of the \pu across the detector.

\minisec{The offset correction in 2010}
%
%
The \pu correction devised for 2010 is derived from MC. The correction
is a function of \itpu, estimated by the number of
reconstructed vertices, \Npv. It is based on calorimeter towers, which are
static, ${\Delta\eta\times\Delta\phi=0.1\times0.1}$, grid elements built directly from calorimeter cells.
Calorimeter towers are calibrated at the \EM scale. As such they are appropriate for use in correcting the
energy of \EMJES calibrated jets.

The calorimeter \textit{tower offset}, $\mathcal{O}_{\mrm{fst\;(tow)}}^{\EM}$,
is determined by measuring the average transverse energy 
of towers, $E^{\rm tower}_{\rm t}$, in events with different values of \Npv. This average is then compared
to the one for events with ${\Npv=\Npv^{\mrm{ref}}=1}$,
\begin{equation}
  \mathcal{O}_{\mrm{fst\;(tow)}}^{\EM}(\eta, \Npv)
  = \langle E^{\rm tower}_{\rm t}(\eta,\Npv)\rangle - \langle E^{\rm tower}_{\rm t}(\eta,N_{\mrm{PV}}^{\mrm{ref}} )\rangle \;.
\end{equation}
The tower offset is an estimation of the local average \pu density. The
\EM jet offset, $\mathcal{O}_{\mrm{fst}}^{\EM}$,
may therefore be approximated as this density,
multiplied by the \textit{effective tower area} of a jet, $\Ajet^{\rm tower}$,
\begin{equation}
  \mathcal{O}_{\mrm{fst}}^{\EM} = \mathcal{O}_{\mrm{fst\;(tow)}}^{\EM} \times \Ajet^{\rm tower} \;.
\end{equation}
The effective tower area stands for the number of towers contained within the jet.
In order to derive $\Ajet^{\rm tower}$ for a given jet, calorimeter cells belonging to the constituent \topos are used.

\minisec{The offset correction in 2011}
%
%
With the increased instantaneous luminosity and the shorter bunch spacing,
the \LC offset correction in 2011, $\mathcal{O}_{\mrm{fst}}^{\LC}$, 
depends on \otpu as well as on \itpu.
These two components are respectively parametrized as 
proportional to the number of reconstructed vertices, \Npv, and to the
average number of interactions, \Mu. The offset in derived from MC simulation,
\begin{eqnarray}
  \mathcal{O}_{\mrm{fst}}^{\LC} \left( N_{\mrm{PV}},\mu,\eta \right) & = &
        p_{\mrm{t}}^{\mrm{\LC}} \left( N_{\mrm{PV}},\mu,\eta \right) - p_{\mrm{t}}^{\mrm{truth(\LC)}} \nonumber \\
  & = & \frac{\partial p_{\mrm{t}}^{\mrm{\LC}}}{\partial N_{\mrm{PV}}}
              \left( \eta \right) \times \left(  N_{\mrm{PV}}- N_{\mrm{PV}}^{\mrm{ref}} \right) +
        \frac{\partial p_{\mrm{t}}^{\mrm{\LC}}}{\partial \mu}
                \left( \eta \right) \times \left(  \mu-\mu^{\mrm{ref}} \right) \nonumber \\
  & = & \alpha \left( \eta \right) \times \left(  N_{\mrm{PV}}- N_{\mrm{PV}}^{\mrm{ref}} \right) +
         \beta \left( \eta \right) \times \left(  \mu-\mu^{\mrm{ref}} \right) \;.
\label{eqDefinitionDefaultOffset} \end{eqnarray}
The offset is defined for a jet with transverse momentum at the \LC scale, $p_{\mrm{t}}^{\mrm{\LC}}$, which is associated with a truth jet with
transverse momentum (at the \LC scale), $p_{\mrm{t}}^{\mrm{truth(\LC)}}$. The reference values for \Npv and \Mu,
$N_{\mrm{PV}}^{\mrm{ref}}=1$ and $\mu^{\mrm{ref}}=8$, are set such that
$\mathcal{O}_{\mrm{fst}}^{\LC} \left( N_{\mrm{PV}}^{\mrm{ref}},\mu^{\mrm{ref}},\eta \right) = 0$.

\minisec{Using the offset to correct the energy scale of jets}
%
%
The offset is determined in MC by matching calorimeter jets to truth jets.
In this way, the average tower transverse energy (for 2010) and the 
coefficients $\alpha \left( \eta \right)$ and $\beta \left( \eta \right)$ (for 2011), are derived.

It is assumed that the \pu contamination affects the momentum and the energy of jets in
a similar manner\footnote{ This is not necessarily true for the effect of \pu on the mass of jets. For the mass,
sophisticated methods may be needed, such as jet grooming techniques aiming at removing jet
constituents associated with \pu~\cite{ATLAS:2012kla}. Such procedures are, however, beyond the scope of this analysis.
}.
Jets in the data and in the MC are corrected for \pu by estimating the percentage of
the jet transverse energy or transverse momentum originating from \pu,
\begin{equation}
  \renewcommand{\arraystretch}{1.9} 
  \wp_{\mrm{t}}^{\;\mathcal{O}_{\mrm{fst}}^{\Scale}} 
       = \left\{
         \begin{array}{cl}
            \scalemath{1.02}{\frac{\mathcal{O}_{\mrm{fst}}^{\EM} \left( N_{\mrm{PV}},\eta \right)}{E_{\mrm{t}}^{\mrm{\EM}}}}
            & \quad \mathrm{(2010)} \\
            \scalemath{1.02}{\frac{\mathcal{O}_{\mrm{fst}}^{\LC} \left( N_{\mrm{PV}},\mu,\eta \right)}{p_{\mrm{t}}^{\mrm{\LC}}}}
            & \quad \mathrm{(2011)}
         \end{array}\right. .
\label{eqPuCorrection0} \end{equation}

The four-momentum of a jet is modified for \pu by the transformation,
\begin{equation}{
    E^{\mrm{det}} \rightarrow E^{\mrm{\Scale}} \times \left( 1-\wp_{\mrm{t}}^{\;\mathcal{O}_{\mrm{fst}}^{\Scale}} \right) \quad,\quad
    \eta \rightarrow \eta \quad,\quad \phi \rightarrow \phi \quad,\quad
    p_{\mrm{t}}^{\mrm{det}} \rightarrow p_{\mrm{t}}^{\mrm{\Scale}} \times \left( 1-\wp_{\mrm{t}}^{\;\mathcal{O}_{\mrm{fst}}^{\Scale}} \right)\; \;,
}\label{eqPuCorrection1} \end{equation}
where $E^{\mrm{\Scale}}$ ($p_{\mrm{t}}^{\mrm{\Scale}}$) and $E^{\mrm{det}}$ ($p_{\mrm{t}}^{\mrm{det}}$) are
the \Scale-scale energy (transverse momentum) of a jet, respectively before and after \pu subtraction,
and $\eta$ and $\phi$ are respectively its pseudo-rapidity and azimuthal angle. 
Here, as defined previously, \Scale stands for
the \EM scale or the \LC scale for jets in 2010 or 2011 respectively.
Following the \pu subtraction, the \JES correction, $\mathcal{F}^{{\rm calib}}_{\mrm{\Scale}}$,
is applied, taking the jet from the \EM or \LC scale to the hadronic energy-scale (see \autoref{eqCalibJetEnergy}).
The fully calibrated jet has transverse momentum,
\begin{equation}{
  p_{\mrm{t}}^{\rm calib} = p_{\mrm{t}}^{\rm det} \times \mathcal{F}^{{\rm calib}}_{\mrm{\Scale}}\left(p_{\mrm{t}}^{\rm det},\eta\right) \; .
}\label{eqPuCorrection2} \end{equation}
%
%

\section{A \pu correction using the jet area/median method}
%
%
A novel approach to estimate the \pu energy content of jets, the \textit{jet area/median method},
has been proposed by Cacciari, Salam and Soyez~\cite{Cacciari:2008gn,Cacciari:2007fd}. Therein, under the assumption that \pu is  
distributed uniformly in a given event, the \pu transverse momentum density, \Rho, may be calculated, and then subtracted from the jets which
originate from the hard scatter. In order to do this, one must first define the \textit{jet area}, $\Ajet$.
One possible definition of the area, used in the following, is the so-called 
\textit{active area}. The area is calculated by adding to the event a dense coverage of infinitely soft
(zero momentum) \textit{ghost particles} and counting how many
are clustered inside a given jet. It can be understood as a measure of the susceptibility of a jet
to contamination from an underlying distribution of particles with uniform, diffuse structure.
A possible way to calculate the \pt-density in a given event,
is by taking the median of the distribution of the \pt-density of all jets in the event,
\begin{equation}{
  \rho_{\mrm{evt}} = \mathop{\mrm{median}}_{j \in \mrm{jets}} 
    \left[ \left\{ \frac{p_{\mrm{t},j}}{A_{\mrm{jet},j}}\right\}\right] \; .
}\label{eqRhoDefinition} \end{equation}
\noindent The logic behind the use of the median is that it is much
less susceptible to contamination from outliers (i.e. hard perturbative jets) than the mean.
A \pu correction to the transverse momentum of a jet could then take the form
\begin{equation}{
    p_{\mrm{t}}^{\mrm{det}} = p_{\mrm{t}}^{\mrm{\LC}} - \Ajet \cdot \rho_{\mrm{evt}} \; ,
}\label{eqPuEvtCorrection} \end{equation}
where $p_{\mrm{t}}^{\mrm{\LC}}$ and $p_{\mrm{t}}^{\mrm{det}}$ are the \LC scale
transverse momenta of a jet with area $\Ajet$, respectively before and after \pu subtraction.

It is important to note that according to this definition, $\rho_{\mrm{evt}}$ encompasses contributions from the underlying event (UE) as well
as from \pu. An alternative approach, which only subtracts \pu energy, is to parametrize the dependence of $\rho_{\mrm{evt}}$
on the amount of \pu by determining an average response as a function of the number of interactions.
The drawback of this approach as opposed to using $\rho_{\mrm{evt}}$ directly,
is that one has to determine beforehand the average \pt-density, a quantity which depends on specific trigger and selection cuts.
A combination of the average and the event-by-event approaches is used in this analysis.

Two important choices which have to be considered are those of the jet algorithm and the jet radius, $R$, which
are used to calculate $\rho_{\mrm{evt}}$.
Both the \KT~\cite{Catani:1993hr,Ellis:1993tq} and the \CAalg~\cite{Dokshitzer:1997in,Wobisch:1998wt}
algorithms are suitable options, because they produce jets whose area
distribution is representative of the jet particle content. In contrast, algorithms that give mostly conical
jets (like \AKT~\cite{Cacciari:2008gp}  and, to a lesser extent, \SISCONE~\cite{Salam:2007xv} ) tend not to be
suitable, because they fill in the holes
between the cones with jets with very small areas, which can have unrepresentative \pt-density
values. The question of what $R$ value to use is one of the freedoms of the method.
The effect on the performance of changing the baseline parameters of the median
calculation are studied in \autoref{sectStabilityOfMedianForDifferentAlgoParam}.

In the following, unless otherwise stated, the median is calculated using \KT jets with $R = 0.4$.
The constituents which compose these jets are \topos, which are calibrated at the \LC energy scale.

\subsection{Parametrization of the average \pu energy\label{sectParametrizationOfAvgPU}}

In order to take into account the rapidity-dependence of the calorimeter-response and of the
rate of scattering due to multiple \pp collisions, the \pt-density of jets is calculated in bins of \eta. 
\Autoref{eqRhoDefinition} is therefore modified,
such that the median, \Rho, is calculated using jets within a limited \eta-window of size $2\delta \eta = 0.2$,
\begin{equation}
  \rho \equiv \rho(\left|\eta\right|) = \mathop{\mrm{median}}_{j \in \mrm{jets},\;
  |\eta_j - \eta|< \delta \eta} \left[ \left\{ \frac{\pt^j}{\Ajet^j}\right\}\right]\,.
\label{eqMedianInEta} \end{equation}

The median is calculated on an event-by-event basis from distributions of the \pt-density of
jets, calibrated at the \LC scale.
\Autoref{FIGptOverAdist1} shows distributions in
data\footnote{ Here and in the following, events in data are selected after having passed one of the jet triggers.
The random trigger will be addressed in \autoref{sectSystematicChecksMedianCorrection}.}
and in MC of the density, $\pt/\Ajet$.
The corresponding distributions of \Rho are presented in \autoref{FIGptOverAdist2}.
\begin{figure}[htp]
\begin{center}
\subfloat[]{\label{FIGptOverAdist1}\includegraphics[trim=5mm 14mm 0mm 10mm,clip,width=.52\textwidth]{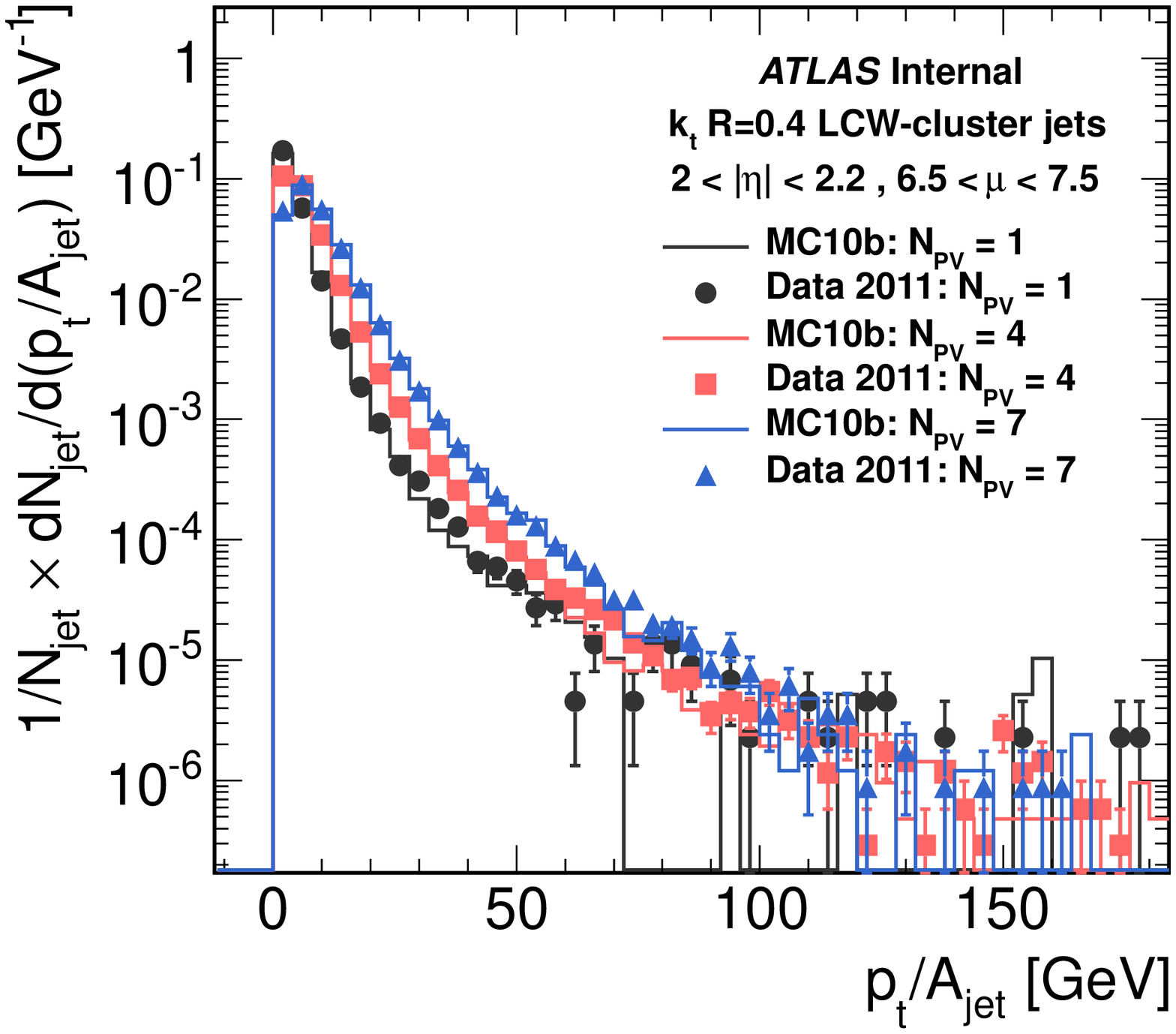}}
\subfloat[]{\label{FIGptOverAdist2}\includegraphics[trim=5mm 14mm 0mm 10mm,clip,width=.52\textwidth]{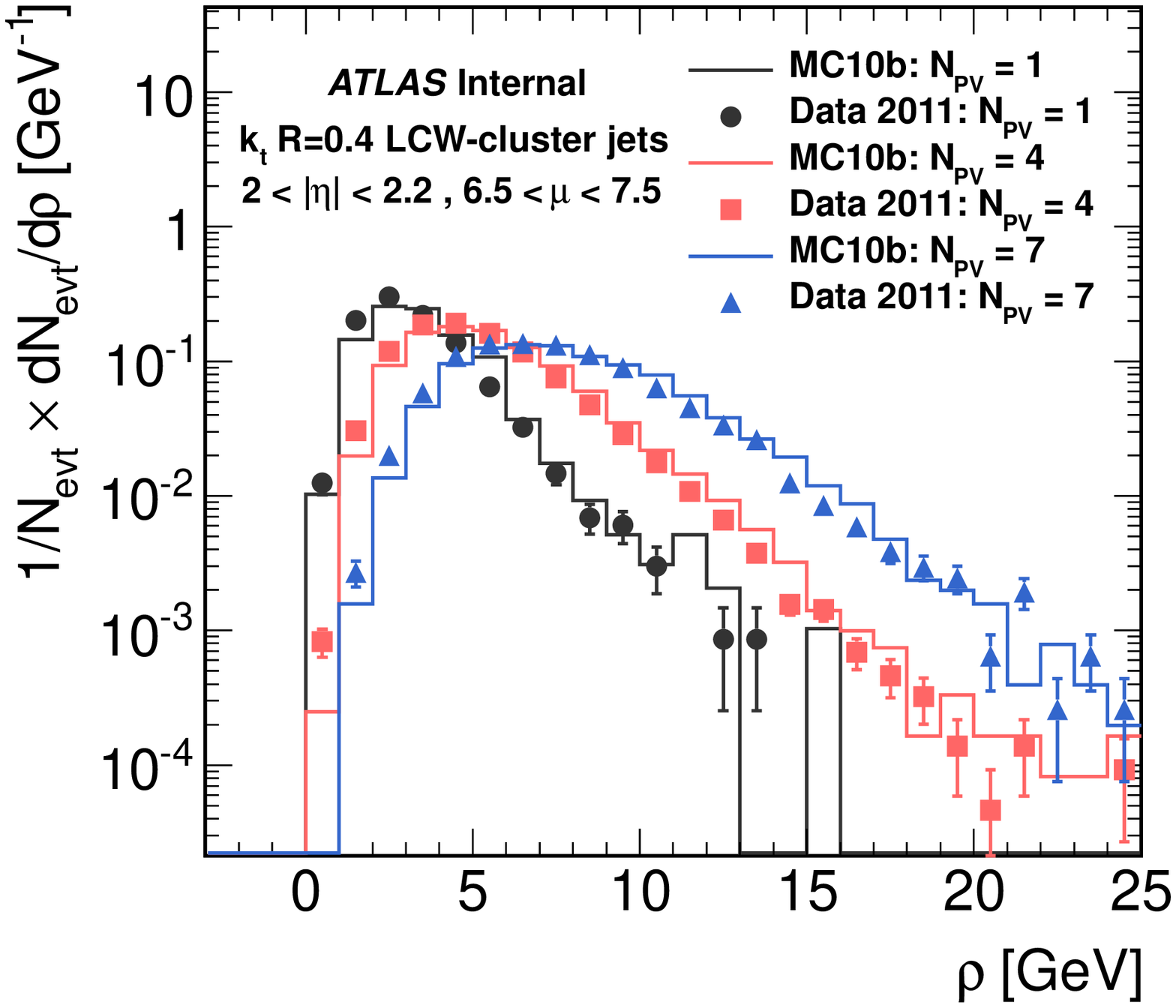}}
\caption{\label{FIGptOverAdist}Distributions of the \pt-density of jets, $\pt/\Ajet$
  \Subref{FIGptOverAdist1} and of the corresponding median, \Rho, \Subref{FIGptOverAdist2}, in data and in MC10b, as indicated in the figures.
  Jets are constructed with the \KT algorithm with a size parameter, $R = 0.4$, and have pseudo-rapidity, \etaRange{2}{2.2}.
  Events with an average number of interactions, \muRange{6.5}{7.5}, and several values of reconstructed vertices, \Npv, are used.
}
\end{center}
\end{figure} 
The agreement between the data and the MC is very good. The observed trend is for the \pt-density to increase with \Npv. 
The median of the distribution of the density also grows
in magnitude as \pu increases.
The distribution of the median has long tails at high values of \Rho. Despite this fact, it will be shown in the following that
the average value of \Rho is a good estimator for the contribution of \pu in an event.

The average median for events with $N_{\mrm{PV}} = 1$ and different \Mu ranges is shown in \autoref{FIGavgRhoEtaInMuBins}
for data and for MC.
\begin{figure}[htp]
\begin{center}
\subfloat[]{\label{FIGavgRhoEtaInMuBins1}\includegraphics[trim=5mm 10mm 65mm 10mm,clip,width=.478\textwidth]{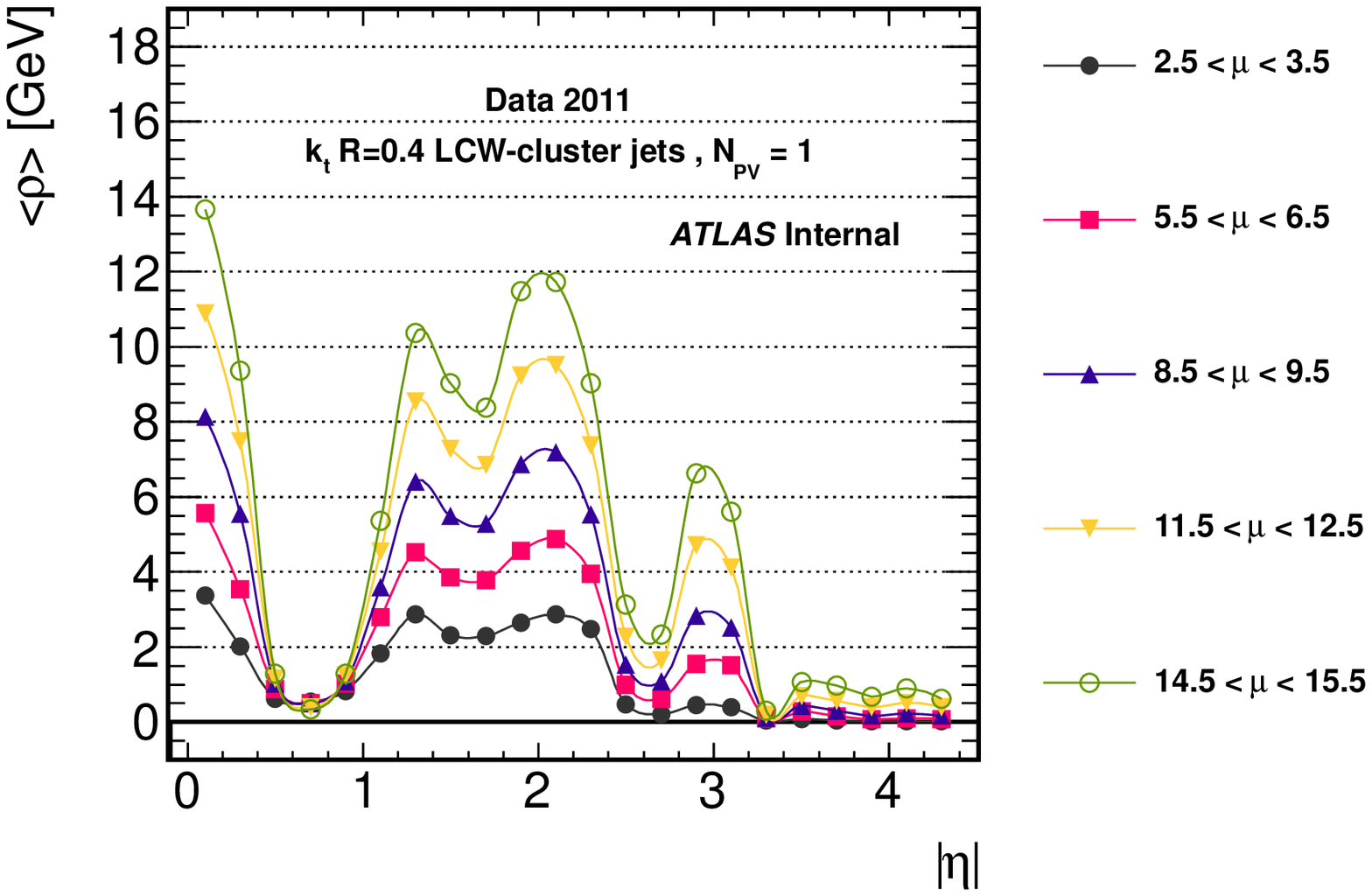}} 
\subfloat[\figQaud]{\label{FIGavgRhoEtaInMuBins2}\includegraphics[trim=5mm 10mm 0mm 10mm,clip,width=.715\textwidth]{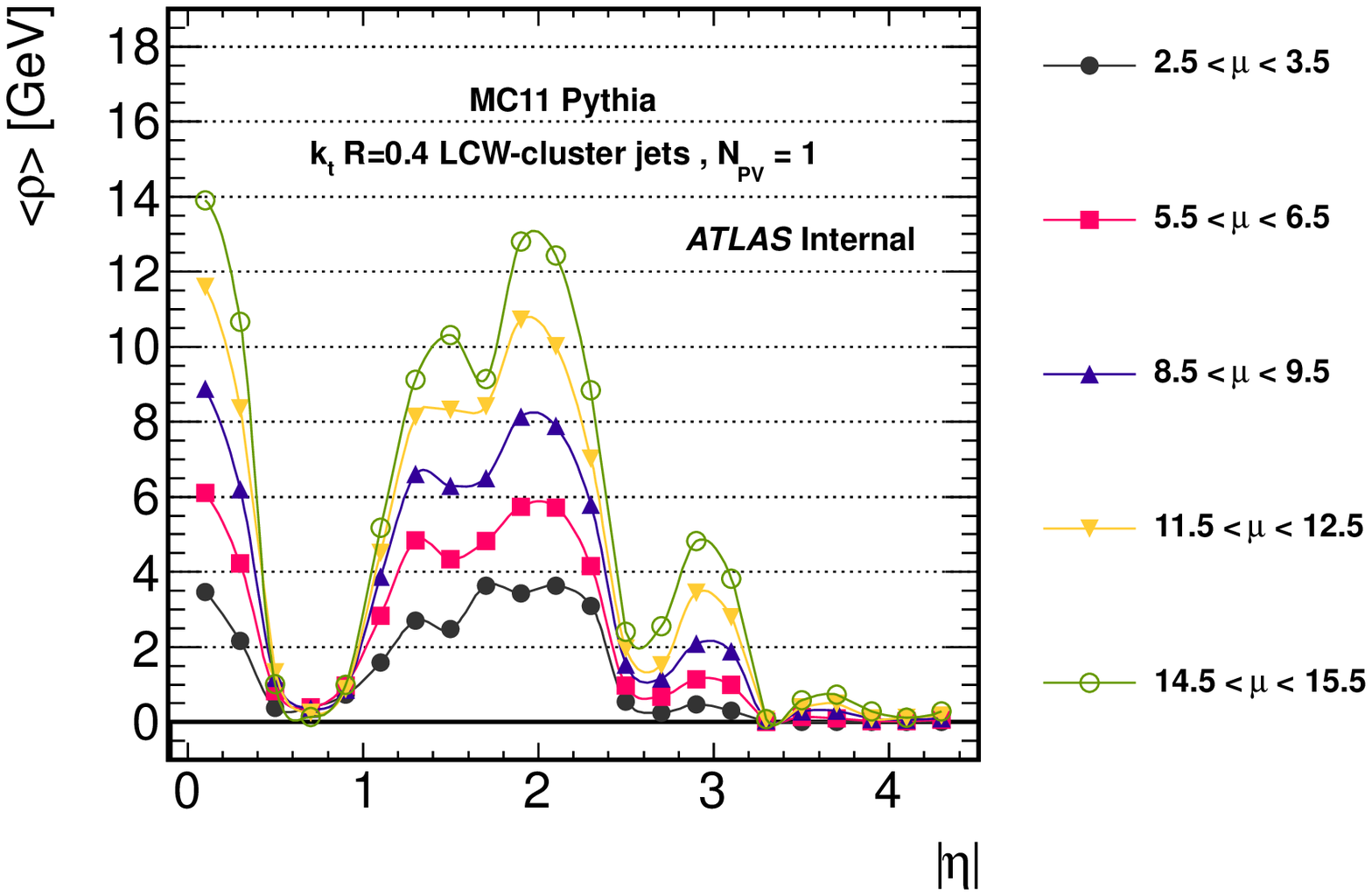}} 
  \caption{\label{FIGavgRhoEtaInMuBins}Dependence of the average median, $\left<\rho\right>$, on pseudo-rapidity, \Eta, for
    events with a single reconstructed vertex, \Npv, and different ranges of the average number of interactions, \Mu, as indicated in the figures,
    for data \Subref{FIGavgRhoEtaInMuBins1} and for MC \Subref{FIGavgRhoEtaInMuBins2}. The connecting lines are intended to guide the eye.
  }
\end{center}
\end{figure} 
The dips in the value of $\left<\rho\right>$ around $|\eta|=0.8,1.8~\mrm{and}~2.8$ correspond to transition regions
between elements of the calorimeter, where the jet response is in general lower
(see also \autoref{chapJetEnergyCalibration}, \autoref{FIGjesCorrection2}).
For \etaHigher{3.2}, $\left<\rho\right>$ is very small, which has to do with the increase
in \otpu in forward rapidities. The high rate of \pu in this region
results in a uniform flux of particles which flood the calorimeter. The integrated positive- and negative-energy
signals cancel each other out. The subsequent occupancy of positive-energy clusters in the forward calorimeter is thus very low.

As \Mu increases in the data, so does the average median. While the MC describes this general feature of the data,
the magnitude of $\left<\rho\right>$ in a given \Mu bin in the MC is slightly different. In addition, compared to the data, the dependence on \Eta
in MC is different in the transition region between the barrel and the endcap ($1.2 < \eta < 1.8$).
These differences stem from two sources. The first is miss-modelling of the distribution of dead material around the crack areas.
The second is the difference in the energy flow of the \pu; the MC does not describe well the low-\pt region
of the \pu particle spectrum.

The dependence of the average median on \Npv for a limited range of \Mu values, and its dependence on \Mu
for a fixed $N_{\mrm{PV}} = 1$ is shown in \autoref{FIGavgRhoInEtaMuBins}. The median is calculated using jets within
limited regions in $\left|\eta\right|$ in a given event.
Additional rapidity bins are shown in \autoref{chapJetAreaMethodApp}, \autorefs{FIGavgRhoNpvInEtaMuBinsApp}~-~\ref{FIGavgRhoMuInEtaMuBinsApp}.
\begin{figure}[htp]
\begin{center}
\subfloat[]{\label{FIGavgRhoInEtaMuBins1}\includegraphics[trim=5mm 10mm 65mm 10mm,clip,width=.43\textwidth]{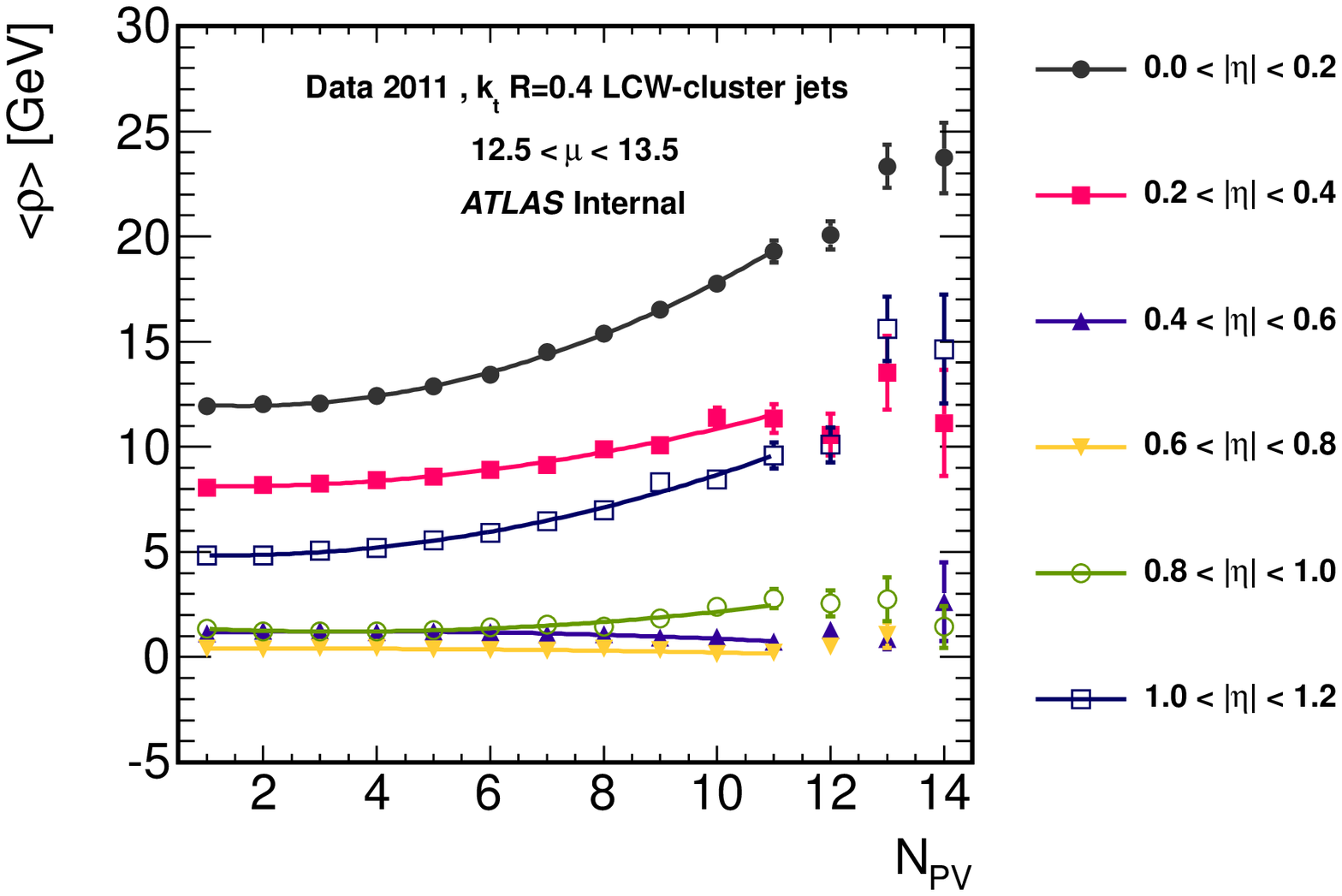}} 
\subfloat[\figQaud]{\label{FIGavgRhoInEtaMuBins2}\includegraphics[trim=5mm 10mm 0mm 10mm,clip,width=.643\textwidth]{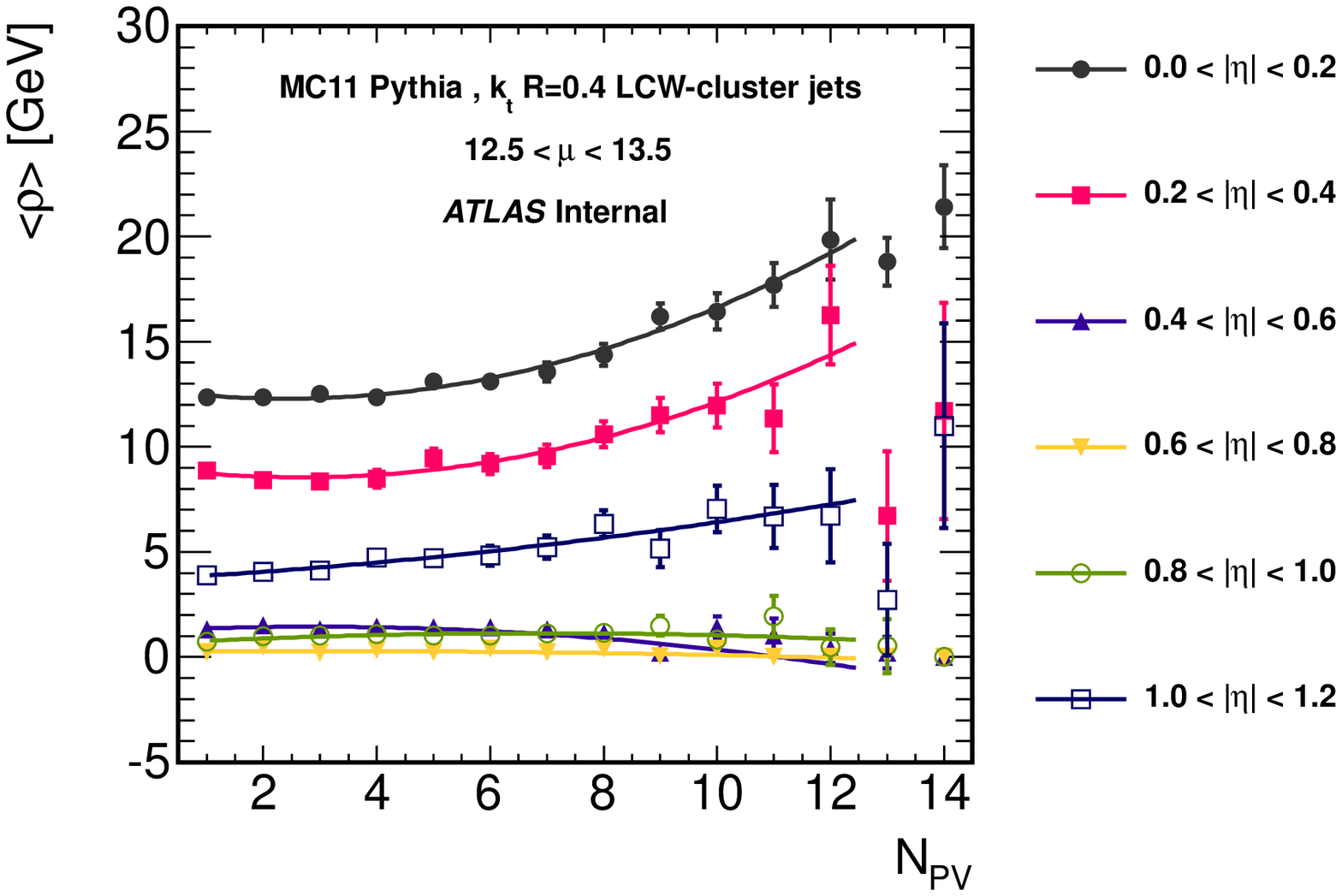}} \\
\subfloat[]{\label{FIGavgRhoInEtaMuBins3}\includegraphics[trim=5mm 10mm 65mm 10mm,clip,width=.43\textwidth]{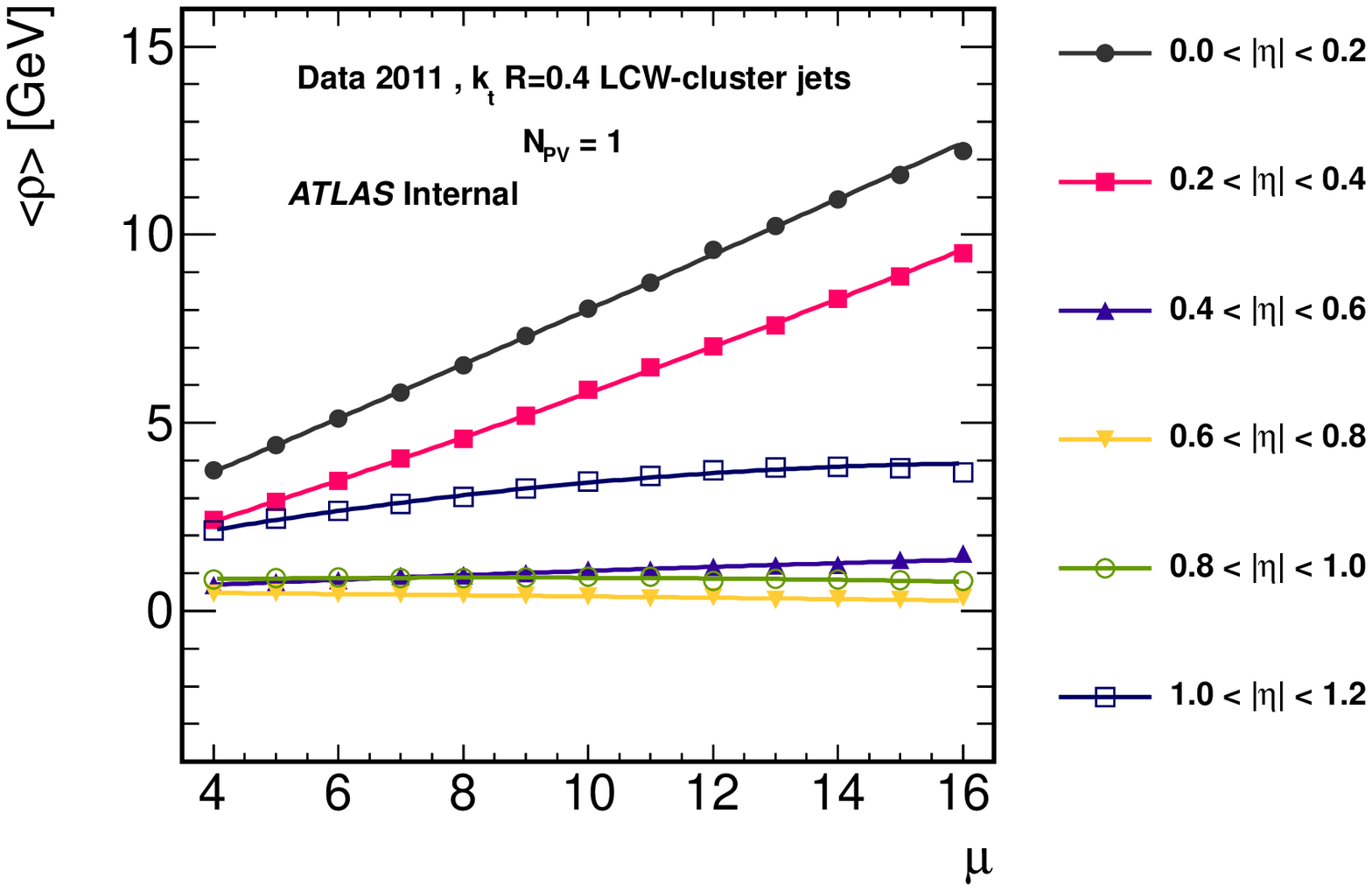}} 
\subfloat[\figQaud]{\label{FIGavgRhoInEtaMuBins4}\includegraphics[trim=5mm 10mm 0mm 10mm,clip,width=.643\textwidth]{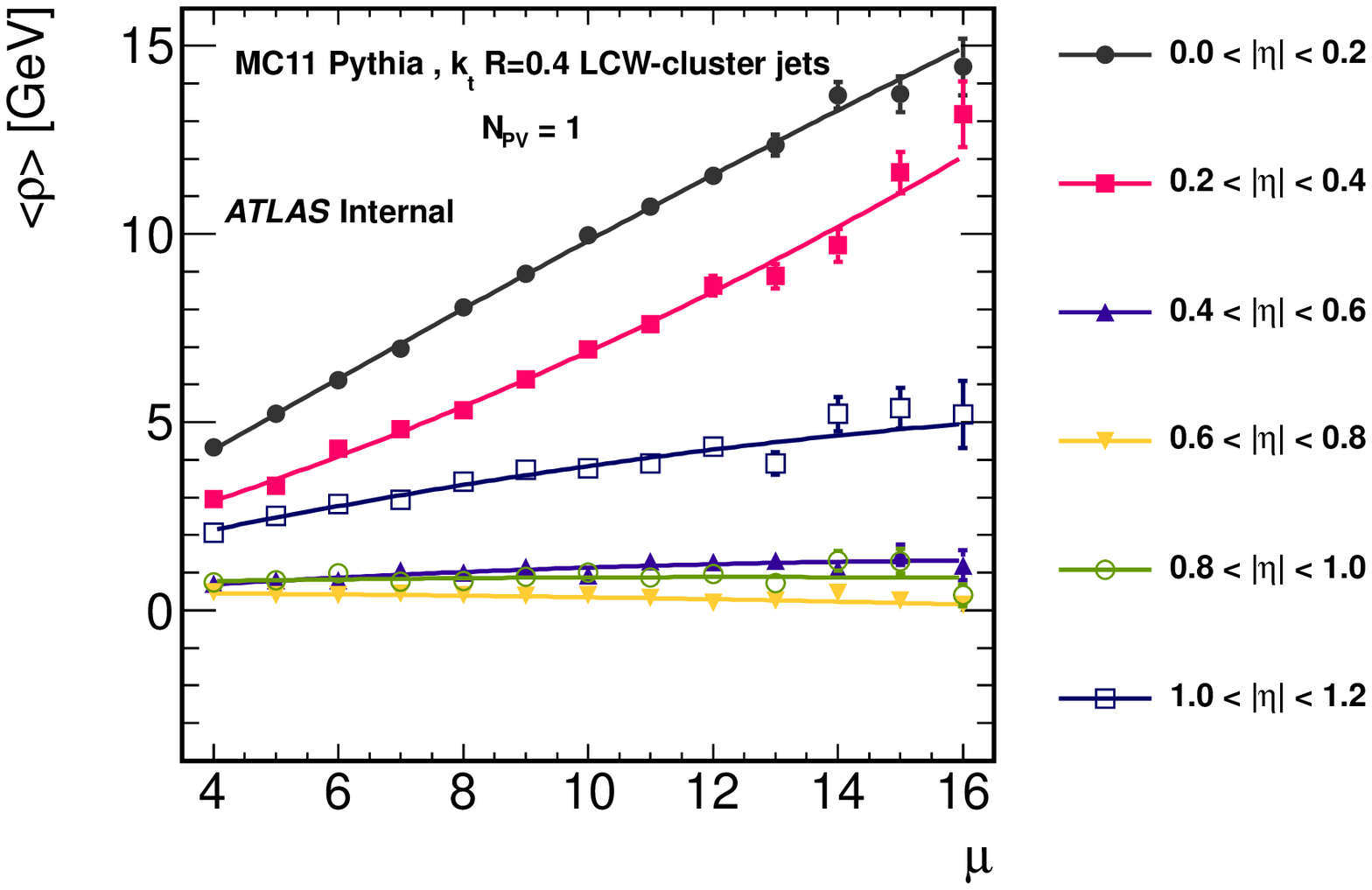}}
\caption{\label{FIGavgRhoInEtaMuBins}\Subref{FIGavgRhoInEtaMuBins1}-\Subref{FIGavgRhoInEtaMuBins2} Dependence of the average median,
  $\left<\rho\right>$, on the number of reconstructed vertices, \Npv, for events with an average number of interactions, \muRange{12.5}{13.5}. \\
  \Subref{FIGavgRhoInEtaMuBins3}-\Subref{FIGavgRhoInEtaMuBins4} Dependence of $\left<\rho\right>$ on \Mu for events with $\Npv = 1$. \\
  Several pseudo-rapidity, \Eta, bins are shown for data and for MC, as indicated in the figures.
  The lines represent polynomial fits to the points.
}
\end{center}
\end{figure} 
A strong dependence on \Npv and on \Mu is observed. 
The magnitude of $\left<\rho\right>$ generally increases with \Npv. The dependence is moderate
below $\Npv\sim4$, above which a sharp rise is observed in most \Eta regions. 
The dependence on \Mu is linear in the range of \Mu values considered.
A strong dependence on rapidity is also observed, as expected.
The rapidity regions in which the dependence of $\left<\rho\right>$ on \Npv and on \Mu is weak, correspond to the
transition regions between the different elements of the calorimeter, as discussed above.

The dependence on \Npv and on \Mu is parametrized using second-degree polynomials. These
parametrizations are used in the following in order to estimate
the \pu \pt-density as a function of \Npv, \Mu and \Eta in a given event.

\subsection{\Pu subtraction with the median}

As discussed above, it is possible to estimate the amount of \pu in an event
using the \pt-density.
The four-momentum of a jet can then be corrected by rescaling it according to the density in its region.
Given a measured density, \Rho, around a jet with area $\Ajet$,
the percentage of the momentum of the jet originating from \pu can be estimated as the ratio
\begin{equation}{
    \wp_{\mrm{t}}^{\;\rho} = \frac{\rho \times \Ajet}{p_{\mrm{t}}^{\mrm{\LC}}} \; ,
}\label{eqPuCorrection3} \end{equation}
where $p_{\mrm{t}}^{\mrm{\LC}}$ is the transverse momentum of the jet before any \pu correction.
The four momentum is then rescaled as in \autoref{eqPuCorrection1} with
the substitution $\wp_{\mrm{t}}^{\;\mathcal{O}_{\mrm{fst}}^{\LC}} \rightarrow \wp_{\mrm{t}}^{\;\rho}$.
Since the \JES correction is derived for jets which have been recalibrated using the
offset correction, a different correction using the median can cause a bias. An additional scaling function,
$f(p_{\mrm{t}}^{\rm calib},\eta)$ may therefore be required after the calibration step from \LC- to
hadronic-scale,
\begin{equation}{
  p_{\mrm{t}}^{\rm calib} = p_{\mrm{t}}^{\rm det} \times \mathcal{F}^{{\rm calib}}_{\mrm{\LC}}\left(p_{\mrm{t}}^{\rm det},\eta\right) \;,
  \quad  p_{\mrm{t}}^{\rm calib} \rightarrow p_{\mrm{t}}^{\rm calib} \times \left(1+ f\left(p_{\mrm{t}}^{\rm calib},\eta\right) \right) \; .
  }\label{eqPuCorrection4} \end{equation}

The appropriate value of the median can be calculated in several ways.
One has the option of either using the parametrization of $\left<\rho\right>$ derived
in \autoref{sectParametrizationOfAvgPU}, or of calculating the \pt-density
from event-level observables. For the latter, there is freedom in choosing the
area over which the median is calculated. One may choose \eg to include in the median
only jets from the vicinity of the jet which is to be calibrated. Another option would be to
estimate the \pu contamination in the event, in a region of the calorimeter where the \pu is stable
over time.
Several such schemes have been explored. The notation $\epsilon_{i}$
is used to define the different correction procedures,
\begin{list}{-}{}
\mynobreakpar
\item
  $\bm{\epsilon_{\textbf{0}}}$\headFont{- no correction};
\item
  $\bm{\epsilon_{\textbf{1}}}$\headFont{- default offset correction - }
  the current correction method used in \ATLAS. (See \autoref{sectDefaultOffsetCorrection});
\item
  $\bm{\epsilon_{\textbf{2}}}$\headFont{- central rapidity correction - }
  the median is calculated on an even-by-event basis, as an average within the central rapidity region, $|\eta|<1.8$.
  The average is calculated using the value of \Rho in 18 pseudo-rapidity bins of 0.2 units width;
\item
  $\bm{\epsilon_{\textbf{3}}}$\headFont{- local correction - }
  the median is calculated on an even-by-event basis,
  as an average within a range of $\pm1$ pseudo-rapidity units around the jet for which the correction is made.
  The average is calculated using the value of \Rho in six pseudo-rapidity bins of 0.2 units width;
\item
  $\bm{\epsilon_{\textbf{4}}}$\headFont{- average correction - }
  the average \pu \pt-density is parametrized as a function of \Npv, \Mu and \Eta. The correction
  is applied with regard to a reference value, $N_{\mrm{PV}} = 1$;
\item
  $\bm{\epsilon_{\textbf{5}} = \epsilon_{\textbf{3}} \oplus \epsilon_{\textbf{4}}}$\headFont{- local/average combination - }
  a combination of the local ($\epsilon_{3}$) and of the average ($\epsilon_{4}$) median calculations. The relative weights 
  between the two elements depend on rapidity (at central rapidity, higher significance is given to
  the local median correction).
\end{list}
For all of the above ($\epsilon_{0}$-$\epsilon_{5}$), jets undergo \pu subtraction using the respective
calculation of \Rho, followed by the original \JES calibration (\autoref{eqPuCorrection2}).
A special configuration, incorporating the correction introduced in \autoref{eqPuCorrection4}, is
\begin{list}{-}{}
\mynobreakpar
\item
  $\bm{\epsilon_{\textbf{6}} = \epsilon_{\textbf{5}} \otimes} \bm{f}\left(\bm{p_{\mrm{\textbf{t}}}},\pmb{\eta}\right)$\headFont{- local/average combination with \JES factor - }
  the combination of the local and of the average median calculations, $\epsilon_{5}$, is modified by introducing 
  a \pt dependant \JES scaling function, $f\left(p_{\mrm{t}},\eta\right)$,
  in additional to the baseline \JES calibration.
\end{list}
As will be shown in the following, the best performance is achieved using $\epsilon_{6}$. This is therefore defined as
the nominal correction.

It should be emphasized that unlike the offset correction (\autoref{eqDefinitionDefaultOffset}),
the average correction defined in $\maybebm{\epsilon_{4}}$ and used in
$\maybebm{\epsilon_{5}}$ and $\maybebm{\epsilon_{6}}$ uses a reference value for \Npv, but does not use
one for \Mu. Instead, the dependence on \Npv is binned in \Mu.

Due to the discrepancy between data and MC (observed \eg in \autoref{FIGavgRhoEtaInMuBins}),
the MC is not used for calculation of the average \pu correction for data.
Instead, the median (\autoref{eqPuCorrection1}) is taken directly from the data.
The MC is used in order to derive $f\left(p_{\mrm{t}},\eta\right)$, used in $\epsilon_{6}$. An associated uncertainty
for the $f\left(p_{\mrm{t}},\eta\right)$ correction is introduced, as discussed below.

\subsection{Performance of the median correction in MC}
%
%
Using truth jets, it is possible to test the performance of different \pu subtraction methods in MC.
This may be done using the \textit{relative transverse momentum offset}, $O_{p_{\mrm{t}}}$, which is
the relative difference between the transverse momentum
of truth and reconstructed calorimeter jets, respectively denoted by $p_{\mrm{t}}^{\mrm{truth}}$ and
$p_{\mrm{t}}^{\mrm{rec}}$. It has previously been defined in \autoref{eqJetPtOffsetDef} as,
\begin{equation*}
  O_{p_{\mrm{t}}} = \frac{p_{\mrm{t}}^{\mrm{rec}}-p_{\mrm{t}}^{\mrm{truth}}}{p_{\mrm{t}}^{\mrm{truth}}} \,.
\end{equation*}
Truth and calorimeter jets are matched within
${\Delta R_{\mrm{match}} = 0.5 \cdot R}$ of the truth jet, where ${R = 0.6}$ is the size parameter
of the jets.
In addition, truth jets are isolated from their counterparts within ${\Delta R_{\mrm{iso}} = 2 \cdot R}$.
No such isolation condition is required of reconstructed jets.

A distribution of the relative \pt offset is shown in \autoref{FIGClosureOffsetPtBins} for jets within 
\etaRange{0.2}{0.4}, calibrated using $\epsilon_{6}$.
\begin{figure}[htp]
\begin{center}
  \qquad\qquad\qquad\qquad\includegraphics[trim=5mm 10mm 0mm 10mm,clip,width=.643\textwidth]{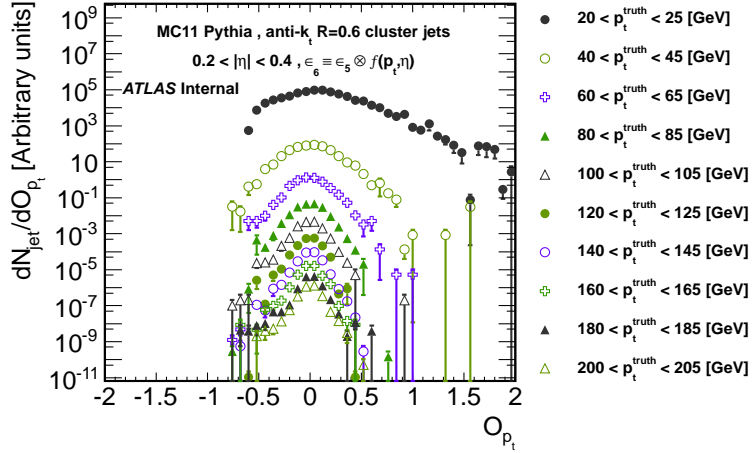}
  \caption{\label{FIGClosureOffsetPtBins}Distributions of the relative transverse momentum offset, $O_{p_{\mrm{t}}}$, for detector jets,
  calibrated using $\epsilon_{6}$, within pseudo-rapidity, $0.2 < |\eta| < 0.4$, and in
  several ranges of truth jet transverse momentum, $p_{\mrm{t}}^{\mrm{truth}}$, as indicated in the figure.
  }
\end{center}
\end{figure} 
A tail towards high values of $O_{p_{\mrm{t}}}$ is observed in the first \pt bin, indicating that the
low-\pt jets tend to be over-calibrated.
As \pt increases, the intrinsic energy resolution improves.
In addition, the fractional magnitude of the \pu decreases as the \pt of the jet increases.
As a result, the distribution of $O_{p_{\mrm{t}}}$ becomes narrower.

The relative \pt offset is used to asses the quality of the different \pu subtraction methods.
\Autoref{FIGmeanClosureOffsetNpvMu} shows the dependence of the average $O_{p_{\mrm{t}}}$
on \Npv and on \Mu for jets within $|\eta| < 1.8$ and different \pt ranges. 
Two correction schemes are compared, $\epsilon_{0}$ (no \pu subtraction) and $\epsilon_{6}$ (nominal correction).
\begin{figure}[htp]
\begin{center}
\subfloat[]{\label{FIGmeanClosureOffsetNpvMu1}\includegraphics[trim=5mm 10mm 65mm 10mm,clip,width=.43\textwidth]{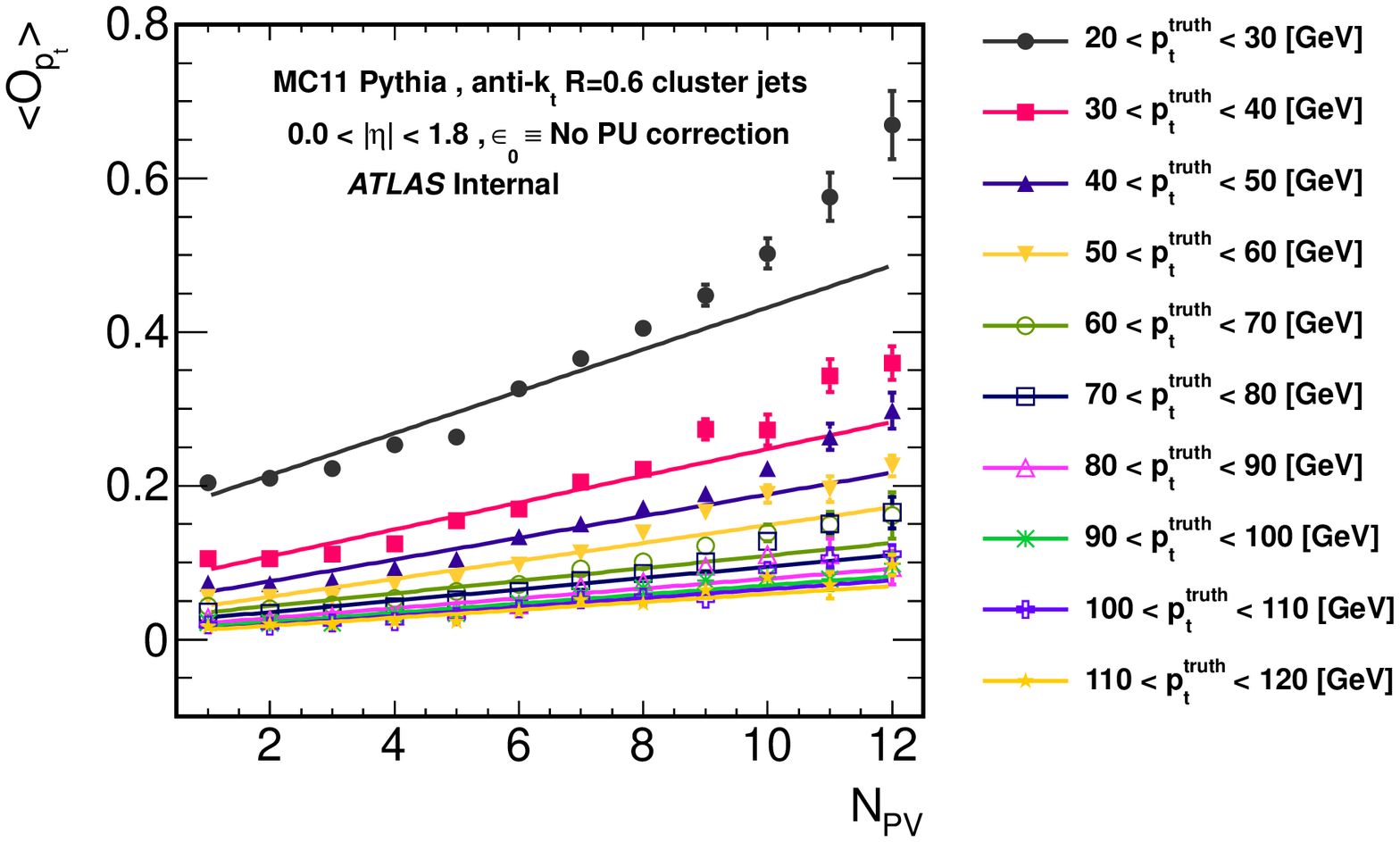}} 
\subfloat[\figQaud]{\label{FIGmeanClosureOffsetNpvMu2}\includegraphics[trim=5mm 10mm 0mm 10mm,clip,width=.643\textwidth]{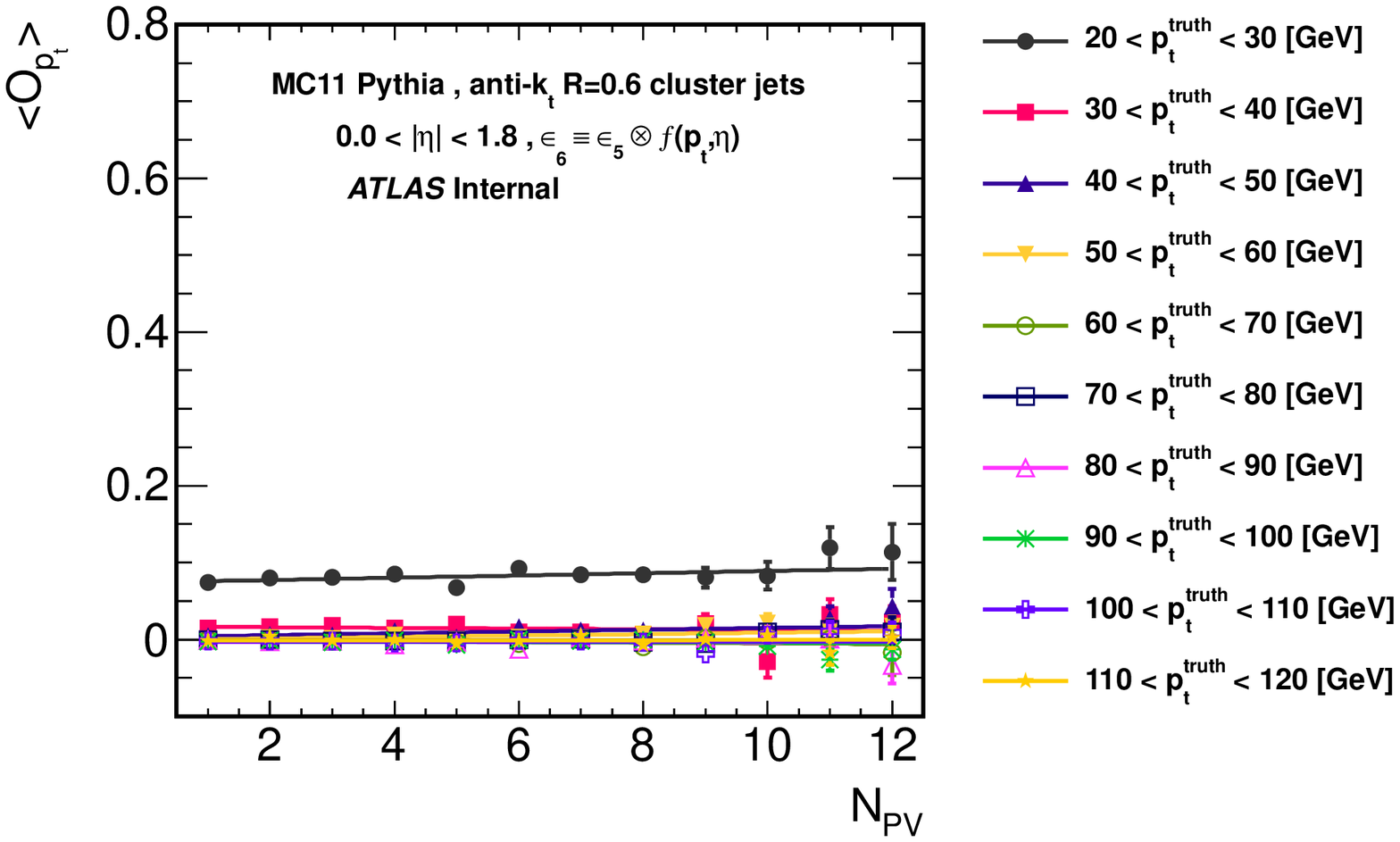}} \\
\subfloat[]{\label{FIGmeanClosureOffsetNpvMu3}\includegraphics[trim=5mm 10mm 65mm 10mm,clip,width=.43\textwidth]{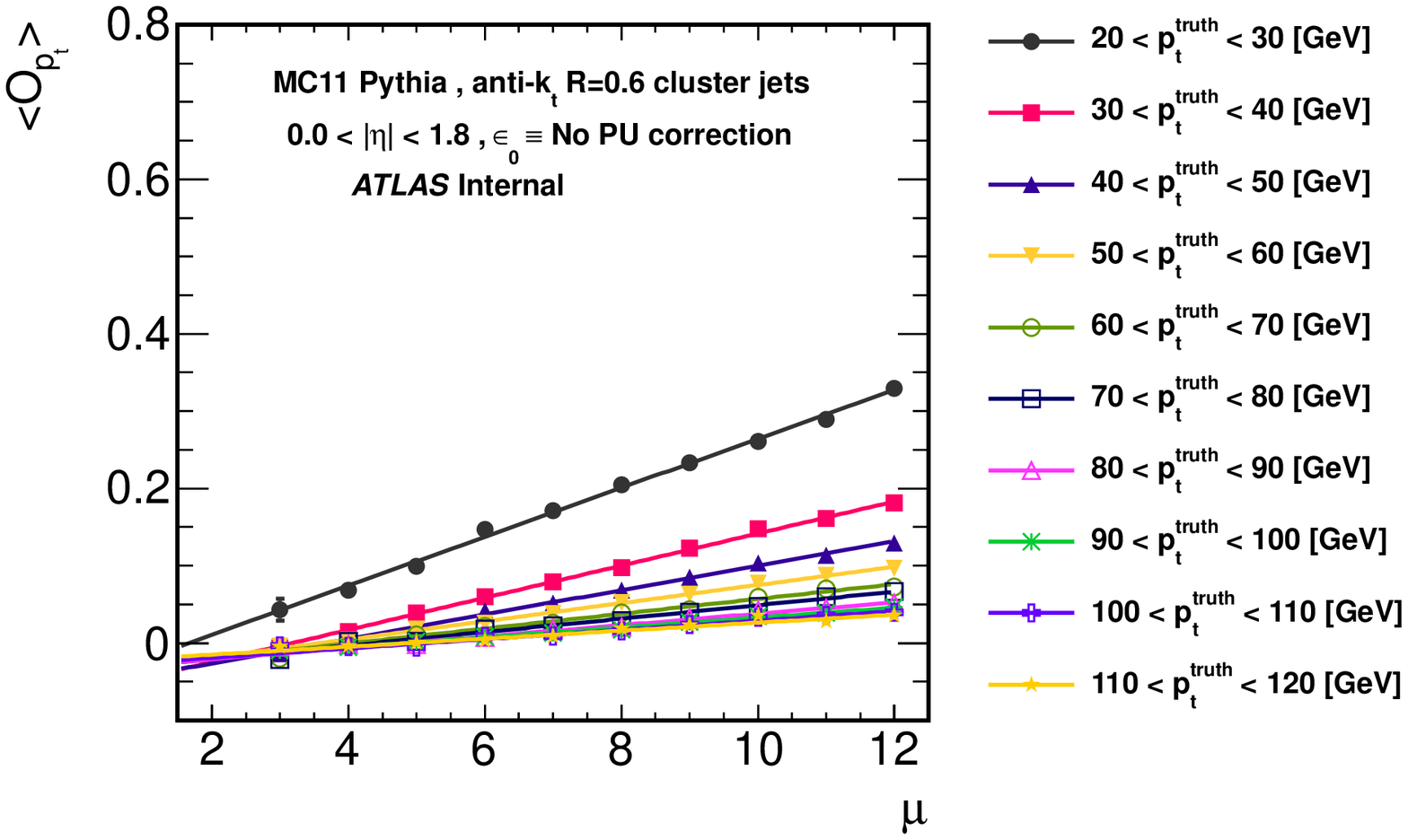}} 
\subfloat[\figQaud]{\label{FIGmeanClosureOffsetNpvMu4}\includegraphics[trim=5mm 10mm 0mm 10mm,clip,width=.643\textwidth]{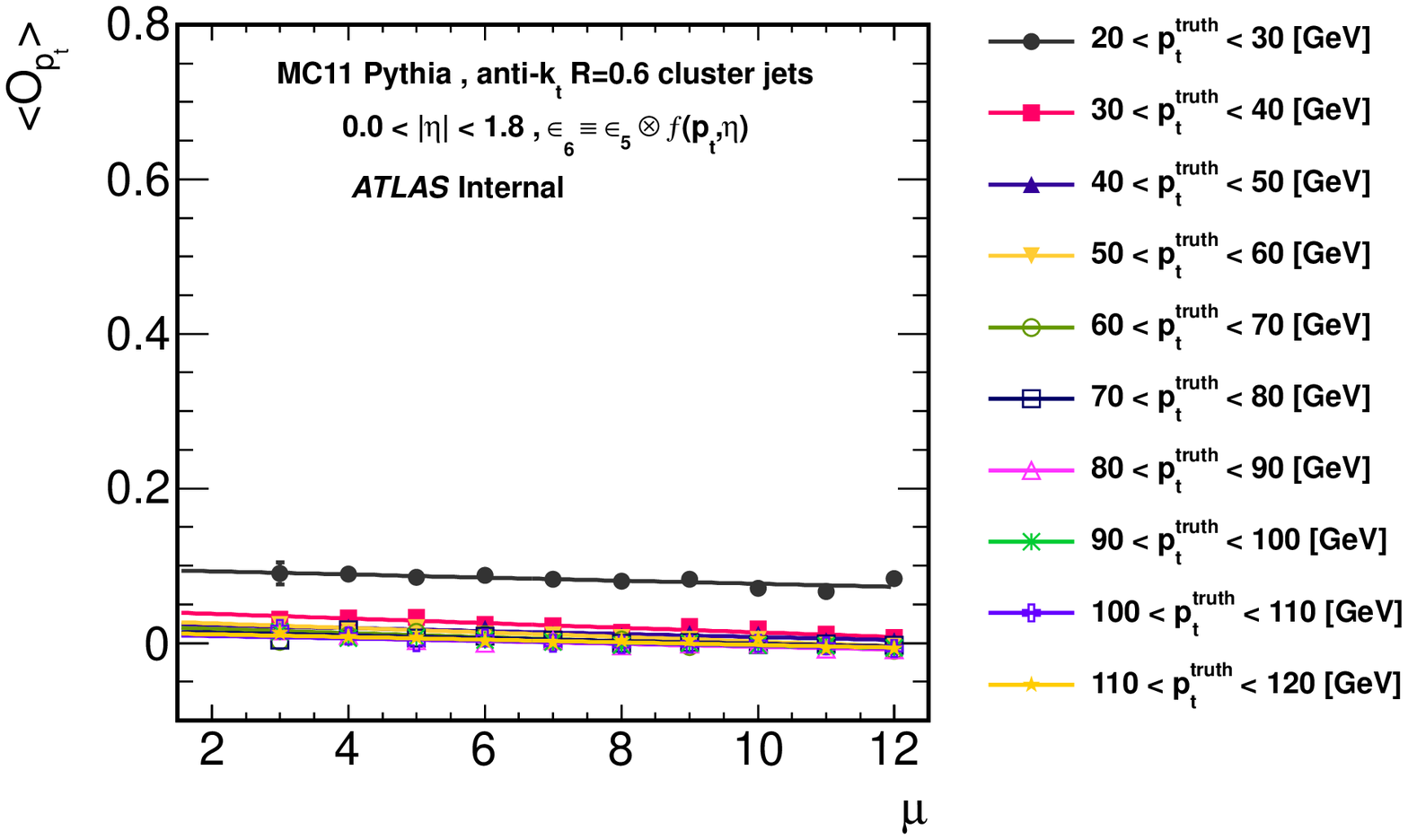}} 
  \caption{\label{FIGmeanClosureOffsetNpvMu}Dependence of the average relative
    transverse momentum offset, $<O_{p_{\mrm{t}}}>$, on the number of
    reconstructed vertices, \Npv, (\Subref{FIGmeanClosureOffsetNpvMu1}, \Subref{FIGmeanClosureOffsetNpvMu2}) and on the average number of interactions, \Mu,
    (\Subref{FIGmeanClosureOffsetNpvMu3}, \Subref{FIGmeanClosureOffsetNpvMu4}), using
    jets with pseudo-rapidity, $|\eta| < 1.8$, corrected for \pu with $\epsilon_{0}$ or $\epsilon_{6}$, for
    several ranges of truth jet transverse momentum, $p_{\mrm{t}}^{\mrm{truth}}$, as indicated.
    The lines represent linear fits to the points.
  }
\end{center}
\end{figure} 
Jets which are not corrected for \pu (\autorefs{FIGmeanClosureOffsetNpvMu1} and~\ref{FIGmeanClosureOffsetNpvMu3})
show a strong \Npv and \Mu dependence. The dependence is especially strong for low-\pt
jets, as the \pu constitutes a significant percentage of the reconstructed jet \pt. Following the nominal \pu subtraction
procedure (\autorefs{FIGmeanClosureOffsetNpvMu2} and~\ref{FIGmeanClosureOffsetNpvMu4}), most of the dependence is removed.
The residual dependence on \Mu is stronger compared to the respective dependence on \Npv.
Jets with $20 < p_{\mrm{t}}^{\mrm{truth}} < 30$\GeV exhibit a bias of $\sim150$\MeV compared to higher \pt values.

In order to quantify the residual dependence following the \pu correction, $O_{p_{\mrm{t}}}$ is parametrized as
\begin{equation}{
  O_{p_{\mrm{t}}}\left(\Npv\right)  = \alpha_{\pt}^{\Npv} + \beta_{\pt}^{\Npv} \cdot \Npv \quad \mrm{and} \quad
  O_{p_{\mrm{t}}}\left(\mu\right)   = \alpha_{\pt}^{\Mu}  + \beta_{\pt}^{\Mu} \cdot \mu \; .
}\label{eqOffsetFitForm1} \end{equation}
The slope parameters, $\beta_{\pt}^{\Npv}$ and $\beta_{\pt}^{\Mu}$, represent respectively the per-\Npv and per-\Mu dependence of the offset.
In the case of complete \pu subtraction, both would be zero.
It is also helpful to inspect the ratio of slope parameters between that of a given correction, $\beta_{\pt}^{\Npv}\left(\epsilon_{i}\right)$, (with $i \in$1-6)
and the case where no \pu correction is applied, $\beta_{\pt}^{\Npv}\left(\epsilon_{0}\right)$.
Accordingly, the \textit{slope ratio parameters} are defined as
\begin{equation}{
    \mathcal{R}_{\beta}^{\Npv}  = \frac{\beta_{\pt}^{\Npv}\left(\epsilon_{0}\right)-\beta_{\pt}^{\Npv}\left(\epsilon_{i}\right)}{\beta_{\pt}^{\Npv}\left(\epsilon_{0}\right)}
    \quad \mrm{and} \quad
    \mathcal{R}_{\beta}^{\mu}  = \frac{\beta_{\pt}^{\mu}\left(\epsilon_{0}\right)-\beta_{\pt}^{\mu}\left(\epsilon_{i}\right)}{\beta_{\pt}^{\mu}\left(\epsilon_{0}\right)} \;,
}\label{eqOffsetFitForm2} \end{equation}
where in the case of a perfect correction, $\mathcal{R}_{\beta}^{\Npv} = \mathcal{R}_{\beta}^{\mu} = 1$. The slope ratio parameters represent the
residual fractional dependence, of the mean relative \pt offset on \Npv and on \Mu.

A linear fit is performed on the mean of $O_{p_{\mrm{t}}}$ for the different correction
methods, $\epsilon_{0}$-$\epsilon_{6}$.
The values of the slope parameters from the fits are presented in \autoref{FIGmeanClosureOffsetNpvMuSlope}
as a function of jet \pt for jets\footnote{
The first three \pt bins in \autoref{FIGmeanClosureOffsetNpvMu1} exhibit a strong dependence on \Npv, not captured by the linear fit.
The respective slopes are therefore slightly underestimated. This feature is unique to $\epsilon_{0}$, and so does not
affect the interpretation of the result.} within \etaLower{1.8}.
Additional rapidity bins are shown in \autoref{chapJetAreaMethodApp}, \autoref{FIGmeanClosureOffsetNpvMuSlopeApp}.
The fit results for $\mathcal{R}_{\beta}^{\Npv}$ and for $\mathcal{R}_{\beta}^{\mu}$ are presented in \autoref{FIGmeanClosureOffsetNpvMuSlopeRatio}
as a function of jet \pt for jets within $|\eta| < 1.8$.
\begin{figure}[htp]
\begin{center}
\subfloat[]{\label{FIGmeanClosureOffsetNpvMuSlope1}\includegraphics[trim=5mm 10mm 65mm 10mm,clip,width=.43\textwidth]{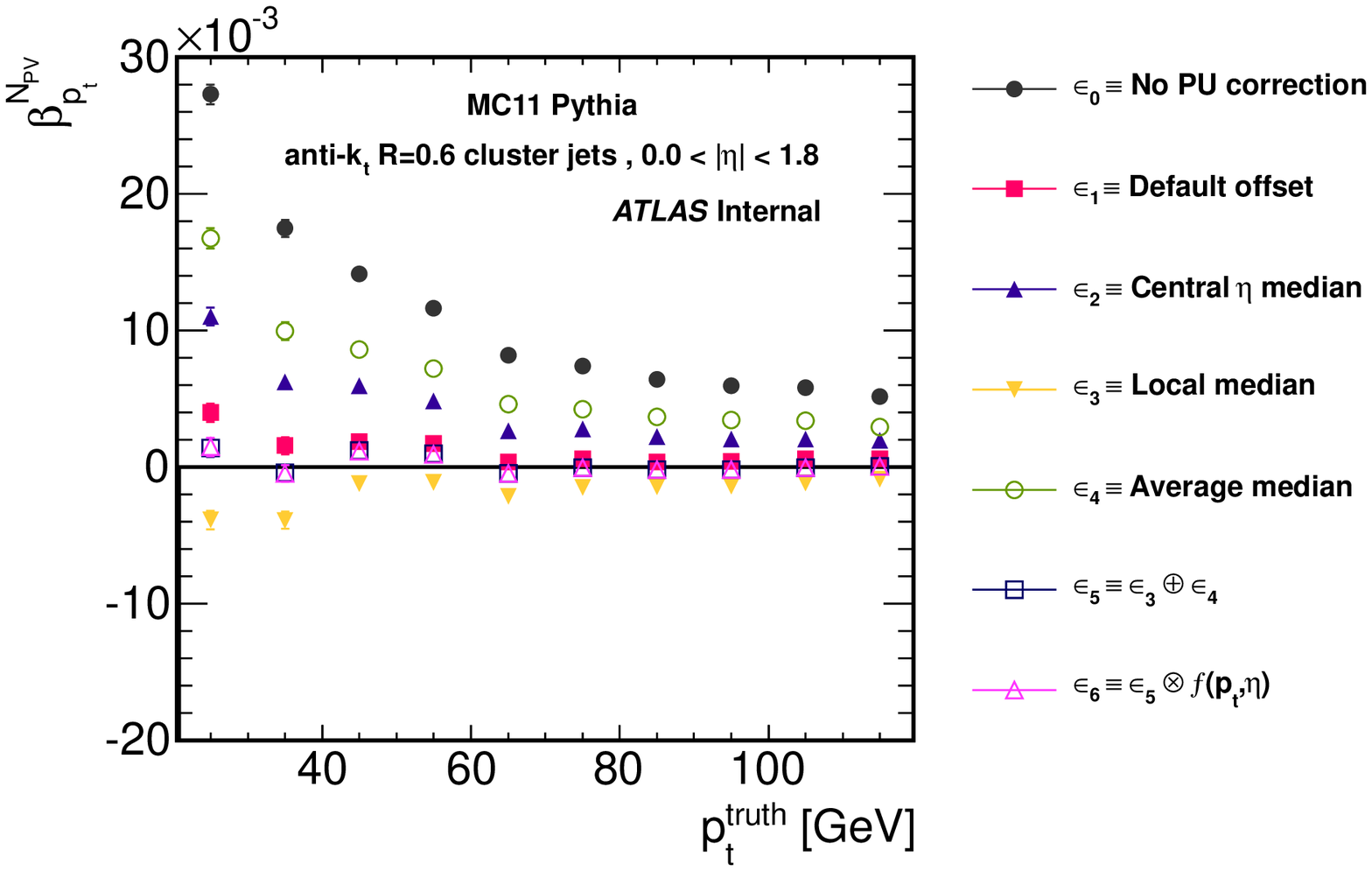}} 
\subfloat[\figQaud]{\label{FIGmeanClosureOffsetNpvMuSlope2}\includegraphics[trim=5mm 10mm 0mm 10mm,clip,width=.643\textwidth]{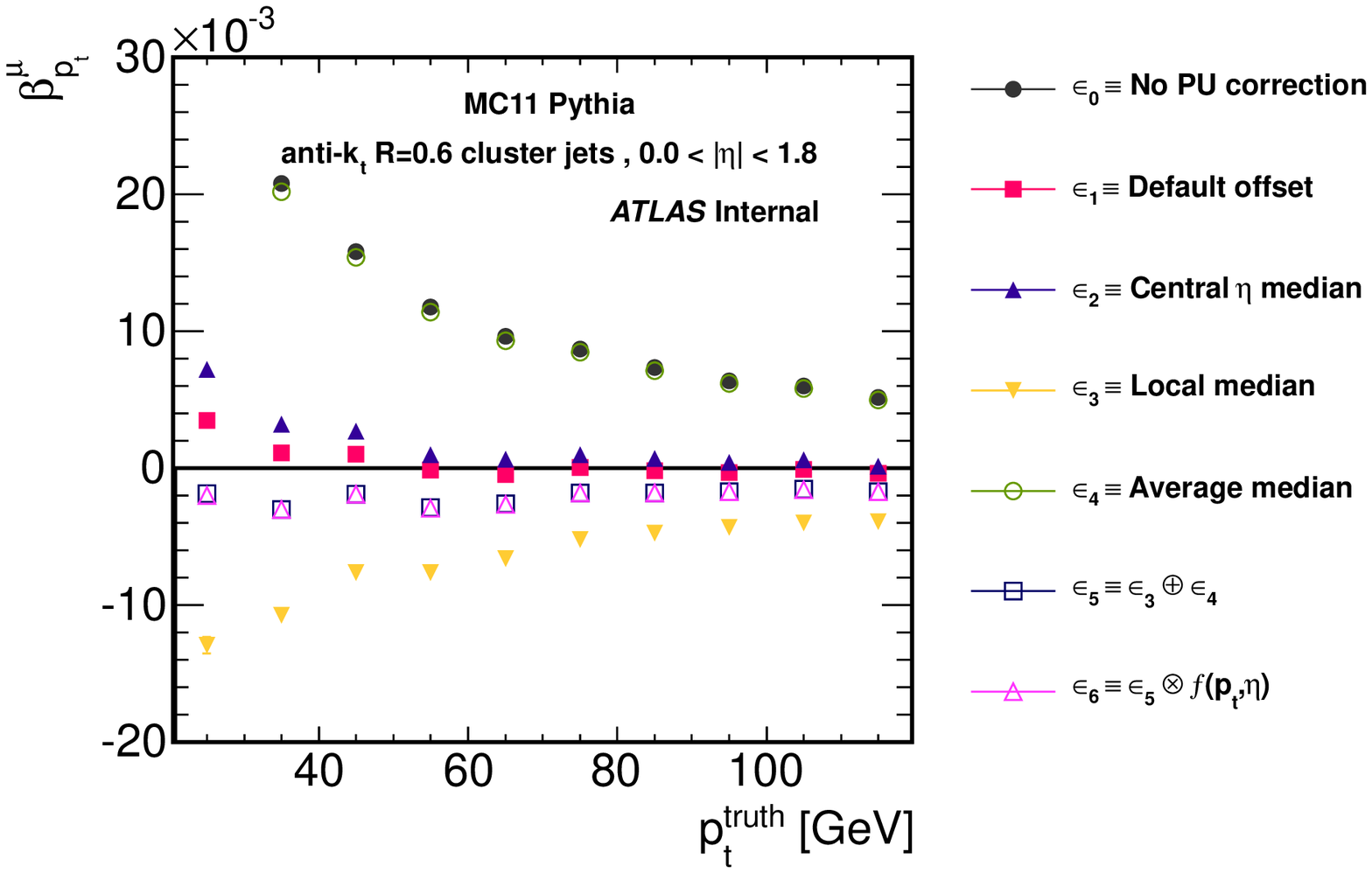}}
  \caption{\label{FIGmeanClosureOffsetNpvMuSlope}Dependence on truth jet transverse momentum, $p_{\mrm{t}}^{\mrm{truth}}$, of the parameters
    defined in \autoref{eqOffsetFitForm1}, $\beta_{\pt}^{\Npv}$ \Subref{FIGmeanClosureOffsetNpvMuSlope1}
    and $\beta_{\pt}^{\Mu}$ \Subref{FIGmeanClosureOffsetNpvMuSlope2},
    using jets with  pseudo-rapidity, $|\eta| < 1.8$, corrected for \pu with $\epsilon_{0}$-$\epsilon_{6}$,
    as indicated in the figures.
  }
\end{center}
\end{figure} 
\begin{figure}[htp]
\begin{center}
\subfloat[]{\label{FIGmeanClosureOffsetNpvMuSlopeRatio1}\includegraphics[trim=5mm 10mm 65mm 10mm,clip,width=.43\textwidth]{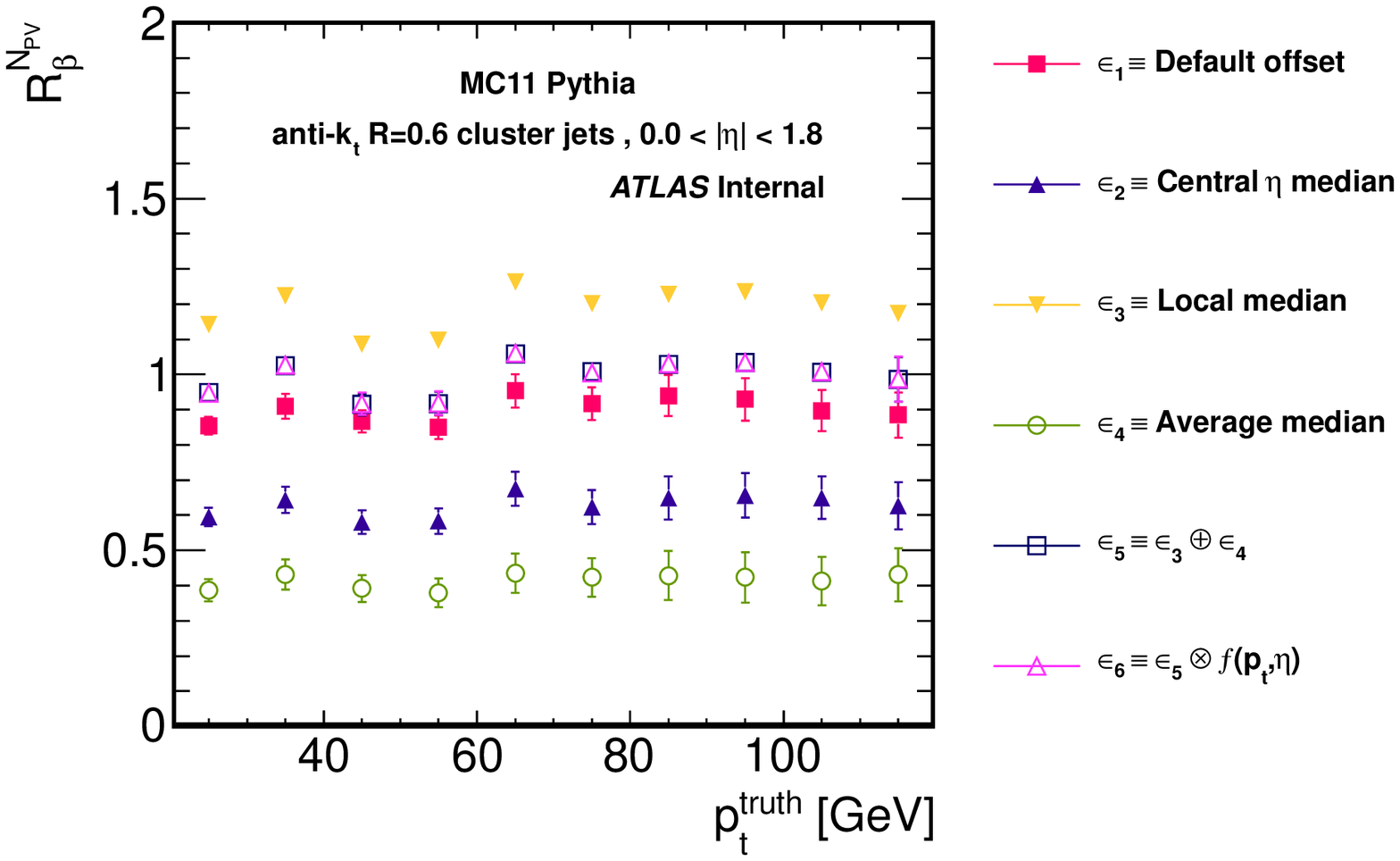}} 
\subfloat[\figQaud]{\label{FIGmeanClosureOffsetNpvMuSlopeRatio2}\includegraphics[trim=5mm 10mm 0mm 10mm,clip,width=.643\textwidth]{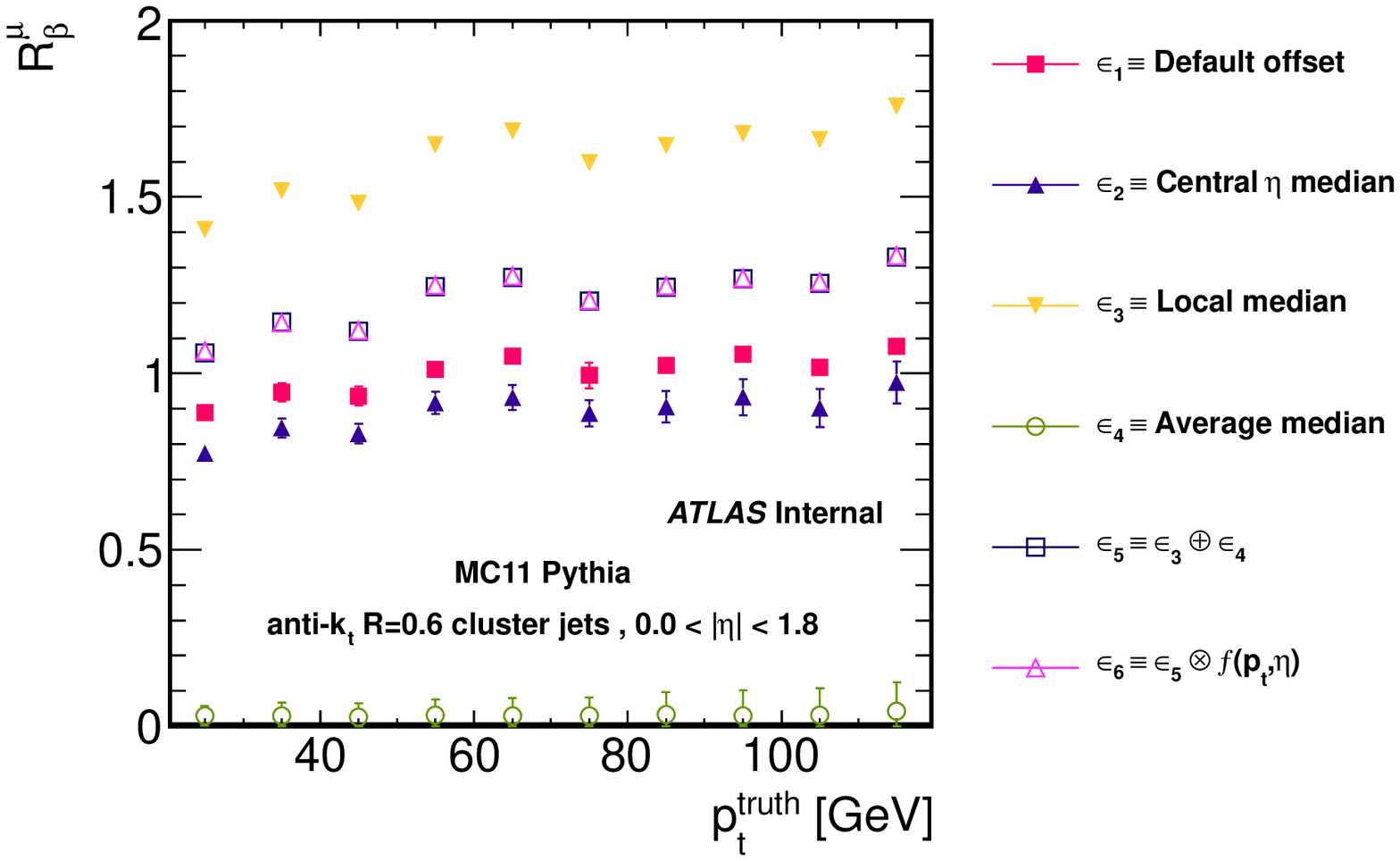}}
  \caption{\label{FIGmeanClosureOffsetNpvMuSlopeRatio}Dependence on truth jet transverse momentum, $p_{\mrm{t}}^{\mrm{truth}}$, of the parameters
    defined in \autoref{eqOffsetFitForm2}, $\mathcal{R}_{\beta}^{\Npv}$ \Subref{FIGmeanClosureOffsetNpvMuSlopeRatio1}
    and $\mathcal{R}_{\beta}^{\Mu}$ \Subref{FIGmeanClosureOffsetNpvMuSlopeRatio2},
    for jets with pseudo-rapidity, $|\eta| < 1.8$, corrected for \pu with $\epsilon_{1}$-$\epsilon_{6}$, as indicated in the figures.
  }
\end{center}
\end{figure} 

All correction methods tend to decrease the dependence on \Npv and on \Mu. The average correction ($\epsilon_{4}$) tends to underestimate the
amount of \pt. Among the event-by-event corrections, the central rapidity correction ($\epsilon_{2}$) underestimates the \pu contamination,
while the local correction ($\epsilon_{3}$) overestimates it. The corrections $\epsilon_{5}$ and $\epsilon_{6}$, which combine the average and the local
methods, achieve the best closure. The residual dependence of the latter two on \Npv fluctuates around zero. The respective residual dependence
on \Mu is of the order of $2 \cdot 10^{-3}$. For reference, in absolute scale the bias in \Mu is of the order of $40\MeV/ \mu$ for jets with $\pt = 20$\GeV.

As mentioned previously, the average correction does not use a reference value for \Mu, therefore there is no improvement between
$\epsilon_{0}$ and $\epsilon_{4}$ in \autoref{FIGmeanClosureOffsetNpvMuSlope2}. Several alternative correction methods were tested. These included
a reference value in \Mu instead of, or in addition to, the reference value in \Npv. The combined residual dependence on \Npv and on \Mu
was found to be greater for such configurations. The \Mu dependence is thus not addressed directly within the scope of the average correction.
However, taking into account the combined local and average calculation of \Rho, the remaining dependence on \Mu is negligible.

The dependencies of the mean of the distributions of $\left(O_{p_{\mrm{t}}} \times p_{t}^{\mrm{truth}}\right)$
and of $O_{p_{\mrm{t}}}$ on \pt are presented in \autoref{FIGmeanClosureOffsetPtCompCalib}. The former
(\autorefs{FIGmeanClosureOffsetPtCompCalib1}~-~\ref{FIGmeanClosureOffsetPtCompCalib2}) are used to illustrate in absolute units
the bias in \pt for the different \pu subtraction schemes. The latter (\autorefs{FIGmeanClosureOffsetPtCompCalib3}~-~\ref{FIGmeanClosureOffsetPtCompCalib4})
illustrate the significance of the bias with regard to the \pt-scale of jets.
Corrections $\epsilon_{0}$-$\epsilon_{6}$ are applied for jets in two \Eta regions.
\begin{figure}[htp]
\begin{center}
\subfloat[]{\label{FIGmeanClosureOffsetPtCompCalib1}\includegraphics[trim=5mm 10mm 65mm 10mm,clip,width=.43\textwidth]{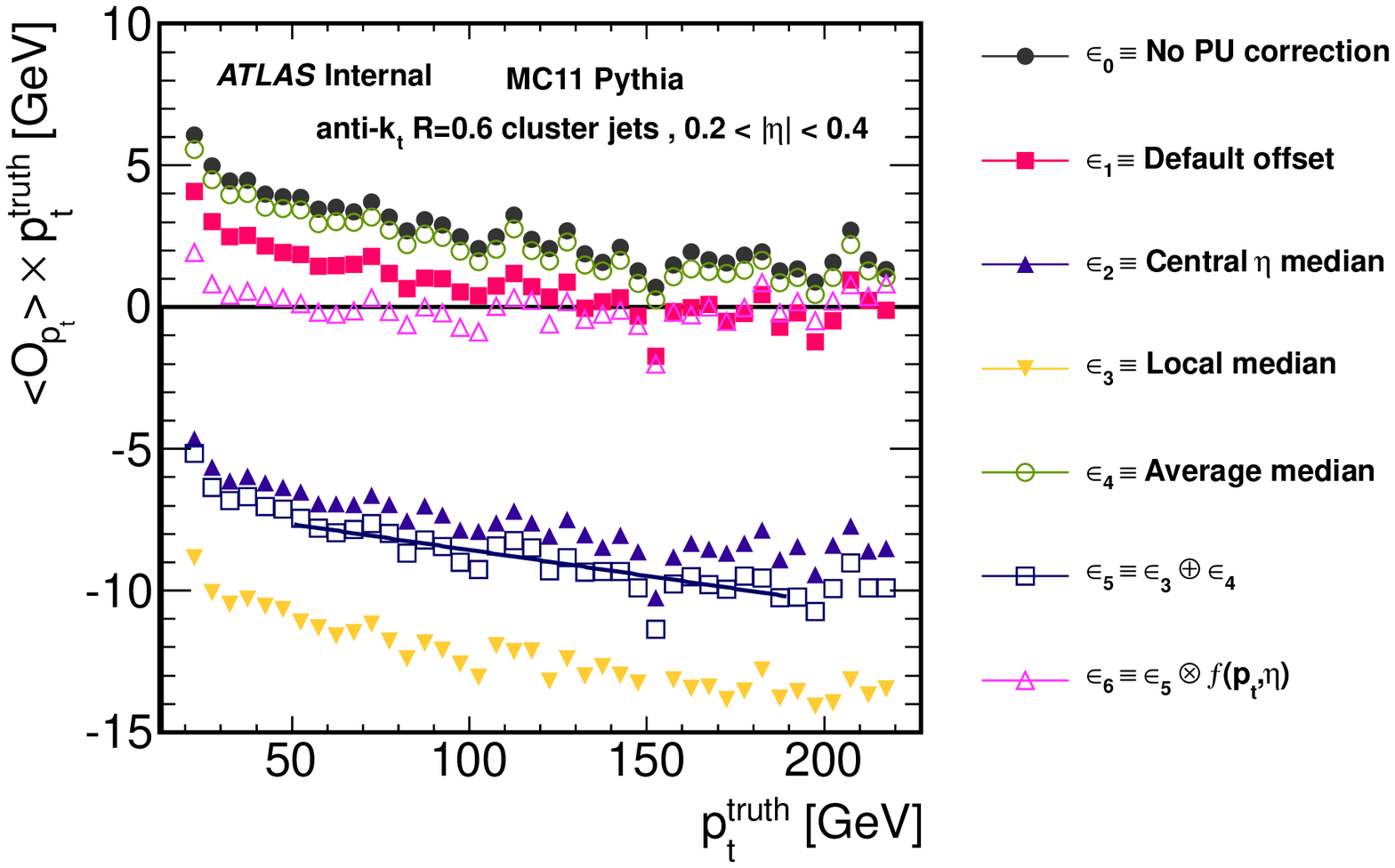}} 
\subfloat[\figQaud]{\label{FIGmeanClosureOffsetPtCompCalib2}\includegraphics[trim=5mm 10mm 0mm 10mm,clip,width=.643\textwidth]{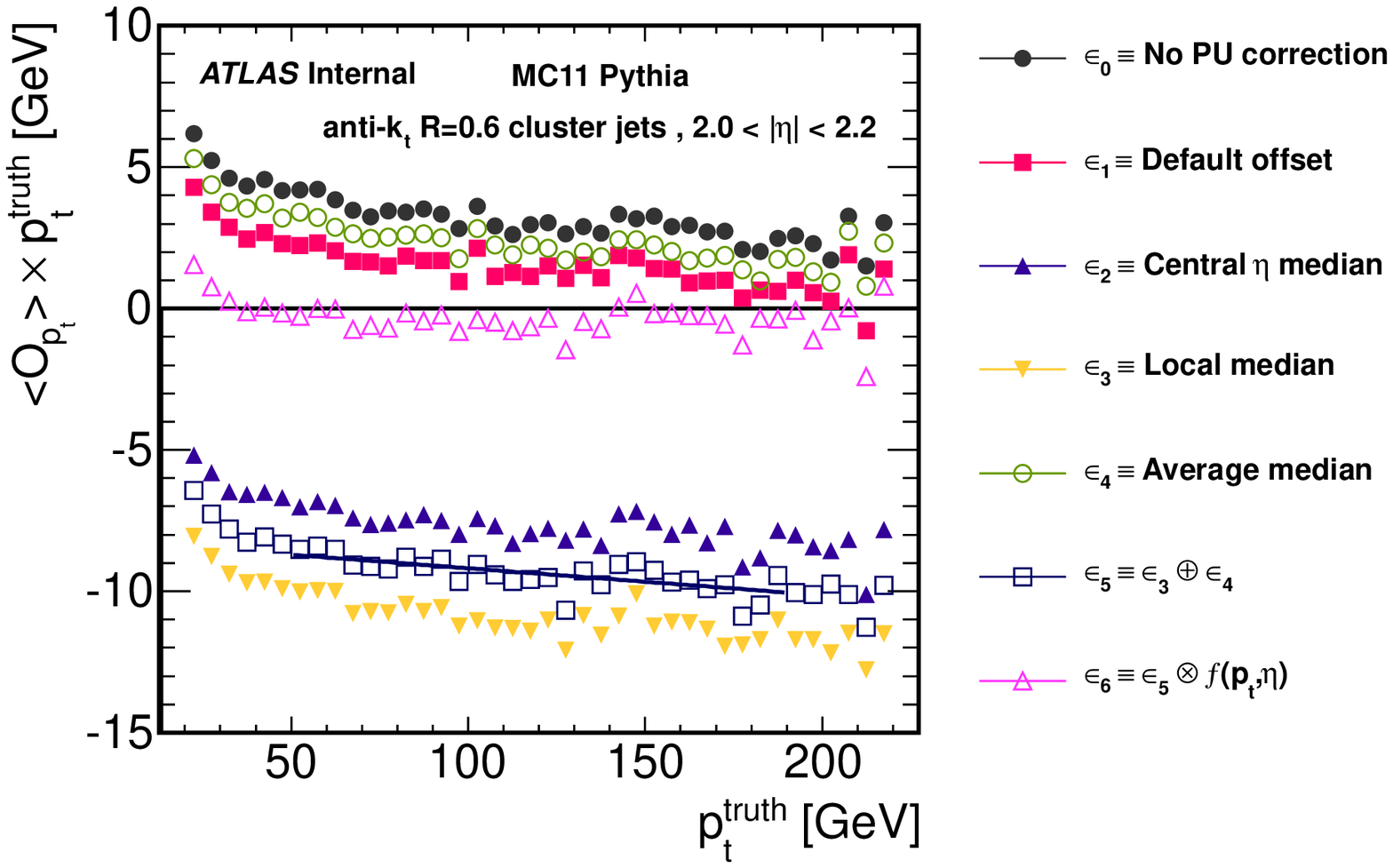}} \\
\subfloat[]{\label{FIGmeanClosureOffsetPtCompCalib3}\includegraphics[trim=5mm 10mm 65mm 10mm,clip,width=.43\textwidth]{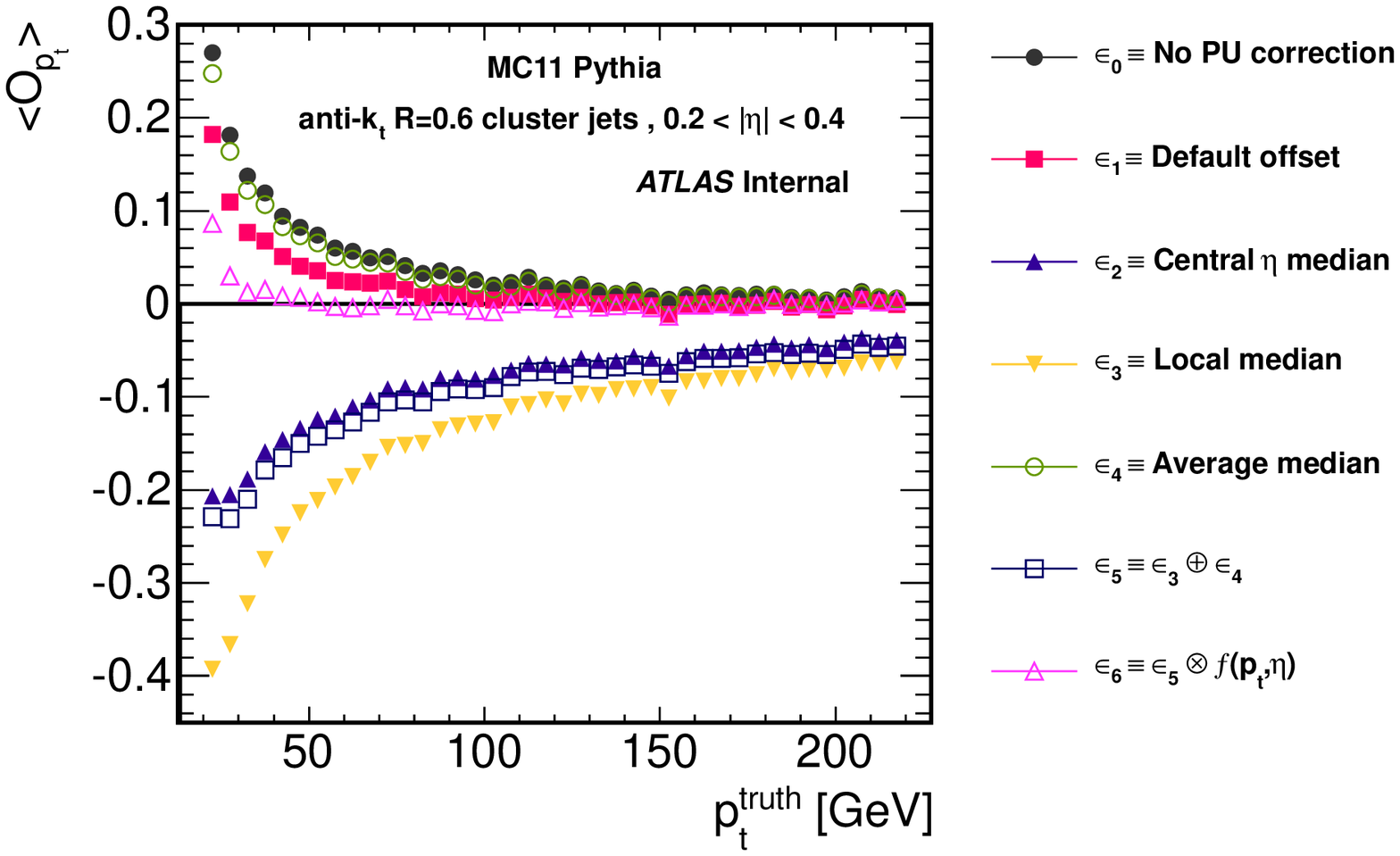}} 
\subfloat[\figQaud]{\label{FIGmeanClosureOffsetPtCompCalib4}\includegraphics[trim=5mm 10mm 0mm 10mm,clip,width=.643\textwidth]{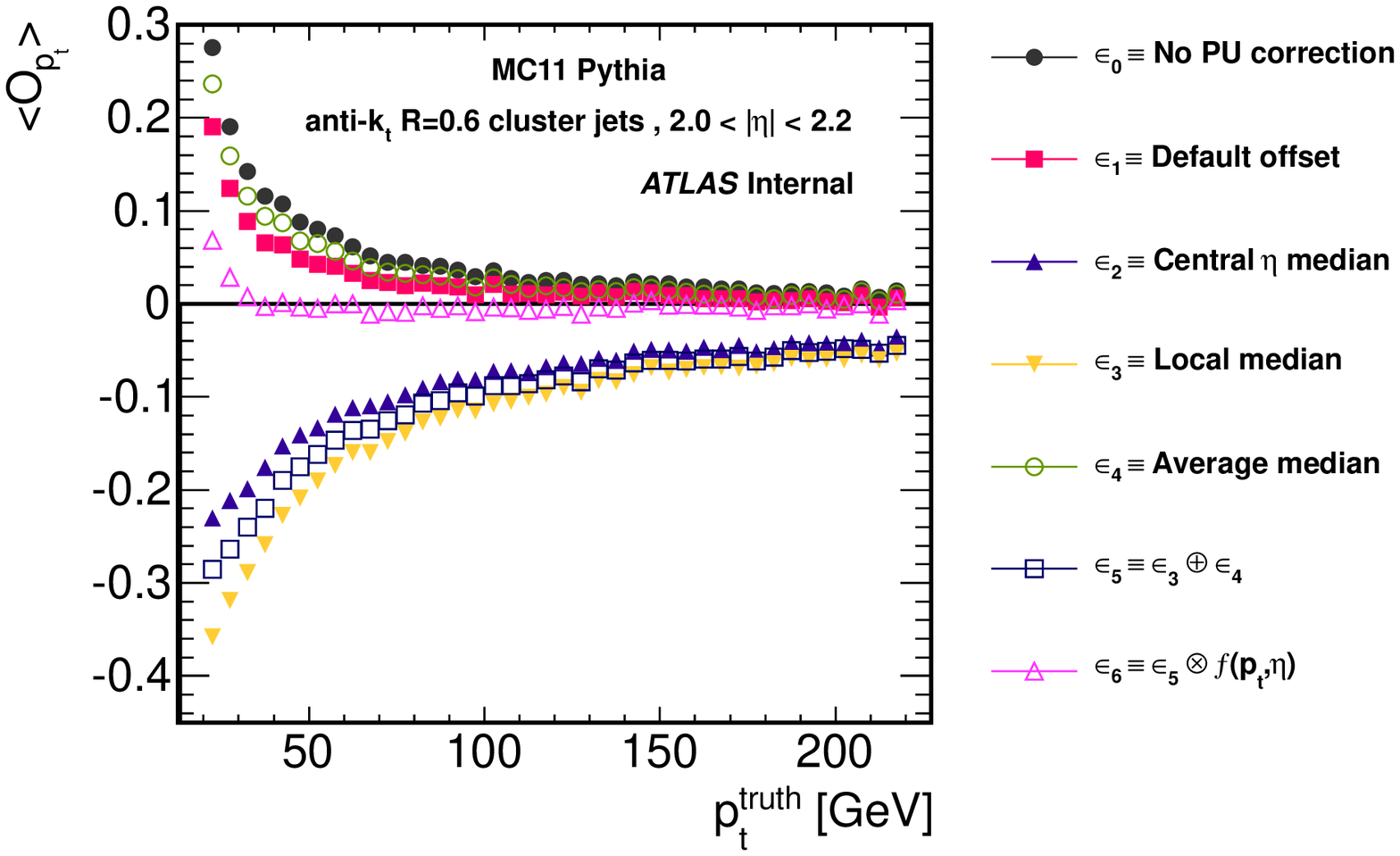}}
  \caption{\label{FIGmeanClosureOffsetPtCompCalib}Dependence on truth jet transverse momentum, $p_{\mrm{t}}^{\mrm{truth}}$,
    of $<\left(O_{p_{\mrm{t}}} \times p_{\mrm{t}}^{\mrm{truth}}\right)>$ (\Subref{FIGmeanClosureOffsetPtCompCalib1}, \Subref{FIGmeanClosureOffsetPtCompCalib2})
    and of $<O_{p_{\mrm{t}}}>$ (\Subref{FIGmeanClosureOffsetPtCompCalib3}, \Subref{FIGmeanClosureOffsetPtCompCalib4}),
    where $O_{p_{\mrm{t}}}$ is the relative transverse momentum offset, using jets within two pseudo-rapidity, \Eta, regions,
    corrected for \pu with $\epsilon_{0}$-$\epsilon_{6}$,
    as indicated in the figures.
    The lines in \Subref{FIGmeanClosureOffsetPtCompCalib1} and \Subref{FIGmeanClosureOffsetPtCompCalib2} represent linear
    fits to points in $\epsilon_{5}$.
  }
\end{center}
\end{figure} 
For the case of $\epsilon_{5}$,
a linear fit is performed within ${50 < p_{\mrm{t}}^{\mrm{truth}} < 250\GeV}$. The results are extrapolated to higher \pt values
and used to derive the \JES scalling function, $f\left(p_{\mrm{t}},\eta\right)$, (see \autoref{eqPuCorrection4})
which is used for $\epsilon_{6}$.
From \autoref{FIGmeanClosureOffsetPtCompEta}, one may compare the performance of the default offset correction, $\epsilon_{1}$,
with that of the nominal median-based correction, $\epsilon_{6}$, in different \Eta bins.
\begin{figure}[htp]
\begin{center}
\subfloat[]{\label{FIGmeanClosureOffsetPtCompEta1}\includegraphics[trim=5mm 10mm 65mm 10mm,clip,width=.478\textwidth]{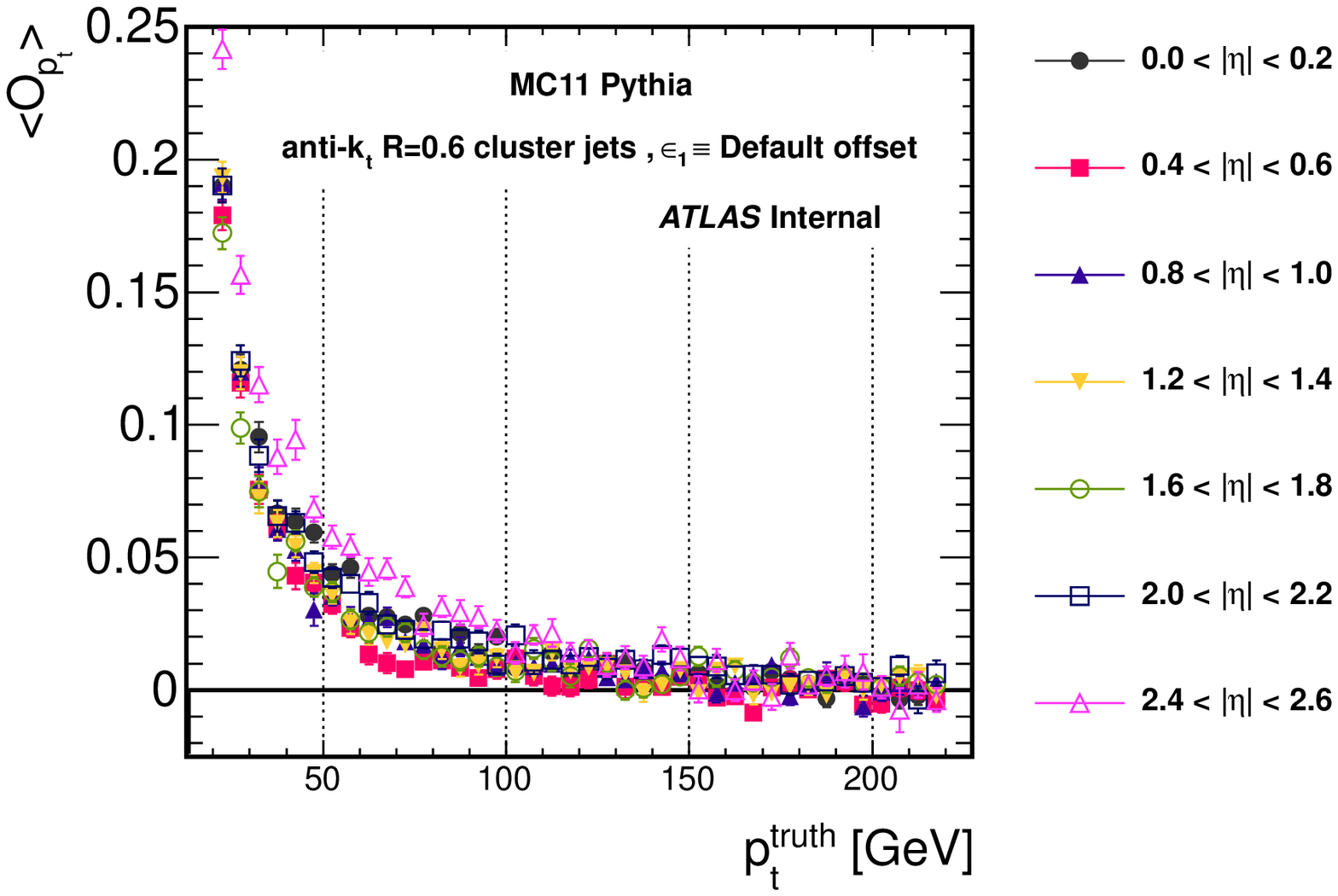}} 
\subfloat[\figQaud]{\label{FIGmeanClosureOffsetPtCompEta2}\includegraphics[trim=5mm 10mm 0mm 10mm,clip,width=.715\textwidth]{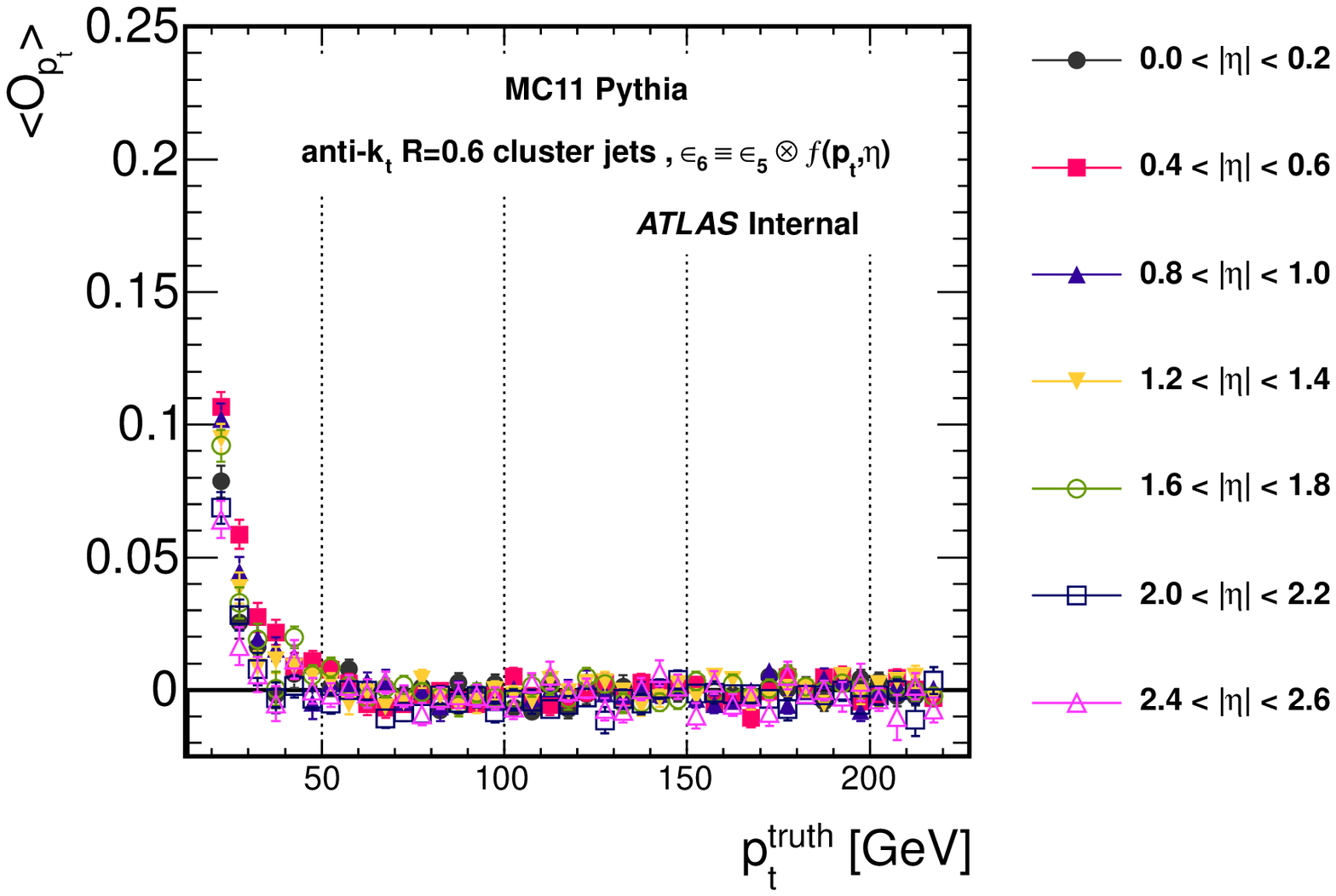}} 
  \caption{\label{FIGmeanClosureOffsetPtCompEta}Dependence on truth jet transverse momentum, $p_{\mrm{t}}^{\mrm{truth}}$,
    of the mean transverse momentum offset, $<O_{p_{\mrm{t}}}>$, for two correction
    methods, $\epsilon_{1}$ \Subref{FIGmeanClosureOffsetPtCompEta1} and $\epsilon_{6}$ \Subref{FIGmeanClosureOffsetPtCompEta2},
    for jets within several pseudo-rapidity, \Eta, regions, as indicated in the figures.
  }
\end{center}
\end{figure} 

The closure using $\epsilon_{6}$ is comparable to that which is achieved using $\epsilon_{1}$
above~100\GeV. At low-\pt, the reach of jet reconstruction using $\epsilon_{6}$ is extended down to~30\GeV and the overall performance
is improved.

Another important figure of merit is the jet \pt-resolution. The resolution is estimated by the standard deviation of the
distribution of the \pt offset, $O_{p_{\mrm{t}}}$. The standard deviation for jets which are calibrated with $\epsilon_{0}$-$\epsilon_{6}$
within a limited \Eta range is compared in \autoref{FIGsdClosureOffsetPtCompEta1}. Different \Eta bins are shown in
\autoref{FIGsdClosureOffsetPtCompEta2}, where jets are calibrated by $\epsilon_{6}$.
\begin{figure}[htp]
\begin{center}
\subfloat[]{\label{FIGsdClosureOffsetPtCompEta1}\includegraphics[trim=5mm 14mm 0mm 10mm,clip,width=.52\textwidth]{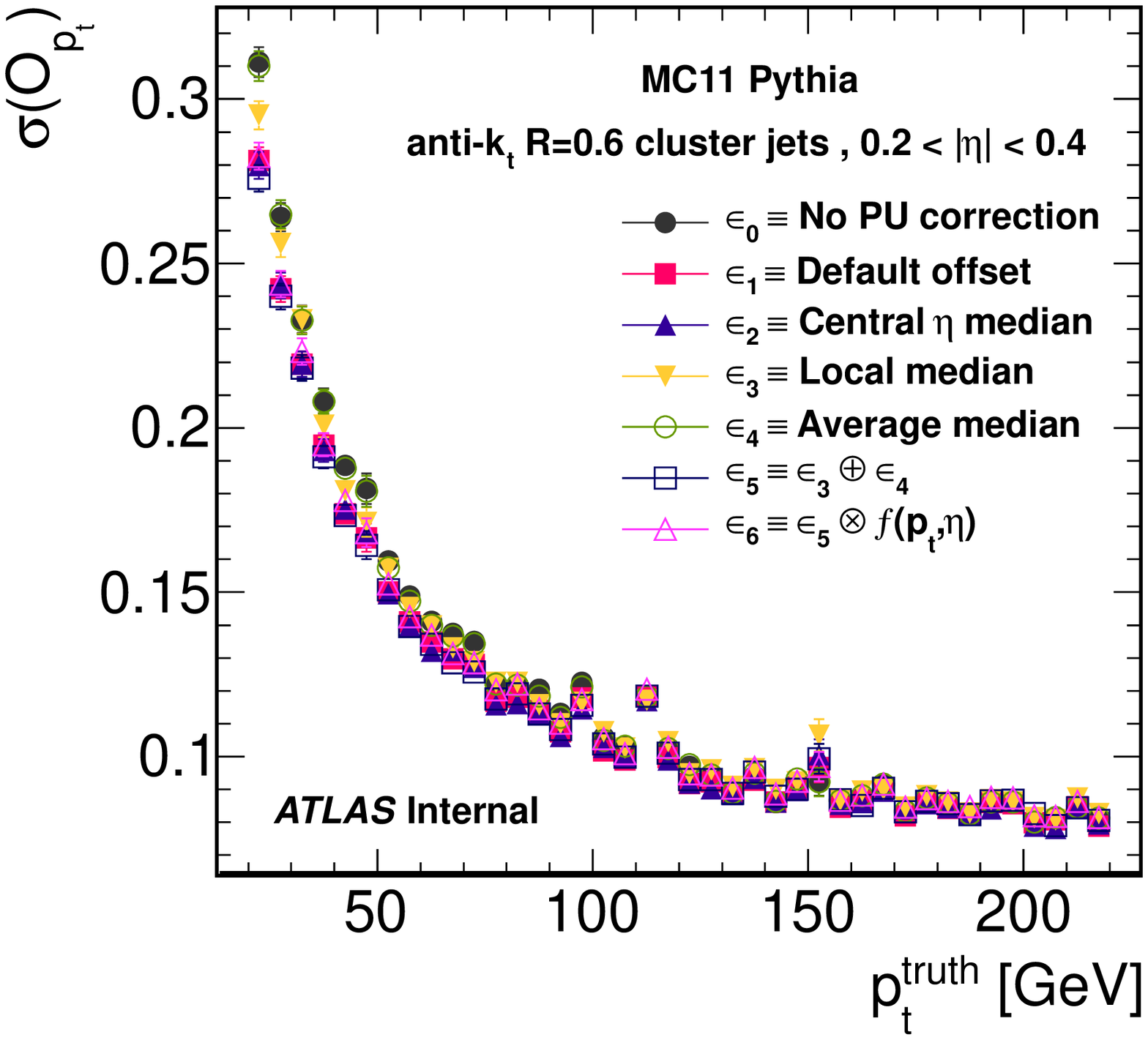}}
\subfloat[]{\label{FIGsdClosureOffsetPtCompEta2}\includegraphics[trim=5mm 14mm 0mm 10mm,clip,width=.52\textwidth]{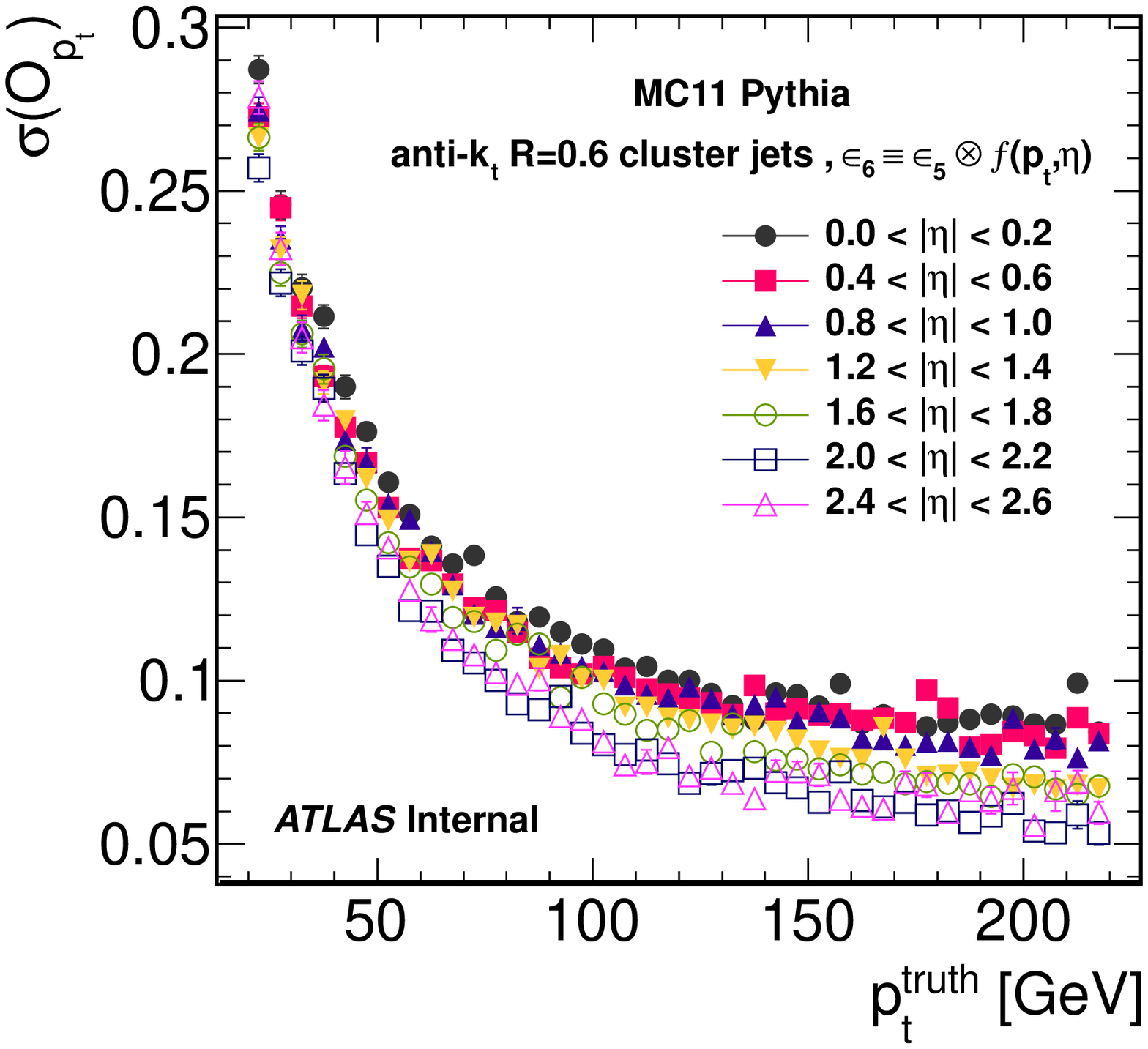}}
\caption{\label{FIGsdClosureOffsetPtCompEta}Dependence on truth jet transverse momentum, $p_{\mrm{t}}^{\mrm{truth}}$,
    of the standard deviation of the transverse momentum offset, $\sigma(O_{p_{\mrm{t}}})$,
    for jets with pseudo-rapidity, \etaRange{0.2}{0.4}, corrected for \pu with $\epsilon_{0}$-$\epsilon_{6}$ \Subref{FIGsdClosureOffsetPtCompEta1},
    and for jets within different \Eta regions, corrected for \pu with $\epsilon_{6}$ \Subref{FIGsdClosureOffsetPtCompEta2},
    as indicated in the figures.
}
\end{center}
\end{figure} 
The resolution using $\epsilon_{6}$ is comparable to that which is achieved using the default offset correction ($\epsilon_{1}$) across \Eta.

\section{Systematic checks of the median correction\label{sectSystematicChecksMedianCorrection}}
%
%
\subsection{Stability in MC for different parameters of the jet algorithm\label{sectStabilityOfMedianForDifferentAlgoParam}}
%
%
A basic assumption of the jet area/median method to correct for \pu the jets of the hard interaction, is that
the \pt-density of accompanying jets provides a good handle on the \pu in that event.
Implicit in this assumption, is that \Rho is independent of the type of jets used to calculate it.
Another requirement of the method, is that the correction is independent of the type of hard jets.
The validity of these two properties is investigated in the following.

\Autorefs{FIGjetClosureSystemJetsA1} and~\ref{FIGjetClosureSystemJetsA2} respectively show the dependence of the slope parameter, $\beta_{\pt}^{\Mu}$,
(see \autoref{eqOffsetFitForm1}) and of the slope ratio parameter, $\mathcal{R}_{\beta}^{\Npv}$, (see \autoref{eqOffsetFitForm2})
on truth jet transverse momentum\footnote{ The following study was performed using MC10b. The values of the fit results presented,
are therefore slightly different than those shown in previous sections. However, the conclusions regarding jet parameters
and the subsequent performance are a feature of the correction method and hold true.}.
The median used for the \pu correction is calculated from \KT jets with size parameters, $0.3 < R < 0.6$.
Here, the average correction method, $\epsilon_{4}$, is used to subtract \pu from \AKT jets with $R = 0.6$.
A stable result, independent of $R$, is achieved in the range, ${R > 0.2}$.

\Autoref{FIGjetClosureSystemJetsB1} illustrates the change in performance, according to
the choice of algorithm used for reconstructing the jets from which \Rho is calculated.
The best results are achieved for the \KT algorithm,
with slightly worse performance of the \CAalg algorithm. A drop of roughly 10\% on average is exhibited
by \AKT jets in comparison, as expected.
In \Autoref{FIGjetClosureSystemJetsB2}
the nominal definition of \Rho (calculated from \KT jets with $R = 0.4$) is used. Here the local ($\epsilon_{3}$) and
the average ($\epsilon_{4}$) \pu correction methods are applied to correct \AKT jets with two different size parameters.
The difference in performance for the different jet sizes is small.
\begin{figure}[htp]
\begin{center}
\vspace{-20pt}
\subfloat[]{\label{FIGjetClosureSystemJetsA1}\includegraphics[trim=5mm 14mm 0mm 10mm,clip,width=.52\textwidth]{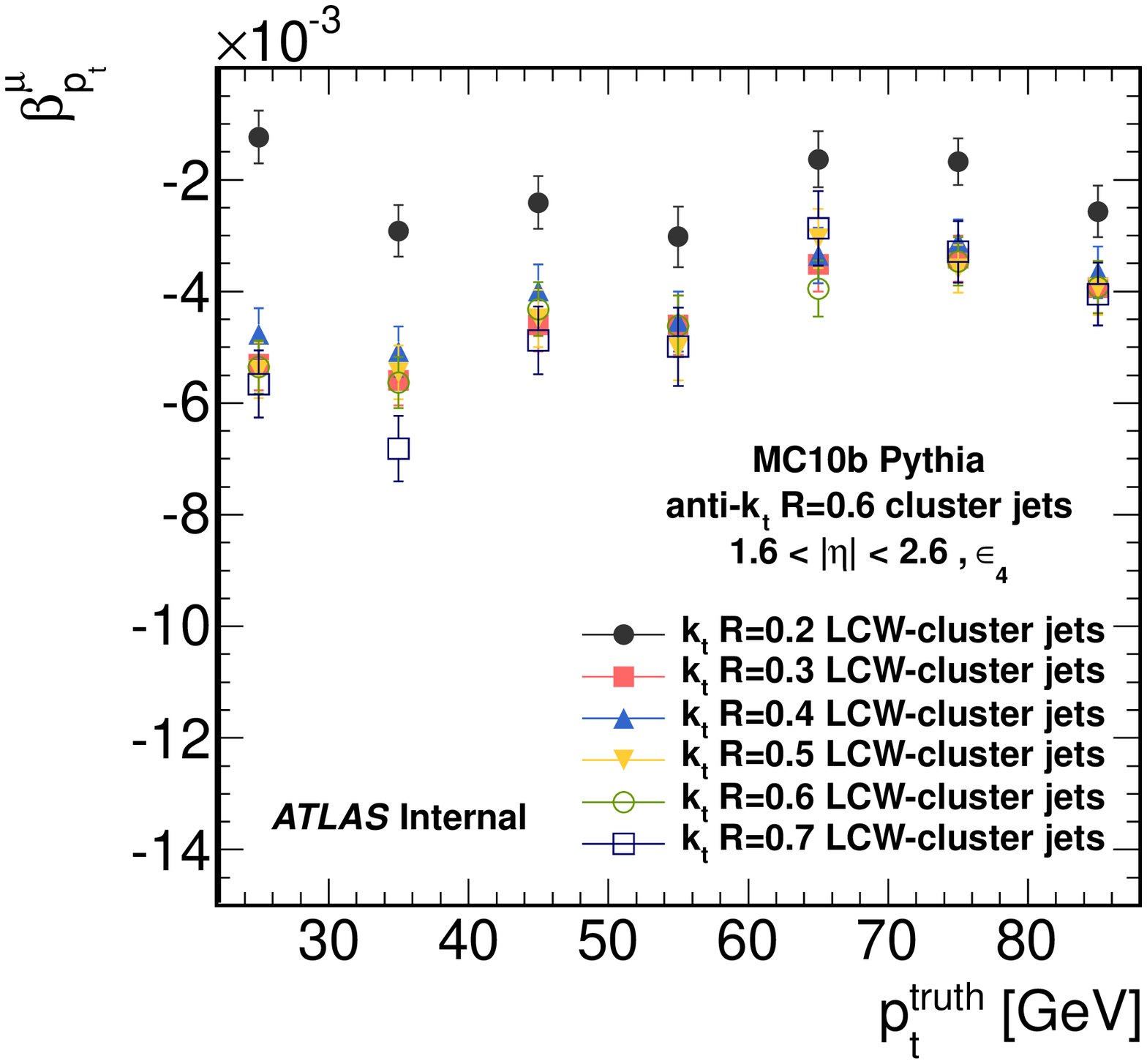}}
\subfloat[]{\label{FIGjetClosureSystemJetsA2}\includegraphics[trim=5mm 14mm 0mm 10mm,clip,width=.52\textwidth]{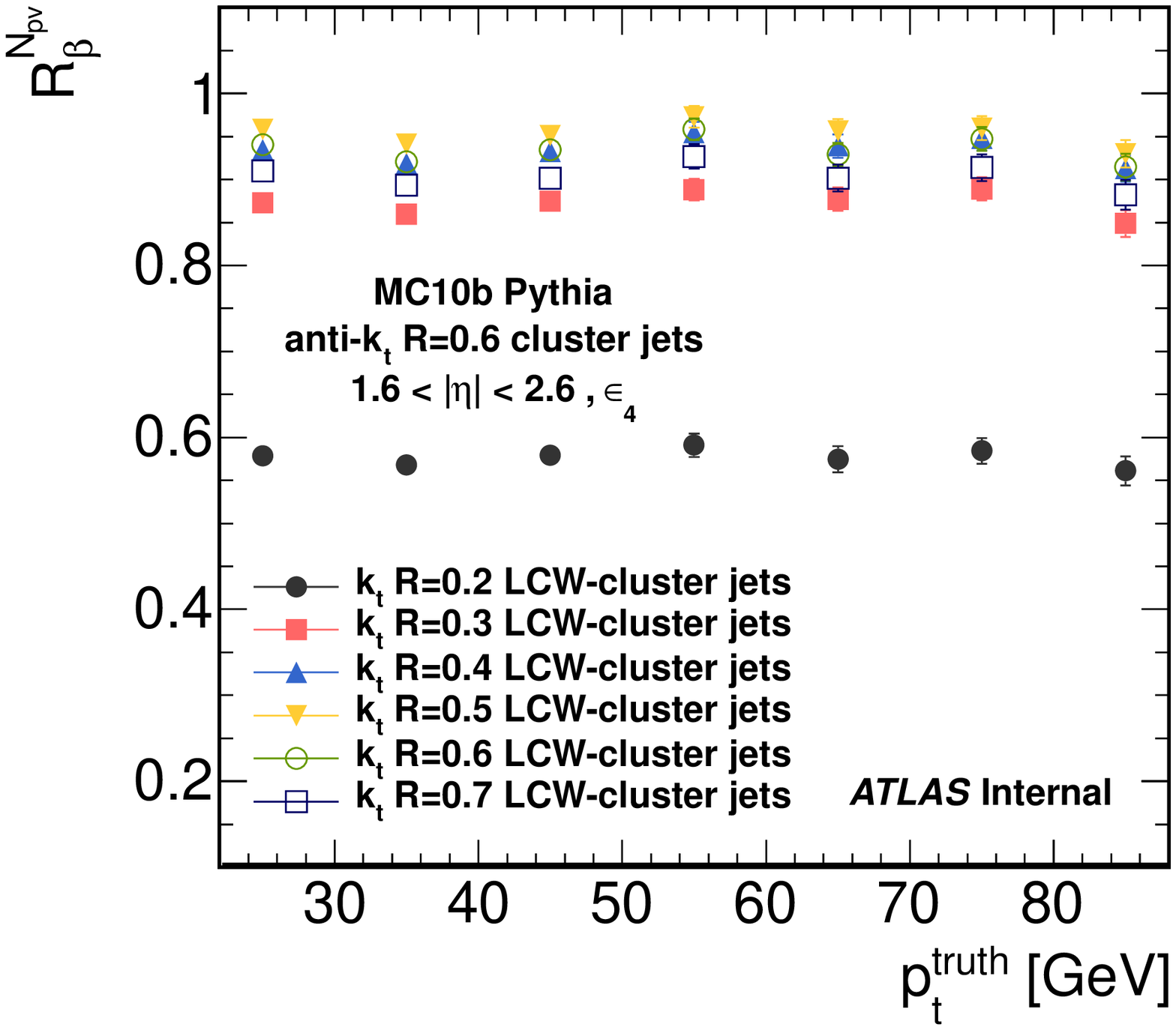}}
  \caption{\label{FIGjetClosureSystemJetsA}Dependence on truth jet transverse momentum, $p_{\mrm{t}}^{\mrm{truth}}$, of the
  parameters $\beta_{\pt}^{\Mu}$ \Subref{FIGjetClosureSystemJetsA1}
  and $\mathcal{R}_{\beta}^{\Npv}$ \Subref{FIGjetClosureSystemJetsA2}, defined in \autorefs{eqOffsetFitForm1}~and~\ref{eqOffsetFitForm2},
  for jets in MC10b with size parameter, $R = 0.6$, and pseudo-rapidity, $1.6 < |\eta| < 2.6$. Jets are
  corrected for \pu with $\epsilon_{4}$, which is calculated using \KT jets with different size parameters, as indicated in the figures.
  }
\end{center}
%
%
\begin{center}
\subfloat[]{\label{FIGjetClosureSystemJetsB1}\includegraphics[trim=5mm 14mm 0mm 10mm,clip,width=.52\textwidth]{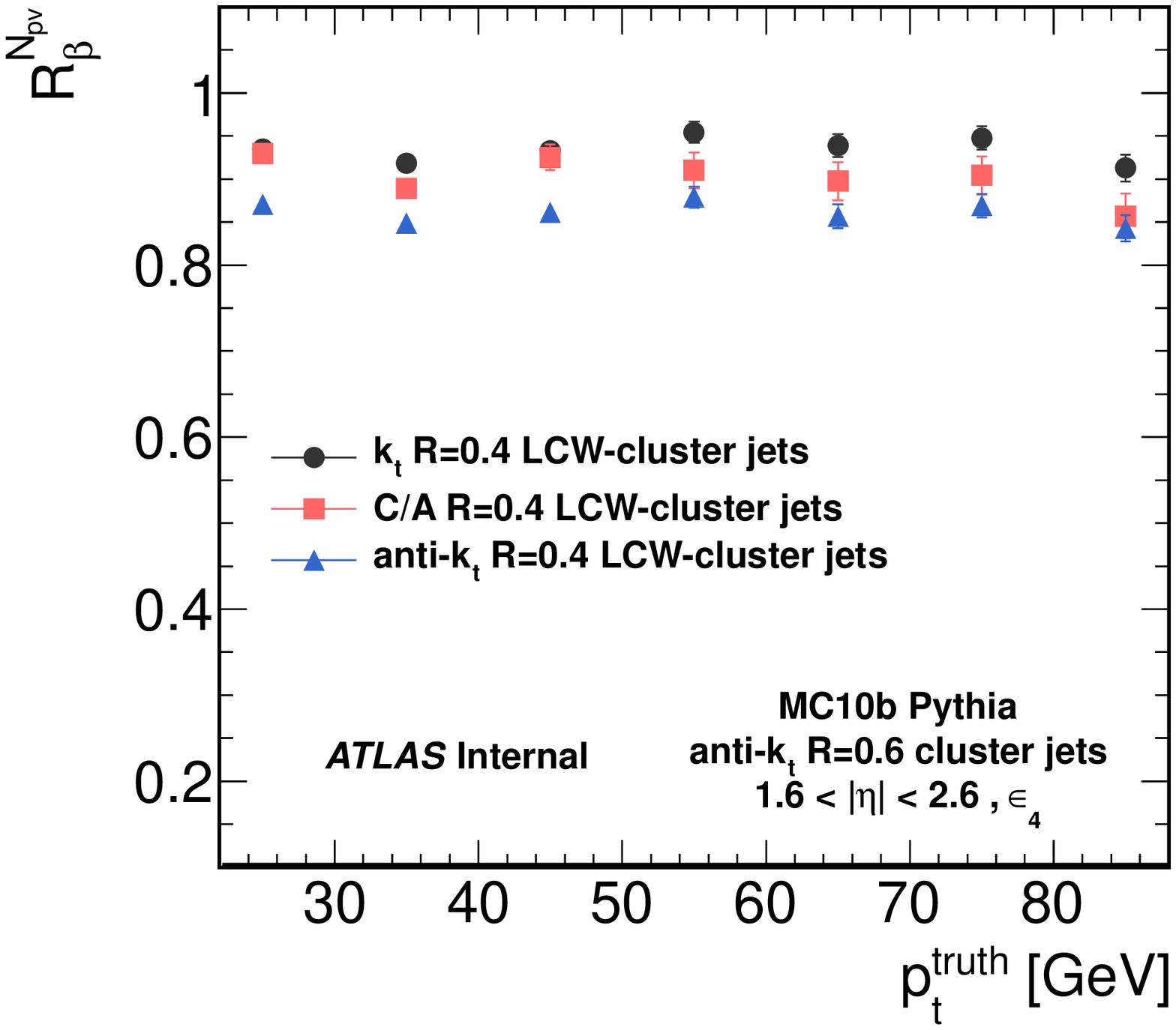}}
\subfloat[]{\label{FIGjetClosureSystemJetsB2}\includegraphics[trim=5mm 14mm 0mm 10mm,clip,width=.52\textwidth]{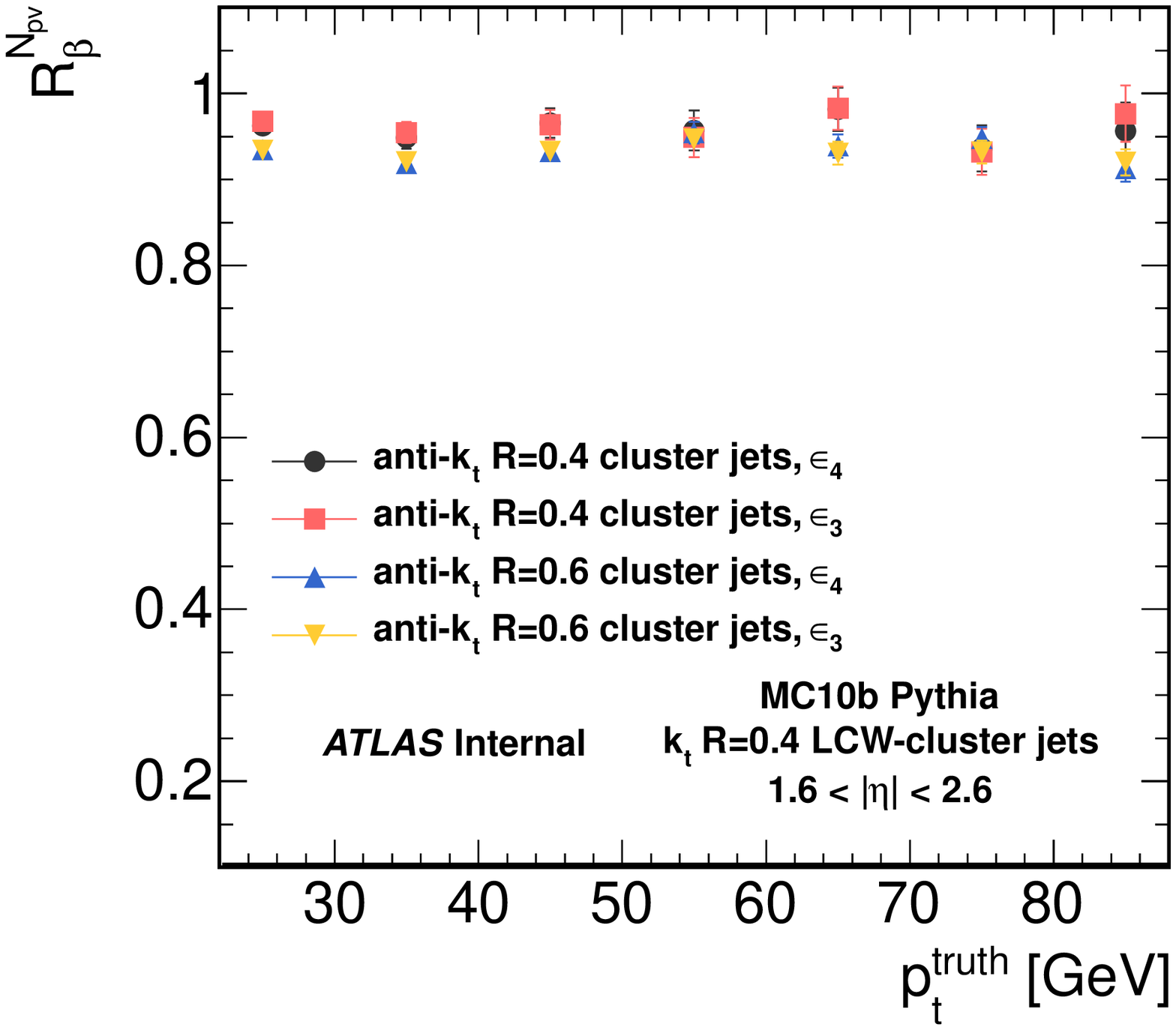}}
  \caption{\label{FIGjetClosureSystemJetsB}Dependence in MC10b on truth jet transverse momentum, $p_{\mrm{t}}^{\mrm{truth}}$, of the
  parameter $\mathcal{R}_{\beta}^{\Npv}$ defined in \autoref{eqOffsetFitForm1}. \\
  \Subref{FIGjetClosureSystemJetsB1} Jets of type \AKT with size parameter, $R = 0.6$, and pseudo-rapidity, $1.6 < |\eta| < 2.6$, are corrected for \pu
  with $\epsilon_{4}$. The correction is calculated from jets which are reconstructed using different algorithms, as indicated.
  \\
  \Subref{FIGjetClosureSystemJetsB2} Jets of type \AKT with size parameters, $R = 0.4$ and $R = 0.6$, and pseudo-rapidity, $1.6 < |\eta| < 2.6$,
  are corrected for \pu with $\epsilon_{3}$ and $\epsilon_{4}$, calculated using
  $R = 0.4$ \KT jets (the nominal configuration of the correction), as indicated in the figure.
}
\end{center}
\end{figure} 
%
%

\subsection{Stability in data}
%
%
Stability of the average median in the data is validated by comparing sub-samples of events.
The following possible sources of bias are investigated:
\begin{list}{-}{}
\mynobreakpar
\item \headFont{dependence on the data-taking period -} the median may depend on changing conditions in the
                                                        calorimeter over time (due to \eg radiation damage and faulty electronics);
\item \headFont{dependence on the trigger -}            the median may depend on the \pt scale of the
                                                        hard interaction in an event, which is correlated with the trigger selection;
\item \headFont{dependence on the position of the 
      bunch-crossing in 
      the bunch train -}                                \otpu depends on the history of collisions within
                                                        any~$\sim400\ns$ window; accordingly, so may the median.
\end{list}

\minisec{Dependence on the data-taking period}
%
%
In \autoref{FIGavgRhoChecksPariods} the dependence of the average median, $\left<\rho\right>$, on \Npv is compared
between data taken in different periods, and with the average over all periods (inclusive sample).
The response is very stable; no dependence on the date-taking period is observed.
\begin{figure}[htp]
\begin{center}
\subfloat[]{\label{FIGavgRhoChecksPariods0}\includegraphics[trim=5mm 14mm 0mm 10mm,clip,width=.52\textwidth]{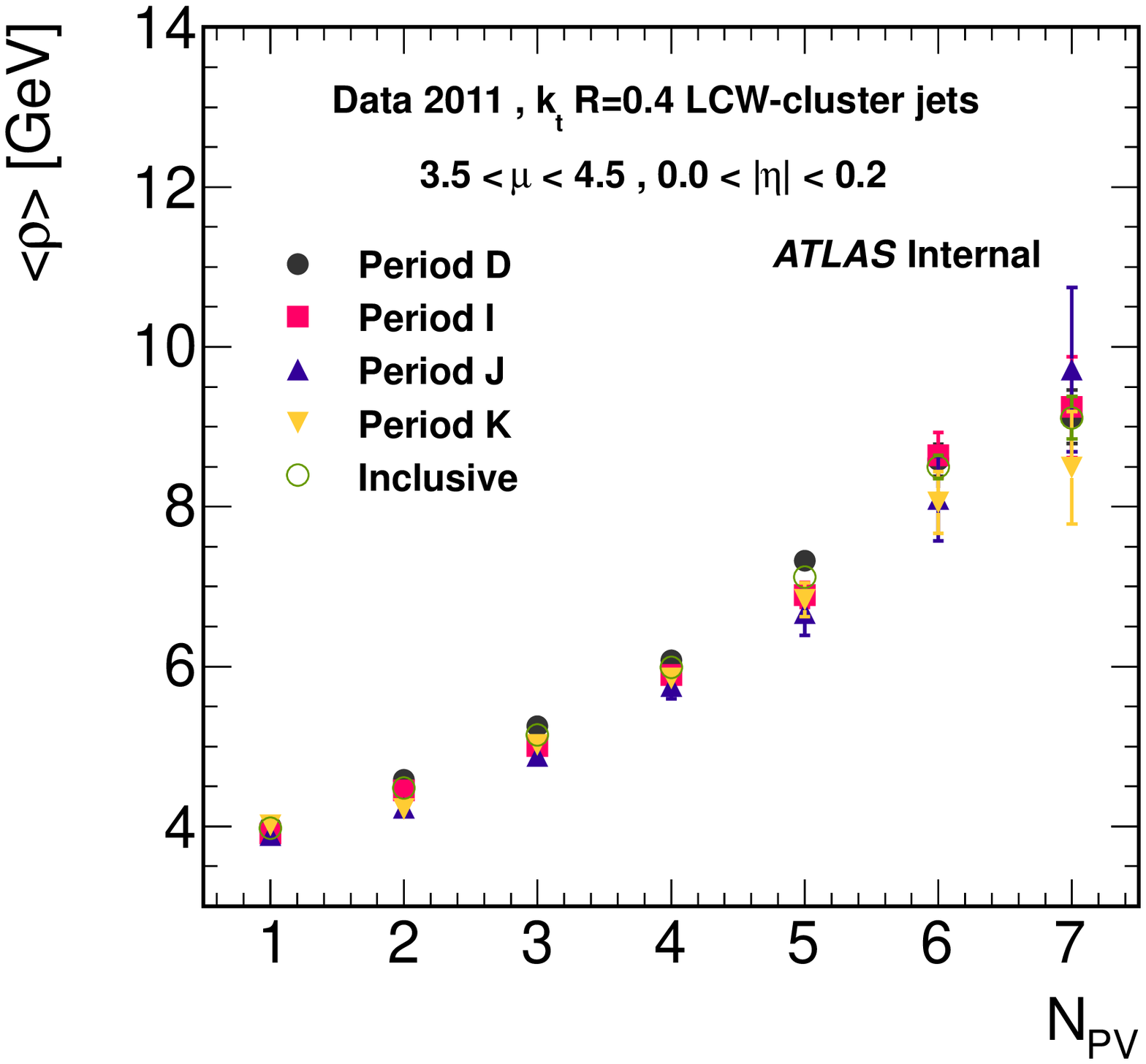}}
\subfloat[]{\label{FIGavgRhoChecksPariods1}\includegraphics[trim=5mm 14mm 0mm 10mm,clip,width=.52\textwidth]{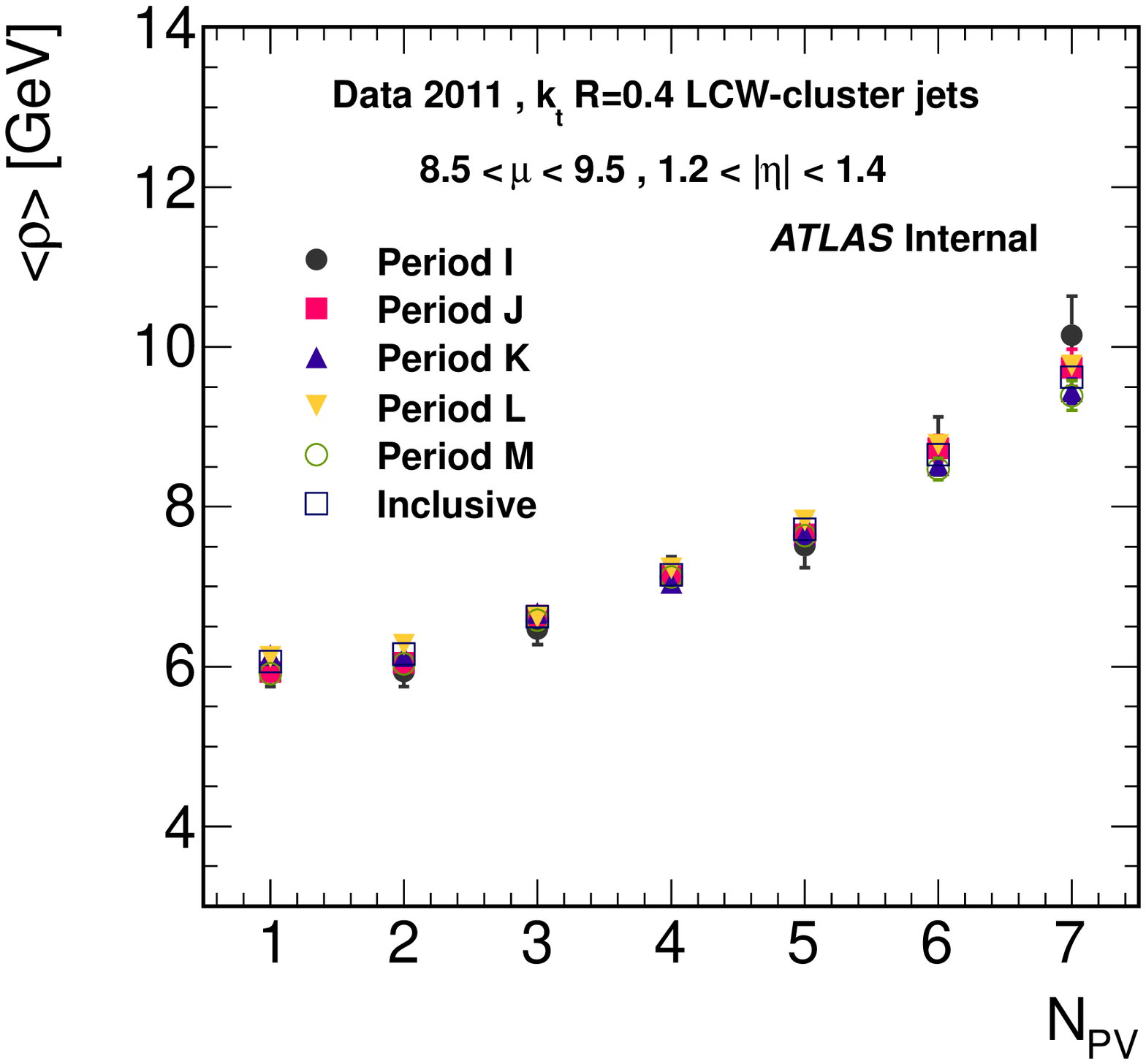}}
  \caption{\label{FIGavgRhoChecksPariods}Dependence 
    of the average median, $\left<\rho\right>$, on the number of reconstructed vertices, \Npv,
    for jets with pseudo-rapidity, \etaLower{0.2}, in events with an average number of interactions, \muRange{3.5}{4.5} \Subref{FIGavgRhoChecksPariods0},
    and for jets with \etaRange{1.2}{1.4} in events with \muRange{8.5}{9.5} \Subref{FIGavgRhoChecksPariods1}.
    Different data-taking periods and an average over all periods (inclusive) are compared, as indicated in the figures.
  }
\end{center}
\end{figure} 

\minisec{Dependence on the trigger}
%
%
\Autoref{FIGavgRhoChecksTriggers} shows the dependence of $\left<\rho\right>$ on \Npv using data
taken with different triggers\footnote{ Using the two-trigger selection scheme, these refer to the triggers which are associated with the highest-\pt
jet in the event.}.
The random trigger, \ttt{EF\ul rd0}, and the first five (ordered by trigger threshold)
jet triggers, \ttt{EF\ul J10}~-~\ttt{EF\ul J40}, are compared in \autoref{FIGavgRhoChecksTriggers0};
the random trigger and an inclusive sample, consisting of all eleven jet triggers, \ttt{EF\ul J10}~-~\ttt{EF\ul J240},
(see \autoref{TBLcentralTrigNames2011}), are compared in \autoref{FIGavgRhoChecksTriggers1}.
\begin{figure}[htp]
\begin{center}
\subfloat[]{\label{FIGavgRhoChecksTriggers0}\includegraphics[trim=5mm 14mm 0mm 10mm,clip,width=.52\textwidth]{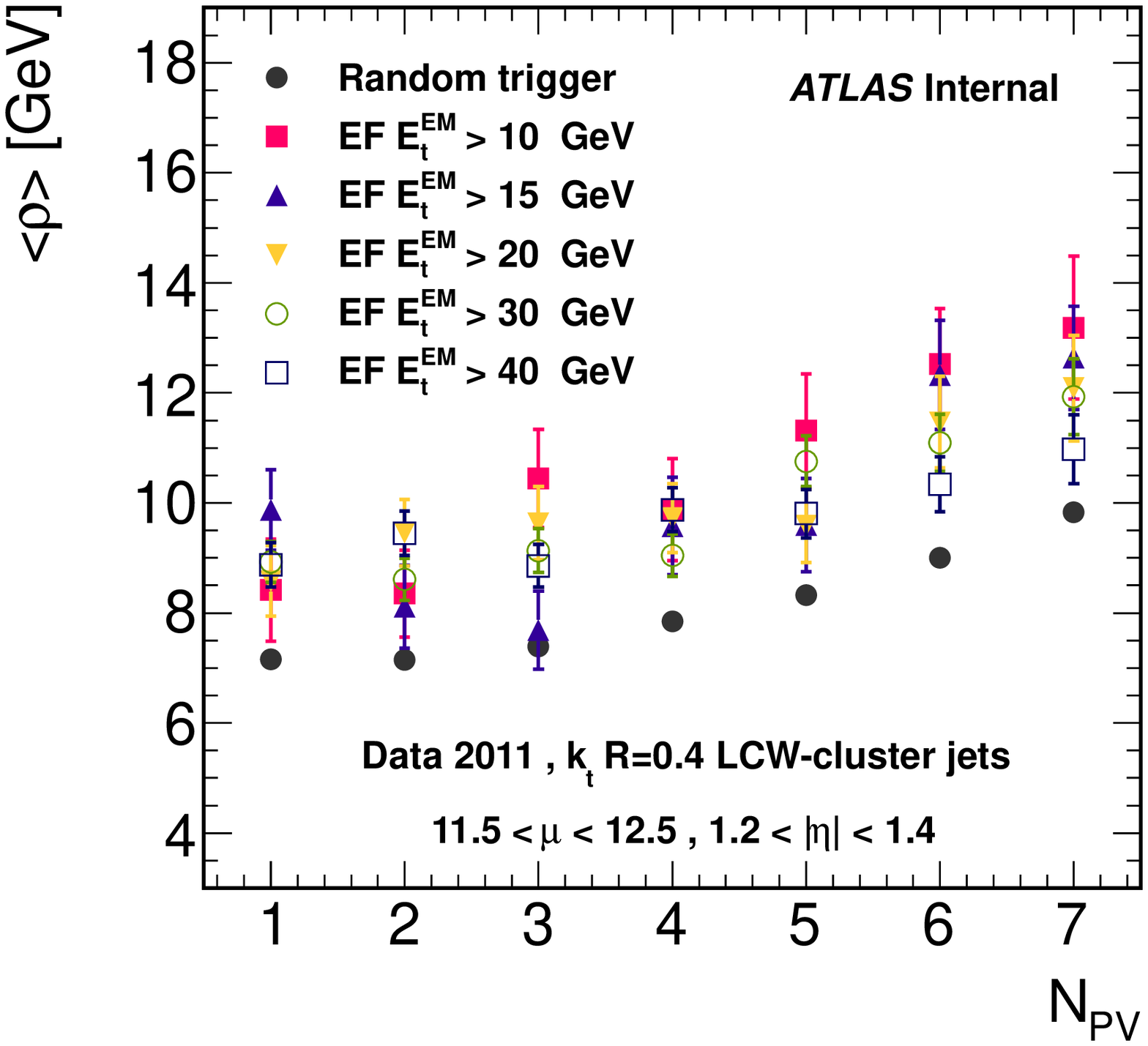}}
\subfloat[]{\label{FIGavgRhoChecksTriggers1}\includegraphics[trim=5mm 14mm 0mm 10mm,clip,width=.52\textwidth]{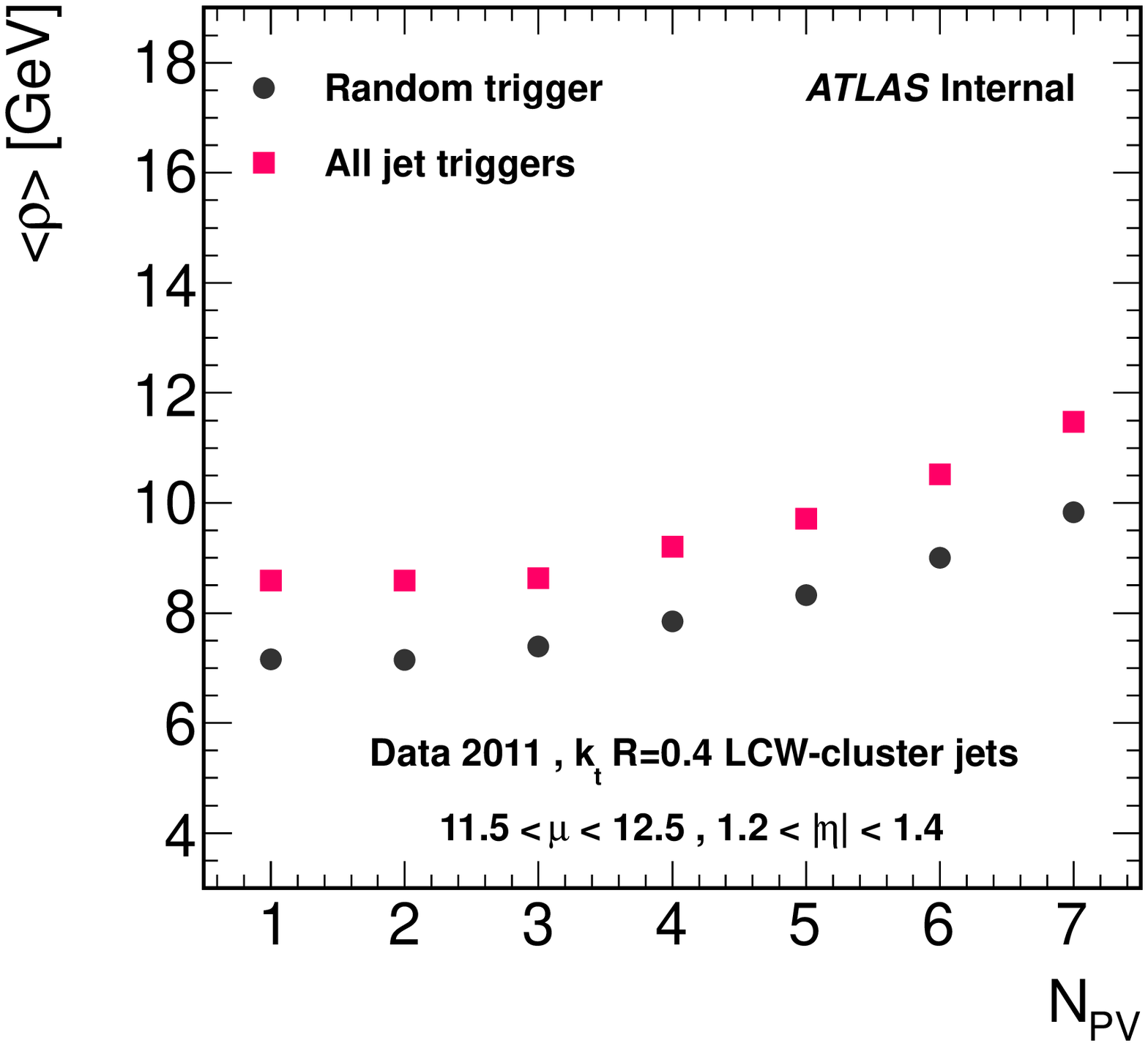}}
  \caption{\label{FIGavgRhoChecksTriggers}Dependence 
    of the average median, $\left<\rho\right>$, on the number of reconstructed vertices, \Npv,
    for jets with pseudo-rapidity, \etaRange{1.2}{1.4}, in events with an average number of interactions, \muRange{11.5}{12.5}.
    Data taken with different triggers are compared, as indicated in the figures.
    The random trigger, \ttt{EF\ul rd0}, and the several jet triggers, \ttt{EF\ul J10}~-~\ttt{EF\ul J40}, are compared in
    \Subref{FIGavgRhoChecksTriggers0}. The random trigger and an inclusive sample, consisting of all eleven
    jet triggers, \ttt{EF\ul J10}~-~\ttt{EF\ul J240}, are compared in \Subref{FIGavgRhoChecksTriggers1}.
  }
\end{center}
\end{figure} 
Data taken using the different jet triggers present a consistent response within errors; data taken with
the random trigger is characterized by smaller values of $\left<\rho\right>$ in comparison.

The difference can be explained by the fact that
events taken with the random trigger generally have lower jet multiplicity in comparison with those taken using
jet triggers. This is due to the increase in \ue activity with \pt for the low-\pt events, which are accessible by \ttt{EF\ul rd0}.
Beyond a certain threshold in the scale of the hard interaction (roughly 30\GeV), the energy density of the \ue does
not significantly increase~\cite{:2012fs}. The median is therefore constant on average for events picked up by the jet triggers.
The averaging of \Rho in a given region in \Eta also takes into account
events where no jets are reconstructed within that region ($\rho=0$).
Consequently, for the random trigger, more events with a low or a null value of the median
are included in the calculation of $\left<\rho\right>$.

In order to account for the different response between trigger streams, the average median is parametrized in data
which are taken exclusively with jet triggers. 
A correction factor (generally between~0.6 and~1) is applied, when using this parametrization for
correcting jets in random-triggered data.

\minisec{Dependence on the position of the bunch-crossing in the bunch train}
%
%
As discussed in \autoref{chapInAndOutOfTimePU},
the positive- and negative-energy components of \otpu depend on the number of overlapping bunch-crossings, which are integrated in the span of
calorimeter signal shaping. As a result, events at the beginning of a bunch train have different \otpu composition compared with later events. 
Roughly 400\ns after the beginning of a bunch train, the typical shaping time of calorimeter elements, this dependence is expected to lessen.

In order to asses this effect, the average median is studied in events with different values of $\Delta_{\mrm{bc}}$, which is defined as the amount of time
separating an event from the beginning of its respective bunch train. Several examples are shown in \autoref{FIGavgRhoChecksDistBX}.
\begin{figure}[htp]
\begin{center}
\subfloat[]{\label{FIGavgRhoChecksDistBX0}\includegraphics[trim=5mm 14mm 0mm 10mm,clip,width=.52\textwidth]{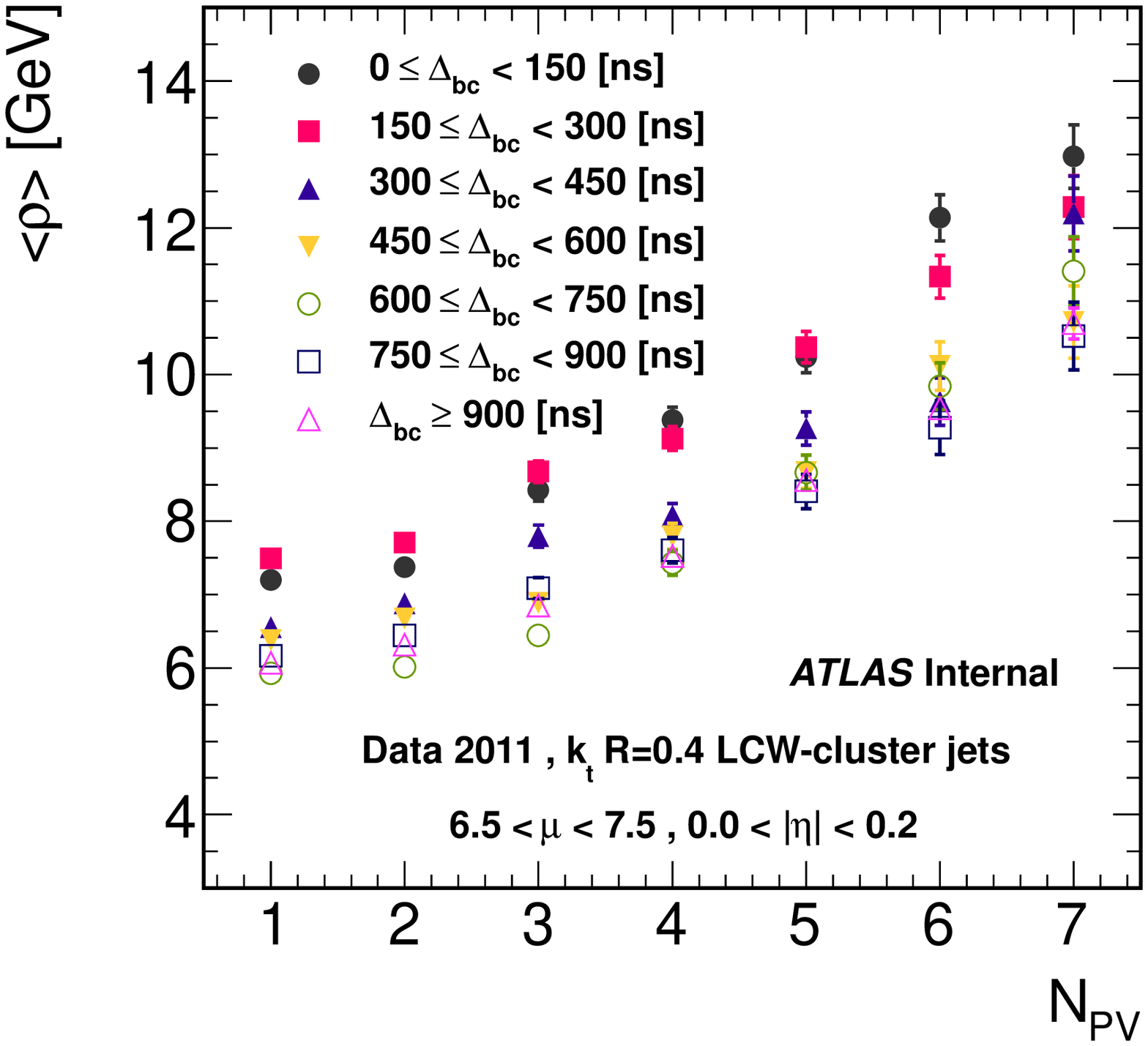}}
\subfloat[]{\label{FIGavgRhoChecksDistBX1}\includegraphics[trim=5mm 14mm 0mm 10mm,clip,width=.52\textwidth]{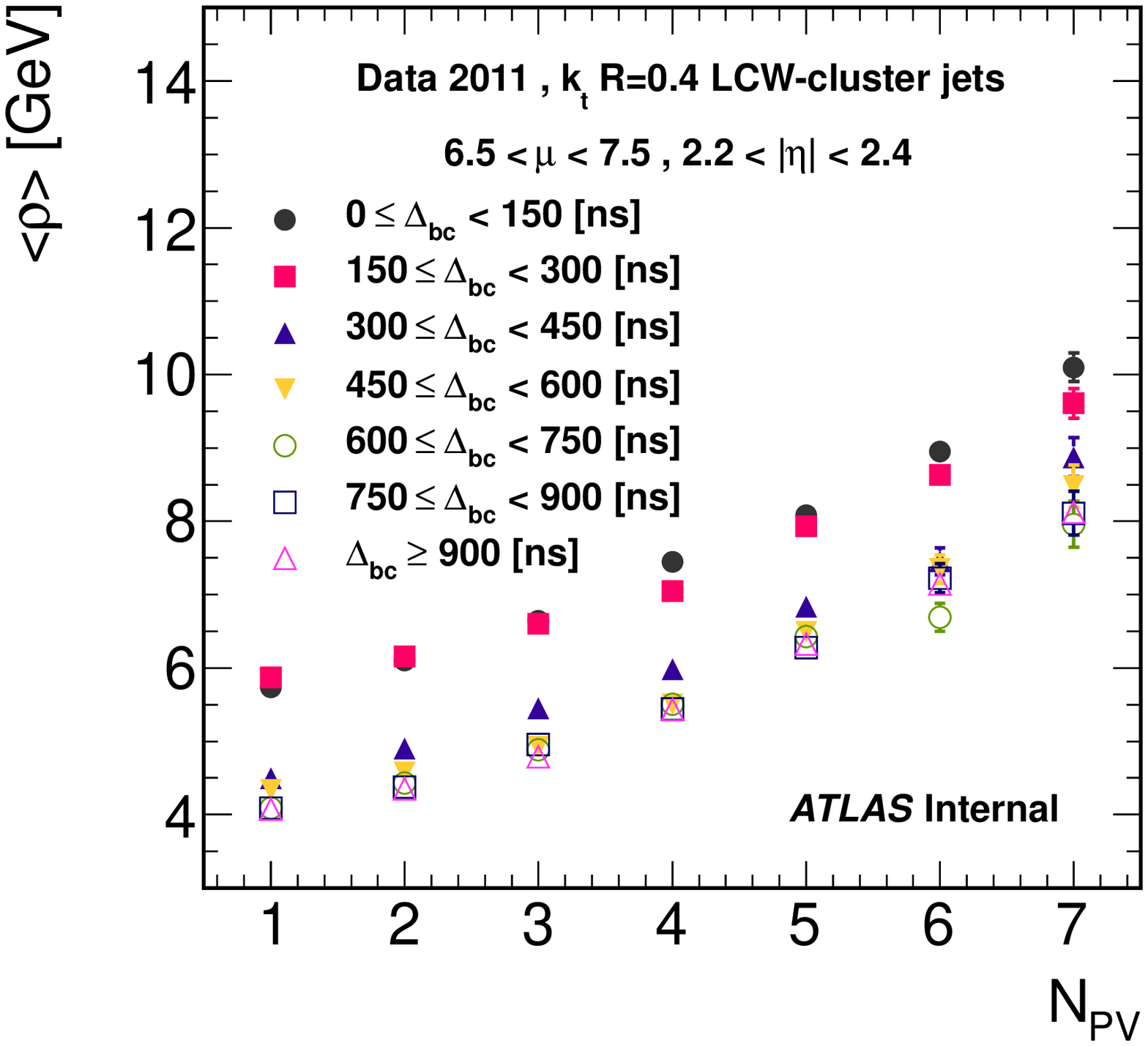}}
  \caption{\label{FIGavgRhoChecksDistBX}Dependence 
    of the average median, $\left<\rho\right>$, on the number of reconstructed vertices, \Npv,
    for jets with pseudo-rapidity, \etaLower{0.2}, \Subref{FIGavgRhoChecksDistBX0} or with
    \etaRange{2.2}{2.4}  \Subref{FIGavgRhoChecksDistBX1}, in events with an average number of interactions, \muRange{6.5}{7.5},
    and different values of $\Delta_{\mrm{bc}}$, the distance of an event from the beginning of the bunch train, as indicated in the figures.
  }
\end{center}
\end{figure} 
The values of \Rho measured in events at the beginning of the bunch train ($\Delta_{\mrm{bc}} < 300$\ns) are on average higher
than those which follow in time, as the negative component of the \otpu is absent. For $\Delta_{\mrm{bc}} > 300$\ns the
differences between $\Delta_{\mrm{bc}}$ classes are mostly within statistical errors.
In order to account for this dependence, the average \Rho is parametrized separately for events
with $\Delta_{\mrm{bc}} < 150$, $150 < \Delta_{\mrm{bc}} < 300$ and $\Delta_{\mrm{bc}} > 300$\ns.

\subsection{Associated systematic uncertainty of the jet energy scale\label{chapSystematicUncertMedianPU2011}}
%
%
The effects of the different components of the
\JES uncertainties (see \autoref{chapSystematicUncertaintiesJES})
on the transverse momentum of jets, are estimated by introducing positive or negative 
variations to the energy scale of jets in the data. The variations are randomly distributed
according to a Gaussian function, with the respective source of uncertainty as its width.
After the energy of jets is shifted, they are re-sorted according to \pt, and the leading jet is redefined
as the jet with the highest (shifted) \pt in the event. The \pt spectrum of the leading jet is remeasured.
The relative changes per-bin of the \pt distribution are taken as a measure of the uncertainty.
The different sources of uncertainty are individually varied, and are considered uncorrelated.
The total uncertainty is thus the quadratic sum of the uncertainties derived from the different sources.

All of the components of the \JES uncertainty are used in this analysis, with the exception of the in- and \otpu components,
which were derived for the default offset correction.
The residual \Npv and \Mu dependencies of the median correction, expressed respectively through $\beta_{\pt}^{\Npv}$
and $\beta_{\pt}^{\Mu}$ in \autoref{eqOffsetFitForm1},
are used to estimate the \pu uncertainty on the final energy scale of jets.
A variation on the fully calibrated transverse momentum of jets is also performed,
using the relative non-closure in MC. The latter is estimated by the average 
transverse momentum offset, $<O_{p_{\mrm{t}}}>$. Any difference of $<O_{p_{\mrm{t}}}>$ from zero indicates 
a bias in the final energy scale correction, $f\left(p_{\mrm{t}},\eta\right)$, introduced in \autoref{eqPuCorrection4}.
The relative non-closure ranges between \pow{0.2}{-3} and \pow{5}{-3} for jets with \etaLower{2.8}.
The three sources of uncertainty (variations in \Npv, \Mu and jet \pt) together make up the total
systematic \JES uncertainty associated with \pu.

A breakdown of the different relative \pu uncertainties is shown in
\autoref{FIGrelUncertLeadingJetPtNominalPu1} for jets with rapidity, \yLower{0.3}, calibrated under $\epsilon_{6}$.
The corresponding comparison of the complete set of systematic uncertainties originating from
all \insitu sources as well as that of \pu, is presented in \autoref{FIGrelUncertLeadingJetPtNominalPu2}.
\begin{figure}[htp]
\begin{center}
  \vspace{-40pt}
  \subfloat[]{\label{FIGrelUncertLeadingJetPtNominalPu1}\includegraphics[trim=30mm 4mm 0mm 10mm,clip,width=.405\textwidth]{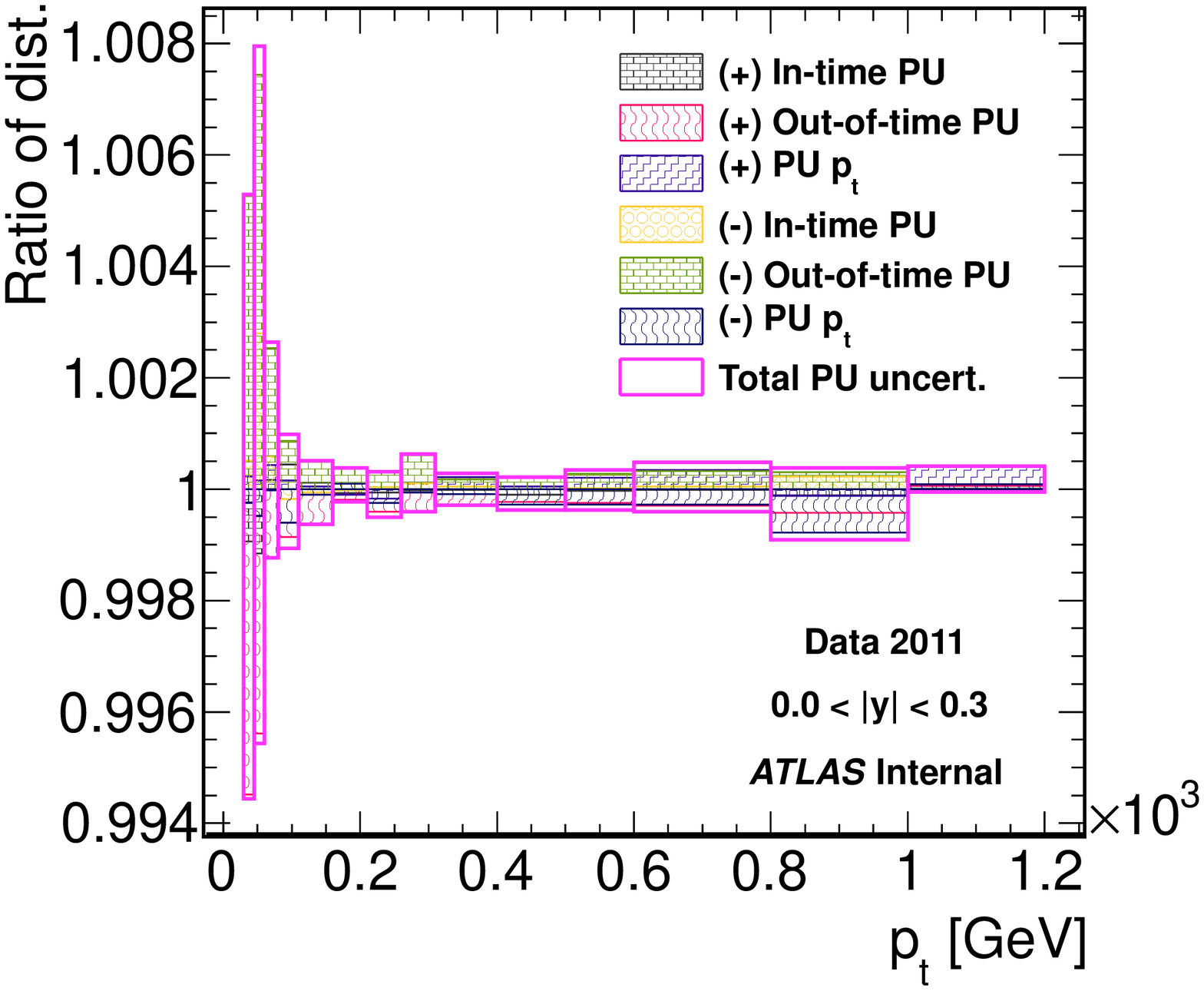}}
  \subfloat[\figQaud]{\label{FIGrelUncertLeadingJetPtNominalPu2}\includegraphics[trim=5mm  1.5mm 0mm 0mm,clip,width=.643\textwidth]{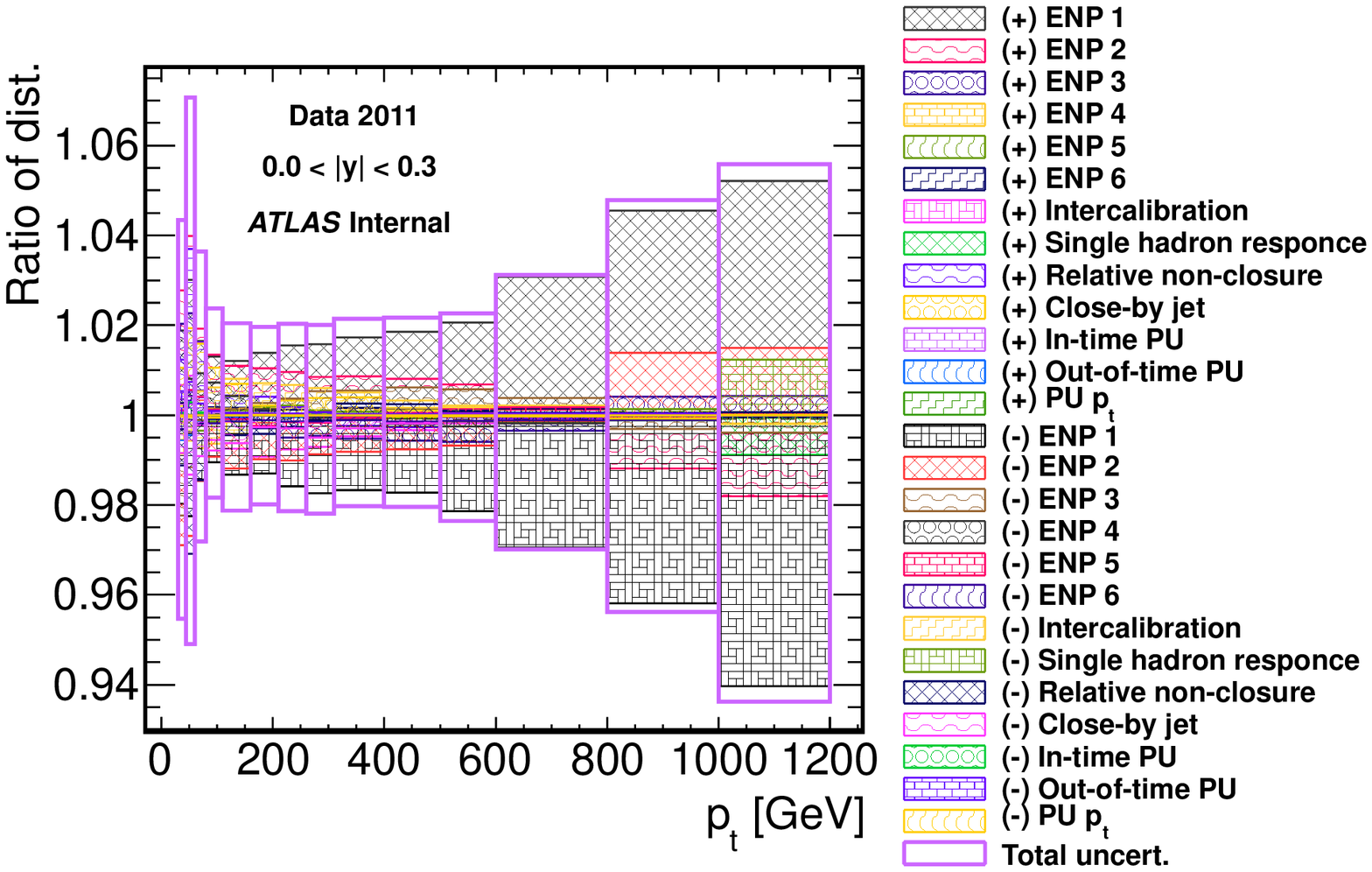}}
  \caption{\label{FIGrelUncertLeadingJetPtNominalPu}Dependence on the transverse momentum, \pt, of jets,
  of the ratio between the \pt distributions of jets with and without shifts of the jet energy scale (\JES), for jets with rapidity, \yLower{0.3},
  corrected for \pu with $\epsilon_{6}$ in the 2011 data.
    Positive $(+)$ and negative $(-)$ shifts of the energy scale are made,
    as indicated in the figures. \\
    \Subref{FIGrelUncertLeadingJetPtNominalPu1} Breakdown of the systematic sources of uncertainty associated with \pu; \itpu (variations in \Npv),
    \otpu (variations in \Mu) and PU \pt (variations in jet \pt due to non-closure in MC), as well as the total uncertainty
    from these three sources (total PU uncert.\/). \\
    \Subref{FIGrelUncertLeadingJetPtNominalPu2} Breakdown of all sources of uncertainty associated with the energy scale of jets,
    including those listed in \autoref{chapSystematicUncertaintiesJES}, those associated with \pu, and the total \JES systematic uncertainty
    (total uncert.\/).
  }
\end{center}
%
%
%
\begin{center}
\subfloat[]{\label{FIGrelUncertLeadingJetPtNominalDefaulPuComp1}\includegraphics[trim=5mm 14mm 0mm 10mm,clip,width=.52\textwidth]{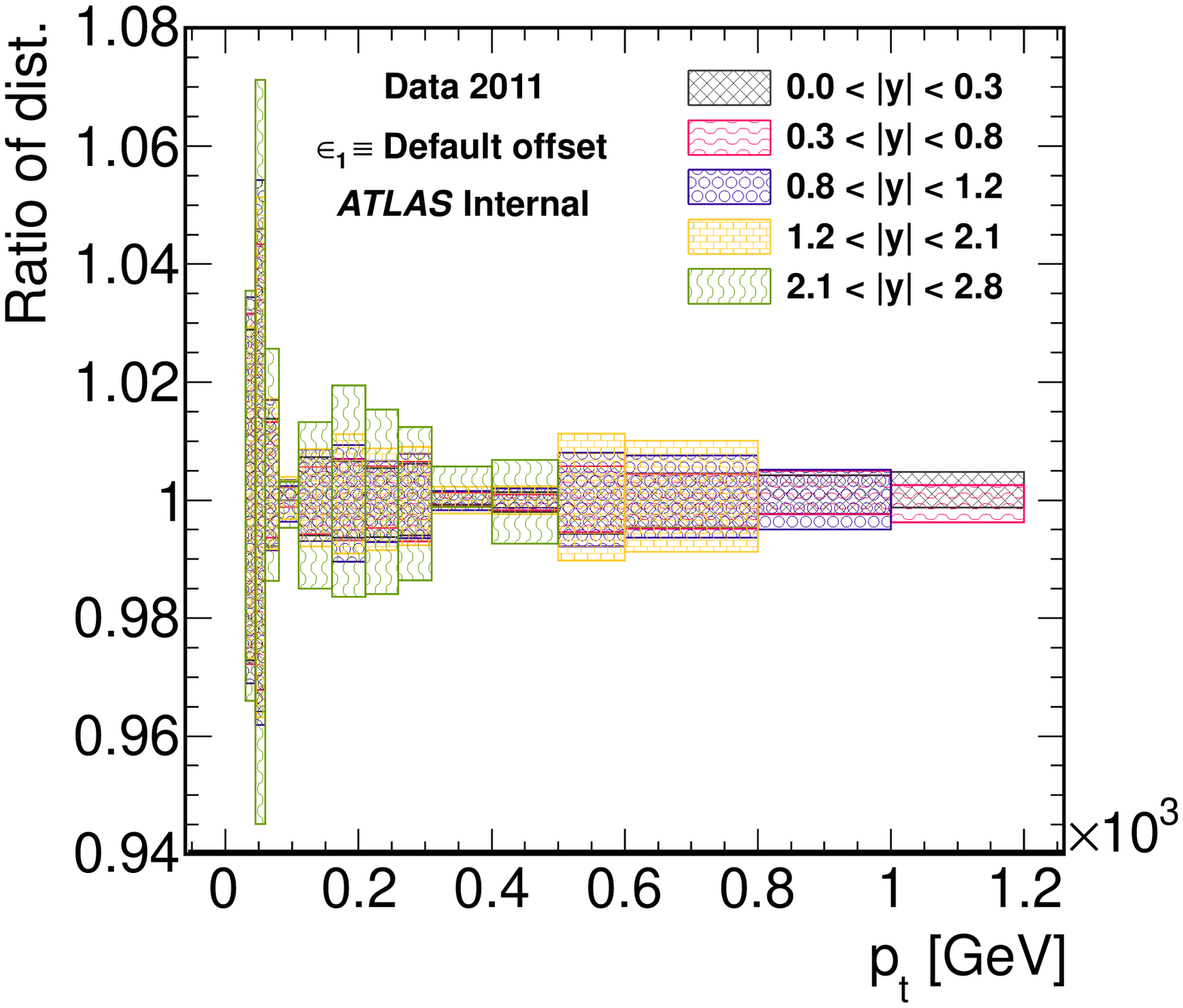}}
\subfloat[]{\label{FIGrelUncertLeadingJetPtNominalDefaulPuComp2}\includegraphics[trim=5mm 14mm 0mm 10mm,clip,width=.52\textwidth]{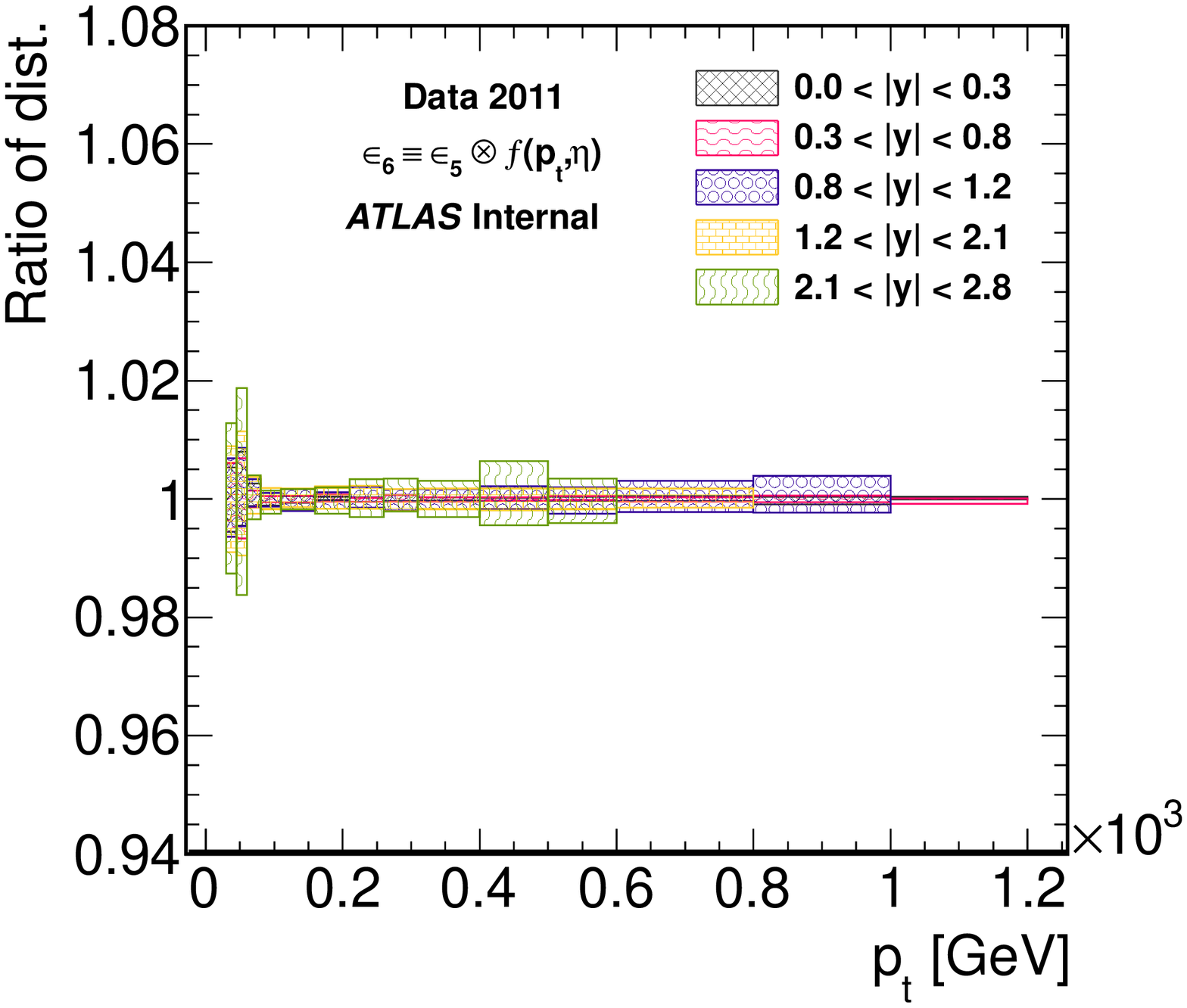}}
  \caption{\label{FIGrelUncertLeadingJetPtNominalDefaulPuComp}Dependence on the transverse momentum, \pt, of jets,
  of the ratio between the \pt distributions of jets with and without shifts of the jet energy scale associated with \pu,
  for jets within different rapidity, $y$, regions, as indicated in the figures. Jets in the 2011 data are
    corrected for \pu using the default offset correction, $\epsilon_{1}$, \Subref{FIGrelUncertLeadingJetPtNominalDefaulPuComp1}
  or the nominal \pu correction method, $\epsilon_{6}$, \Subref{FIGrelUncertLeadingJetPtNominalDefaulPuComp2}.
  }
\end{center}
\end{figure} 
The relative \pu uncertainty for jets calibrated with the nominal correction method, $\epsilon_{6}$,
is compared with the respective uncertainty for jets calibrated with the default offset correction, $\epsilon_{1}$,
in \autoref{FIGrelUncertLeadingJetPtNominalDefaulPuComp}.

To summarize, the total relative \JES uncertainty on the \pt spectrum of the leading jet in an event is roughly~7\%
at \ptLower{50}. As \pt increases the uncertainty decreases down to 2\%; the low uncertainty is kept up to 600\GeV, beyond which
it rises up to $\sim5\%$ in the \pt range considered.
In absolute terms the uncertainty on \pu following the median correction is roughly 1\% at low
transverse momenta, dropping to 0.1\% for \ptHigher{80}.
The corresponding relative contribution of the \pu uncertainties to the total \JES uncertainty is roughly 15\% at low
\pt and 0.5\% otherwise. 
The uncertainties associated with the median correction are much smaller in magnitude compared to those of the offset correction.
This has to do with the fact that the median \pu correction is completely data-driven; the offset correction on the
other hand is based on MC, and has a relatively large uncertainty due to mis-modeling of the effects of \pu on
simulated jets~\cite{ATLAS-CONF-2012-064}.

\section{In-situ measurements}
%
In order to test the performance of the median correction in data, several \insitu
measurements are presented in the following.

\subsection{Dijet balance}
%
%
A common method to measure \insitu the \pt resolution of jets is the \textit{dijet balance method}. The method is 
based on momentum conservation in the transverse plane. It works properly in the ideal case, in which there are
two particle-level jets in an event, which are reconstructed as exactly two calorimeter-jets. Due to radiation
from the hard interaction and to \pu, one is unlikely to observe an event with exactly two jets, balanced in \pt.
A \pt cut is therefore made on a possible third jet; events with exactly two jets above threshold, regardless
of the number of reconstructed jets below threshold, are then defined as dijet events. The two selected
jets are referred to as the leading jets in the event.
In order to further enhance the signal, a restriction is also placed on the difference in azimuth
between the two selected jets, $\Delta\phi_{1,2}$, such that the pair of jets is \textit{back-to-back},
\begin{equation}
  \Delta\phi_{1,2} > 2.8 \,.
\label{eqJetPtBalanceDef0} \end{equation}

The asymmetry between the transverse momenta of the two leading jets in a given event is defined as
\begin{equation}
  B_{1,2} = \frac{p_{\mrm{t,1}}-p_{\mrm{t,2}}}{p_{\mrm{t,1}}+p_{\mrm{t,2}}} \,,
\label{eqJetPtBalanceDef1} \end{equation}
where $p_{\mrm{t,1}}$ and $p_{\mrm{t,2}}$ are the transverse momenta of the two jets.
\Autoref{FIGdijetBalanceAllPt1} shows distributions of
$B_{1,2}$ for jet pairs with different average transverse momenta,
$\bar{p}_{\mrm{t}} = \frac{1}{2} \left( p_{\mrm{t,1}}+p_{\mrm{t,2}} \right)$, where
the nominal \pu correction method, $\epsilon_{6}$, is used.
As the average \pt of jet pairs increases, the asymmetry
distributions become more narrow, due to the improvement in resolution.

The fitted Gaussian width of the asymmetry distribution, $\sigma\left(B_{1,2}\right)$,
is used to quantify the resolution.
Assuming transverse momentum balance and requiring the jets to be in the same \Eta region,
the relation between $\sigma\left(B_{1,2}\right)$ and the relative jet resolution is
\begin{equation}
  \sigma\left(B_{1,2}\right) = \frac{\sqrt{ \left(\sigma_{p_{\mrm{t,1}}}\right)^{2} + \left(\sigma_{p_{\mrm{t,2}}}\right)^{2} }}{ \left<  p_{\mrm{t,1}}+p_{\mrm{t,2}} \right>} \approx \frac{\sigma_{p_{\mrm{t}}}}{\sqrt{2}p_{\mrm{t}}} \;,
\label{EQjetPtBalanceRes} \end{equation}
where the balance in transverse momentum implies that $\left<p_{\mrm{t,1}}\right> \approx \left<p_{\mrm{t,2}}\right> \equiv p_{\mrm{t}}$.

The fit results for the width of the asymmetry distribution as a function of the average momenta of the jet pairs
are presented in \autoref{FIGdijetBalanceAllPt2}. Different \pu subtraction methods are compared,
including dijet events where a third possible jet has transverse momentum, ${p_{\mrm{t}}^{\mrm{3^{rd}\;jet}} < 10\GeV}$.
In each case, $\bar{p}_{\mrm{t}}$ is computed with
jets which are corrected for \pu with the same method as used for $B_{1,2}$.
\begin{figure}[htp]
\begin{center}
  \subfloat[\qquad \qquad \qquad \qquad \quad]{\label{FIGdijetBalanceAllPt1}\includegraphics[trim=5mm 10mm 0mm 10mm,clip,width=.62\textwidth]{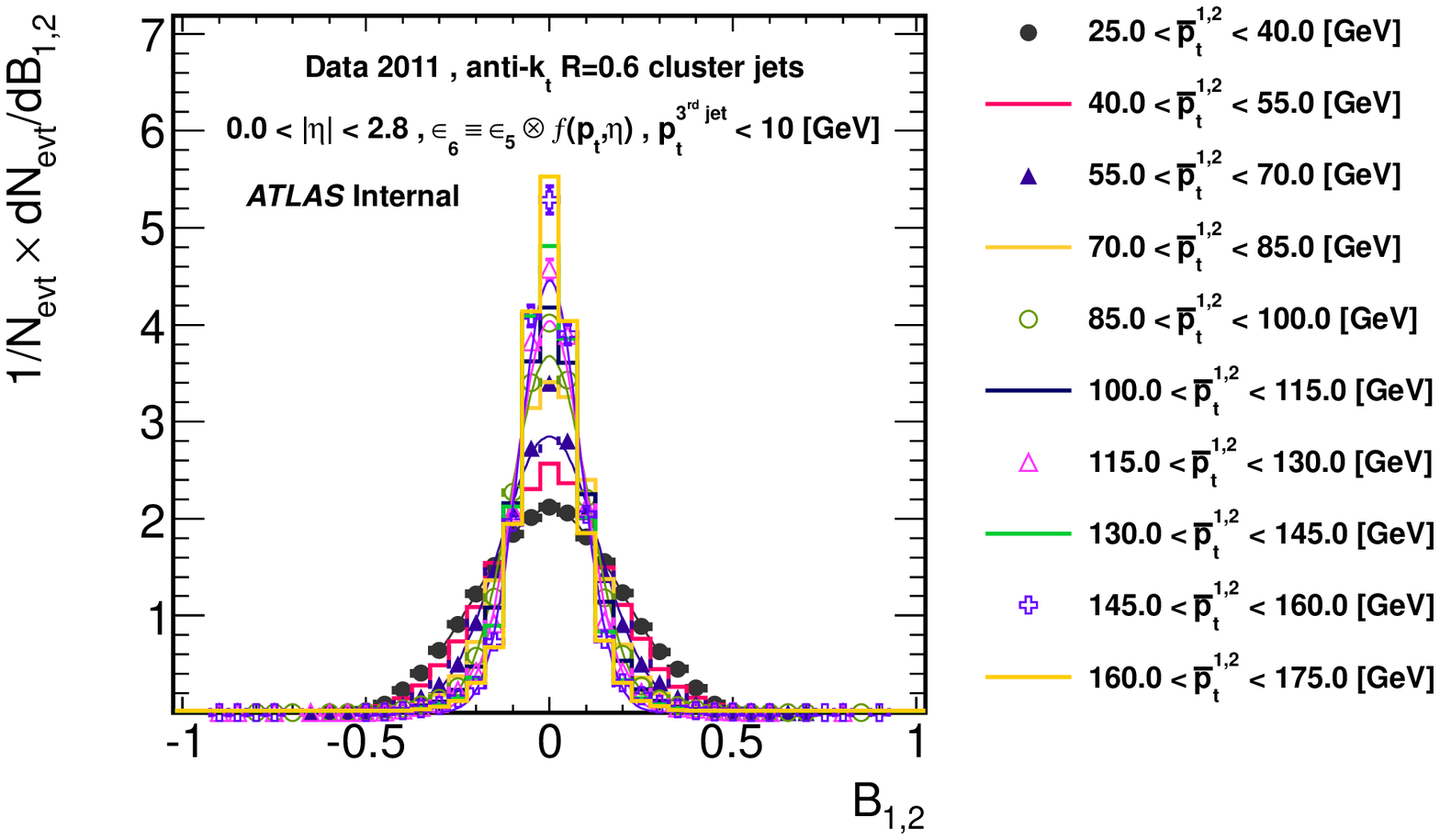}}
  \subfloat[]{\label{FIGdijetBalanceAllPt2}\includegraphics[trim=5mm 10mm 0mm 10mm,clip,width=.435\textwidth]{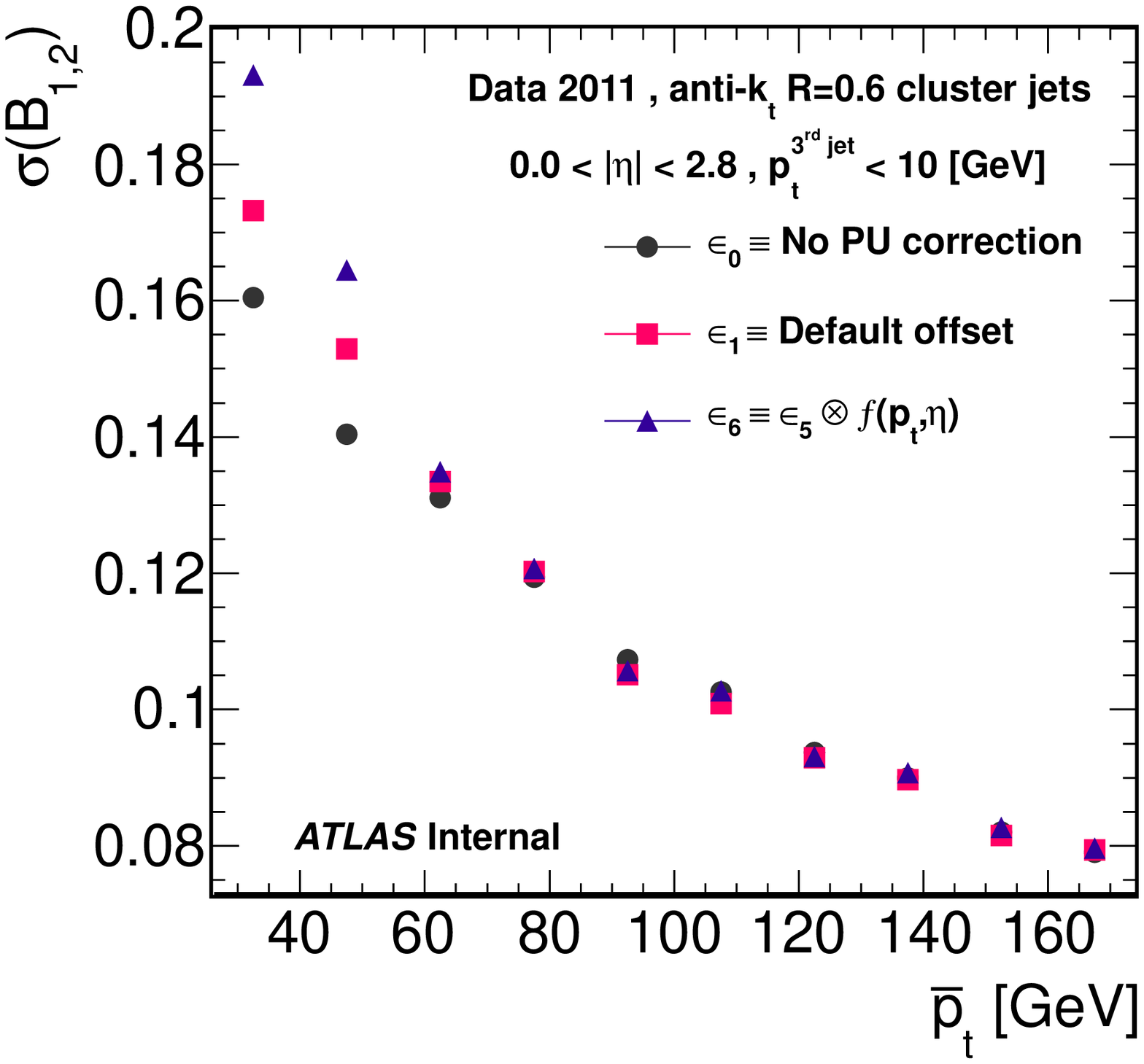}}
  \caption{\label{FIGdijetBalanceAllPt}\Subref{FIGdijetBalanceAllPt1} Normalized distributions of the
    dijet asymmetry, $B_{1,2}$, for jet pairs with different average transverse momenta,
    $\bar{p}_{\mrm{t}}$, as indicated, using jets with pseudo-rapidity, $|\eta| < 2.8$, corrected for \pu
    with $\epsilon_{6}$. \\
    \Subref{FIGdijetBalanceAllPt2} Dependence
    on  $\bar{p}_{\mrm{t}}$ of the width of the $B_{1,2}$ distribution, $\sigma\left(B_{1,2}\right)$, using jets with pseudo-rapidity, $|\eta| < 2.8$,
    corrected for \pu with $\epsilon_{0}$, $\epsilon_{1}$ or $\epsilon_{6}$, as indicated in the figure.\\    
    In both figures, a third possible jet in the event has transverse momentum, $p_{\mrm{t}}^{\mrm{3^{rd}\;jet}} < 10$\GeV.   
  }
\end{center}
\end{figure} 

Above~60\GeV in \pt, $\sigma\left(B_{1,2}\right)$ is independent on the correction method. Below,
$\sigma\left(B_{1,2}\right)$ is smaller in magnitude for jets which are not corrected for \pu,
compared to those which are.
This can be explained by the following.
The \pu subtraction decreases the energy of jets, but does not change their resolution. Accordingly, following the correction,
the balance in \pt of the dijet system (the nominator in \autoref{EQjetPtBalanceRes}) is unchanged, while the average 
(the denominator in \autoref{EQjetPtBalanceRes}) decreases. The value of $\sigma\left(B_{1,2}\right)$ therefore increases
following the \pu correction.
For larger values of $\bar{p}_{\mrm{t}}$, this effect becomes smaller, as the relative magnitude of the \pu energy compared
to the jet energy decreases.


In \autoref{FIGdijetBalanceSigmaDataMC1} the dependence of $\sigma\left(B_{1,2}\right)$ on $\bar{p}_{\mrm{t}}$ is compared for
different \pu correction schemes in data and in MC.
Good agreement is observed above~60\GeV. Below, however, some differences are observed.
These may be attributed to the miss-match of the \pt distributions of jets in MC compared to those in data,
as observed in \autoref{FIGdijetBalanceSigmaDataMC2}.  At low \pt, the data to MC ratio
of the \pt spectrum of the leading jet is different than that of the
sub-leading jets by up to~50\%.
The MC should reflect the physics of jet production, the performance of the detector and the \pu conditions in the data.
Correct description of the \pt distributions of the three leading jets in MC is essential, in order to
reproduce both the balance in \pt between the two leading jets, and the dependence on $p_{\mrm{t}}^{\mrm{3^{rd}\;jet}}$,
observed in data. The differences in the \pt distributions, may therefore serve to explain the
variance in $\sigma\left(B_{1,2}\right)$ at low values of $\bar{p}_{\mrm{t}}$.
\begin{figure}[ht]
\begin{center}
  \subfloat[]{\label{FIGdijetBalanceSigmaDataMC1}\includegraphics[trim=10mm 2mm 0mm 20mm,clip,width=.52\textwidth]{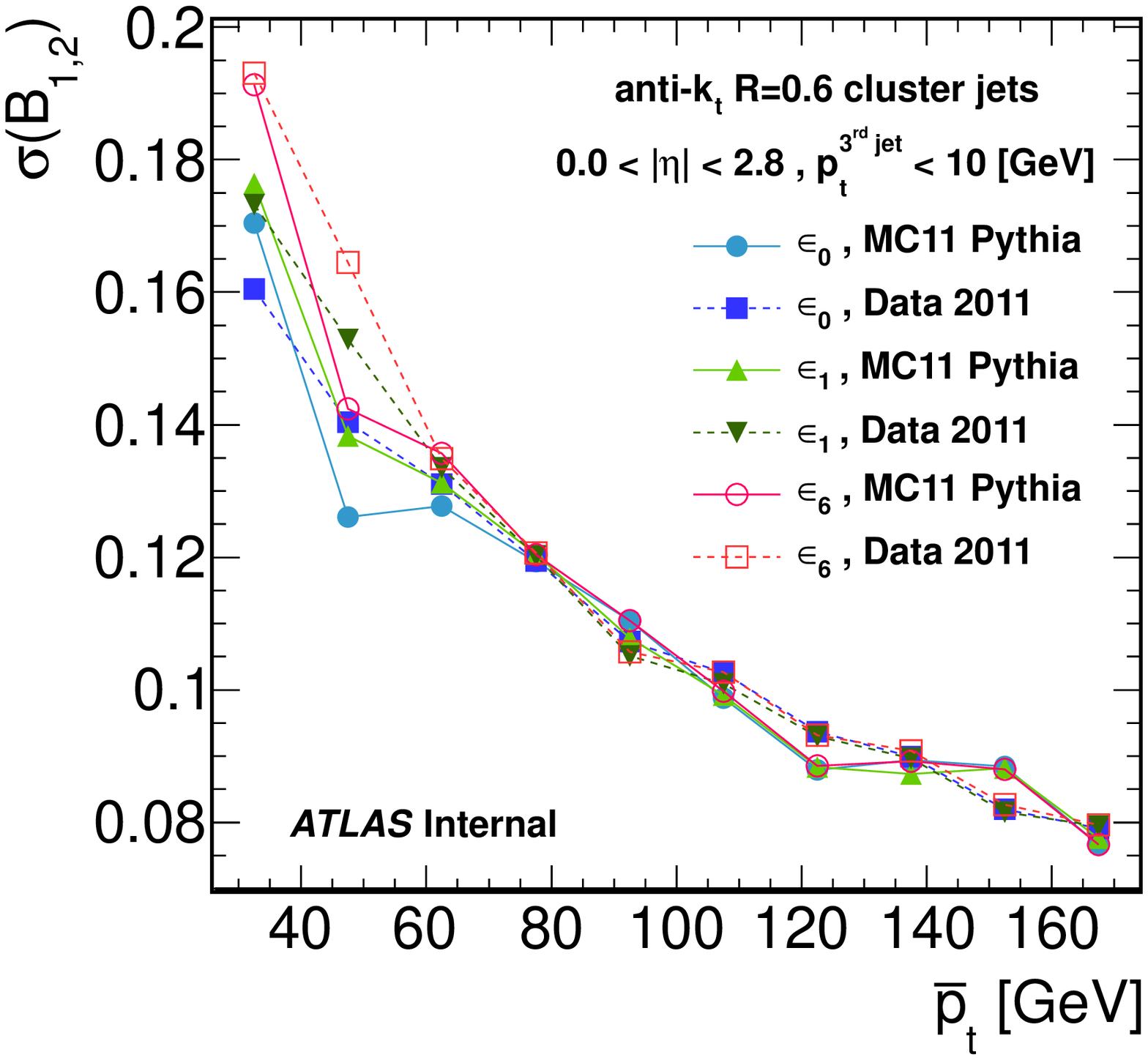}}
  \subfloat[]{\label{FIGdijetBalanceSigmaDataMC2}\includegraphics[trim=10mm 5mm 0mm 15mm,clip,width=.52\textwidth]{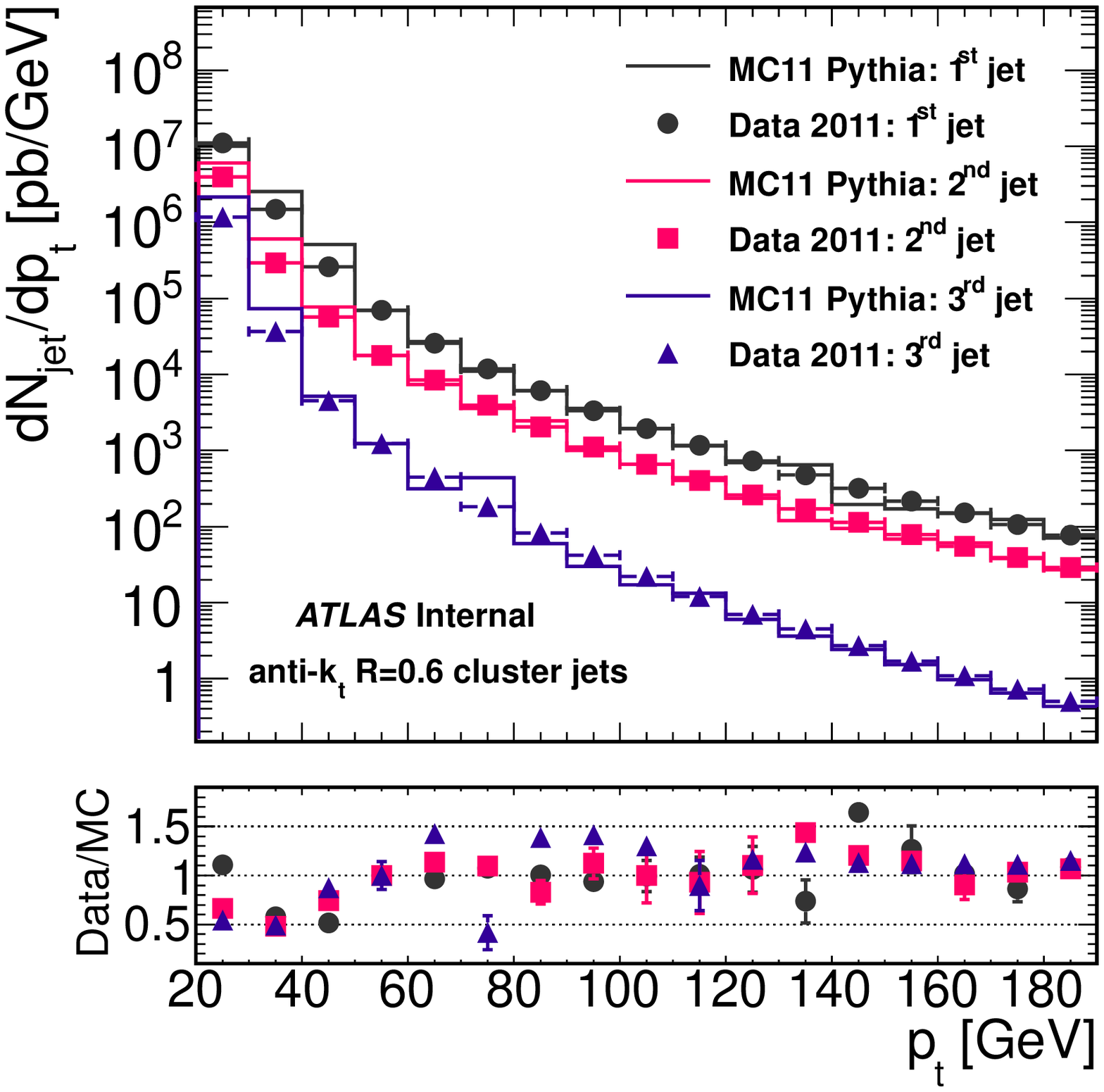}}
  \caption{\label{FIGdijetBalanceSigmaDataMC}\Subref{FIGdijetBalanceSigmaDataMC1} Dependence on the average
  momenta of jet pairs, $\bar{p}_{\mrm{t}}$, of the width of the \pt asymmetry
  distribution in dijet events, $\sigma\left(B_{1,2}\right)$, using jets with pseudo-rapidity, $|\eta| < 2.8$, corrected
  for \pu with $\epsilon_{0}$ (no \pu subtraction), $\epsilon_{1}$ (default offset correction) or $\epsilon_{6}$ (the nominal \pu correction),
  in data and in MC, as indicated in the figure.
  A third possible jet in the event has transverse momentum, $p_{\mrm{t}}^{\mrm{3^{rd}\;jet}} < 10$\GeV.
  The connecting lines are intended to guide the eye.\\
  \Subref{FIGdijetBalanceSigmaDataMC2} Distributions of the transverse momentum, \pt, of the
  three jets with the highest-\pt in an event, in data and in MC, as indicated in the figure, 
  where the distributions in MC are rescaled such that they agree in normalization with the data at $\pt = 50\GeV$.
  The bottom panel shows the ratio of data to MC.
  }
\end{center}
\end{figure} 
%
%

\subsection{Jet \pt and dijet invariant mass spectra \label{chapJetPtDijetMassSpectra}}
%
In order to validate the performance of the \pu correction, differential spectra of the transverse momentum of the leading
jet, and of the invariant mass of the two leading jets, are measured for events with different numbers of reconstructed vertices.

The transverse momentum of the leading jet in an event is measured in bins of the absolute value of the
rapidity of the jet, $\left|y\right|$, extending up to~2.8.
The mass of the two leading jets, $m_{12}$, also includes jets with rapidity ${\left|y\right| < 2.8}$. The transverse momenta of the leading and
of the sub-leading (second highest-\pt) jets in an event are respectively constrained to be, ${p_{\mrm{t,1}} > 40}$ and ${p_{\mrm{t,2}} > 30\GeV}$.
Expressing the rapidities of the leading and sub-leading jets
as $y_{1}$ and $y_{2}$ respectively,
the rapidity of one of the outgoing partons associated with one of the jets in the two-parton \com frame is 
${\ystr = \left| y_{1}-y_{2} \right| / 2}$.

Luminosity in this study is calculated, as described in \autoref{chapTwoTriggerLumiCalcScheme}, using a two-trigger selection
procedure. Accordingly, luminosity classes are defined by the \pt and rapidity of the two leading jets in an event.
The vertex distribution and trigger prescales
are correlated, as prescale factors were modified with the increase in luminosity during data taking.
Accordingly, the baseline luminosity classes are modified to account for event samples with any given value of \Npv\footnote{ This
is done by extracting the probability for an event to have a given number of reconstructed vertices
for each trigger bin; a trigger bin refers to the classification of an event according to the triggers associated with the two leading jets.
%
}.

The \pt spectrum of the leading jet within $\left|y\right| < 0.3$ and the dijet
mass spectrum for $\ystr < 0.5$ are shown in \autoref{FIGinSituMassPt} for events with fixed values of \Npv. 
In order to quantify the performance of the \pu correction, the standard deviation with regard to single-vertex events, $\sigma\left(\Npv\right)$, is calculated.
This is done by using the distributions with ${\Npv=1}$
as reference, and calculating bin-by-bin the differences \wrt respective multi-vertex (${\Npv>1}$) events,
\begin{equation}{
  \sigma\left(\Npv\right) = \sqrt{ \frac{1}{n_{\Npv}-1}\sum_{i=2}^{n_{\Npv}}{ \left( \mathfrak{b}_{1} - \mathfrak{b}_i \right)^{2}}  }
 \;.
}\label{eqStandardDeviationNpvDef} \end{equation}
Here $ n_{\Npv}$ represents the maximal number of reconstructed vertices, and $\mathfrak{b}_i$ stands for
the bin content of a given distribution, which has $\Npv=i$ reconstructed vertices.

Results for $\sigma\left(\Npv\right)$ are given in \autoref{FIGinSituMassPtStdNpv} for the first $y$ and \ystr bins of the
respective \pt and mass spectra. Additional rapidity and \ystr bins are
shown in \autoref{chapJetAreaMethodApp}, \autorefs{FIGinSituPtStdNpvApp}~-~\ref{FIGinSituMassStdNpvApp}.
\begin{figure}[htp]
\begin{center}
\vspace{-20pt}
\subfloat[]{\label{FIGinSituMassPt1}\includegraphics[trim=5mm 14mm 0mm 10mm,clip,width=.52\textwidth]{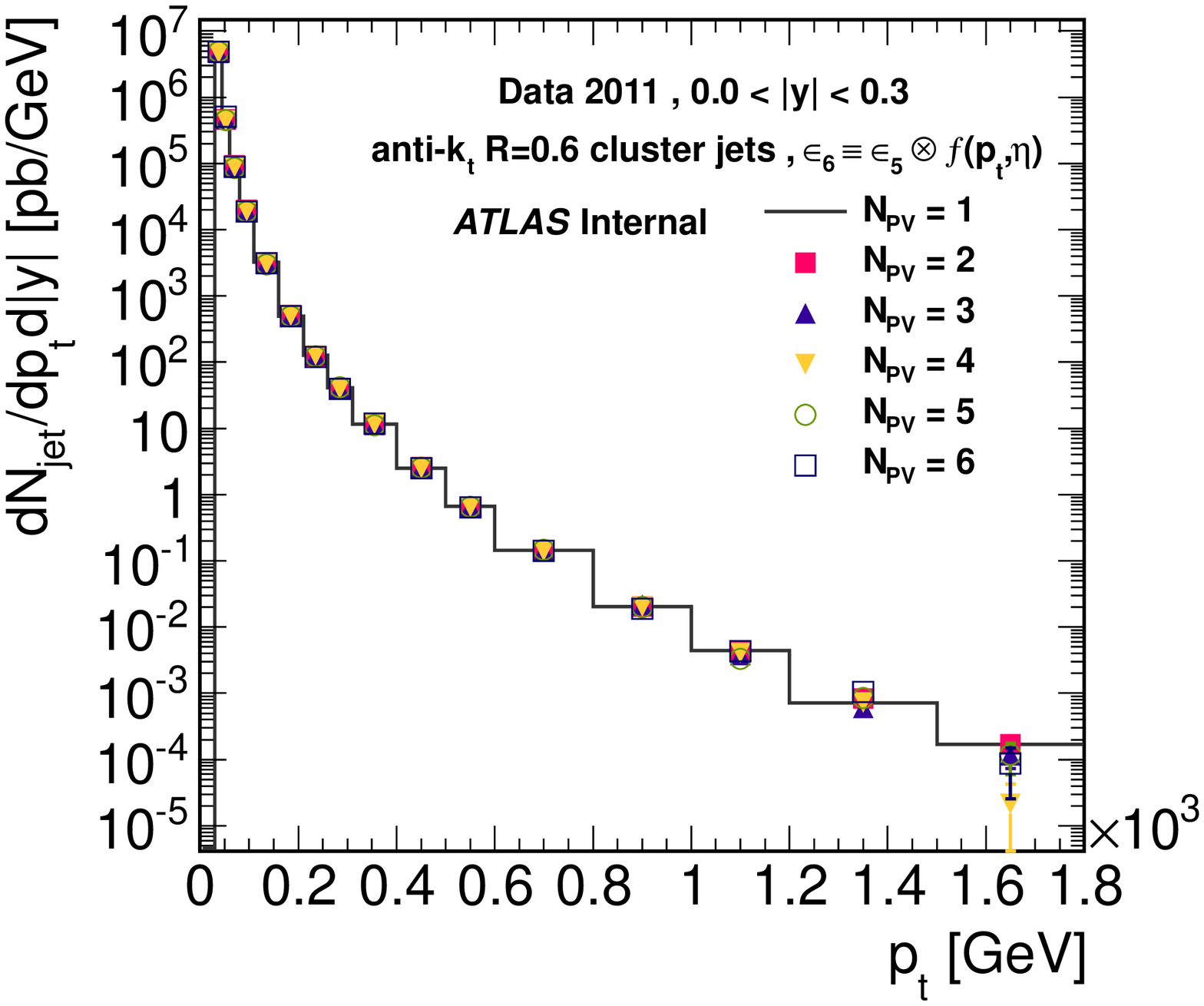}}
\subfloat[]{\label{FIGinSituMassPt2}\includegraphics[trim=5mm 14mm 0mm 10mm,clip,width=.52\textwidth]{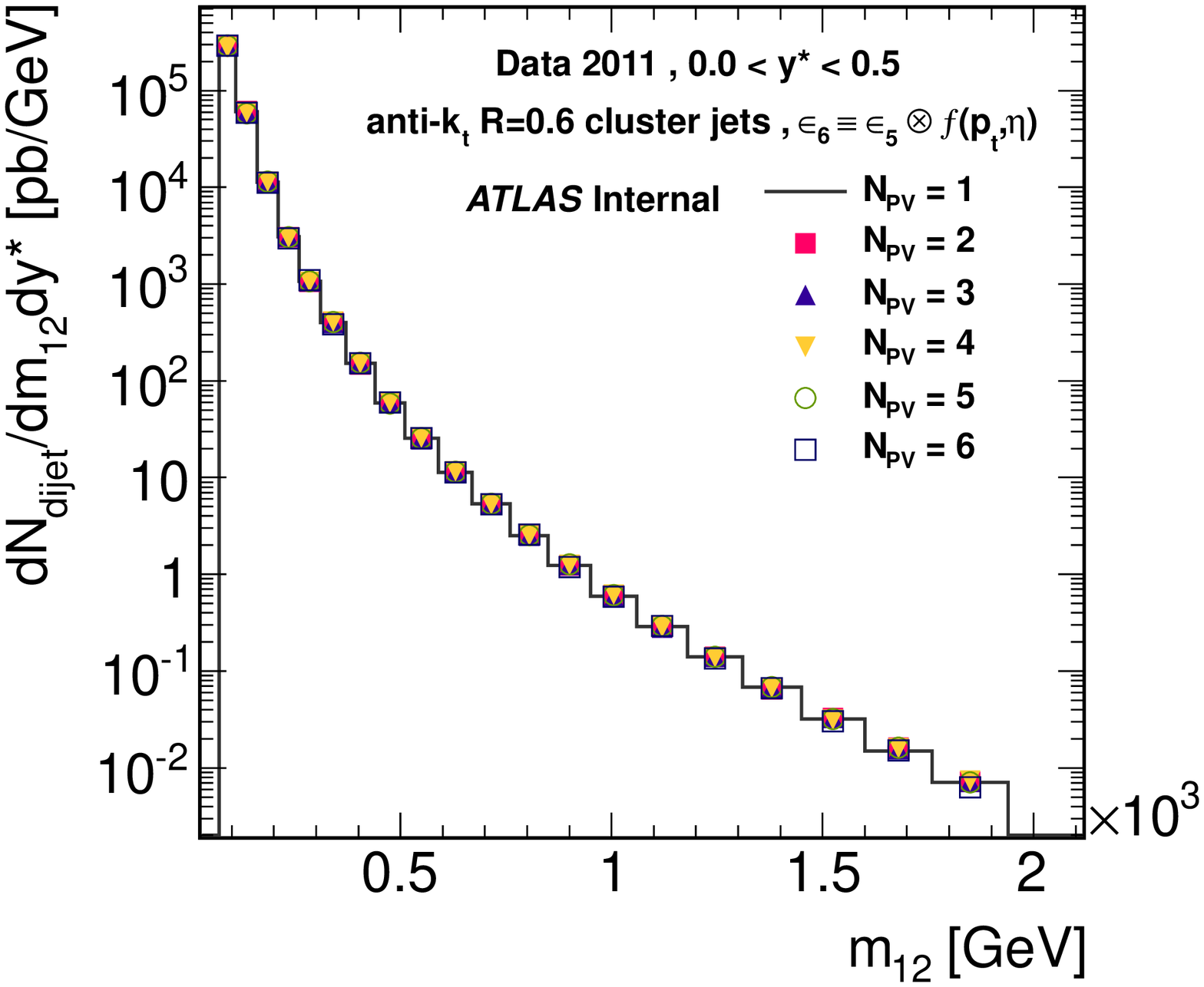}}
  \caption{\label{FIGinSituMassPt}\Subref{FIGinSituMassPt1} Differential transverse momentum, \pt, spectra
    of the highest-\pt jet in an event, using jets with rapidity, $\left|y\right| < 0.3$. \\
    \Subref{FIGinSituMassPt2} Differential spectra of the invariant mass distribution, $m_{12}$, of the two highest-\pt jets in an event,
    for \com jet rapidity, $\ystr < 0.5$. \\
    In both figures jets are calibrated using the nominal \pu subtraction method, $\epsilon_{6}$, and
    events have a fixed number of reconstructed vertices, \Npv, as indicated.
  }
\end{center}
%
%
\begin{center}
\subfloat[]{\label{FIGinSituMassPtStdNpv1}\includegraphics[trim=5mm 14mm 0mm 10mm,clip,width=.52\textwidth]{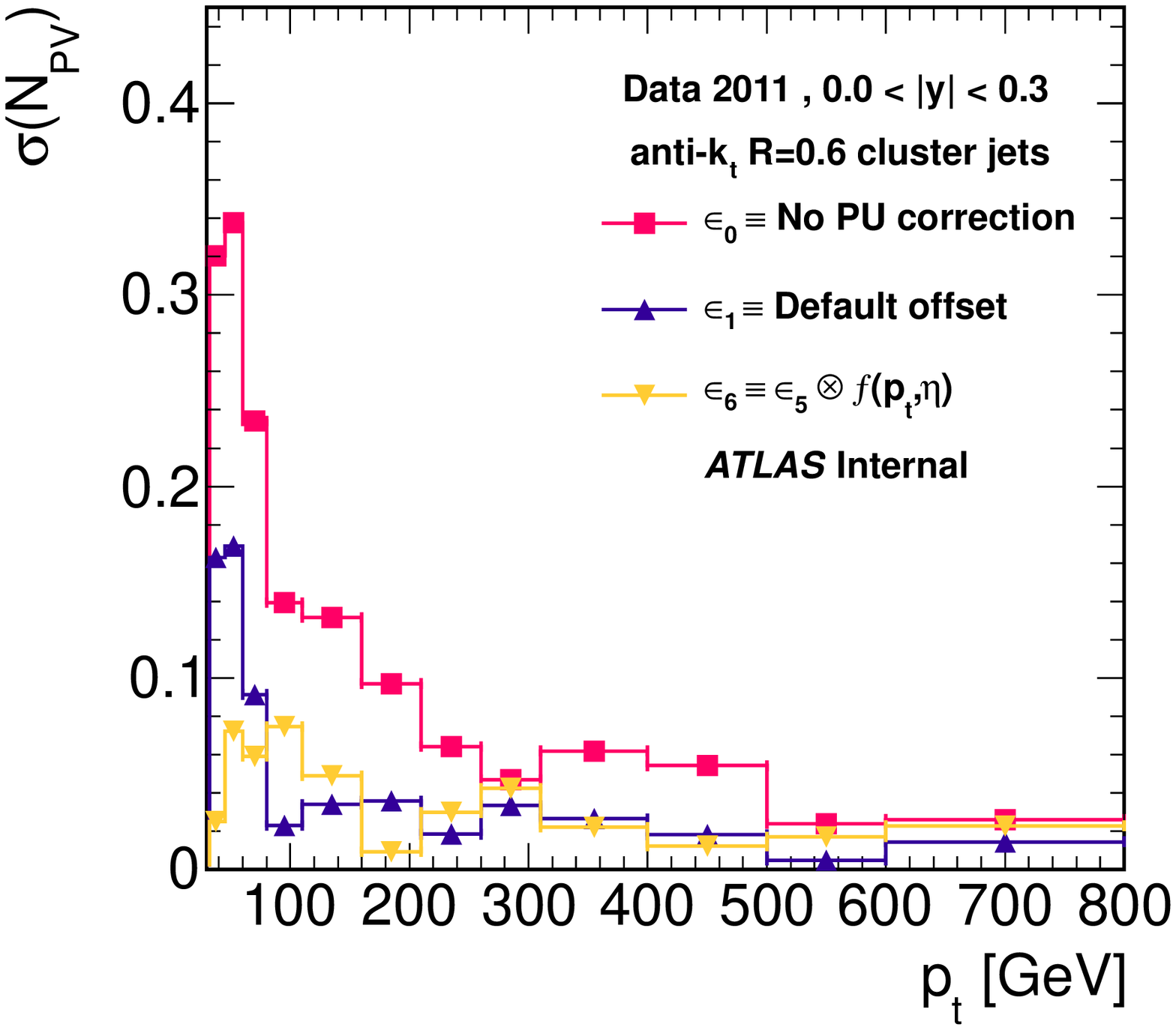}}
\subfloat[]{\label{FIGinSituMassPtStdNpv2}\includegraphics[trim=5mm 14mm 0mm 10mm,clip,width=.52\textwidth]{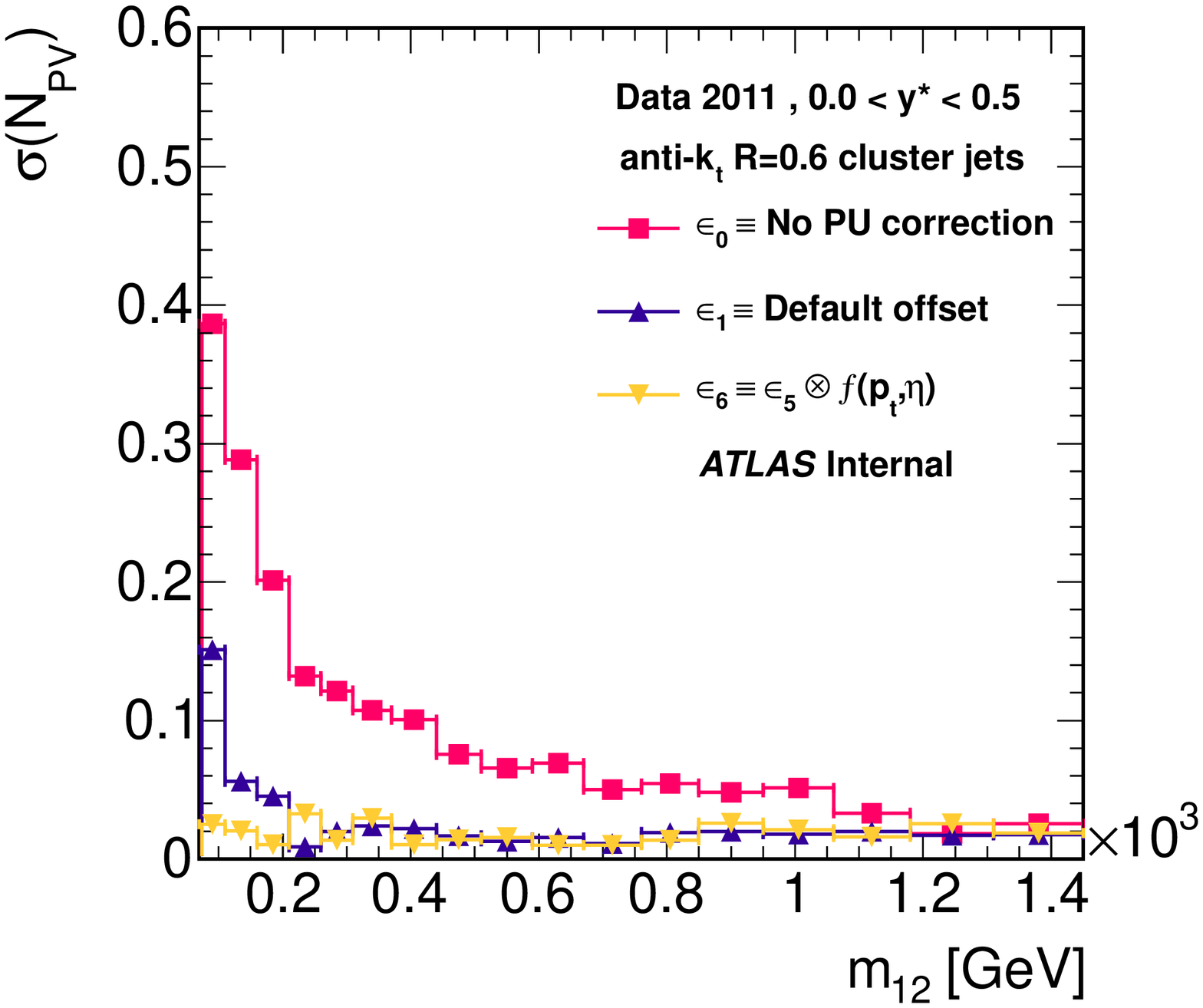}}
  \caption{\label{FIGinSituMassPtStdNpv}The standard deviation with regard to single-vertex events, $\sigma\left(\Npv\right)$,
    (see \autoref{eqStandardDeviationNpvDef}) for two observables,
    the differential transverse momentum, \pt, spectrum
    of the highest-\pt jet in an event, using jets with rapidity, $\left|y\right| < 0.3$, \Subref{FIGinSituMassPtStdNpv1} and
    the differential mass, $m_{12}$, spectrum of the two highest-\pt jets in an event, for
    \com jet rapidity, $\ystr < 0.5$ \Subref{FIGinSituMassPtStdNpv2}.
    Distributions for jets reconstructed with several \pu correction schemes
    ($\epsilon_{0}$, $\epsilon_{1}$ and $\epsilon_{6}$) are compared, as indicated in the figures.
  }
\end{center}
\end{figure} 
As expected, both the offset correction ($\epsilon_{1}$) and the median \pu correction ($\epsilon_{6}$) decrease the dependence on \Npv, expressed through smaller
values of $\sigma\left(\Npv\right)$. As \pt and $m_{12}$ increase, the relative effect of \pu decreases and so the \Npv dependence lessens. Some residual
statistical fluctuations remain, generally of the order of 2\% for $\pt>300\GeV$ and $m_{12}>600\GeV$.
The median correction ($\epsilon_{6}$) generally performs better
than the default offset correction ($\epsilon_{1}$) at low \pt and mass.
At respective higher values the performance is comparable between the two methods.
Similar behaviour is observed for higher rapidities, 
though the magnitude of the fluctuations at high transverse momenta and masses are larger
due to insufficient statistics.

\vspace{-5pt}
\section{Summary of the performance of the median \pu correction}
%
%
The jet area/median method is used to devise a data-driven \pu subtraction scheme. The method combines
parametrization of the average \pu contribution to jets, as well as
event-by-event estimates of the local \pu \pt-density around a given jet.
The correction is used to subtract \pu from jets with \ptHigher{30} and \etaLower{2.8}
in data taken during 2011.

Several performance tests are conducted using simulated events.
The residual \Npv dependence of the \pt of jets following the correction is shown to be
insignificant. The corresponding \Mu dependence is of the order of~${0.2\% / \mu}$ for low-\pt jets.
The \pt dependence is likewise insignificant within this kinematic range,
achieving \pt-closer better than~0.1\% on average at~30\GeV.
This extends the reach of the standard \pu correction in \atlas, the offset correction, for which 
the relative bias in \pt becomes negligible (roughly~0.2\%) at~60\GeV.
The \pt resolution is unchanged \wrt that achieved using the offset correction.
The median method is shown to be robust under changes of jet algorithm and size parameter.

In data, the median correction is parametrized separately under different restrictions. These
account for different \pt scales of jets which undergo the calibration,
as well as for different restrictions on the history of previous collisions.
The \pu uncertainty on the jet energy scale following the median correction is estimated in data,
and found to be smaller than the corresponding uncertainty for the offset correction.

\Insitu measurements are performed of the \pt balance of dijets, where good agreement is observed between data and MC.
Measurements are also made of the \pt spectra of jets in different rapidity regions
and of the differential dijet mass spectra in different \ystr bins.
Following the median correction, the dependence on \Npv is shown to be
alleviated \wrt events which are not corrected for \pu; the achieved performance
is on a par with or better than that of the offset correction.

The median method is used for \pu subtraction in the dijet differential \xsec measurement,
which is the subject of the next chapter.
              }{}
\ifthenelse{\boolean{do:jetMass}}           { 
\chapter{Dijet mass distribution\label{chapMeasurementOfTheDijetMass}}
%
\section{\Xsec definition}
%
%
The dijet \xsec is defined using \AKT jets with size parameter, $R = 0.6$.
The measurement is corrected for all experimental effects, and thus is defined for the
particle-level final state of a \pp collision~\cite{Buttar:2008jx}.
The \xsec is measured as
a function of the invariant mass of the leading (highest $\pt$) and sub-leading (second highest-\pt)
jets; the mass is defined as ${m_{12} = \sqrt{ (E_{1}+E_{2})^{2} -
(\vec{p}_{1}+\vec{p}_{2})^2}}$, where $E_{1}$ and $E_{2}$ ($\vec{p}_{1}$ and $\vec{p}_{2}$) are
the energies (momenta) of the leading and sub-leading jets respectively.
The rapidities of the two leading jets are similarly defined as $y_{1}$ and $y_{2}$.
The \xsec is binned in the variable, ${\ystr = \left| y_{1}-y_{2} \right| / 2}$. 
For massless partons, \ystr is determined by the polar scattering angle, $\theta^{*}$, in the \com moving along the beamline,
\begin{equation}
\ystr = \frac{1}{2}
\ln{\left(\frac{1+|\cos{\theta^{*}}|}{1-|\cos{\theta^{*}}|} \right) } \;.
\label{eqYstar}
\end{equation}
It stands for the rapidity of one of the outgoing partons in the two-parton \com frame.

The following phase-space regions
are used to define the \xsec measurement for data taken during 2010 and during 2011:
\begin{equation}
  \renewcommand{\arraystretch}{1.9} 
         \begin{array}{cccc}
            \mathrm{(2010)} & p_{t,1} > 30\GeV \;, & p_{t,2} > 20\GeV \;, & |y_{1,2}| < 4.4 \;,   \\
            \mathrm{(2011)} & p_{t,1} > 40\GeV \;, & p_{t,2} > 30\GeV \;, & |y_{1,2}| < 2.8 \;,
         \end{array}
\label{eqMassMeasurementPhasespace} \end{equation}
where $p_{t,1}$ and $p_{t,2}$ are the transverse momenta of the leading and sub-leading
jets respectively.
Requiring the leading jet to have \pt higher than that of the sub-leading jet improves the stability of the NLO
calculation~\cite{Frixione:1997ks}.
Due to the increase in \pu in 2011 compared to 2010, both the rapidity coverage and the \pt coverage are 
more restricted in 2011.

In leading order (LO) approximation, the fractions of the proton momenta carried by the interacting partons, $x$, can
be reconstructed from the two outgoing jets.
%
The momentum fraction of the two interacting partons, denoted by $(+)$ or $(-)$, can be estimated as
\begin{equation}
x_{\pm} = \frac{1}{\sqrt{s}} \left( E_{\mrm{t,1}} e^{\pm\eta_{1}} + E_{\mrm{t,2}} e^{\pm\eta_{2}} \right) \;,
\label{eqMedianInEta} \end{equation}
where the \cms energy of the collision is \sqs, and
$E_{\mrm{t,i}}$ and $\eta_{i}$ respectively stand for the transverse energy and pseudo-rapidity
of jet $i$.
The momentum transfer of the interaction, $Q$, is usually taken as the mass of the dijet system.

The $(x,Q^2)$ kinematic plane for the 2010 and 2011 datasets is presented in \autoref{FIGdijetMassXQ2Plane}. 
The results are equivalent for the momentum fractions of the two partons, and so only $x_{+}$ is shown.
The data included in the figure are not corrected for detector effects, other than the \JES calibration of jets.
\begin{figure}[ht]
\begin{center}
  \subfloat[]{\label{FIGdijetMassXQ2Plane1}\includegraphics[trim=5mm 14mm 30mm 10mm,clip,width=.48\textwidth]{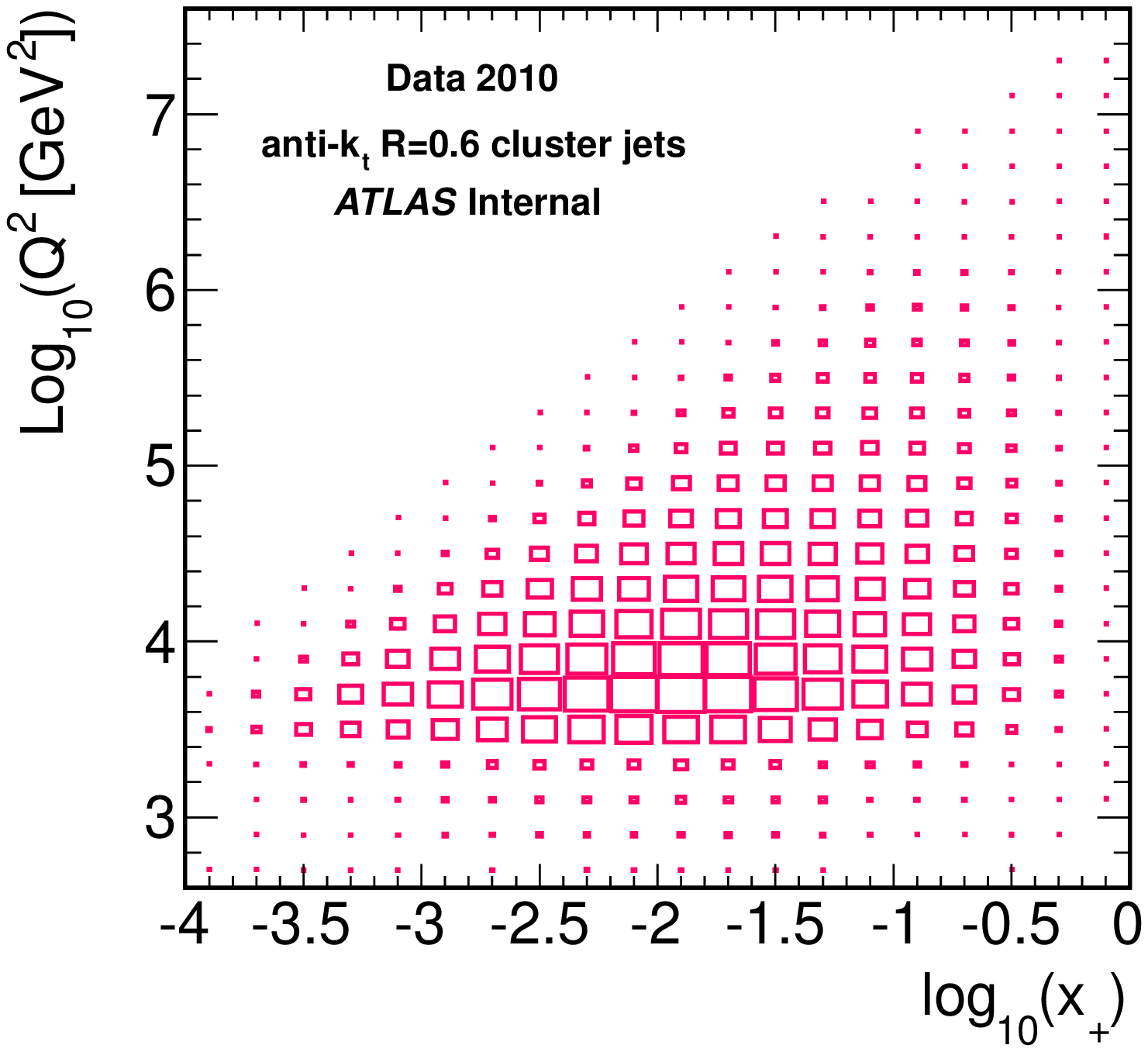}} \quad
  \subfloat[]{\label{FIGdijetMassXQ2Plane2}\includegraphics[trim=5mm 14mm 30mm 10mm,clip,width=.48\textwidth]{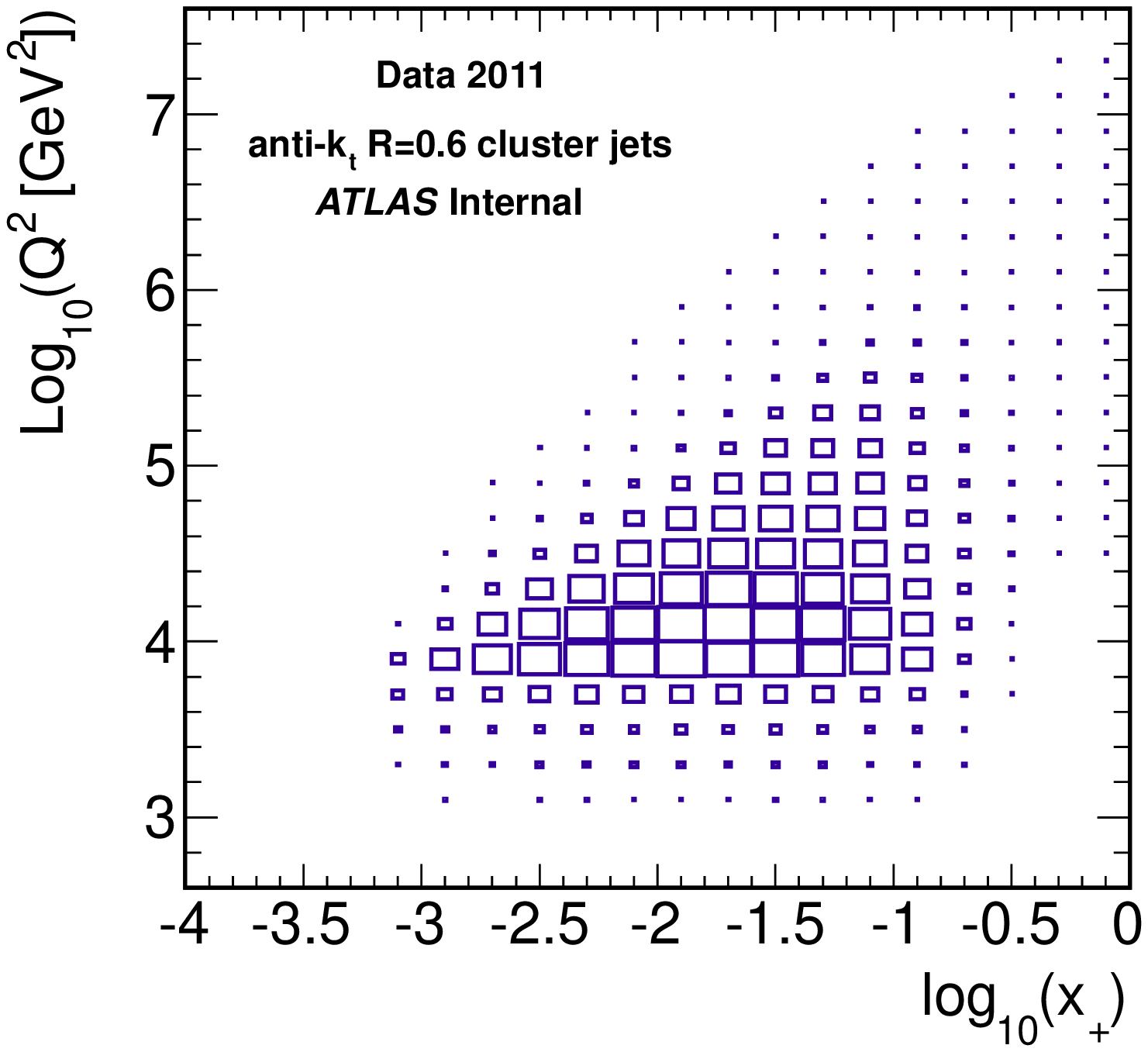}}
  \caption{\label{FIGdijetMassXQ2Plane}The $(x,Q^2)$ kinematic plane for the dijet mass measurement, where $x$ denotes (to leading order) the
  fraction of the momentum of the proton carried by an interacting parton, and $Q^2$ stands
  for the momentum transfer of the interaction. Shown are data collected during 2010 \Subref{FIGdijetMassXQ2Plane1}
  and during 2011 \Subref{FIGdijetMassXQ2Plane2} within the corresponding phase-space regions,
  as detailed in \autoref{eqMassMeasurementPhasespace}.
  Box sizes in the figures indicate the relative number of entries in a given bin,
  compared to other bins in the same figure. 
  }
\end{center}
\end{figure} 
The measurements cover masses from~70\GeV up to several\TeV, and probe proton momentum
fractions from~$10^{-4}$ up to about~1. The 2010 dataset reaches lower values of $x$ due to
the extended range in rapidity and \pt.

\section{Theoretical Predictions}
%
%
The measured jet \xsec is compared to fixed-order NLO pQCD
predictions, which are corrected for non-perturbative effects.
Theory calculations are performed in the same kinematic range as the measurement in data.

\subsection{NLO calculation}
%
%
The \nlojet (version 4.1.2)
package~\cite{Nagy:2003tz} is used with the
CT10 NLO parton distribution functions~\cite{Lai:2010vv}.
The same value, which corresponds to the transverse momentum of the leading jet, $p_{t,1}$,
is used for the normalisation and factorisation scales, respectively denoted by $\mu_{\rm R}$ and $\mu_{\rm F}$.
The forward dijet \xsec in \nlojet is numerically stable if
instead of a scale fixed entirely by $\pt$, a scale that depends on
the rapidity separation between the two jets is used\footnote{If a scale fixed with jet \pt is used,
\nlojet predicts an unstable or even negative \xsec for large ($\gtrsim 3$) rapidity
separations between the two jets.}.  The values
chosen for each bin in $\ystr$ follow the formula,
\begin{equation}
\mu = \mu_{\rm R} = \mu_{\rm F} = \pt \, e^{0.3 \ystr} \;,
  \label{eqDijetscale}
\end{equation}
which is illustrated in \autoref{FIGjetMassRenormFactorScales}.
This choice is motivated by the formula,
\begin{equation}
  \mu = \mu_{\rm R} = \mu_{\rm F} = \frac{m_{\rm 12}}{2\cosh(0.7 \ystr)} \;,
  \label{eqMassscale}
\end{equation}
that is suggested in~\cite{Ellis:1992en}.  
\begin{figure}[htp]
\begin{center}
\includegraphics[trim=0mm 0mm 0mm 0mm,clip,width=.57\textwidth]{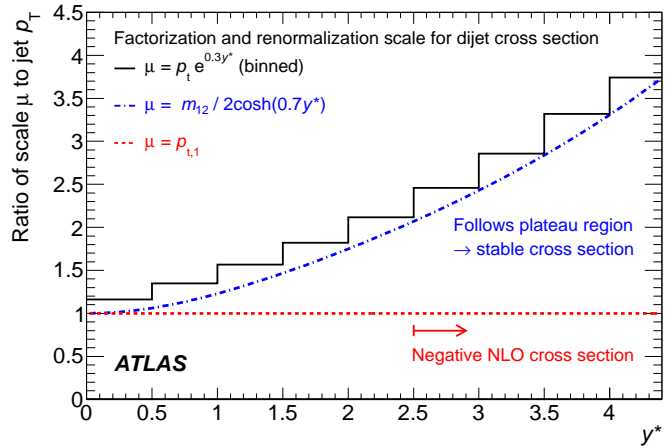}
\caption{\label{FIGjetMassRenormFactorScales}Ratio of the renormalisation and factorisation scale,
  \Mu, to the jet transverse momentum, $p_{\mrm{T}}$, used for the dijet predictions obtained with \nlojet, as a function of 
  \com jet rapidity, \ystr, using several scale parametrizations; the chosen
  parametrization, given in \autoref{eqDijetscale} (full black line);
  the formula (\autoref{eqMassscale}) suggested in~\protect\cite{Ellis:1992en} (dashed blue line);
  and a fixed scale proportional to the transverse momentum of the leading jet, $p_{\mrm{t,1}}$ (dashed red line).
  (Figure taken from~\protect\cite{Aad:2011fc}.)
}
\end{center}
\end{figure} 

\subsection{Non-perturbative corrections\label{chapNonPerturbativeCorrections}}
%
%
The fixed-order NLO calculations predict parton-level \xsecs,
which must be corrected for non-perturbative effects in order to be compared
with data.  This is done by using leading-logarithmic parton shower generators.
Several combinations of generators, PDF sets, and underlying
event tunes are used (see \autoref{chapEventGenerators}). The corrections are derived with
\begin{list}{-}{}
  \item \pythia~6.425~\cite{pythia} with the MRST~LO*~\cite{PDF-MRST} PDF set and the AUET2B~\cite{ATL-PHYS-PUB-2011-009} tune,
\end{list}
referred to as the nominal simulation setup in the following.
Additional configurations are,
\begin{list}{-}{}
  \item \pythia~6.425 with the CTEQ6L1~\cite{PDF-CTEQ} PDF set and the AUET2B and AMBT2B~\cite{ATL-PHYS-PUB-2011-009} tunes;
  \item \pythia~8 (v150)~\cite{Sjostrand:2007gs} with the MRST~LO** PDF set and 4C~\cite{Corke:2010yf} tune; and
  \item \herwigpp~\cite{Herwigpp} (v2.5.1) with the MRST~LO* and the CTEQ6L1 PDF sets and the UE7000-3~\cite{ATL-PHYS-PUB-2011-009} tune.
\end{list}

The hadron-level \xsec is computed in simulation samples where hadronization and
the underlying event are switched on, $\sigma_{\left(\mrm{+HU}\right)}$, or off, $\sigma_{\left(\mrm{\textrm{-}HU}\right)}$.
The bin-wise ratio of \xsecs,
\begin{equation}
  \mathcal{R}_{\mrm{HU}}(m_{12},\ystr) = \frac{\sigma_{\left(\mrm{+HU}\right)}(m_{12},\ystr)}{\sigma_{\left(\mrm{\textrm{-}HU}\right)}(m_{12},\ystr)} \,,
\label{eqHadronisationCorrections0} \end{equation}
is computed for each such configuration as a function of $m_{12}$ and \ystr.
The non-perturbative correction factors
are taken as the values of a functional fit to the ratio of \xsecs in the nominal setup.
The correction factors take the parton-level \xsec, $\sigma_{\mrm{part}}$,
to the hadron-level \xsec, $\sigma_{\mrm{had}}$,
\begin{equation}
  \sigma_{\mrm{had}}(m_{12},\ystr) 
   =  \mathcal{R}_{\mrm{HU}} (m_{12},\ystr) \times \sigma_{\mrm{part}}(m_{12},\ystr) \,.
\label{eqHadronisationCorrections1} \end{equation}

\Autoref{FIGhadronicCorrections} shows the correction factors, $\mathcal{R}_{\mrm{HU}}$, for
a single \ystr bin for the 2010 and the 2011 phase-space definitions of the \xsec.
\begin{figure}[htp]
\begin{center}
  \subfloat[]{\label{FIGhadronicCorrections1}\includegraphics[trim=5mm 14mm 0mm 10mm,clip,width=.52\textwidth]{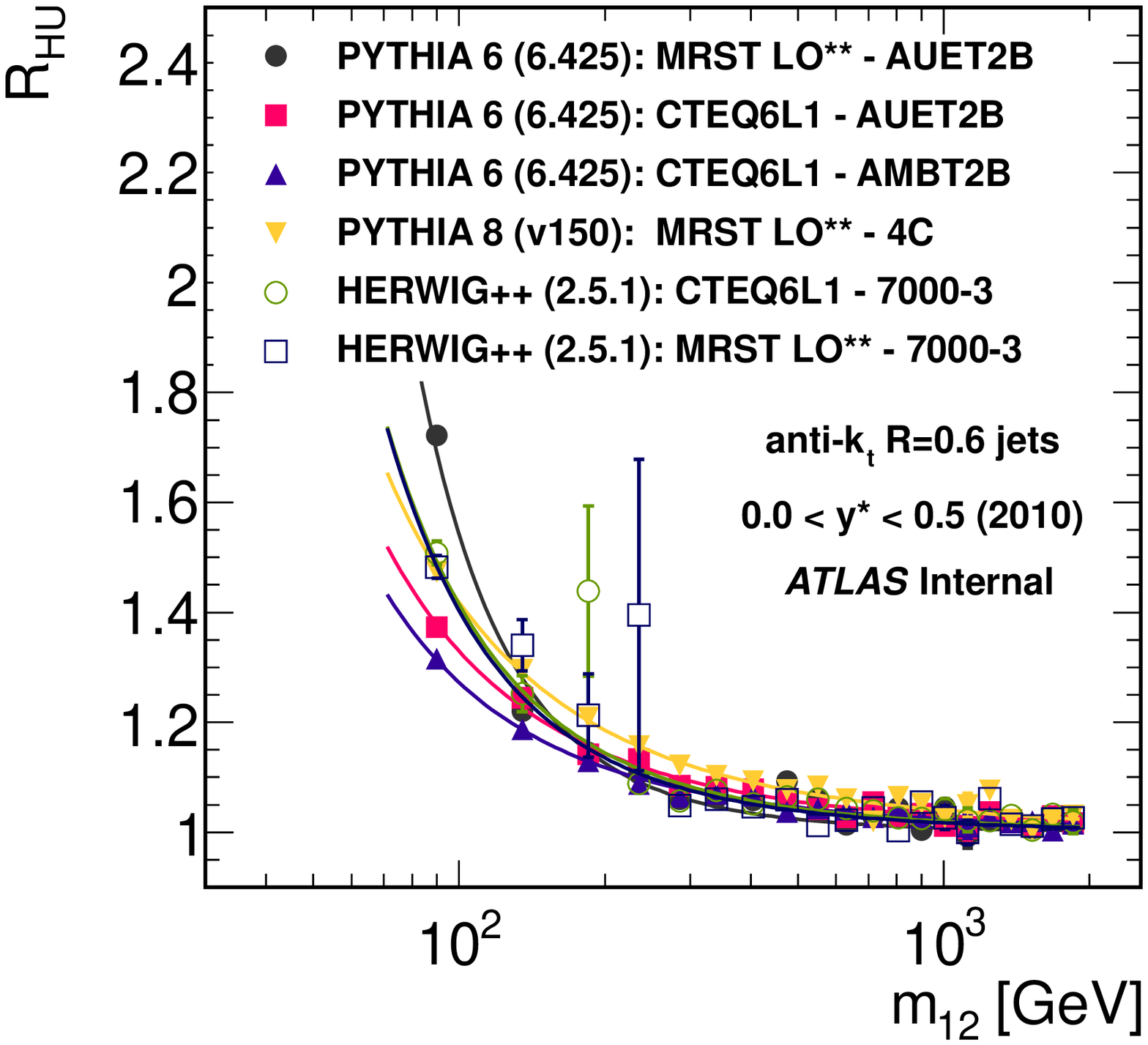}}
  \subfloat[]{\label{FIGhadronicCorrections2}\includegraphics[trim=5mm 14mm 0mm 10mm,clip,width=.52\textwidth]{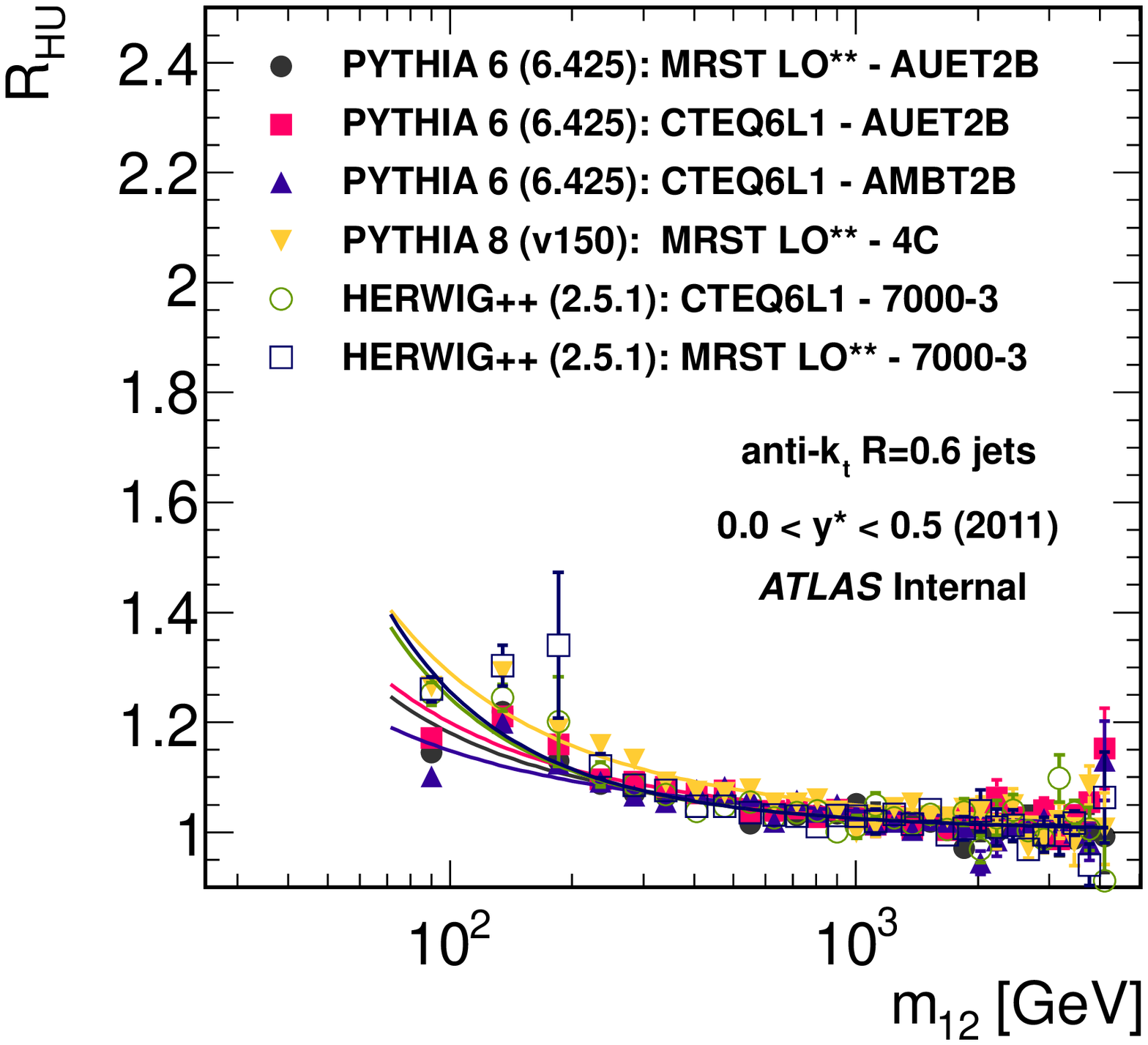}}
  \caption{\label{FIGhadronicCorrections}Non-perturbative correction factors, $\mathcal{R}_{\mrm{HU}}$, as a function
  of the invariant mass, $m_{12}$, of the two jets with the highest transverse momentum in an event, for \com jet rapidity, ${\ystr < 0.5}$, for several
  combinations of generators, PDF sets, and underlying event tunes, as indicated in the figure and explained in the text.
  The lines represent fits to the various MC samples, using the 2010 \Subref{FIGhadronicCorrections1} and the 2011 \Subref{FIGhadronicCorrections2}
  phase-space definitions of the measurement.
  }
\end{center}
\end{figure} 
The lines represent fits to the different MC samples.
Additional \ystr bins using the 2011 phase-space definition
are shown in \autoref{chapMeasurementOfTheDijetMassApp}, \autoref{FIGhadronicCorrections0App}.
Similar values of $\mathcal{R}_{\mrm{HU}}$ (not shown) were found for the 2010 measurement.
The correction factors have a significant influence (up to~80\%) at low dijet mass.

The uncertainty on the non-perturbative correction is taken as the 
spread of the fit results for the various MCs, compared to the nominal setup.
Additional sources of uncertainty on the theoretical prediction are discussed in the following.

\subsection{Uncertainties on the NLO prediction}
%
%
In addition to the uncertainty associated with the hadronization and UE corrections,
the main uncertainties on the NLO prediction come from several sources; the
uncertainties on the PDFs; the choice of factorisation and
renormalization scales; and the uncertainty on the value of the strong
coupling constant, $\alpha_{\rm s}$.  To allow for fast and flexible
evaluation of PDF and scale uncertainties, the
\applgrid~\cite{Carli:2010rw} software was interfaced with \nlojet in
order to calculate the perturbative coefficients once and store them
in a look-up table.  The PDF uncertainties are defined at 68\%~CL and
evaluated following the prescriptions given for each PDF set. These
account for the experimental uncertainties, tension between input data sets,
parametrisation uncertainties, and various theoretical uncertainties
related to PDF determination.

To estimate the uncertainty on the NLO prediction due to neglected
higher-order terms, each observable was recalculated while varying the
renormalization scale by a factor of two with respect to the default
choice.  Similarly, to estimate the sensitivity to the choice of scale
where the PDF evolution is separated from the matrix element, the
factorisation scale was separately varied by a factor of two.  Cases
where the two scales are simultaneously varied by a factor 2 in
opposite directions were not considered due to the presence of
logarithmic factors in the theory calculation that become large in
these configurations.  The envelope of the variation of the
observables was taken as a systematic uncertainty.

The effect of the
uncertainty on the value of the strong coupling constant, $\alpha_{\rm
s}$, is evaluated following the recommendation of the CTEQ
group~\cite{Lai:2010nw}, in particular by using different PDF sets
that were derived using the positive and negative variations of the
coupling from its best estimate.
Electroweak corrections were not included in the theory predictions~\cite{Moretti:2006ea}.

The total relative uncertainty on the NLO prediction and the breakdown to its components is shown in \autoref{FIGtheoryUncertainty}
for a single \ystr bin, using the 2010 and the 2011 phase-space definitions of the \xsec.
Additional \ystr bins of the measurement
are shown in \autoref{chapMeasurementOfTheDijetMassApp}, \autorefs{FIGtheoryUncertainty0App}~-~\ref{FIGtheoryUncertainty2App}.
\begin{figure}[ht]
\begin{center}
  \subfloat[]{\label{FIGtheoryUncertainty1}\includegraphics[trim=5mm 14mm 0mm 10mm,clip,width=.52\textwidth]{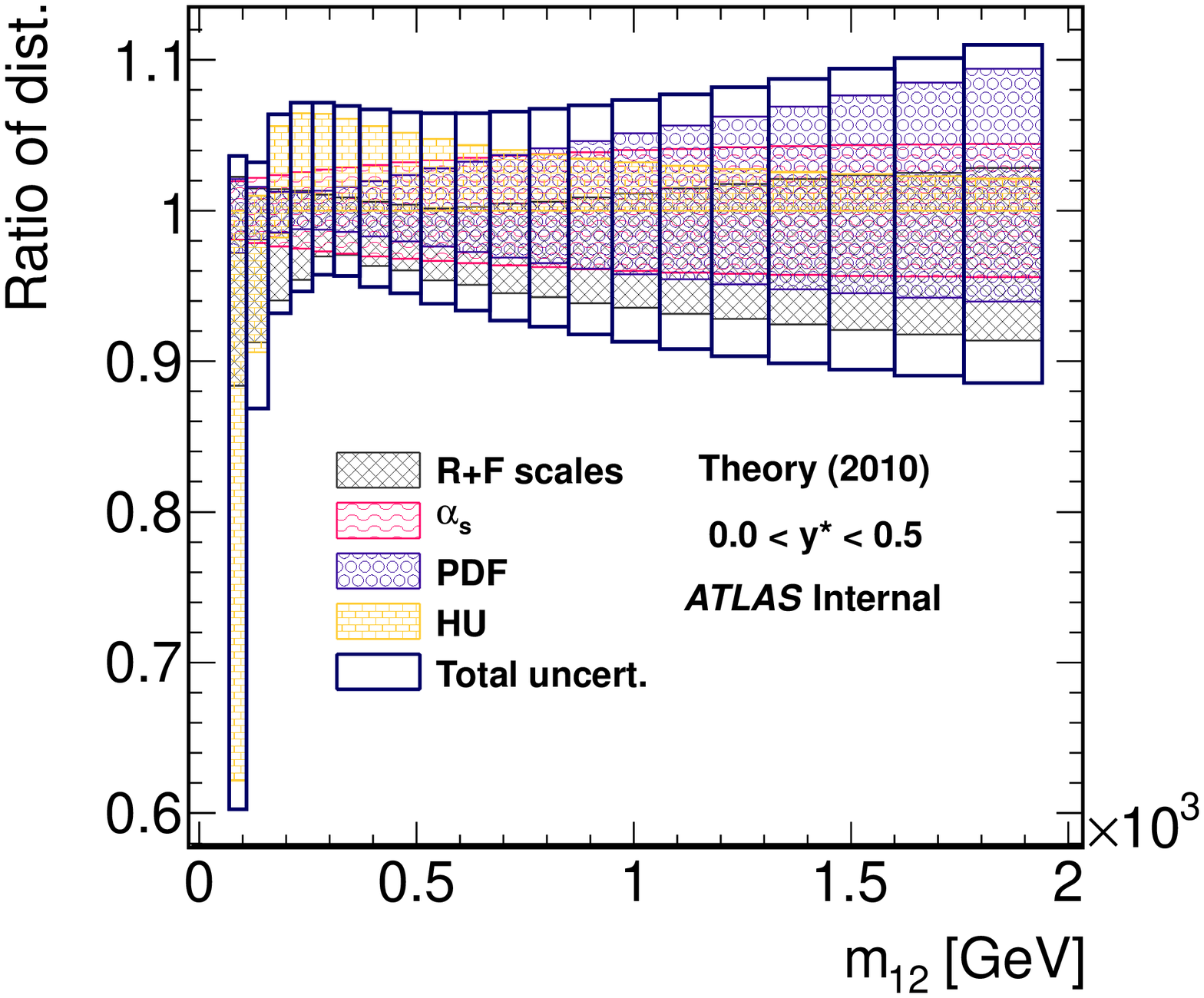}}
  \subfloat[]{\label{FIGtheoryUncertainty2}\includegraphics[trim=5mm 14mm 0mm 10mm,clip,width=.52\textwidth]{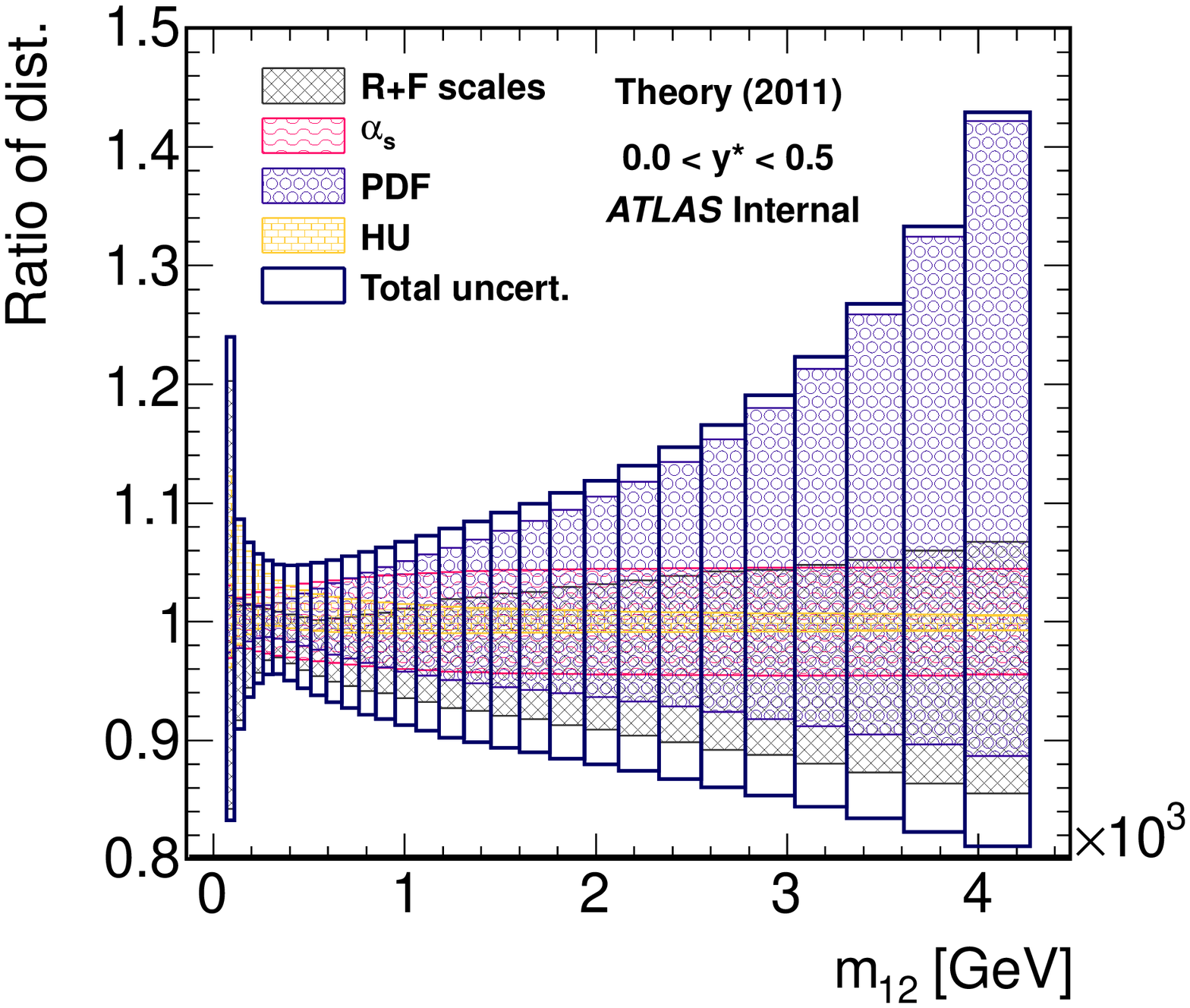}}
  \caption{\label{FIGtheoryUncertainty}The ratio of \nlojet expectations obtained under different assumptions relative to
  the nominal calculation, as a function
  of the invariant mass, $m_{12}$, of the two highest-\pt hadron-level jets with \com rapidity, ${\ystr < 0.5}$,
  for the 2010 \Subref{FIGtheoryUncertainty1} and for the 2011 \Subref{FIGtheoryUncertainty2} measurements.
  The variations on the nominal expectation include the uncertainty on the
  renormalization and factorisation scales (R+F scales), the uncertainty on the value of the strong
  coupling constant ($\alpha_{\rm s}$), use of different parton density functions (PDF), the uncertainty on the
  hadronization and UE corrections (HU) and the total uncertainty on all the latter (total uncert.\/).
  }
\end{center}
\end{figure} 
The uncertainty on the hadronization and UE corrections dominates at low values of $m_{12}$, while 
at high masses the uncertainty associated with the PDFs is most prominent. For intermediate mass values
(roughly between 0.5~and~1\TeV) the total uncertainty is smaller than 10\% in all \ystr bins.

\section{Bias and resolution in the invariant mass}
%
%
The offset in \ystr, $O_{\ystr}$, and in $m_{12}$, $O_{m_{12}}$, of the dijet system 
are defined as
\begin{equation}
  O_{\ystr}  = \ystr^{\mrm{rec}}-\ystr^{\mrm{truth}} \quad \mrm{and} \quad
  O_{m_{12}} = \frac{m_{12}^{\mrm{rec}}-m_{12}^{\mrm{truth}}}{m_{12}^{\mrm{truth}}} \;,
\label{eqJetAngularOffsetDef} \end{equation}
where  $\ystr^{\mrm{truth}}$ ($m_{12}^{\mrm{truth}}$) and $\ystr^{\mrm{rec}}$ ($m_{12}^{\mrm{rec}}$) are
the \ystr (invariant mass) of the dijet system, determined at the hadron-level and at the detector-level, respectively.
The two parameters respectively represent the bias in the rapidity
and the fractional bias in the mass of the dijet system.
The width of the distributions of $O_{\ystr}$ and $O_{m_{12}}$ represent the corresponding 
resolution of the two parameters, denoted as $\sigma(O_{\ystr})$ and $\sigma\left(O_{m_{12}}\right)$.
The parameters are determined in MC by selecting the two leading jets at the hadron and at the detector
levels in a given event, without requiring exact matching between the jets.

The average $O_{\ystr}$, the average $O_{m_{12}}$ and the corresponding resolution parameters are
calculated in MC10 and in MC11 for dijet systems with different \ystr values.
\Autorefs{FIGdijetMassOffsetResYstr}~-~\ref{FIGdijetMassOffsetResMass} show the dependence
of the parameters on $m_{12}^{\mrm{truth}}$.
\begin{figure}[htp]
\begin{center}
  \subfloat[]{\label{FIGdijetMassOffsetResYstr1}\includegraphics[trim=5mm 14mm 0mm 10mm,clip,width=.52\textwidth]{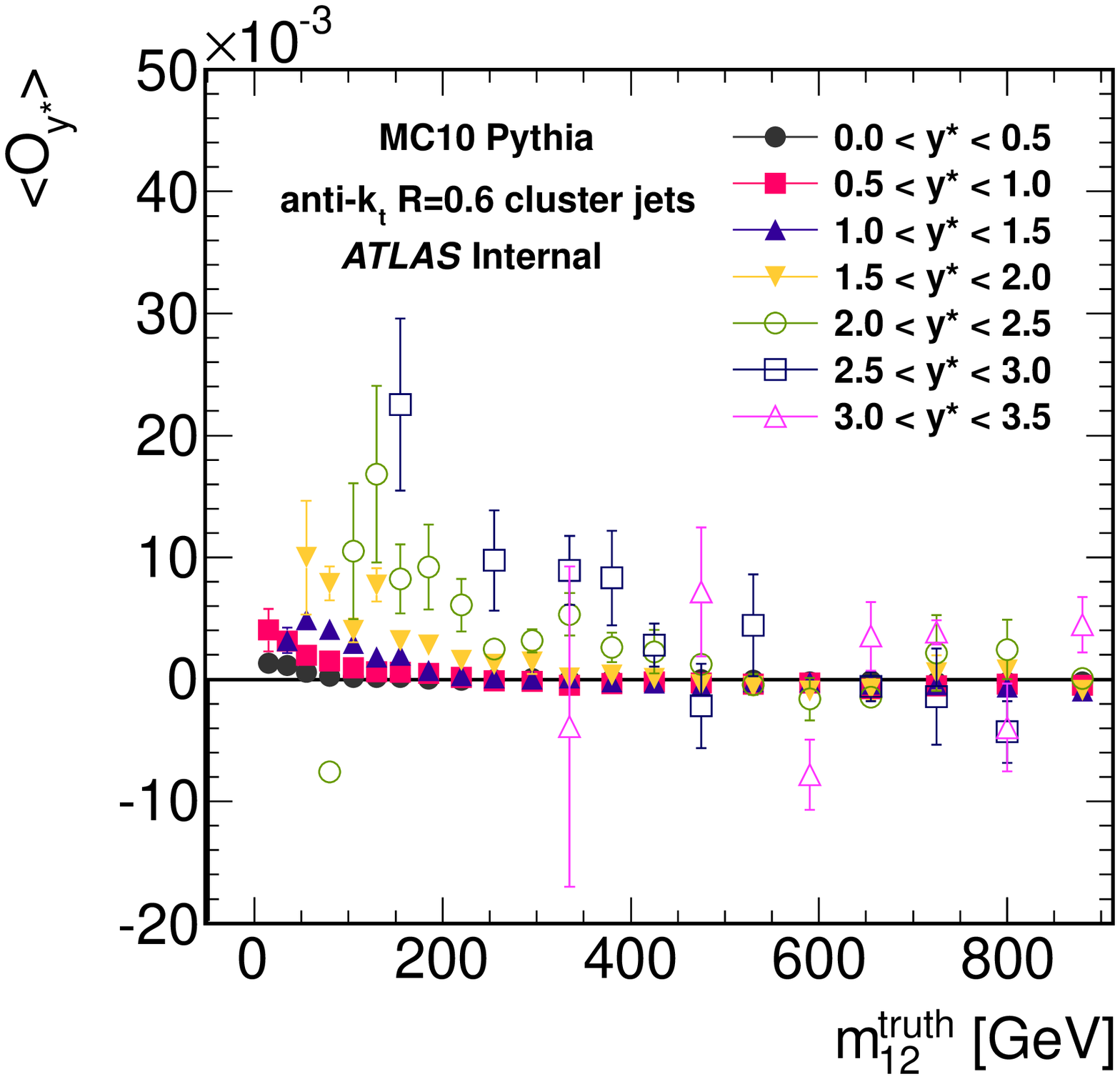}}
  \subfloat[]{\label{FIGdijetMassOffsetResYstr2}\includegraphics[trim=5mm 14mm 0mm 10mm,clip,width=.52\textwidth]{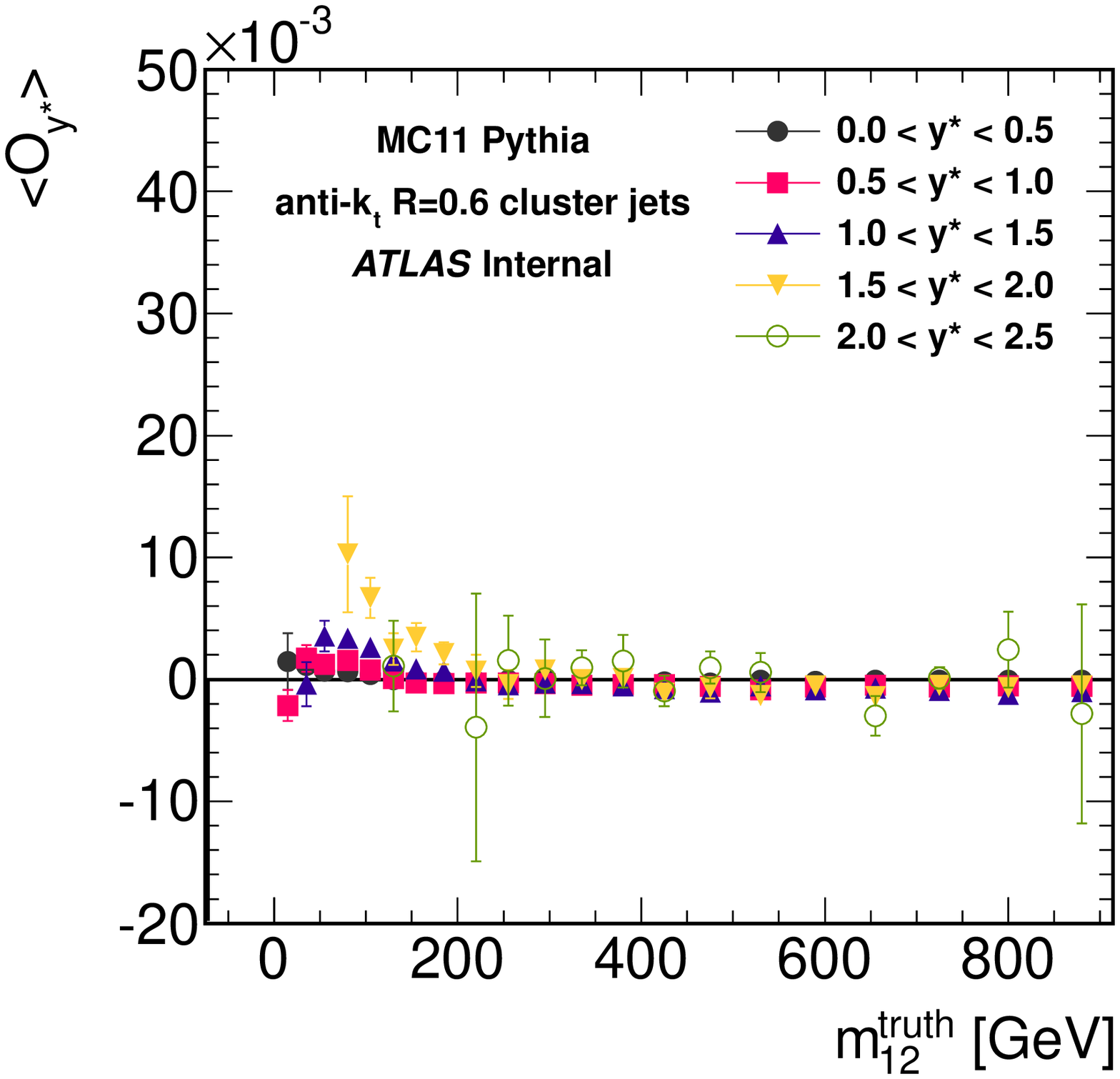}} \\
  \subfloat[]{\label{FIGdijetMassOffsetResYstr3}\includegraphics[trim=5mm 14mm 0mm 10mm,clip,width=.52\textwidth]{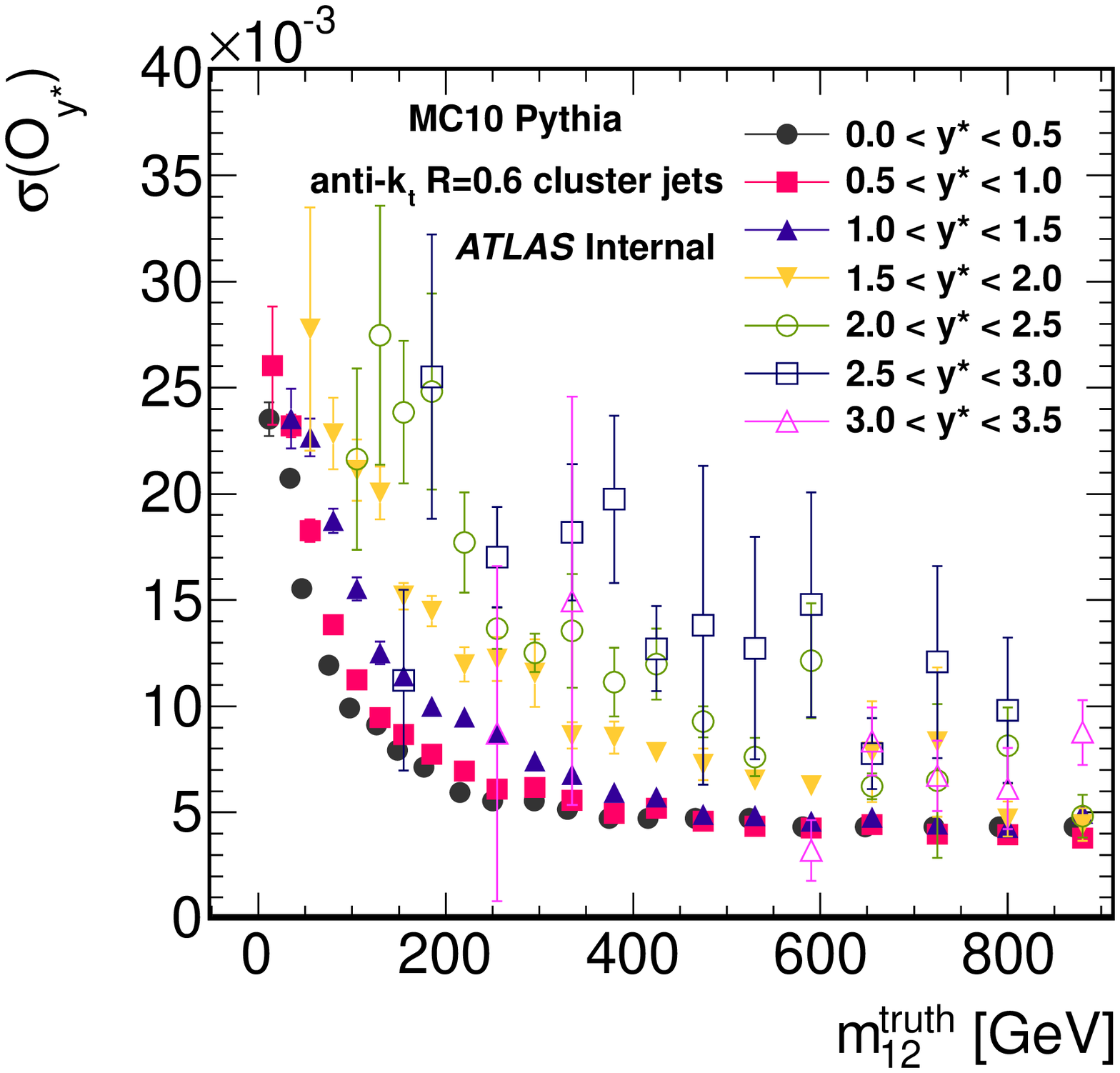}}
  \subfloat[]{\label{FIGdijetMassOffsetResYstr4}\includegraphics[trim=5mm 14mm 0mm 10mm,clip,width=.52\textwidth]{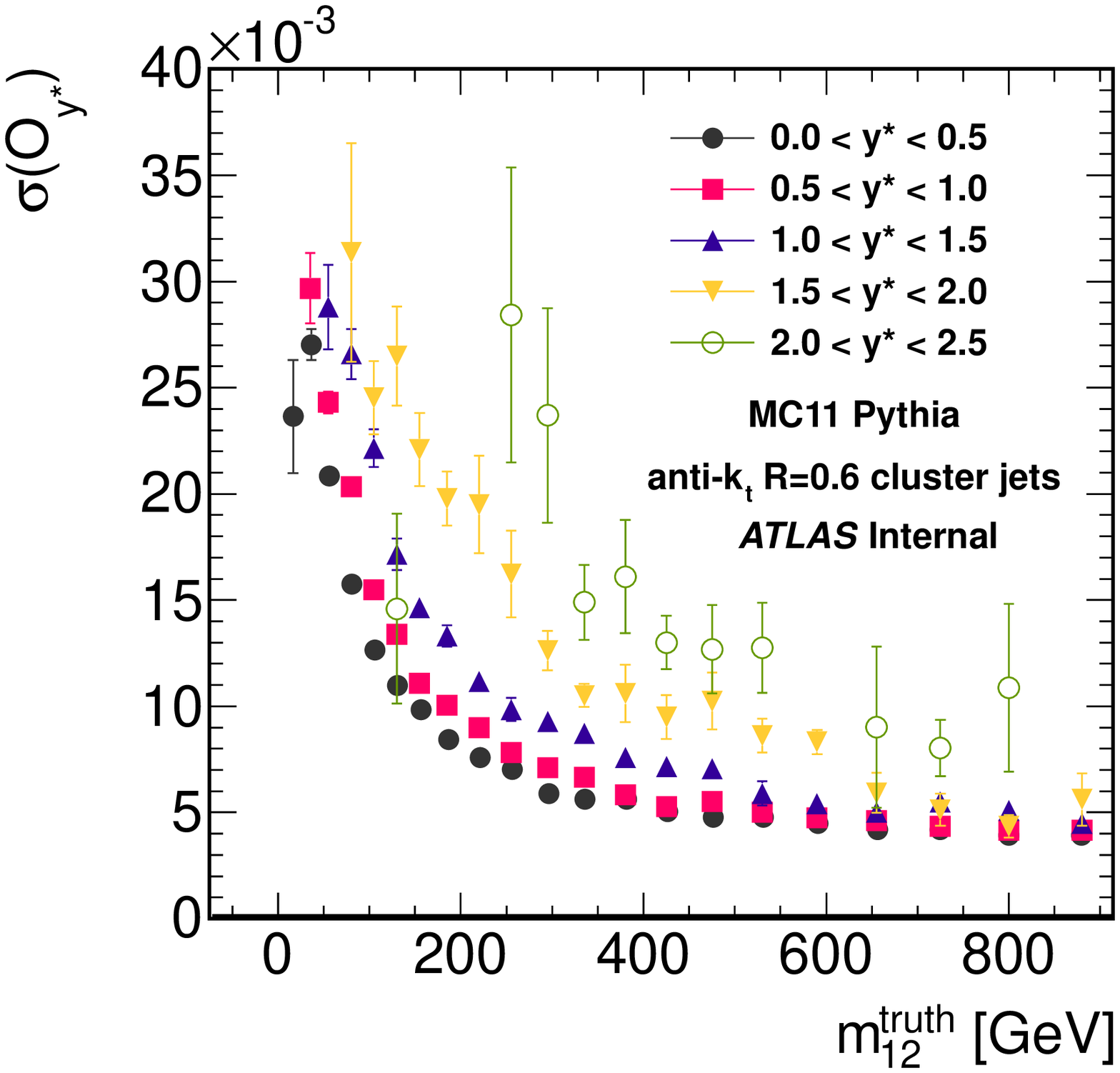}}
  \caption{\label{FIGdijetMassOffsetResYstr}Dependence of the average
    rapidity offset, $<O_{\ystr}>$, (\Subref{FIGdijetMassOffsetResYstr1} and \Subref{FIGdijetMassOffsetResYstr2})
    and of the width of the distribution of the offset, $\sigma(O_{\ystr})$,
    (\Subref{FIGdijetMassOffsetResYstr3} and \Subref{FIGdijetMassOffsetResYstr4}) on the
    invariant mass, $m_{12}$, of the two jets with the highest transverse momentum in an event, for different \com jet rapidities, \ystr,
    in MC10 and in MC11, as indicated in the figures.
  }
\end{center}
\end{figure} 
The \ystr offset is generally smaller in magnitude than~2\%, decreasing as $m_{12}$ increases. For mass values above~$\sim200\GeV$,
$O_{\ystr}$ is mostly consistent with zero. Similar behaviour is exhibited in MC10 and in MC11.
The resolution, $\sigma(O_{\ystr})$, decreases with $m_{12}$ from about~3\% at low masses to below~1.5\% above~600\GeV.

\begin{figure}[htp]
\begin{center}
  \subfloat[]{\label{FIGdijetMassOffsetResMass1}\includegraphics[trim=5mm 14mm 0mm 10mm,clip,width=.52\textwidth]{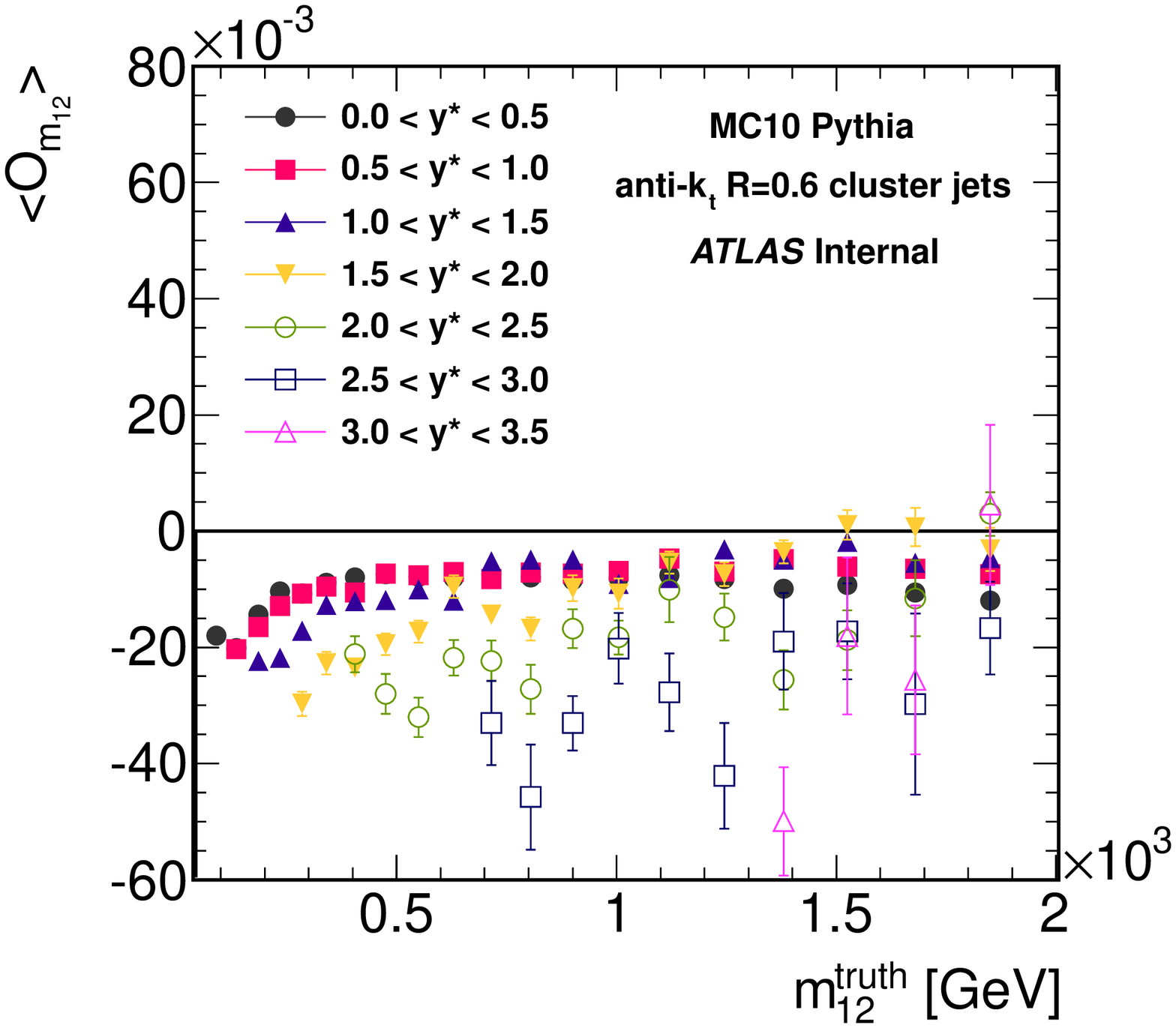}}
  \subfloat[]{\label{FIGdijetMassOffsetResMass2}\includegraphics[trim=5mm 14mm 0mm 10mm,clip,width=.52\textwidth]{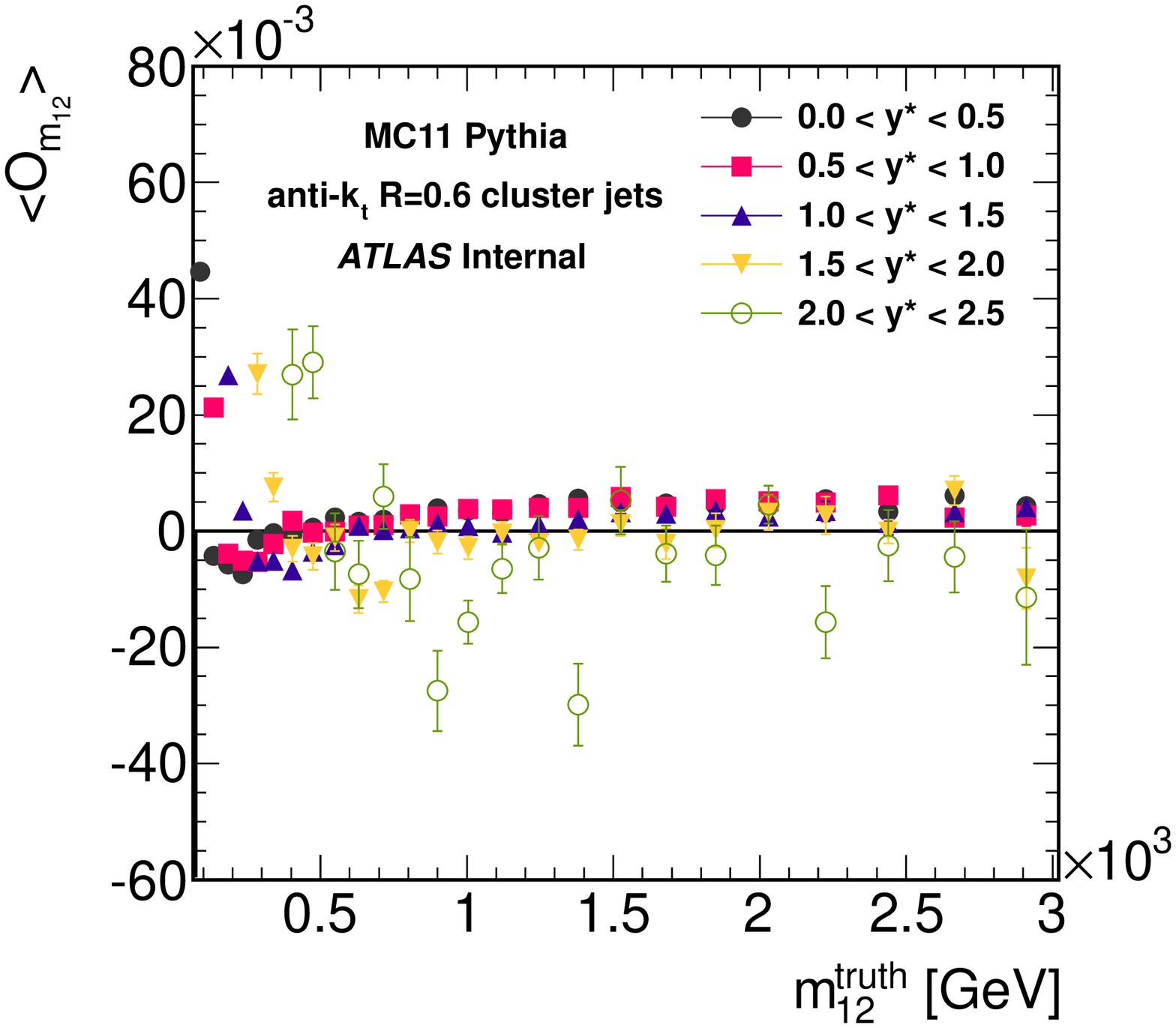}} \\
  \subfloat[]{\label{FIGdijetMassOffsetResMass3}\includegraphics[trim=5mm 14mm 0mm 10mm,clip,width=.52\textwidth]{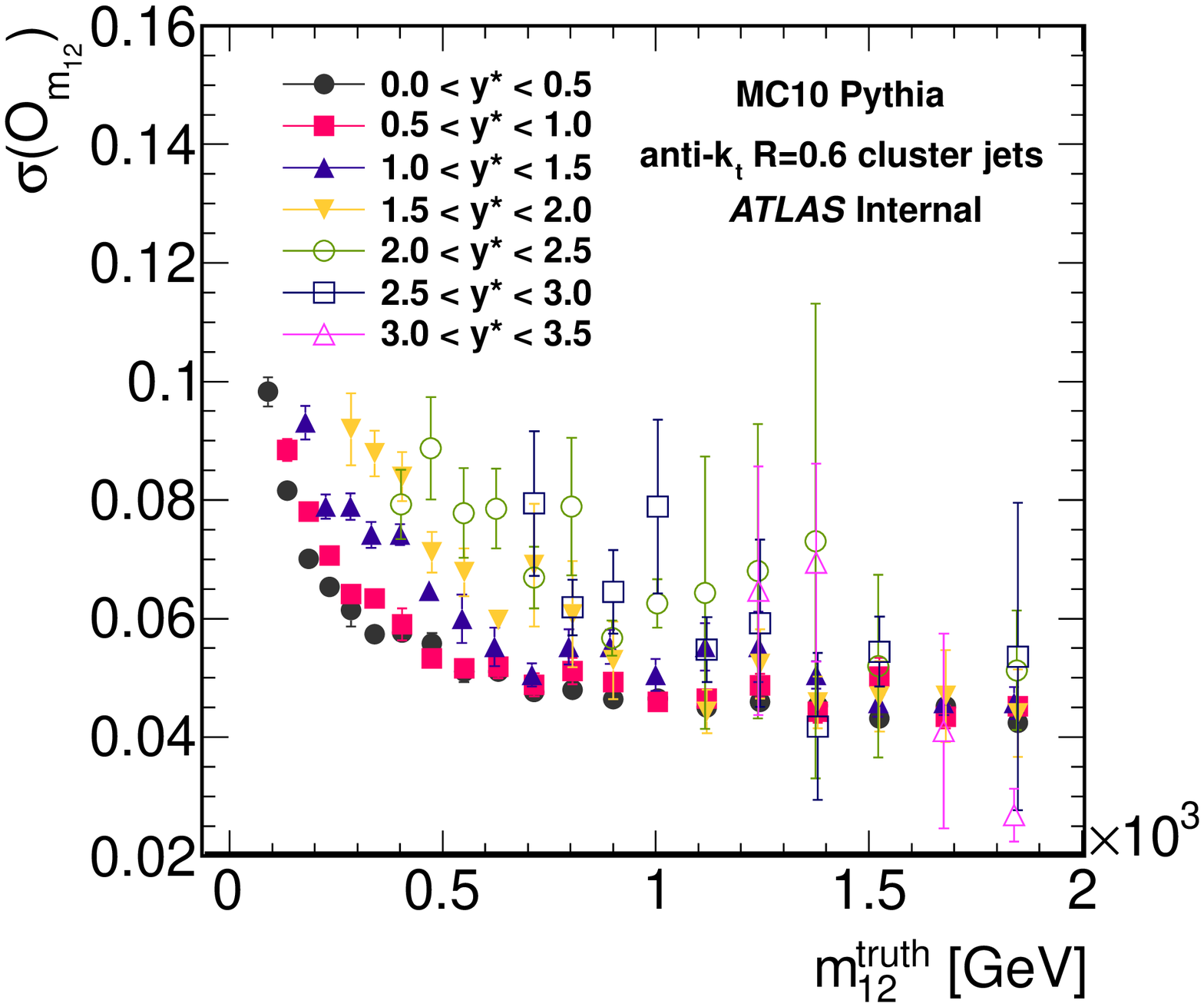}}
  \subfloat[]{\label{FIGdijetMassOffsetResMass4}\includegraphics[trim=5mm 14mm 0mm 10mm,clip,width=.52\textwidth]{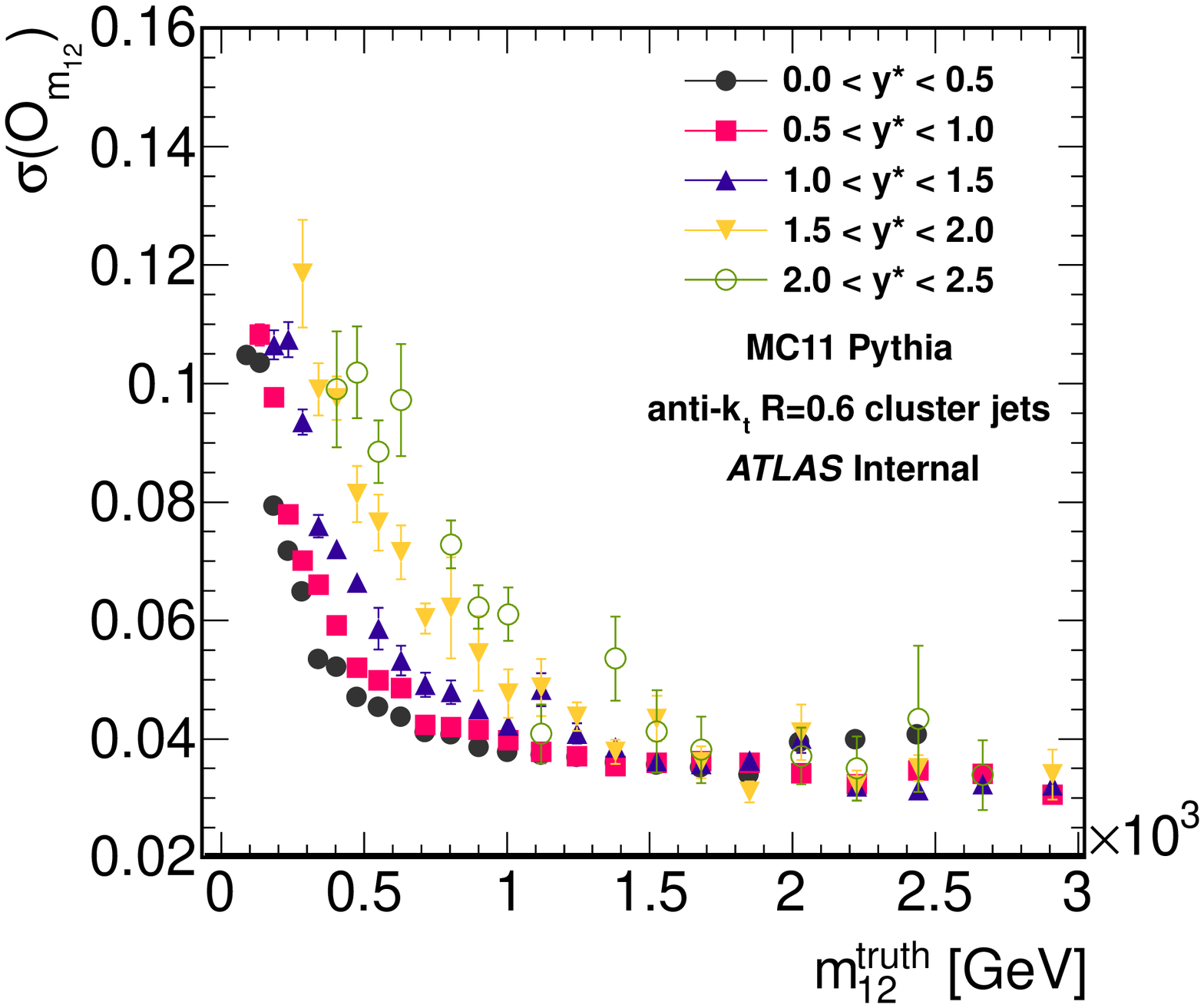}}
  \caption{\label{FIGdijetMassOffsetResMass}Dependence of the average
    mass offset, $<O_{m_{12}}>$, (\Subref{FIGdijetMassOffsetResMass1} and \Subref{FIGdijetMassOffsetResMass2})
    and of the width of the distribution of the offset, $\sigma(O_{m_{12}})$,
    (\Subref{FIGdijetMassOffsetResMass3} and \Subref{FIGdijetMassOffsetResMass4}) on the
    invariant mass, $m_{12}$, of the two jets with the highest transverse momentum in an event, for different \com jet rapidities, \ystr,
    in MC10 and in MC11, as indicated in the figures.
  }
\end{center}
\end{figure} 
For dijet systems with $\ystr < 2.5$, the mass offset, $O_{m_{12}}$, is typically smaller than~30\% in 2010 and~10\% in 2011.
The exception is the low mass region (up to~100\GeV) in 2011, for which a bias of up
to~40\% is observed. For the last two bins in \ystr in 2010, a slightly larger offset of up to~50\% in absolute value
is observed, decreasing with increasing $m_{12}$.
The degraded performance at high \ystr in 2010 and at low mass in 2011 is due to \pu; the latter is
dominant for jets in the forward region (which populate the high \ystr bins in 2010), as well as for low energy jets (which
comprise low dijet masses in 2011).

The relative dijet mass resolution, $\sigma\left(O_{m_{12}}\right)$, decreases as $m_{12}$ increases.
This trend extends up to roughly 500\GeV
for low values of \ystr or to~${1-1.5\TeV}$ for high values of \ystr. At high mass values the resolution is typically constant,
consistent with the expected behaviour of the energy resolution of jets, also seen
in \autoref{chapJetReconstruction}, \autoref{FIGjetResolutionPt}.
In general, $\sigma\left(O_{m_{12}}\right)$ decreases from~10\% to~4\% in~2010 and from~13\% to~3\% in~2011.
Comparable resolution is observed for low \ystr and low mass as for dijets with higher \ystr and higher mass.
The reason for this is that 
for a given average \pt of the two jets, higher values of \ystr are associated with higher dijet masses.


In order to extract the invariant mass distribution at the particle-level, all the detector effects, such as
efficiency and resolution, have to be corrected for.
Aside from the \JES correction, all other corrections are performed using an iterative unfolding procedure,
discussed in the following.

\section{Unfolding\label{chapDijetMassUnfolding}}
%
%
Unfolding of the dijet mass differential
distributions is performed using the \textit{Iterative Bayes} algorithm
provided by the \RooUnfold software package~\cite{RooUnfold}. The
unfolding algorithm uses the method described by ${\mrm{D'\kern -0.1em Agostini}}$~\cite{D'Agostini1995487}, and 
previously used in~\cite{Aad:2011sc}.
This procedure takes as input two elements, a measured distribution and a response matrix obtained from MC.
For the case of the dijet mass measurement, the former is the detector-level two-dimensional distribution
of the differential dijet mass \xsec, binned in $m_{12}$ and in \ystr.
The second element, the response matrix, provides a mapping between the kinematic variations of the
reconstructed objects, and those obtained directly from the event generator. 
%

\minisec{Adjusting the kinematic distributions of jets in the MC}
%
%
Due to the finite binning of the unfolding procedure and the very steep kinematic distributions that need to
be unfolded, the response matrix depends on the shape of the input distribution in the MC.
It is therefore advantageous to start the unfolding procedure with input distributions which are
as close as possible to the corresponding distributions in the data.

For the case of MC10, the \pt spectrum of the two leading jets without any correction is presented in \autoref{FIGleadingJetPtRW2010_1}.
The distributions in MC are rescaled such that they agree in normalization with the data
at ${\pt = 50\GeV}$. At low \pt, the MC underestimates the data by a factor
of~2.3. As \pt increases, the difference between MC and data increases up to~50\% at ${\pt = 600\GeV}$.
The main origin of the weights in MC10 has been identified as the result of mismatch between the trigger in the data
and in the MC.

To reach a better agreement, the \pt distribution in the MC is reweighted. 
Transverse momentum reweighting is performed in bins of rapidity.
The weights are estimated from the ratio of the data distribution to the MC distribution
at detector-level, which is then parametrized by a smooth function. This function is
applied as a weight in the MC at the particle-level, based on truth jet \pt.
The \pt distributions of the two leading jets in reweighted events are shown in \autoref{FIGleadingJetPtRW2010_2}.
\begin{figure}[htp]
\begin{center}
  \subfloat[]{\label{FIGleadingJetPtRW2010_1}\includegraphics[trim=5mm 14mm 0mm 10mm,clip,width=.52\textwidth]{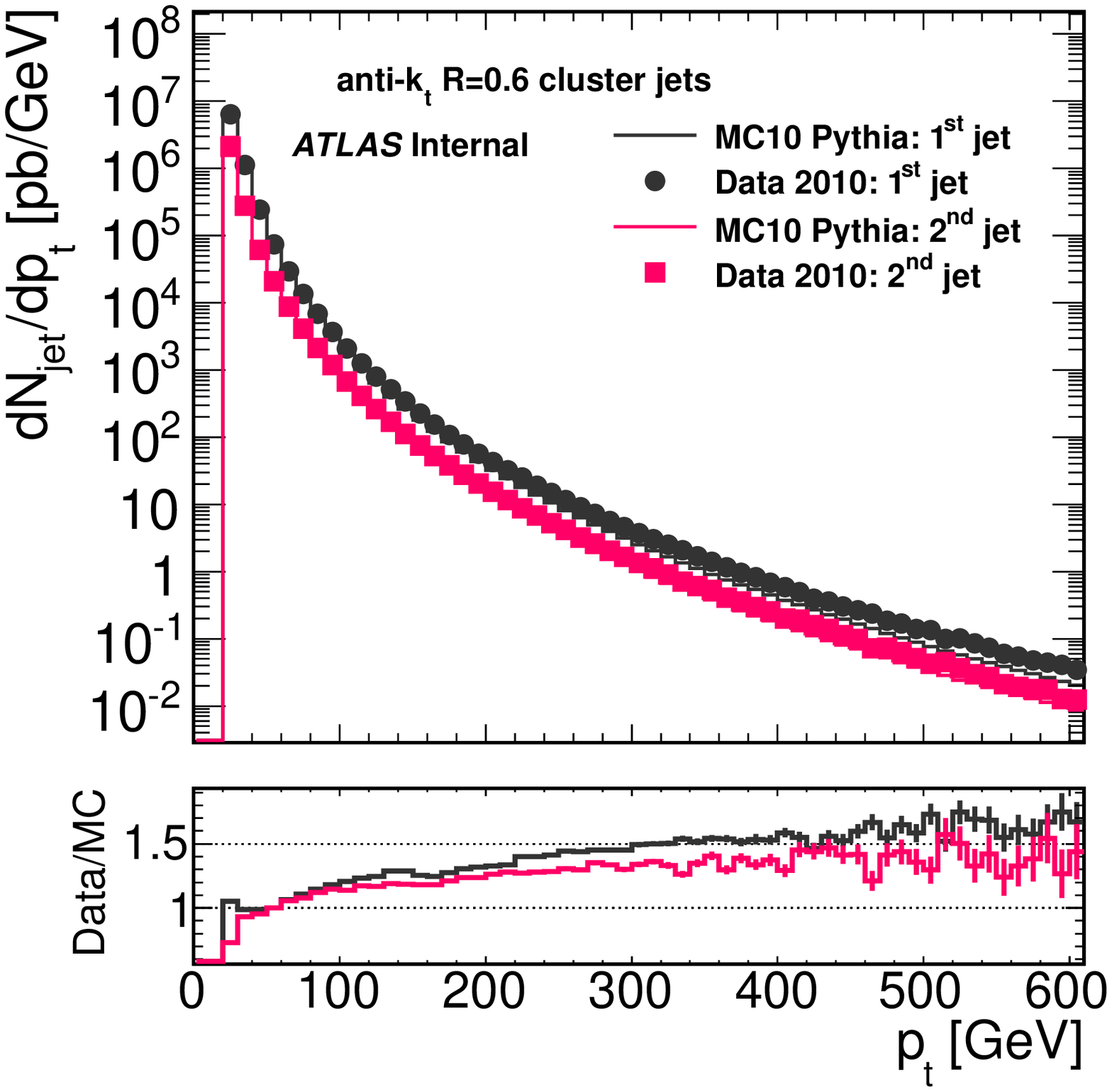}} 
  \subfloat[]{\label{FIGleadingJetPtRW2010_2}\includegraphics[trim=5mm 14mm 0mm 10mm,clip,width=.52\textwidth]{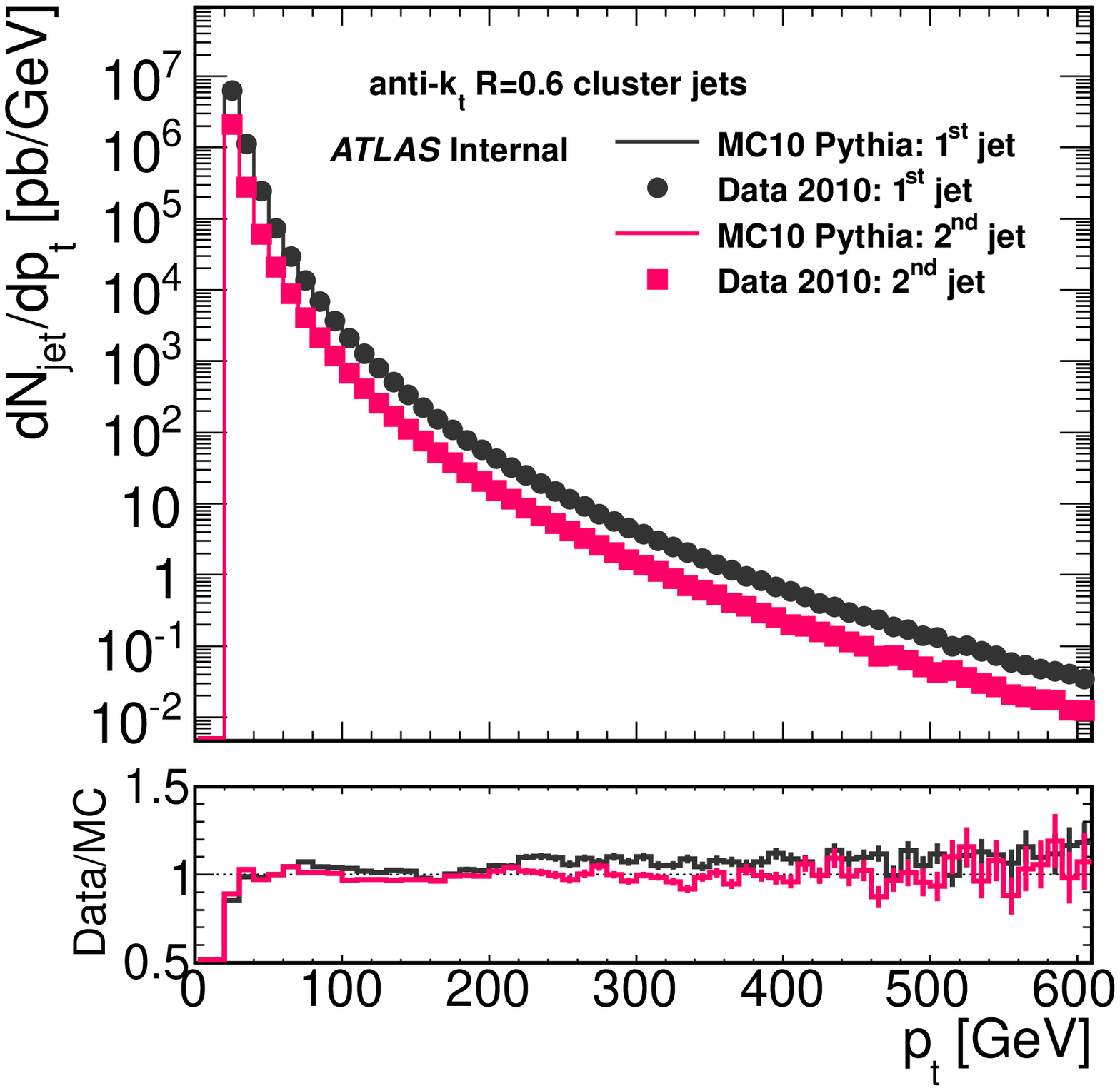}}
  \caption{\label{FIGleadingJetPtRW2010}Transverse momentum, \pt, spectra of the two highest-\pt jets
  in an event, denoted as $1^{\mrm{st}}$ and $2^{\mrm{nd}}$~jet in the figures, for events before \Subref{FIGleadingJetPtRW2010_1}
  and after \Subref{FIGleadingJetPtRW2010_2} the reweighting procedure described in the text, using
  MC10 and the 2010 data, where the distributions in MC are rescaled such that they agree in normalization with the data
  at $\pt = 50\GeV$.
  The bottom panels show the ratio of data to MC of the various distributions.
  }
\end{center}
\end{figure} 
The difference between MC and data is reduced compared to that before reweighting; this approximate
procedure therefore seems adequate. The residual differences, at the level of~${10-20\%}$, are taken care of by the unfolding.
Following the \pt reweighting, the multiplicity distribution of the reconstructed vertices, \Npv, in the MC is matched to that in the data
(the reweighted vertex distribution is shown in \autoref{chapJetAreaMethod}, \autoref{FIGNpvDist2011DataMc1}.)
The vertex reweighting does not bias the \pt spectrum, as there is no correlation between \Npv and
jet \pt in MC10.

Due to improvements in the simulation, no \pt reweighting based on truth jets is required for MC11.
To achieve adequate agreement between the 2011 data and MC11, it is enough to
constrain the spectrum of jets in MC, by matching the trigger conditions with those in the data. Trigger assignment is made according to the \pt of
the two leading jets in an event (see \autoref{chapTwoTriggerLumiCalcScheme}). Accordingly, events in MC11 are given weights such that
the number of weighted events in each trigger bin\footnote{ The expression
trigger bin refers here to the classification of an event according to the triggers associated with the two leading jets.
Associations are made according to jet \pt and \Eta, as specified
in \autoref{chapDataSelection}, \autorefs{TBLcentralTrigNames2010}~-~\ref{TBLcentralTrigNames2011}.}
matches the corresponding number in data.
As mentioned in \autoref{chapSimulationOfTheATLASdetector}, MC11 is divided into several ``periods'', which correspond to different
\pu conditions and trigger prescales.
In order to avoid biasing the vertex distribution, trigger matching is performed separately for events with any given value of \Npv.
Distributions of the number of reconstructed vertices in events assigned to several different trigger bins are shown
in \autoref{FIGleadingJetPtRW2011_1}. The \pt spectra of the two leading jets from the same events are shown in \autoref{FIGleadingJetPtRW2011_2}.
\begin{figure}[htp]
\begin{center}
  \subfloat[\qquad\qquad\qquad\qquad\qquad\qquad\qquad]
             {\label{FIGleadingJetPtRW2011_1}\includegraphics[trim=5mm 10mm -20mm 10mm,clip,width=.63\textwidth]{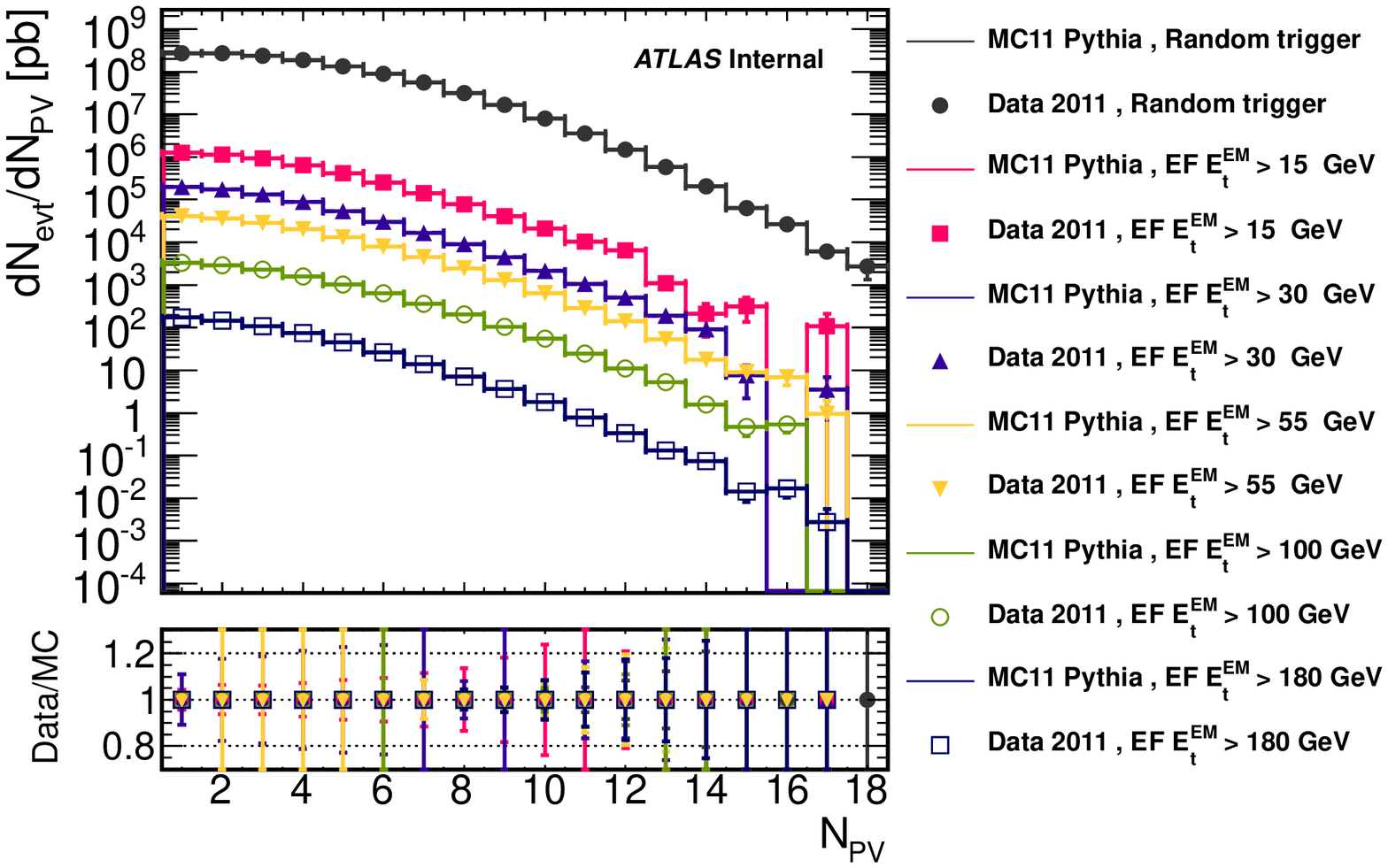}}
  \subfloat[]{\label{FIGleadingJetPtRW2011_2}\includegraphics[trim=20mm 14mm 0mm 10mm,clip,width=.38\textwidth]{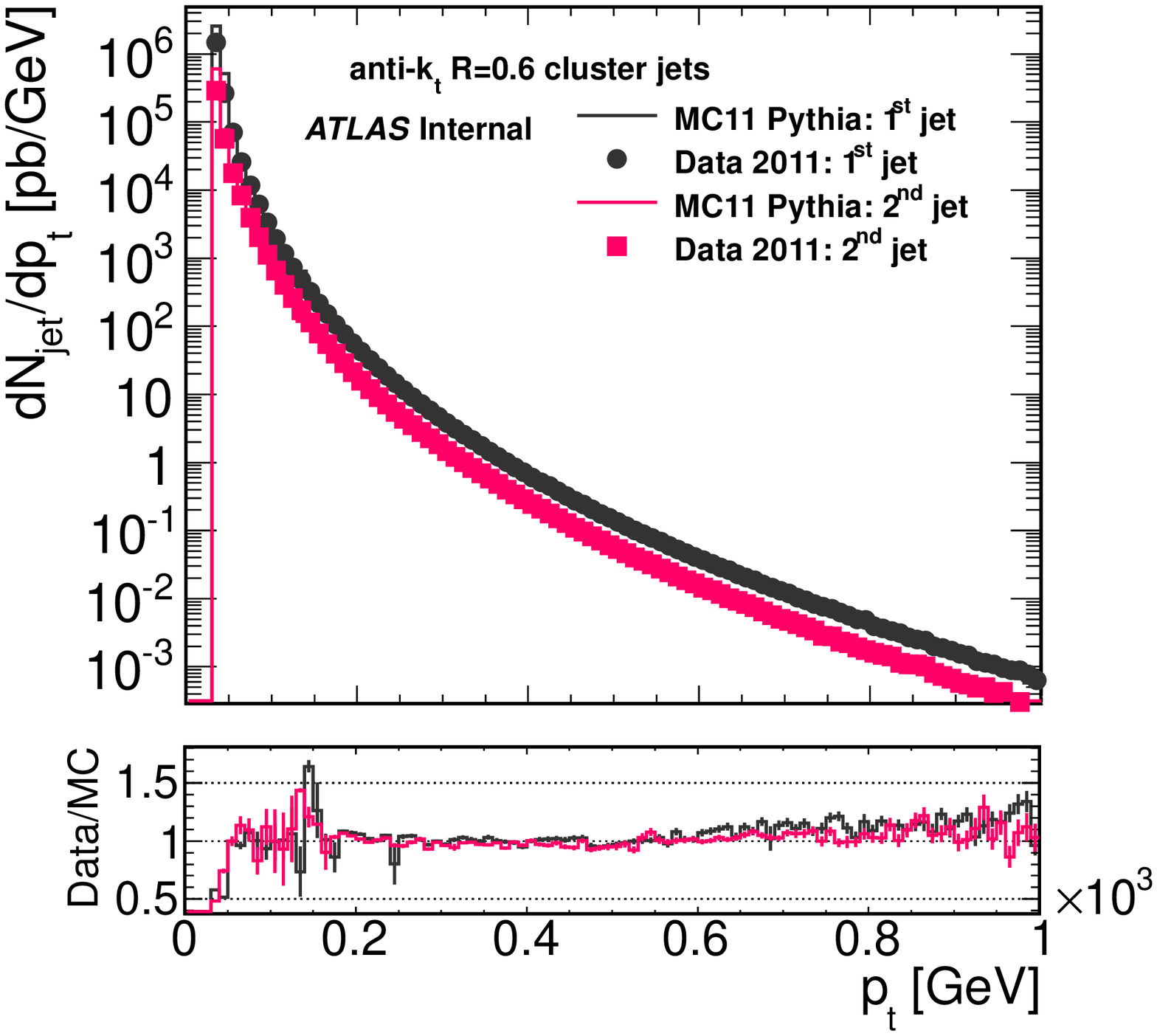}}
  \caption{\label{FIGleadingJetPtRW2011}\Subref{FIGleadingJetPtRW2011_1} Distributions of the number of reconstructed vertices, \Npv,
      in the 2011 data and in MC11, for events in which the jet with the highest transverse momentum is associated with different triggers,
      including a random trigger and jet triggers, where the latter are 
      ordered by trigger thresholds at the \EM scale, $E_{\mrm{t}}^{\mrm{EM}}$, as indicated in the figure. \\
      \Subref{FIGleadingJetPtRW2011_2} Transverse momentum, \pt, spectra of the two highest-\pt jets
      in an event, denoted as $1^{\mrm{st}}$ and $2^{\mrm{nd}}$~jet in the figure, in the 2011 data and in MC11,
      where the distributions in MC are rescaled such that they agree in normalization with the data at $\pt = 50\GeV$.
      \\
      The bottom panels in both figures show the ratio of data to MC.
      }
\end{center}
\end{figure} 
The distributions in MC are rescaled such that they agree in normalization with the data at $\pt = 50\GeV$.
The \pt spectrum in MC11 differs from data by up to~20\% for \ptLower{50}, by up to 10\% within \ptRange{50}{500} and by up
to~20\% within \ptRangeT{0.5}{1}.

As already mentioned, the steeply falling distribution of jet \pt has a significant effect on migrations between mass bins in the
unfolding response matrix.
To a lesser degree, discrepancies in the relatively flat rapidity distribution of jets may also affect the unfolding.
A comparison between the pseudo-rapidity of the two leading jets in MC10 and in MC11 with the corresponding
data is shown in \autoref{FIGleadingJetEtaRW}. The simulation samples are \pt-reweighted (MC10) or trigger-corrected (MC11), as
discussed above.
\begin{figure}[htp]
\begin{center}
  \subfloat[]{\label{FIGleadingJetEtaRW_1}\includegraphics[trim=5mm 14mm 0mm 10mm,clip,width=.52\textwidth]{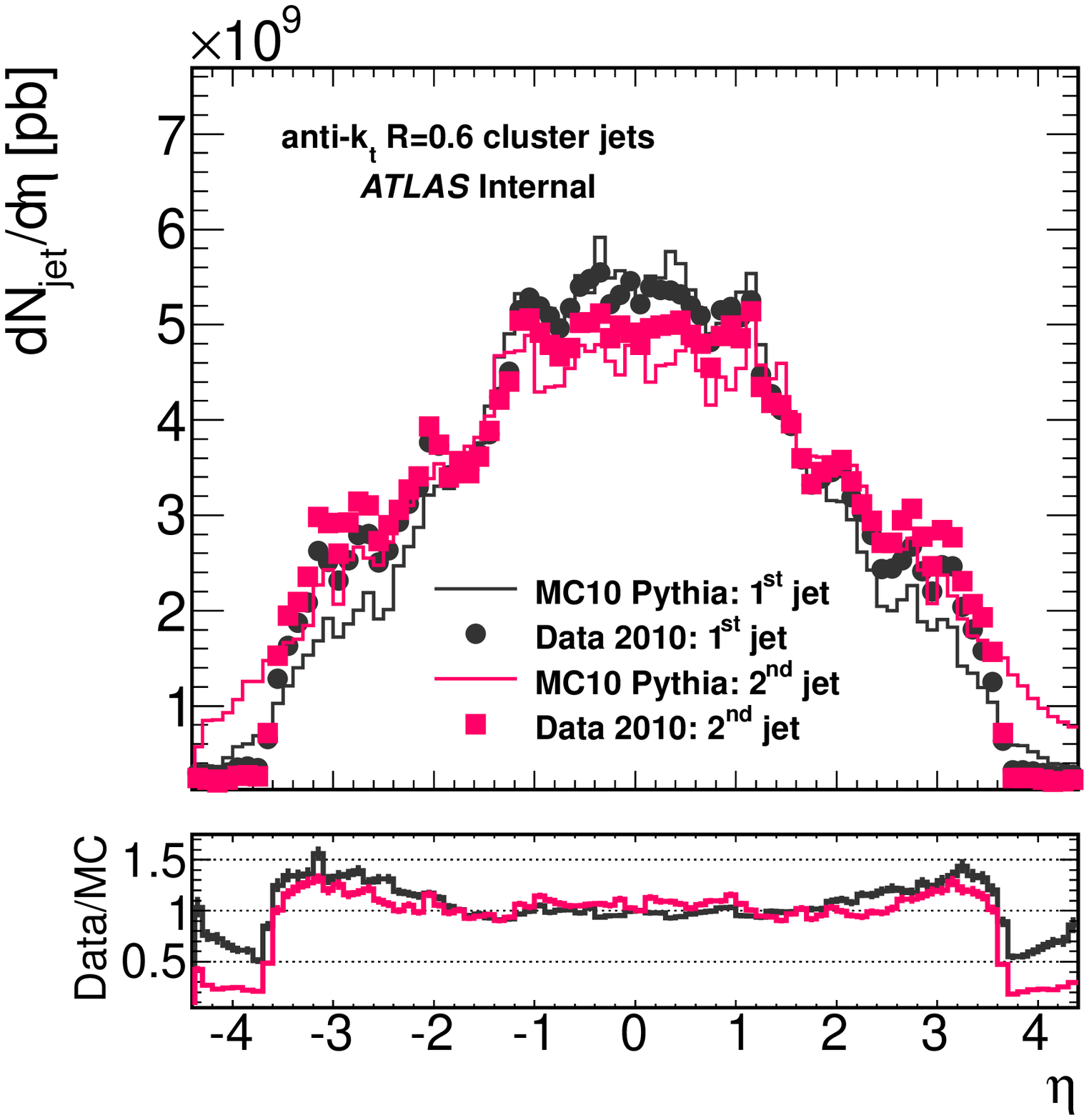}} 
  \subfloat[]{\label{FIGleadingJetEtaRW_2}\includegraphics[trim=5mm 14mm 0mm 10mm,clip,width=.52\textwidth]{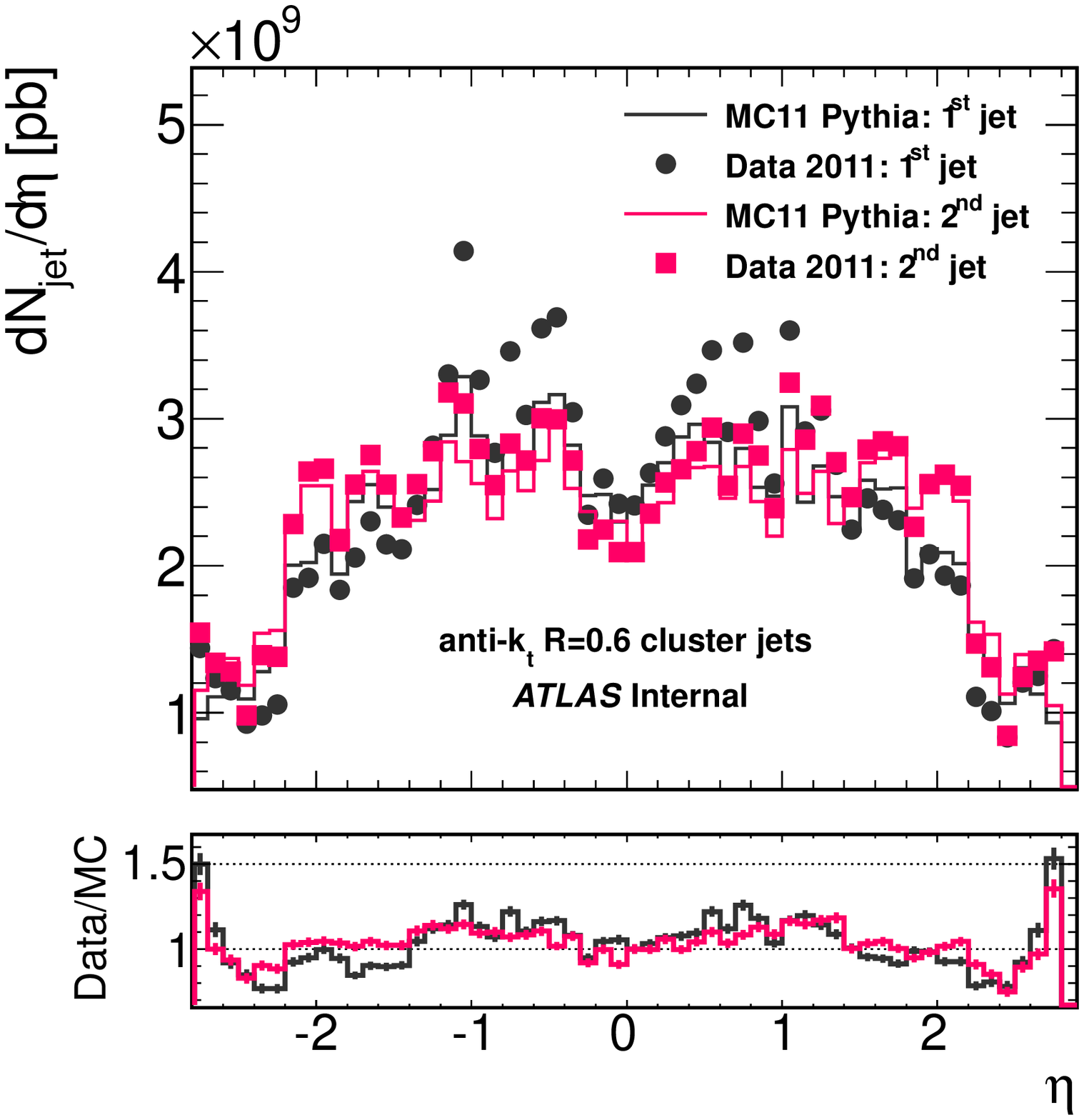}}
  \caption{\label{FIGleadingJetEtaRW}Pseudo-rapidity, \Eta, distributions of the two jets with the highest
  transverse momentum
  in an event, denoted as $1^{\mrm{st}}$ and $2^{\mrm{nd}}$~jet in the figure, using
  the 2010 data with MC10 \Subref{FIGleadingJetEtaRW_1} or the 2011 data with MC11 \Subref{FIGleadingJetEtaRW_2}, as indicated.
  The bottom panels show the ratio of data to MC.
  }
\end{center}
\end{figure} 
The differences between data and MC are smaller than~25\% within \etaLower{2.5} and within roughly~50\% for
\etaRange{2.5}{4.4}.
In principle, it is possible to further improve the agreement in the \Eta distributions of the MC with data.
However, this comes at the
cost of degrading the agreement in the \pt distributions, as the momentum and rapidity distributions of jets are correlated.
No further correction for the rapidity of jets is therefore made.

\minisec{The number of unfolding iterations}
%
%
In order to unfold the data, the response matrix has to be inverted. In this case, the inversion of the matrix
is notoriously difficult, as it is an ill-posed problem.
The usual approach is to apply an iterative procedure with a smoothing function~\cite{D'Agostini1995487}.
The latter prevents oscillations in
the solution, which are generated by inherent correlations in the data.
Such an approach is implemented as part of the unfolding algorithm in this analysis.
The number of iterations used in the procedure, denoted by $\kappa_{\mrm{unf}}$, can not be determined beforehand,
and must be optimized for a given measurement.


The value of $\kappa_{\mrm{unf}}$ used in the analysis is determined from simulation.
Half of the available MC statistics are
used to reconstruct a response matrix; the other half, a statistically independent sample,
is ``unfolded'' using this matrix, as would the data. One can then compare the ``unfolded'' MC distribution of
the differential \xsec with the respective truth-level distribution from the same events.
The optimal value of $\kappa_{\mrm{unf}}$ is one which minimizes the differences between these two distributions.

\Autoref{FIGunfoldingIterations1} shows the first step in the optimization procedure of $\kappa_{\mrm{unf}}$.
\begin{figure}[htp]
\begin{center}
  \hspace{23pt}
  \subfloat[\qquad\qquad]{\label{FIGunfoldingIterations1}\includegraphics[trim=30mm 5mm 0mm  5mm,clip,width=.405\textwidth]{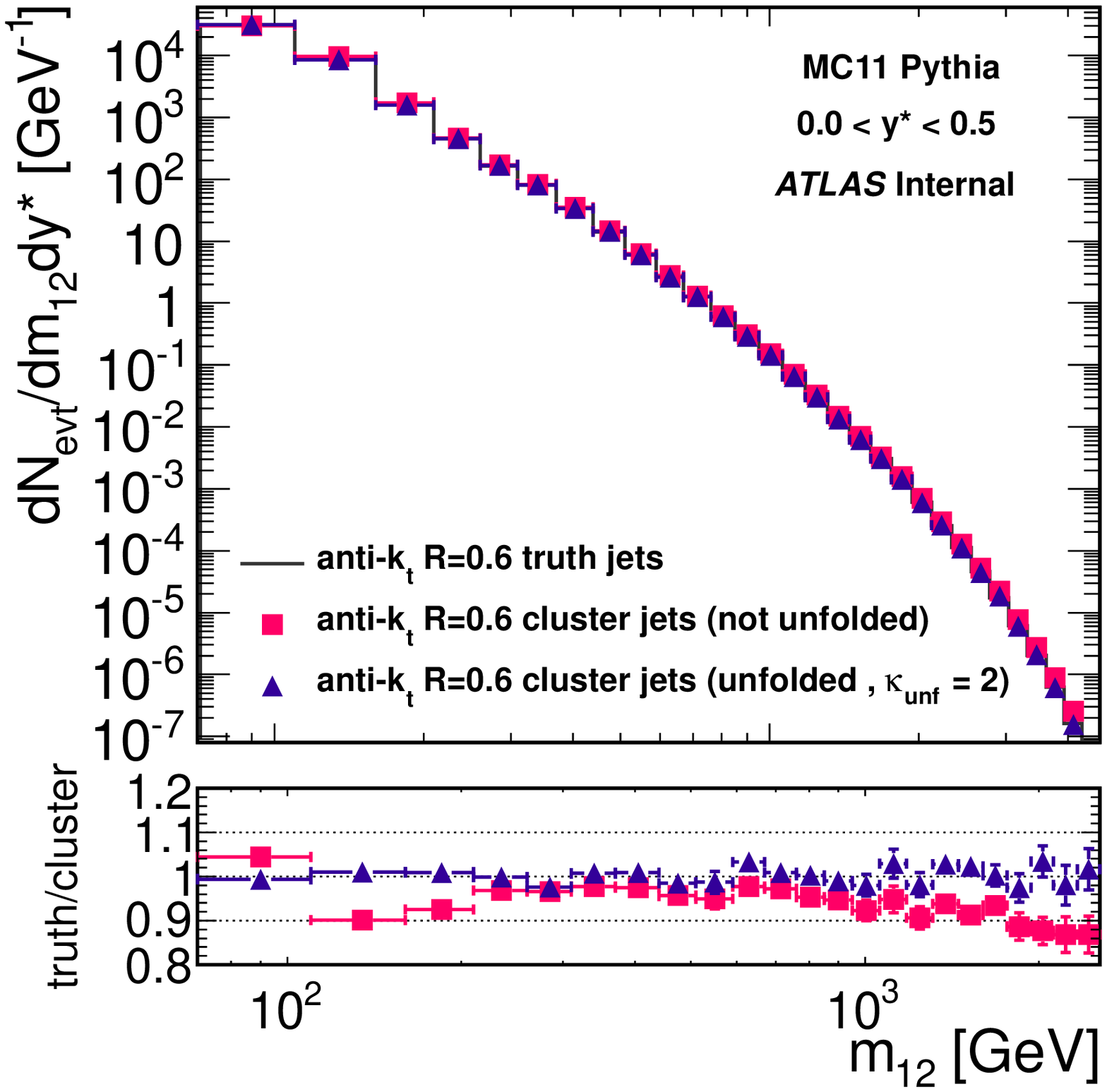}}
  \hspace{3pt}
  \subfloat[]{\label{FIGunfoldingIterations2}\includegraphics[trim=10mm 2mm 0mm 10mm,clip,width=.46\textwidth]{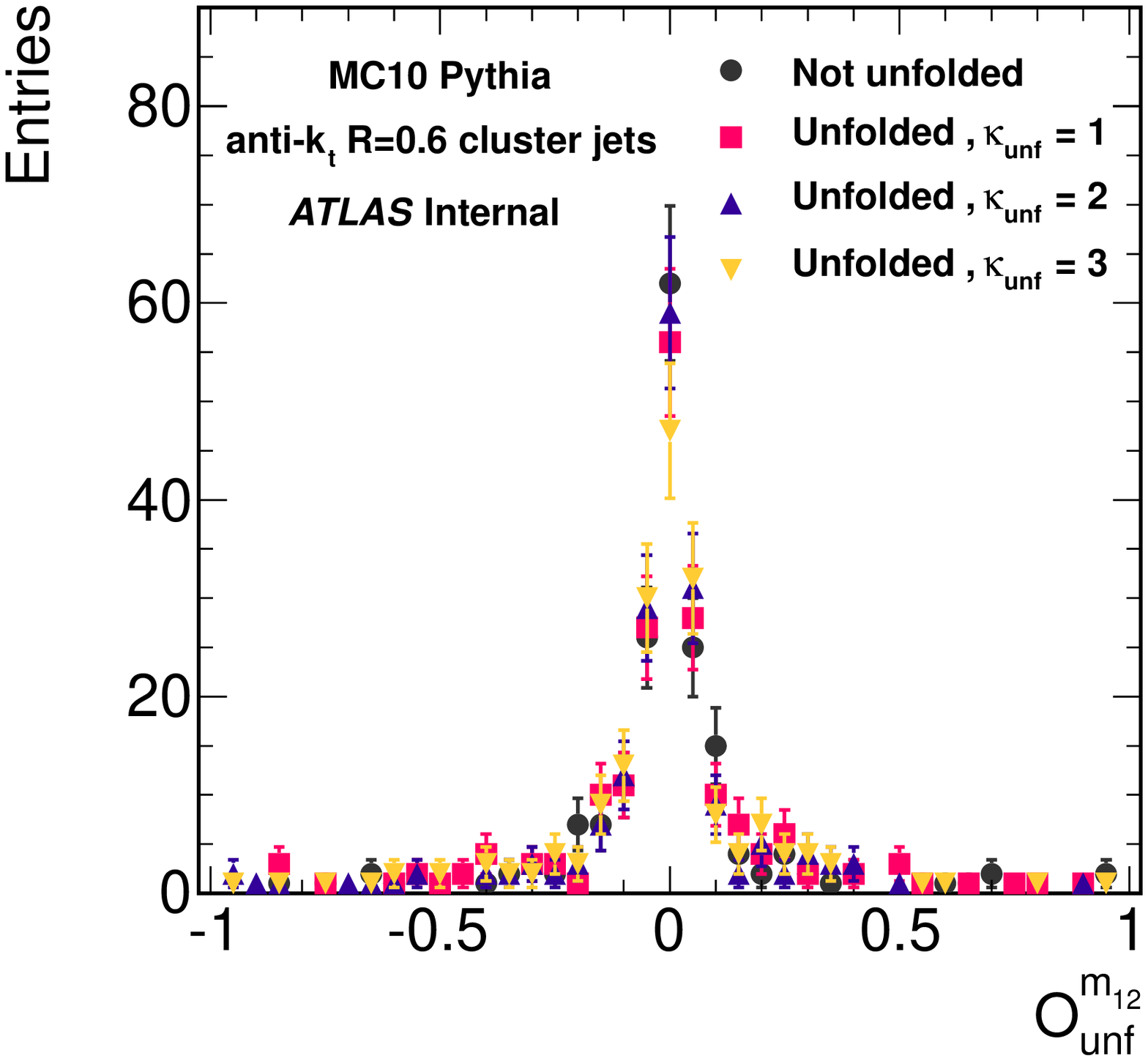}} \\
  \subfloat[]{\label{FIGunfoldingIterations3}\includegraphics[trim=10mm 2mm 0mm 10mm,clip,width=.46\textwidth]{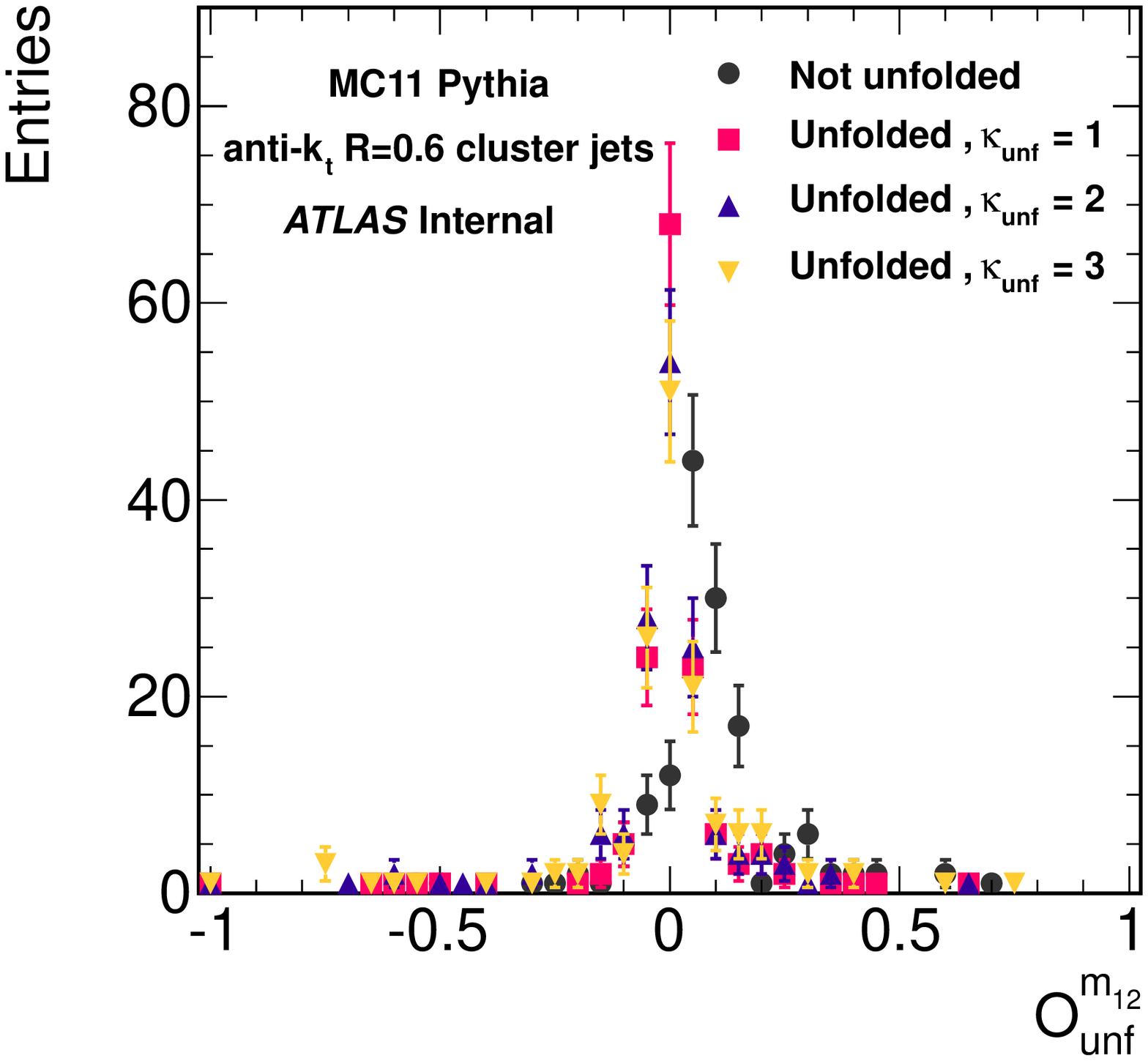}}
  \subfloat[]{\label{FIGunfoldingIterations4}\includegraphics[trim=10mm 2mm 0mm 10mm,clip,width=.46\textwidth]{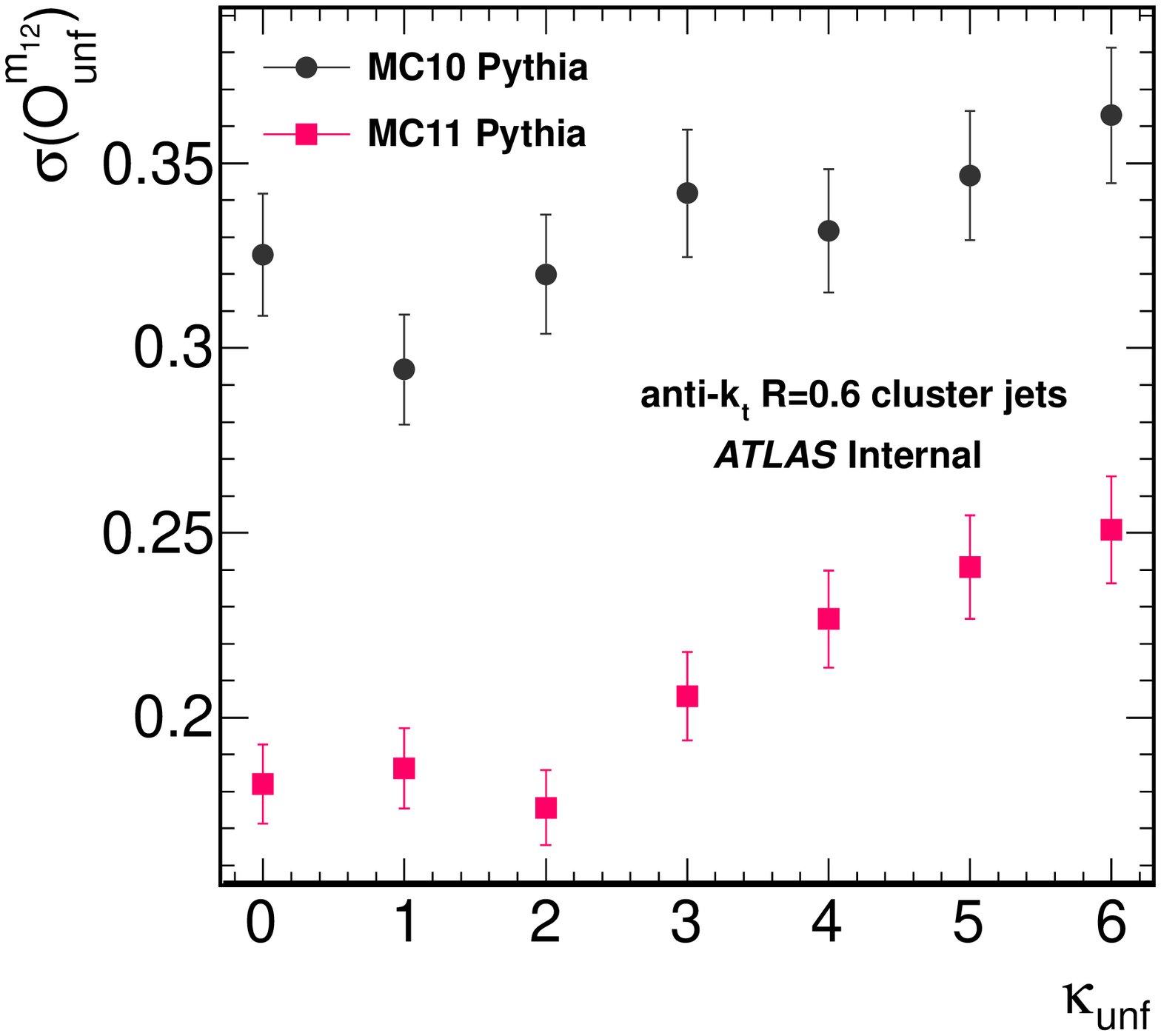}}
  \caption{\label{FIGunfoldingIterations}\Subref{FIGunfoldingIterations1} Invariant mass, $m_{12}$, distribution of the two
    jets with the highest transverse momentum in an event in MC11,
    for \com jet rapidity, ${\ystr < 0.5}$, for truth jets, calorimeter jets without unfolding and calorimeter jets with unfolding
    (using two iterations, $\kappa_{\mrm{unf}}$), as indicated in the figure.
    The bottom panel shows the ratio between the truth- and the calorimeter-jet distributions. \\
    \Subref{FIGunfoldingIterations2}-\Subref{FIGunfoldingIterations3} Distributions of the
    unfolding offset of the invariant mass distribution, $O_{\mrm{unf}}^{m_{12}}$,
    for the case where no unfolding is performed and for unfolded distributions using several values of $\kappa_{\mrm{unf}}$,
    in MC10 and in MC11, as indicated. \\
    \Subref{FIGunfoldingIterations4} Dependence of the width of the unfolding offset, $\sigma(O_{\mrm{unf}}^{m_{12}})$,
    on $\kappa_{\mrm{unf}}$, in MC10 and in MC11,
    where $\kappa_{\mrm{unf}} = 0$ indicates the case for which no unfolding is performed.
  }
\end{center}
\end{figure} 
The dijet mass is constructed from truth jets in a single \ystr bin in MC11, drawn as a function of $m_{12}$ (the histogram is the figure.)
The same variable is then constructed from calorimeter jets and drawn before (red squares) and after (blue triangles) unfolding.
The agreement of the calorimeter-jet distribution with the truth-jet distribution is improved after unfolding.

The unfolding offset is defined as
\begin{equation}
  O_{\mrm{unf}}^{m_{12}} = \frac{\mathfrak{b}_{\mrm{calo}}^{m_{12},\ystr} - \mathfrak{b}_{\mrm{truth}}^{m_{12},\ystr}}{\mathfrak{b}_{\mrm{truth}}^{m_{12},\ystr}} \,,
\label{eqUnfoldingOffsetDef} \end{equation}
where $\mathfrak{b}_{\mrm{calo}}^{m_{12},\ystr}$ ($\mathfrak{b}_{\mrm{truth}}^{m_{12},\ystr}$) stands for the number
of entries in a bin of the double-differential dijet mass distribution, constructed from calorimeter (truth) jets.
The unfolding offset, derived from MC10 and from MC11, is presented in \autorefs{FIGunfoldingIterations2} and~\ref{FIGunfoldingIterations3}.
Several distributions are shown, corresponding to $O_{\mrm{unf}}^{m_{12}}$ without unfolding, and to $O_{\mrm{unf}}^{m_{12}}$
with unfolding and using several values of $\kappa_{\mrm{unf}}$.
In MC11, the mean value of $O_{\mrm{unf}}^{m_{12}}$ is non-zero before unfolding, suggesting that
the unfolding correction is significant. Otherwise, it is difficult to tell by eye which value of $\kappa_{\mrm{unf}}$
gives the best result, either for MC10 or for MC11.
The width of the distributions of the unfolding offset, denoted by $\sigma(O_{\mrm{unf}}^{m_{12}})$,
is therefore used in order to quantify the performance.

The dependence of $\sigma(O_{\mrm{unf}}^{m_{12}})$ on the number of unfolding iterations is
shown in \autoref{FIGunfoldingIterations4}.
In MC10, the value of $\sigma(O_{\mrm{unf}}^{m_{12}})$ decreases between
$\kappa_{\mrm{unf}} = 0$ (no unfolding) and $\kappa_{\mrm{unf}} = 1$. As $\kappa_{\mrm{unf}}$ is increased above unity,
the width of the unfolding offset grows.
For MC11, $\sigma(O_{\mrm{unf}}^{m_{12}})$ of $\kappa_{\mrm{unf}} = 1$ is larger compared to
the case of no unfolding. However, as mentioned \wrt \autoref{FIGunfoldingIterations3}, the mean of
$O_{\mrm{unf}}^{m_{12}}$ is smaller after unfolding, indicating an improved performance nonetheless.
As $\kappa_{\mrm{unf}}$ is increased above unity, the magnitude of $\sigma(O_{\mrm{unf}}^{m_{12}})$ reaches a minimum at
$\kappa_{\mrm{unf}} = 2$.
The values of the width of the offset in MC11 are generally smaller compared to the corresponding values
in MC10.

The minimal values of $\sigma(O_{\mrm{unf}}^{m_{12}})$ represent the best agreement
between the unfolded and corresponding truth-level distributions.
The optimal values for MC10 and for MC11 are therefore chosen as
$\kappa_{\mrm{unf}}^{2010} = 1$ and $\kappa_{\mrm{unf}}^{2011} = 2$ respectively.
%

\section{Systematic uncertainties\label{chapDijetMassSystematicUncertainties}}
%
%
Several sources of systematic uncertainty are considered for the measurement, discussed in the following sections.
The uncertainty in each case is studied by changing the properties of jets, calculating the dijet mass \xsec,
and comparing the result to the nominal measurement. The difference in the distributions is taken as an estimate of the
uncertainty. The variations in the properties of jets are performed either in the data or in the MC. For the latter,
the effect of the variations on the measurement is expressed through the changes induced in the unfolding response matrix,
or in the way that it is used to unfold the data.

\subsection{Systematic uncertainty on the jet energy scale\label{ENUMdijetMassUncertainty1}}
%
%
The \JES is the dominant source of uncertainty in the measurement in most $m_{12},\ystr$ regions.
The following components, in total seven for the~2010 measurement and thirteen for~2011, are considered in the analysis:
\begin{list}{-}{}
\mynobreakpar
\item \headFont{2010 -} single hadron response; cluster thresholds; Perugia 2010 and Alpgen+Herwig+Jimmy; intercalibration; relative non-closure; \itpu;
\item \headFont{2011 -} ENP 1-6; intercalibration; single hadron response; relative non-closure; close-by jet; in-time PU, out-of-time PU and PU \pt.
\end{list}
\mynobreakpar
The different components of the uncertainty are discussed in detail in \autoref{chapSystematicUncertaintiesJets},
with the exception of the \pu uncertainty for 2011, which is discussed in \autoref{chapSystematicUncertMedianPU2011}.

The effects of the different components of the
\JES uncertainties on the dijet mass measurement are estimated by introducing positive or negative 
variations to the energy scale of jets in data.
After the energy of jets is varied, they are re-sorted according to \pt, and the two leading jets
are chosen as the dijet pair. The dijet mass is reconstructed
using the modified jets, and the differential \xsec, $\partial^{2}\sigma/\partial m_{12}\partial\ystr$, is
unfolded. The mass spectra are finally compared to the nominal sample, which consists of an unfolded measurement
of the differential \xsec, performed without any \JES shifts.
The ratio between the shifted and the nominal distributions
is taken as the relative \JES uncertainty on the measurement.
The total uncertainty is the quadratic sum of the uncertainties derived from the different sources.

The effect on the measurement due to
the different sources of \JES uncertainty, derived from positive or negative shifts of the energy
scale of jets, may be inferred from \autoref{FIGdijetMassUncertaintyJES} for
dijet systems with $\ystr<0.5$.
\begin{figure}[htp]
\begin{center}
  \hspace{50pt} \subfloat[\quad\qquad\qquad\qquad\qquad\qquad]{\label{FIGdijetMassUncertaintyJES1}\includegraphics[trim=0mm 0mm 0mm 0mm,clip,width=.857\textwidth]{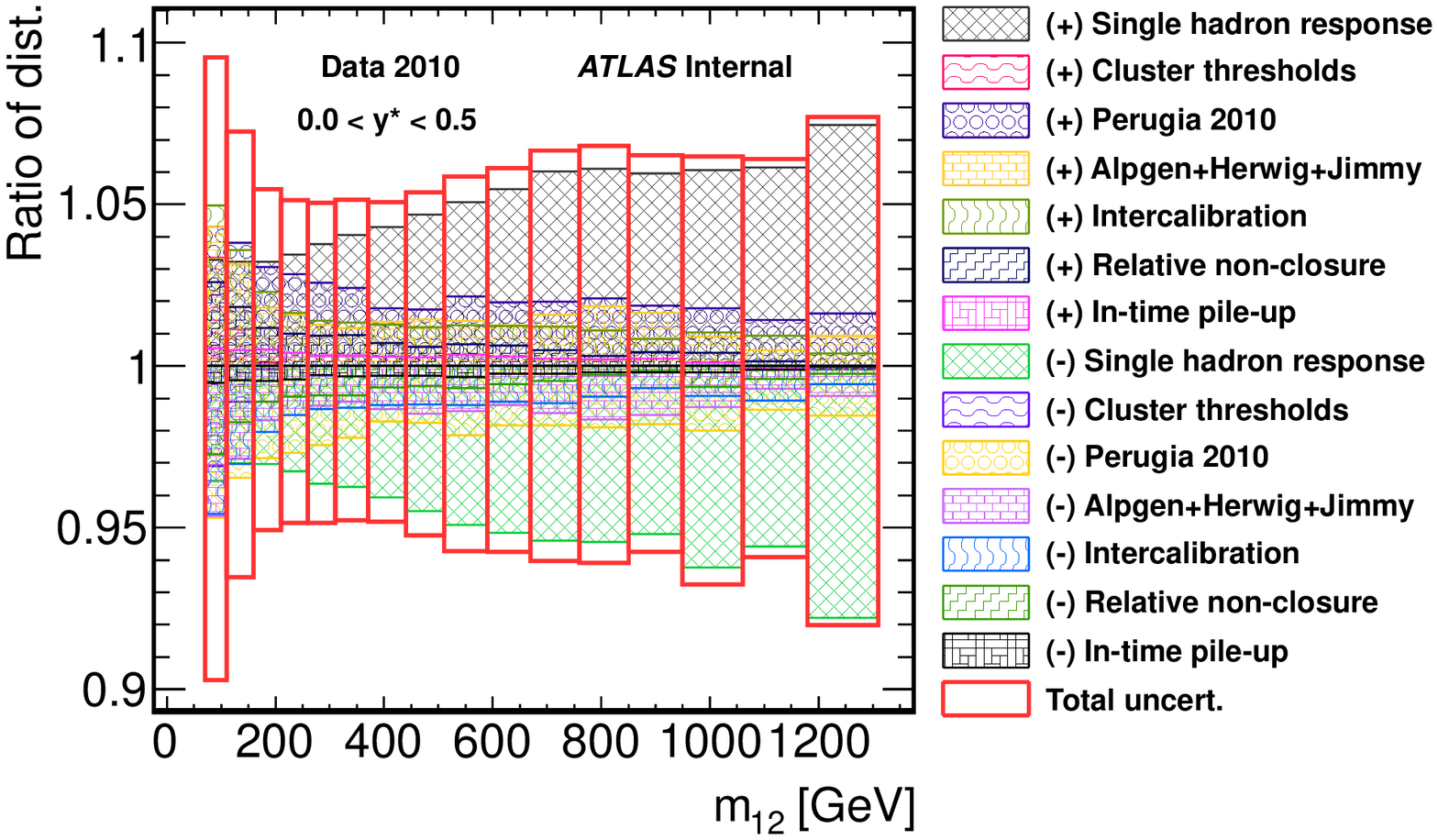}} \\
  \hspace{50pt} \subfloat[\quad\qquad\qquad\qquad\qquad\qquad]{\label{FIGdijetMassUncertaintyJES2}\includegraphics[trim=0mm 0mm 0mm 0mm,clip,width=.857\textwidth]{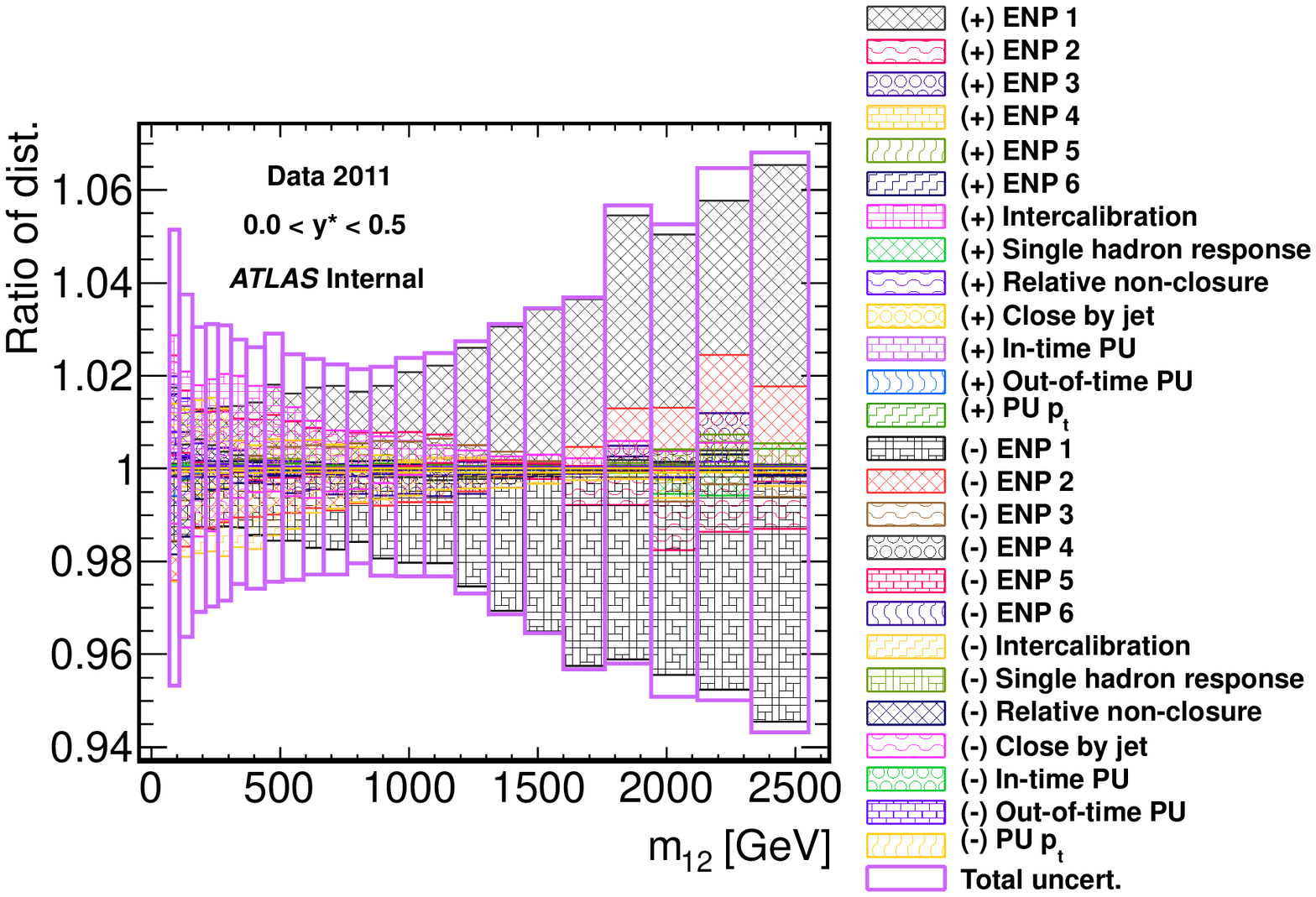}}
  \caption{\label{FIGdijetMassUncertaintyJES}Dependence on the invariant mass, $m_{12}$, of the two jets
  with the highest transverse momentum in an event,
  of the ratio between the $m_{12}$ distributions
  with and without systematic variations, for \com jet rapidity, $\ystr < 0.5$, for the
  2010 \Subref{FIGdijetMassUncertaintyJES1} and for the 2011 \Subref{FIGdijetMassUncertaintyJES2} measurements,
  where positive $(+)$ and negative $(-)$ variations
  are performed according to the different components (described in the text) of the uncertainty on the energy scale of jets (\JES),
  and the total \JES uncertainty is denoted by ``total uncert.\/''
  }
\end{center}
\end{figure} 
The total \JES uncertainty for $\ystr<0.5$ in 2010 is roughly~10\% at low mass values, decreasing to between~5-6\% for $m_{12}>200\GeV$.
In 2011 the total uncertainty for $\ystr<0.5$ is smaller in the mid mass range (${0.5<m_{12}<1.5\TeV}$)
compared to 2010; this is due to improvement in the understanding
of the detector and in the techniques used to estimate the uncertainty. At low masses the uncertainty in 2011 is comparable
to 2010, at around~5\%; this is due to the increase in \pu in 2011, the effect of which is relatively strong at low-\pt. 
For ${m_{12} > 1.5\TeV}$, the uncertainty increases up to~7\%, and is kept fixed at this value for lack
of statistics.

\subsection{Other sources of uncertainty\label{ENUMdijetMassUncertainty2}}
%
%
In addition to the \JES, the following sources of uncertainty on the measurement are studied:
\begin{list}{-}{}
\mynobreakpar
\item \headFont{jet energy and angular resolutions -}   the properties of jets in the MC are varied according to the uncertainty on
                                                        the jet energy and angular resolutions (see \autoref{chapJetPtAngularResolution}),
                                                        leading to a change in the unfolding response matrix;
\item \headFont{physics model used for the unfolding -} the unfolding response matrix is computed using \alpgen, coupled to \herwig and \jimmy (\AHJ)
                                                        (see \autoref{chapEventGenerators}), where the \pt and \Npv distributions in the MC are
                                                        reweighted to match those in the data, following the procedure discussed
                                                        in \autoref{chapJetAreaMethod}, \autoref{FIGNpvDist2011DataMc1};
\item \headFont{shape of the dijet mass spectrum
                in the unfolding -}                     the unfolding response matrix is constructed using a MC
                                                        in which the shape of the dijet invariant mass distribution is distorted;
\item \headFont{the number of unfolding iterations -}   the unfolding procedure is performed using different values of $\kappa_{\mrm{unf}}$.
\end{list}
The uncertainty derived from the various sources is summarized in the next section.

\subsection{Total systematic uncertainty}
%
%
The total relative systematic uncertainty on the dijet mass \xsec measurement for ${\ystr<0.5}$,
as well as the different sources which comprise it, are depicted in
\autoref{FIGdijetMassTotalSystematicUncertainty}. The uncertainties are expressed through changes in the invariant
mass distribution, induced by variation of the different sources of uncertainty.
\begin{figure}[htp]
\begin{center}
  \subfloat[]{\label{FIGdijetMassTotalSystematicUncertainty1}\includegraphics[trim=5mm 14mm 0mm 10mm,clip,width=.52\textwidth]{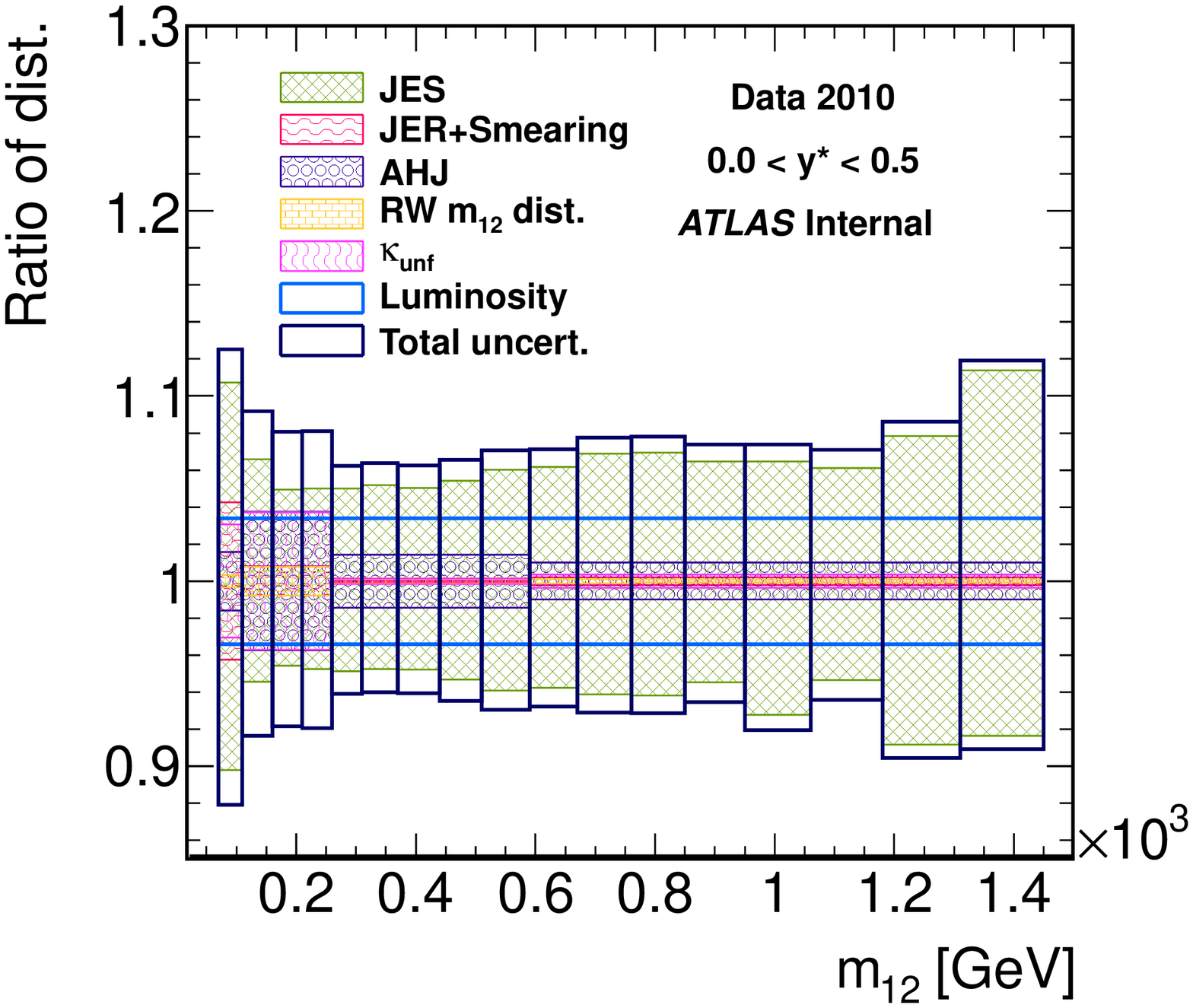}}
  \subfloat[]{\label{FIGdijetMassTotalSystematicUncertainty2}\includegraphics[trim=5mm 14mm 0mm 10mm,clip,width=.52\textwidth]{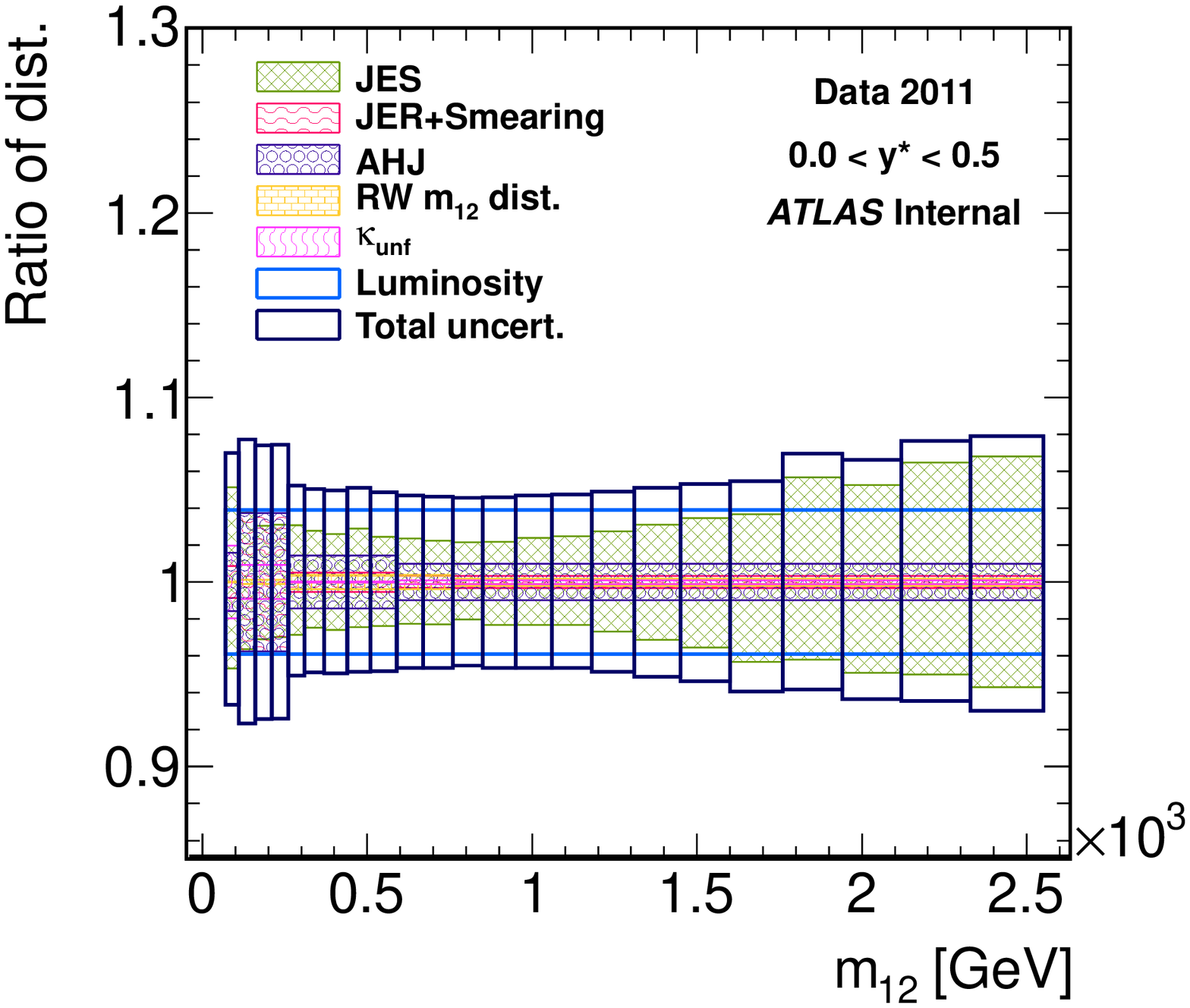}}
  \caption{\label{FIGdijetMassTotalSystematicUncertainty}Dependence on the invariant mass, $m_{12}$, of the two jets with
  the highest transverse momentum in an event,
  of the ratio between the $m_{12}$ distributions
  with and without systematic variations (described in the text), for \com jet rapidity, $\ystr < 0.5$, 
  in the 2010 \Subref{FIGdijetMassTotalSystematicUncertainty1} and in the 2011 \Subref{FIGdijetMassTotalSystematicUncertainty2} measurements,
  including the uncertainty on the jet energy scale (\JES);
  the uncertainty on the jet energy and angular resolutions (\JES + smearing);
  the uncertainty associated with the choice of physics generator (\AHJ); 
  the uncertainty associated with variation in the simulated shape of the $m_{12}$ spectrum (RW $m_{12}$ dist.\/);
  the choice of the number of iterations used in the unfolding procedure ($\kappa_{\mrm{unf}}$);
  the uncertainty on the luminosity (luminosity); and
  the total uncertainty on all the latter (total uncert.\/).
  }
\end{center}
\end{figure} 
The various components of the uncertainty, described in the previous sections, are
the jet energy scale; the jet energy and angular resolution;
the uncertainty associated with the choice of
physics generator; the uncertainty associated with the simulated shape of the dijet mass differential spectrum; and the
choice of the number of iterations used in the unfolding procedure.
In addition, the uncertainty on the luminosity~(3.4\% for 2010 and~3.9\% for 2011) is also taken into account.

The total relative uncertainty for ${\ystr<0.5}$ is between~${6-12\%}$ in 2010 and between~${5-8\%}$ in 2011.
Additional \ystr bins are presented
in \autoref{chapMeasurementOfTheDijetMassApp}, \autorefs{FIGdijetMassTotalSystematicUncertainty2010AppA}~-~\ref{FIGdijetMassTotalSystematicUncertainty2011App}.

\section{Results\label{chapDijetMassResults}}
%
%
The dijet double-differential \xsec, for \AKT ${R=0.6}$ jets,
is measured as a function of the dijet invariant mass for
various values of \ystr.
The 2010 measurements include dijet masses between~$70\GeV$ and~${3.04\TeV}$, with 
values of \ystr up to~3.5, using jets within pseudo-rapidity, \etaLower{4.4}.
The 2011 measurements extend the mass range to~${4.27\TeV}$, but for a smaller \ystr range, ${\ystr<2.5}$,
and for jets with \etaLower{2.8}. In the 2010 measurements, the jets are required to have
\ptHigher{30} for the leading jet, and \ptHigher{20} for the sub-leading jet. In 2011, these limits are
higher by~10\GeV.

The measured \xsecs have been corrected for all detector effects.
The systematic uncertainties include variations of the jet energy scale and the
jet energy and angular resolutions, several checks of the unfolding procedure, and the uncertainty on
the luminosity.
The measurements are compared to fixed-order NLO pQCD predictions from \nlojet, using the CT10 PDF set, and
corrected for non-perturbative effects. The theoretical uncertainties take
into account scale variations, change in the parton distribution functions,
and stability of the non-perturbative corrections.

The \xsec results are presented in \autorefs{FIGdijetMassResult2010} and~\ref{FIGdijetMassResult2011}.
\begin{figure}[htp]
\begin{center}
 \vspace{-20pt}
 \subfloat[\qquad\qquad\qquad\qquad\qquad\qquad\qquad\qquad]{\label{FIGdijetMassResult2010a}\includegraphics[trim=15mm 1mm 10mm 10mm,clip,width=.99\textwidth]{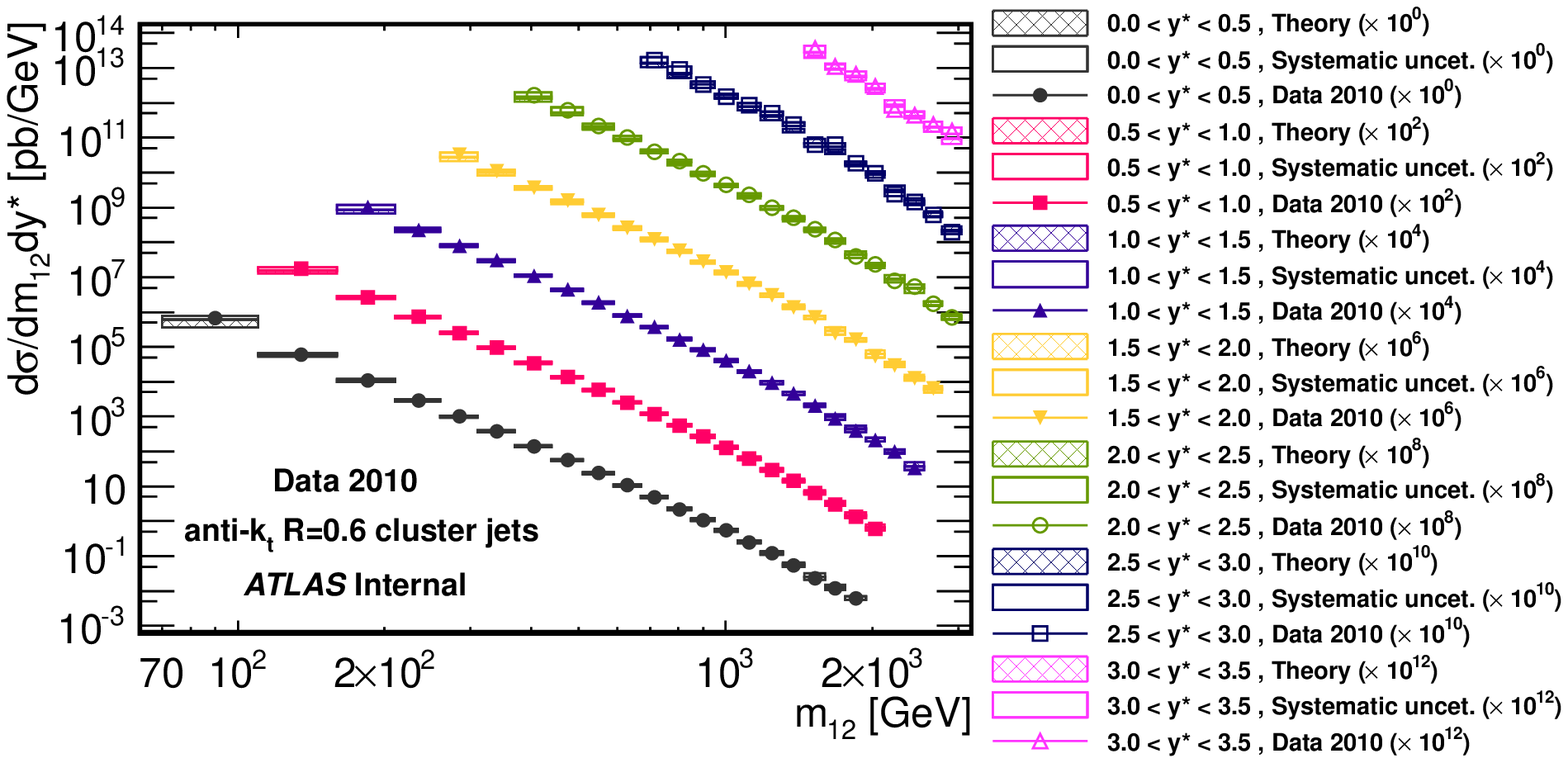}} \\
 \vspace{-7pt}
  \subfloat[]{\label{FIGdijetMassResult2010b}\includegraphics[trim=15mm 10mm 5mm 15mm,clip,width=.99\textwidth]{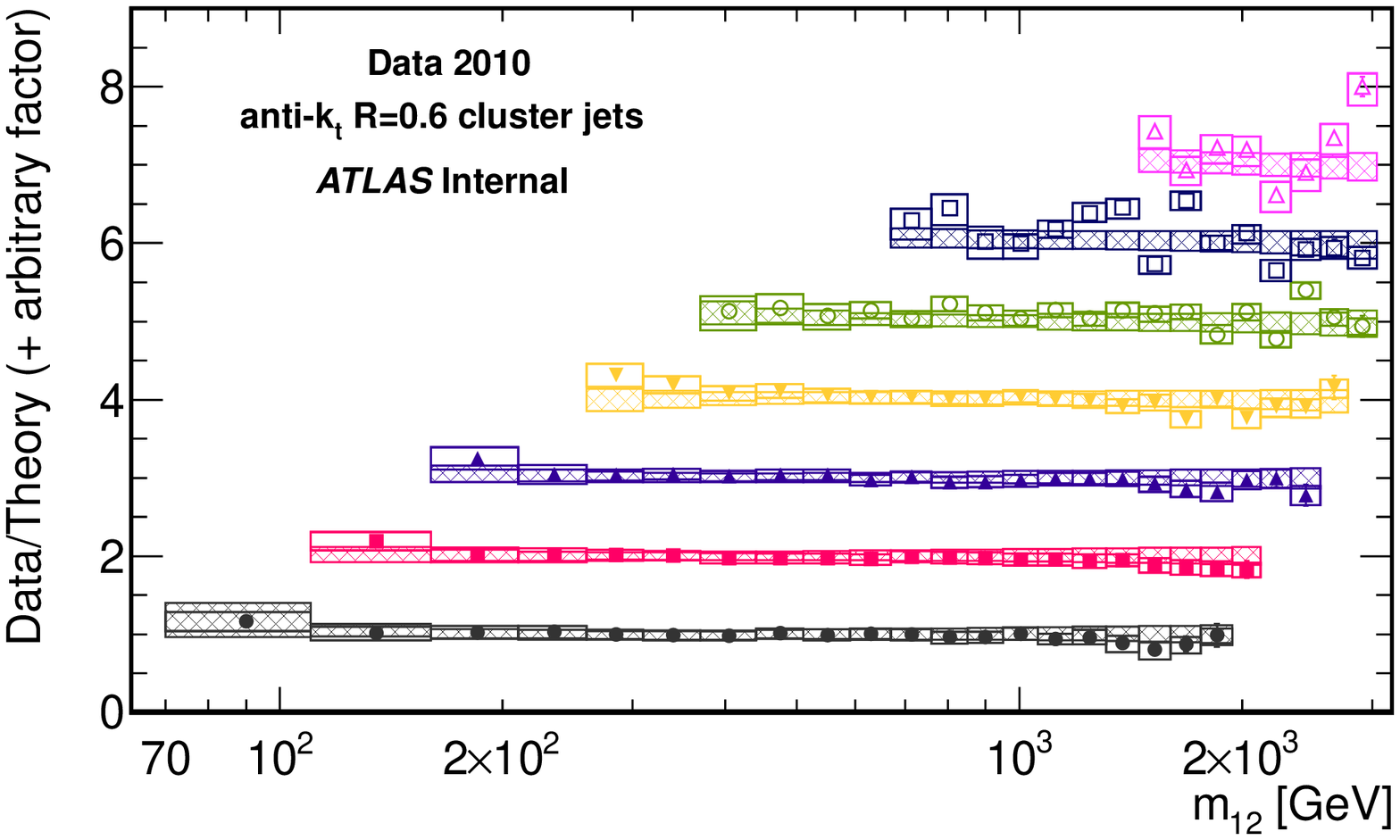}}
  \caption{\label{FIGdijetMassResult2010}Double-differential dijet \xsec as a function of the invariant mass, $m_{12}$,
  binned in \com jet rapidity, \ystr, in the 2010 data, compared to \nlojet pQCD calculations using the CT10 PDF set,
  which are corrected for non-perturbative effects.
  The uncertainty on the theoretical calculations, the systematic uncertainty on the data and the statistical
  uncertainty on the data (generally too small to discern) are shown as well. \\
  \Subref{FIGdijetMassResult2010a} Invariant mass distributions, multiplied by the factors specified in the legend. \\
  \Subref{FIGdijetMassResult2010b} Ratio of the data to the theoretical predictions, where the results for different \ystr bins
  are separated by arbitrary constants for convenience.
  }
\end{center}
\end{figure} 
\begin{figure}[htp]
\begin{center}
 \vspace{-20pt}
 \subfloat[\qquad\qquad\qquad\qquad\qquad\qquad\qquad\qquad]{\label{FIGdijetMassResult2011a}\includegraphics[trim=15mm 1mm 10mm 10mm,clip,width=.99\textwidth]{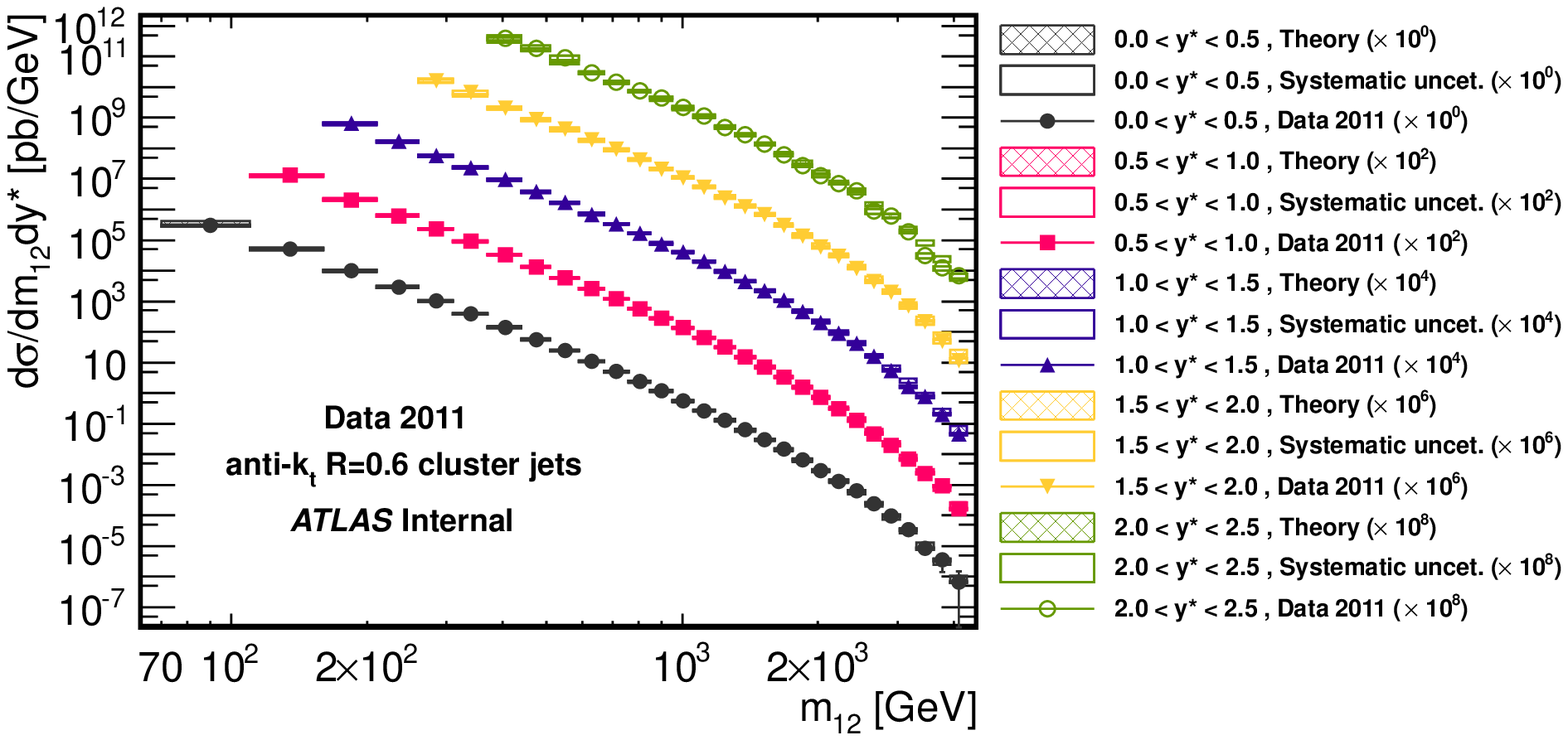}} \\
 \vspace{-7pt}
  \subfloat[]{\label{FIGdijetMassResult2011b}\includegraphics[trim=15mm 10mm 5mm 15mm,clip,width=.99\textwidth]{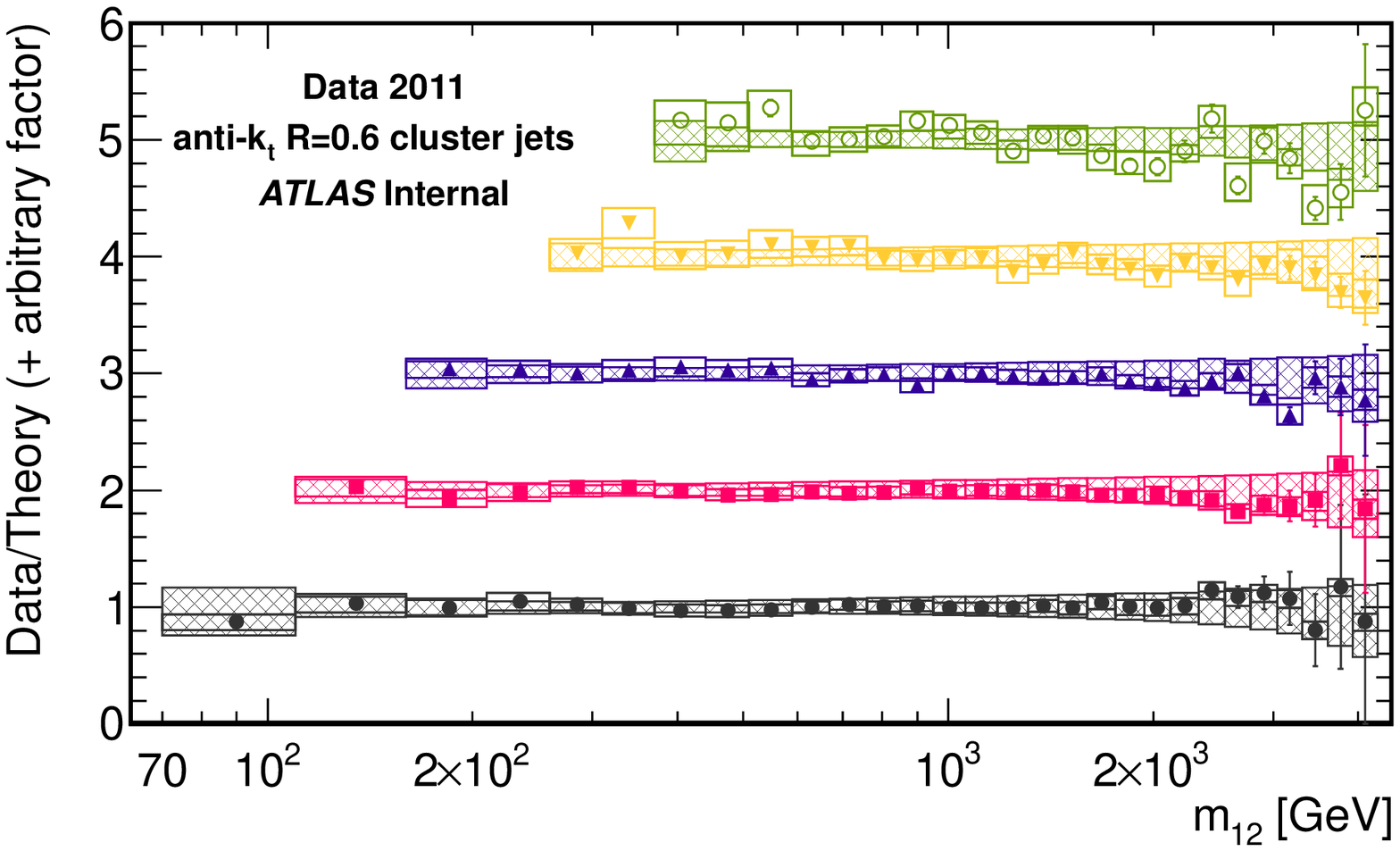}}
  \caption{\label{FIGdijetMassResult2011}Double-differential dijet \xsec as a function of the invariant mass, $m_{12}$,
  binned in \com jet rapidity, \ystr, in the 2011 data, compared to \nlojet pQCD calculations using the CT10 PDF set,
  which are corrected for non-perturbative effects.
  The uncertainty on the theoretical calculations, the systematic uncertainty on the data and the statistical
  uncertainty on the data (generally too small to discern) are shown as well. \\
  \Subref{FIGdijetMassResult2010a} Invariant mass distributions, multiplied by the factors specified in the legend. \\
  \Subref{FIGdijetMassResult2010b} Ratio of the data to the theoretical predictions, where the results for different \ystr bins
  are separated by arbitrary constants for convenience.
  }
\end{center}
\end{figure} 
Within the uncertainties (statistical, systematic and theoretical),
the fixed-order NLO pQCD predictions give a good description of the measurements over~12 orders of magnitude.
The good agreement between the measurements performed in 2010 and in 2011, under very different beam conditions,
indicate that \itpu and \otpu are experimentally well under control.
                }{}
\ifthenelse{\boolean{do:fourJetDPS}}        { 
\chapter{Hard double parton scattering in four-jet events\label{chapDoublePartonScattering}}
%
The dominant mechanism for the production of events containing four high-\pt jets
at the LHC is double gluon bremsstrahlung. This process is described quantitatively
(up to LO) by pQCD. A few Feynman diagrams depicting such events are shown in \autoref{FIGfourJetGlounBremsstrahlung}.
\begin{figure}[htp]
\vspace{10pt}
\begin{center}
\begin{overpic}[trim=0mm 90mm 0mm 0mm,clip,width=.47\textwidth]{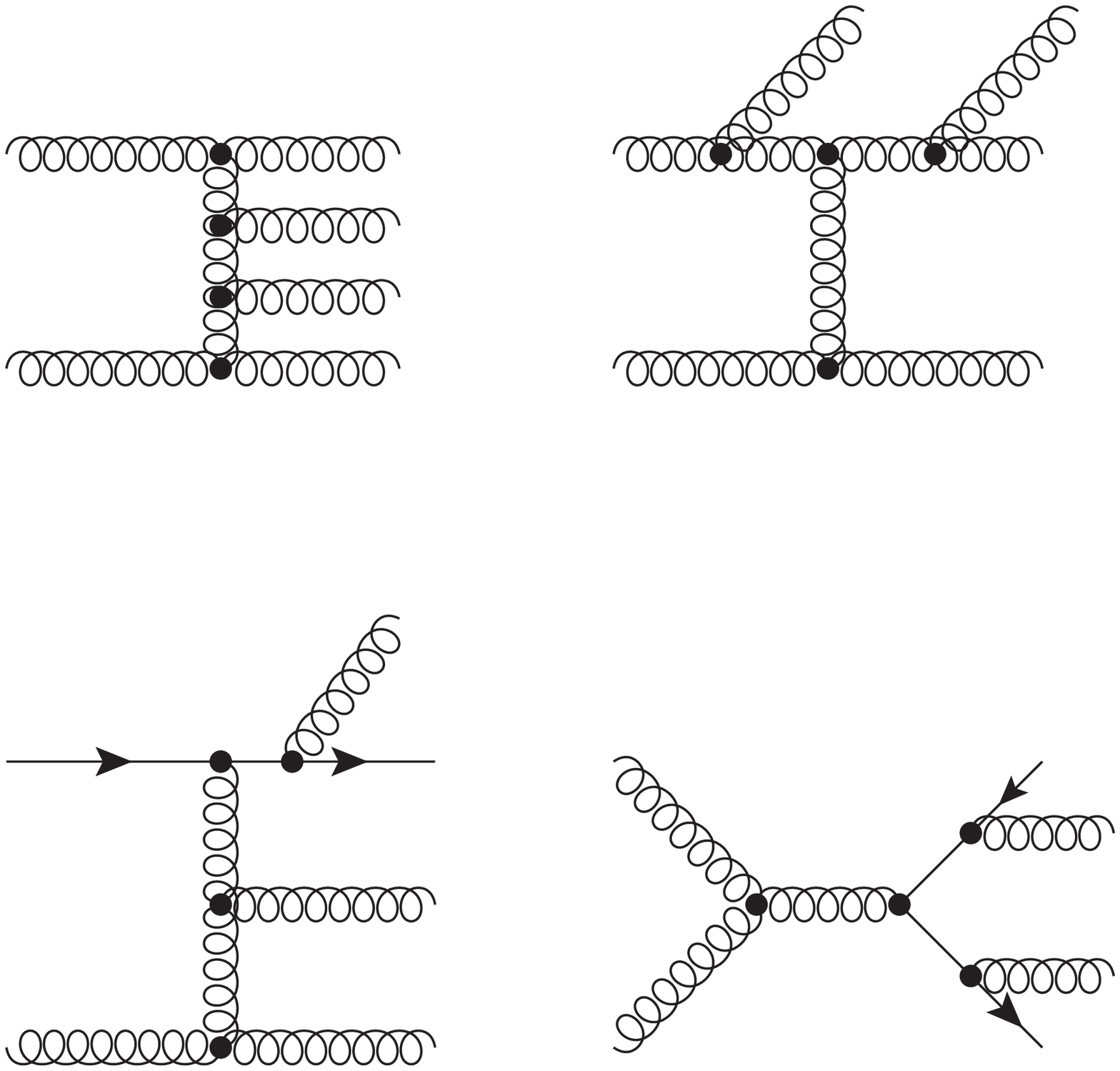}
\put(6,0){$gg \longrightarrow gggg$}
\put(60,0){$gg \longrightarrow gggg$}
\end{overpic}
\;\;
\begin{overpic}[trim=0mm -12mm 0mm 100mm,clip,width=.47\textwidth]{figures/fourJetDPS/FeynmanSPS} 
\put(7,0){$gq \longrightarrow gqgg$}
\put(60,0){$gg \longrightarrow ggq\bar{q}$}
\end{overpic}
\caption{\label{FIGfourJetGlounBremsstrahlung}Four of the Feynman diagrams which contribute to
the leading-order matrix element expression for the double gluon bremsstrahlung process.
}
\end{center}
\end{figure}
This topology is characterized by two partons in the initial state and four in the final state, denoted as \twofour.
Alternatively, four-jet final states can also be produced through the double parton scattering (DPS) mechanism,
which is discussed in \autoref{chapMultiPartonInteractionsInGenerators}.
In the case of DPS, two pairs of dijet events occur simultaneously in a single $pp$ collision.
The process of DPS is denoted in the following as \dtwot.

Recently it has been advocated that in addition to the two processes discussed above, a third process should 
contribute, in which a parton from one of the protons, splits perturbatively into two partons,
which then interact with two partons from the other proton.
This configuration, denoted as \threefour, leads to slightly different topologies of the
four jets than expected in the \dtwot and \twofour processes, although its contribution may
be substantial~\cite{Blok:2012mw}.
In this analysis, no attempt is made to differentiate between the \dtwot and the \threefour topologies,
as the differences may be too small for the typical jet energy resolution, for jets with $\pt\sim20\GeV$.

On an event-by-event basis, it is impossible to determine whether a \twofour or a \dtwot interaction
describes an event. However, several kinematic features distinguish the two processes on average.
The natural approach would have been to devise a strategy of extracting the DPS signal, based on a
sample of MC generated events, which contained multi-parton interactions and full detector simulation.
Unfortunately, a MC simulating four-jet final states through matrix-elements, such as \alpgen, coupled to
\herwig and \jimmy (\AHJ), or \sherpa, turned out not to be adequate. The available \AHJ sample, in which
the user can identify jets which originate from extra parton scatterings in the event-record,
did not have enough hard double-scattering events passing the experimental selection cuts.
On the other hand, in \sherpa, the information on the extra scatters in the event is not available to the user.
A different strategy was therefore devised, that of using a neural network.

In the following analysis, the rate of DPS in four jet events in the 2010 \atlas data 
is extracted using the neural network approach. The respective value of \sigeff
(defined in \autoref{chapMultiPartonInteractionsInGenerators}, \autoref{EQsigmaEff}) is subsequently measured.

\section{Strategy of the analysis\label{chap}}
%
\minisec{Phase-space and data sample of the measurement}
%
%
The rate of hard double parton scattering is measured in events which have
exactly four jets in the final state.
Events are selected using the two-trigger selection scheme described in \autoref{chapTwoTriggerLumiCalcScheme}.
The measurement also uses exclusive dijet events as part of the calculation of \sigeff. Dijet
events are defined in a similar manner as four-jet events, in that they have exactly
two jets in the final state.
Jets are reconstructed using the \AKT algorithm with size parameter, $R = 0.6$.
They are defined as having transverse momentum, \ptHigher{20}, and pseudo-rapidity, \etaLower{4.4}.
Additional requirements on jet-\pt may be imposed for four-jet or dijet event selection, as discussed in the following.
Jets in dijet and in four-jet events are ordered in descending order of their transverse-momenta.
That is, denoting by $p_{t,i}$ the transverse momentum
of the $i^{\mrm{th}}$ jet in an event, the \pt of jets in \eg a four-jet event, fulfil the condition,
\begin{equation*}
{p_{t,1}>p_{t,2}>p_{t,3}>p_{t,4}} \;.
\end{equation*}

\Pu constitutes a major source of background to the measurement. It introduces fake
four-jet or dijet events and removes legitimate events. 
There are two main sources of miss-counting of four-jet events (with similar considerations for dijet events).
The first are three-jet events in which the \pt of a fourth jet
should be smaller than~20\GeV, but is increased above threshold due to \pu.
The second are five-jet events, in which one of the jets is split due to \pu,
and the two split jets are both below threshold. 
It is also possible for additional hard jets,
which originate from a different $pp$ collision, to be counted as part of the main interaction.
Thus an event may move in to or out of the accepted phase-space.
In order to avoid such uncertainty, only singe-vertex events taken during 2010 are
used in the analysis.

In addition to the requirements specified above, the leading jet in a four-jet event
is required to have \ptHigher{42.5}. 
Restricting the \pt of the leading jet to be higher than the minimal threshold
for reconstruction, creates a scale separation between the jets. This helps to combine the
four jets in DPS events into two pairs, each originating from a different scattering, as discussed in the
next sections. 
The restriction on the leading jet is also motivated by the trigger strategy of the analysis.
It amounts to the requirement that at least one jet in the event fires
a jet trigger (see \autoref{TBLcentralTrigNames2010}). Four-jet events in which the two leading jets are
associated with the \ttt{MBTS} trigger are not included. Such events constitute a very small
fraction of the overall statistics, but dominate completely the phase-space region in which
both leading jets have ${20<\pt<42.5\GeV}$. Since they are associated with very high luminosity weights, they tend
to introduce large statistical fluctuations into the measurement.

To summarize, the measurement is performed for exclusive four-jet events which have
\begin{equation}
  \renewcommand{\arraystretch}{1.75}
  \begin{array}{ccccc}
  \Npv = 1 \;,& N_{\rm jet} = 4 \;,& p_{t,1} > 42.5\GeV \;,& p_{t,2-4} > 20\GeV  \;,& |\eta_{1-4}| < 4.4 \;,
  \end{array}
\label{EQphaseSpaceOfMeasurementDPS1}
\end{equation}
where \Npv is the number of reconstructed vertices in the event,
$N_{\rm jet}$ is the number of reconstructed \AKT, $R = 0.6$ jets;
and $\eta_{1-4}$ stands for the pseudo-rapidity of the four leading jets.

The added restriction on jet-\pt may also be imposed on dijet events.
Two classes of exclusive dijet events are defined, denoted by $a$ and $b$. These are,
\begin{equation}
  \renewcommand{\arraystretch}{1.75}
  \begin{array}{cccccc}
  (a) & \Npv = 1 \;,& N_{\rm jet} = 2 \;,& \multicolumn{2}{c}{p_{t,1-2} > 20\GeV \;, } &  |\eta_{1,2}| < 4.4 \;, \\
  (b) & \Npv = 1 \;,& N_{\rm jet} = 2 \;,& p_{t,1} > 42.5\GeV \;,& p_{t,2} > 20\GeV  \;,& |\eta_{1,2}| < 4.4 \;.
  \end{array}
\label{EQphaseSpaceOfMeasurementDPS2}
\end{equation}
Here $p_{t,2}$ stands for the transverse momentum of the sub-leading jet in the event,
and $\eta_{1,2}$  stands for the pseudo-rapidity  of the two jets.

For completeness, the \xsec for the four-jet final state is denoted by $\sigma_{4j}^{b}$, and those
of the two classes of dijet final states are denoted by $\sigma_{2j}^{a}$ and $\sigma_{2j}^{b}$.
The superscripts $a$ and $b$ indicate the phase-space restrictions on the respective \xsecs.
The former ($a$) represents a \xsec which includes the restriction that all jets have \ptHigher{20}.
The latter ($b$) stands for \xsecs which include events where the leading jet has \ptHigher{42.5} and
all other jets have \ptHigher{20}.

\minisec{Definition of the observable}
%
%
The cornerstone of the analysis is the extraction of \fDPS, the fraction of four-jet events in which
the jets originate from a hard DPS, and the measurement of the ratio of dijet and four-jet \xsecs.
These are subsequently used to determine the value
of \sigeff (see \autoref{chapMultiPartonInteractionsInGenerators}) in the four-jet final state.

The simplified form,
\begin{equation}
\sigeff = 
\frac{m}{2}\frac{ \sigma_{2j}^{a} \sigma_{2j}^{b} }{ \sigma_{\rm DPS}^{b} } = \frac{m}{2\fDPS}\frac{ \sigma_{2j}^{a} \sigma_{2j}^{b} }{ \sigma_{4j}^{b} } \;,
\label{EQsimpleSigmaDPSFourJet1}
\end{equation}
is used, where the double parton scattering \xsec,
\begin{equation*}
\sigma_{\rm DPS}^{b} = \fDPS \cdot \sigma_{4j}^{b} \;, 
\end{equation*}
is extracted using the measured \xsec for four-jet events, $\sigma_{4j}^{b}$. The latter
includes all events with four-jet final states, that is both \twofour and \dtwot topologies.
Four-jet final states of type $b$ in DPS originate from two interactions, of which only one is restricted
to have at least one jet with \ptHigher{42.5}. The expression for \sigeff therefore includes one
instance of the dijet \xsec of type $a$ and one of type $b$.

In principle, the two \xsecs, $\sigma_{2j}^{a}$ and~$\sigma_{2j}^{b}$, overlap in the region where the leading
jet has \ptHigher{42.5}. However, due to the steep fall of the dijet \xsec with \pt, the overlapping
events constitute roughly~3.5\% of $\sigma_{2j}^{a}$. The two dijet final states are, therefore, considered
to a good approximation as different, such that the symmetry factor in \autoref{EQsimpleSigmaDPSFourJet1} is
\begin{equation*}
m=2\;.
\end{equation*}
The small overlap also helps to correctly combine the four jets in DPS events into two pairs. The probability for a jet from a
scattering of type $a$ to have ${\pt\sim40\GeV}$ (or higher) is small. The two leading jets are therefore likely to have
originated from one scattering and the next two from the other scattering. This distinction helps to construct kinematic
variables which differentiate between DPS and non-DPS events, as discussed in \autoref{chapExtractionOfFdpsWithNN}.

It is important to emphasise that the present determination of \sigeff
is based on exclusive dijet and four-jet \xsecs, while \autoref{EQsimpleSigmaMPI} is defined for inclusive \xsecs.
This choice greatly facilitates the separation between DPS and direct four-jet production, though
it may not be theoretically sound. However, due to the~20\GeV cut on jet-\pt,
the difference between the inclusive and the exclusive \xsecs was found to have a negligible effect on the analysis.

In general, the \xsecs for dijet and four-jet production take the form
\begin{equation}
\sigma_{nj} = \frac{N_{nj}}{ \mathcal{A}_{nj} \epsilon_{nj} \mathcal{L}_{nj} } 
            = \frac{\hat{\sigma}_{nj}}{ \mathcal{A}_{nj}  }\;,
\label{EQmeasureXS}
\end{equation}
where the subscript, $nj$, denotes either dijet ($2j$) or four-jet ($4j$) topologies. For each $n$ channel,
$N_{nj}$ is the number of measured events, $\mathcal{A}_{nj}$ is the geometrical acceptance of the
measurement, $\epsilon_{nj}$ is the efficiency for reconstructing the event\footnote{ In
the 2010 data there is need to differentiate between the acceptance and the efficiency, as the
non-geometrical inefficiencies of the detector (\eg faulty electronic channels) were not properly simulated as part of the
MC10 campaign.}, and $\mathcal{L}_{nj}$ is the luminosity.
Finally, $\hat{\sigma}_{nj} = N_{nj}/(\epsilon_{nj} \mathcal{L}_{nj})$ is the \textit{observed \xsec} at the detector-level.
Using this identity, \autoref{EQsimpleSigmaDPSFourJet1} may be written as
\begin{equation}
\sigeff = 
          \frac{1}{\fDPS}
          \frac{ \hat{\sigma}_{2j}^{a} \hat{\sigma}_{2j}^{b} }{ \hat{\sigma}_{4j}^{b} }
          \frac{ \mathcal{A}_{4j}^{b}  }{ \mathcal{A}_{2j}^{a} \mathcal{A}_{2j}^{b} }
          \;.
\label{EQsimpleSigmaDPSFourJet2}
\end{equation}
%
%


The expression for \sigeff is simplified by defining the \textit{acceptance ratio},
\begin{equation}
\alpha_{2j}^{4j} = \frac{ \mathcal{A}_{4j}^{b} }{ \mathcal{A}_{2j}^{a} \mathcal{A}_{2j}^{b} } \;.
\label{EQsimpleSigmaDPSFourJet3}
\end{equation}
This factor is used to absorb the difference in the acceptance of the final states into a single quantity.
Naively, one may expect that the acceptance factor for each jet in an event should be the same, which would result
in ${\mathcal{A}_{4j}^{b} = \mathcal{A}_{2j}^{a} \mathcal{A}_{2j}^{b}}$ and ${\alpha_{2j}^{4j}=1}$.
In reality, the cancellation of efficiencies is not complete, as one has to take into account the effects
of close-by-jets.
Any departure of $\alpha_{2j}^{4j}$ from unity, is mainly due to configurations where a pair of jets overlap.
Such topologies reduce the classification efficiency of an event as a $2j$- or a $4j$-event, and affect
each channel in different proportions.

Following these considerations, \autoref{EQsimpleSigmaDPSFourJet2} reduces to
\begin{equation}
\sigeff = \frac{ \mathcal{S}_{4j}^{2j} \; \alpha_{2j}^{4j} }{\fDPS} \;,
\label{EQsimpleSigmaDPSFourJet4}
\end{equation}
with
%
\begin{equation}
\mathcal{S}_{4j}^{2j}
        = \frac{ \hat{\sigma}_{2j}^{a} \hat{\sigma}_{2j}^{b} }{ \hat{\sigma}_{4j}^{b} } \;,
\label{EQsimpleSigmaDPSFourJet5}
\end{equation}
where one may identify that ${\mathcal{S}_{4j}^{2j} \; \alpha_{2j}^{4j} \;=\; \sigma_{2j}^{a} \; \sigma_{2j}^{b} \; / \; \sigma_{4j}^{b}}$.

To summarize, three elements, $\mathcal{S}_{4j}^{2j}$, $\alpha_{2j}^{4j}$ and \fDPS, 
enter the determination of \sigeff. These are discussed in turn in the following sections.

\minisec{Uncertainty on the measurement}
%
%
The systematic uncertainty on the measurement originates from uncertainties on the measured properties 
of jets in data, and on the simulated properties in the MC. The uncertainty in the data is mainly due to
that on the jet energy scale (\JES). In the MC, the uncertainty on the energy and angular resolutions and
the dependence on the shape of the input distributions, all contribute to the total uncertainty.

The fraction of DPS events is extracted from the data using a neural network, which is based on the MC. The statistical
as well as the \JES uncertainties on the data, therefore, affect
the fit-result for \fDPS. Similarly, the uncertainties on $\alpha_{2j}^{4j}$ and on \fDPS
are correlated as well, since both elements depend on the simulated properties of jets in the MC.
The different correlations between the elements of the
measurement are taken into account when propagating the combined uncertainty to \sigeff,
as discussed in \autoref{chapDpsUncertainty}.

\section{Measurement of the ratio of dijet and four-jet \xsecs\label{chap}}
%

\minisec{The ratio of observed \xsecs}
%
The ratio, $\mathcal{S}_{4j}^{2j}$, 
is estimated in data by counting the number of dijet and four-jet events,
as defined in \autorefs{EQphaseSpaceOfMeasurementDPS1} and~\ref{EQphaseSpaceOfMeasurementDPS2}.
Events are weighted by the appropriate luminosity factors and are corrected for trigger efficiency.
In addition, jet reconstruction efficiency corrections are applied for each jet, depending on its transverse momentum
and pseudo-rapidity. That is,
a dijet (four-jet) event is given a weight according to the efficiency for
reconstructing the two (four) jets.

The numbers of events used for the calculations of the \xsecs are
\begin{equation}
  \renewcommand{\arraystretch}{1.75}
  \begin{array}{cccc}
   N_{2j}^a = 1087776 \;, & N_{2j}^b = 871830  & \mrm{and} & N_{4j}^b = 307236 \;.
  \end{array}
\label{EQphaseSpaceOfMeasurementDPS3}
\end{equation}
The combination of these numbers with the corresponding luminosity and efficiency factors, leads to determination
of the observed \xsecs,
\begin{equation}
        \renewcommand{\arraystretch}{1.75}
        \begin{array}{llll}
          \hat{\sigma}_{2j}^{a} & = & 2.103\cdot10^{8}\pm5.3\cdot10^{5}~\mrm{(stat.)}~\mrm{pb} &, \\
          \hat{\sigma}_{2j}^{b} & = & 1.562\cdot10^{7}\pm2.7\cdot10^{4}~\mrm{(stat.)}~\mrm{pb} &, \\
          \hat{\sigma}_{4j}^{b} & = & 2.221\cdot10^{6}\pm9.7\cdot10^{3}~\mrm{(stat.)}~\mrm{pb} &. \\
        \end{array}
\label{eqObservedDpsXS} \end{equation}
Finally, the respective value of the ratio of observed \xsecs comes out to be
\begin{equation}
\mathcal{S}_{4j}^{2j} = 1.48 \pm 0.04~\mrm{(stat.)}~\mrm{mb} \;.
\label{EQsimpleSigmaDPSFourJet7}
\end{equation}
The quoted errors on the observed \xsecs and on $\mathcal{S}_{4j}^{2j}$
are derived from the statistical uncertainties on the respective data samples.
These are propagated to \sigeff, along with additional systematic uncertainties, as discussed in \autoref{chapDpsUncertainty}.

\minisec{The acceptance ratio}
%
The second component of the measurement is the acceptance ratio, $\alpha_{2j}^{4j}$,
which is determined in MC.
In order to calculate $\alpha_{2j}^{4j}$, the acceptance for each class of events
($\mathcal{A}_{2j}^{a}$, $\mathcal{A}_{2j}^{b}$ or $\mathcal{A}_{4j}^{b}$) is individually estimated.
Counting the number of events which pass the selection cuts for class $n$ using truth (particle) jets, $N_{nj}^{a,b\;\mrm{truth}}$,
or using detector jets, $N_{nj}^{a,b\;\mrm{calo}}$, the acceptance is defined as
%
\begin{equation}
\mathcal{A}_{nj}^{a,b} = \frac{N_{nj}^{a,b\;\mrm{calo}}}{N_{nj}^{a,b\;\mrm{truth}}+N_{nj}^{a,b\;\mrm{calo}}} \;,
\label{EQacceptanceDefinition}
\end{equation}
where the same restrictions on the phase-space of jets
(\autorefs{EQphaseSpaceOfMeasurementDPS1}~and~\ref{EQphaseSpaceOfMeasurementDPS2})
are used for truth jets, as for detector jets.

The acceptance is sensitive to the migration of events in to and out of the
phase-space of the measurement.
The MC is therefore reweighted such that the \pt spectra of the four leading jets
match the corresponding distributions in data. 
The weights are derived according to the \pt of the leading jet (as discussed in \autoref{chapDijetMassUnfolding}),
which also serves to correct the distributions of the other three jets.
The reweighted MC may be compared to data in \autoref{FIGfourJetPtSpectrumDataPythia}.
\begin{figure}[ht]
\begin{center}
  \includegraphics[trim=5mm 5mm 0mm 16mm,clip,width=.6\textwidth]{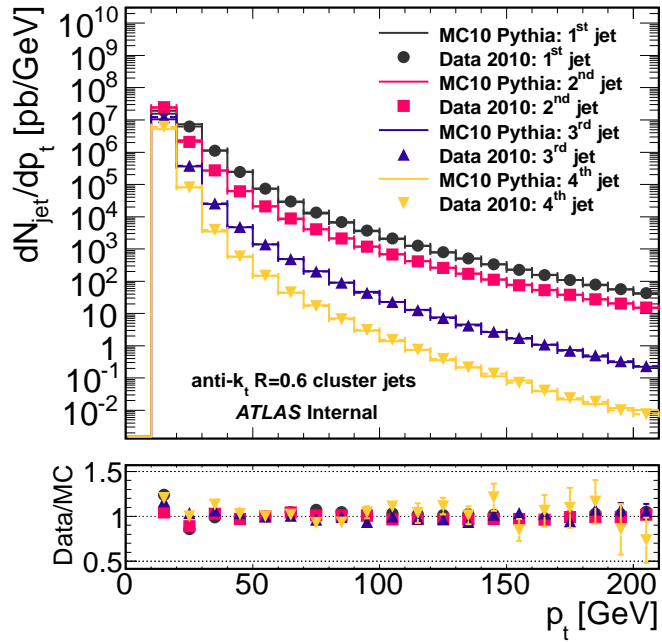}
  \caption{\label{FIGfourJetPtSpectrumDataPythia}Transverse momentum, \pt, spectra of the four highest-\pt jets
  in an event, denoted as $1^{\mrm{st}}$, $2^{\mrm{nd}}$, $3^{\mrm{rd}}$ and $4^{\mrm{th}}$~jet in the figure,
  in the 2010 data and in MC, 
  where the distributions in MC are rescaled such that they agree in normalization with the data at $\pt = 50\GeV$.
  The bottom panel shows the ratio of data to MC.
  }
\end{center}
\end{figure} 
The distributions in MC are rescaled such that they agree in normalization with the data at ${\pt = 50\GeV}$.
The event samples in data and in MC are inclusive; the only requirement imposed is that in each
event, at least two jets with \ptHigher{20} are reconstructed.
The differences between data and MC are at most of the order of~20\% for 
each of the four leading jets, for \ptLower{50}. For higher values of transverse momentum, the deviations of the
MC with respect to data are generally below~10\%. The dependence of the acceptance ratio on the shape of the
\pt spectra in MC is estimated in the next section.

The values of the acceptance factors which are derived using the reweighted MC are
\begin{equation*}
        \renewcommand{\arraystretch}{1.75}
        \begin{array}{cccc}
          \mathcal{A}_{2j}^{a} = 1.11 \pm 0.01  \;, &
          \mathcal{A}_{2j}^{b} = 1.13 \pm 0.01  & \mrm{and} &
          \mathcal{A}_{4j}^{b} = 1.10 \pm 0.03  \;,
        \end{array}
\label{eqObservedDpsAccept} \end{equation*}
where the errors are due to the limited MC statistics.
The corresponding value of the acceptance ratio is
%
\begin{equation}
\alpha_{2j}^{4j} = 0.88 \pm 0.03  \;.
\label{EQsimpleSigmaDPSFourJet8}
\end{equation}
%
%
%
The statistical uncertainty 
on $\alpha_{2j}^{4j}$ is propagated to \sigeff in the following as a systematic uncertainty.
Additional uncertainties are derived for $\alpha_{2j}^{4j}$, and subsequently propagated to \sigeff, in \autoref{chapDpsUncertainty}.

\section{Extraction of the fraction of DPS events using a neural network\label{chapExtractionOfFdpsWithNN}}
%
An artificial neural network (NN) is a powerful method of solving complex
problems, such as pattern recognition~\cite{citeulike:308856}.
Neural networks use \textit{supervised learning}, a machine-learning task of inferring a function from a set of training examples.
Each example consists of an input object and a desired output value.
Neural networks are organized into at least three layers; the input layer, the hidden layer, and the output layer.
The three layers consist of various numbers of neurons.
There can be more than one hidden layer; this
does not necessarily coincide with an increase in performance, but instead raises the complexity of the network.

The analysis uses a Multi-Layer Perceptron, one of the better known artificial NN models.
Learning occurs in the perceptron by changing connection weights after each piece
of data is processed, based on the amount of error in the output compared to the predicted result. 
The learning process is carried out through backpropagation~\cite{Rumelhart:1988:LRB:65669.104451},
a generalization of the least mean squares algorithm in the linear perceptron.

\subsection{Input samples to the neural network\label{chapInputSamplesNeuralNetwork}}
%
Two input sets are used in order to train the NN, referred to as the \textit{NN background sample}
and the \textit{NN signal sample}.
The background sample represents four-jet events in which no DPS occurs, and all four jets originate from a single
interaction. This sample is constructed using MC events, where the option of turning MPI off is available.
The signal sample represents DPS four-jet events, in which the two sets of jets originate from two
separate interactions. Since the available MCs for the analysis do not describe well the
features of hard DPS in four-jet events, this sample is constructed by an overlay of dijet events. These represent
events in which two completely uncorrelated hard interactions take place in a given event.

The background and signal samples for the analysis are described in the following.

\minisec{Background sample for the NN}
%
%
The background input to the NN, standing for \twofour pQCD events, is generated with \sherpa.
The generator is run with multiple interactions turned off, using the \ttt{MI\ul HANDLER=None} setting.
The CKKW matching scale of \sherpa is set at~15\GeV.
This implies that partons with transverse momentum above~15\GeV
in the final state, necessarily originate from matrix elements, and not from the parton shower.
All jets passing selection (having \ptHigher{20}), are therefore associated with the hard interaction.

The \sherpa sample consists of truth (particle) jets, which are not passed through the full detector simulation. In order to stand on
equal footing with data, the  energy and momentum of truth jets are
smeared according to the parametrized energy resolution of corresponding
detector jets (see \autoref{chapJetPtAngularResolution}). The energy smearing
is performed as for the systematic study discussed in \autoref{ENUMdijetMassUncertainty2}.

The \pt-spectra of the four leading jets in an event in the \sherpa sample, before and after the smearing procedure,
are presented in \autoref{FIGsmearedSherpaPtAndRes1}. Following the smearing, the peak at low values of
transverse momentum is enhanced. Due to the shape of the \pt distribution, more jets are pushed
into the phase-space of the measurement than out of it. Overall, following the smearing procedure,
an additional~4\% of events are included in the phase-space of the measurement.

The relative \pt resolution of jets in the smeared \sherpa MC may be compared to that of detector jets in the nominal \pythia MC in
\autoref{FIGsmearedSherpaPtAndRes2}.
The relative \pt resolution is derived in \pythia by matching truth jets with detector jets. 
The width of the distribution of the variable,
\begin{equation*}
  O_{p_{\mrm{t}}} = \frac{p_{\mrm{t}}^{\mrm{rec}}-p_{\mrm{t}}^{\mrm{truth}}}{p_{\mrm{t}}^{\mrm{truth}}} \;,
\end{equation*}
is taken as the relative resolution. For the \pythia sample, $p_{\mrm{t}}^{\mrm{truth}}$ and $p_{\mrm{t}}^{\mrm{rec}}$
represent the transverse momenta of truth and detector jets, respectively.
For the \sherpa sample, smeared truth jets take the role of detector jets,
and are matched to truth jets which are not smeared.
\begin{figure}[ht]
\begin{center}
\subfloat[]{\label{FIGsmearedSherpaPtAndRes1}\includegraphics[trim=10mm 5mm 0mm  5mm,width=.52\textwidth]{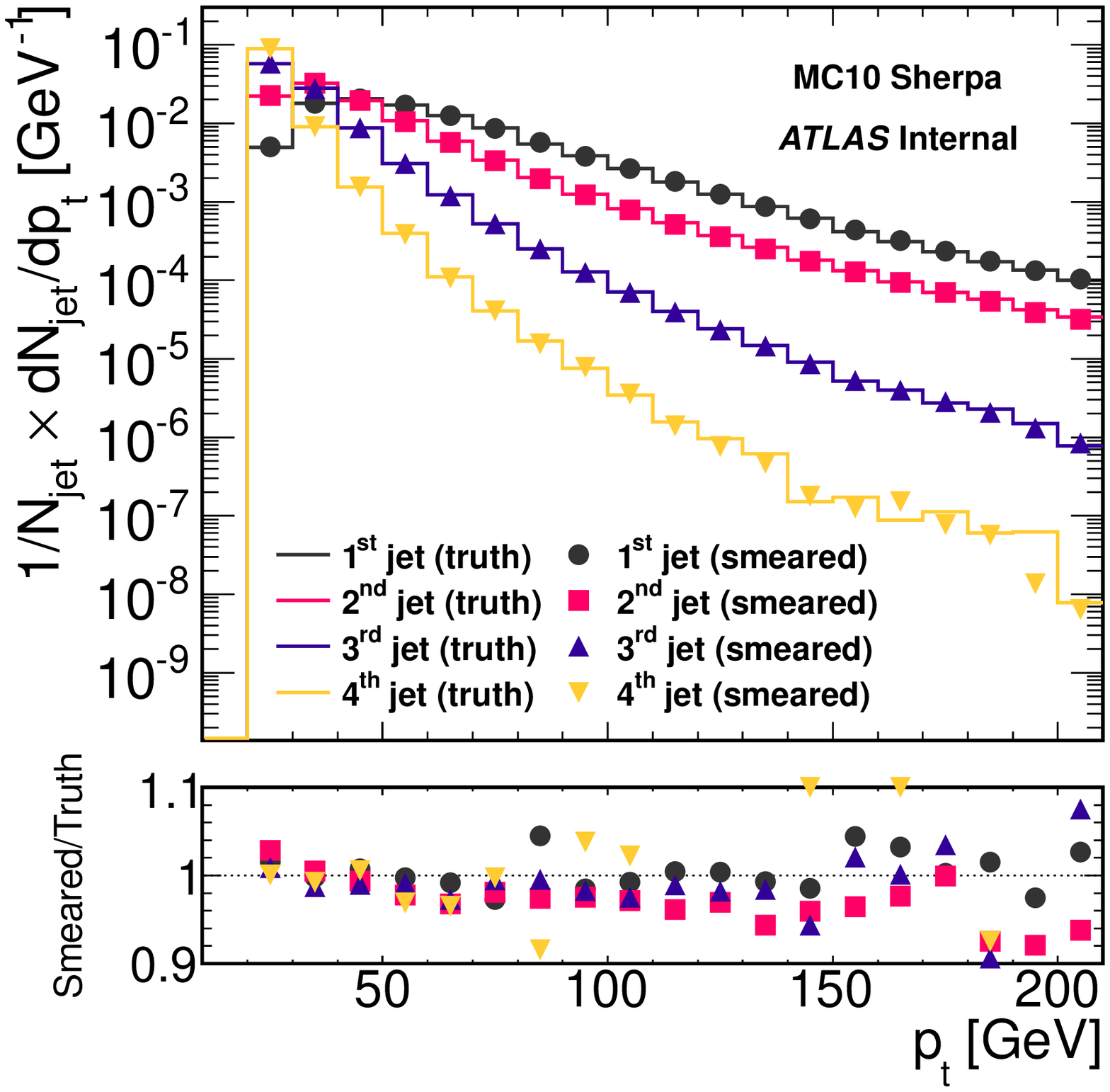}}
\subfloat[]{\label{FIGsmearedSherpaPtAndRes2}\includegraphics[trim=10mm 2mm 0mm 10mm,clip,width=.52\textwidth]{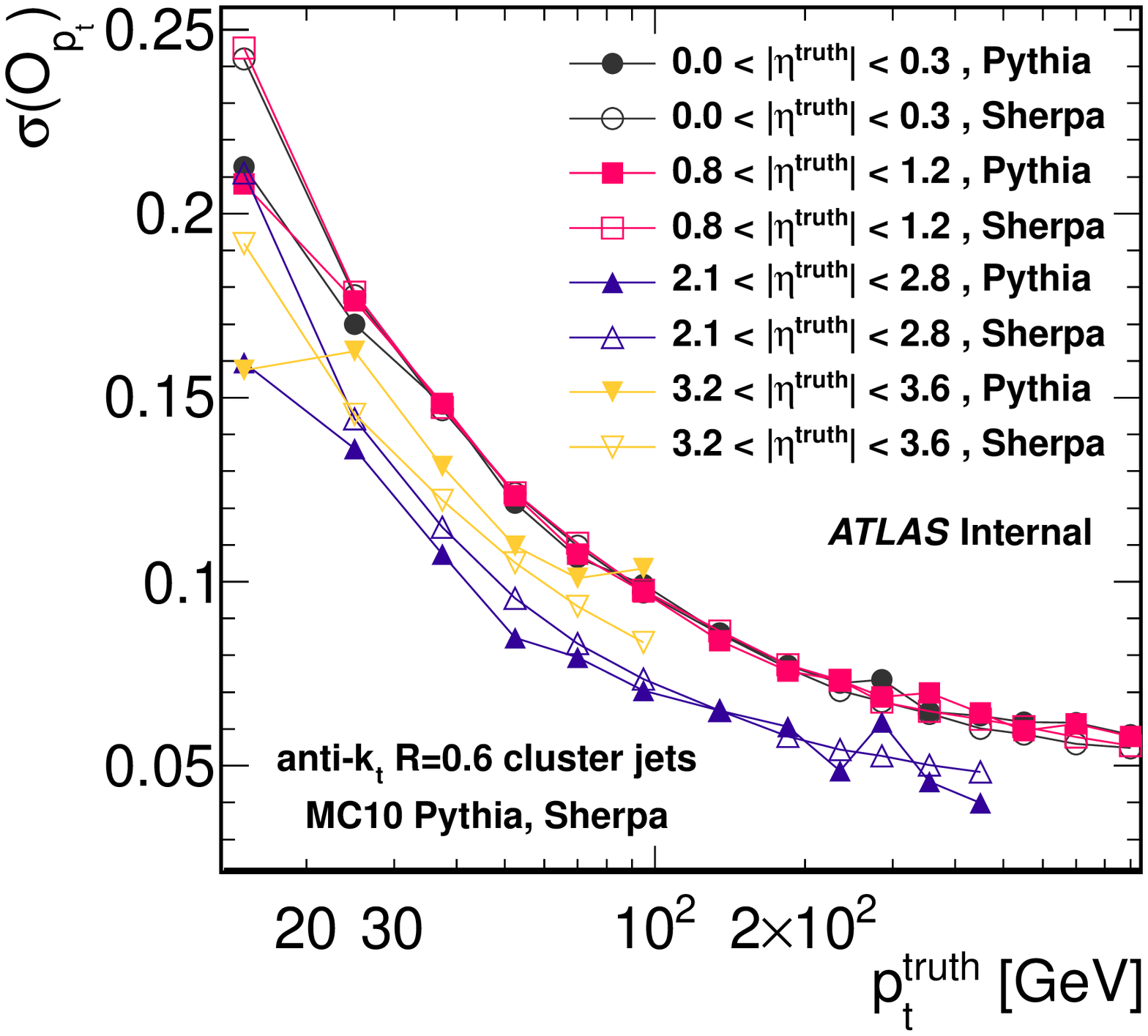}}
  \caption{\label{FIGsmearedSherpaPtAndRes}\Subref{FIGsmearedSherpaPtAndRes1} Transverse momentum, \pt,
  spectra of truth jets and smeared truth jets in \sherpa, for the
  four highest-\pt jets in an event, denoted as $1^{\mrm{st}}$, $2^{\mrm{nd}}$, $3^{\mrm{rd}}$ and $4^{\mrm{th}}$~jet.
  The bottom panel shows the ratio between the spectra for each jet separately, as indicated.\\
  \Subref{FIGsmearedSherpaPtAndRes2} Relative transverse momentum resolution, $\sigma(O_{p_{\mrm{t}}})$, of jets in the
  smeared \sherpa MC and in the nominal \pythia MC, as a function of the
  transverse momentum of truth jets, $p_{\mrm{t}}^{\mrm{truth}}$, for jets
  within different ranges of pseudo-rapidity, $\eta^{\rm truth}$, as indicated in the figure. 
  }
\end{center}
\end{figure} 
The first bin (${10 < p_{\mrm{t}}^{\mrm{truth}} < 20\GeV}$) in \autoref{FIGsmearedSherpaPtAndRes2}
shows that the resolution is over-estimated for smeared jets in \sherpa by~20\%,
relative to that of detector jets in \pythia. The discrepancy is due to the fact that the parametrized resolution is
derived from \insitu techniques using jets with \ptHigher{20}, and extrapolated to lower transverse momenta~\cite{:2012ag}.
Above~20\GeV, good agreement is observed between the two MC samples.
The dependence of the results of the DPS measurement on the \pt resolution of jets in \sherpa is estimated in
the following. A corresponding systematic uncertainty is assigned to the measurement.

To summarize, the NN background sample consists of exclusive four-jet events.
Events are selected by requiring the existence of exactly four smeared truth jets with
transverse momentum, \ptHigher{20}, where the leading jet is further required to
have \ptHigher{42.5}. This is consistent with the corresponding phase-space of the
measurement, as defined in \autoref{EQphaseSpaceOfMeasurementDPS1}.

\minisec{Signal sample for the neural network}
%
%
The NN sample which describes DPS events is composed of overlaid dijet events.
The natural choice would have been to use dijets from data in order to construct this sample. However,
the number of events with two low-\pt jets in the data is too small in order to construct a statistically robust
overlay sample.
The dijets are therefore taken from the nominal \pythia simulation.
In order to illustrate that \pythia describes well the chrechtaristics of dijet events in data,
distributions of the dijet invariant mass, $m_{12}$, are used. The measurement of the invariant mass
spectra is discussed in detail in \autoref{chapMeasurementOfTheDijetMass}.
Using the 2010 dataset, data unfolded to the hadron-level are compared to \pythia events
in \autoref{FIGsmearedSherpaPtAndRes}, where truth (particle) jets are
used to calculate $m_{12}$ in \pythia.
\begin{figure}[ht]
\begin{center}
  \includegraphics[trim=8mm 0mm 3mm 5mm,width=1.02\textwidth]{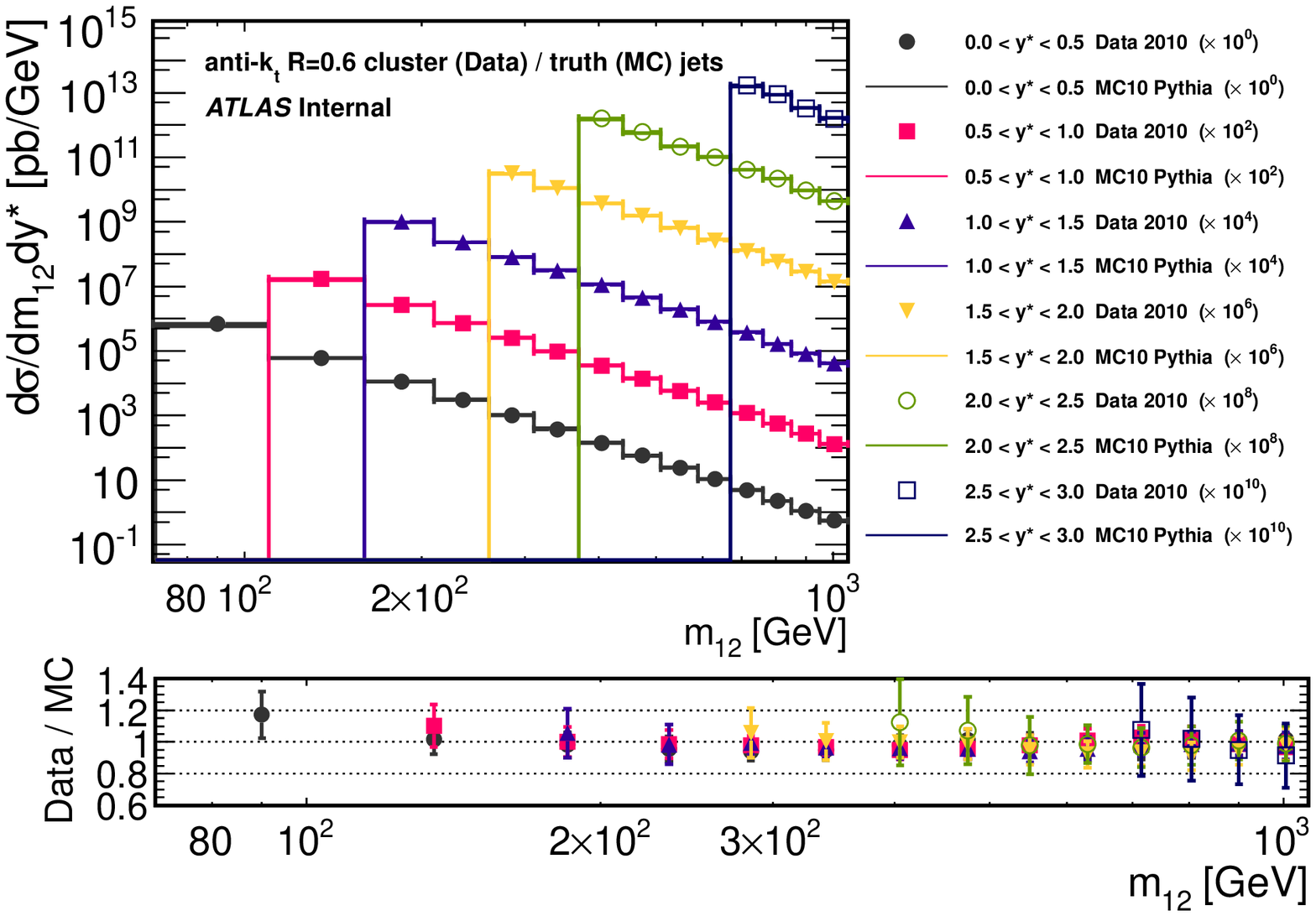}
  \caption{\label{FIGsmearedSherpaPtAndRes}Double-differential dijet \xsec as a function of the invariant mass, $m_{12}$,
  binned in \com jet rapidity, \ystr, in the 2010 data, compared to the $m_{12}$ distributions of
  particle (truth) jets in the nominal \pythia MC, where distributions are multiplied by the factors specified in the legend,  
  the error bars represent the statistical and systematic uncertainties, added in quadrature,
  and the bottom panel shows the ratio of distributions between the data and the MC.
  }
\end{center}
\end{figure} 
The \pythia MC is consistent with the data within the uncertainties.

The first step in constructing the overlaid sample is selection of exclusive dijet events in the MC.
The latter refer to events in which exactly two detector jets are reconstructed, each with
transverse momentum, \ptHigher{20}.
Once a list of dijet events is assembled, pairs of events are overlaid into four-jet events. The conditions
which must be fulfilled in order for a given pair of events to be overlaid are the following:
\begin{list}{-}{}
\item none of the four jets overlaps in $\eta,\phi$ by a distance of less than 0.6~units;
\item the vertex-$z$ coordinates of the vertices in the two overlaid events are no more than 10~mm apart;
\item at least one of the four jets has \ptHigher{42.5};
\item only single-vertex \pythia events are used.
\end{list}
The first condition allows for the four-jet event to be reconstructed in the calorimeter to begin with;
the second condition avoids possible bias, due to events where two jet pairs originate from separate vertices;
the third condition is set in order to match the corresponding phase-space of the measurement;
finally, the fourth condition is imposed in order to match the conditions in the data.

Once a pair of dijet events are overlaid into a four-jet event, the jets are re-ordered in \pt. As a result, the sub-leading
jet in the four-jet overlay may have originally been either the sub-leading jet in the first dijet pair, or the leading jet in the
second dijet pair. This reflects the signature of a four-jet DPS event in data; the ordering in \pt of the four reconstructed
jets does not necessarily correspond to the two highest-\pt jets originating from one interaction and the second 
pair from the other interaction.
The significance of the ordering in \pt of the four jets comes into play in the variables which constitute
the input parameters to the NN. The latter are
the subject of the following section.

\subsection{Input variables to the neural network}
%
%
The underlying assumption driving the analysis involving the NN is that for the case of DPS, two
dijet interactions occur simultaneously and independently. On the other hand, in a \twofour QCD
four-jet interaction, there is correlation between all four jets in the final state.

The correlations in a \twofour interaction have several characteristics; one would
\eg observe many events in which the third and fourth jets by order of \pt (assumed
here to be higher-order radiative corrections to the hard scattering) are collinear in azimuth
with either of the leading pair.
Additionally, radiative jets are more likely to span the area in rapidity between the two hard scatters,
or between one of the latter and the beamline (the proton remnants).
Up to detector effects (calibration and resolution)
and additional energy flow (\eg jets with \ptLower{20}), the four jets
in the final state are expected to all balance together in \pt.
Finally, no pair-wise balance in \pt and no pair-wise back-to-back symmetry in azimuth are expected.

In a DPS event, under the assumption of uncorrelated \dtwot QCD interactions,
different topologies should be prominent. For instance, one would expect that there would be pair-wise
balance in \pt between the two pairs corresponding to the two interactions. Each pair should
be back-to-back in the transverse place, and the azimuthal angle between the axes of interaction should have a random
distribution.
In addition, the rapidity differences between the two pairs of jets is expected to have a random distribution as well.
%

These potentially discriminating features motivate the choice of variables which are selected as input for the NN.
Several combinations of different parameter sets have been explored.
The best performance in discriminating between signal and background is achieved using the following variables:
\begin{equation}
\renewcommand{\arraystretch}{1.5} 
\begin{array}{ccccccc}
\multicolumn{3}{c}{ \Delta^{\pt}_{12} = \dfrac{\left|\vec{p}_{t,1}+\vec{p}_{t,2}\right|}{p_{t,1}+p_{t,2}} \quad , }  &
\multicolumn{3}{c}{ \Delta^{\pt}_{34} = \dfrac{\left|\vec{p}_{t,3}+\vec{p}_{t,4}\right|}{p_{t,3}+p_{t,4}} \quad , }  &  \\
\Delta^\phi_{12} = \phi_{1}-\phi_{2}  \quad , & \Delta^\phi_{13} = \phi_{1}-\phi_{3} &  , &
\Delta^\phi_{23} = \phi_{2}-\phi_{3}  \quad , & \Delta^\phi_{34} = \phi_{3}-\phi_{4} &  , & \\
\Delta^\eta_{13} = \eta_{1}-\eta_{3}  \quad , & \Delta^\eta_{14} = \eta_{1}-\eta_{4} &  , &
\Delta^\eta_{23} = \eta_{2}-\eta_{3}  \quad , & \Delta^\eta_{24} = \eta_{2}-\eta_{4} &  , &
\end{array}
\label{eqMassMeasurementPhasespace} \end{equation}
where $p_{t,i}$, $\eta_{i}$ and $\phi_{i}$ stand for the transverse momentum, pseudo-rapidity and azimuthal angle
of jet $i$, respectively, with $i=1,2,3,4$. 

The first two variables, $\Delta^{\pt}_{12}$ and $\Delta^{\pt}_{34}$, represent the normalized \pt balance between the two jets of the
leading jet-pair and between the two jets of the sub-leading jet-pair, respectively.
The \pt balance is normalized to the respective scalar sum
of transverse momenta in order to avoid two types of bias.
The fist type has to do with the shape of the
\pt spectrum of jets; one would like the NN to be insensitive to the shape
in the MC, as it does not perfectly describe the data.
A second possible source of bias is the uncertainty on the \JES; dealing with normalized variables reduces the
uncertainty, as the absolute energy scale does not play a direct role.

In the NN signal sample, there is no
way of knowing which combination of two jet-pairs corresponds to the original configuration of the event.
For instance, for a given event it is
possible that jets~1 and~2 originated from one \twotwo interaction and jets~3 and~4 from the other; alternatively,
it is also possible that jets~1 and~3 and jets~2 and~4 form the respective pairs.
In order to resolve the ambiguity of jet pairing and be consistent with data,
the information about which jet originated from which interaction is not used
with the NN signal sample. The variables $\Delta^{\pt}_{12}$ and $\Delta^{\pt}_{34}$
are therefore constructed from the \pt-ordered jets.

Normalized distributions of $\Delta^{\pt}_{12}$ and of $\Delta^{\pt}_{34}$ in the NN signal and background samples
are shown in \autoref{FIGnnVarShapeDeltaPt}.
\begin{figure}[htp]
\begin{center}
\subfloat[]{\label{FIGnnVarShapeDeltaPt1}\includegraphics[width=.52\textwidth]{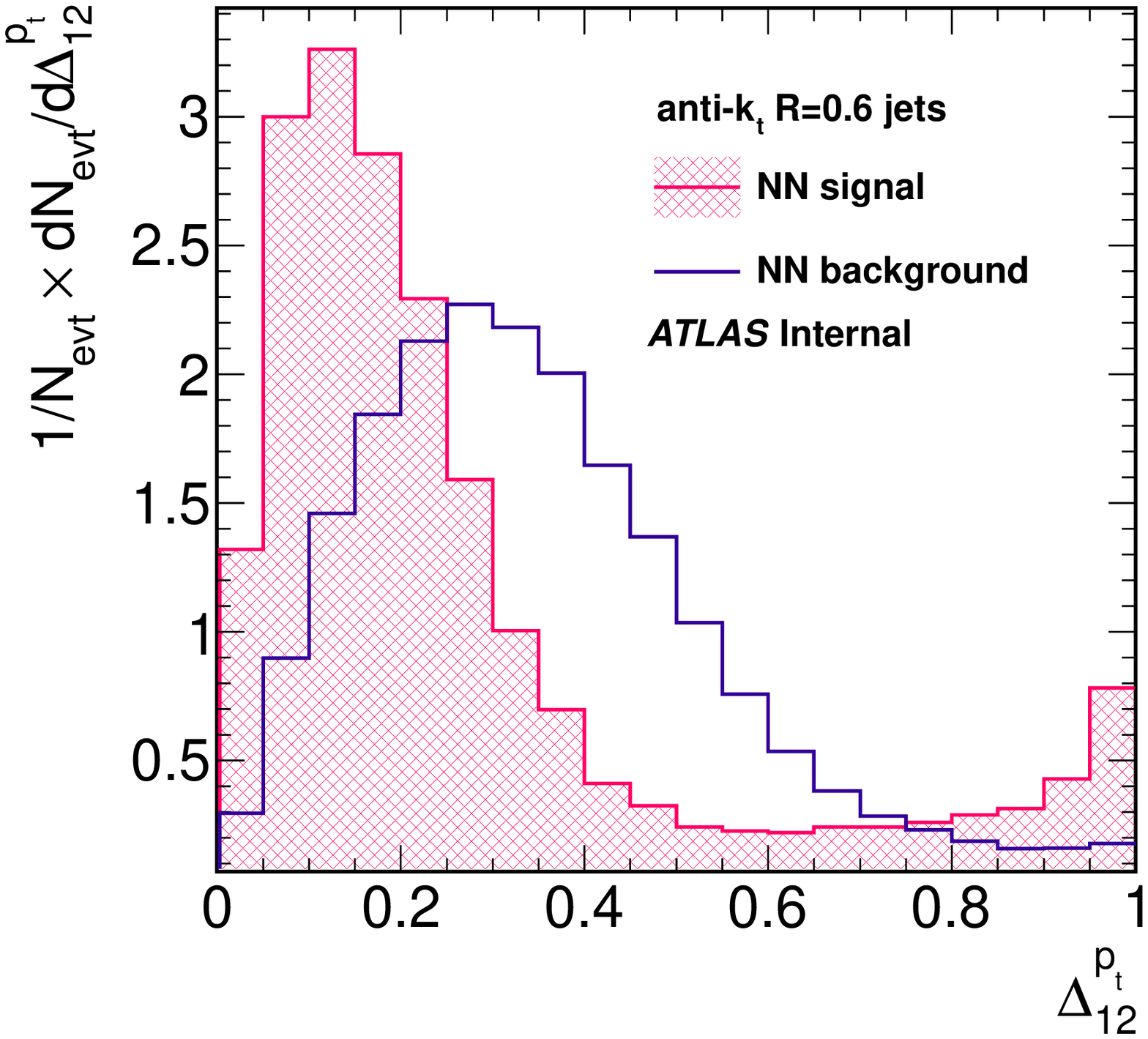}}
\subfloat[]{\label{FIGnnVarShapeDeltaPt2}\includegraphics[width=.52\textwidth]{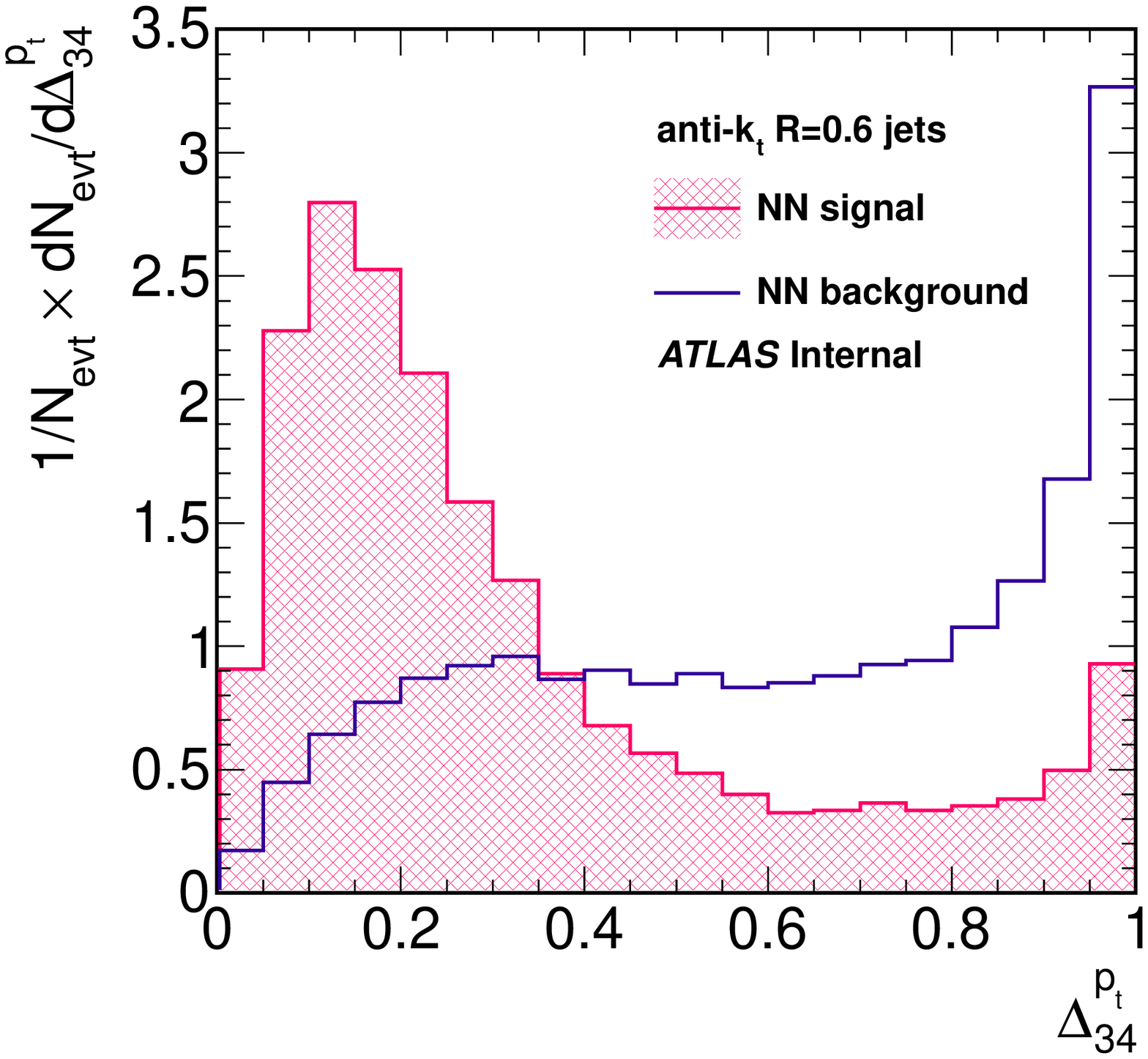}}
  \caption{\label{FIGnnVarShapeDeltaPt}Normalized differential distributions of the variables, $\Delta^{\pt}_{12}$ \Subref{FIGnnVarShapeDeltaPt1}
  and $\Delta^{\pt}_{34}$ \Subref{FIGnnVarShapeDeltaPt2}, which are defined in \autoref{eqMassMeasurementPhasespace},
  for the signal and background input samples of the NN, as indicated.
  }
\end{center}
\end{figure} 
For the NN signal sample, both distributions peak at low values,
indicating that both the leading and the sub-leading jet pairs are balanced in \pt. The small peak around
unity likely indicates events in which the re-ordering of jet-\pt switches the positions of jets~2 and~3.
For the NN background sample, the leading jet-pair exhibits a wider peak at low values of $\Delta^{\pt}_{12}$
compared to that in the signal sample.
This is due to the fact that the two leading jets can not balance well in \pt, as the second
pair of jets carries some of the momentum of the same hard scatter. The distribution of $\Delta^{\pt}_{34}$ in
the background sample peaks around unity, indicating that in most events jets~3 and~4 are both collinear with
a single one of the two leading jets. The tail toward lower values of $\Delta^{\pt}_{34}$ corresponds to events in which
each of the leading jets is collinear with one of the two sub-leading jets. In such events, a partial balance
in \pt is possible, as the two leading jets are in most cases approximately back-to-back.

The second set of input variables, $\Delta^\phi_{ij}$ $\left(ij=12,13,23~\mrm{and}\;34\right)$, represents
the ``back-to-back''-ness of the different jet-pair combinations, and the relative angles between the two axes of interaction for each
combination of jet-pairs.
Here information regarding all possible combinations is preserved (either explicitly, or through the
difference between a pair of $\Delta^\phi_{ij}$ variables); pairing ambiguity is thus avoided.
Similarly, the variables $\Delta^\eta_{ij}$ $\left(ij=13,14,23~\mrm{and}\;24\right)$
represent the full information of the relative pseudo-rapidity
between any pair of jets of the four-jet system. 

Normalized distributions of the different $\Delta^\phi_{ij}$ and $\Delta^\eta_{ij}$ variables 
in the NN signal and background samples are shown in \autorefs{FIGnnVarShapeDeltaPhi}~and~\ref{FIGnnVarShapeDeltaEta}.
\begin{figure}[htp]
\begin{center}
\subfloat[]{\label{FIGnnVarShapeDeltaPhi1}\includegraphics[width=.52\textwidth]{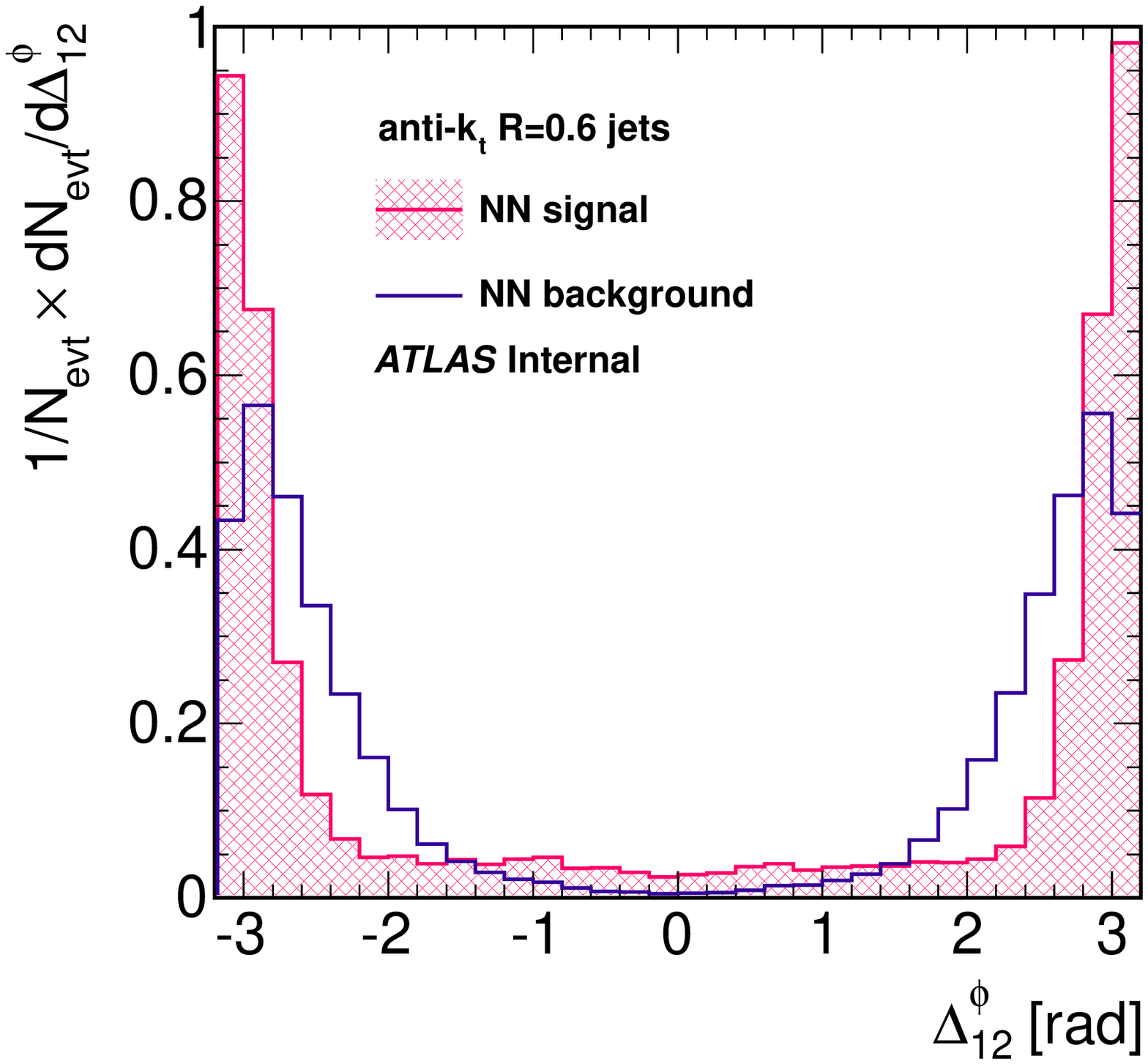}}
\subfloat[]{\label{FIGnnVarShapeDeltaPhi2}\includegraphics[width=.52\textwidth]{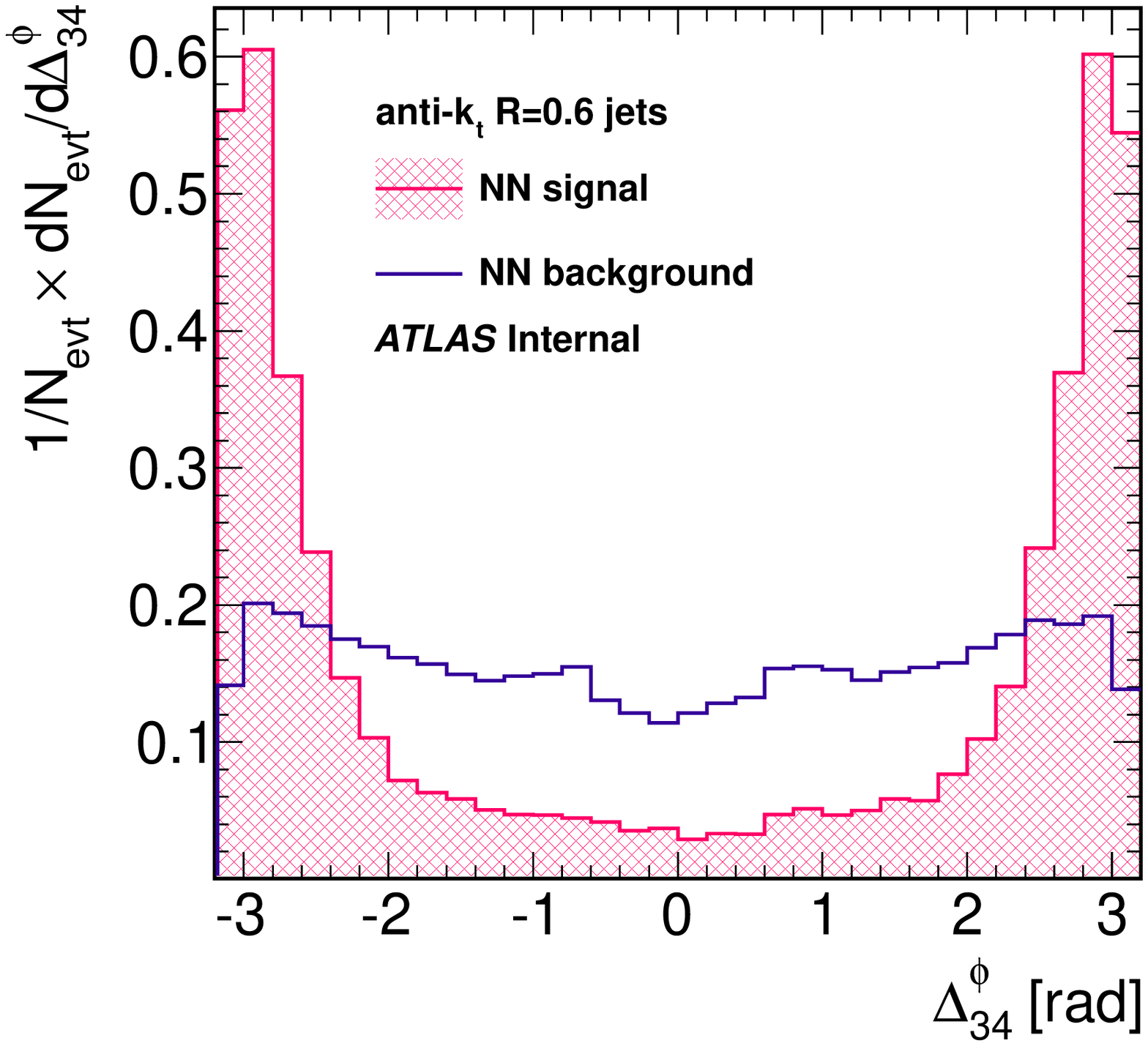}} \\
\subfloat[]{\label{FIGnnVarShapeDeltaPhi3}\includegraphics[width=.52\textwidth]{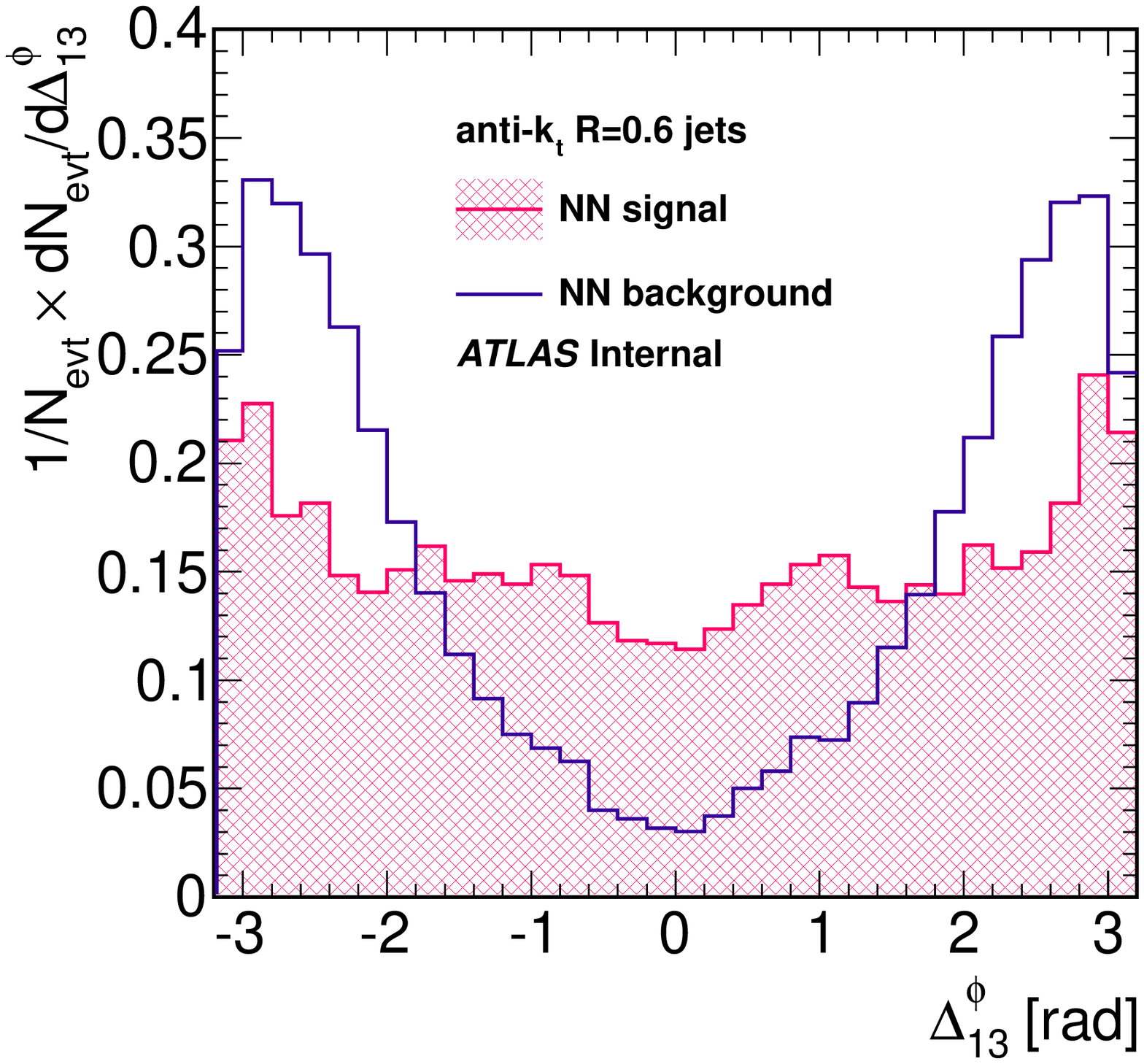}}
\subfloat[]{\label{FIGnnVarShapeDeltaPhi4}\includegraphics[width=.52\textwidth]{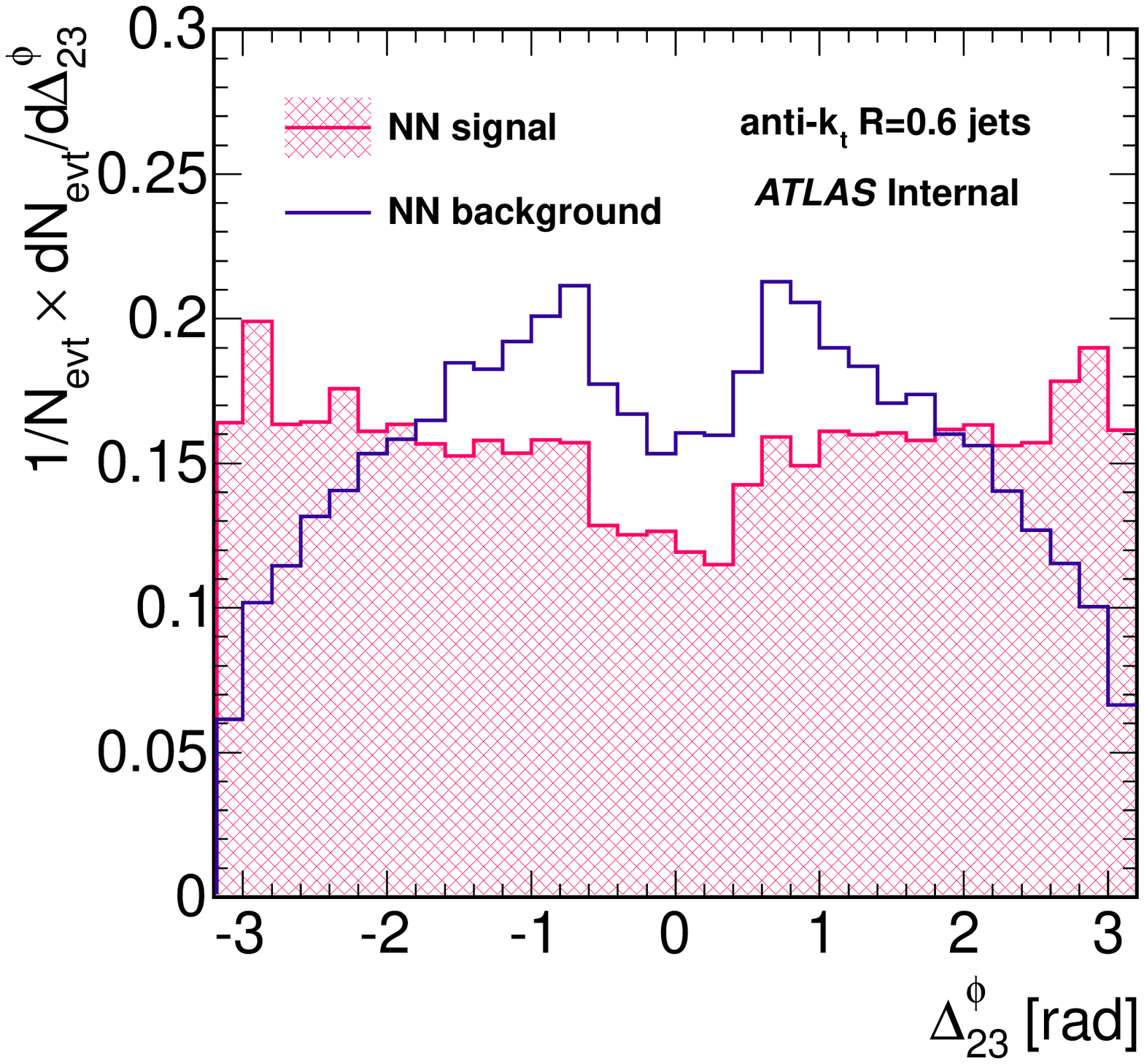}}
  \caption{\label{FIGnnVarShapeDeltaPhi}Normalized differential distributions of the variables, $\Delta^{\phi}_{12}$ \Subref{FIGnnVarShapeDeltaPhi1},
  $\Delta^{\phi}_{34}$ \Subref{FIGnnVarShapeDeltaPhi2}, $\Delta^{\phi}_{13}$ \Subref{FIGnnVarShapeDeltaPhi3}
  and $\Delta^{\phi}_{23}$ \Subref{FIGnnVarShapeDeltaPhi4}, which are defined in \autoref{eqMassMeasurementPhasespace},
  for the signal and background input samples of the NN, as indicated.
  }
\end{center}
\end{figure} 
\begin{figure}[htp]
\begin{center}
\subfloat[]{\label{FIGnnVarShapeDeltaEta1}\includegraphics[width=.52\textwidth]{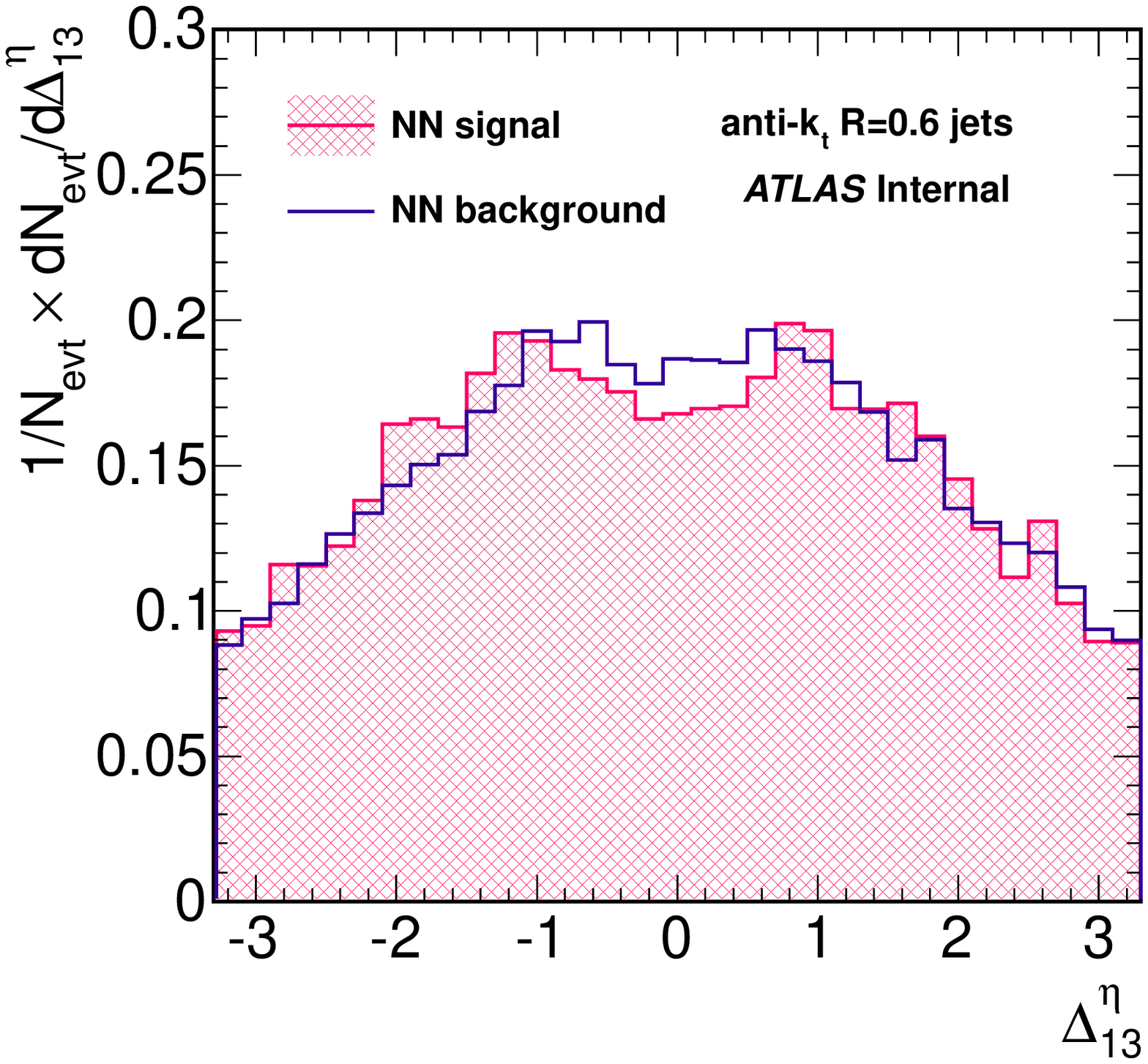}}
\subfloat[]{\label{FIGnnVarShapeDeltaEta2}\includegraphics[width=.52\textwidth]{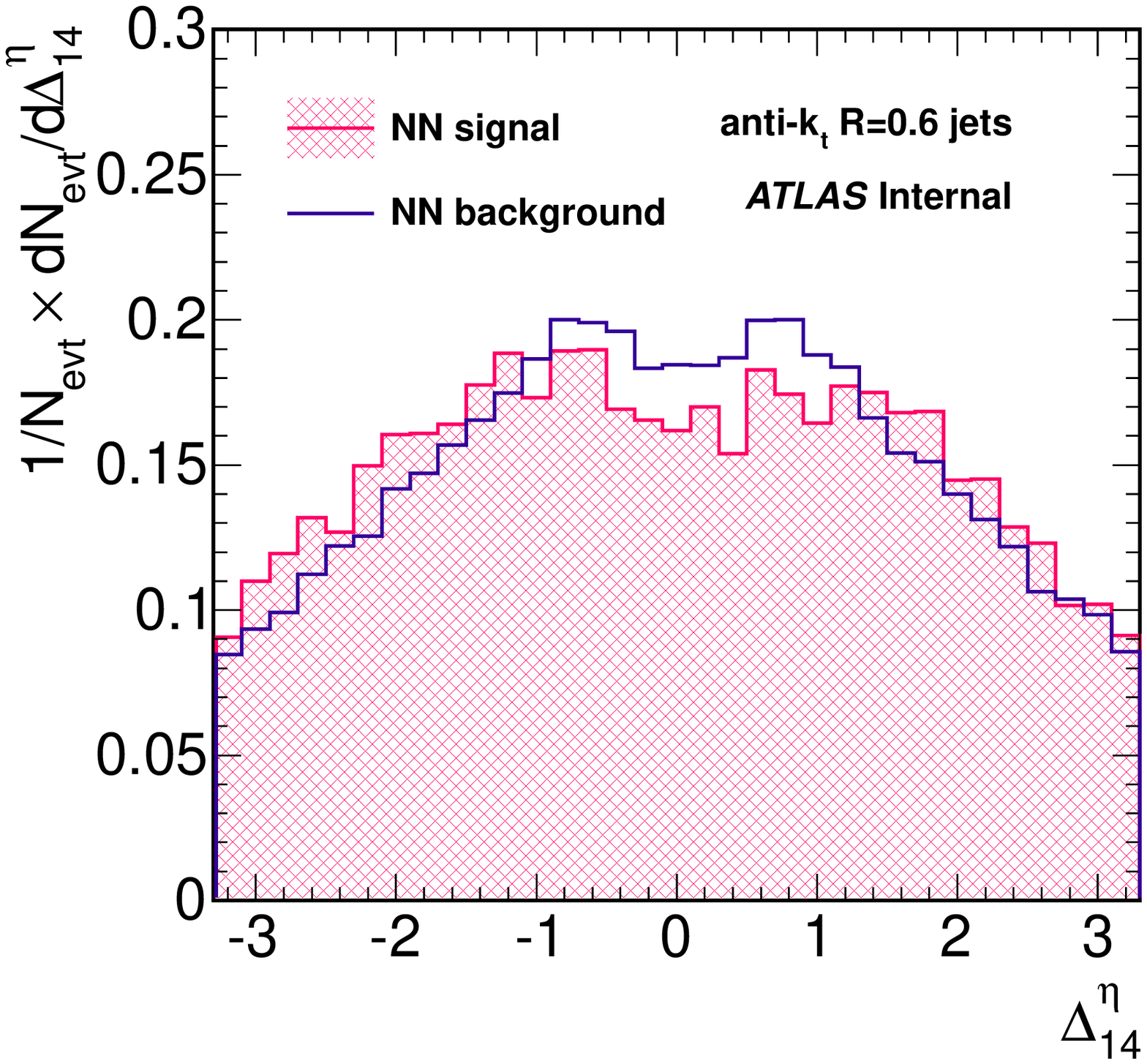}} \\
\subfloat[]{\label{FIGnnVarShapeDeltaEta3}\includegraphics[width=.52\textwidth]{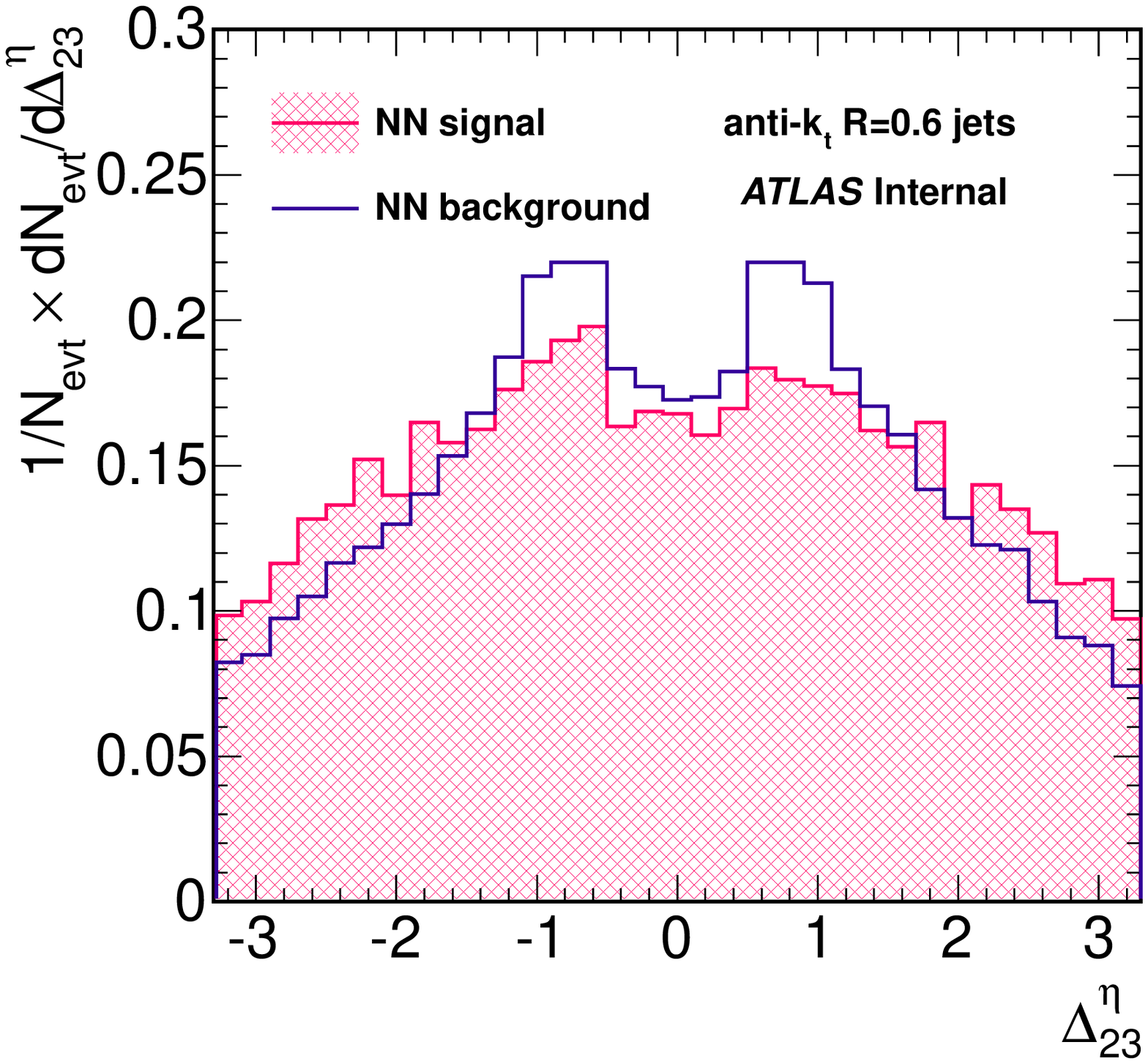}}
\subfloat[]{\label{FIGnnVarShapeDeltaEta4}\includegraphics[width=.52\textwidth]{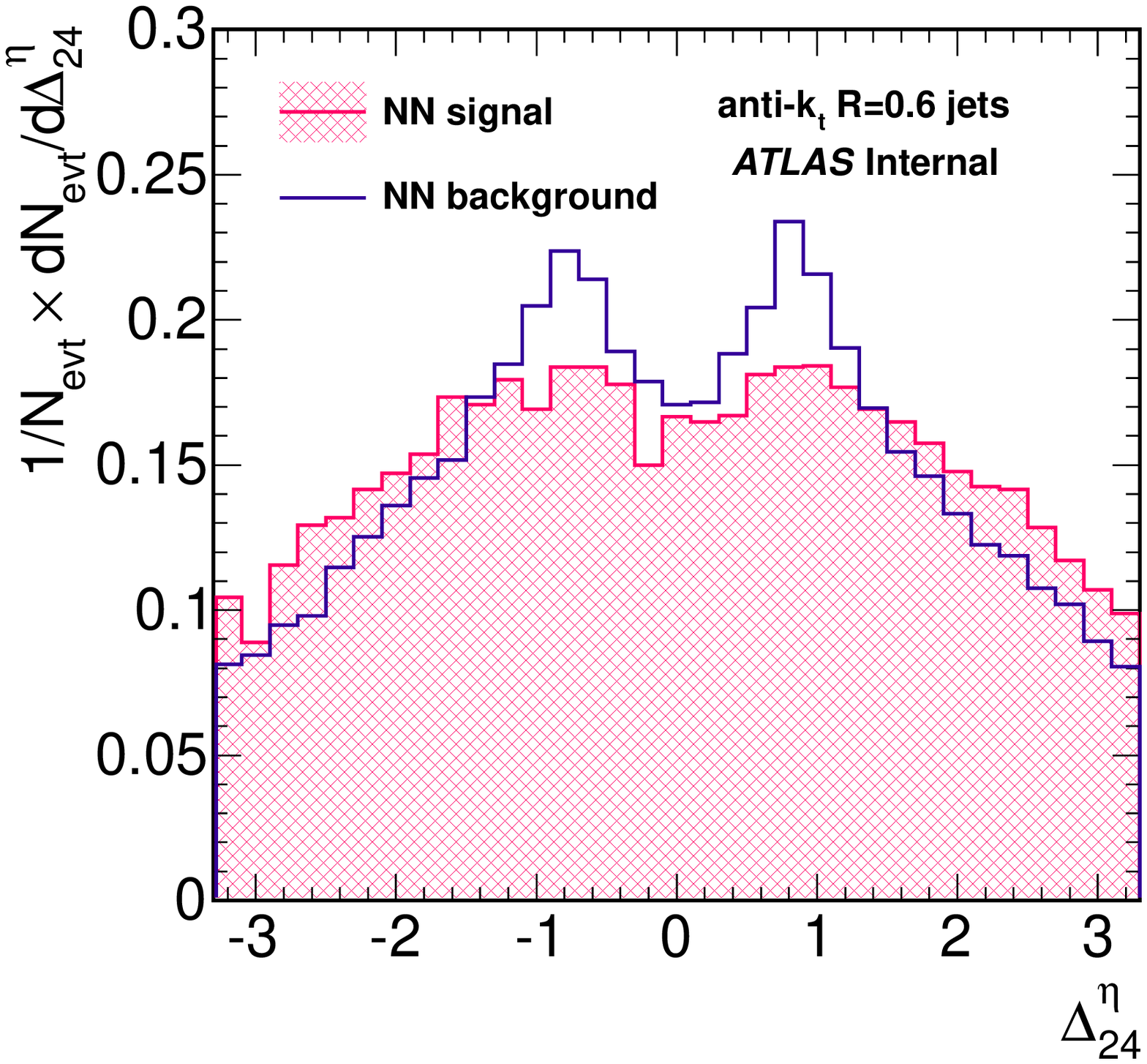}}
  \caption{\label{FIGnnVarShapeDeltaEta}Normalized differential distributions of the variables, $\Delta^{\eta}_{13}$ \Subref{FIGnnVarShapeDeltaEta1},
  $\Delta^{\eta}_{14}$ \Subref{FIGnnVarShapeDeltaEta2}, $\Delta^{\eta}_{23}$ \Subref{FIGnnVarShapeDeltaEta3}
  and $\Delta^{\eta}_{24}$ \Subref{FIGnnVarShapeDeltaEta4}, which are defined in \autoref{eqMassMeasurementPhasespace},
  for the signal and background input samples of the NN, as indicated.
  }
\end{center}
\end{figure} 
As expected, the $\Delta^\phi_{ij}$ variables show that in the NN signal sample the two pairs
of jets are in most cases back-to-back. The corresponding distributions for the azimuthal differences between
jets~1 and~3 and between jets~2 and~3 are relatively flat, as these differences are random by construction. The small enhancements
around~$\pm\pi$ represent events in which the re-ordering of jets in \pt had switched the
order between jets~2 and~3.
In the NN background sample, the two leading jets are only approximately back-to-back, resulting
in wider peaks around~$\pm\pi$ in $\Delta^\phi_{12}$, compared to the signal sample.
The peaks around~$\pm\pi$ in $\Delta^\phi_{13}$ correspond to events in which jet~2 emits collinear radiation, and thus the leading
jet approximately remains back-to-back with both jets~2 and~3. As expected, the distribution of $\Delta^\phi_{23}$ features peaks close
to zero. The angle between jets~3 and~4 in the background sample is mostly random, though small
enhancements around~$\pm\pi$ indicate events in which each of the leading jets is collinear with a jet of the sub-leading pair.

Inspection of the distributions of the $\Delta^\eta_{ij}$ variables shows significant differences between
the signal and background samples, only for $\Delta^\eta_{23}$ and for $\Delta^\eta_{24}$. The differences are characterized by
peaks around small $\Delta^\eta_{23}$ and $\Delta^\eta_{24}$ values in the background sample, which are not present in the NN signal.
As expected, these indicate that QCD radiation tends to fill-in the rapidity region between
the hardest scattering objects (or between the latter and the proton remnants).

\subsection{Training and output of the neural network\label{chapTrainingOutputOfNN}}
%
Once the signal and background samples for the NN are prepared, events from each are divided into two (statistically
independent) sub-samples, the \textit{training sample} and the \textit{test sample}. 
As the names imply, the former is used to train the NN, and
the latter to test the robustness of the result.
The same number of events from the signal and background samples is used for the training of the NN.
In all subsequent figures, only the test signal and background samples are shown.

The final structure of the NN consists of two hidden layers. The input layer has ten neurons, the first and second hidden
layers have eight and four neurons, respectively, and the output of the network is a single neuron (a single number).
These choices represent the product of an optimization study on the performance of the NN, and balance the complexity
of the network with the computation time of the training.

During the training phase of the NN, weights between the various input-layer neurons are modified,
changing the output of the network.
The result is quantified by an error function, $E_{\rm NN}$, which measures the agreement of the response of the network with the desired result.
\Autoref{FIGnnLayerStructureSignificance1} shows the value of $E_{\rm NN}$ as a function of the number of training cycles of the network.
The training should be stopped when $E_{\rm NN}$ deteriorates for the test sample relative to the training sample.
Here, training is stopped after 2000~epochs, as $E_{\rm NN}$ converges to a stable value for the training sample,
and there is an ever so slight indication of deterioration for the test sample.
Attempts to use a higher number of training cycles did not achieve improved performance in $E_{\rm NN}$.
The chosen structure of the NN and the final set of weights after completion
of the training are illustrated in \autoref{FIGnnLayerStructureSignificance2}.
The circles represent neurons, where each column stands for a layer of the network,
and the width of a connecting line between neurons is proportional to the magnitude of the respective weight.
\begin{figure}[ht]
\begin{center}
\subfloat[]{\label{FIGnnLayerStructureSignificance1}\includegraphics[trim=5mm 14mm 10mm 0mm,clip,width=.52\textwidth]{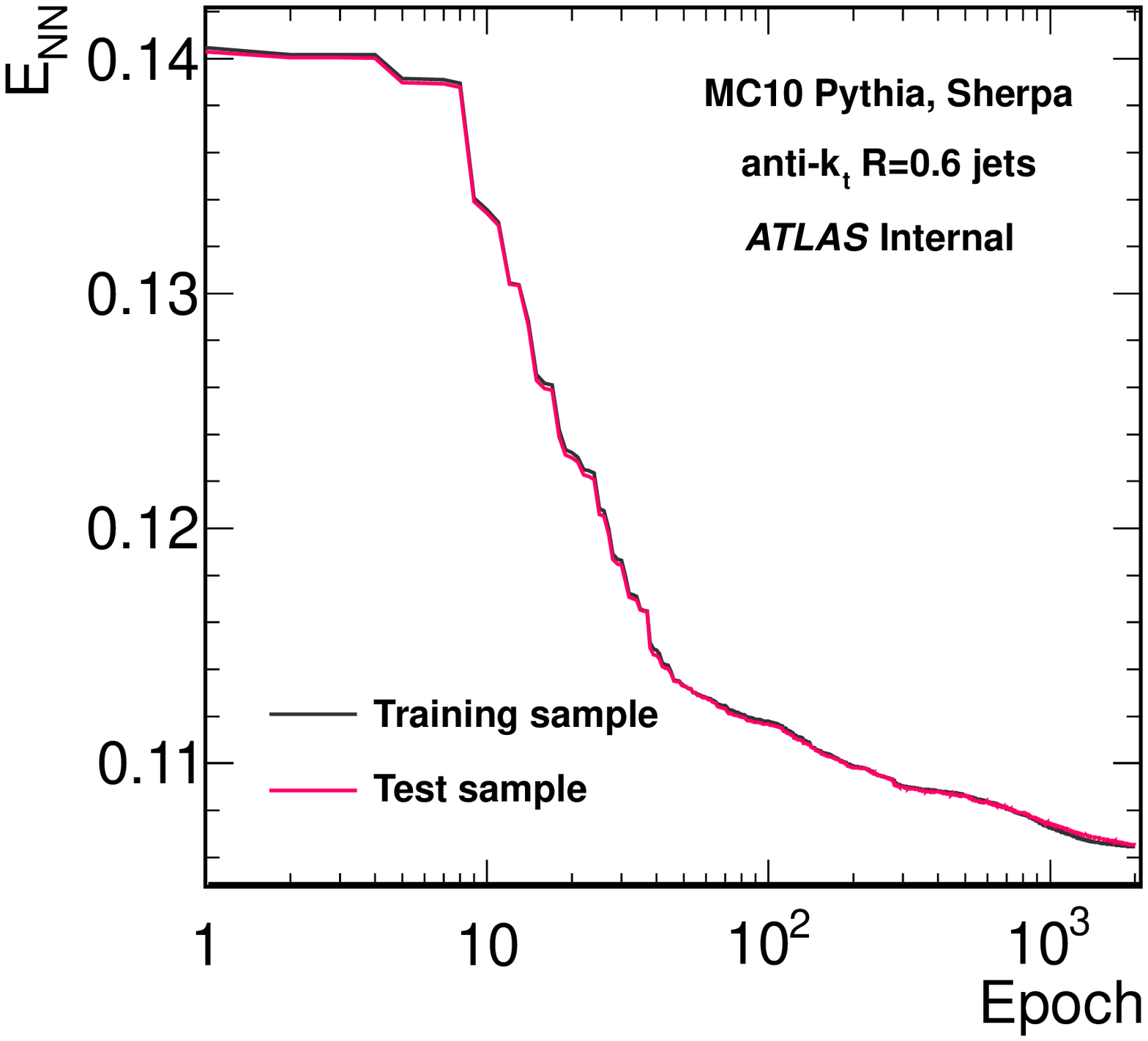}}
\subfloat[]{\label{FIGnnLayerStructureSignificance2}\includegraphics[trim=15mm 0mm 25mm 0mm,clip,width=.52\textwidth]{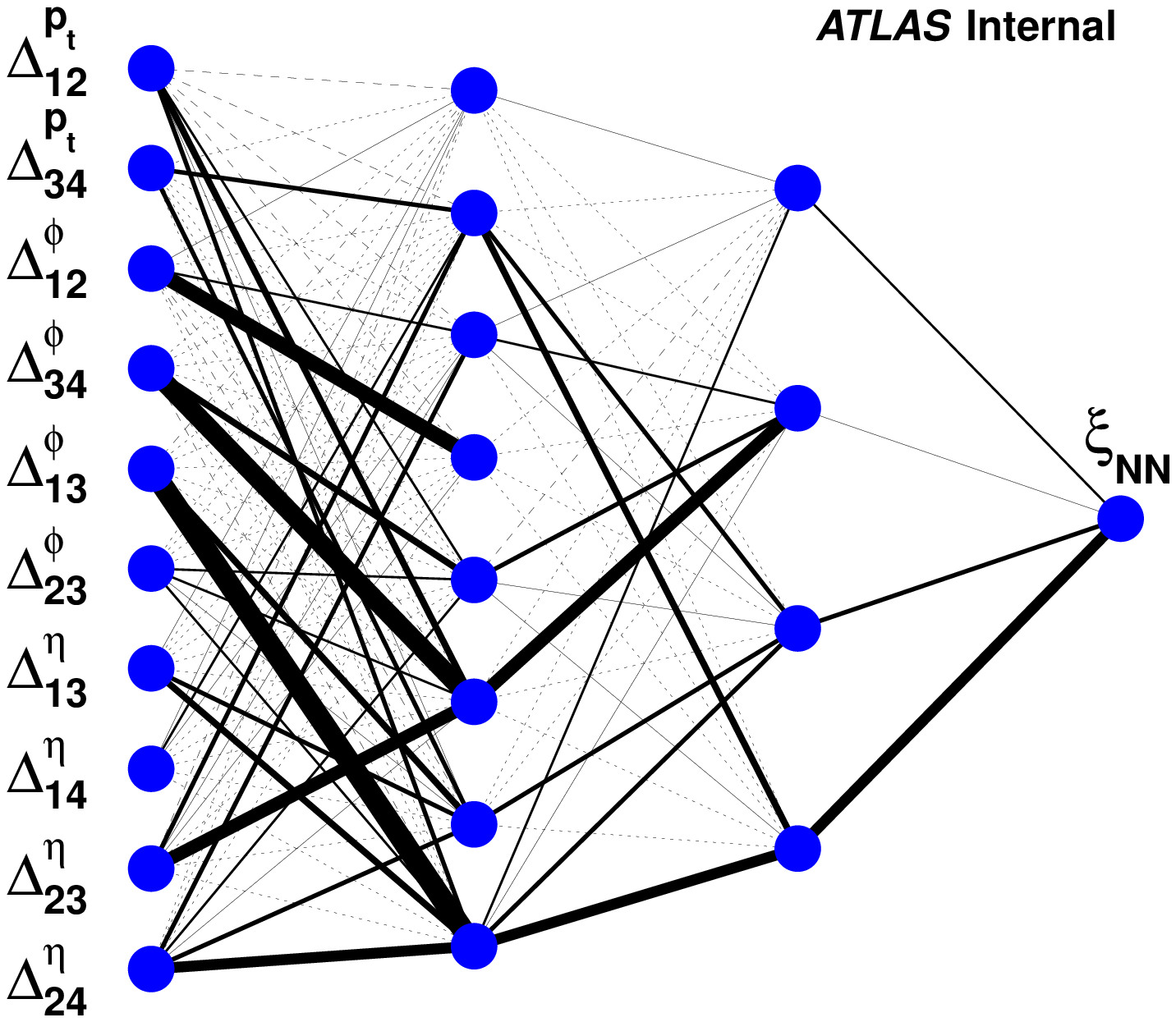}}
  \caption{\label{FIGnnLayerStructureSignificance}\Subref{FIGnnLayerStructureSignificance1} Dependence
  of the value of the error function of the NN, $E_{\rm NN}$, on the number
  of training cycles (epoch) of the network, for the NN training and
  test samples, as indicated in the figure. \\
  \Subref{FIGnnLayerStructureSignificance2} Schematic depiction of the structure of the NN, where
  neurons are denoted by blue circles, layers are denoted by columns of circles,
  and lines represent the weights which connect neuron pairs, where
  the thickness of a line is proportional to the relative magnitude of the corresponding weight.
  Each input variable, depicted by the various $\Delta$ symbols
  (see \autoref{eqMassMeasurementPhasespace}), corresponds to a neuron in the first layer of the network.
  The output of the NN is denoted by $\xi_{\rm NN}$.
  }
\end{center}
\end{figure} 
%
%
The larger the weight of a given neuron, the more significant the contribution of the
corresponding variable.

The output of the NN is quantified by the parameter, $\xi_{\rm NN}$, which typically takes
values in the range~$[0,1]$. An output close to zero is classified by the NN as more likely to belong
to the background sample, while values close to unity are associated with signal events.
Normalized distributions of the output of the NN for the signal and background training samples
are shown in \autoref{FIGnnBackgroundSignalResult1}. There is good separation between the two inputs,
with prominent peaks of the background around zero and of the signal around unity, as required.

The NN is used in order to determine the fraction of DPS events in four-jet events in data.
A distribution of $\xi_{\rm NN}$ in data, normalized to one ($\mathcal{D}$), is fitted to a combination of the
distributions of the signal ($\mathcal{S}$) and background ($\mathcal{B}$) samples, each normalized to one.
The result of the fit in the form,
\begin{equation}
\mathcal{D} = (1-\fDPS) \cdot \mathcal{B} + \fDPS \cdot \mathcal{S} \;,
\label{EQfDpsResultFromNN1}
\end{equation}
is shown in \autoref{FIGnnBackgroundSignalResult2}.
The fitted value is
\begin{equation}
\fDPS = 0.081 \pm 0.004 \;,
\label{EQfDpsResultFromNN2}
\end{equation}
where the goodness-of-fit, $\chi^{2}$, divided by the number of degrees of freedom of the fit (NDF) yields
\begin{equation*}
\chi^{2}/\mrm{NDF} = 1.2 \;.
\label{EQfDpsResultFromNNchi2}
\end{equation*}
\begin{figure}[htp]
\begin{center}
\subfloat[]{\label{FIGnnBackgroundSignalResult1}\includegraphics[trim=5mm 14mm 10mm 25mm,clip,width=.52\textwidth]{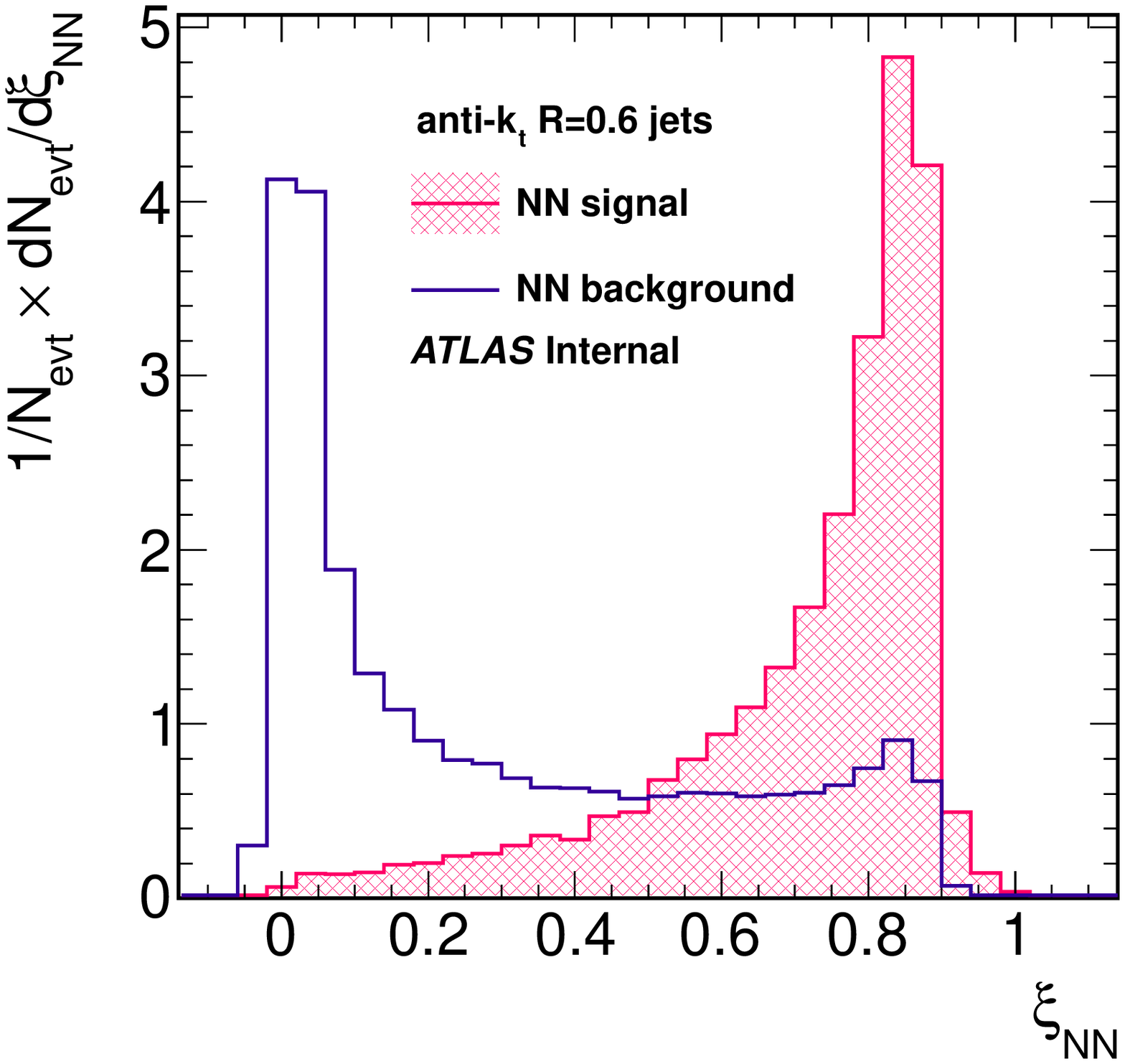}}
\subfloat[]{\label{FIGnnBackgroundSignalResult2}\includegraphics[trim=5mm 14mm 10mm 25mm,clip,width=.52\textwidth]{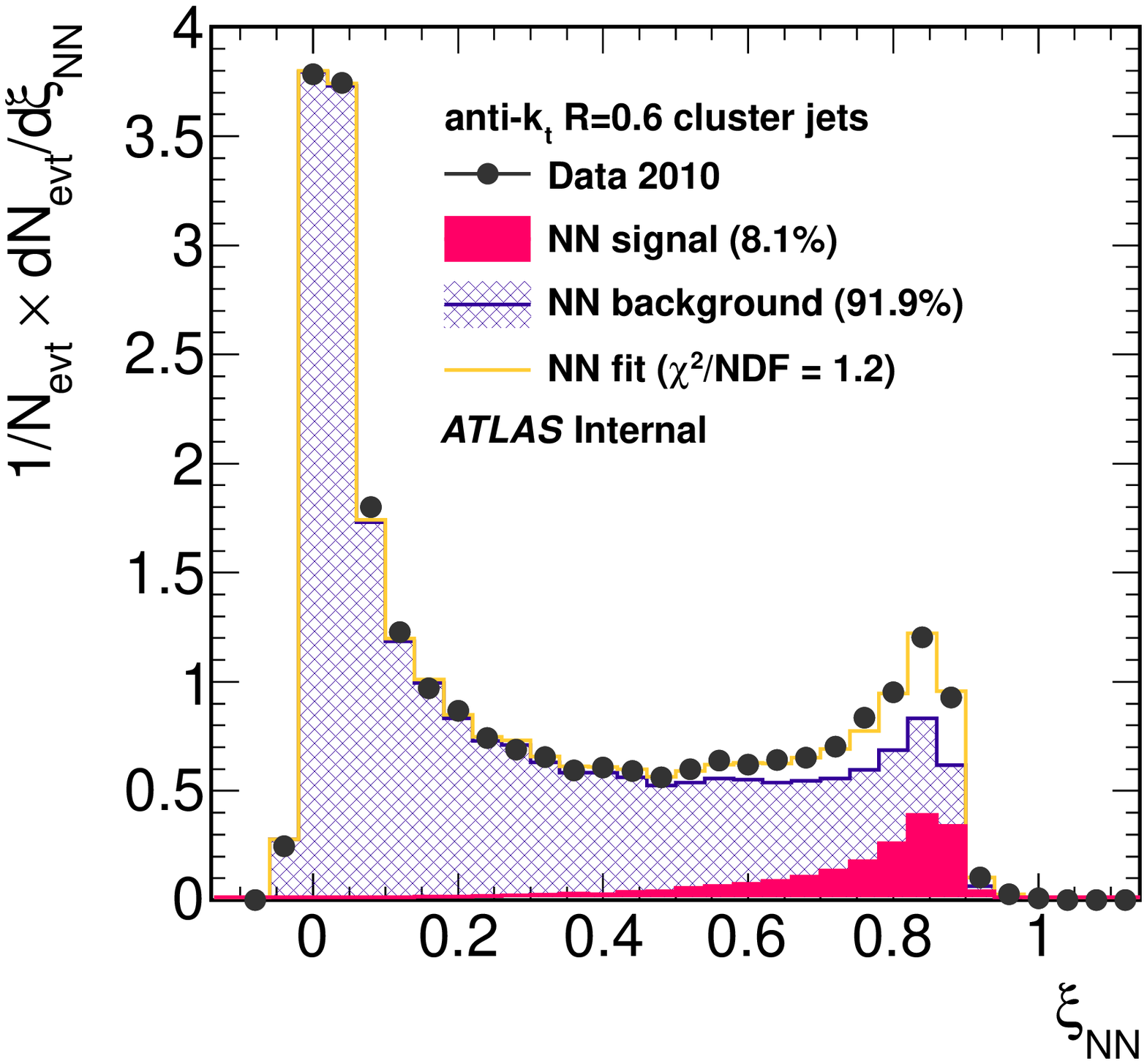}}
  \caption{\label{FIGnnBackgroundSignalResult}\Subref{FIGnnBackgroundSignalResult1} Differential distributions,
  normalized to unity, of the output parameter of the NN, $\xi_{\rm NN}$,
  for the signal and background input samples, as indicated in the figure.\\
  \Subref{FIGnnBackgroundSignalResult2} Differential distributions of $\xi_{\rm NN}$ for
  four-jet events from the 2010 data, for the signal and background input samples of the NN and
  for the sum of the two input samples (denoted by ``NN fit''), where the signal and background samples
  are each normalized according to the fit of the fraction of DPS events, as explained in the text.
  The goodness-of-fit and the number of degrees of freedom
  of the fit are respectively denoted by $\chi^{2}$ and NDF.
  }
\end{center}
\end{figure} 

In order to get a qualitative assessment of the fitted value of \fDPS, normalized distributions of the input variables
to the NN are used.
The data are compared to the expectation, based on a combination of the signal and background samples,
in proportion according to $\fDPS=8.1\%$. The results are shown in
\autoref{chapDoublePartonScatteringApp}, \autorefs{FIGnnVarFitDeltaPtApp}~-~\ref{FIGnnVarFitDeltaEtaApp}.
A good description of the data in most regions of phase-space is achieved.
The difficulty in extracting the signal from an individual distribution can be inferred from this
comparison. It is only the correlations between the variables, captured by the NN, that allow to improve
the sensitivity to the presence of the signal.

Using the output of the NN, it is possible to construct samples of depleted ($\xi_{\rm NN} \sim 0$) and
enriched ($\xi_{\rm NN} \sim 1$) DPS events. 
The momentum fraction of the two interacting partons, $x$, is computed, as in \autoref{eqMedianInEta},
for the leading and sub-leading jet pairs in four-jet events in the data. 
The distributions of $x$ for depleted and enriched DPS events are compared in \autoref{FIGnnJetPairMomentumFraction}.
\begin{figure}[htp]
\begin{center}
\subfloat[]{\label{FIGnnJetPairMomentumFraction1}\includegraphics[trim=5mm 14mm 10mm 25mm,clip,width=.52\textwidth]{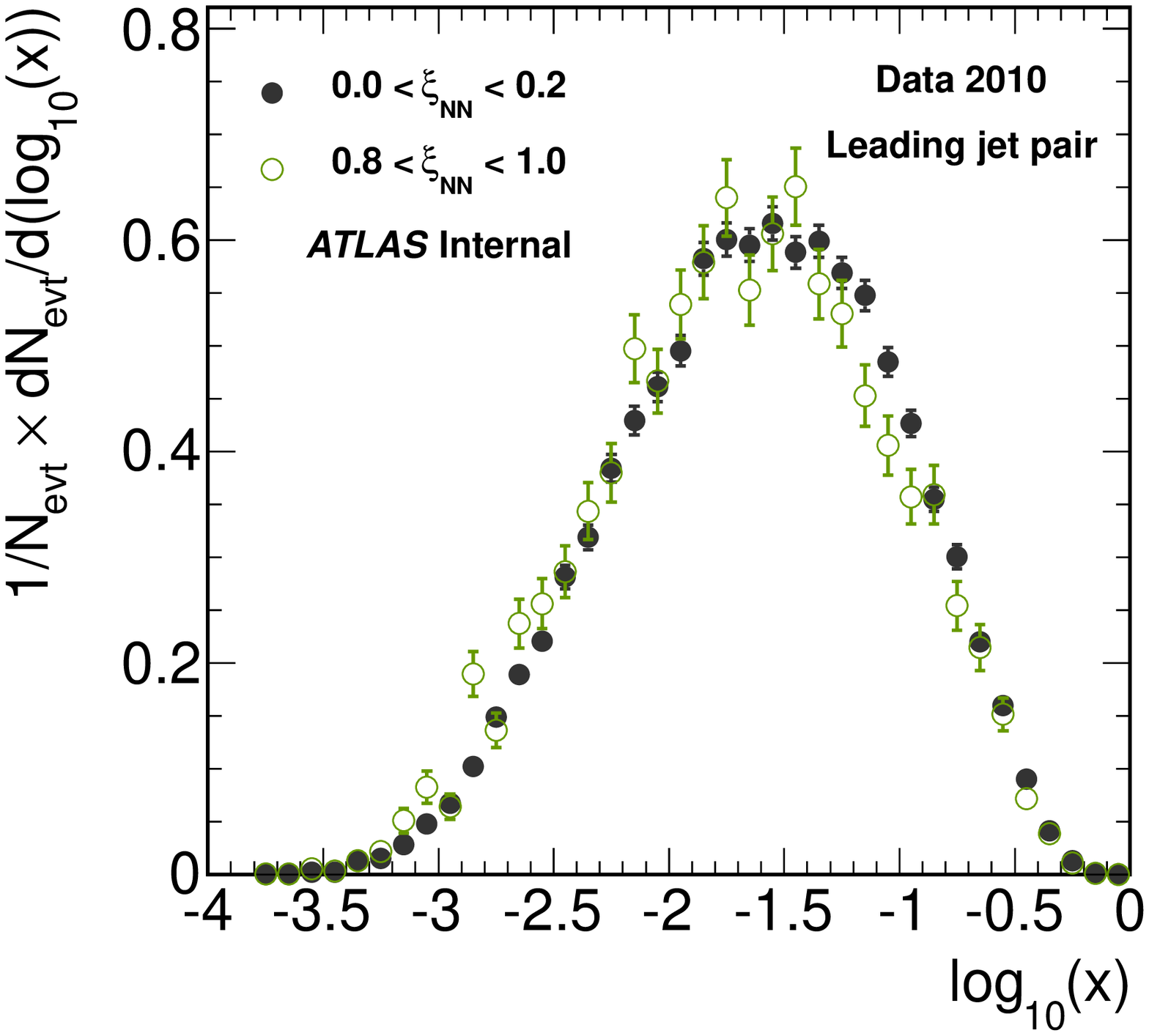}}
\subfloat[]{\label{FIGnnJetPairMomentumFraction2}\includegraphics[trim=5mm 14mm 10mm 25mm,clip,width=.52\textwidth]{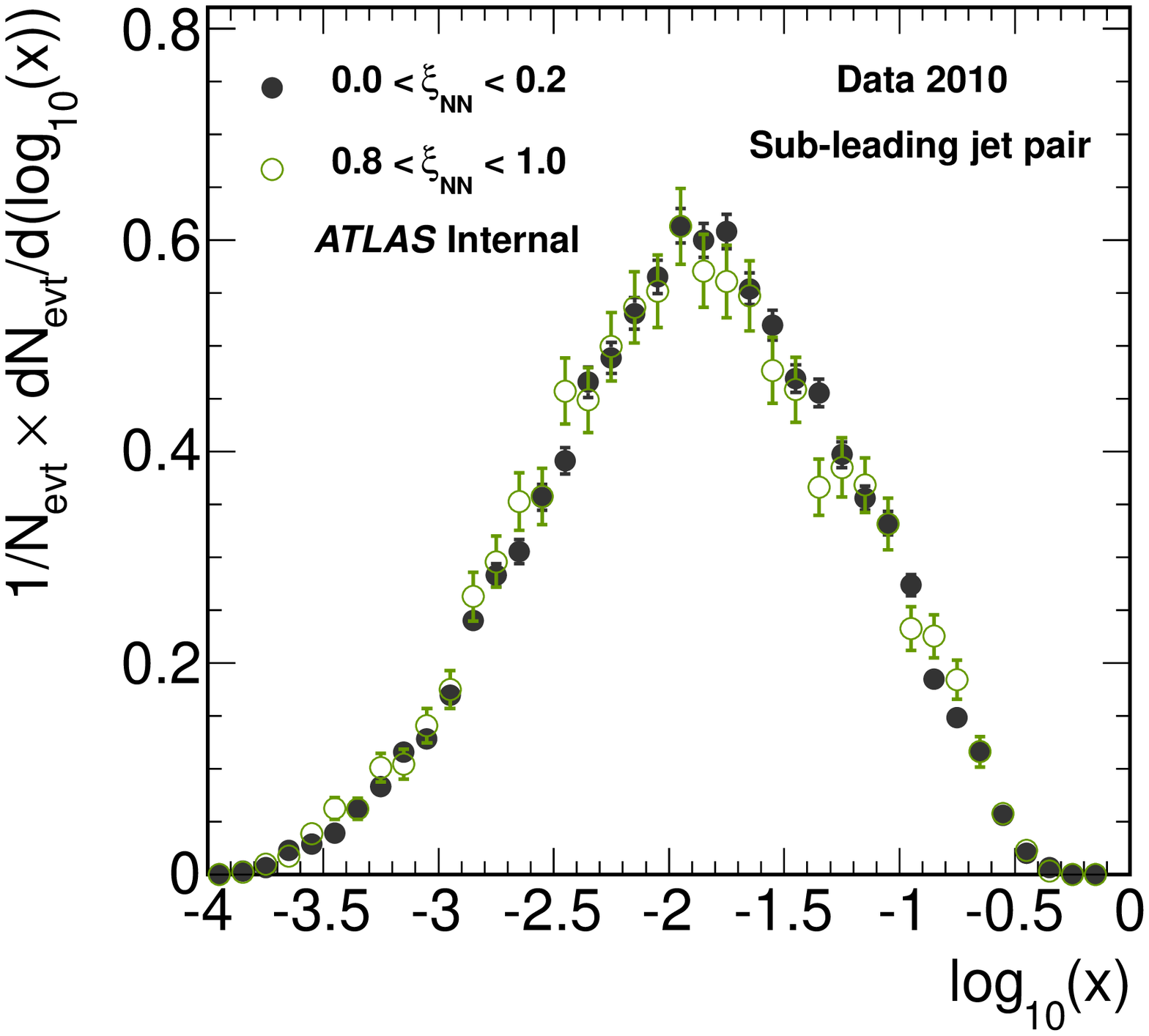}}
  \caption{\label{FIGnnJetPairMomentumFraction}Normalized differential distributions of the
  momentum fraction of the two interacting partons, $x$, associated with the leading \Subref{FIGnnJetPairMomentumFraction1}
  and with the sub-leading \Subref{FIGnnJetPairMomentumFraction2} jet pairs in four-jet events in the 2010 data, for events
  with different values of the output parameter of the NN, $\xi_{\rm NN}$, as indicated in the figures.
  }
\end{center}
\end{figure} 
The four-jet events in the data reach down to momentum fractions, ${x\sim10^{-4}}$.
The sub-leading pair of jets is on average
associated with partons with lower values of $x$ compared to the leading pair, due to the difference in jet \pt.
The sample of enriched DPS events is charechtarised by slightly lower values of $x$ compared to the depleted sample,
as expected.

\section{Systematic and statistical uncertainties\label{chapDpsUncertainty}}
%
%
The measurement of \sigeff is sensitive to several sources of uncertainty, which are correlated in ways which
are hard to evaluate.
The primary uncertainty on \sigeff originates from that on the jet energy scale (\JES), which
influences distributions in the data. The \JES uncertainty
affects both the value of $\mathcal{S}_{4j}^{2j}$, which is directly measured in data,
as well as that of \fDPS, since the fit on the value of the latter is performed using four-jet events from the data.

Additional sources of systematics have to do with the uncertainty in the simulation
of the resolution of jets in energy, in azimuth and in rapidity.
These types of uncertainties affect the value of $\alpha_{2j}^{4j}$, which is directly evaluated in MC, as
well as that of \fDPS, as the MC is used to construct the input samples of the NN.
Estimation of the uncertainties in this case requires for the NN
to be re-trained several times. The subsequent fit for the value of \fDPS on a re-trained network is subject
not only to the uncertainty associated with the MC, but also to fluctuations in the data.
Since the correlations are difficult to disentangle, a conservative approach is taken,
in which they are considered uncorrelated.

\minisec{Uncertainty on the jet energy scale}
%
The procedure for evaluating the uncertainty on the \JES follows that
which is described in \autoref{chapDijetMassSystematicUncertainties}
for the measurement of the dijet mass \xsec. Six sources of uncertainty (see \autoref{chapSystematicUncertaintiesJets})
are considered; single hadron response, cluster thresholds, Perugia 2010, Alpgen+Herwig+Jimmy, intercalibration and relative non-closure.

The effects of the different components of the
\JES uncertainties on the observed \xsecs and on the fitted value of \fDPS are estimated by introducing positive and negative 
variations to the energy scale of jets. The variations are randomly distributed
according to a Gaussian function, with the respective source of uncertainty as its width.
After the energy of jets is varied, they are re-sorted according to \pt, and the event is either rejected, or accepted. Accepted
events fall under one of the classifications of observed \xsecs,
a dijet event of type $a$ or $b$, or a four-jet event of type $b$.
The systematic uncertainty on the \xsecs is estimated as the
difference between the integrated observed \xsecs with and without the \JES variations.

For each source of uncertainty, the value of 
\fDPS is estimated. This is done by fitting a combination of the nominal signal and background samples of the NN
to four-jet data, where the data had been subjected to \JES variations.
The uncertainty on \fDPS is taken as the difference between the result of this fit and the nominal value.

The value of \sigeff is recalculated for every change of the ratio $\mathcal{S}_{4j}^{2j} / \fDPS$, that is
for every component of the \JES uncertainty.
The contributions to the uncertainty on \sigeff from the different sources are considered uncorrelated, and so the total
uncertainty is computed as the quadratic sum of all components.

The uncertainties are listed in \autoref{TBLfullSustUncertDPS}, where those on $\mathcal{S}_{4j}^{2j}$ and on \fDPS
are given separately, for completeness. The \JES uncertainties on \sigeff come out asymmetric, and amount to~$^{+\;5.1}_{-18.4}\%$.
Most of the downward uncertainties on \sigeff originate from the increase of the energy scale in the data, relative to the MC.
The increase in the energy scale results in increasing the number of low-\pt topologies in the data. This leads
to an increase of \fDPS, which turns out to be stronger than the increase in the value of $\mathcal{S}_{4j}^{2j}$.

\minisec{Uncertainty on the properties of jets in the MC}
%
The main source of uncertainty on jets in the MC is that on the jet energy resolution.
The uncertainty on the resolution directly affects $\alpha_{2j}^{4j}$, as it influences the migration of events in and out
of the accepted phase-space of the measurement. 
The response of the NN is less sensitive to the energy resolution in comparison, 
since $\Delta^{\pt}_{12}$ and $\Delta^{\pt}_{34}$ do not have high NN-significance.
On the other hand, the various $\Delta^\phi$ variables, which carry high weights in the NN,
cause the network to be sensitive to the uncertainty on the azimuthal angle of jets in the input samples.

In order to asses the effect of the MC uncertainties on the measurement, the properties of jets in the simulation are distorted.
This is done, either according to the uncertainty on the various resolution
parameters, or by re-weighing the \pt distribution.
The distorted MC samples are used to estimate the acceptance for each class of events,
$\mathcal{A}_{2j}^{a}$, $\mathcal{A}_{2j}^{b}$ and $\mathcal{A}_{4j}^{b}$.
The combination of these leads to a value of $\alpha_{2j}^{4j}$ which is different than the nominal value.
The difference between the two is taken as the associated uncertainty.

In addition, the distorted MCs serve as input for training of the NN. Using the re-trained network,
\fDPS is extracted by fitting  a combination of the distorted signal and background samples
to the nominal sample of four-jet data events. This value of \fDPS is compared to the nominal one
in order to estimate the associated uncertainty.

The difference in the ratio $\alpha_{2j}^{4j} / \fDPS$ compared to the nominal value
is computed for each source of uncertainty and translated into a change in \sigeff.
The different sources are considered as uncorrelated,
and so the quadratic sum of all such elements is taken as the combined uncertainty on \sigeff.

As mentioned above, the result of the fit for \fDPS is sensitive to fluctuations in the four-jet data sample.
These fluctuations are not addressed each time the network is retrained, as the same data are compared to
different combinations of MC samples (also used to train the network).
The assumption that the different elements of the uncertainty are uncorrelated is, therefore, a
conservative estimation of the uncertainty.
The uncertainties are listed in \autoref{TBLfullSustUncertDPS}, where those on $\alpha_{2j}^{4j}$ and on \fDPS
are given separately, for completeness. The MC uncertainties on \sigeff amount to~$^{+\;9.6}_{-11.5}\%$.

\minisec{Statistical uncertainty on the data}
%
The statistical uncertainty on the four-jet data sample and on the signal and background input samples
to the NN, all affect the result of the fit for \fDPS. However, the statistical uncertainty on the data also
affects the value of the observed four-jet \xsec, and accordingly also the value of $\mathcal{S}_{4j}^{2j}$.
%
Instead of calculating the correlation analytically, the combined uncertainty on \sigeff is estimated numerically.
This is done by
smearing the values of each data point in the distributions of $\xi_{\rm NN}$ of the NN inputs, and of the four-jet
data sample. The shifts in value are performed within one standard deviation of the nominal values,
where the magnitude of the deviations is derived from the respective statistical uncertainty.
The smeared samples are then used to perform the fit for the fraction of DPS events, denoted as $\tilde{f}_\mathrm{DPS}$.
The ratio of observed \xsecs is similarly treated (varying $\sigma_{2j}^{a}$ and $\sigma_{2j}^{b}$ in addition to $\sigma_{4j}^{b}$),
resulting in the ``smeared'' ratio of \xsecs, $\tilde{\mathcal{S}}_{4j}^{2j}$.

Using \autoref{EQsimpleSigmaDPSFourJet4} together with the nominal value of the acceptance ratio, one may define the quantity,
\begin{equation}
\tilde{\sigma}_\mathrm{eff} = \frac{ \tilde{\mathcal{S}}_{4j}^{2j} \; \alpha_{2j}^{4j} }{ \tilde{f}_\mathrm{DPS} } \;,
\label{EQsimpleSigmaDPSFourJetSmeared}
\end{equation}
which represents the value of \sigeff, following the smearing procedure.
The fractional bias in the value of \sigeff due to the smearing is then defined as
\begin{equation}
  O_{\tilde{\sigma}_\mathrm{eff}} = \frac{ \tilde{\sigma}_\mathrm{eff} - \sigeff }{ \sigeff } \,.
\label{eqSigmaEffStatBiasDef} \end{equation}
The smearing and fitting steps are performed many times, resulting in a distribution of values of $O_{\tilde{\sigma}_\mathrm{eff}}$,
shown in \autoref{FIGstatisticalSmaringOfSigmaEff}.
The distribution is asymmetric. The final relative uncertainty on \sigeff of~$^{+\;2.9}_{-5.1}\%$, is thus
obtained by finding the ${\pm~35\%}$ confidence region around zero.
\begin{figure}[htp]
\begin{center}
  \includegraphics[trim=5mm 10mm 10mm 25mm,clip,width=.52\textwidth]{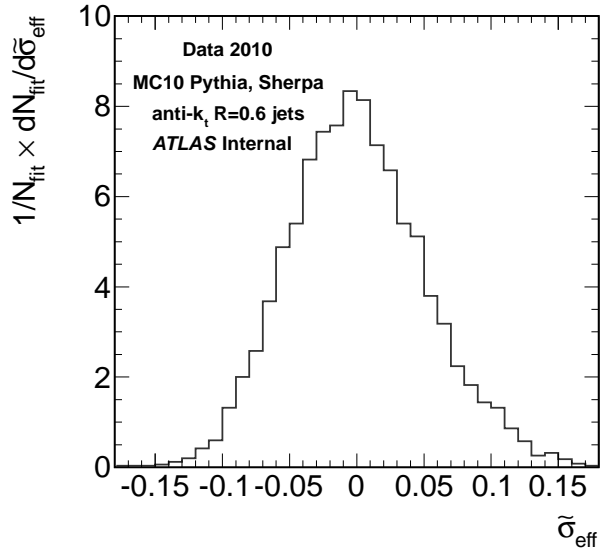}
  \caption{\label{FIGstatisticalSmaringOfSigmaEff}Normalized differential distribution of $O_{\tilde{\sigma}_\mathrm{eff}}$,
  the relative bias in the effective \xsec for DPS, due to the statistical uncertainty on the latter.
  }
\end{center}
\end{figure} 

\minisec{Summary of the uncertainties}
%
%
A summary of the systematic and statistical
uncertainties on the various elements of the measurement is shown in \autoref{TBLfullSustUncertDPS}.
\begin{table}[htp] \begin{center}
\begin{Tabular}[1.8]{ |c|c|c|c|c| } \hline
Source of uncert.\/ in [\%]       &
                             ~~~~~~~$\mathcal{S}_{4j}^{2j}$~~~~~~~ 
                                                            & 
                                                        ~~~~~~~$\alpha_{2j}^{4j}$~~~~~~~
                                                                                 &
                                                                               ~~~~~~~$\fDPS$~~~~~~~
                                                                                                  &
                                                                                               ~~~~~~~$\sigeff$~~~~~~~ 
                                                                                                                 \\ \hhline{=====}
Jet energy scale                  & +10.2 / -8.2            &  --                & +28.9 / -12.8  & +5.1 / -18.4  \\ \hline
Luminosity                        & +3.5 / -3.3             &  --                & --             & +3.5 / -3.3   \\ \hline 
Jet properties in MC              & --                      &  $\pm7$            & $\pm11$        & +9.6 / -11.5  \\ \hline
Limited MC statistics             & --                      &  $\pm3$            & --             & $\pm3$        \\ \hline
Total relative syst.\/ uncert.\/  & +11 / -9                &  $\pm7.5$          & +31 / -17      & +12 / -22     \\ \hhline{=====}
Relative stat.\/ uncert.\/        & $\pm3$                  &  --                & $\pm5$         &  +3 / -5      \\ \hline
\cline{2-5}
\end{Tabular} \end{center}
\caption{\label{TBLfullSustUncertDPS}Summary of the relative systematic and statistical uncertainties
  on the various elements of the DPS measurement;
  the ratio of observed \xsecs, $\mathcal{S}_{4j}^{2j}$,
  the acceptance ratio, $\alpha_{2j}^{4j}$,
  the fraction of DPS events, \fDPS,
  and the effective \xsec for DPS, \sigeff.
  The individual components of the uncertainty are described in the text.
}
\end{table}
The uncertainties on $\mathcal{S}_{4j}^{2j}$, $\alpha_{2j}^{4j}$ and $\fDPS$ are individually derived for completeness.
The uncertainty on $\alpha_{2j}^{4j}$ due to limited MC statistics is propagated to \sigeff as a systematic uncertainty.
In addition to the uncertainties discussed above, the uncertainty on the luminosity is considered as well.

The uncertainties on \sigeff are calculated by taking into account the respective
combination of elements for each source, as previously discussed.
The largest uncertainty on \sigeff is due to that on the \JES.
The final relative uncertainty on $\sigeff$ comes out as~$^{+\;3}_{-5}~\mrm{(stat.)}~^{+\;12}_{-22}~\mrm{(syst.)}~\%$.

\newpage
\section{Results and summary of the measurement\label{chapDpsResultsSummary}}
%
A measurement of hard double parton scattering in four-jet events is performed. The measurement
utilizes a sub-sample of the 2010~\atlas dataset, consisting of single-vertex events
of proton-proton collisions at \com energy, \sqs. Four-jet events are defined
as those in which exactly four jets with transverse momentum, \ptHigher{20}, and
pseudo-rapidity, \etaLower{4.4}, are reconstructed. Events are further constrained such that the
highest-\pt jet has \ptHigher{42.5}.
The rate of DPS events is extracted from data using a neural network, assuming that the topology of the DPS
events is the same as for a random combination of exclusive dijet production.
Complemented by measurements of the dijet and four-jet \xsecs in the appropriate phase-space regions,
the effective \xsec for DPS, \sigeff, is calculated.

The ratio of observed \xsecs,
the acceptance ratio, and the fraction of DPS events, are respectively found to be
\begin{equation*}
        \renewcommand{\arraystretch}{1.75}
        \begin{array}{lclllll}
         \mathcal{S}_{4j}^{2j} &=& 1.48  & \pm~0.04            & \mrm{(stat.)} & ^{+\;0.16}_{-0.13}   & \mrm{(syst.)} \;\; \mrm{mb}  \;\;, \\
         \alpha_{2j}^{4j}      &=& 0.88  &                     &               & \pm~0.07             & \mrm{(syst.)}                                                    \;\;, \\
         \fDPS                 &=& 0.081 & \pm~0.004           & \mrm{(stat.)} & ^{+\;0.025}_{-0.014} & \mrm{(syst.)}             \;\;.
        \end{array}
\end{equation*}
The combination of these yields
\begin{equation}
\sigeff = 16.0~^{+\;0.5}_{-0.8}~\mrm{(stat.)}~^{+\;1.9}_{-3.5}~\mrm{(syst.)}~\mrm{mb}  \;.
\label{EQsimpleSigmaDPSFourJetSmeared}
\end{equation}
This result is consistent with previous measurements, performed in \atlas and 
in other experiments~\cite{Åkesson:173908,Alitti1991145,PhysRevD.47.4857,PhysRevD.56.3811,Abazov:2009gc,Sadeh:1498427},
all of which are summarized in \autoref{FIGallMeasurementsSigmaEff}.

The measured value of \sigeff is smaller than would have been expected from the gluon form factor of the proton.
Following the suggestion in~\cite{Blok:2012mw}, an attempt in the signal-enhanced region of the NN ($\xi_{\rm NN} > 0.8$), was made to look
for the \threefour topology discussed in \autoref{chapMultiPartonInteractionsInGenerators},
in which the balance in \pt should be correlated between the two pairs of jets.
No correlation was observed between $\Delta^{\pt}_{12}$ and $\Delta^{\pt}_{34}$ in this region, which does not
preclude the existence of the \threefour process, as it might be obscured by the resolution in the $\Delta^{\pt}$ variables.

\begin{figure}[ht]
\begin{center}
  \includegraphics[trim=15mm 5mm 20mm 20mm,clip,width=.99\textwidth]{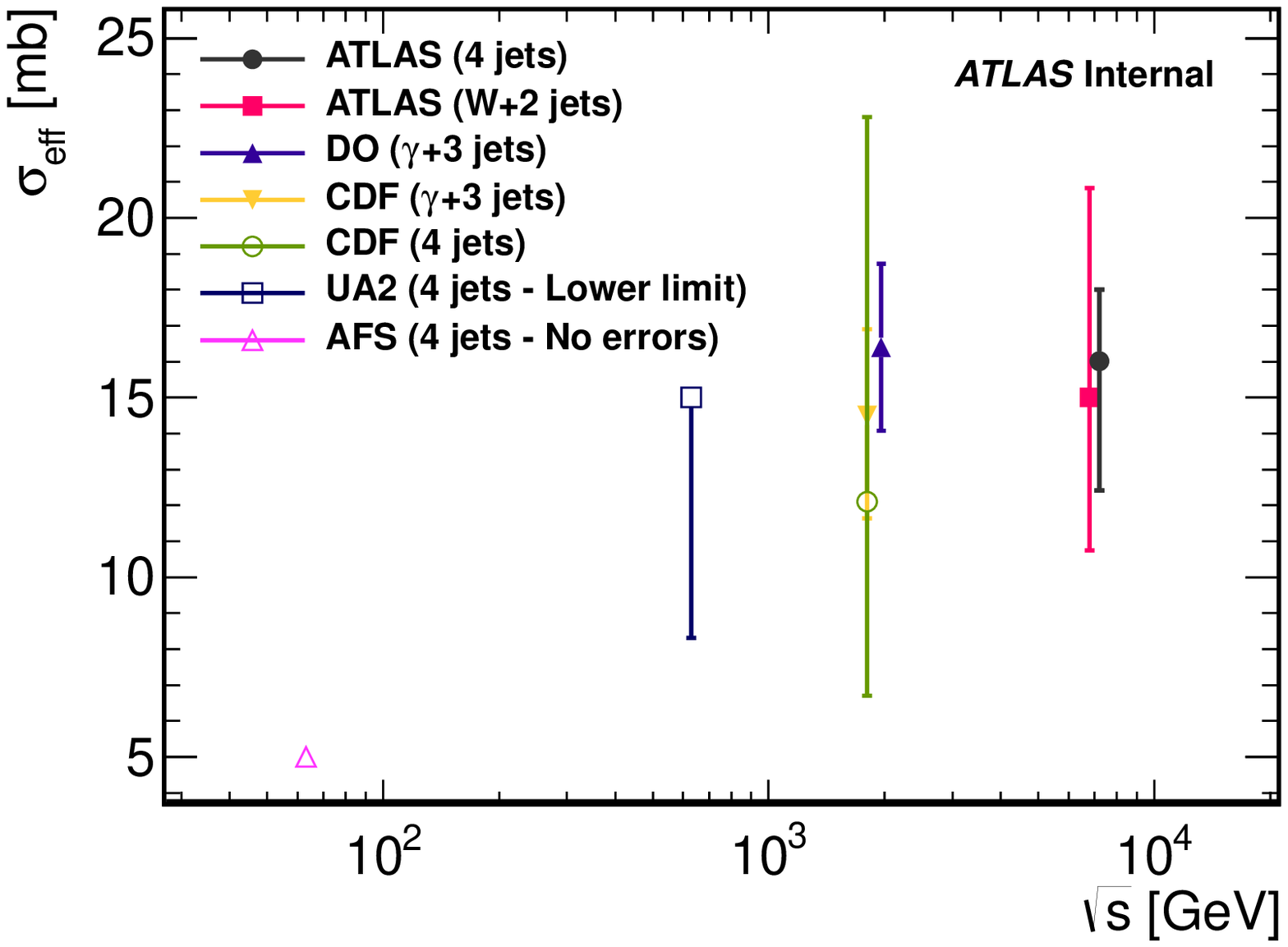}
  \caption{\label{FIGallMeasurementsSigmaEff}Dependence of the effective \xsec for DPS, \sigeff,
  on \com energy, $\sqrt{s}$, in different processes and different
  experiments, including the current measurement (\ATLAS~(4~jets)), as well as those described in \autoref{chapMultiPartonInteractionsInGenerators},
  \autoref{sigmaEffTable}~\protect\cite{Åkesson:173908,Alitti1991145,PhysRevD.47.4857,PhysRevD.56.3811,Abazov:2009gc,Sadeh:1498427}.
  The error bars on the data points represent the statistical and systematic uncertainties, added in quadrature.
  The two \atlas measurements at \sqs are slightly shifted in \com energy, in order to improve
  the visual distinction between the two data points.
  }
\end{center}
\end{figure} 

             }{}
\ifthenelse{\boolean{do:summary}}           { 
\chapter{Summary\label{chapSummary}}
%
%
Jet production is the dominant hard scattering process in the Standard Model.
As such, final states which involve high-\pt jets are an important background
for many searches for new physics, \eg those which involve the decay products of heavy objects.
The study of jets also provides an ideal avenue to probe
QCD and parton distribution functions, which describe the distribution of the momenta of quarks and gluons within a proton.
The two measurements presented in this thesis, of the dijet double-differential \xsec and of double
parton scattering in four-jet events, provide insight into the high-energy regime of jet production.

The dijet double-differential \xsec is measured as a function of the dijet invariant mass, for
various \com jet rapidities, \ystr. 
It uses \AKT, $R=0.6$ jets in data taken during 2010 and during 2011 with the \atlas experiment at a \com energy, \sqs.
Jets in the 2011 dataset are characterized by severe background of additional proton-proton collisions (\pu). A new technique
for \pu subtraction, the jet area/median method, is developed and used in the analysis.
The measurements are sensitive to dijet masses between~$70$ and~${4.27\TeV}$,
including \ystr values up to~3.5. The data are compared to fixed-order NLO pQCD predictions, using \nlojet with the CT10 PDF set, and
corrected for non-perturbative effects. 
The results, which span~12 orders of magnitude, agree well with the predictions within the theoretical and experimental uncertainties.
These represent stringent tests of pQCD in an energy regime previously unexplored.

A measurement of hard double parton scattering in four-jet events is also performed, using
a sub-sample of the 2010~\atlas dataset. Here four-jet events are defined
as those in which exactly four jets with transverse momentum, \ptHigher{20}, and
pseudo-rapidity, \etaLower{4.4}, are reconstructed, where the highest-\pt jet is further constrained to have \ptHigher{42.5}.
The rate of double parton scattering events is estimated using a neural network.
A clear signal is observed, under the assumption
that the DPS signal can be represented by a random combination of exclusive dijet production.
The fraction of DPS candidate events is determined to be
$\fDPS = 0.081~\pm~0.004~\mrm{(stat.)}~^{+\;0.025}_{-0.014}~\mrm{(syst.)}$ in the analyzed phase-space of
four-jet topologies.
Combined with measurements of the dijet and four-jet \xsecs in the appropriate phase-space regions,
the effective \xsec is found to be,
${\sigeff = 16.0~^{+\;0.5}_{-0.8}~\mrm{(stat.)}~^{+\;1.9}_{-3.5}~\mrm{(syst.)}~\mrm{mb}}$.
This result is consistent within the quoted uncertainties with previous measurements of \sigeff
at \com energies between~63\GeV and~7\TeV, using several final states.

                }{}
\ifthenelse{\boolean{do:plotAppendix}}      { \appendix 


%

\ifthenelse {\boolean{do:jetReconstruction}} {
\begin{table}[htp]
%
\chapter{Additional figures and tables\label{sectPlotAppendix}}
%
Auxiliary figures and tables, pertaining to different chapters in the thesis, are presented in the following.

  \section{Jet reconstruction and calibration\label{chapJetReconstructionApp} }
  The quality criteria for jet selection, discussed in \autoref{chapJetReconstruction}, \autoref{chapJetQualitySelection},
  are summarized in \autoref{TBLjetSelectionCriteriaApp}.
  \vspace{-20pt}
\begin{center}
\begin{Tabular}[1.6]{|c||c|c|  } 
\multicolumn{3}{c}{ } \\ [-5pt] \hline
\multicolumn{3}{|c|}{ 2010 } \\
\hline \hline
           & \ttt{Loose} & \ttt{Medium} \\
\hline \hline
\multirow{2}{*}{HEC spikes}
           & $\left(f_{\mrm{HEC}} > 0.5 \;,\;  \left| \mathcal{S}_{\mrm{HEC}}  \right| > 0.5 \right)$
           &  \ttt{Loose} \textbf{OR} \\
           & \textbf{OR} $ \left| E_{\rm neg}  \right| > 60$\GeV 
           & $f_{\mrm{HEC}} > 1-  \left| \mathcal{S}_{\mrm{HEC}}  \right| $  \\
\hline
Coherent  & \multirow{2}{*}{$\left(f_{\mrm{EM}} > 0.95 \;,\; f_{{\rm qlt}} > 0.8 \;,\;  \left| \eta_{\mrm{jet}}  \right| < 2.8\right)$}  & \ttt{Loose} \textbf{OR} \\
\EM noise  &                                                              & $\left(f_{\mrm{EM}} > 0.9 \;,\; \mathcal{S}_{\mrm{EM}}>0.8 \;,\;  \left| \eta_{\mrm{jet}}  \right| < 2.8\right)$ \\
\hline
               & $ \left| t_{\rm jet}  \right| > 25$\ns \textbf{OR}  &  \ttt{Loose} \textbf{OR}\\
Non-collision  & $\left(f_{\mrm{EM}} < 0.05 \;,\; f_{\rm ch} < 0.05 \;,\;  \left| \eta_{\mrm{jet}}  \right| < 2 \right)$  & $ \left| t_{\rm jet}  \right| > 10$\ns \\
background   
               & \textbf{OR} $\left(f_{\mrm{EM}} < 0.05 \;,\;  \left| \eta_{\mrm{jet}}  \right| \geq 2\right)$ & \textbf{OR} $\left(f_{\mrm{EM}} < 0.05 \;,\; f_{\rm ch} < 0.1 \;,\;  \left| \eta_{\mrm{jet}}  \right| < 2\right)$  \\
            
               & \textbf{OR} $\left(f_{\rm max} > 0.99 \;,\;  \left| \eta_{\mrm{jet}}  \right| < 2\right)$  & \textbf{OR} $\left(f_{\mrm{EM}} > 0.95 \;,\; f_{\rm ch} < 0.05 \;,\;  \left| \eta_{\mrm{jet}}  \right| < 2\right)$ \\
\hline 
\end{Tabular}
\begin{Tabular}[1.8]{|c||c|  } 
\multicolumn{2}{c}{ } \\ [-15pt]\hline
\multicolumn{2}{|c|}{ 2011 } \\

\hline \hline
           & \ttt{Loose}  \\
\hline \hline
HEC spikes & 
            $\left(f_{\mrm{HEC}} > 0.5 \;,\;  \left| \mathcal{S}_{\mrm{HEC}}  \right| > 0.5 \right)$ 
            \textbf{OR} $ \left| E_{\rm neg}  \right| > 60$\GeV   \\ \hline
Coherent \EM noise &$\left(f_{\mrm{EM}} > 0.95 \;,\; f_{{\rm qlt}} > 0.8 \;,\;  \left| \eta_{\mrm{jet}}  \right| < 2.8\right)$   \\ \hline
Non-collision background & $ \left| t_{\rm jet}  \right| > 25$\ns \\ \hline
\hline 
\end{Tabular}
 \caption{\ttt{Loose} and \ttt{Medium} selection criteria, used to reject fake jets and non-collision background
 in data taken during 2010 and during 2011, as indicated. The pseudo-rapidity of jets is denoted as $\left|\eta_{\mrm{jet}}\right|$.
 (See also \autoref{chapJetQualitySelection}, \autoref{FIGjetQualityCriteriaEff} and accompanying
 text for definition of the various variables.)
 }
\label{TBLjetSelectionCriteriaApp}
\end{center} 
\end{table}
}{} 

\ifthenelse {\boolean{do:dataSelection}} {
\begin{figure}[htp]
  \vspace{-20pt}
  %
  \section{Data selection\label{chapDataSelectionAndCalibrationApp}}
%
  Additional figures and tables pertaining to \autoref{chapDataSelection} are presented in the following;
  the efficiencies for various jet triggers are shown in \autorefs{FIGTrigEffCentralApp}~-~\ref{FIGTrigEffForwardApp};
  the plateau per-jet trigger efficiency for various data taking periods in 2010 is shown in  \autoref{TBLcentralTrigEfficiency2010App};
  the integrated luminosity for different trigger combinations used in the analysis
  is shown in \autorefs{TBLlumiInTrigBinsApp0}~-~\ref{TBLlumiInTrigBinsApp2}. 
  \vspace{-10pt}
\begin{center}
\subfloat[]{\label{FIGTrigEffCentralApp0}\includegraphics[trim=5mm 14mm 0mm 10mm,clip,width=.52\textwidth]{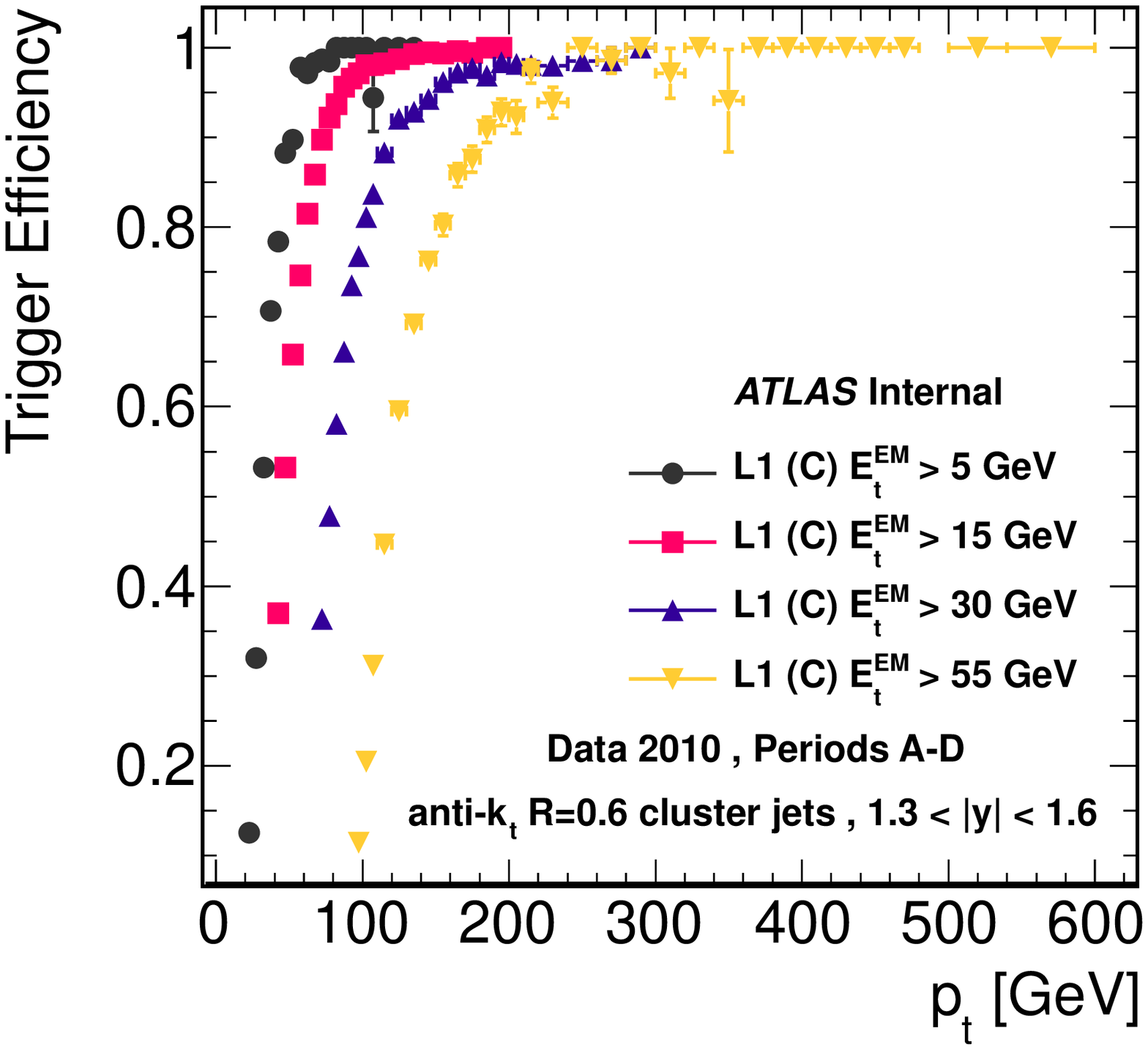}}
\subfloat[]{\label{FIGTrigEffCentralApp1}\includegraphics[trim=5mm 14mm 0mm 10mm,clip,width=.52\textwidth]{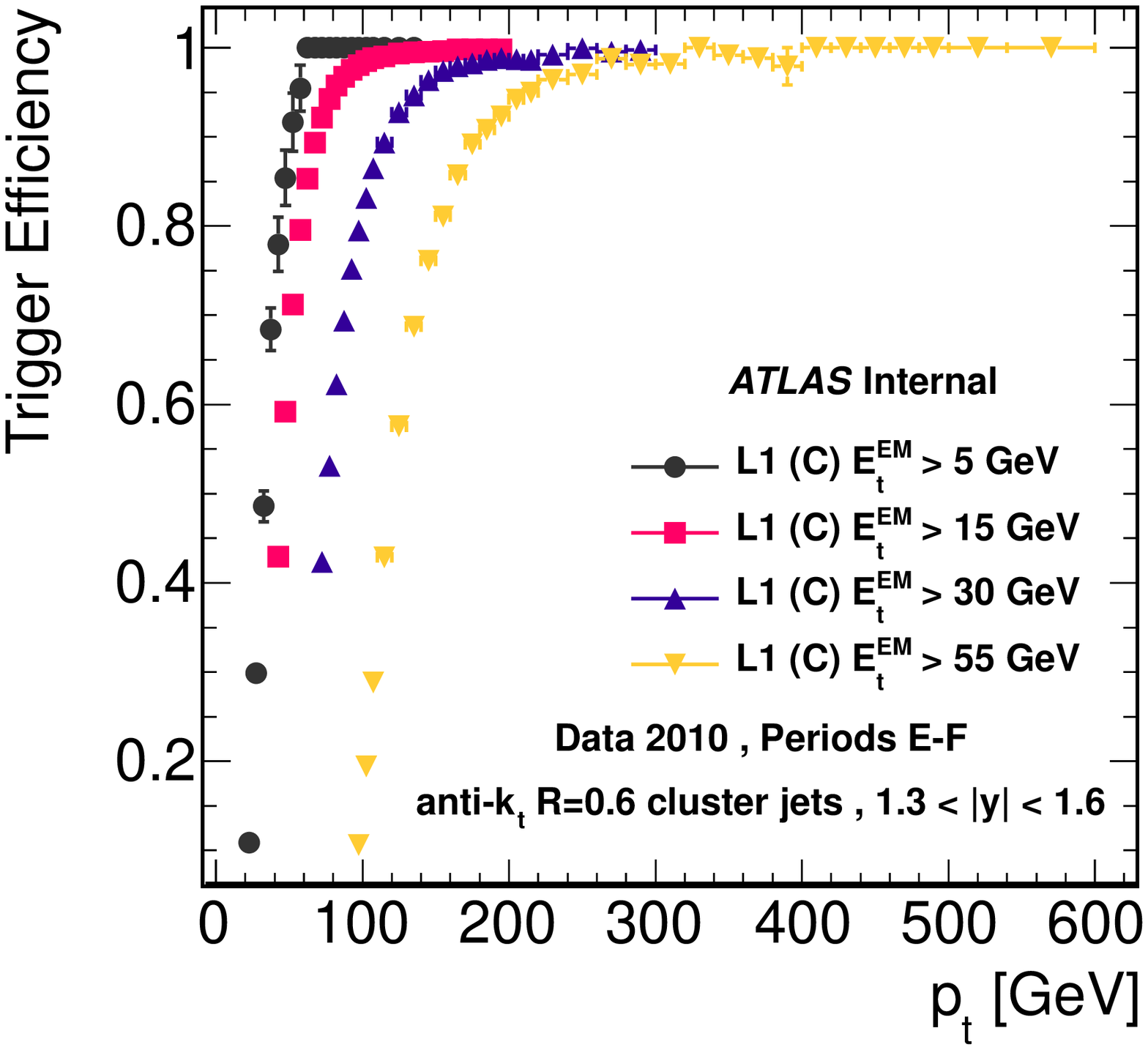}} \\
\subfloat[]{\label{FIGTrigEffCentralApp2}\includegraphics[trim=5mm 14mm 0mm 10mm,clip,width=.52\textwidth]{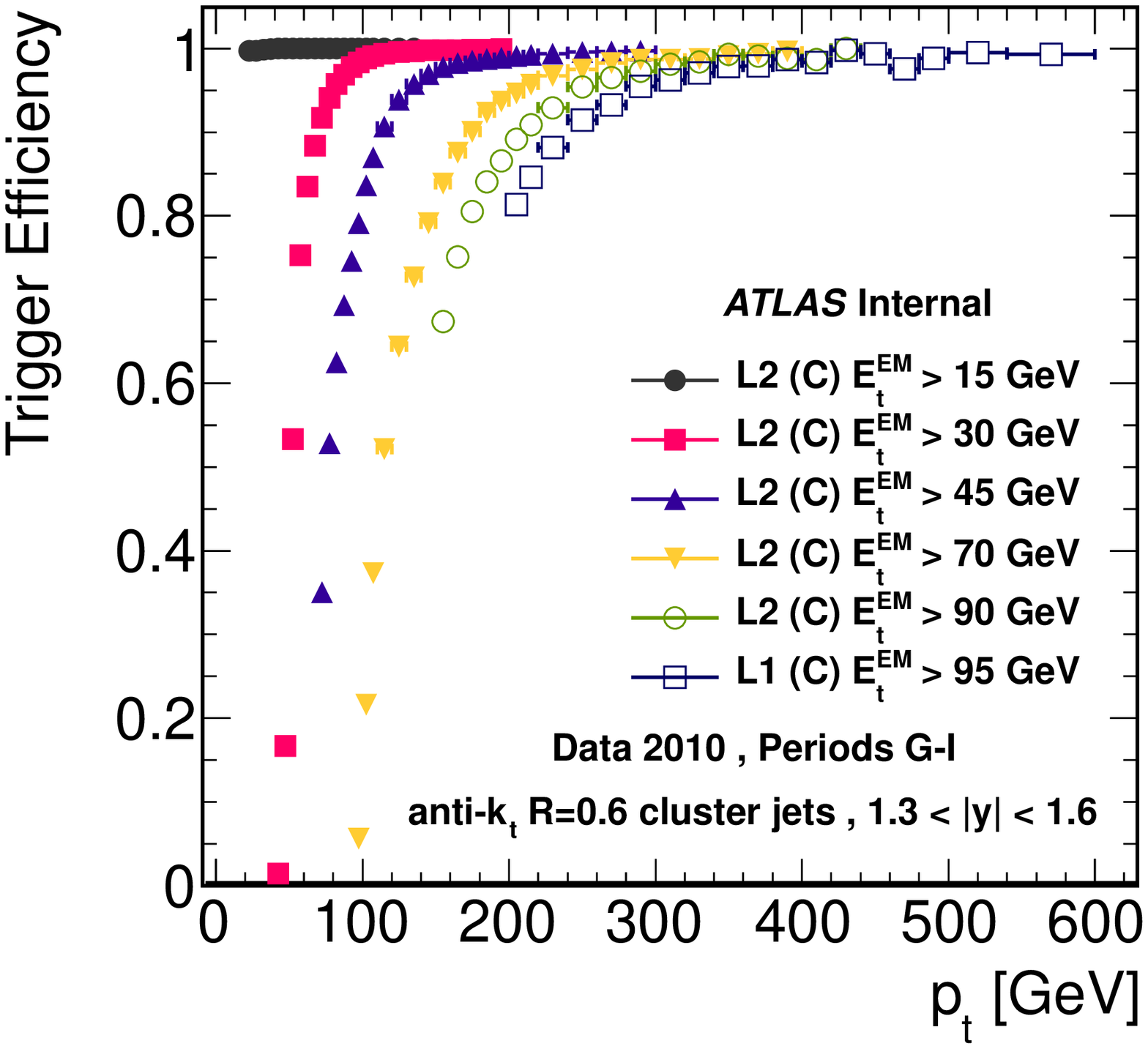}}
\subfloat[]{\label{FIGTrigEffCentralApp3}\includegraphics[trim=5mm 14mm 0mm 10mm,clip,width=.52\textwidth]{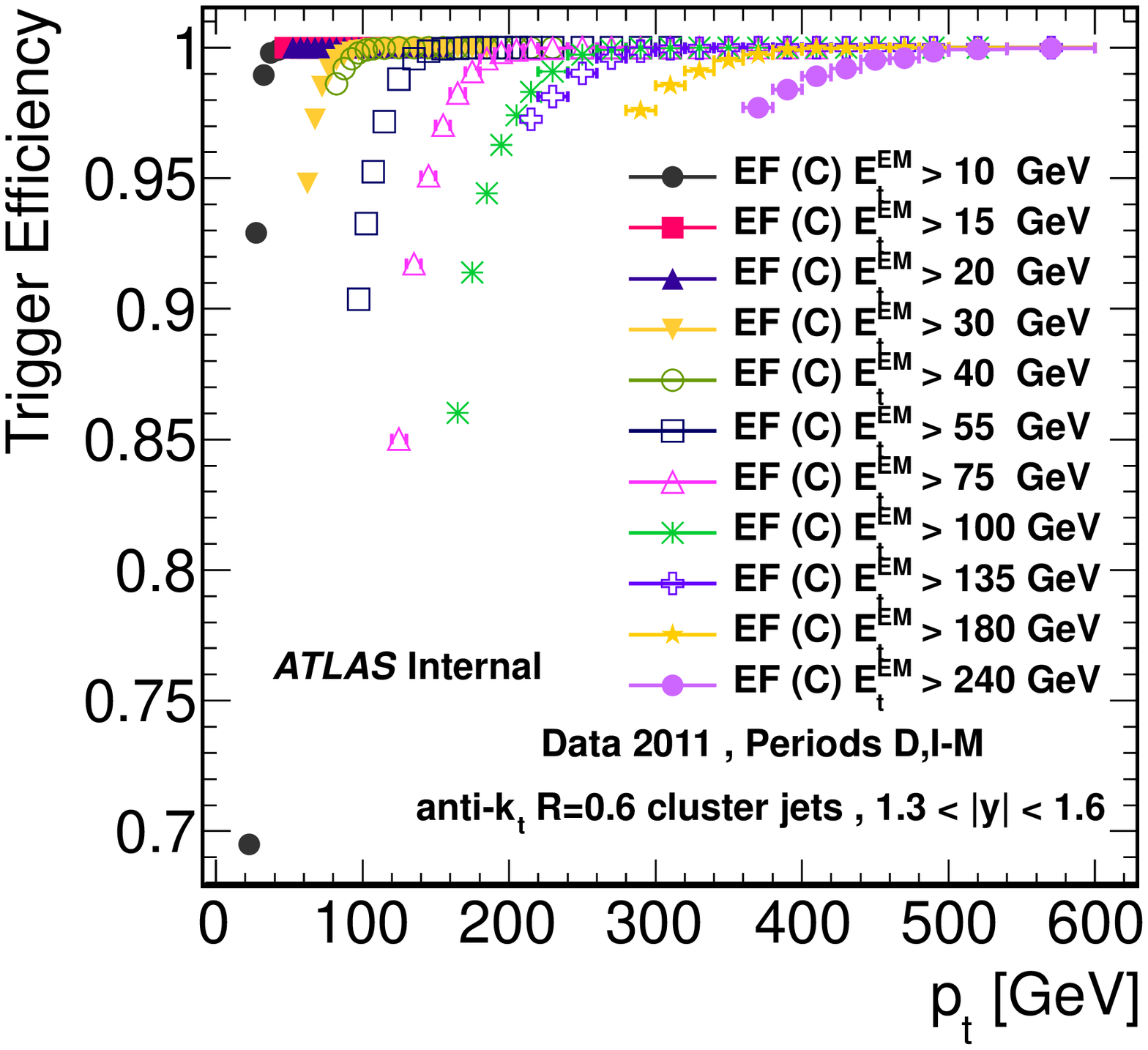}}
\caption{\label{FIGTrigEffCentralApp}Trigger efficiency curves for various Level~1
  (\ttt{L1}), Level~2 (\ttt{L2}) and Event Filter (\ttt{EF}) triggers. Several 2010 (\Subref{FIGTrigEffCentralApp0},
  \Subref{FIGTrigEffCentralApp1} and \Subref{FIGTrigEffCentralApp2}) and 2011 \Subref{FIGTrigEffCentralApp3} data
  taking periods are represented, as indicated in the figures.
  The triggers are associated with jets within rapidity $1.3 < \left|y\right| < 1.6$, belonging to the
  central jet trigger system, denoted by (C).
  Trigger thresholds are denoted by $E_{\mrm{t}}^{\mrm{EM}}$,
  signifying the minimal transverse energy at the electromagnetic scale, of a jet which is required to fire the trigger.
  (See also \autoref{chapDataSelection}, \autoref{FIGTrigEffCentral} and accompanying text.)
}
\end{center}
\end{figure} 
\begin{figure}[htp]
\begin{center}
\subfloat[]{\label{FIGTrigEffForwardApp1}\includegraphics[trim=5mm 14mm 0mm 10mm,clip,width=.52\textwidth]{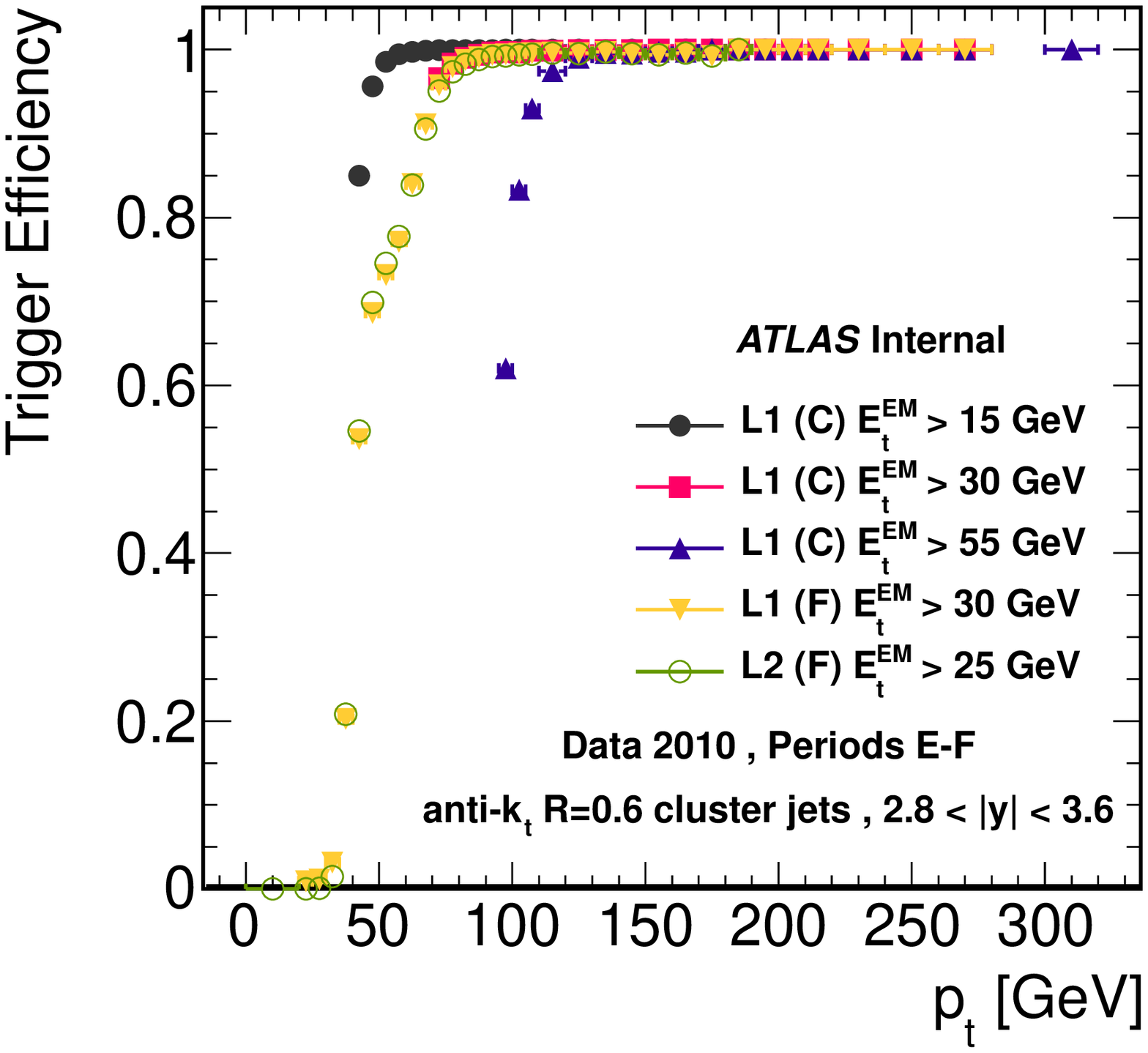}}
\subfloat[]{\label{FIGTrigEffForwardApp2}\includegraphics[trim=5mm 14mm 0mm 10mm,clip,width=.52\textwidth]{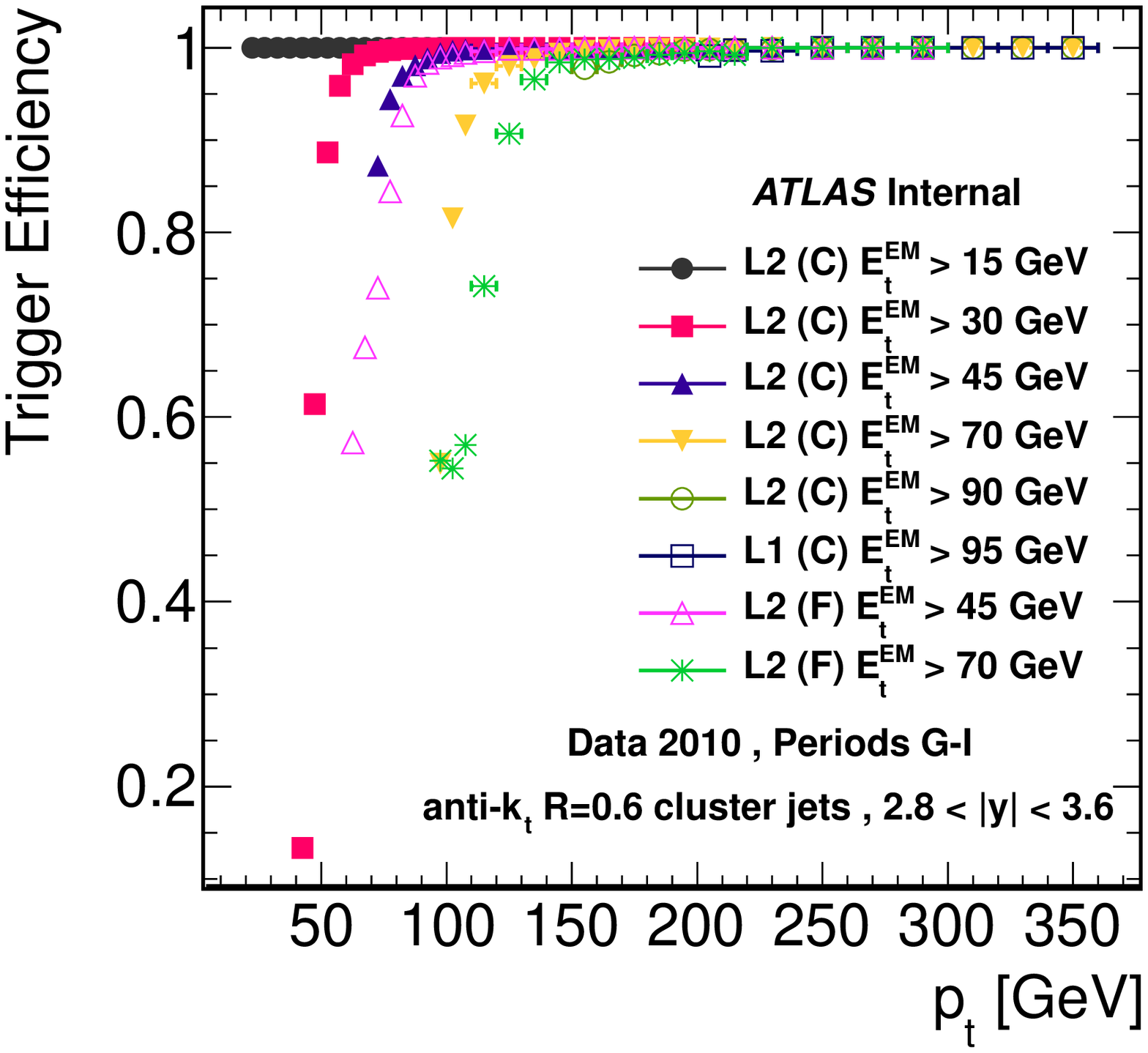}} \\
\subfloat[]{\label{FIGTrigEffForwardApp3}\includegraphics[trim=5mm 14mm 0mm 10mm,clip,width=.52\textwidth]{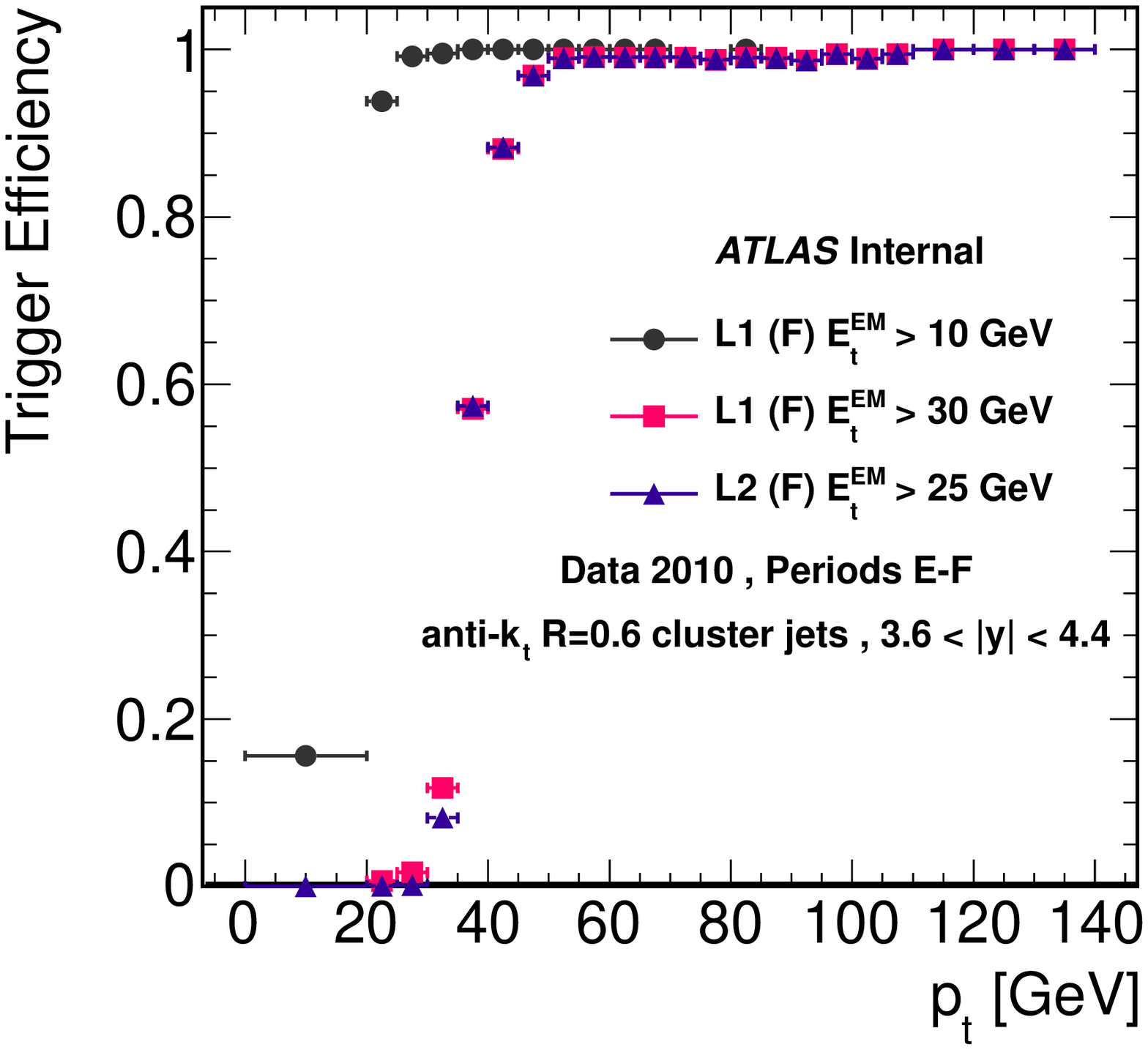}}
\subfloat[]{\label{FIGTrigEffForwardApp4}\includegraphics[trim=5mm 14mm 0mm 10mm,clip,width=.52\textwidth]{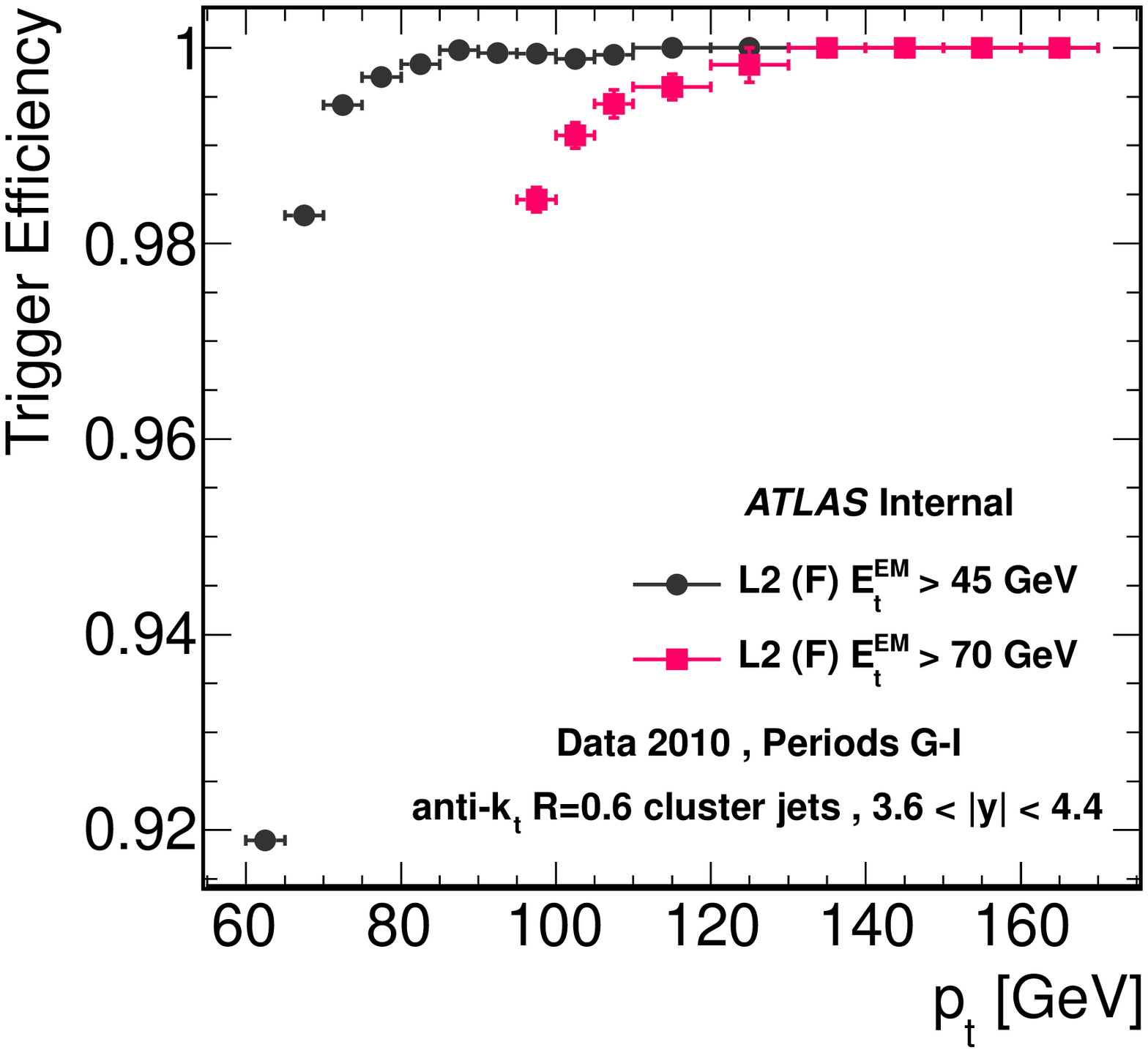}}
\caption{\label{FIGTrigEffForwardApp}Trigger efficiency curves for various Level~1
  (\ttt{L1}) and Level~2 (\ttt{L2}) triggers. Several 2010 data taking periods are represented, as indicated in the figures.
  The triggers are associated with jets within rapidity $2.8 < \left|y\right| < 3.6$ (\Subref{FIGTrigEffForwardApp1} and \Subref{FIGTrigEffForwardApp2})
  and  within rapidity $3.6 < \left|y\right| < 4.4$ (\Subref{FIGTrigEffForwardApp3} and \Subref{FIGTrigEffForwardApp4}),
  belonging to either the central (C) or the forward (F) jet trigger systems.
  Trigger thresholds are denoted by $E_{\mrm{t}}^{\mrm{EM}}$,
  signifying the minimal transverse energy at the electromagnetic scale, of a jet which is required to fire the trigger.
  (See also \autoref{chapDataSelection}, \autoref{FIGTrigEffCentral} and accompanying text.)
}
\end{center}
\end{figure} 
\begin{table}[htp] 
 \begin{center}
\begin{Tabular}[1.3]{ |ccc|c|c|c|c| } \hline \hline
  \multicolumn{3}{|c|}{ \multirow{3}{*}{ \pt~[\GeV] } } & \multicolumn{4}{c|}{Per-jet trigger efficiency} \\ \cline{4-7}
                              &&& Period~A     & Periods~A-D          & Periods~E-F   & \multirow {2}{*}{ Periods~G-I }\\
                              &&& Run $<$ 152777 & (Run $\ge$ 152777) & (excl.~E1-E4) & \\ [2pt]
\hline \hline \hline
\multicolumn{7}{c}{ } \\ [-5pt] \hline
\multicolumn{7}{|c|}{ $1.3 < \left|y\right| < 1.6$ } \\ 
\hline \hline
\multicolumn{1}{|p{24pt}}{\raggedleft20}    &$<\pt\le$& \multicolumn{1}{p{26pt}|}{\raggedright42.5} & 1 & 1    & 1    & 1    \\ \hline
\multicolumn{1}{|p{24pt}}{\raggedleft42.5}  &$<\pt\le$& \multicolumn{1}{p{26pt}|}{\raggedright70}   & 1 & 0.89 & 0.89 & 0.96 \\ \hline
\multicolumn{1}{|p{24pt}}{\raggedleft70}    &$<\pt\le$& \multicolumn{1}{p{26pt}|}{\raggedright97.5} & 1 & 0.88 & 0.88 & 0.87 \\ \hline
\multicolumn{1}{|p{24pt}}{\raggedleft97.5}  &$<\pt\le$& \multicolumn{1}{p{26pt}|}{\raggedright152.5}& 1 & 0.81 & 0.81 & 0.83 \\ \hline
\multicolumn{1}{|p{24pt}}{\raggedleft152.5} &$<\pt\le$& \multicolumn{1}{p{26pt}|}{\raggedright197.5}& 1 & 0.83 & 0.83 & 0.82 \\ \hline
\multicolumn{1}{|p{24pt}}{\raggedleft197.5} &$<\pt\le$& \multicolumn{1}{p{26pt}|}{\raggedright217.5}& 1 & 0.83 & 0.83 & 0.80 \\ \hline
                                     &$\;\;\;\;\pt\ge$& \multicolumn{1}{p{26pt}|}{\raggedright217.5}& 1 & 0.83 & 0.83 & 0.81 \\ \hline
\hline
\multicolumn{7}{c}{ } \\ [-5pt] \hline
\multicolumn{7}{|c|}{ $2.8 < \left|y\right| < 3.6$ } \\
\hline \hline

\multicolumn{1}{|p{24pt}}{\raggedleft20}    &$<\pt\le$& \multicolumn{1}{p{24pt}|}{\raggedright42.5}& 1 & 1 & 1    & 1    \\ \hline
\multicolumn{1}{|p{24pt}}{\raggedleft42.5}  &$<\pt\le$& \multicolumn{1}{p{24pt}|}{\raggedright62.5}& 1 & 1 & 1    & 1    \\ \hline
\multicolumn{1}{|p{24pt}}{\raggedleft62.5}  &$<\pt\le$& \multicolumn{1}{p{24pt}|}{\raggedright72.5}& 1 & 1 & 1    & 0.99 \\ \hline
\multicolumn{1}{|p{24pt}}{\raggedleft72.5}  &$<\pt\le$& \multicolumn{1}{p{24pt}|}{\raggedright95}  & 1 & 1 & 0.97 & 0.99 \\ \hline
\multicolumn{1}{|p{24pt}}{\raggedleft95}    &$<\pt\le$& \multicolumn{1}{p{24pt}|}{\raggedright160} & 1 & 1 & 0.97 & 0.99 \\ \hline
                                     &$\;\;\;\;\pt\ge$& \multicolumn{1}{p{24pt}|}{\raggedright160} & 1 & 1 & 0.97 & 1    \\ \hline

\hline
\multicolumn{7}{c}{ } \\ [-5pt] \hline
\multicolumn{7}{|c|}{$3.6 < \left|y\right| < 4.4$ } \\
\hline \hline

\multicolumn{1}{|p{24pt}}{\raggedleft20}    &$<\pt\le$& \multicolumn{1}{p{24pt}|}{\raggedright42.5}& 1 & 1 & 0.95 & 1    \\ \hline
\multicolumn{1}{|p{24pt}}{\raggedleft42.5}  &$<\pt\le$& \multicolumn{1}{p{24pt}|}{\raggedright50}  & 1 & 1 & 0.95 & 0.99 \\ \hline
\multicolumn{1}{|p{24pt}}{\raggedleft50}    &$<\pt\le$& \multicolumn{1}{p{24pt}|}{\raggedright67.5}& 1 & 1 & 0.95 & 0.99 \\ \hline
\multicolumn{1}{|p{24pt}}{\raggedleft67.5}  &$<\pt\le$& \multicolumn{1}{p{24pt}|}{\raggedright100} & 1 & 1 & 0.95 & 0.97 \\ \hline
                                     &$\;\;\;\;\pt\ge$& \multicolumn{1}{p{24pt}|}{\raggedright100} & 1 & 1 & 0.95 & 0.97 \\ \hline
\hline
\end{Tabular} \end{center}
\caption{\label{TBLcentralTrigEfficiency2010App}The plateau per-jet trigger efficiency for the 2010 data
  as a function of jet transverse momentum, \pt, and rapidity, $y$, in various
  data taking periods, as indicated in the table. Trigger inefficiencies in the crack region,
  within rapidity $1.3 < \left|y\right| < 1.6$, arise due to inhomogeneities in the calorimeter, while those
  in the forward region, $2.8 < \left|y\right| < 4.4$, arise due to a dead FCAL tower.
  (See \autoref{chapTriggerEfficiency} for details.)
}
\end{table}

\begin{table}[p]
\begin{center} \begin{Tabular}[1.3]{ |c||c| }
\multicolumn{2}{c}{ } \\ [-5pt] \hline
\multicolumn{2}{|c|}{ 2010 Period~A , Run $<$ 152777 } \\ 
\hline \hline
Trigger name& \ttt{L1\ul MBTS}\\ \hline \hline
\ttt{L1\ul MBTS} & 1.84$\;\cdot10^{-4}$\\ \hline
\end{Tabular} \end{center}
\begin{center}  \begin{Tabular}[1.3]{ |c||c|c|c|c|c| }
\multicolumn{6}{c}{ } \\ [-5pt] \hline
\multicolumn{6}{|c|}{ 2010 Periods~A-D (Run $\ge$ 152777) } \\ 
\hline \hline
Trigger name      & \ttt{L1\ul MBTS} & \ttt{L1\ul J5} & \ttt{L1\ul J15} & \ttt{L1\ul J30} & \ttt{L1\ul J55} \\ \hline \hline
\ttt{L1\ul MBTS} & 5.62$\;\cdot10^{-4}$& 2.53$\;\cdot10^{-2}$& 2.62$\;\cdot10^{-1}$&  2.62$\;\cdot10^{-1}$& 2.62$\;\cdot10^{-1}$\\ \hline
\ttt{L1\ul J5}&& 2.53$\;\cdot10^{-2}$& 2.62$\;\cdot10^{-1}$&  2.62$\;\cdot10^{-1}$& 2.62$\;\cdot10^{-1}$\\ \hline
\ttt{L1\ul J15}&&& 2.62$\;\cdot10^{-1}$&  2.62$\;\cdot10^{-1}$& 2.62$\;\cdot10^{-1}$\\ \hline
\ttt{L1\ul J30}&&&&  2.62$\;\cdot10^{-1}$& 2.62$\;\cdot10^{-1}$\\ \hline
\ttt{L1\ul J55}&&&&& 2.62$\;\cdot10^{-1}$\\ \hline
\end{Tabular} \end{center}
\begin{center} \begin{Tabular}[1.3]{ |c||c|c|c|c|c|c|c| }
\multicolumn{8}{c}{ } \\ [-5pt] \hline
\multicolumn{8}{|c|}{ 2010 Periods~E-F (excl.~E1-E4) } \\ 
\hline \hline
Trigger name& \ttt{L1\ul MBTS} & \ttt{L1\ul J5} & \ttt{L1\ul J15} & \ttt{L1\ul J30} & \ttt{L1\ul J55} & \ttt{L1\ul FJ10} & \ttt{L1\ul FJ10} \\ \hline \hline
\ttt{L1\ul MBTS} & 7.93$\;\cdot10^{-5}$& 2.46$\;\cdot10^{-3}$& 2.40$\;\cdot10^{-2}$ & 1.10$\;\cdot10^{0}$& 2.17$\;\cdot10^{0}$& 1.55$\;\cdot10^{-2}$& 2.17$\;\cdot10^{0}$\\ \hline
\ttt{L1\ul J5}&& 2.38$\;\cdot10^{-3}$& 2.63$\;\cdot10^{-2}$& 1.10$\;\cdot10^{0}$& 2.17$\;\cdot10^{0}$& 1.78$\;\cdot10^{-2}$& 2.17$\;\cdot10^{0}$\\ \hline
\ttt{L1\ul J15}&&& 2.40$\;\cdot10^{-2}$& 1.10$\;\cdot10^{0}$& 2.17$\;\cdot10^{0}$& 3.87$\;\cdot10^{-2}$& 2.17$\;\cdot10^{0}$\\ \hline
\ttt{L1\ul J30}&&&& 1.10$\;\cdot10^{0}$& 2.17$\;\cdot10^{0}$& 1.10$\;\cdot10^{0}$& 2.17$\;\cdot10^{0}$\\ \hline
\ttt{L1\ul J55}&&&&& 2.17$\;\cdot10^{0}$& 2.17$\;\cdot10^{0}$& 2.17$\;\cdot10^{0}$\\ \hline
\ttt{L1\ul FJ10}&&&&&& 1.54$\;\cdot10^{-2}$& 2.17$\;\cdot10^{0}$\\ \hline
\ttt{L1\ul FJ30}&&&&&&& 2.17$\;\cdot10^{0}$\\ \hline
\end{Tabular} \end{center}
\caption{\label{TBLlumiInTrigBinsApp0}Integrated luminosity in~$\mrm{pb}^{-1}$ for different trigger combinations used in
  several data taking periods in 2010. Data taken up to period~F are included. The uncertainty on the luminosity is 3.4\%.
  (See \autoref{chapTwoTriggerLumiCalcScheme} for details.)
}
\end{table}
\begin{table}[p] \begin{center}
\begin{Tabular}[1.3]{ |c||c|c|c|c|c|c|c|c|c|c| }
\multicolumn{8}{c}{ } \\ [-5pt] \hline
\multicolumn{8}{|c|}{ 2010 Periods~G-I } \\ 
\hline \hline
Trigger name& \ttt{EF\ul MBTS} & \ttt{L2\ul J15} & \ttt{L2\ul J30} & \ttt{L2\ul J45} & \ttt{L2\ul J70} & \ttt{L2\ul J90} & \ttt{L1\ul J95}  \\ \hline \hline
\ttt{EF\ul MBTS} & 9.84$\;\cdot10^{-5}$& 2.41$\;\cdot10^{-3}$&  5.03$\;\cdot10^{-2}$&  2.45$\;\cdot10^{-1}$& 6.27$\;\cdot10^{0}$& 6.00$\;\cdot10^{0}$& 3.44$\;\cdot10^{1}$\\ \hline
\ttt{L2\ul J15}&& 2.33$\;\cdot10^{-3}$&  5.24$\;\cdot10^{-2}$&  2.49$\;\cdot10^{-1}$& 6.27$\;\cdot10^{0}$& 6.00$\;\cdot10^{0}$& 3.44$\;\cdot10^{1}$\\ \hline
\ttt{L2\ul J30}&&&  5.02$\;\cdot10^{-2}$&  2.68$\;\cdot10^{-1}$& 6.29$\;\cdot10^{0}$& 6.02$\;\cdot10^{0}$& 3.44$\;\cdot10^{1}$\\ \hline
\ttt{L2\ul J45}&&&&  2.46$\;\cdot10^{-1}$& 6.33$\;\cdot10^{0}$& 6.06$\;\cdot10^{0}$& 3.44$\;\cdot10^{1}$\\ \hline
\ttt{L2\ul J70}&&&&& 6.27$\;\cdot10^{0}$& 7.72$\;\cdot10^{0}$& 3.44$\;\cdot10^{1}$\\ \hline
\ttt{L2\ul J90}&&&&&& 6.00$\;\cdot10^{0}$& 3.44$\;\cdot10^{1}$\\ \hline
\ttt{L1\ul J95}&&&&&&& 3.44$\;\cdot10^{1}$\\ \hline
\end{Tabular} 
\begin{Tabular}[1.3]{ |c||c|c|c| }
\multicolumn{4}{c}{ } \\ [-5pt] \hline
\multicolumn{4}{|c|}{ 2010 Periods~G-I } \\ 
\hline \hline
Trigger name             & \ttt{L2\ul FJ25} & \ttt{L2\ul FJ45} & \ttt{L2\ul FJ70} \\ \hline \hline
\ttt{EF\ul MBTS} & 1.58$\;\cdot10^{-1}$& 3.73$\;\cdot10^{0}$&  3.44$\;\cdot10^{1}$\\ \hline
\ttt{L2\ul J15}& 1.61$\;\cdot10^{-1}$& 3.73$\;\cdot10^{0}$& 3.44$\;\cdot10^{1}$\\ \hline
\ttt{L2\ul J30}& 1.82$\;\cdot10^{-1}$& 3.75$\;\cdot10^{0}$& 3.44$\;\cdot10^{1}$\\ \hline
\ttt{L2\ul J45}& 3.30$\;\cdot10^{-1}$& 3.81$\;\cdot10^{0}$& 3.44$\;\cdot10^{1}$\\ \hline
\ttt{L2\ul J70}& 6.32$\;\cdot10^{0}$& 7.62$\;\cdot10^{0}$& 3.44$\;\cdot10^{1}$\\ \hline
\ttt{L2\ul J90}& 6.05$\;\cdot10^{0}$& 7.43$\;\cdot10^{0}$& 3.44$\;\cdot10^{1}$\\ \hline
\ttt{L1\ul J95}& 3.44$\;\cdot10^{1}$& 3.44$\;\cdot10^{1}$& 3.44$\;\cdot10^{1}$\\ \hline
\ttt{L2\ul FJ25}& 1.58$\;\cdot10^{-1}$& 3.80$\;\cdot10^{0}$& 3.44$\;\cdot10^{1}$\\ \hline
\ttt{L2\ul FJ45}&& 3.73$\;\cdot10^{0}$& 3.44$\;\cdot10^{1}$\\ \hline
\ttt{L2\ul FJ70}&&& 3.44$\;\cdot10^{1}$\\ \hline
\end{Tabular} \end{center}
\caption{\label{TBLlumiInTrigBinsApp1}Integrated luminosity in~$\mrm{pb}^{-1}$ for the different trigger
  combinations used during periods~G-I in 2010. The uncertainty on the luminosity is 3.4\%.
  (See \autoref{chapTwoTriggerLumiCalcScheme} for details.)
}
\end{table}
\begin{table}[p]
\begin{center} \begin{Tabular}[1.3]{ |c||c|c|c|c|c|c| }
\multicolumn{7}{c}{ } \\ [-5pt] \hline
\multicolumn{7}{|c|}{ 2011 Periods~D,I-M } \\ 
\hline \hline
Trigger name& \ttt{EF\ul rd0} & \ttt{EF\ul J10} & \ttt{EF\ul J15} & \ttt{EF\ul J20} & \ttt{EF\ul J30} & \ttt{EF\ul J40} \\ \hline \hline
\ttt{EF\ul rd0}& 1.48$\;\cdot10^{-3}$&  3.22$\;\cdot10^{-3}$& 9.34$\;\cdot10^{-3}$& 2.58$\;\cdot10^{-2}$& 2.65$\;\cdot10^{-1}$& 2.96$\;\cdot10^{-1}$\\ \hline
\ttt{EF\ul J10}&&  1.74$\;\cdot10^{-3}$& 9.57$\;\cdot10^{-3}$& 2.60$\;\cdot10^{-2}$& 2.64$\;\cdot10^{-1}$& 2.96$\;\cdot10^{-1}$\\ \hline
\ttt{EF\ul J15}&&& 7.83$\;\cdot10^{-3}$& 3.21$\;\cdot10^{-2}$& 2.70$\;\cdot10^{-1}$& 3.02$\;\cdot10^{-1}$\\ \hline
\ttt{EF\ul J20}&&&& 2.43$\;\cdot10^{-2}$& 2.87$\;\cdot10^{-1}$& 3.18$\;\cdot10^{-1}$\\ \hline
\ttt{EF\ul J30}&&&&& 2.62$\;\cdot10^{-1}$& 5.56$\;\cdot10^{-1}$\\ \hline
\ttt{EF\ul J40}&&&&&& 2.94$\;\cdot10^{-1}$\\ \hline
\end{Tabular} \end{center}
\begin{center} \begin{Tabular}[1.3]{ |c||c|c|c|c|c|c| }
\multicolumn{7}{c}{ } \\ [-5pt] \hline
\multicolumn{7}{|c|}{ 2011 Periods~D,I-M } \\ 
\hline \hline
Trigger name    & \ttt{EF\ul J55} & \ttt{EF\ul J75} & \ttt{EF\ul J100} & \ttt{EF\ul J135} & \ttt{EF\ul J180} & \ttt{EF\ul J240} \\ \hline \hline
\ttt{EF\ul rd0}   & 1.05$\;\cdot10^{0}$  & 3.88$\;\cdot10^{0}$  & 1.34$\;\cdot10^{1}$  & 1.24$\;\cdot10^{2}$  & 3.26$\;\cdot10^{2}$ & 3.83$\;\cdot10^{3}$  \\ \hline
\ttt{EF\ul J10}   & 1.05$\;\cdot10^{0}$  & 3.87$\;\cdot10^{0}$  & 1.34$\;\cdot10^{1}$  & 1.24$\;\cdot10^{2}$  & 3.25$\;\cdot10^{2}$ & 3.82$\;\cdot10^{3}$  \\ \hline
\ttt{EF\ul J15}   & 1.06$\;\cdot10^{0}$  & 3.87$\;\cdot10^{0}$  & 1.34$\;\cdot10^{1}$  & 1.24$\;\cdot10^{2}$  & 3.25$\;\cdot10^{2}$ & 3.82$\;\cdot10^{3}$  \\ \hline
\ttt{EF\ul J20}   & 1.07$\;\cdot10^{0}$  & 3.89$\;\cdot10^{0}$  & 1.34$\;\cdot10^{1}$  & 1.24$\;\cdot10^{2}$  & 3.25$\;\cdot10^{2}$ & 3.82$\;\cdot10^{3}$  \\ \hline
\ttt{EF\ul J30}   & 1.31$\;\cdot10^{0}$  & 4.12$\;\cdot10^{0}$  & 1.36$\;\cdot10^{1}$  & 1.24$\;\cdot10^{2}$  & 3.25$\;\cdot10^{2}$ & 3.82$\;\cdot10^{3}$  \\ \hline
\ttt{EF\ul J40}   & 1.34$\;\cdot10^{0}$  & 4.16$\;\cdot10^{0}$  & 1.37$\;\cdot10^{1}$  & 1.24$\;\cdot10^{2}$  & 3.26$\;\cdot10^{2}$ & 3.82$\;\cdot10^{3}$  \\ \hline
\ttt{EF\ul J55}   & 1.05$\;\cdot10^{0}$  & 4.91$\;\cdot10^{0}$  & 1.44$\;\cdot10^{1}$  & 1.25$\;\cdot10^{2}$  & 3.26$\;\cdot10^{2}$ & 3.82$\;\cdot10^{3}$  \\ \hline
\ttt{EF\ul J75}   && 3.86$\;\cdot10^{0}$  & 1.72$\;\cdot10^{1}$  & 1.27$\;\cdot10^{2}$  & 3.28$\;\cdot10^{2}$ & 3.82$\;\cdot10^{3}$  \\ \hline
\ttt{EF\ul J100}  &&& 1.34$\;\cdot10^{1}$  & 1.35$\;\cdot10^{2}$  & 3.35$\;\cdot10^{2}$ & 3.82$\;\cdot10^{3}$  \\ \hline
\ttt{EF\ul J135}  &&&& 1.24$\;\cdot10^{2}$  & 3.61$\;\cdot10^{2}$ & 3.82$\;\cdot10^{3}$  \\ \hline
\ttt{EF\ul J180}  &&&&& 3.25$\;\cdot10^{2}$ & 3.82$\;\cdot10^{3}$  \\ \hline
\ttt{EF\ul J240}  &&&&&& 3.82$\;\cdot10^{3}$ \\ \hline

\end{Tabular} \end{center}
\caption{\label{TBLlumiInTrigBinsApp2}Integrated luminosity in~$\mrm{pb}^{-1}$ for the different trigger combinations
  used throughout 2011 (Periods~D,I-M). The uncertainty on the luminosity is 3.9\%.
  (See \autoref{chapTwoTriggerLumiCalcScheme} for details.)
}
\end{table}
}{} 

\ifthenelse {\boolean{do:jetMedian}} {
  \begin{figure}[htp]
  \vspace{-20pt}
  %
  \section{The jet area/median method for \pu subtraction\label{chapJetAreaMethodApp}}
  %
  Additional figures pertaining to \autoref{chapJetAreaMethod} are presented in the following;
  the dependence of the average median of the transverse momentum density of jets on the number of reconstructed vertices and on the
  average number of interactions is shown in \autorefs{FIGavgRhoNpvInEtaMuBinsApp}~-~\ref{FIGavgRhoMuInEtaMuBinsApp};
  the dependence of the quality criteria for the median \pu correction on the transverse momentum of truth jets
  is shown in \autoref{FIGmeanClosureOffsetNpvMuSlopeApp};
  the \insitu performance qualifiers for the median \pu correction are shown in 
  \autorefs{FIGinSituPtStdNpvApp}~-~\ref{FIGinSituMassStdNpvApp}.
  \vspace{-10pt}
  \begin{center}
  \subfloat[]{\label{FIGavgRhoNpvInEtaMuBinsApp1}\includegraphics[trim=5mm 10mm 65mm 10mm,clip,width=.43\textwidth]{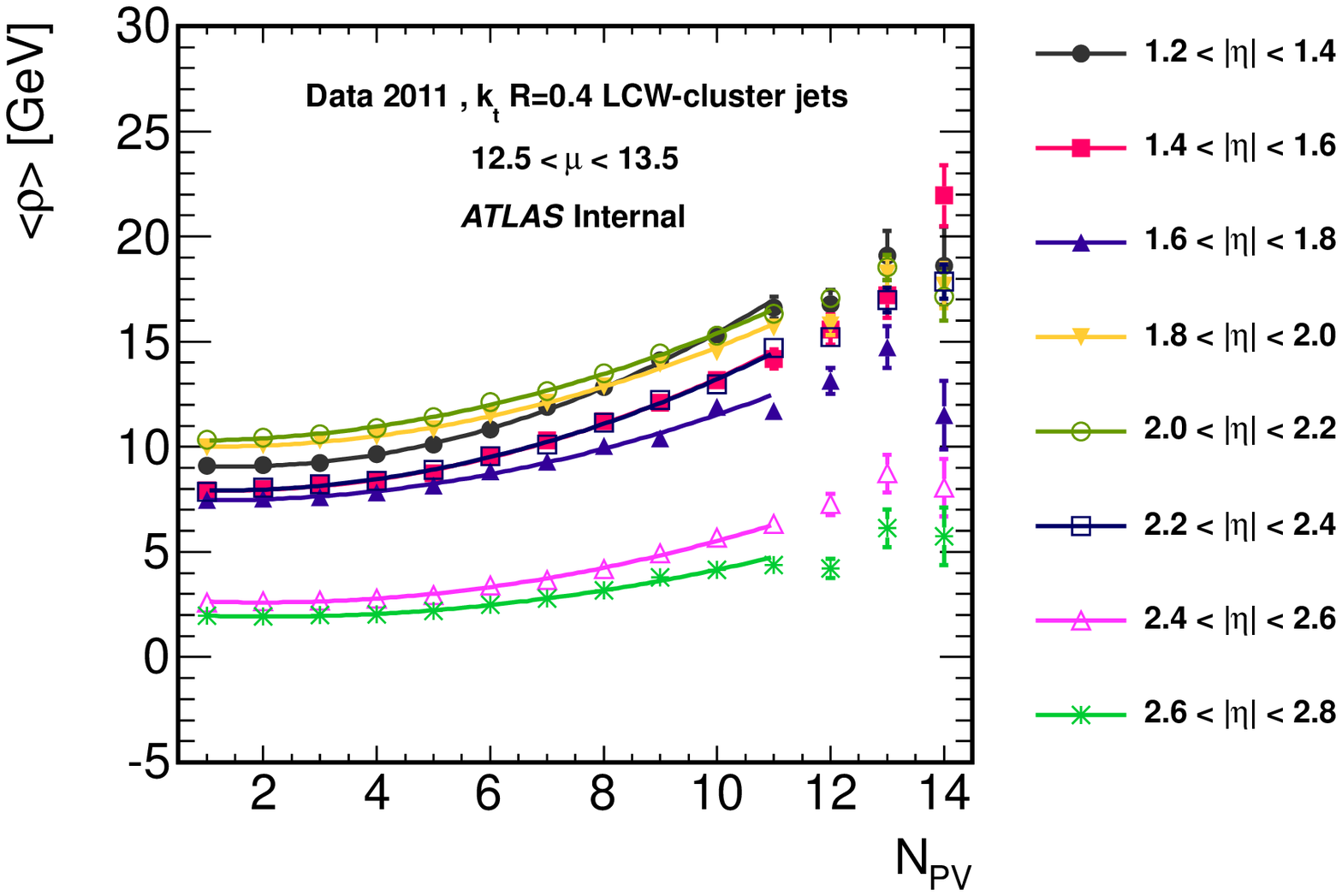}} 
  \subfloat[\figQaud]{\label{FIGavgRhoNpvInEtaMuBinsApp2}\includegraphics[trim=5mm 10mm 0mm 10mm,clip,width=.643\textwidth]{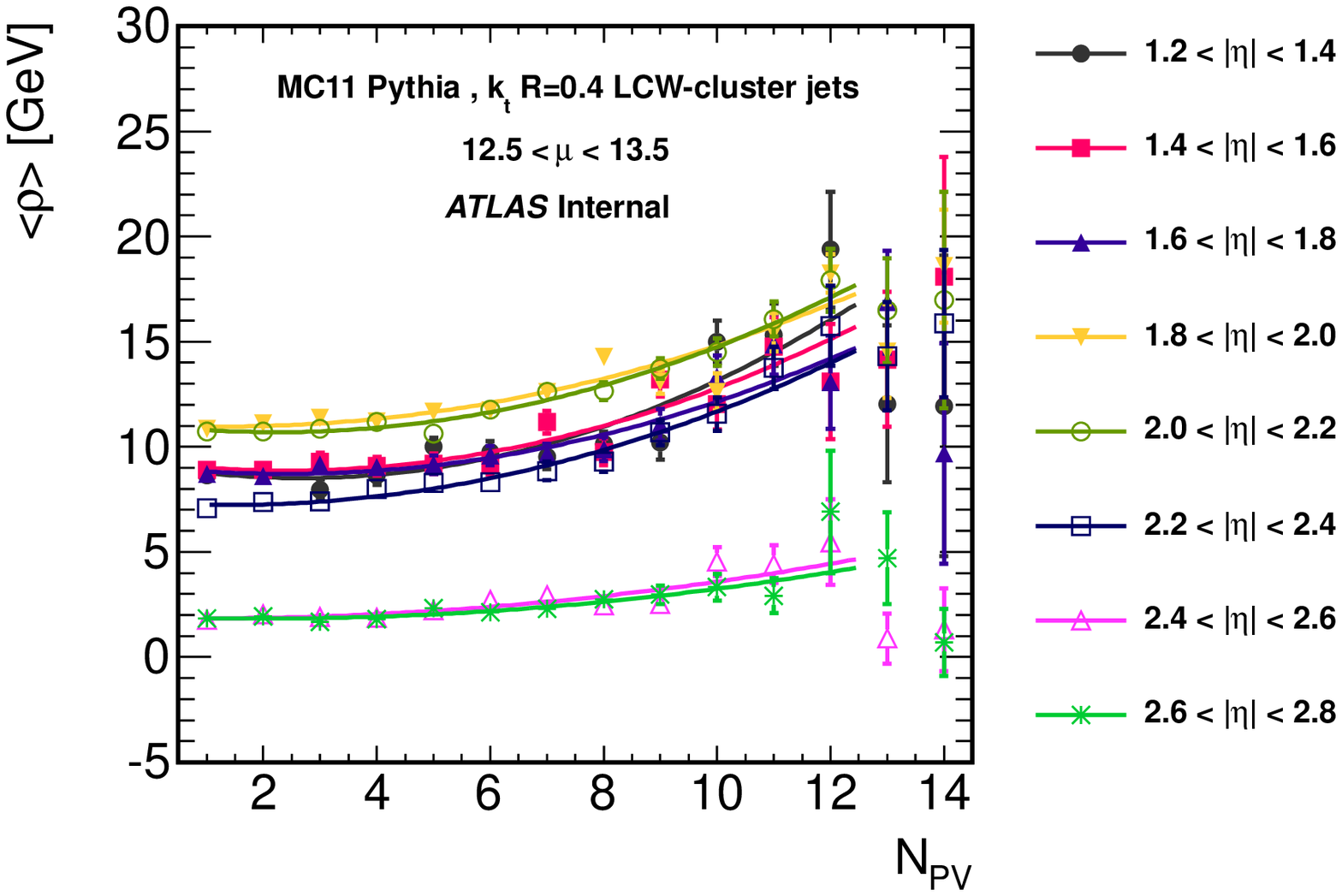}} \\
  \subfloat[]{\label{FIGavgRhoNpvInEtaMuBinsApp3}\includegraphics[trim=5mm 10mm 65mm 10mm,clip,width=.43\textwidth]{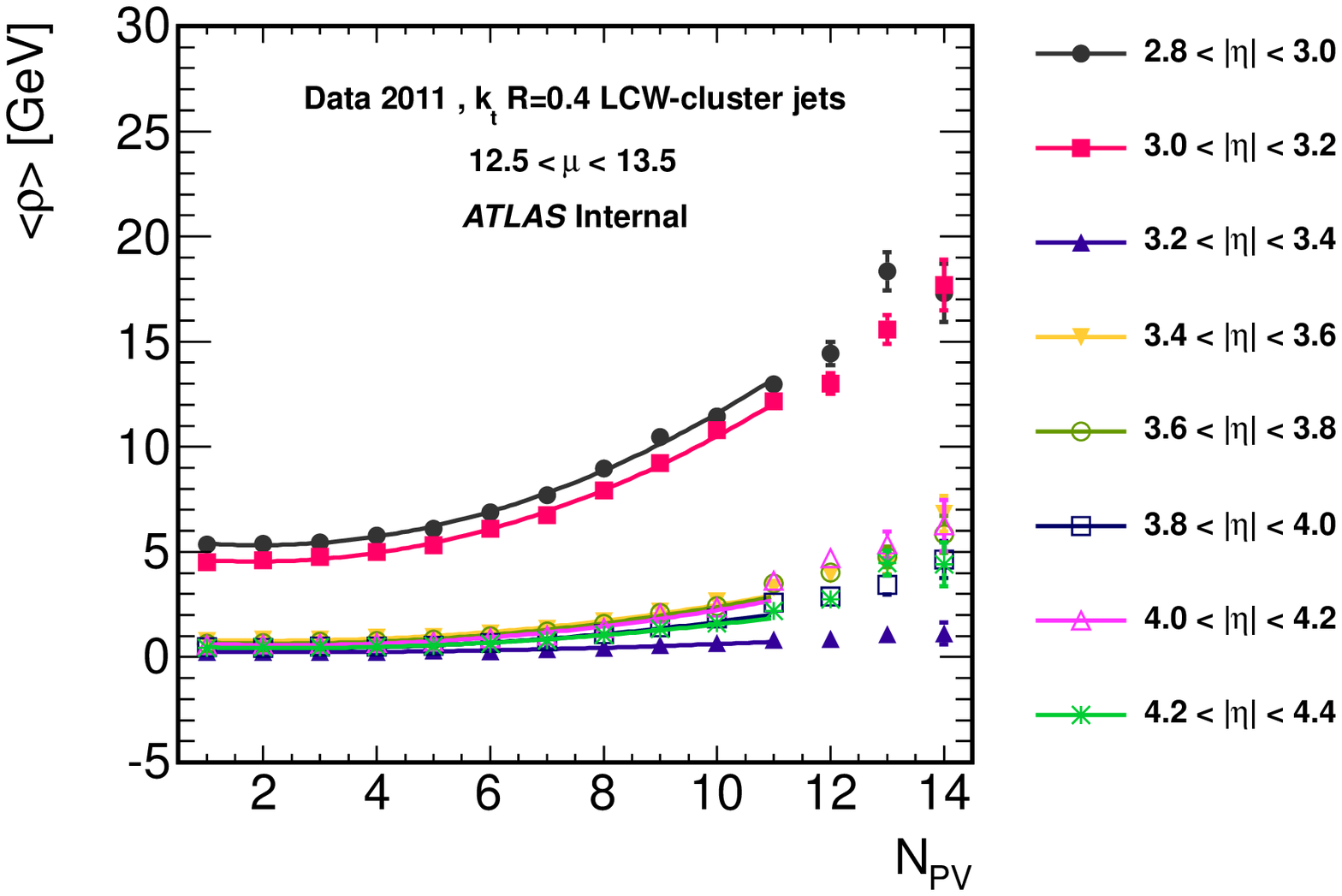}} 
  \subfloat[\figQaud]{\label{FIGavgRhoNpvInEtaMuBinsApp4}\includegraphics[trim=5mm 10mm 0mm 10mm,clip,width=.643\textwidth]{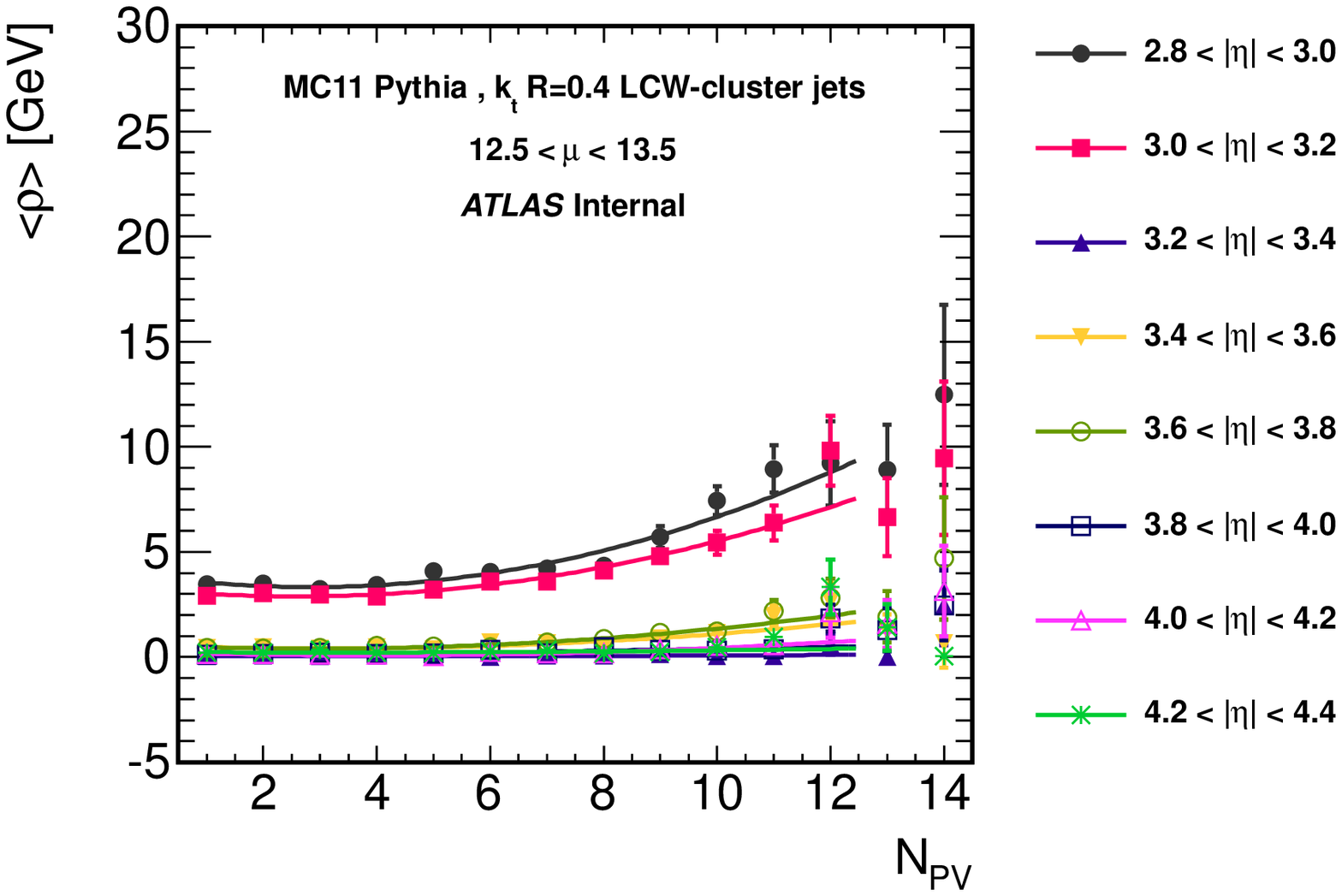}}
  \caption{\label{FIGavgRhoNpvInEtaMuBinsApp}Dependence of the average median, $\left<\rho\right>$, on the number of reconstructed vertices, \Npv,
    for a fixed range of the average number of interactions, \Mu. Several
    pseudo-rapidity bins within $1.2 < \eta < 2.8$ (\Subref{FIGavgRhoNpvInEtaMuBinsApp1}, \Subref{FIGavgRhoNpvInEtaMuBinsApp2}) and
    $2.8 < \eta < 4.4$ (\Subref{FIGavgRhoNpvInEtaMuBinsApp3}, \Subref{FIGavgRhoNpvInEtaMuBinsApp4}) are shown, using data and MC,
    as indicated in the figures.  The lines represent polynomial fits to the points.
    (See also \autoref{chapJetAreaMethod}, \autoref{FIGavgRhoInEtaMuBins} and accompanying text.)
  }
  \end{center}
  \end{figure} 
  \begin{figure}[htp]
  \begin{center}
  \subfloat[]{\label{FIGavgRhoMuInEtaMuBinsApp1}\includegraphics[trim=5mm 10mm 65mm 10mm,clip,width=.43\textwidth]{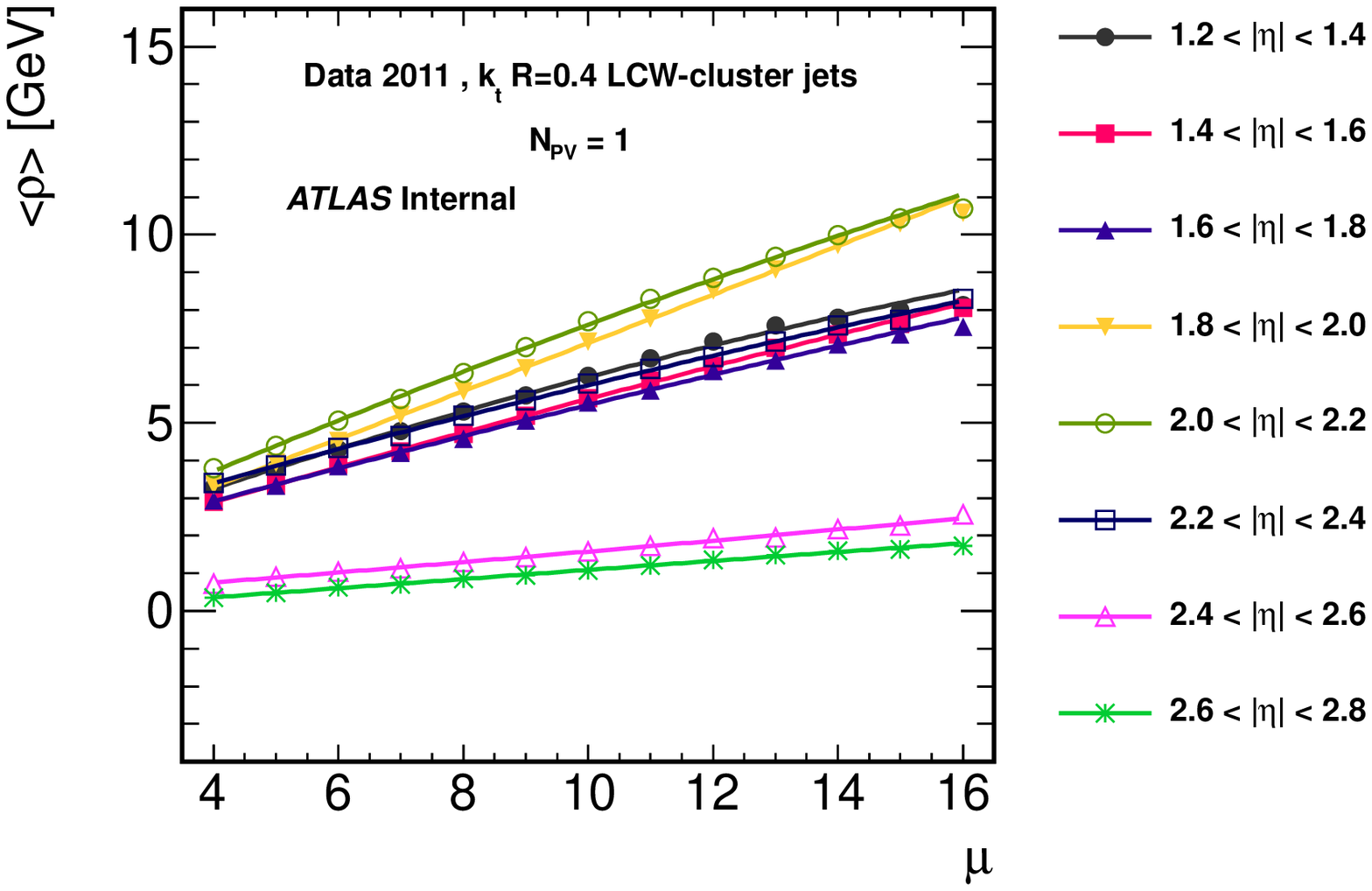}} 
  \subfloat[\figQaud]{\label{FIGavgRhoMuInEtaMuBinsApp2}\includegraphics[trim=5mm 10mm 0mm 10mm,clip,width=.643\textwidth]{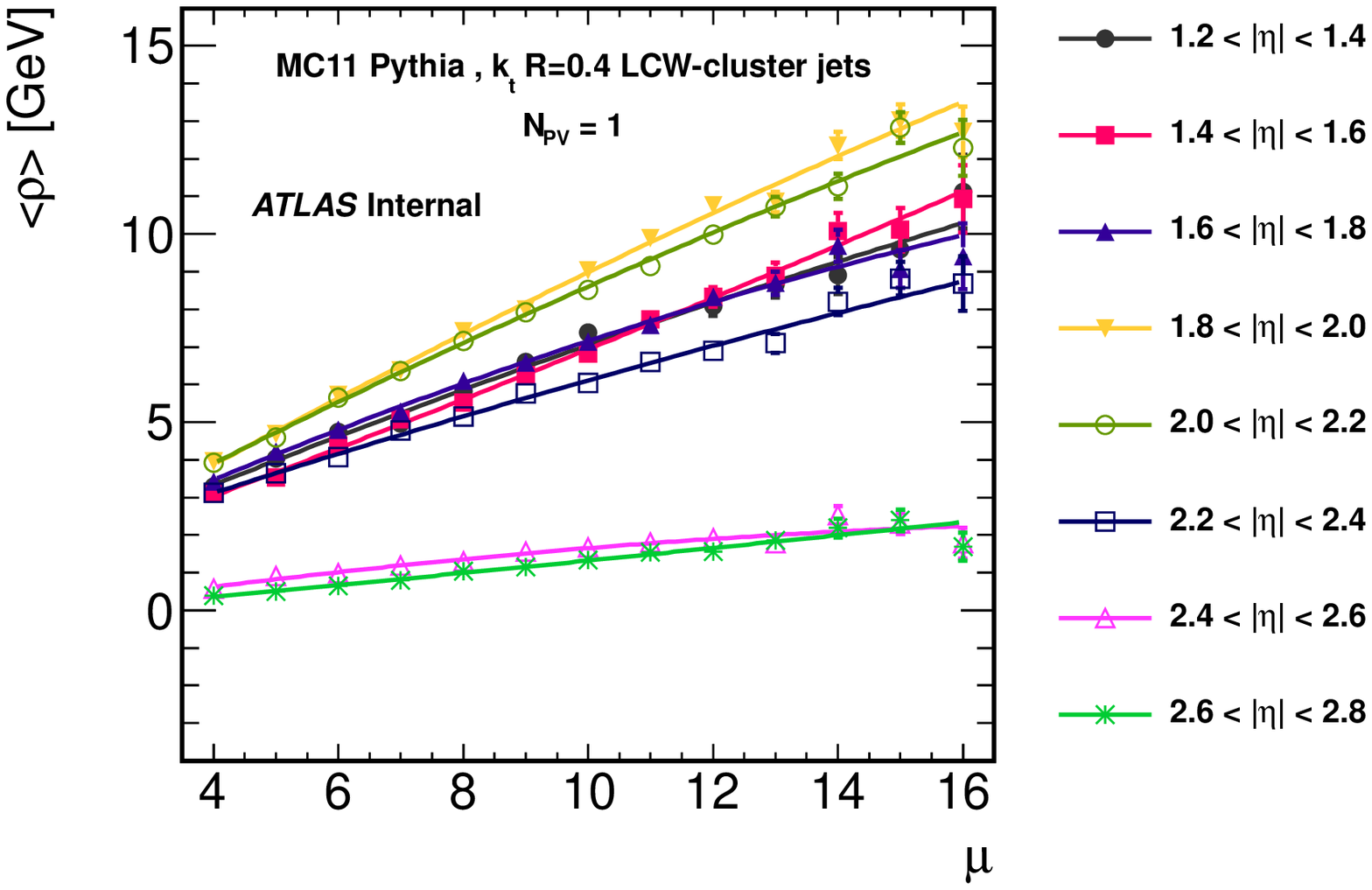}} \\
  \subfloat[]{\label{FIGavgRhoMuInEtaMuBinsApp3}\includegraphics[trim=5mm 10mm 65mm 10mm,clip,width=.43\textwidth]{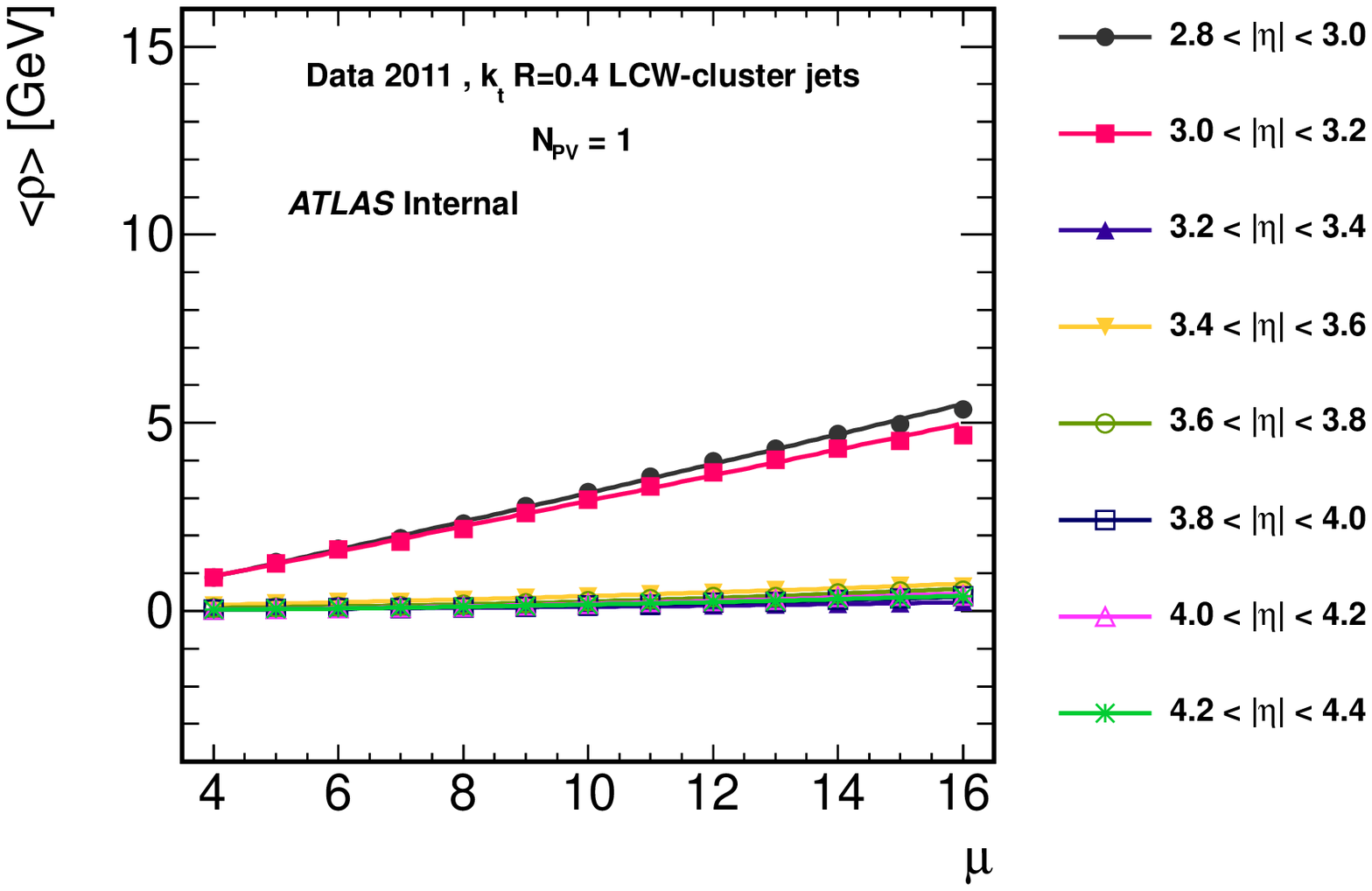}} 
  \subfloat[\figQaud]{\label{FIGavgRhoMuInEtaMuBinsApp4}\includegraphics[trim=5mm 10mm 0mm 10mm,clip,width=.643\textwidth]{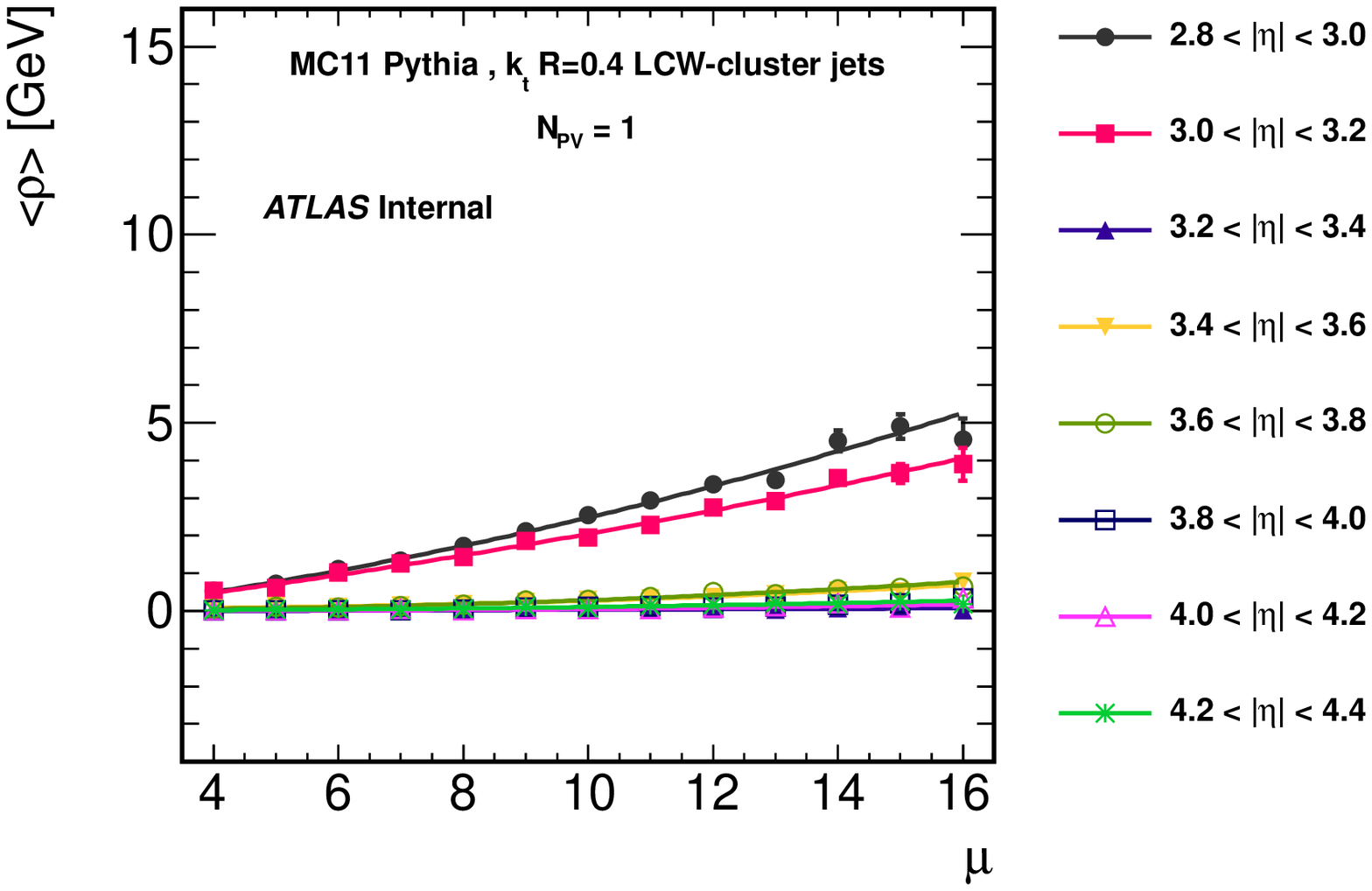}} 
    \caption{\label{FIGavgRhoMuInEtaMuBinsApp}Dependence of the average median, $\left<\rho\right>$,
    on the average number of interactions, \Mu, for events with a single reconstructed vertex, \Npv. Several
    pseudo-rapidity bins within $1.2 < \eta < 2.8$ (\Subref{FIGavgRhoMuInEtaMuBinsApp1}, \Subref{FIGavgRhoMuInEtaMuBinsApp2}) and
    $2.8 < \eta < 4.4$ (\Subref{FIGavgRhoMuInEtaMuBinsApp3}, \Subref{FIGavgRhoMuInEtaMuBinsApp4}) are shown, using data and MC,
    as indicated in the figures.  The lines represent polynomial fits to the points.
    (See also \autoref{chapJetAreaMethod}, \autoref{FIGavgRhoInEtaMuBins} and accompanying text.)
  }
  \end{center}
  \end{figure} 
  \begin{figure}[htp]
  \begin{center}
  \subfloat[]{\label{FIGmeanClosureOffsetNpvMuSlopeApp1}\includegraphics[trim=5mm 10mm 65mm 15mm,clip,width=.43\textwidth]{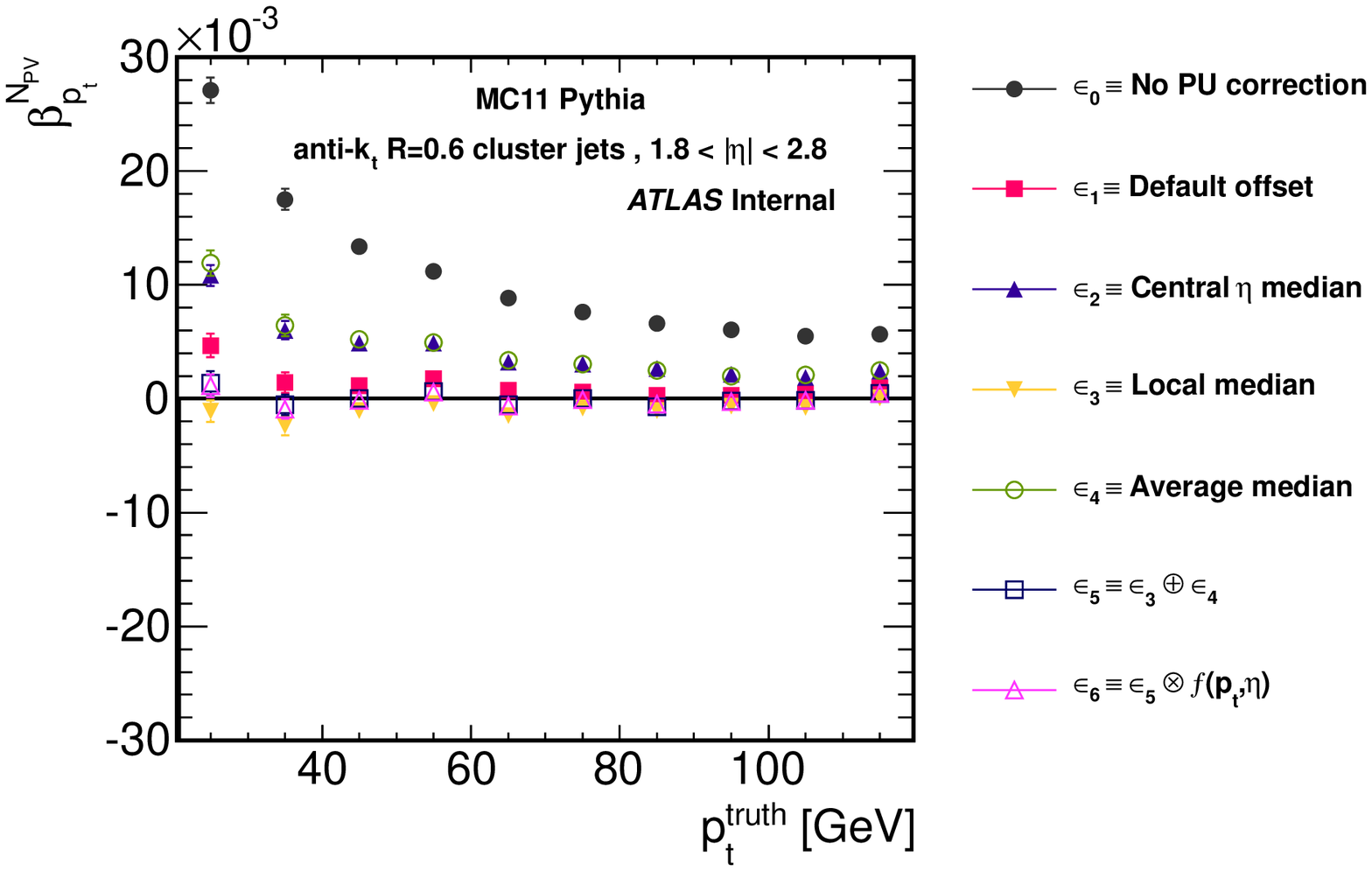}} 
  \subfloat[\figQaud]{\label{FIGmeanClosureOffsetNpvMuSlopeApp2}\includegraphics[trim=5mm 10mm 0mm 15mm,clip,width=.643\textwidth]{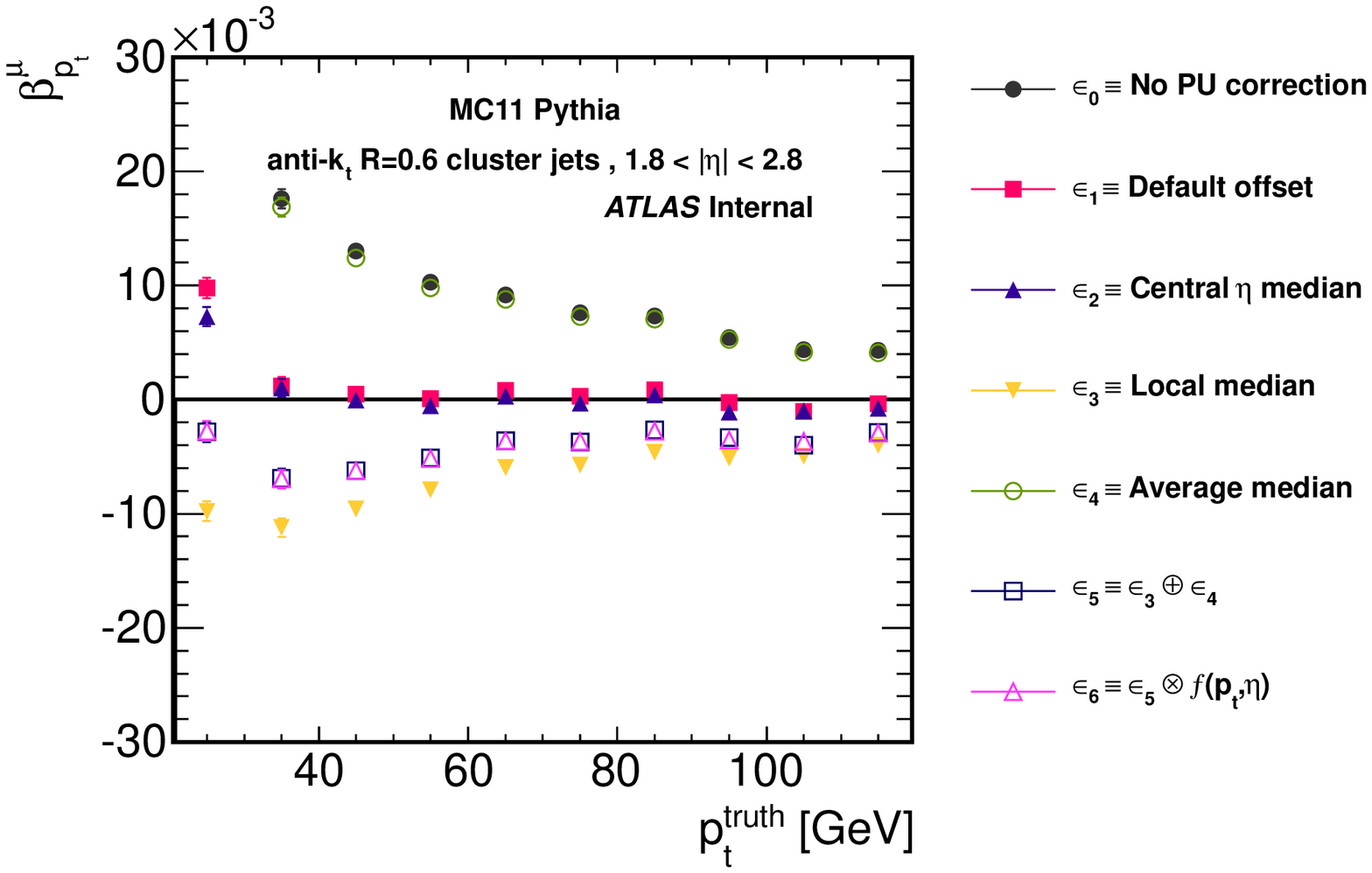}} \\
  \subfloat[]{\label{FIGmeanClosureOffsetNpvMuSlopeApp3}\includegraphics[trim=5mm 10mm 65mm 15mm,clip,width=.43\textwidth]{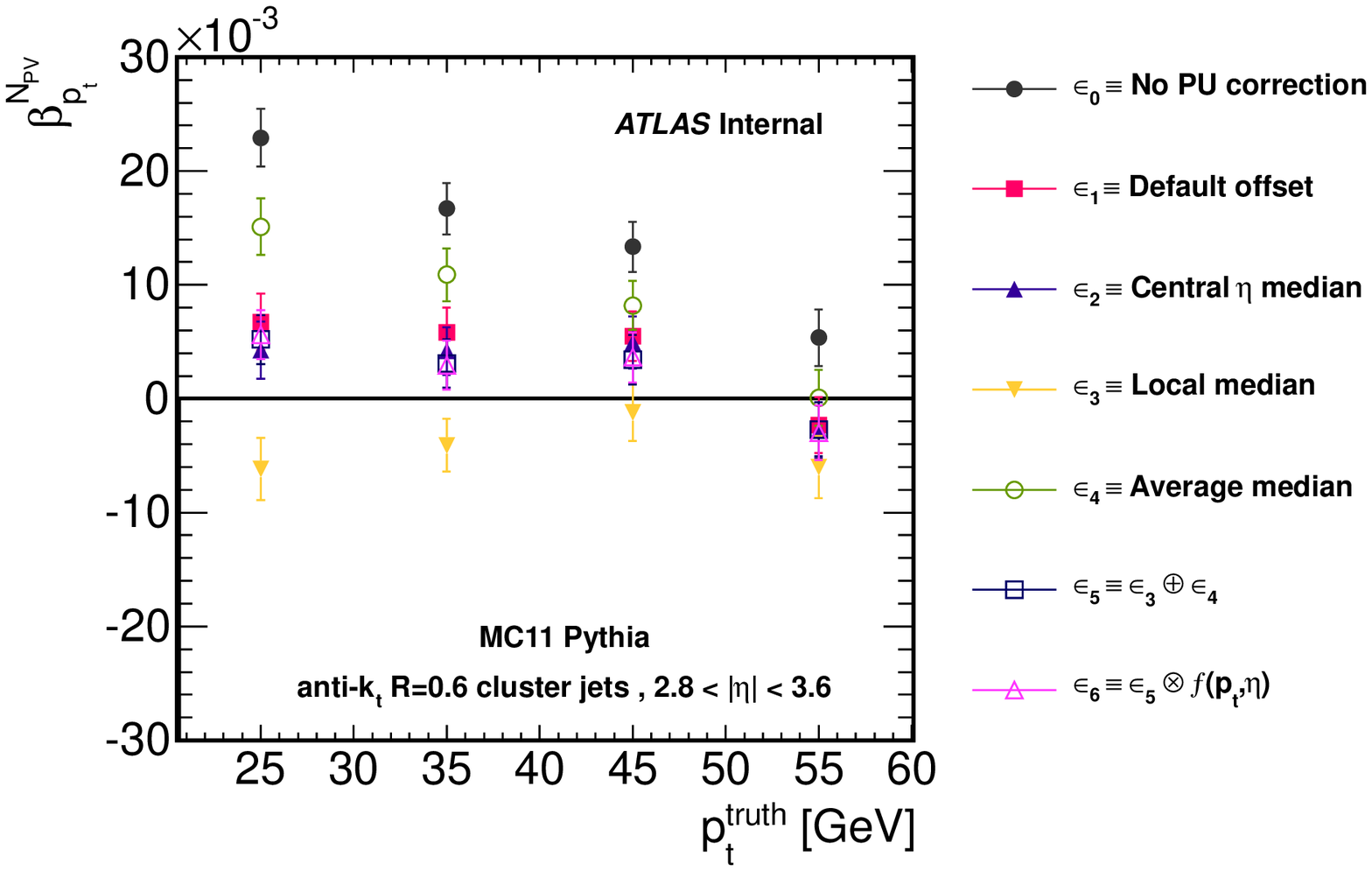}} 
  \subfloat[\figQaud]{\label{FIGmeanClosureOffsetNpvMuSlopeApp4}\includegraphics[trim=5mm 10mm 0mm 15mm,clip,width=.643\textwidth]{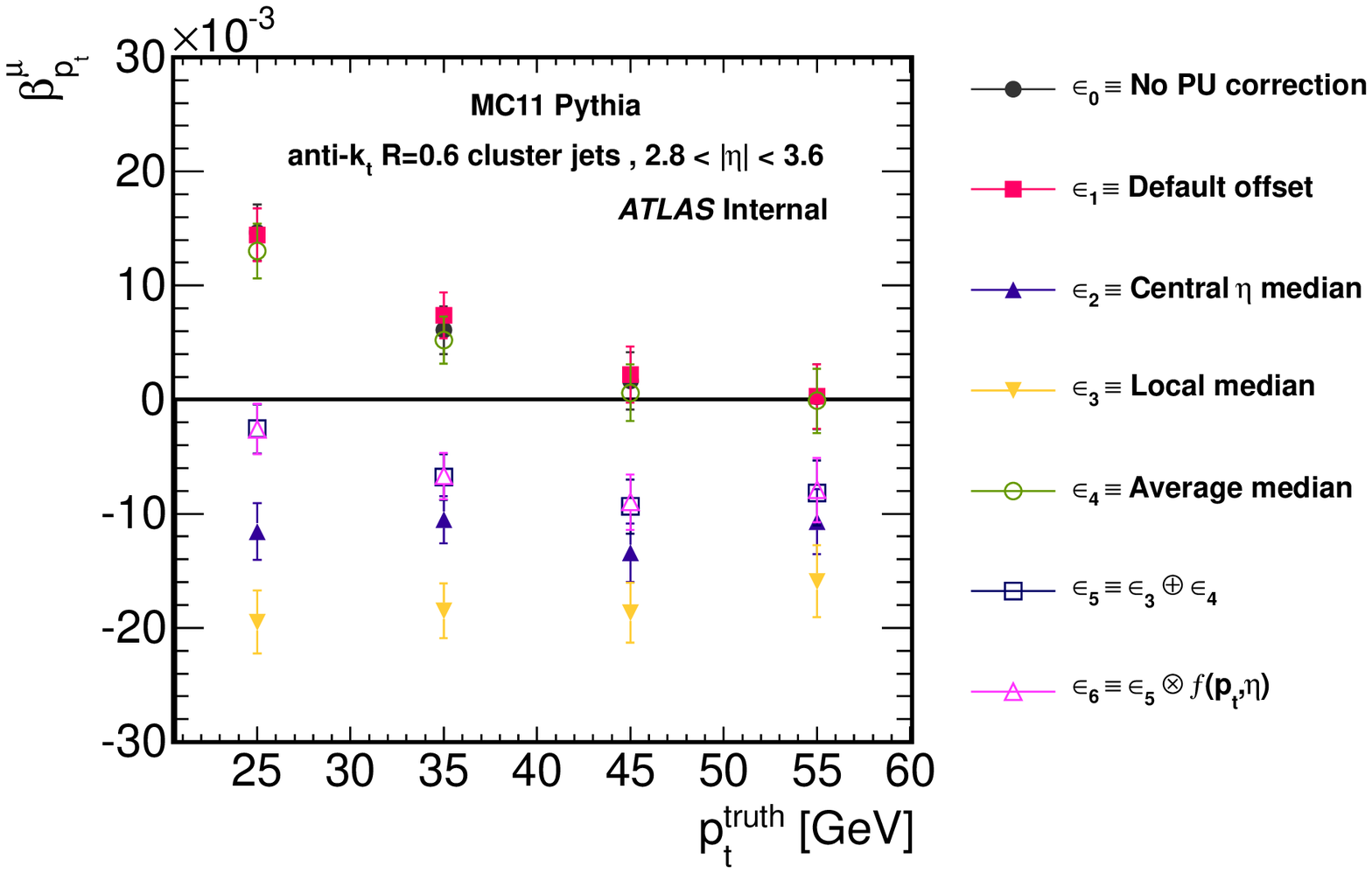}} \\
  \subfloat[]{\label{FIGmeanClosureOffsetNpvMuSlopeApp5}\includegraphics[trim=5mm 10mm 65mm 15mm,clip,width=.43\textwidth]{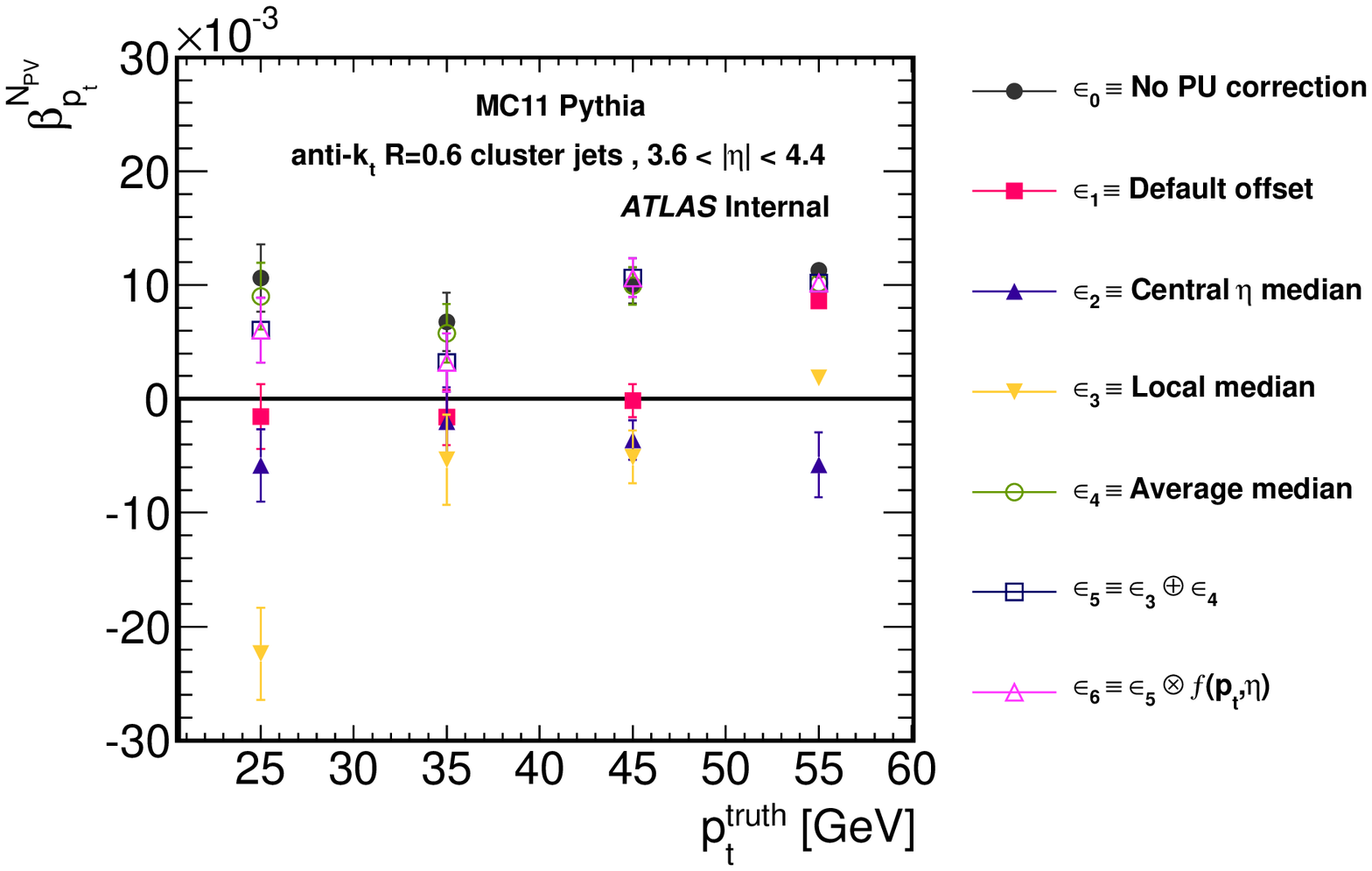}} 
  \subfloat[\figQaud]{\label{FIGmeanClosureOffsetNpvMuSlopeApp6}\includegraphics[trim=5mm 10mm 0mm 15mm,clip,width=.643\textwidth]{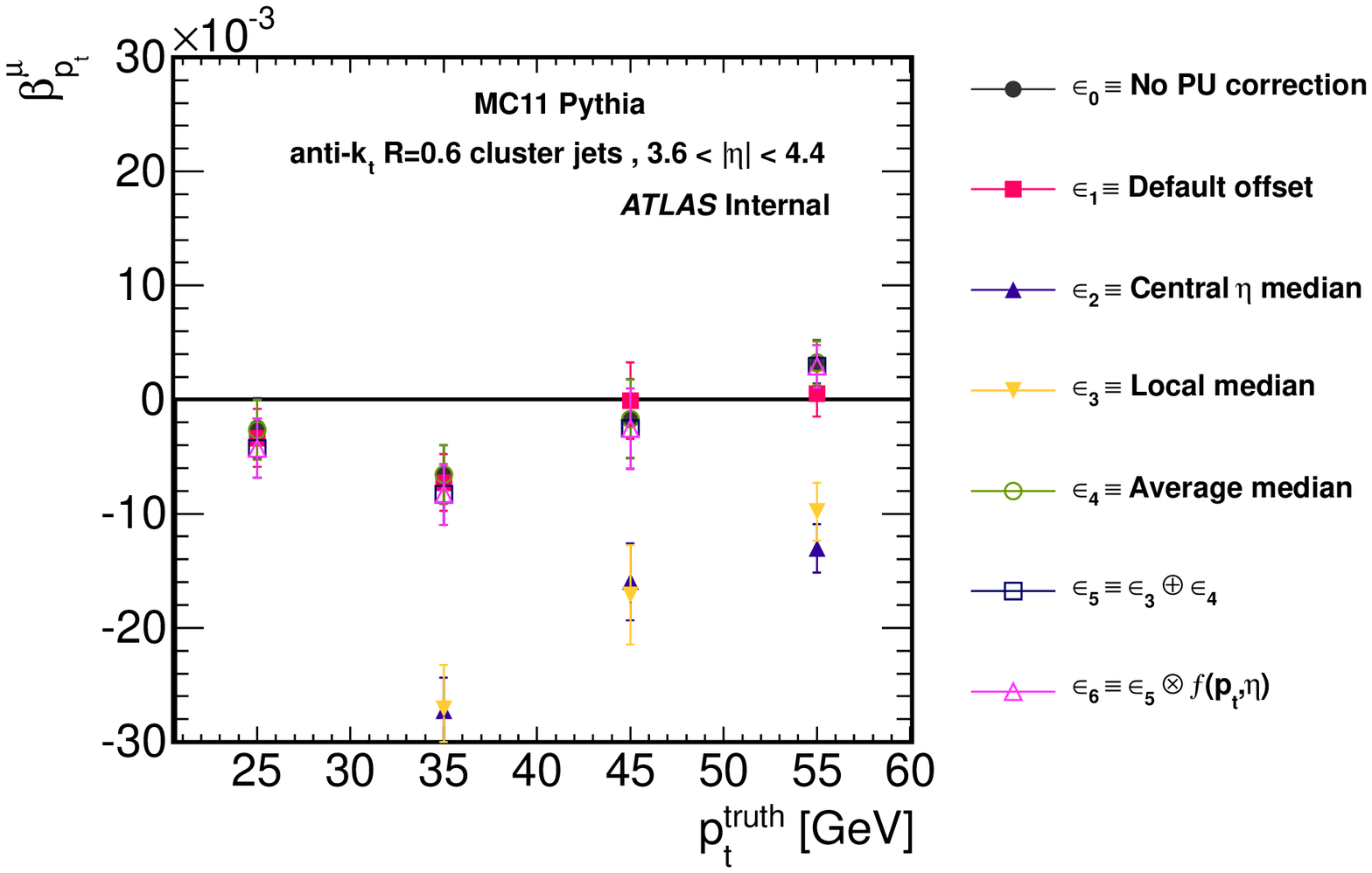}}
    \caption{\label{FIGmeanClosureOffsetNpvMuSlopeApp}Dependence on truth jet transverse momentum, $p_{\mrm{t}}^{\mrm{truth}}$, of the parameters
      defined in \autoref{chapJetAreaMethod}, \autoref{eqOffsetFitForm1},
      $\beta_{\pt}^{\Npv}$ (\Subref{FIGmeanClosureOffsetNpvMuSlopeApp1}, \Subref{FIGmeanClosureOffsetNpvMuSlopeApp3} and \Subref{FIGmeanClosureOffsetNpvMuSlopeApp5})
      and $\beta_{\pt}^{\Mu}$ (\Subref{FIGmeanClosureOffsetNpvMuSlopeApp2}, \Subref{FIGmeanClosureOffsetNpvMuSlopeApp4} and \Subref{FIGmeanClosureOffsetNpvMuSlopeApp6}),
      using jets corrected for \pu with $\epsilon_{0}$-$\epsilon_{6}$,
      within several pseudo-rapidity, \Eta, regions, as indicated in the figures.
      (See also \autoref{chapJetAreaMethod}, \autoref{FIGmeanClosureOffsetNpvMuSlope} and accompanying text.)
    }
  \end{center}
  \end{figure} 
  %
  %


  %
  %
  \begin{figure}[htp]
  \begin{center}
  \subfloat[]{\label{FIGinSituPtStdNpvApp1}\includegraphics[trim=5mm 14mm 0mm 10mm,clip,width=.52\textwidth]{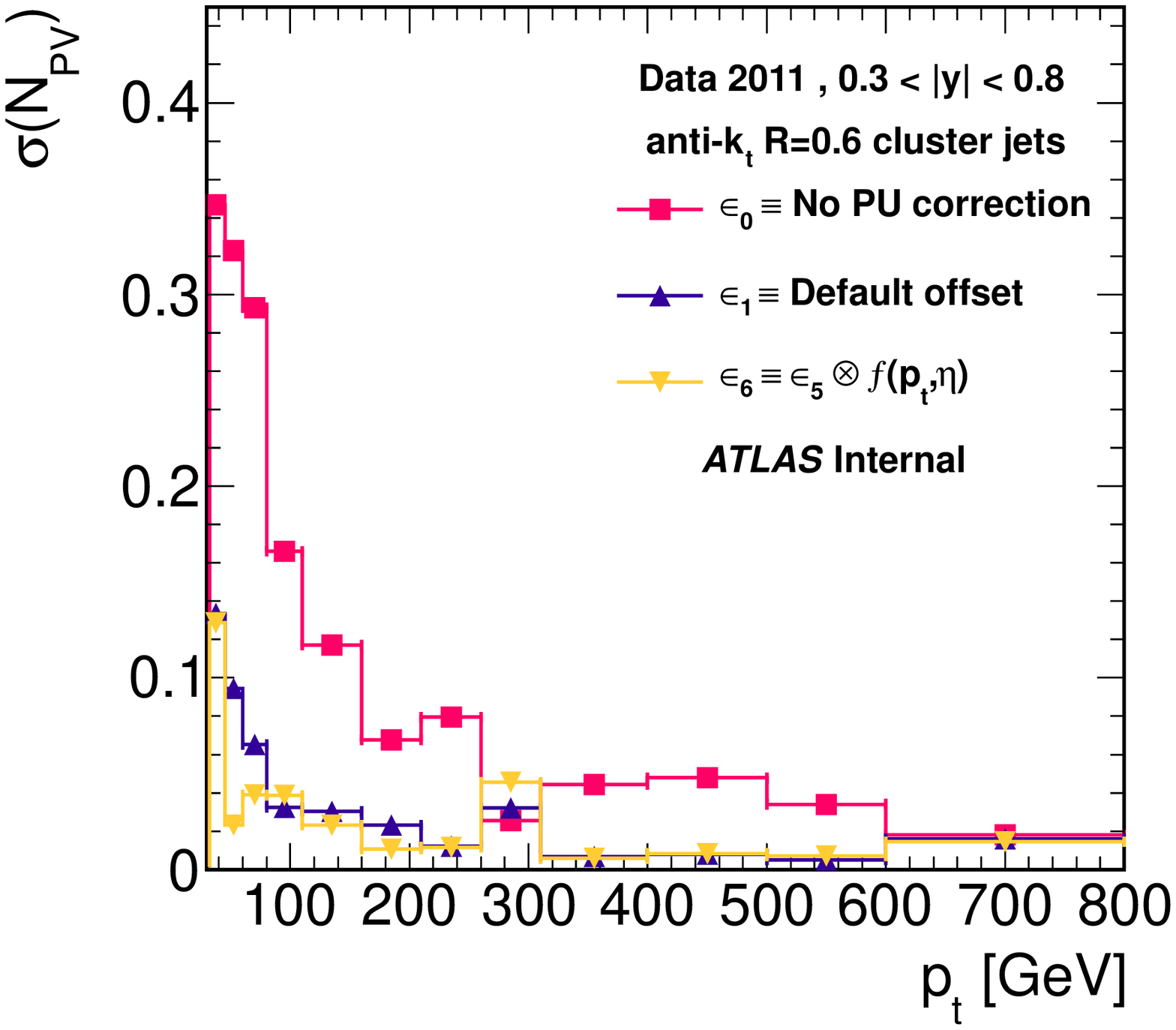}}
  \subfloat[]{\label{FIGinSituPtStdNpvApp2}\includegraphics[trim=5mm 14mm 0mm 10mm,clip,width=.52\textwidth]{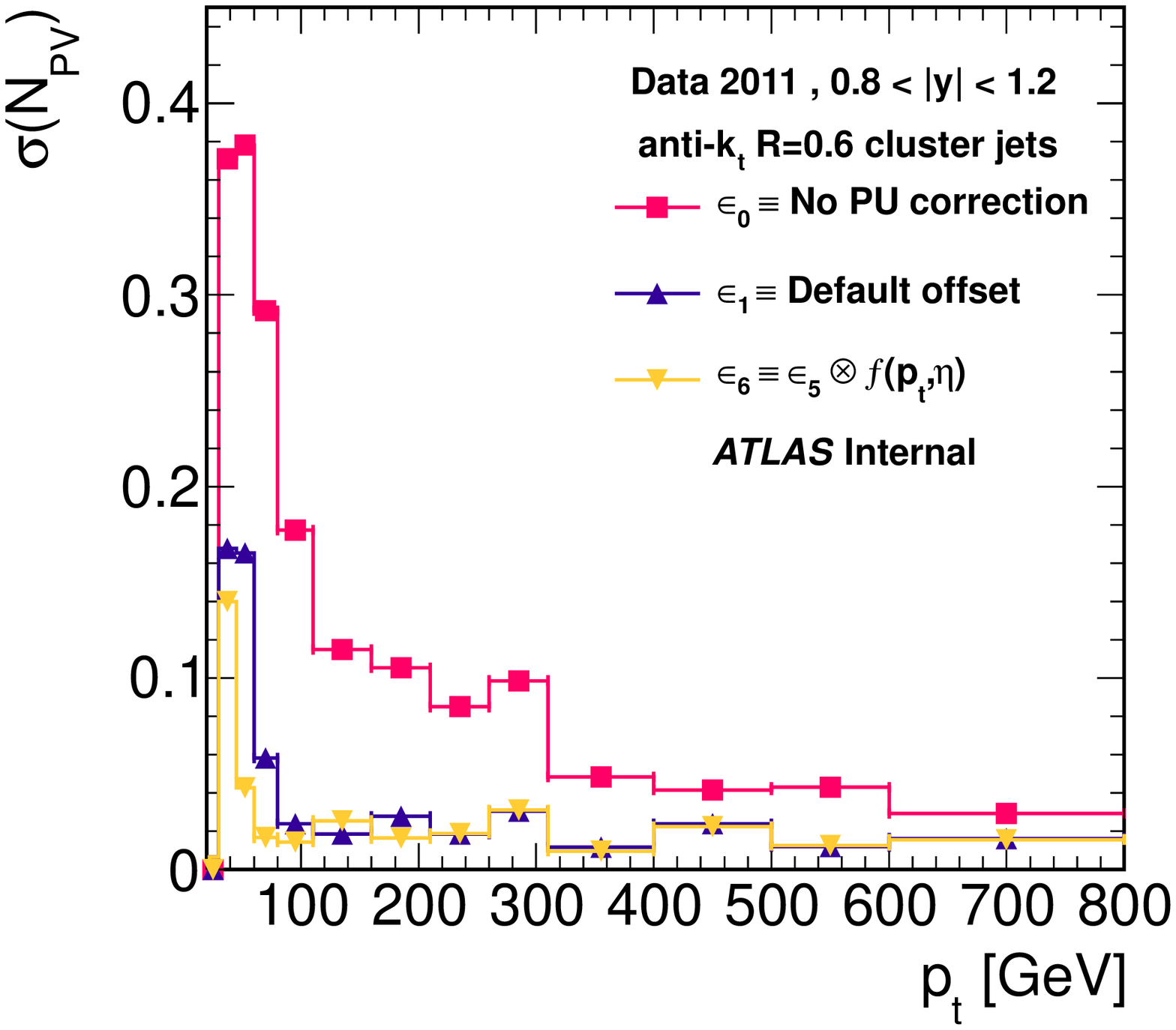}} \\
  \subfloat[]{\label{FIGinSituPtStdNpvApp3}\includegraphics[trim=5mm 14mm 0mm 10mm,clip,width=.52\textwidth]{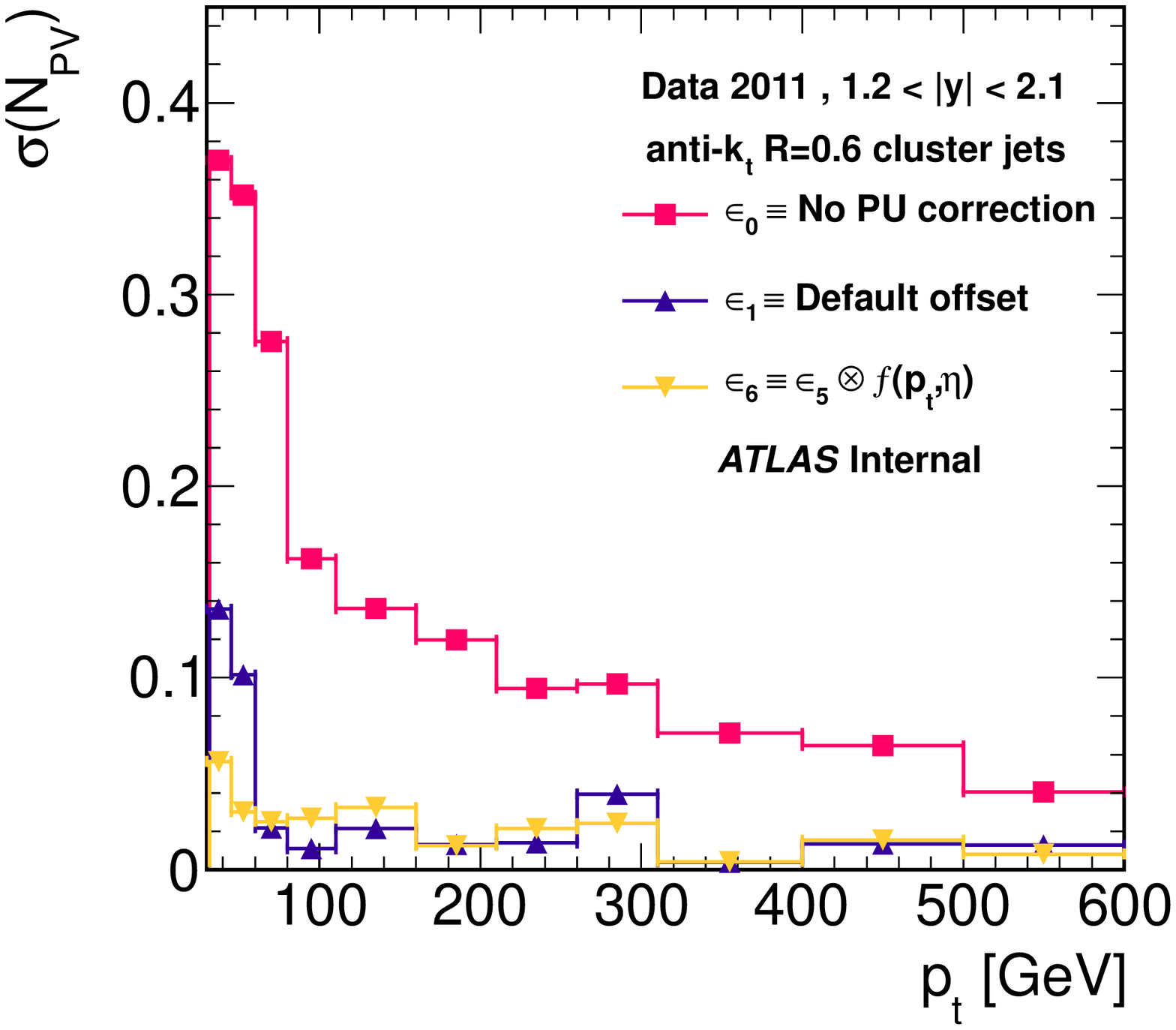}}
  \subfloat[]{\label{FIGinSituPtStdNpvApp4}\includegraphics[trim=5mm 14mm 0mm 10mm,clip,width=.52\textwidth]{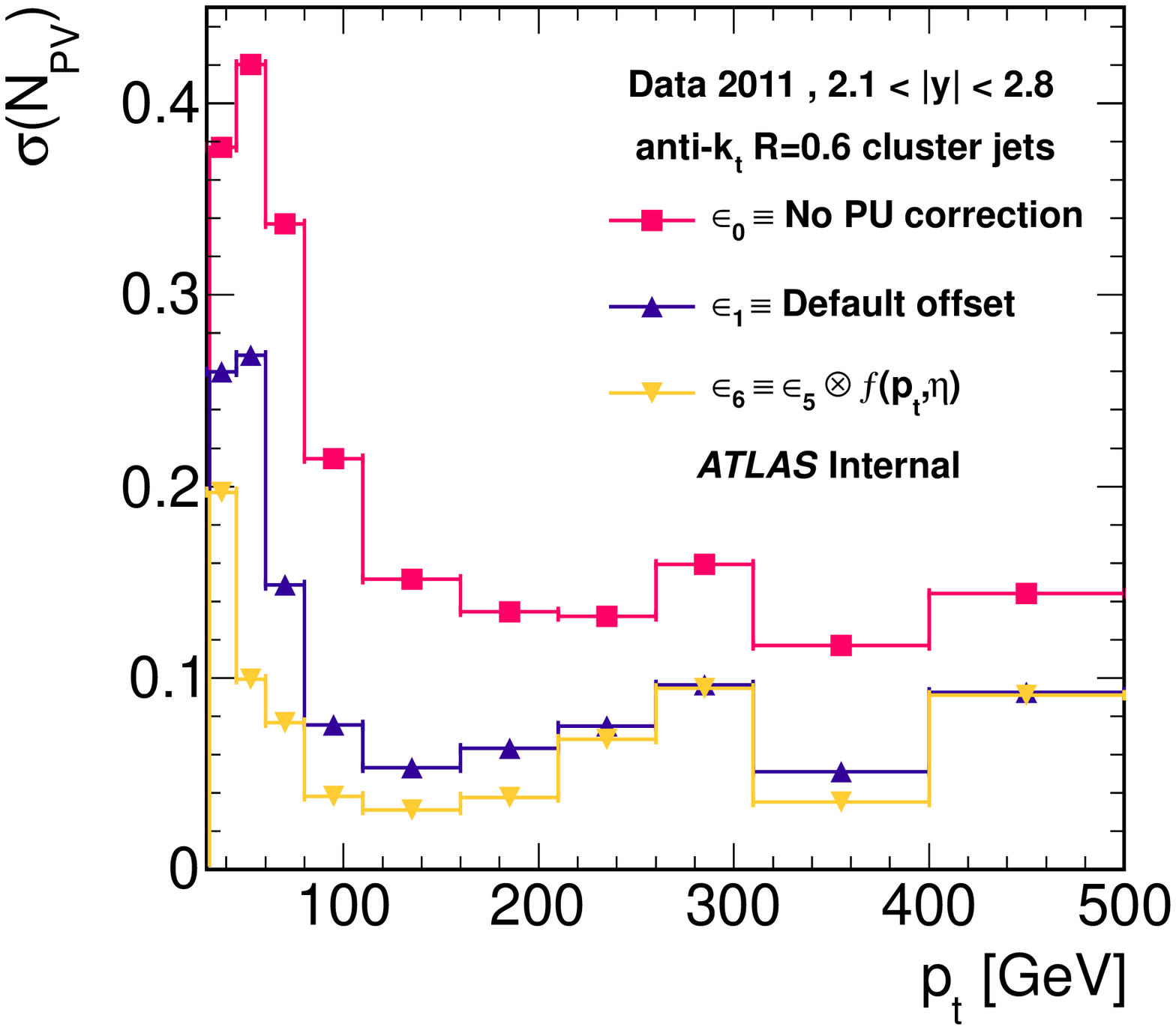}}
    \caption{\label{FIGinSituPtStdNpvApp}Standard deviation with regard to single-vertex events, $\sigma\left(\Npv\right)$,
      of the differential transverse momentum, \pt, spectrum of the highest-\pt jet in an event,
      for jets within several rapidity, $y$, regions, corrected for \pu by
      $\epsilon_{0}$, $\epsilon_{1}$ and $\epsilon_{6}$, as indicated in the figures.
      (See also \autoref{chapJetAreaMethod}, \autoref{eqStandardDeviationNpvDef}, \autoref{FIGinSituMassPtStdNpv} and accompanying text.)
    }
  \end{center}
  \end{figure} 
  \begin{figure}[htp]
  \begin{center}
  \subfloat[]{\label{FIGinSituMassStdNpvApp1}\includegraphics[trim=5mm 14mm 0mm 10mm,clip,width=.52\textwidth]{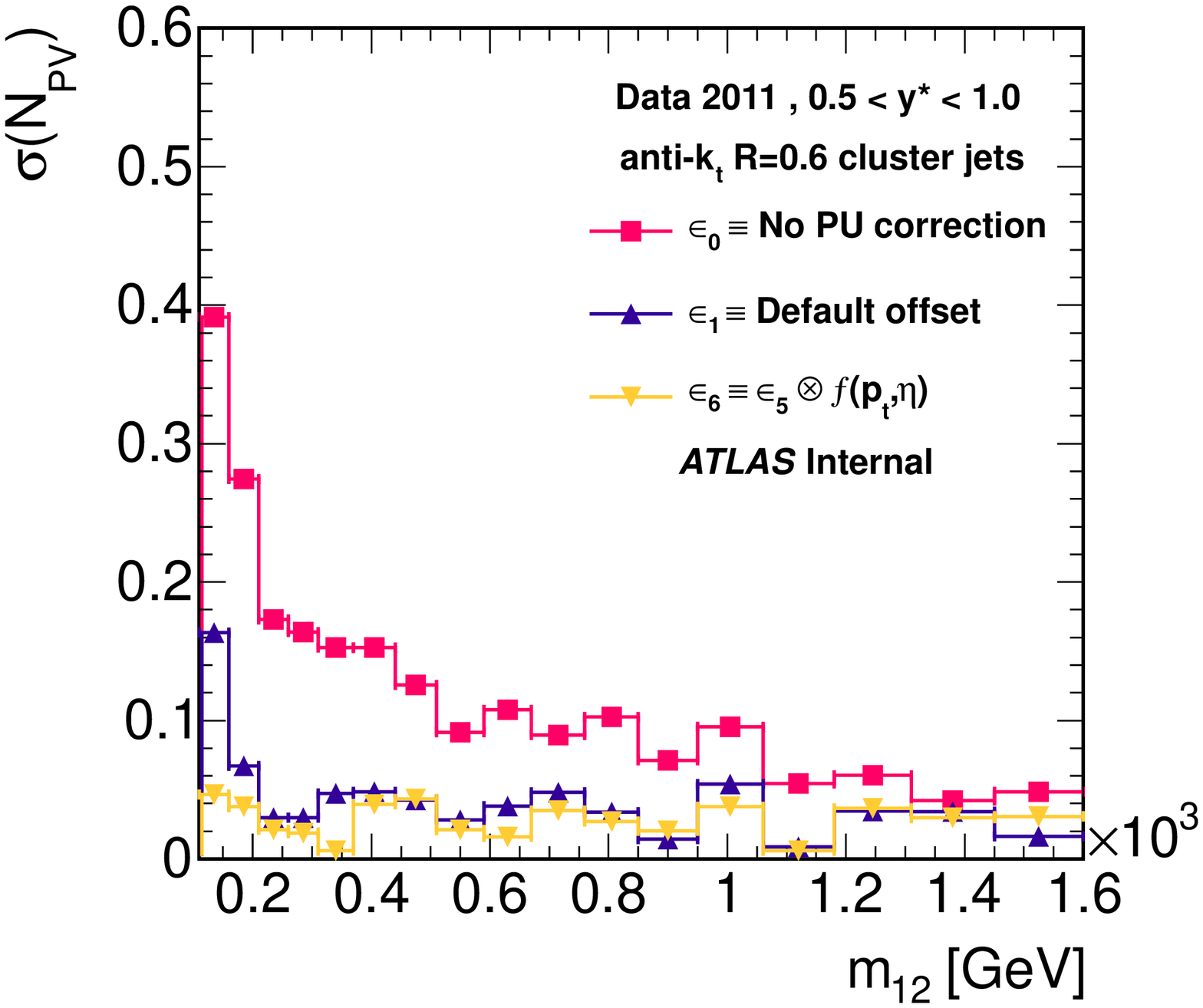}}
  \subfloat[]{\label{FIGinSituMassStdNpvApp2}\includegraphics[trim=5mm 14mm 0mm 10mm,clip,width=.52\textwidth]{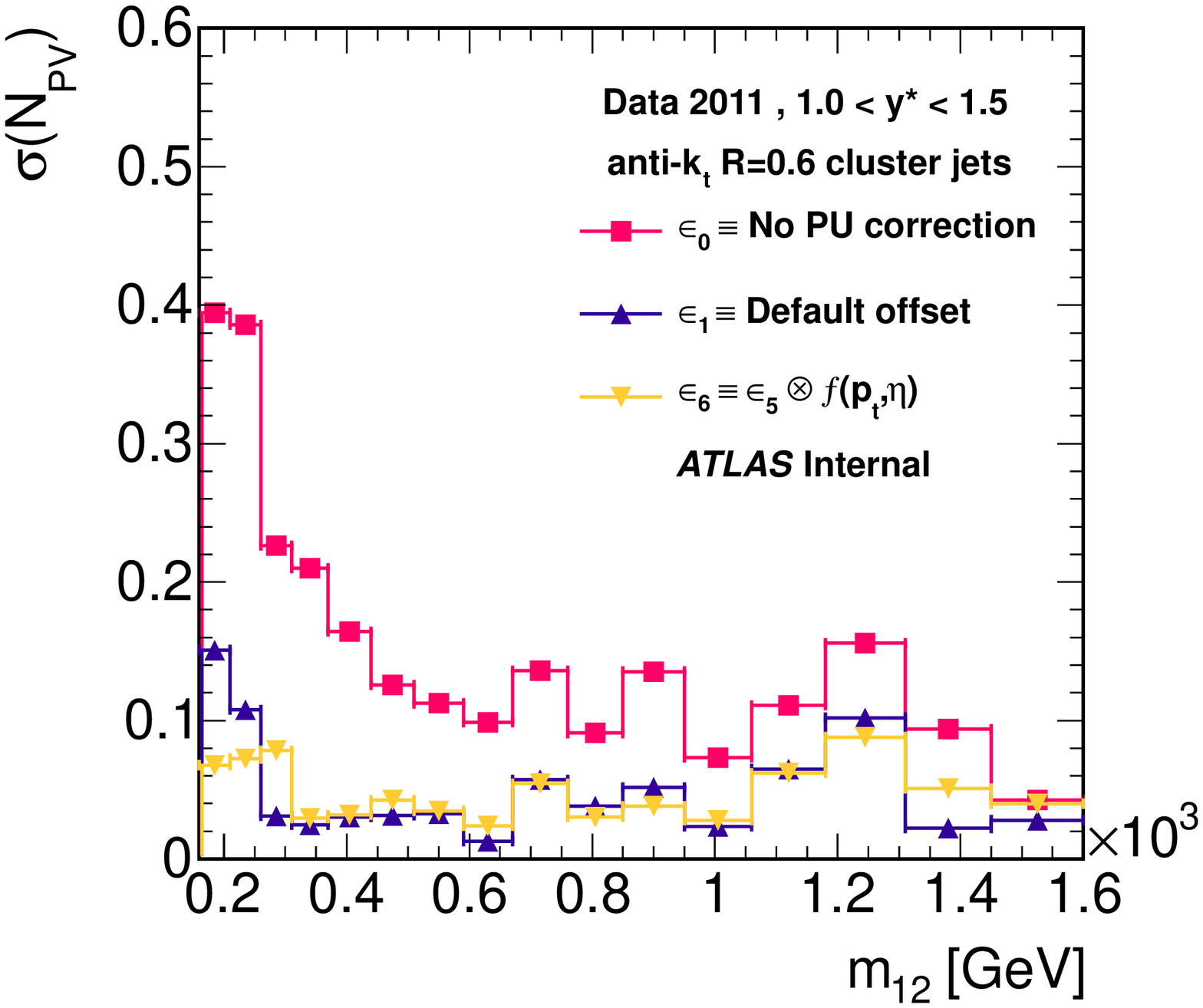}} \\
  \subfloat[]{\label{FIGinSituMassStdNpvApp3}\includegraphics[trim=5mm 14mm 0mm 10mm,clip,width=.52\textwidth]{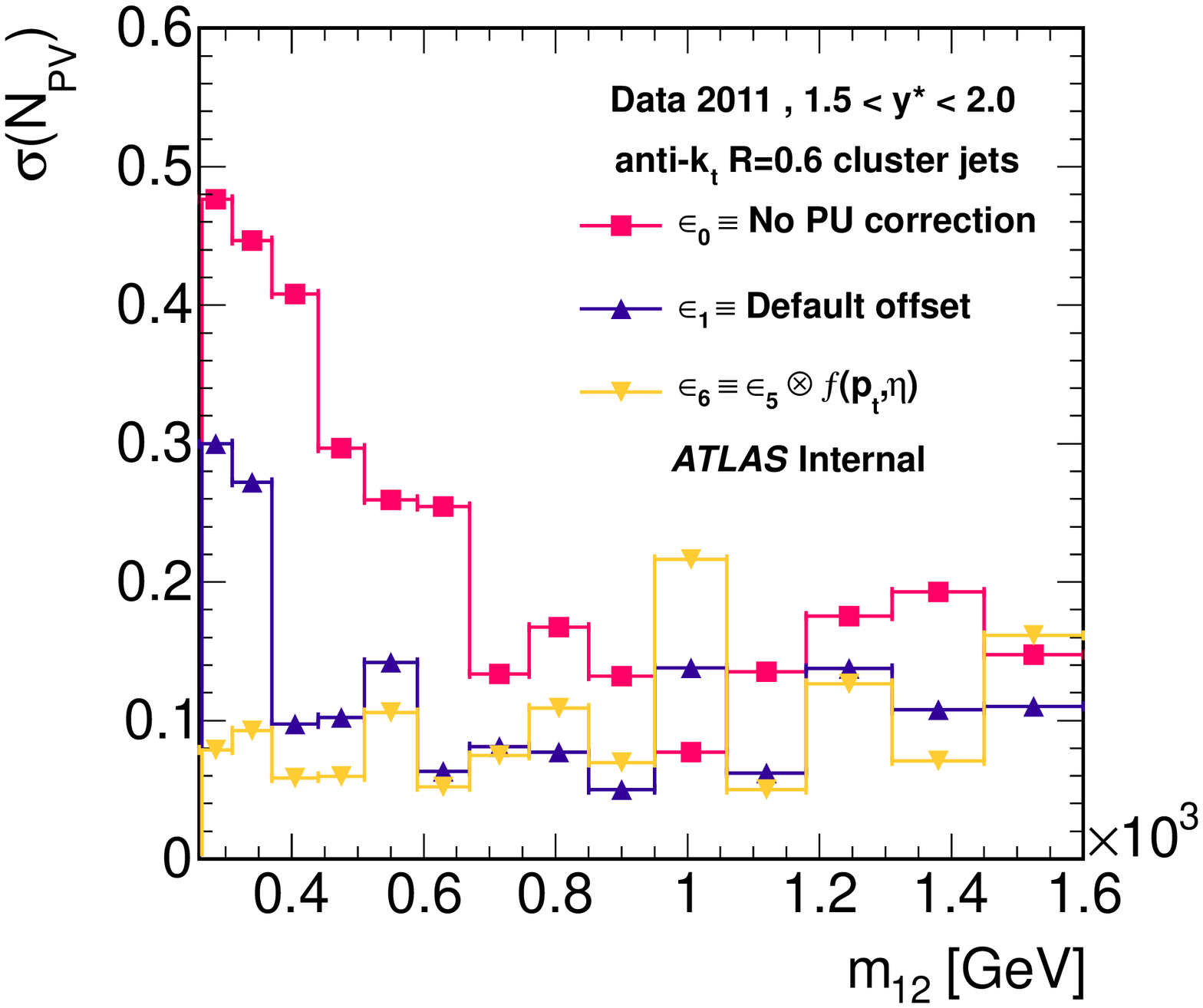}}
  \subfloat[]{\label{FIGinSituMassStdNpvApp3}\includegraphics[trim=5mm 14mm 0mm 10mm,clip,width=.52\textwidth]{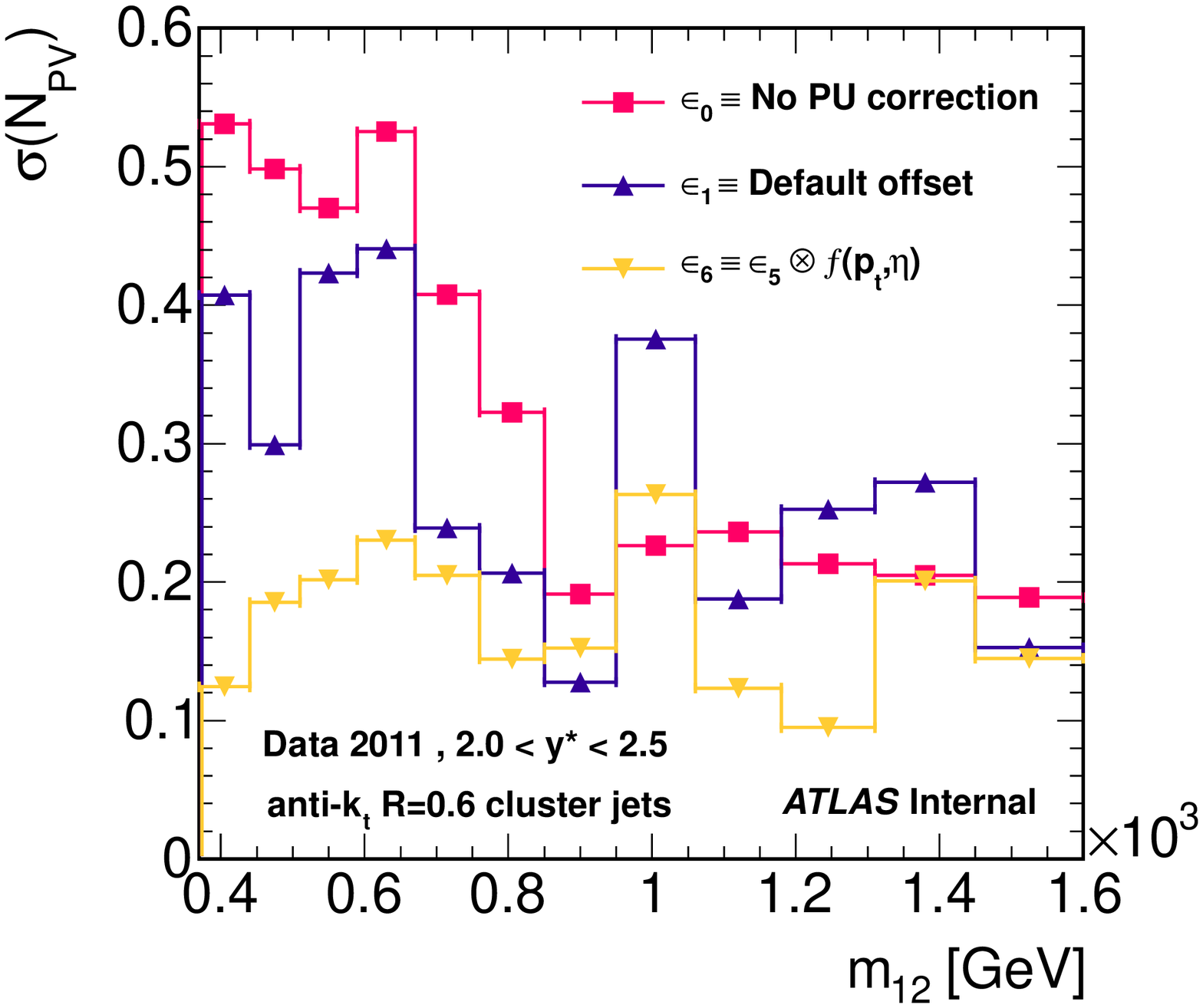}}
    \caption{\label{FIGinSituMassStdNpvApp}Standard deviation with regard to single-vertex events, $\sigma\left(\Npv\right)$,
      of the invariant mass distribution, $m_{12}$, of the two jets with the highest transverse momentum in an event,
      for several \com jet rapidities, \ystr, using jets corrected for \pu by
      $\epsilon_{0}$, $\epsilon_{1}$ and $\epsilon_{6}$, as indicated.
      (See also \autoref{chapJetAreaMethod}, \autoref{eqStandardDeviationNpvDef}, \autoref{FIGinSituMassPtStdNpv} and accompanying text.)
    }
  \end{center}
  \end{figure} 
}{} 

\ifthenelse {\boolean{do:jetMass}} {
  \begin{figure}[htp]
  \vspace{-20pt}
  %
  \section{Dijet mass distribution\label{chapMeasurementOfTheDijetMassApp}}
  %
  Additional figures pertaining to \autoref{chapMeasurementOfTheDijetMass} are presented in the following;
  non-perturbative correction factors for the theoretical calculation of the dijet invariant mass distributions are shown in
  \autoref{FIGhadronicCorrections0App};
  the uncertainties on the theoretical calculation and the systematic uncertainties on the data for the invariant mass measurement may be inferred from
  \autorefs{FIGtheoryUncertainty0App}~-~\ref{FIGdijetMassTotalSystematicUncertainty2011App};
  \vspace{-10pt}
  \begin{center}
    \subfloat[]{\label{FIGhadronicCorrections0App1}\includegraphics[trim=5mm 14mm 0mm 10mm,clip,width=.52\textwidth]{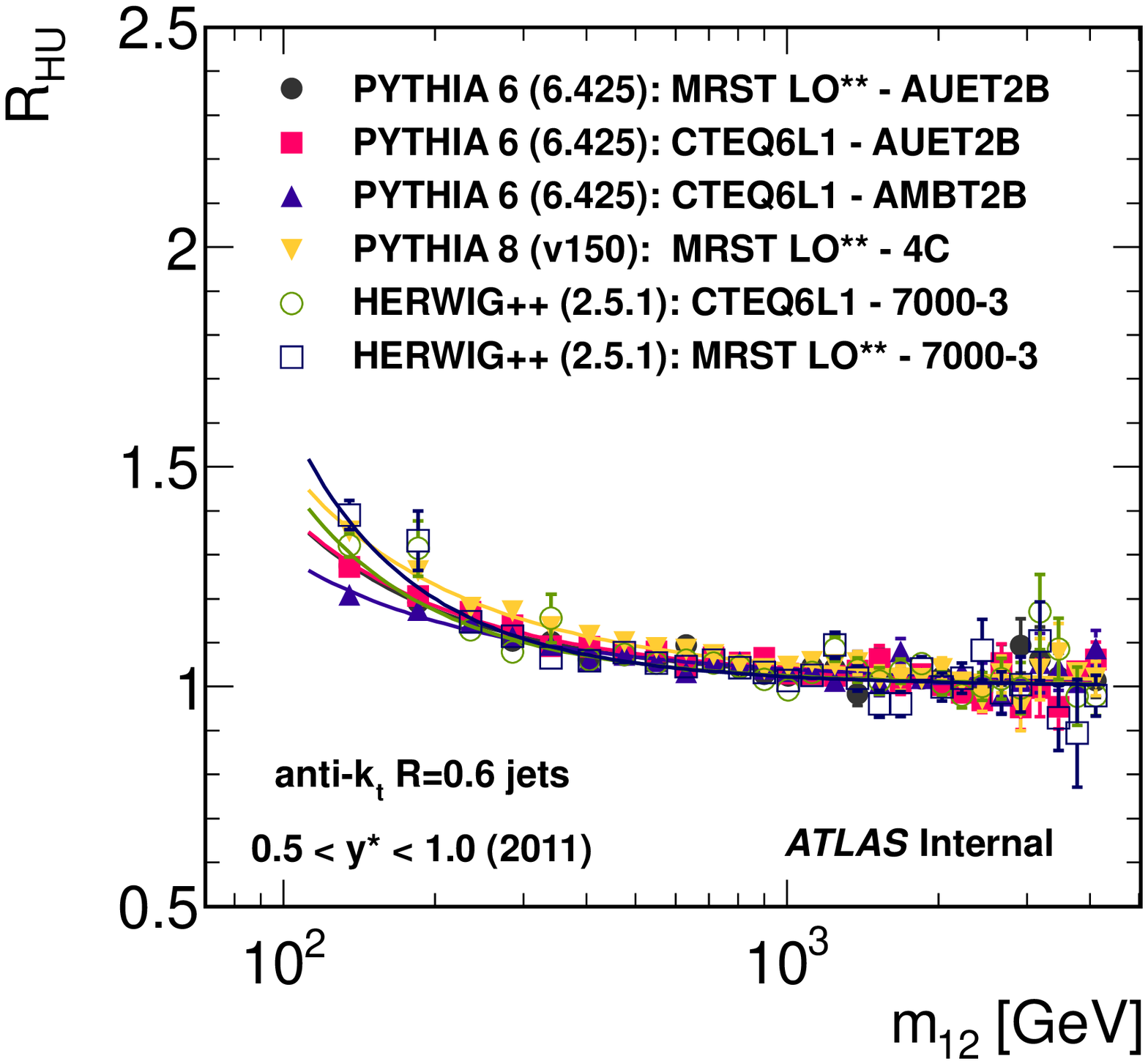}}
    \subfloat[]{\label{FIGhadronicCorrections0App2}\includegraphics[trim=5mm 14mm 0mm 10mm,clip,width=.52\textwidth]{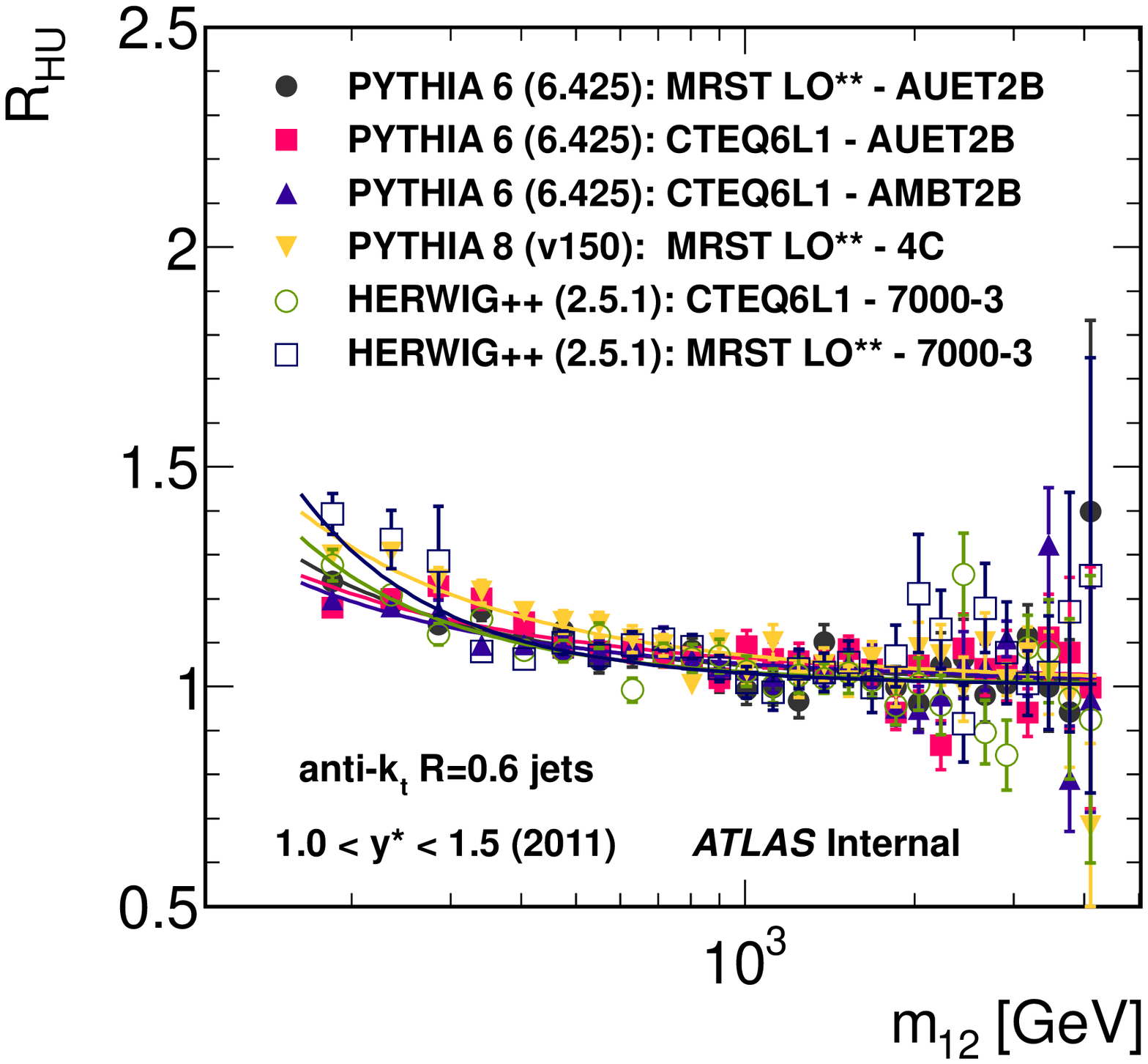}} \\
    \subfloat[]{\label{FIGhadronicCorrections0App3}\includegraphics[trim=5mm 14mm 0mm 10mm,clip,width=.52\textwidth]{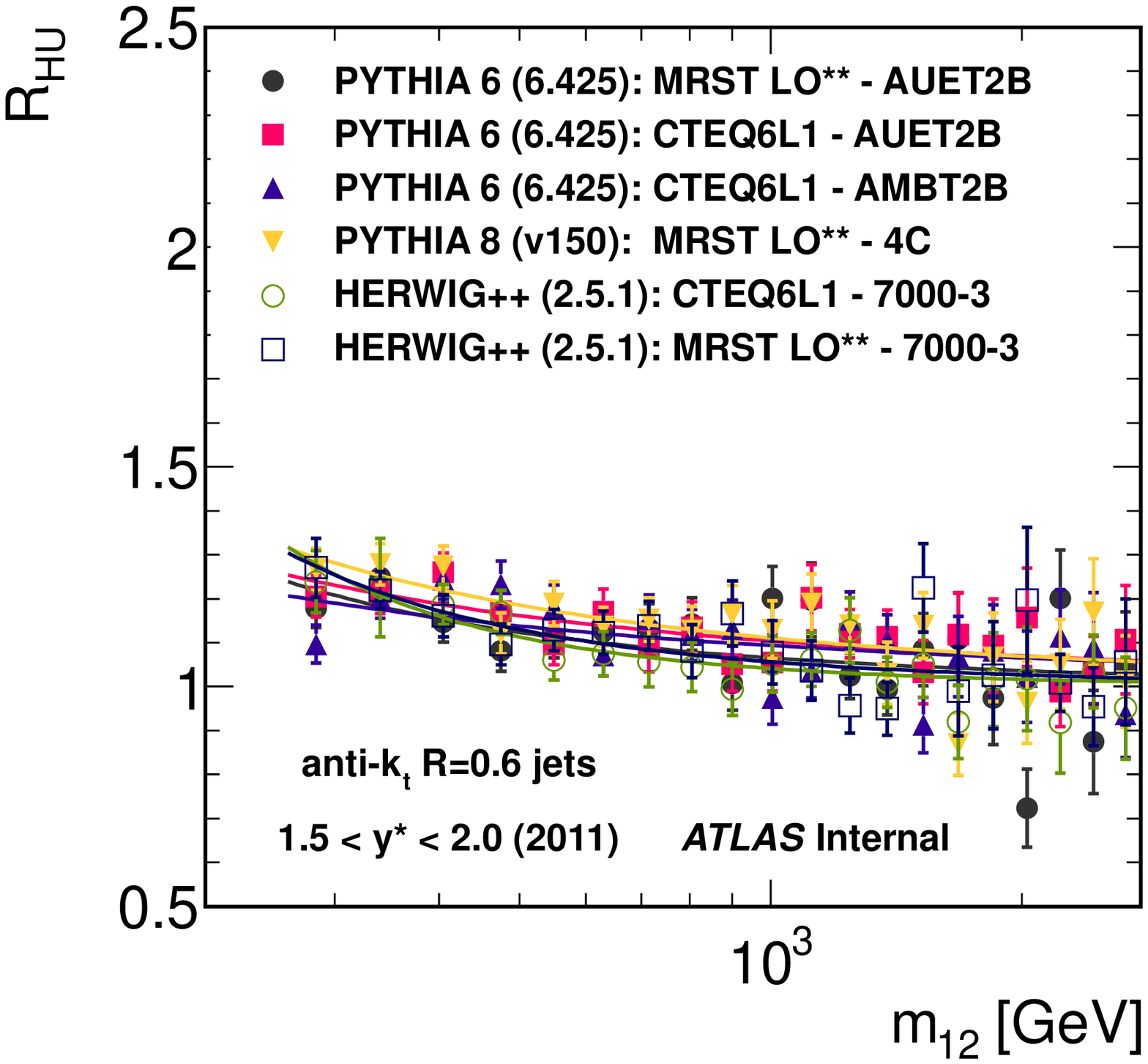}}
    \subfloat[]{\label{FIGhadronicCorrections0App4}\includegraphics[trim=5mm 14mm 0mm 10mm,clip,width=.52\textwidth]{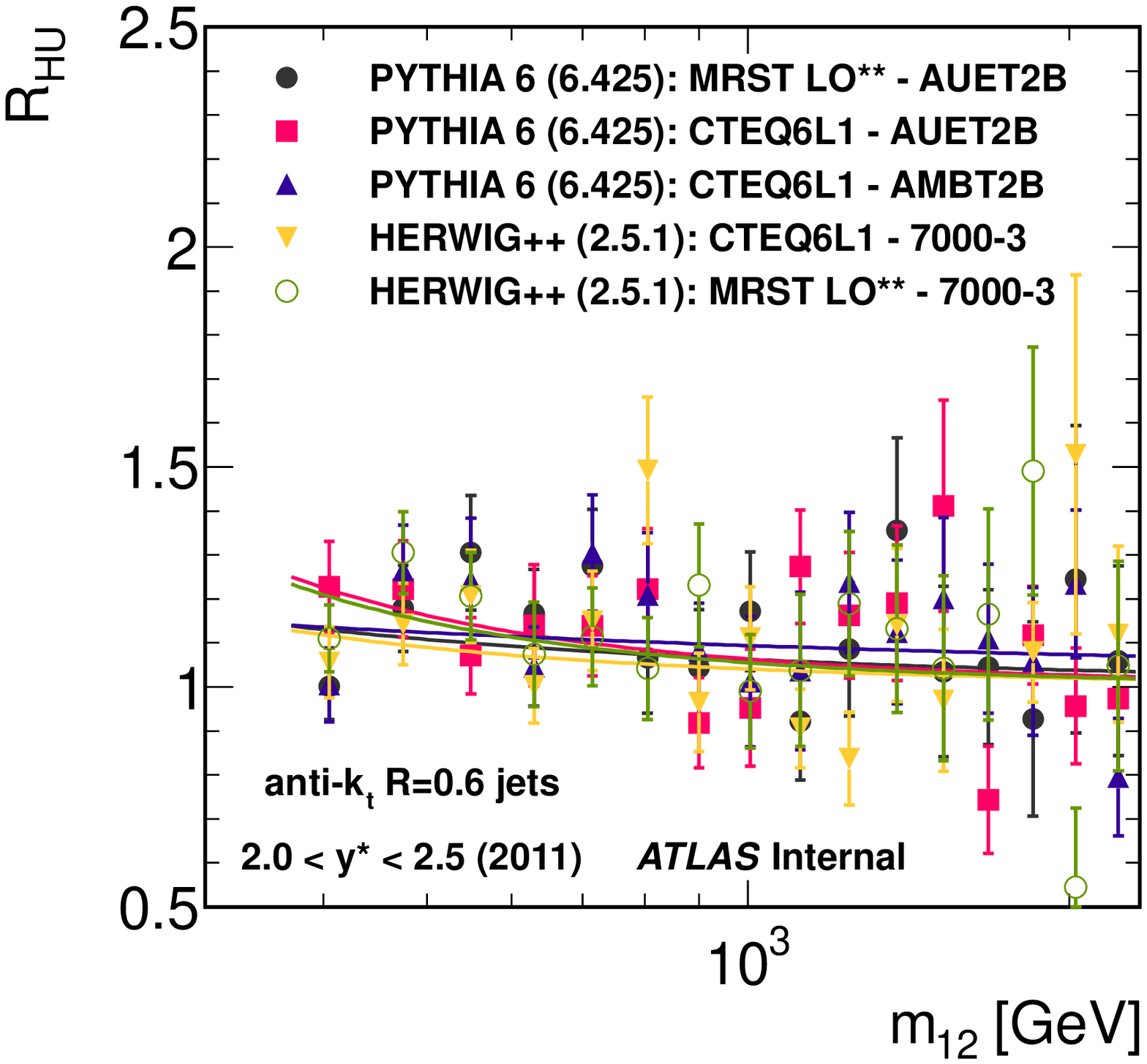}}
    \caption{\label{FIGhadronicCorrections0App}Non-perturbative correction factors, $\mathcal{R}_{\mrm{HU}}$, for the
  invariant mass, $m_{12}$, spectrum of the two jets with the highest transverse momentum in an event, as a function
  of $m_{12}$, for different \com jet rapidities, \ystr, for several
  combinations of generators, PDF sets, and underlying event tunes, as indicated in the figure and explained in \autoref{chapNonPerturbativeCorrections}.
  The lines represent fits to the various MC samples, using the 2011 phase-space definition of the measurement.
    (See also \autoref{chapMeasurementOfTheDijetMass}, \autoref{FIGhadronicCorrections}.)
    }
  \end{center}
  \end{figure} 
  \begin{figure}[htp]
  \begin{center}
    \subfloat[]{\label{FIGtheoryUncertainty0App1}\includegraphics[trim=5mm 14mm 0mm 10mm,clip,width=.52\textwidth]{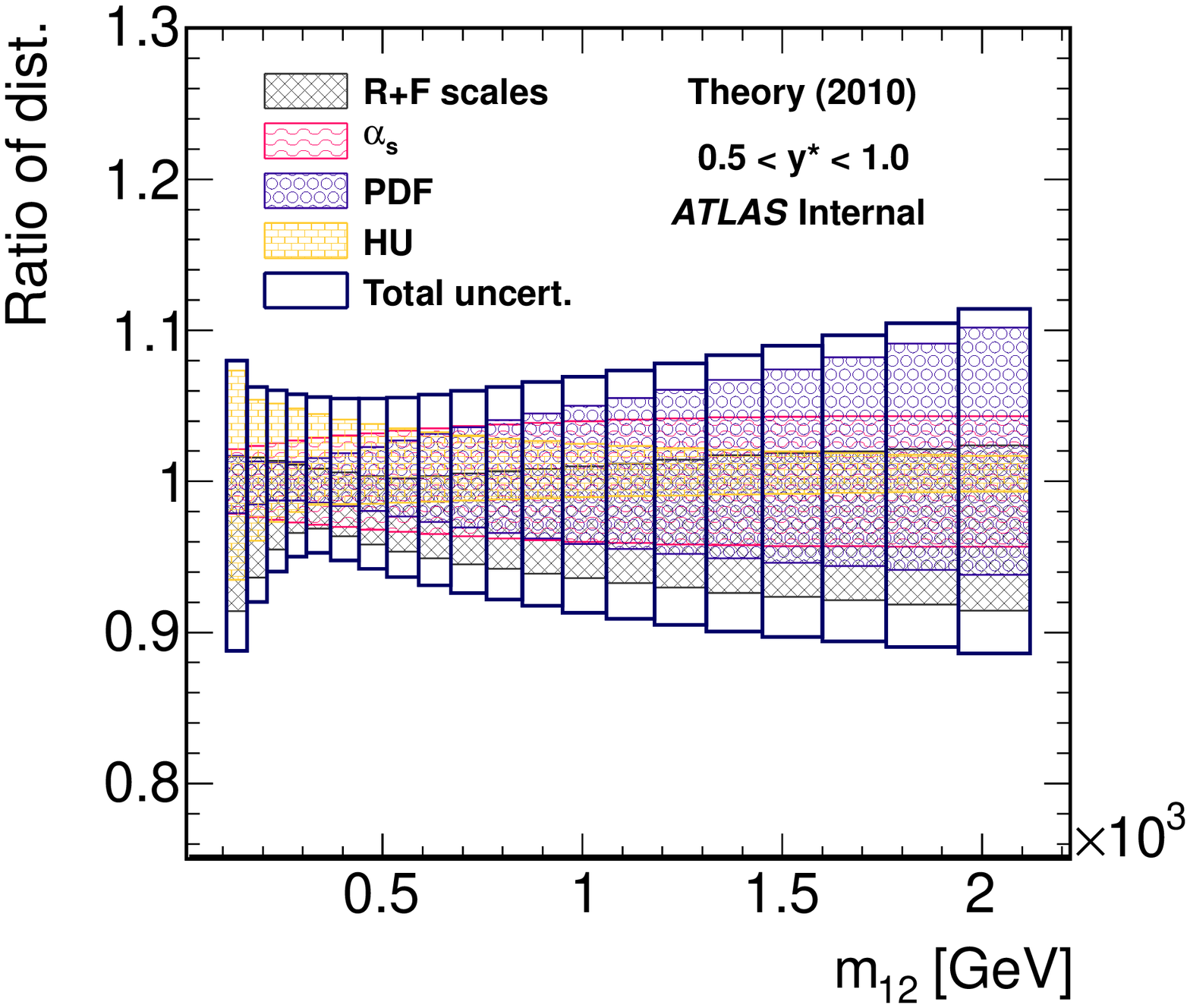}}
    \subfloat[]{\label{FIGtheoryUncertainty0App2}\includegraphics[trim=5mm 14mm 0mm 10mm,clip,width=.52\textwidth]{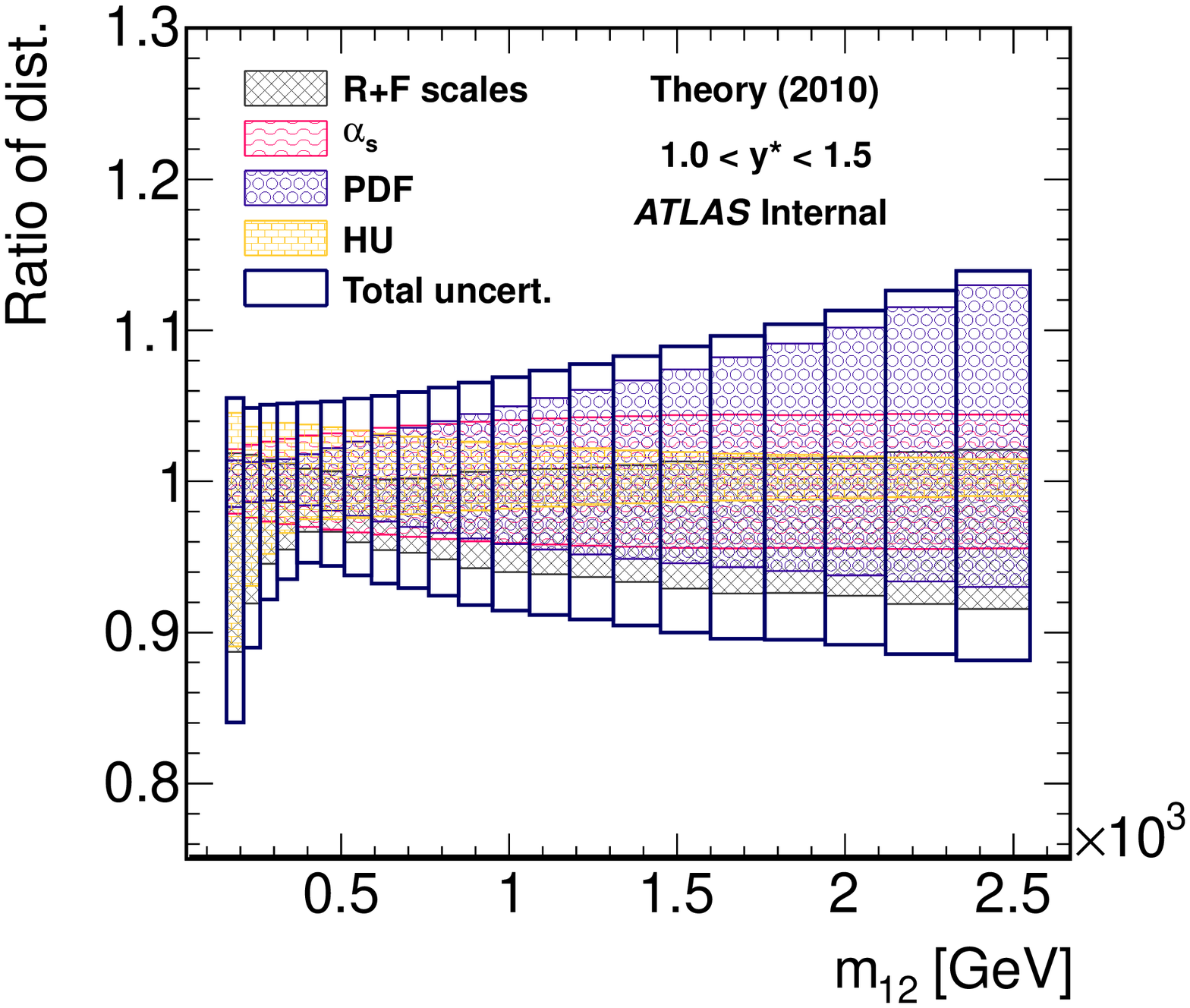}} \\
    \subfloat[]{\label{FIGtheoryUncertainty0App3}\includegraphics[trim=5mm 14mm 0mm 10mm,clip,width=.52\textwidth]{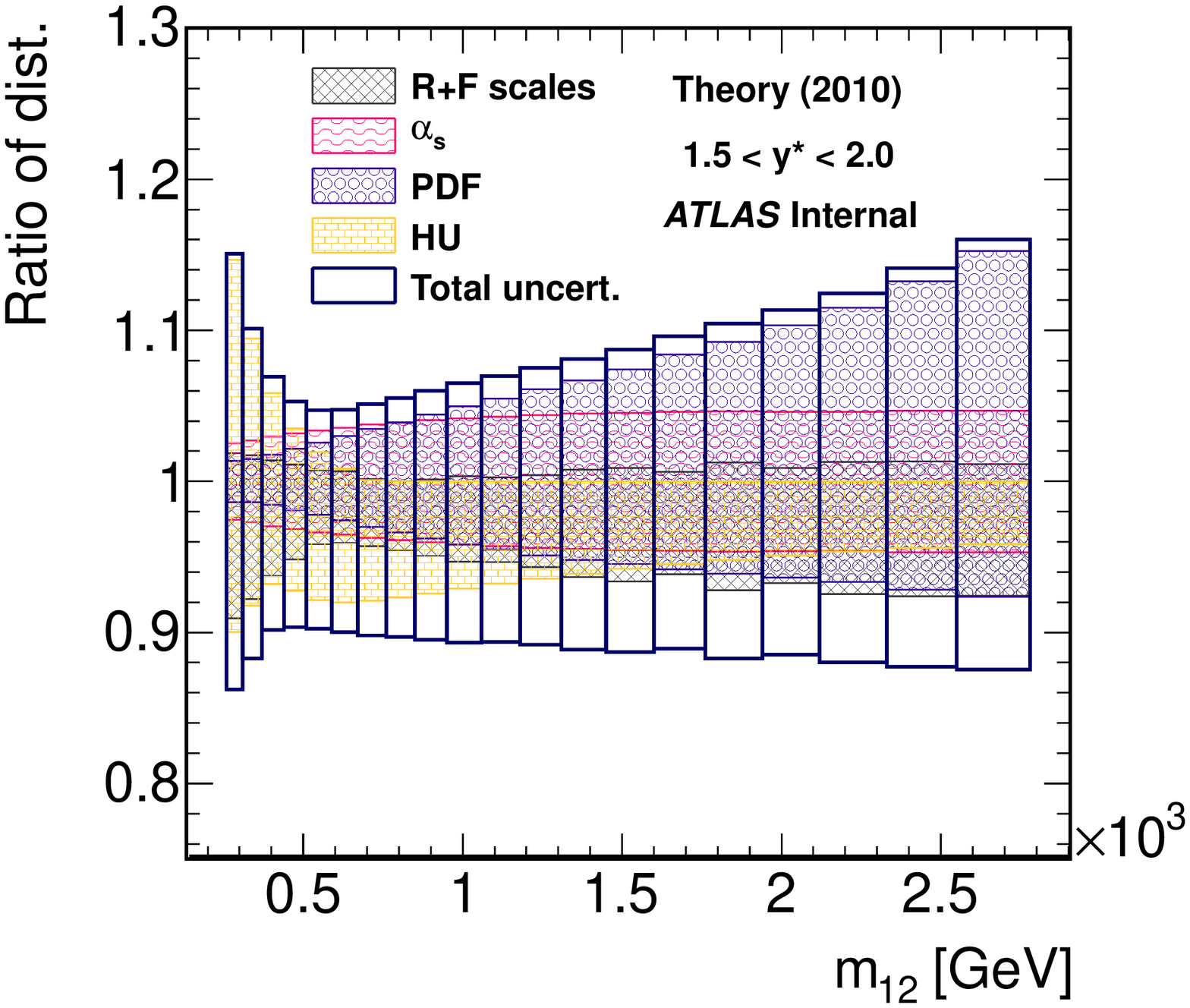}}
    \subfloat[]{\label{FIGtheoryUncertainty0App4}\includegraphics[trim=5mm 14mm 0mm 10mm,clip,width=.52\textwidth]{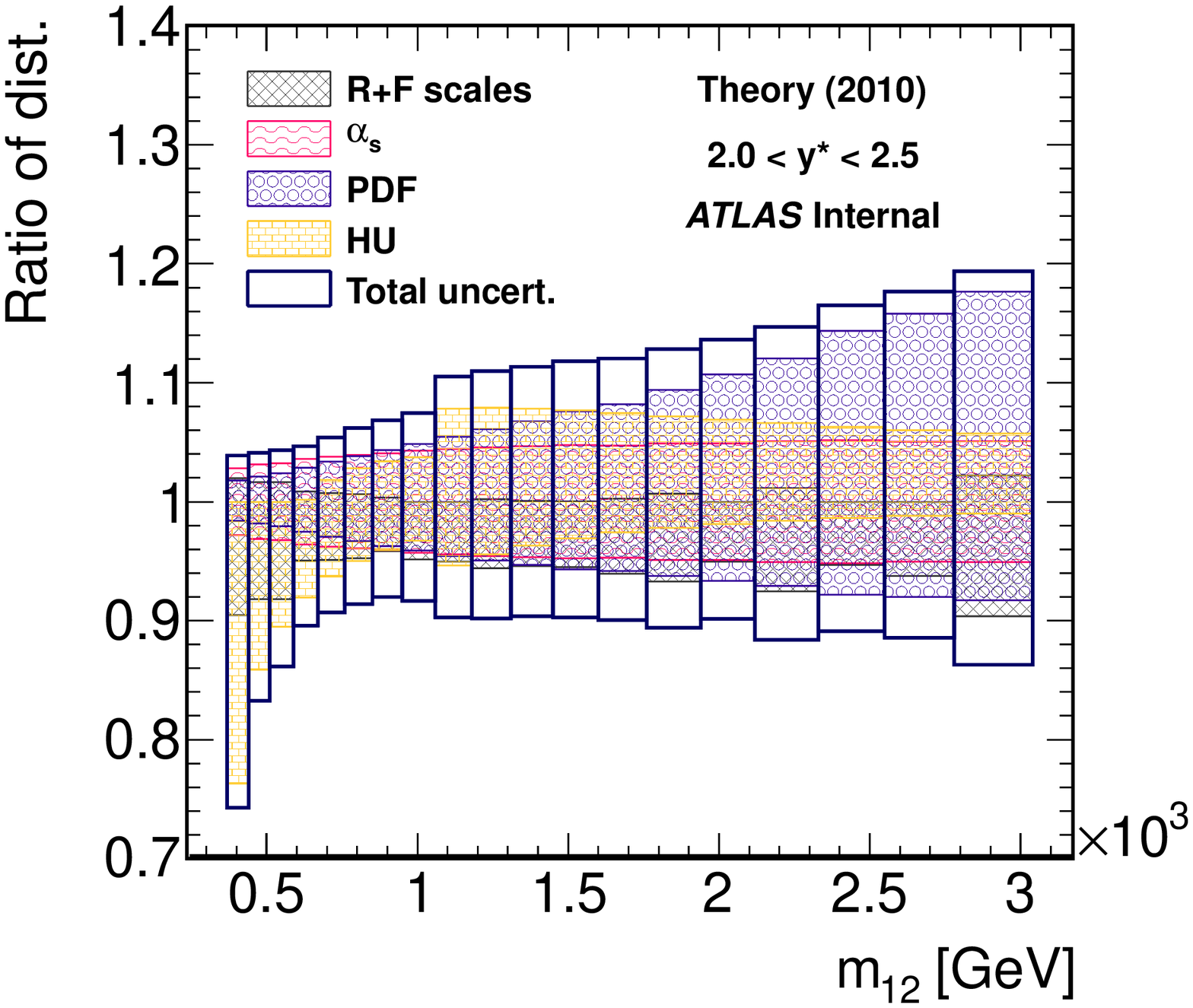}}
    \caption{\label{FIGtheoryUncertainty0App}The ratio of \nlojet expectations obtained under different assumptions relative to
    the nominal calculation, as a function
    of the invariant mass, $m_{12}$, of the two hadron-level jets with the highest transverse momentum, for different \com jet rapidities, \ystr,
    for the 2010 measurement.
    The variations on the nominal expectation include the uncertainty on the
    renormalization and factorisation scales (R+F scales), the uncertainty on the value of the strong
    coupling constant ($\alpha_{\rm s}$), use of different parton density functions (PDF), the uncertainty on the
    hadronization and UE corrections (HU) and the total uncertainty on all the latter (total uncert.\/).
    (See also \autoref{chapMeasurementOfTheDijetMass}, \autoref{FIGtheoryUncertainty} and accompanying text.)
  }
  \end{center}
  \end{figure} 
  \begin{figure}[htp]
  \begin{center}
    \subfloat[]{\label{FIGtheoryUncertainty1App1}\includegraphics[trim=5mm 14mm 0mm 10mm,clip,width=.52\textwidth]{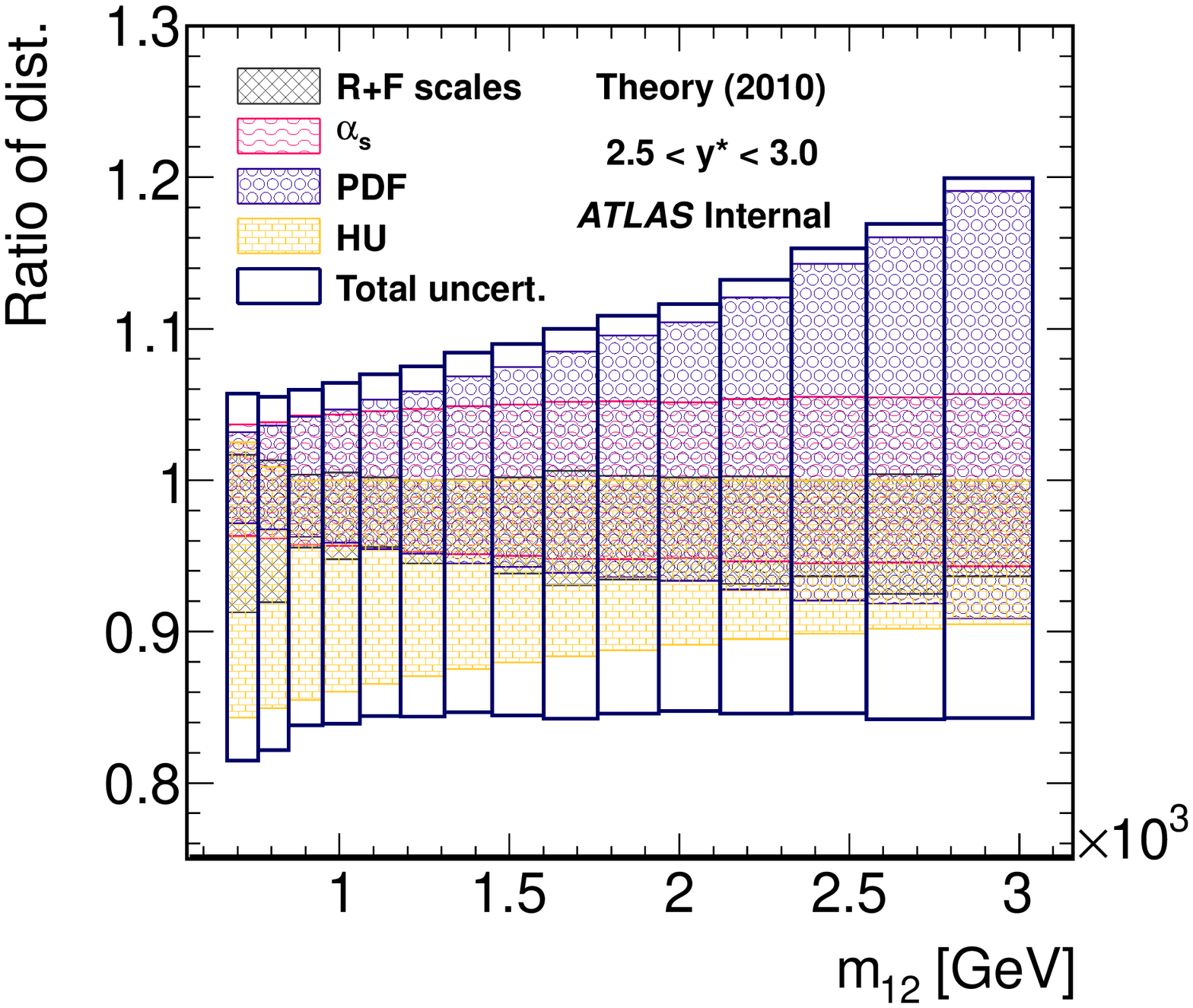}}
    \subfloat[]{\label{FIGtheoryUncertainty1App2}\includegraphics[trim=5mm 14mm 0mm 10mm,clip,width=.52\textwidth]{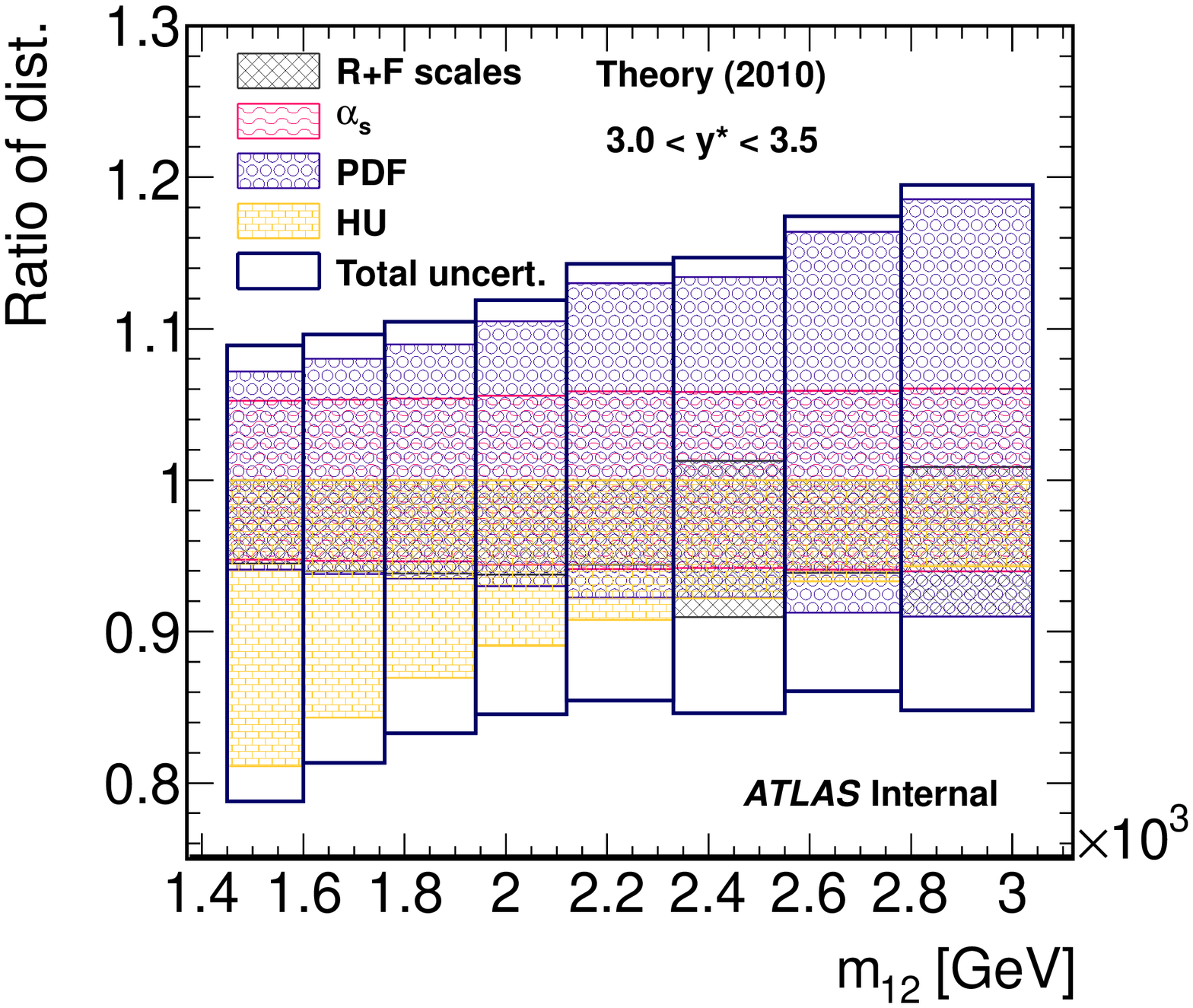}} \\
    \subfloat[]{\label{FIGtheoryUncertainty1App3}\includegraphics[trim=5mm 14mm 0mm 10mm,clip,width=.52\textwidth]{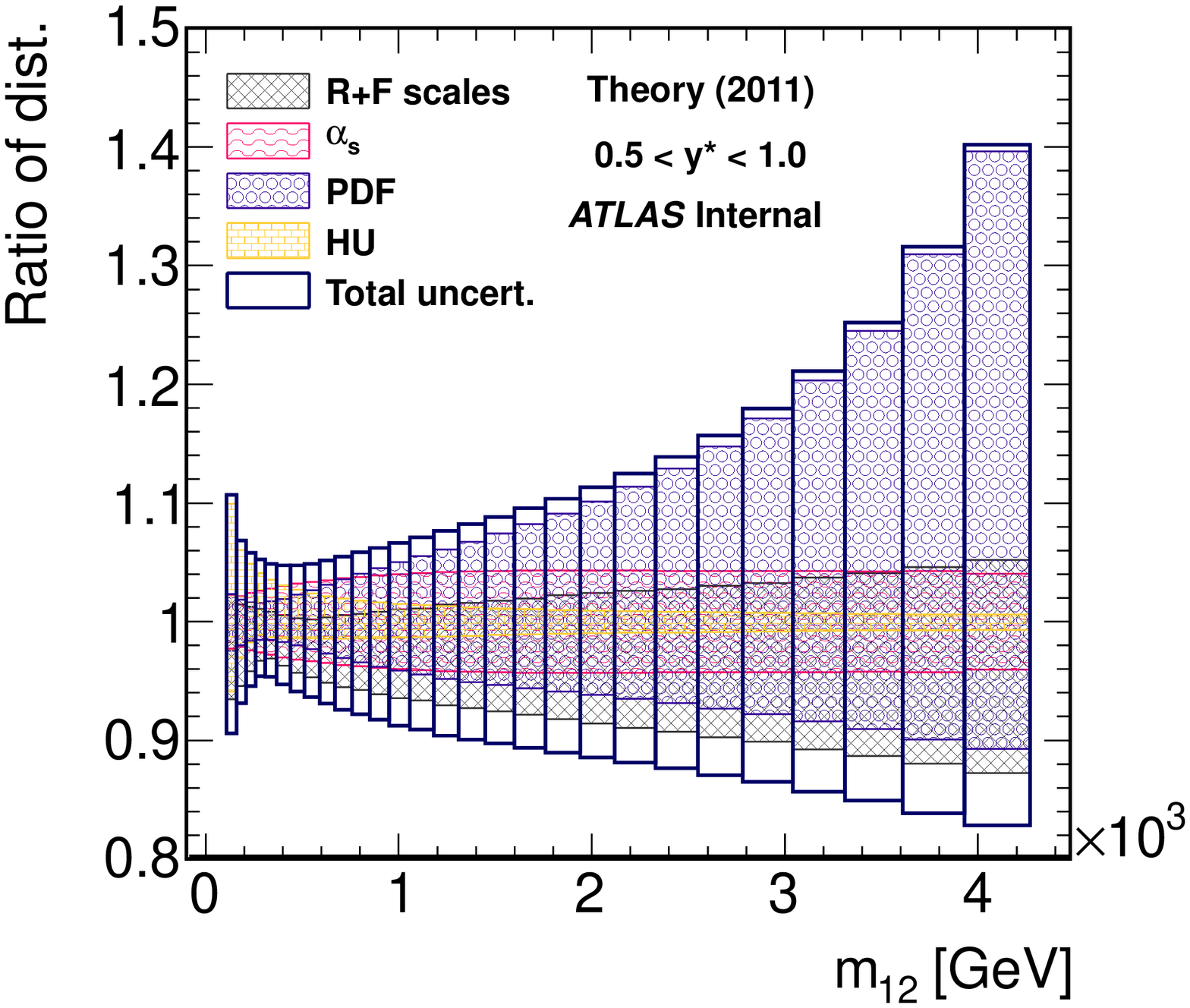}}
    \subfloat[]{\label{FIGtheoryUncertainty1App4}\includegraphics[trim=5mm 14mm 0mm 10mm,clip,width=.52\textwidth]{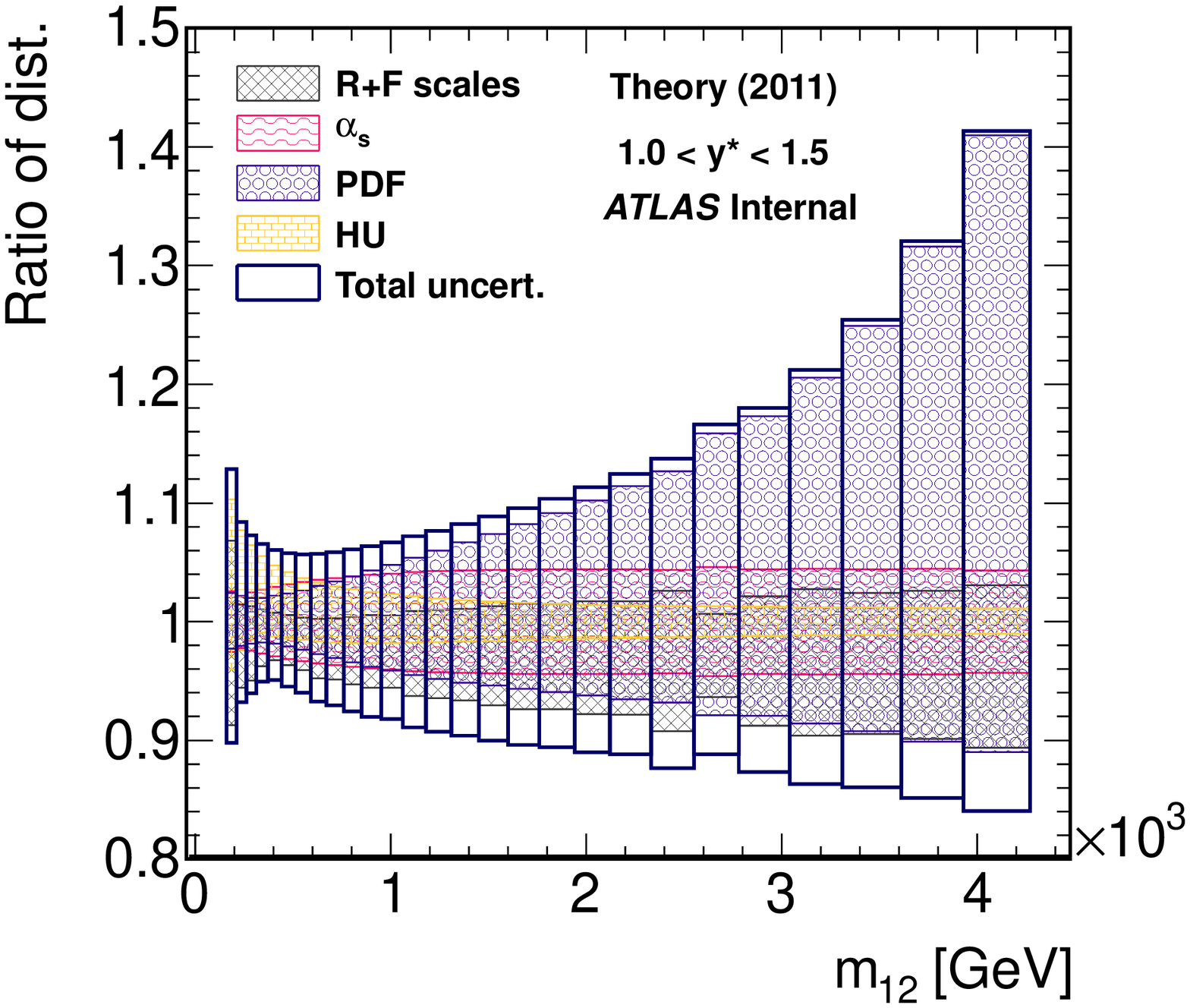}}
    \caption{\label{FIGtheoryUncertainty1App}The ratio of \nlojet expectations obtained under different assumptions relative to
    the nominal calculation, as a function
    of the invariant mass, $m_{12}$, of the two hadron-level jets with the highest transverse momentum, for different \com jet rapidities, \ystr,
    for the 2010 (\Subref{FIGtheoryUncertainty1App1},\Subref{FIGtheoryUncertainty1App2})
    and for the 2011 (\Subref{FIGtheoryUncertainty1App3} and \Subref{FIGtheoryUncertainty1App4}) measurements.
    The variations on the nominal expectation include the uncertainty on the
    renormalization and factorisation scales (R+F scales), the uncertainty on the value of the strong
    coupling constant ($\alpha_{\rm s}$), use of different parton density functions (PDF), the uncertainty on the
    hadronization and UE corrections (HU) and the total uncertainty on all the latter (total uncert.\/).
    (See also \autoref{chapMeasurementOfTheDijetMass}, \autoref{FIGtheoryUncertainty} and accompanying text.)
  }
  \end{center}
  \end{figure} 
  \begin{figure}[htp]
  \vspace{-35pt}
  \begin{center}
    \subfloat[]{\label{FIGtheoryUncertainty2App1}\includegraphics[trim=5mm 14mm 0mm 10mm,clip,width=.52\textwidth]{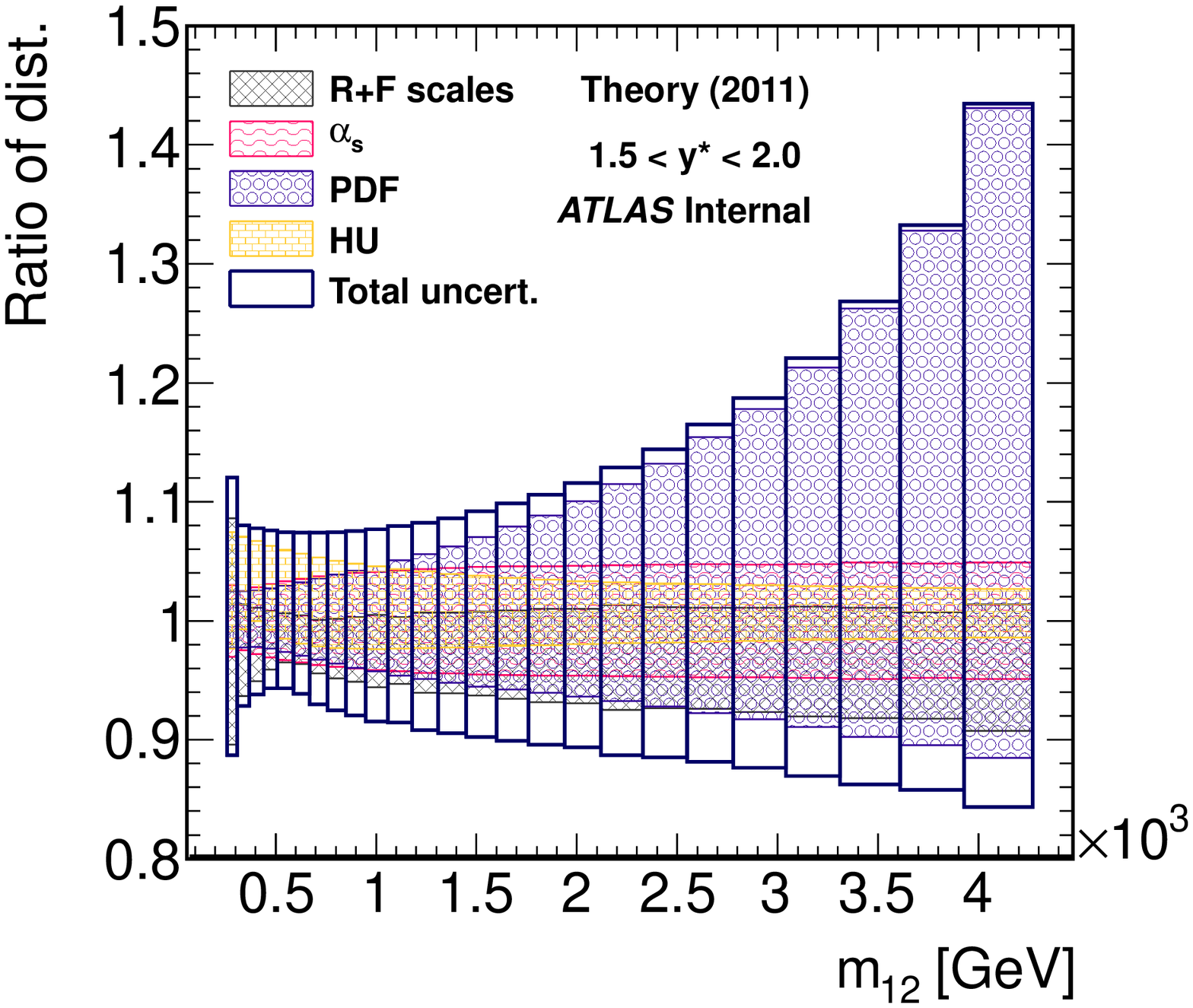}}
    \subfloat[]{\label{FIGtheoryUncertainty2App2}\includegraphics[trim=5mm 14mm 0mm 10mm,clip,width=.52\textwidth]{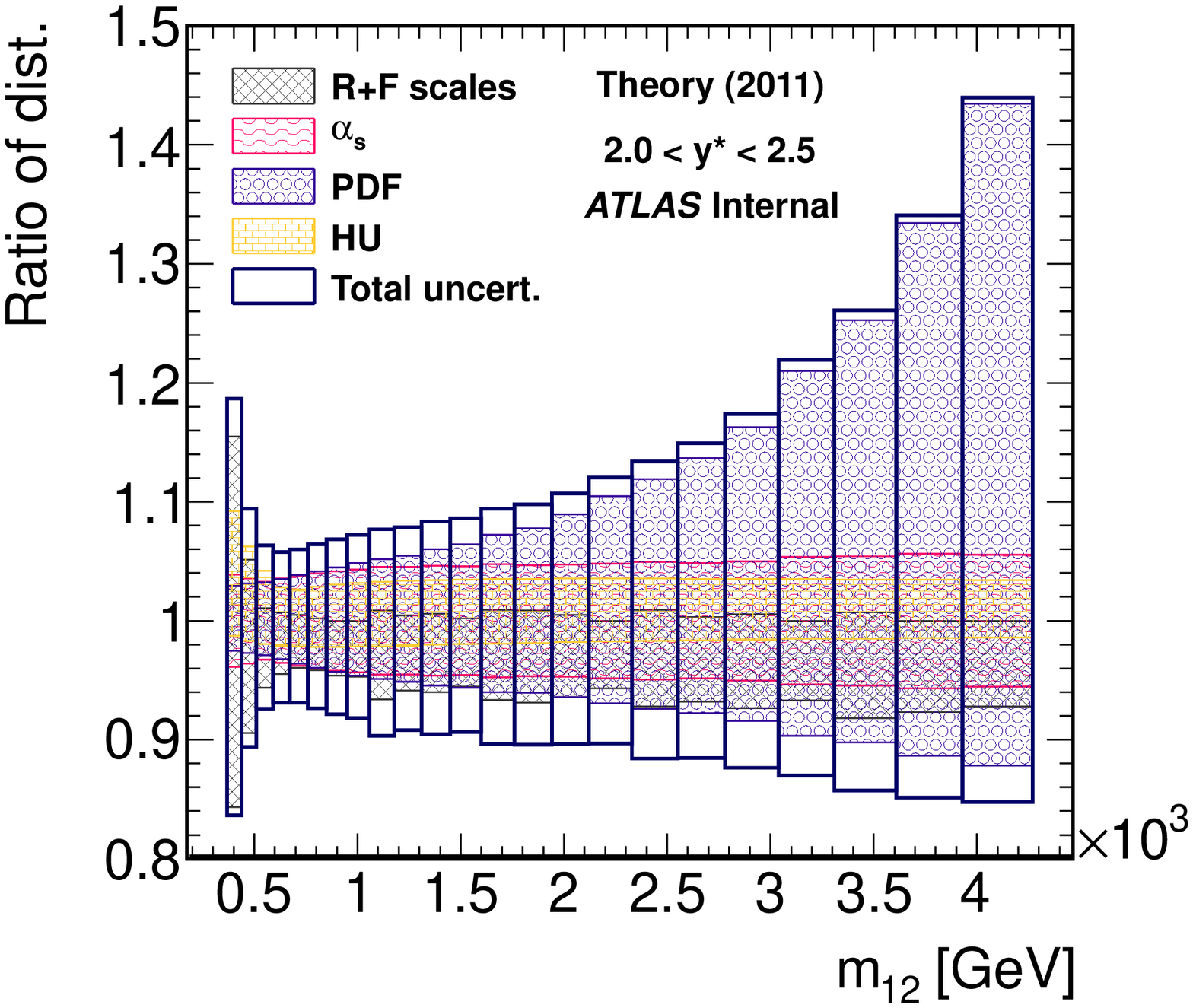}}
    \caption{\label{FIGtheoryUncertainty2App}The ratio of \nlojet expectations obtained under different assumptions relative to
    the nominal calculation, as a function
    of the invariant mass, $m_{12}$, of the two hadron-level jets with the highest transverse momentum, for different \com jet rapidities, \ystr,
    for the 2011 measurement.
    The variations on the nominal expectation include the uncertainty on the
    renormalization and factorisation scales (R+F scales), the uncertainty on the value of the strong
    coupling constant ($\alpha_{\rm s}$), use of different parton density functions (PDF), the uncertainty on the
    hadronization and UE corrections (HU) and the total uncertainty on all the latter (total uncert.\/).
    (See also \autoref{chapMeasurementOfTheDijetMass}, \autoref{FIGtheoryUncertainty} and accompanying text.)
  }
  \end{center}
  %
  %
\begin{center}
  \vspace{-30pt}
  \subfloat[]{\label{FIGdijetMassTotalSystematicUncertainty2010AppA1}\includegraphics[trim=5mm 14mm 0mm 10mm,clip,width=.52\textwidth]{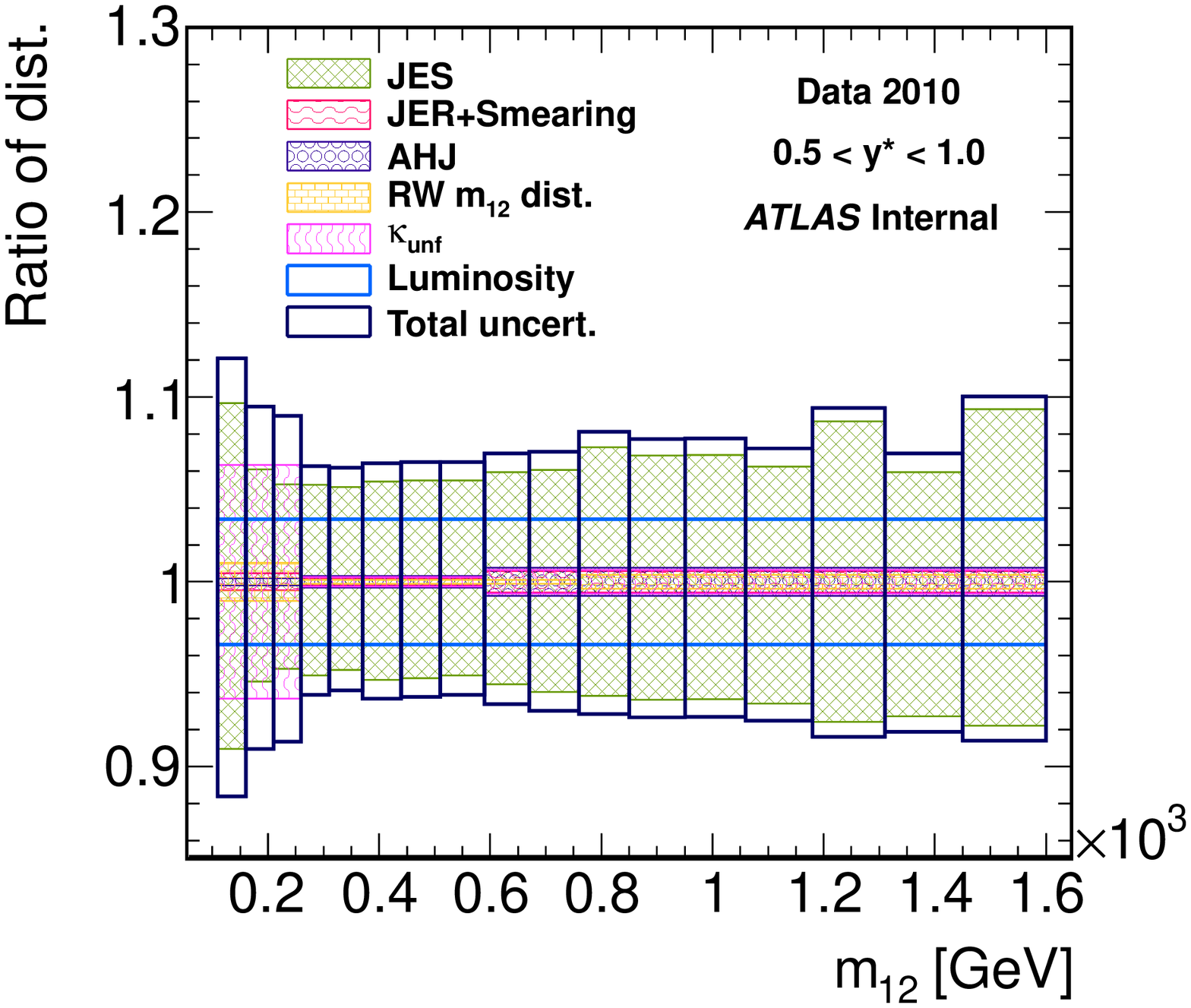}}
  \subfloat[]{\label{FIGdijetMassTotalSystematicUncertainty2010AppA2}\includegraphics[trim=5mm 14mm 0mm 10mm,clip,width=.52\textwidth]{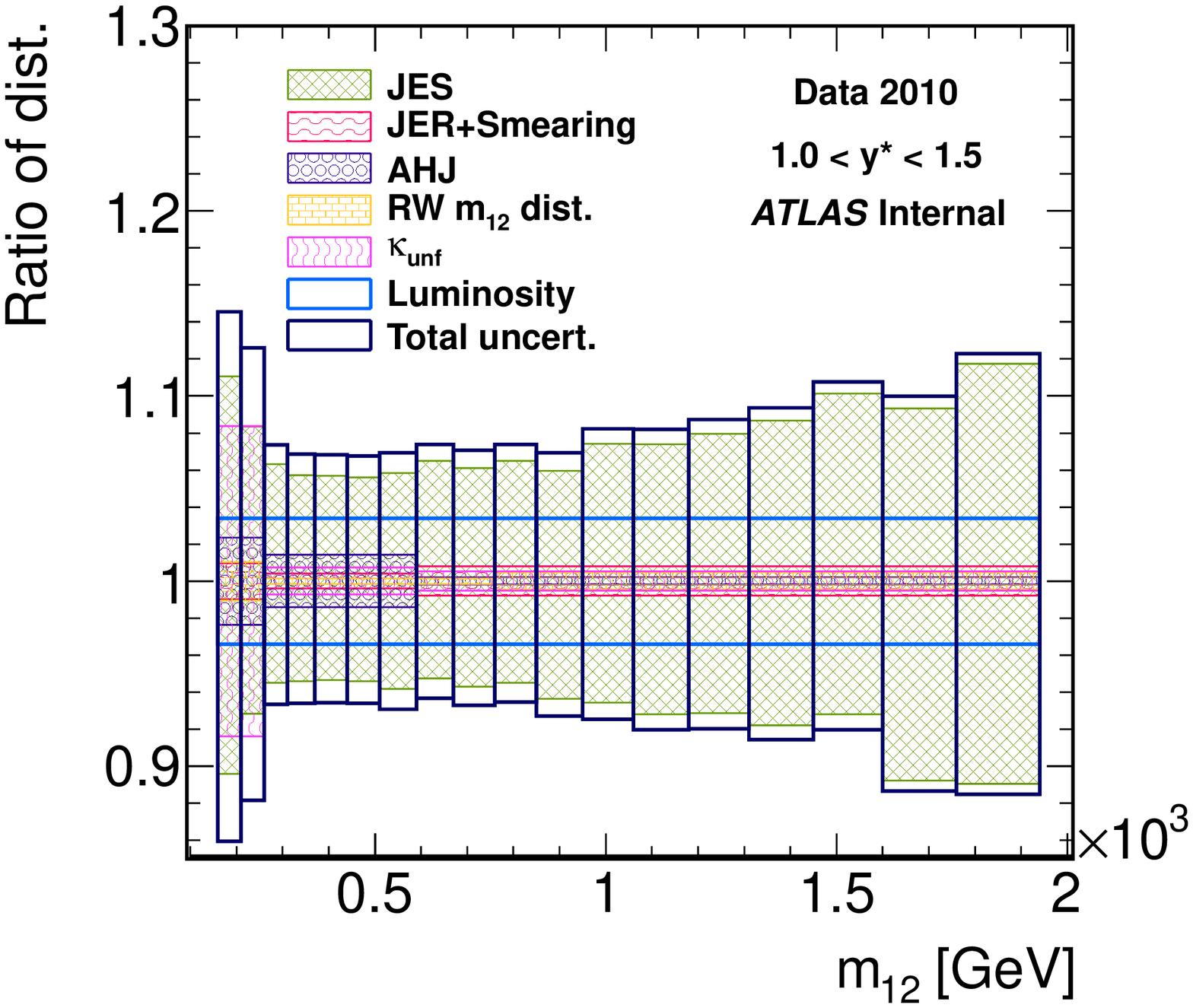}}
  \caption{\label{FIGdijetMassTotalSystematicUncertainty2010AppA}Dependence on the invariant mass, $m_{12}$, of the two jets with the highest transverse momentum in an event,
  of the ratio between the $m_{12}$ distributions
  with and without systematic variations (described in the text), for \com jet rapidity, $\ystr < 0.5$, in the 2010 measurement,
  including the uncertainty on the jet energy scale (\JES);
  the uncertainty on the jet energy and angular resolutions (\JES + smearing);
  the uncertainty associated with the choice of physics generator (\AHJ); 
  the uncertainty associated with variation in the simulated shape of the $m_{12}$ spectrum (RW $m_{12}$ dist.\/);
  the choice of the number of iterations used in the unfolding procedure ($\kappa_{\mrm{unf}}$);
  the uncertainty on the luminosity (luminosity); and
  the total uncertainty on all the latter (total uncert.\/).
  (See also \autoref{chapMeasurementOfTheDijetMass}, \autoref{FIGdijetMassTotalSystematicUncertainty} and accompanying text.)
  }
\end{center}
\end{figure} 

   \clearpage
\begin{figure}[htp]
\begin{center}
  \subfloat[]{\label{FIGdijetMassTotalSystematicUncertainty2010AppB1}\includegraphics[trim=5mm 14mm 0mm 10mm,clip,width=.52\textwidth]{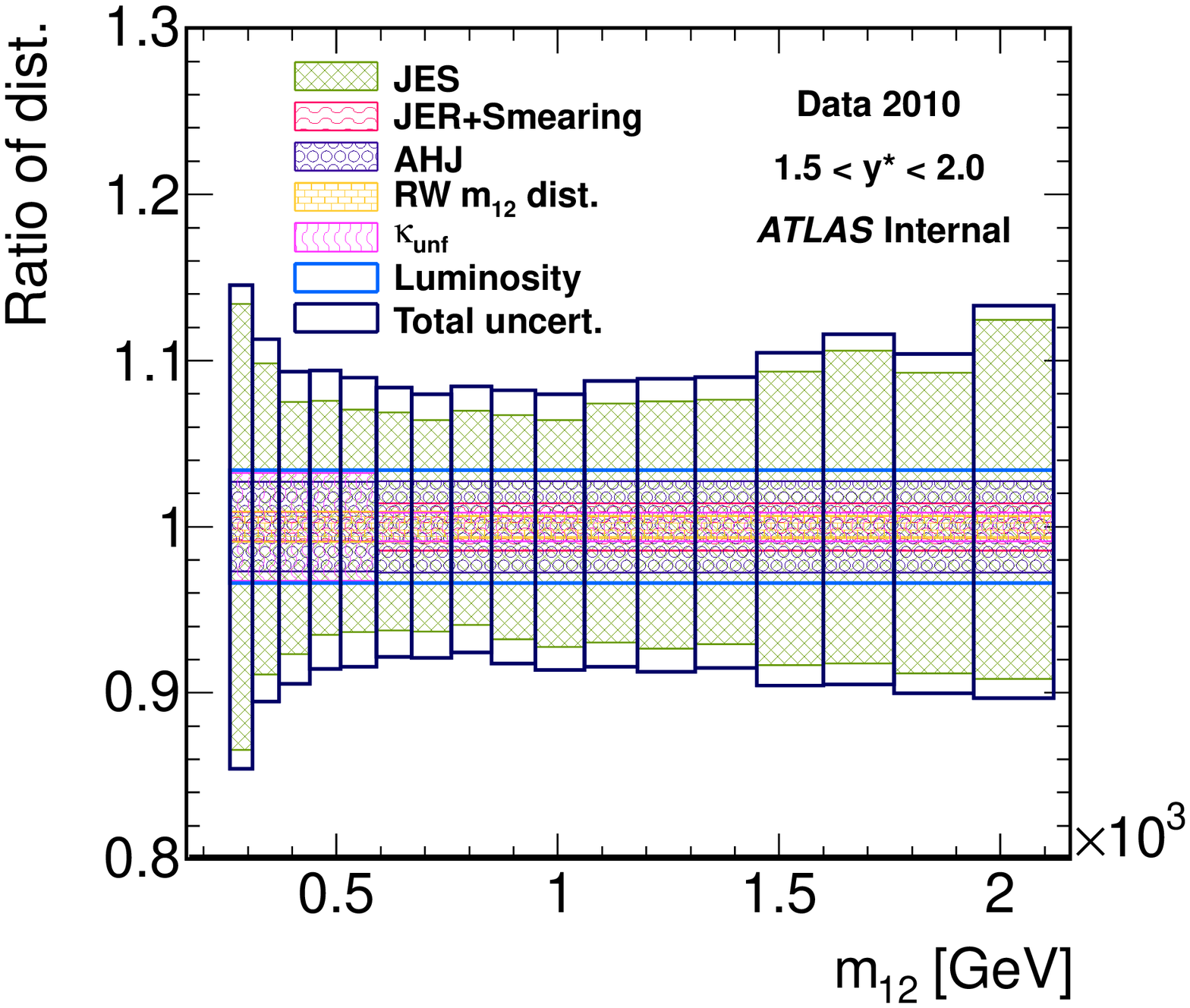}}
  \subfloat[]{\label{FIGdijetMassTotalSystematicUncertainty2010AppB2}\includegraphics[trim=5mm 14mm 0mm 10mm,clip,width=.52\textwidth]{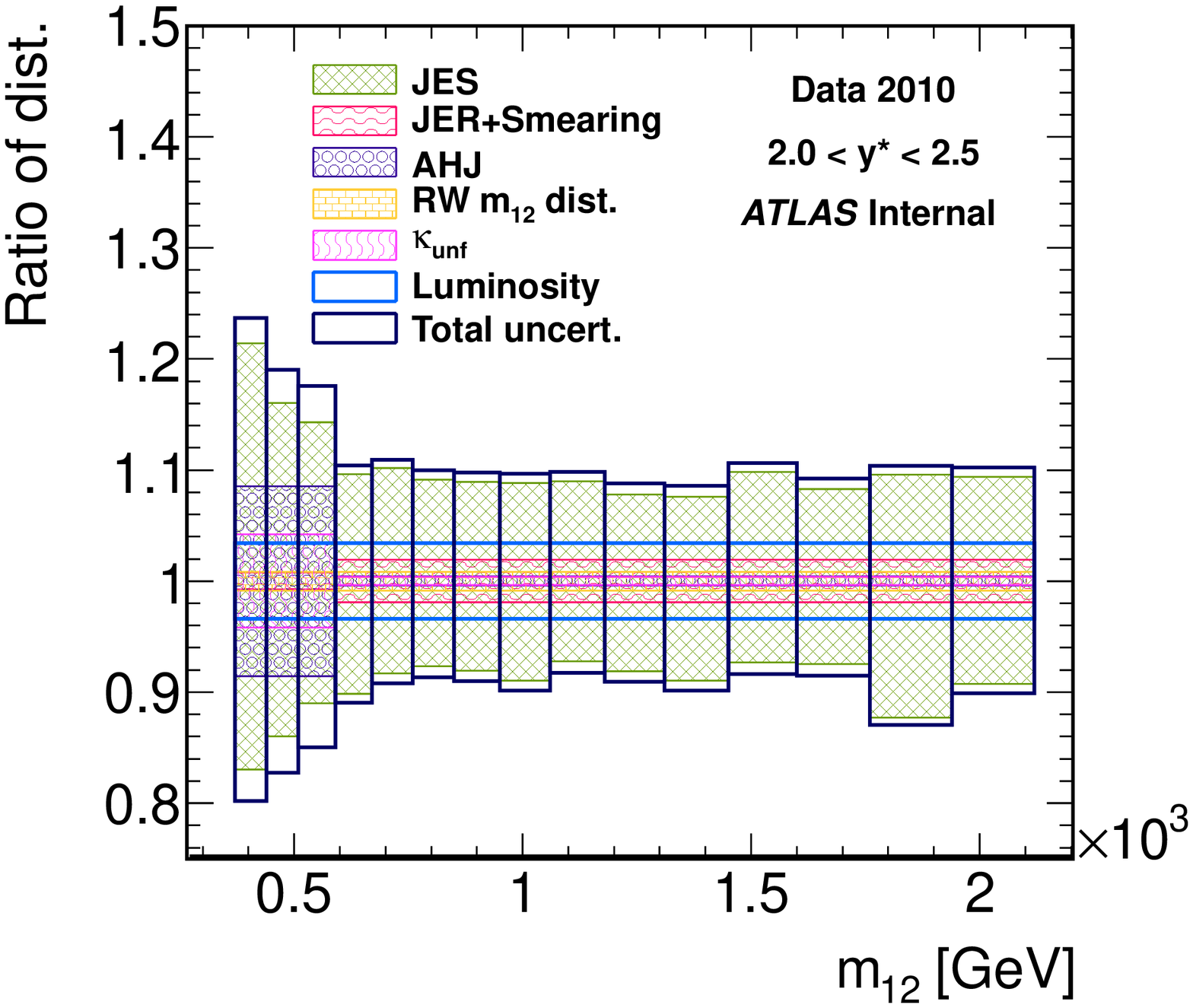}} \\
  \subfloat[]{\label{FIGdijetMassTotalSystematicUncertainty2010AppB3}\includegraphics[trim=5mm 14mm 0mm 10mm,clip,width=.52\textwidth]{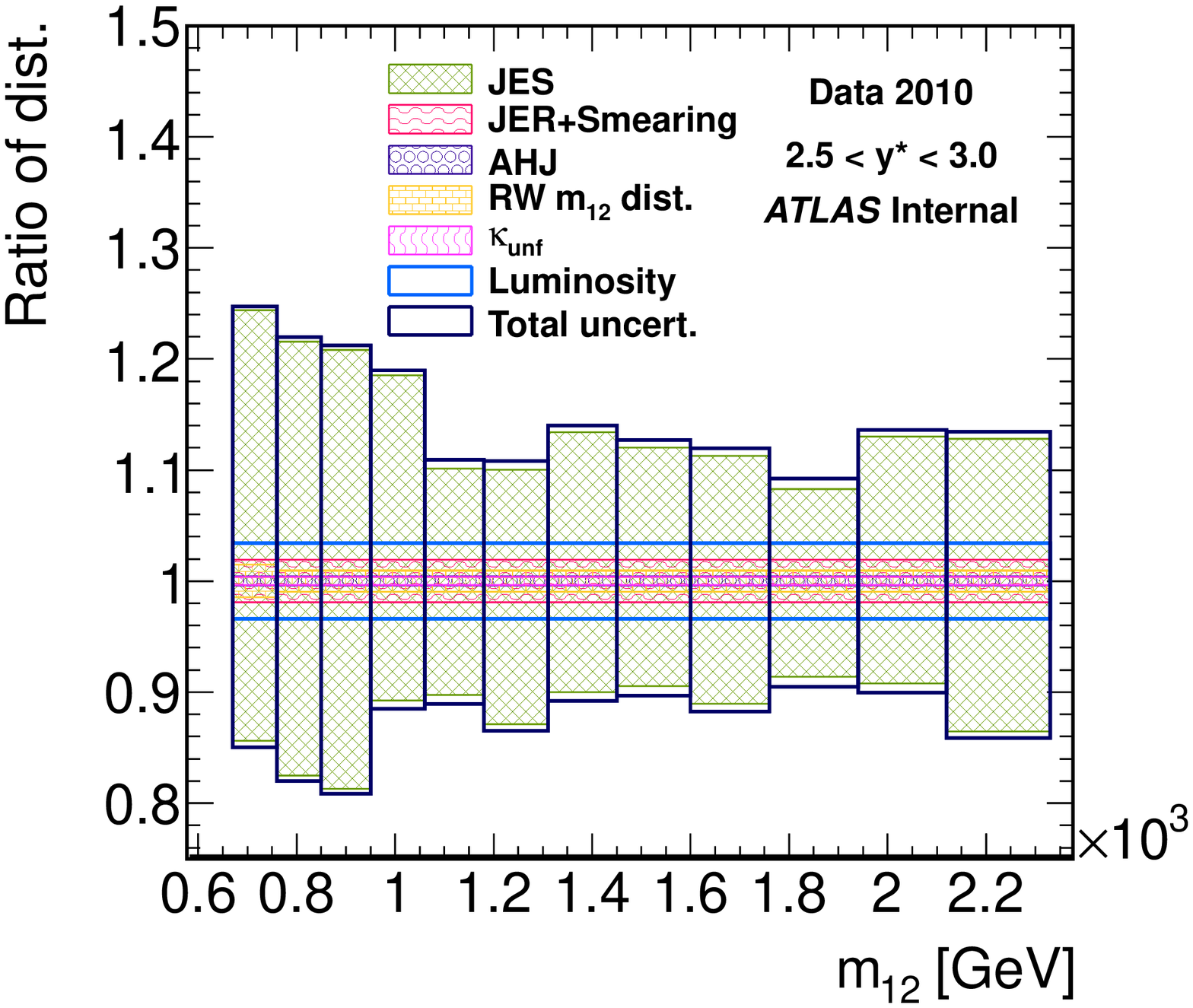}}
  \subfloat[]{\label{FIGdijetMassTotalSystematicUncertainty2010AppB4}\includegraphics[trim=5mm 14mm 0mm 10mm,clip,width=.52\textwidth]{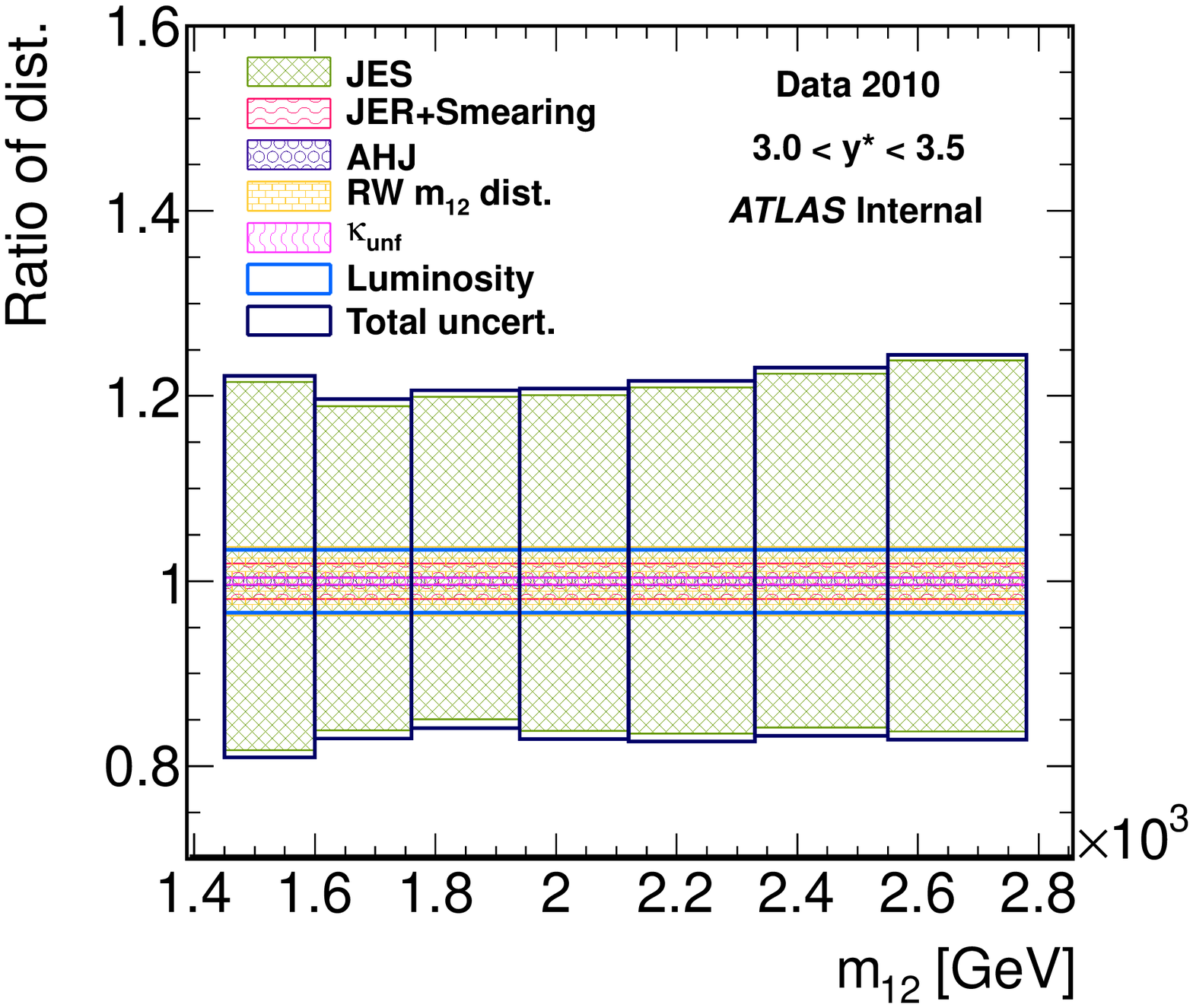}}
  \caption{\label{FIGdijetMassTotalSystematicUncertainty2010AppB}Dependence on the invariant mass, $m_{12}$, of the two jets with the highest transverse momentum in an event,
  of the ratio between the $m_{12}$ distributions
  with and without systematic variations (described in the text), for different \com jet rapidities, \ystr, in the 2010 measurement,
  including the uncertainty on the jet energy scale (\JES);
  the uncertainty on the jet energy and angular resolutions (\JES + smearing);
  the uncertainty associated with the choice of physics generator (\AHJ); 
  the uncertainty associated with variation in the simulated shape of the $m_{12}$ spectrum (RW $m_{12}$ dist.\/);
  the choice of the number of iterations used in the unfolding procedure ($\kappa_{\mrm{unf}}$);
  the uncertainty on the luminosity (luminosity); and
  the total uncertainty on all the latter (total uncert.\/).
  (See also \autoref{chapMeasurementOfTheDijetMass}, \autoref{FIGdijetMassTotalSystematicUncertainty} and accompanying text.)
  }
\end{center}
\end{figure} 
\begin{figure}[htp]
\begin{center}
  \subfloat[]{\label{FIGdijetMassTotalSystematicUncertainty2011App1}\includegraphics[trim=5mm 14mm 0mm 10mm,clip,width=.52\textwidth]{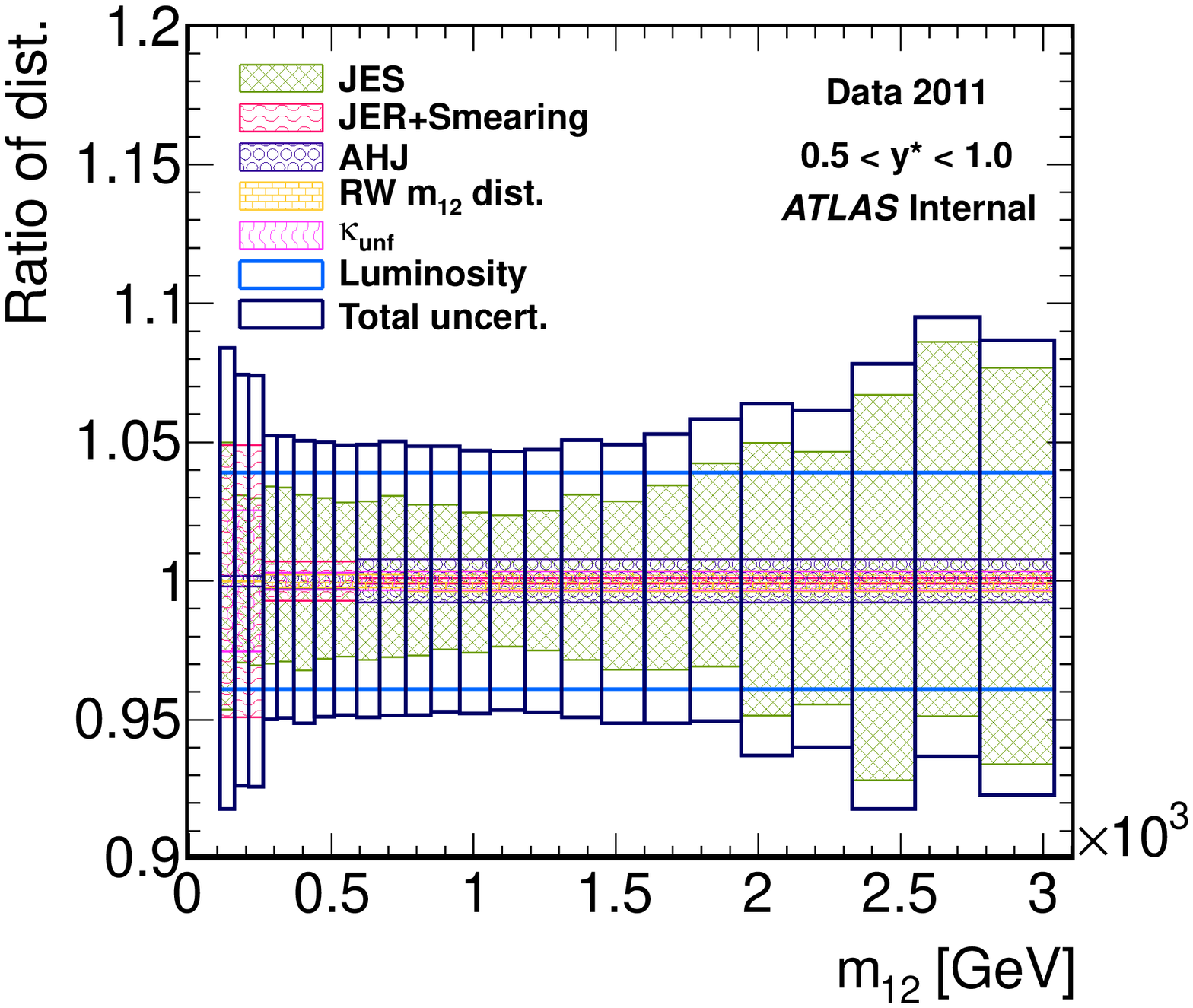}}
  \subfloat[]{\label{FIGdijetMassTotalSystematicUncertainty2011App2}\includegraphics[trim=5mm 14mm 0mm 10mm,clip,width=.52\textwidth]{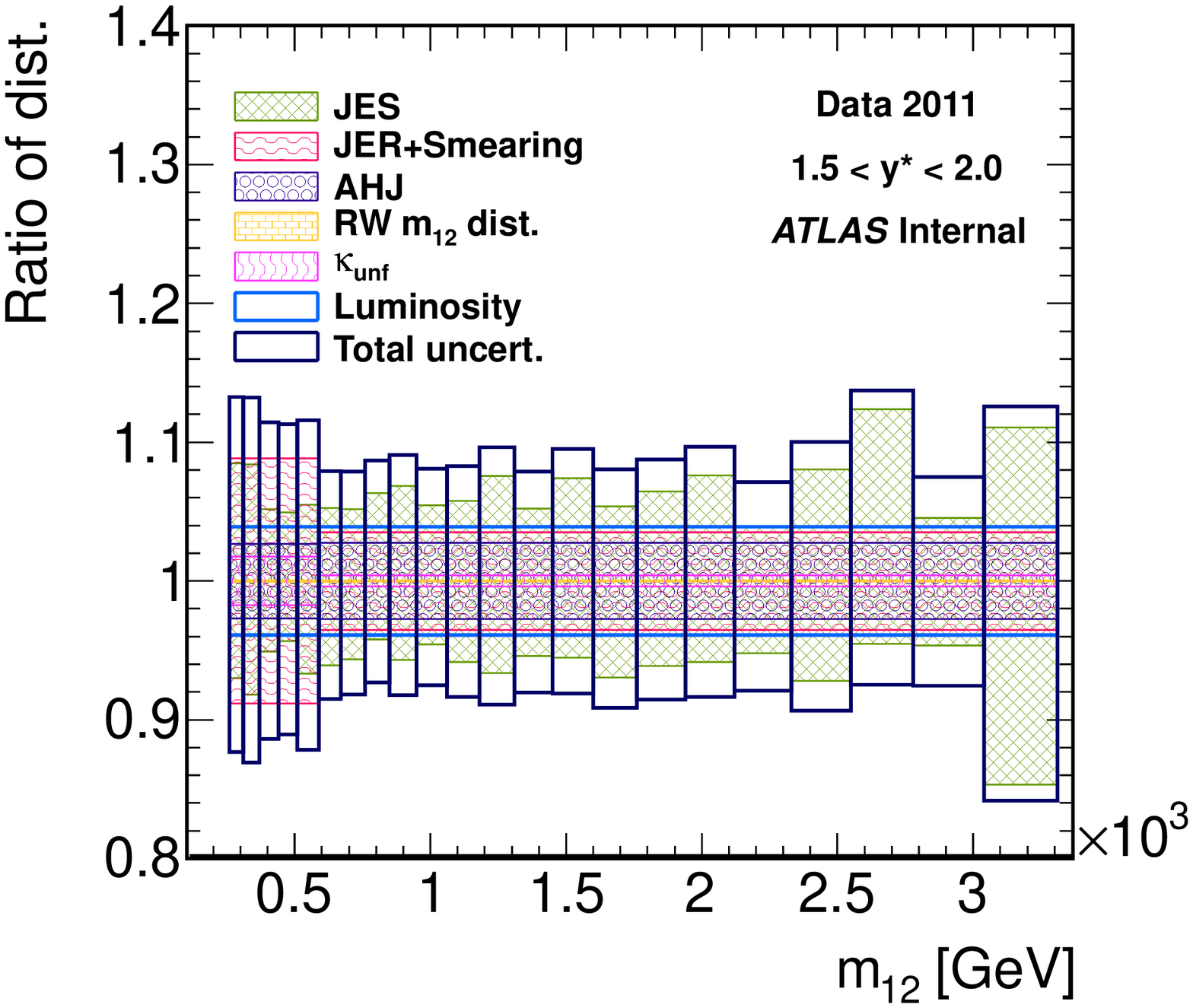}} \\
  \subfloat[]{\label{FIGdijetMassTotalSystematicUncertainty2011App3}\includegraphics[trim=5mm 14mm 0mm 10mm,clip,width=.52\textwidth]{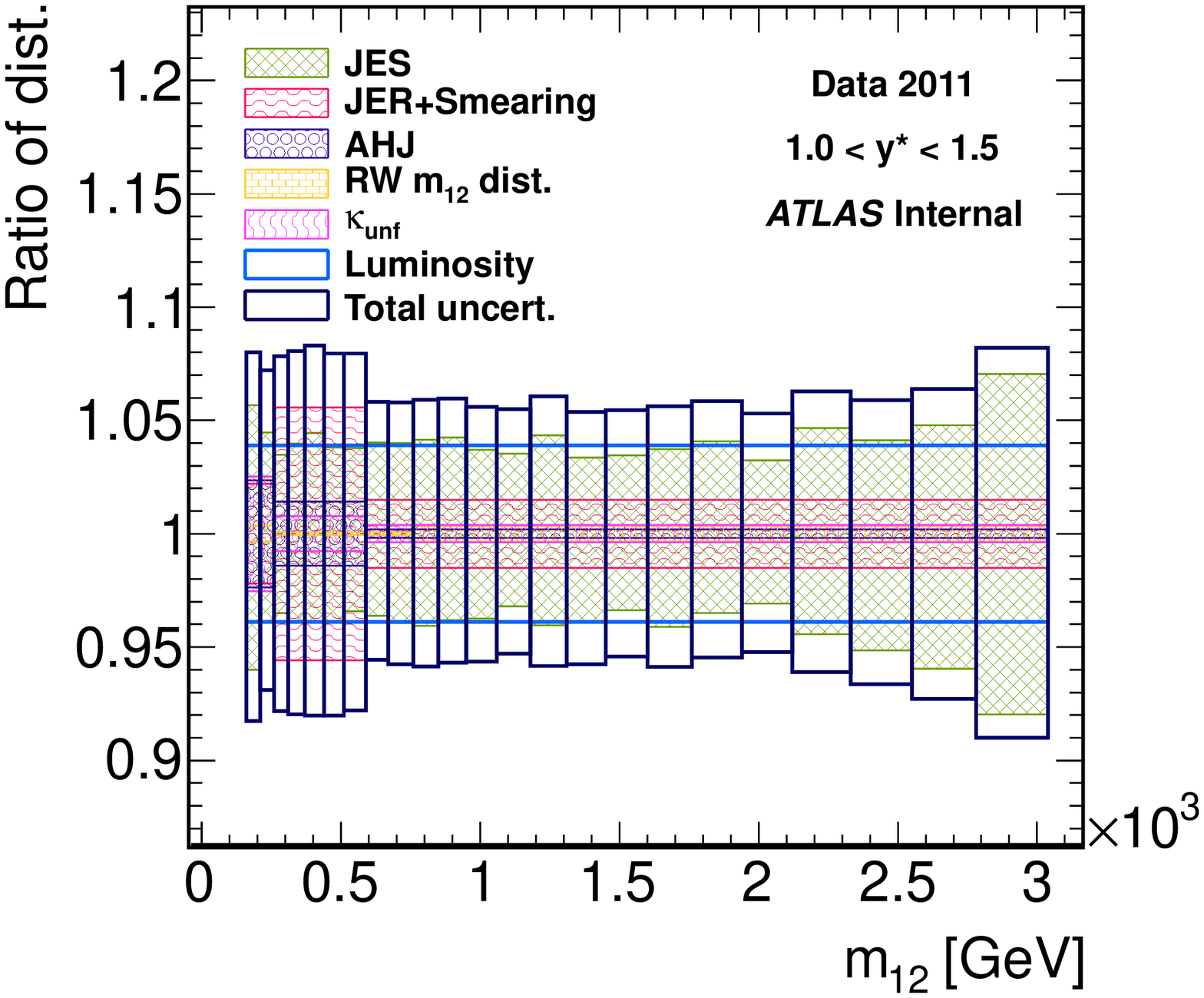}}
  \subfloat[]{\label{FIGdijetMassTotalSystematicUncertainty2011App4}\includegraphics[trim=5mm 14mm 0mm 10mm,clip,width=.52\textwidth]{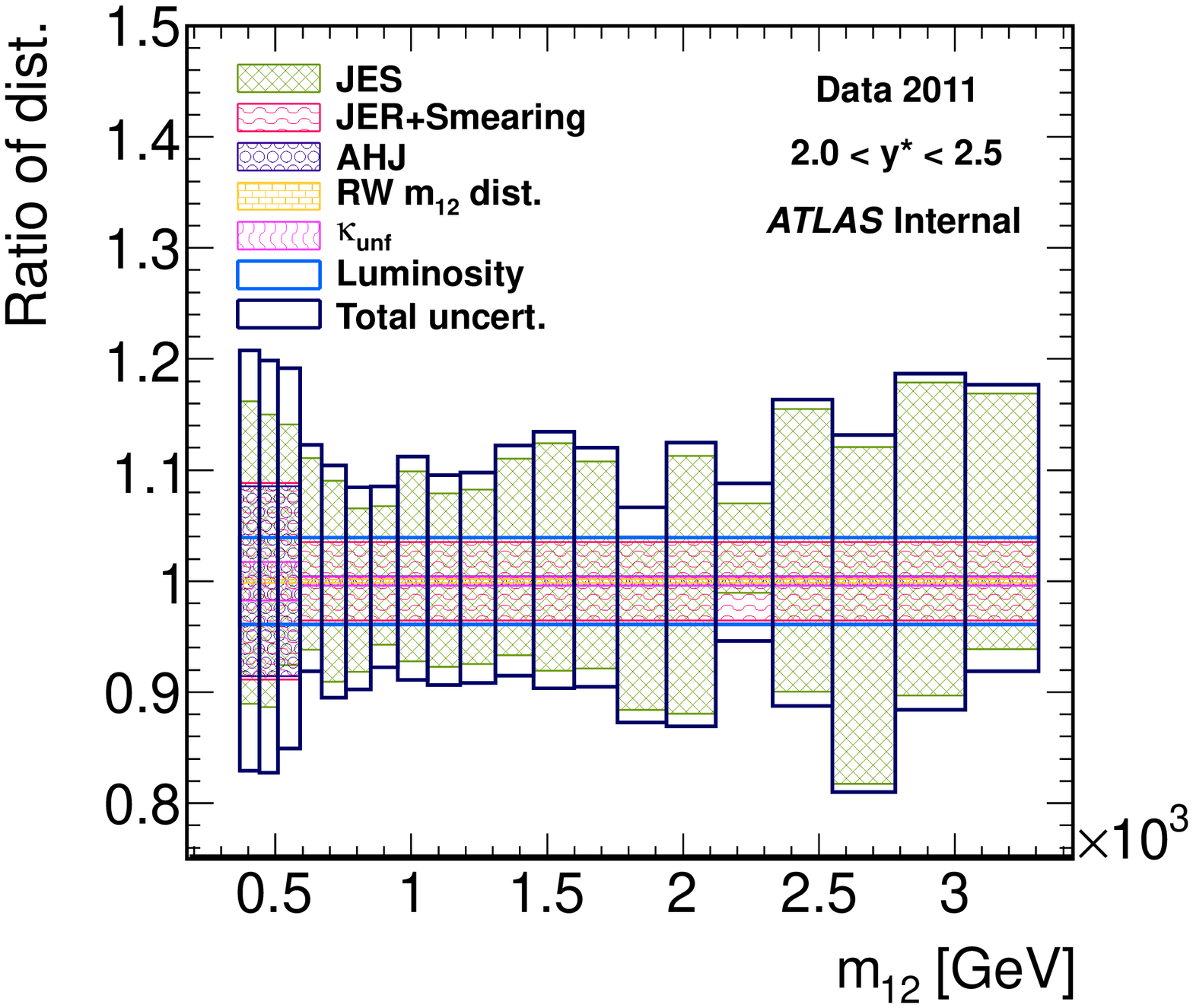}}
  \caption{\label{FIGdijetMassTotalSystematicUncertainty2011App}Dependence on the invariant mass, $m_{12}$, of the two jets with the highest transverse momentum in an event,
  of the ratio between the $m_{12}$ distributions
  with and without systematic variations (described in the text), for different \com jet rapidities, \ystr, in the 2011 measurement,
  including the uncertainty on the jet energy scale (\JES);
  the uncertainty on the jet energy and angular resolutions (\JES + smearing);
  the uncertainty associated with the choice of physics generator (\AHJ); 
  the uncertainty associated with variation in the simulated shape of the $m_{12}$ spectrum (RW $m_{12}$ dist.\/);
  the choice of the number of iterations used in the unfolding procedure ($\kappa_{\mrm{unf}}$);
  the uncertainty on the luminosity (luminosity); and
  the total uncertainty on all the latter (total uncert.\/).
  (See also \autoref{chapMeasurementOfTheDijetMass}, \autoref{FIGdijetMassTotalSystematicUncertainty} and accompanying text.)
  }
\end{center}
\end{figure} 

}{} 

\ifthenelse {\boolean{do:fourJetDPS}} {
  \begin{figure}[htp]
  %
  %
  \section{Hard double parton scattering in four-jet events\label{chapDoublePartonScatteringApp}}
  %
  %
  Additional figures pertaining to \autoref{chapDoublePartonScattering} are presented in the following;
  normalized distributions of the input variables to the NN (defined in \autoref{eqMassMeasurementPhasespace})
  are used in \autorefs{FIGnnVarFitDeltaPtApp}~-~\ref{FIGnnVarFitDeltaEtaApp}
  to compare the data to the expectation. The latter refers to a combination of the signal and background samples,
  normalized in proportion according to the fitted value of the fraction of DPS events, $\fDPS = 8.1\%$.
  \vspace{-10pt}
  \begin{center}
  \subfloat[]{\label{FIGnnVarFitDeltaPtApp1}\includegraphics[width=.52\textwidth]{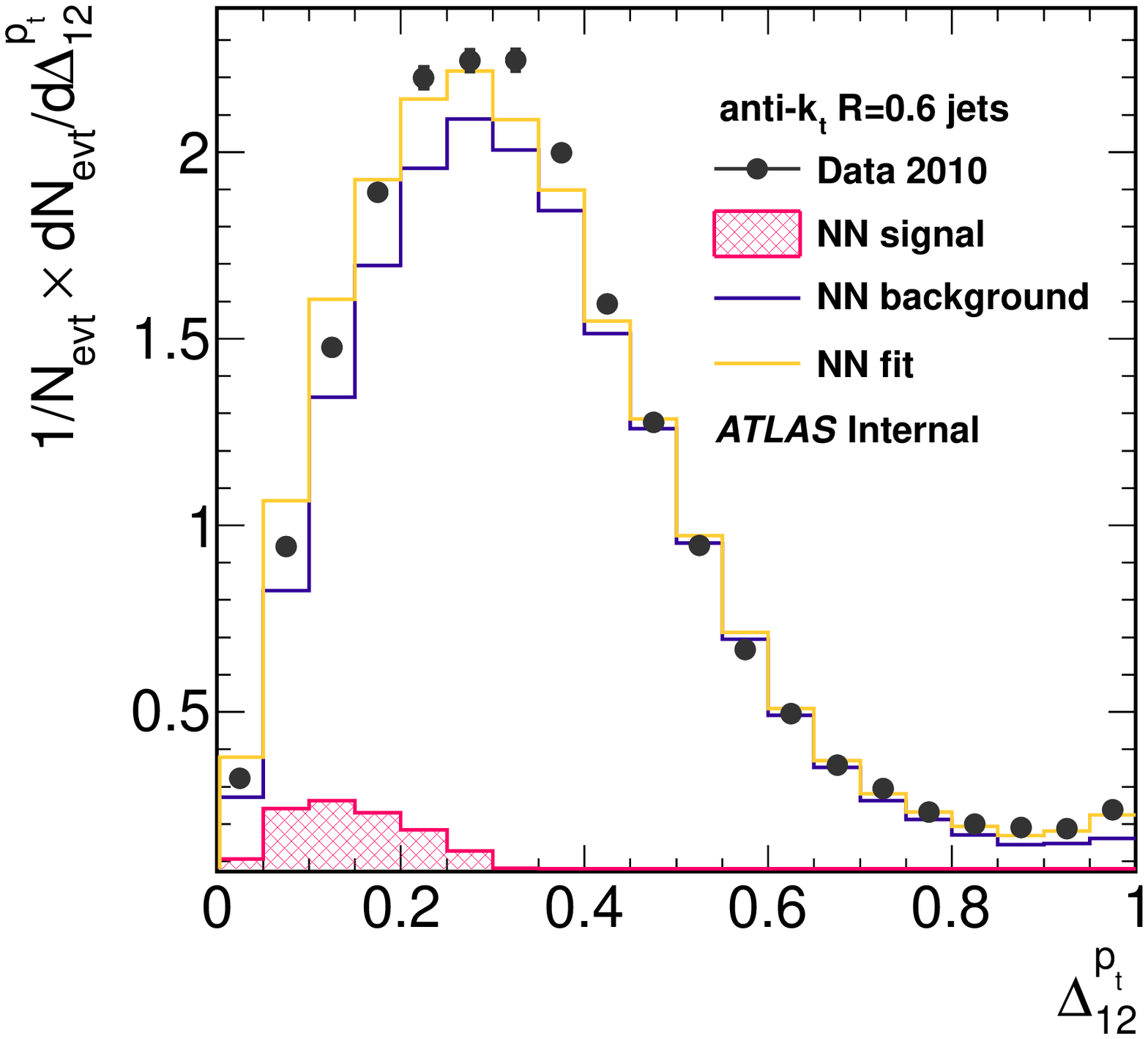}}
  \subfloat[]{\label{FIGnnVarFitDeltaPtApp2}\includegraphics[width=.52\textwidth]{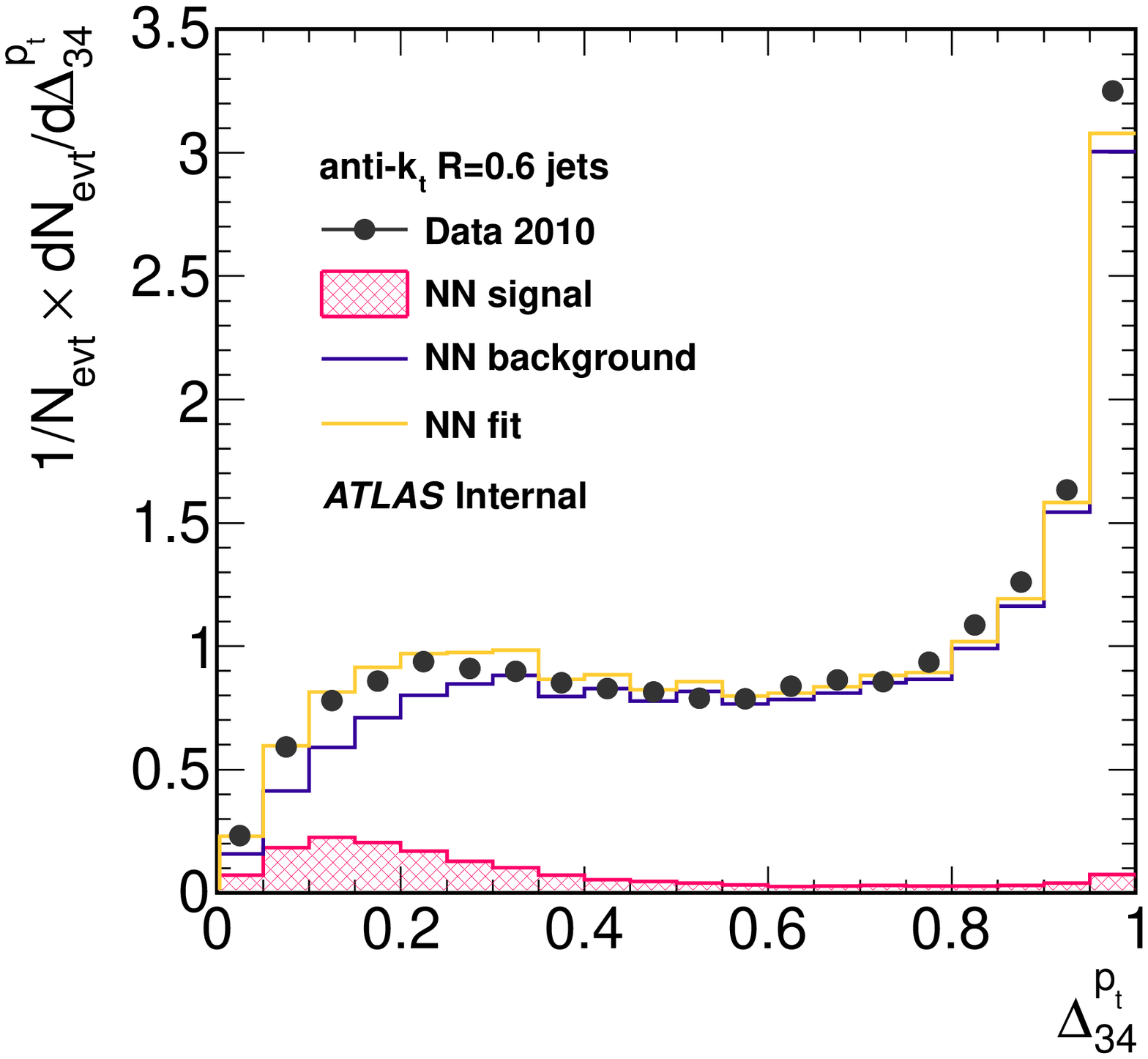}}
    \caption{\label{FIGnnVarFitDeltaPtApp}Differential distributions of the variables, $\Delta^{\pt}_{12}$ \Subref{FIGnnVarFitDeltaPtApp1}
    and $\Delta^{\pt}_{34}$ \Subref{FIGnnVarFitDeltaPtApp2}, which are defined in
    \autoref{chapDoublePartonScattering}, \autoref{eqMassMeasurementPhasespace},
    for four-jet events in the 2010 data, for the signal and background input samples of the NN and
    for the sum of the two NN inputs (denoted by ``NN fit''), where the signal and background samples
    are each normalized according to the fit to the fraction of DPS events, as explained in \autoref{chapTrainingOutputOfNN}.
    }
  \end{center}
  \end{figure} 
  \begin{figure}[htp]
  \begin{center}
  \subfloat[]{\label{FIGnnVarFitDeltaPhiApp1}\includegraphics[width=.52\textwidth]{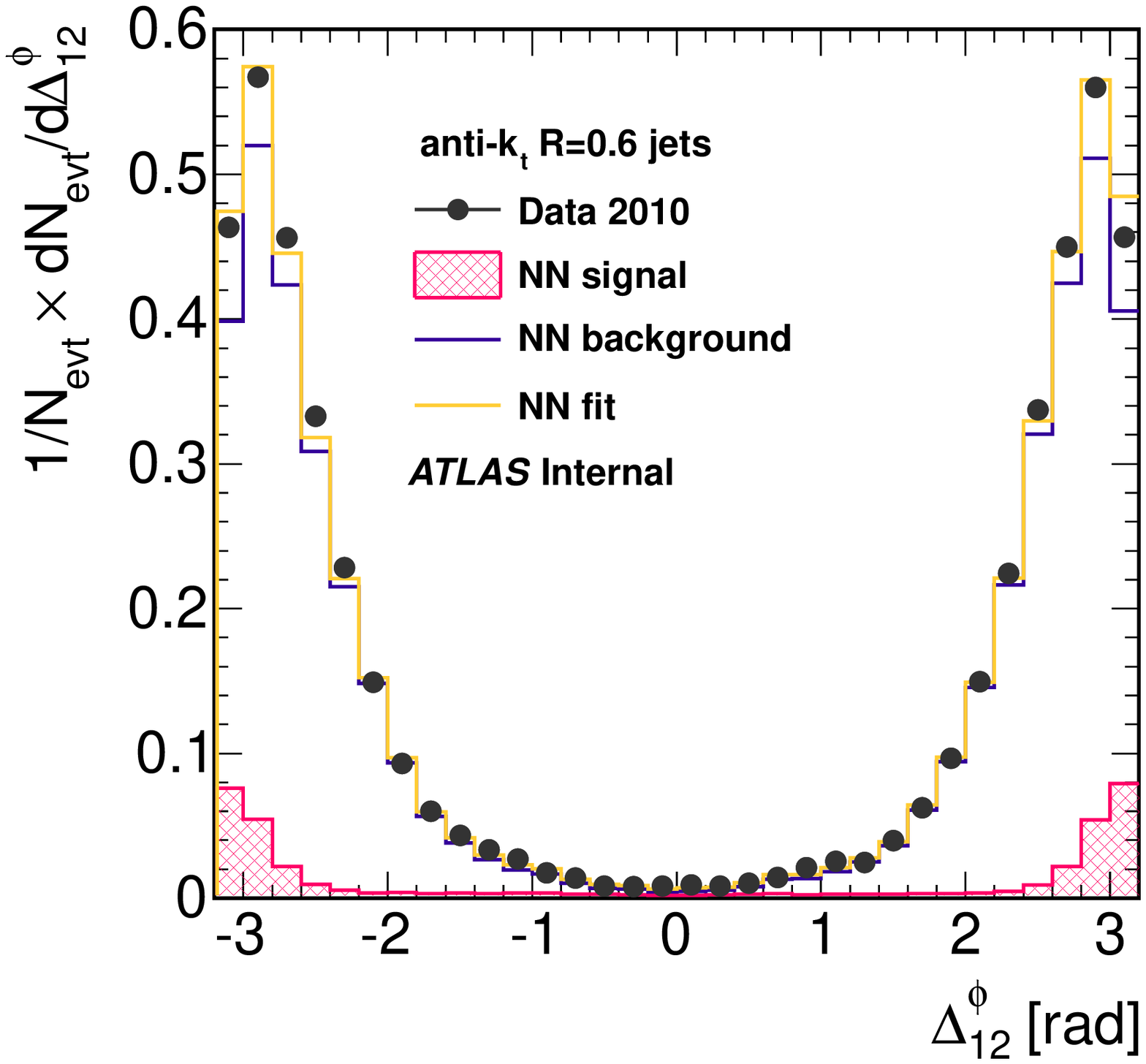}}
  \subfloat[]{\label{FIGnnVarFitDeltaPhiApp2}\includegraphics[width=.52\textwidth]{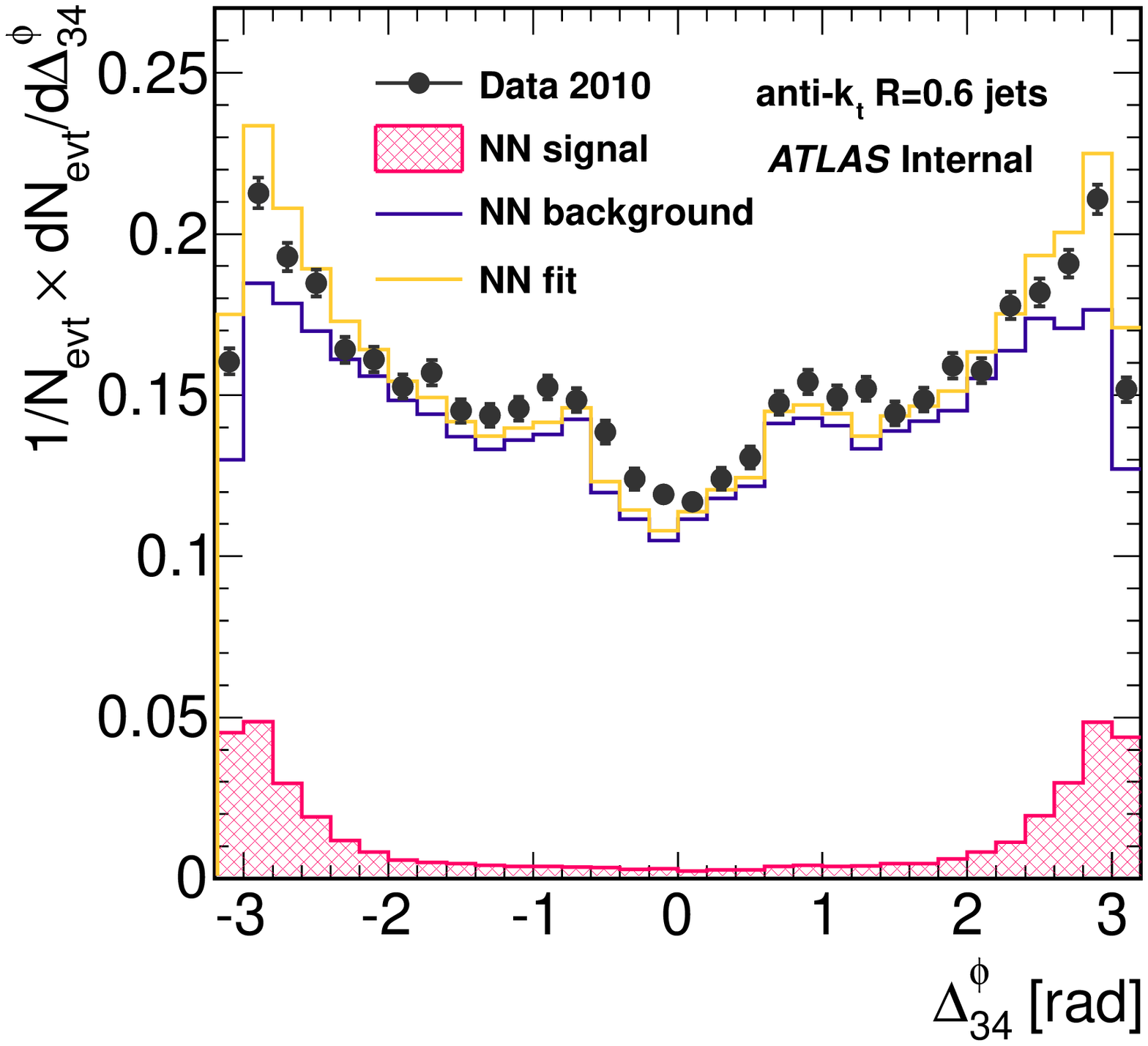}} \\
  \subfloat[]{\label{FIGnnVarFitDeltaPhiApp3}\includegraphics[width=.52\textwidth]{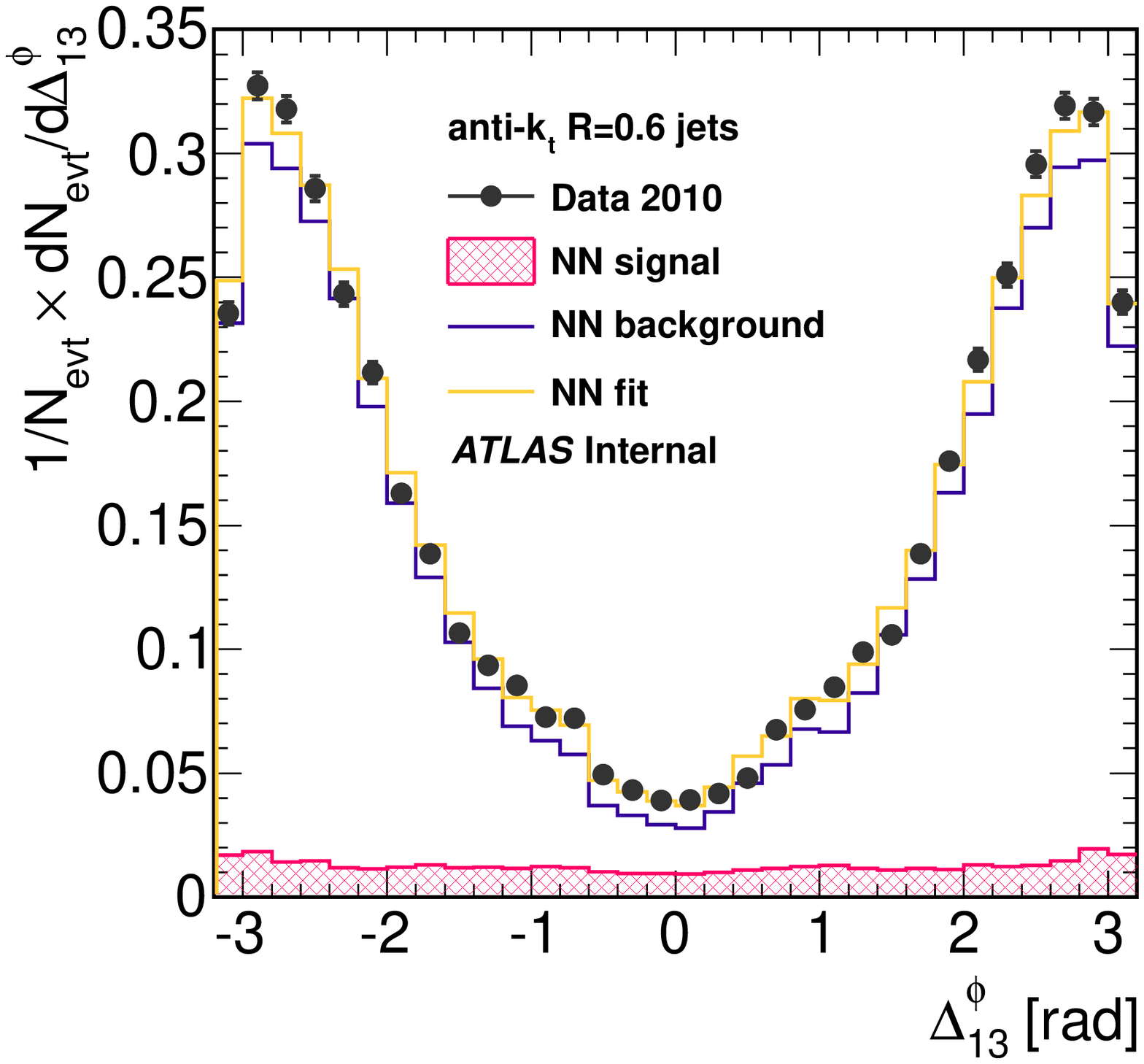}}
  \subfloat[]{\label{FIGnnVarFitDeltaPhiApp4}\includegraphics[width=.52\textwidth]{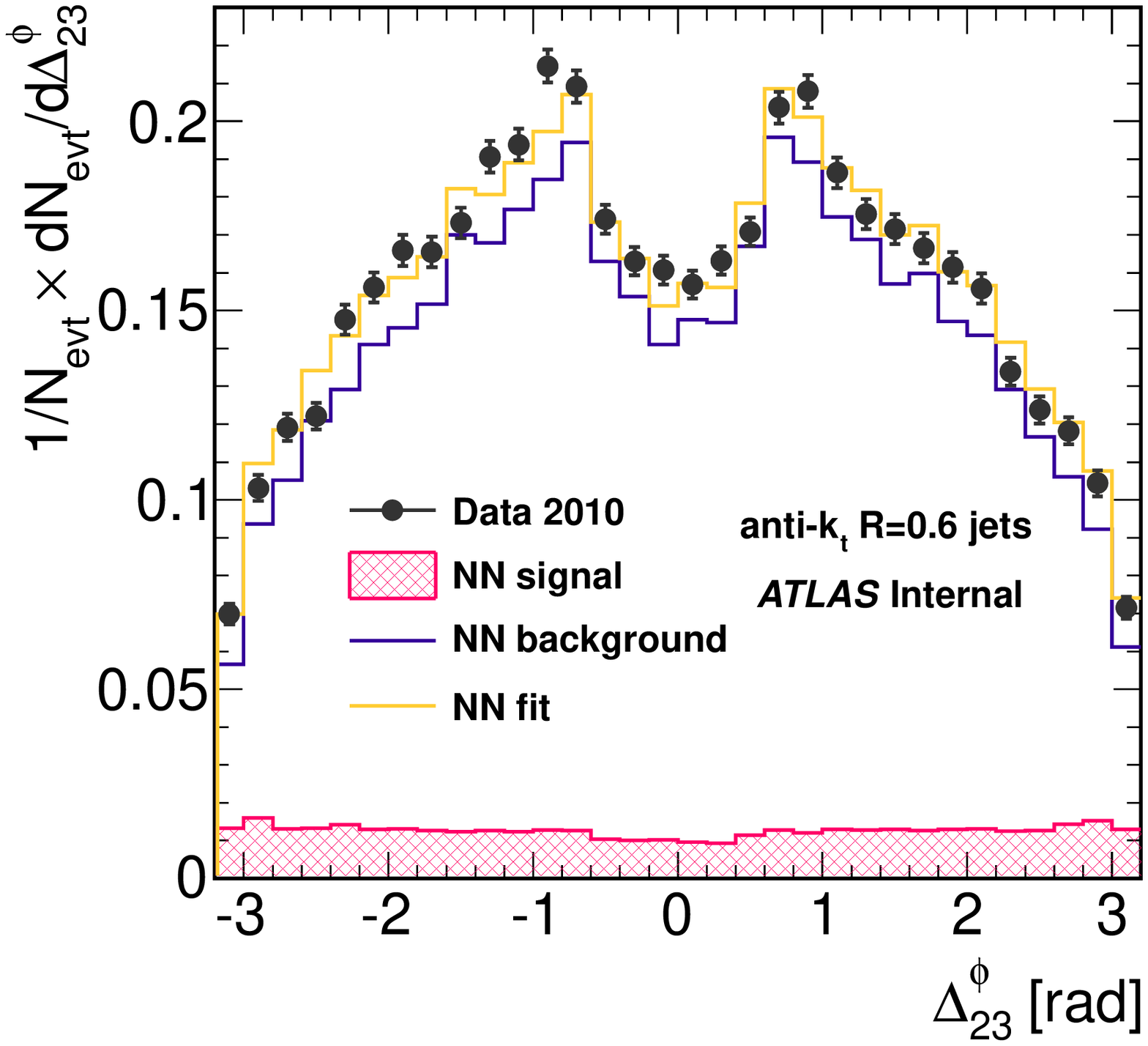}}
    \caption{\label{FIGnnVarFitDeltaPhiApp}Differential distributions of the variables, $\Delta^{\phi}_{12}$ \Subref{FIGnnVarFitDeltaPhiApp1},
    $\Delta^{\phi}_{34}$ \Subref{FIGnnVarFitDeltaPhiApp2}, $\Delta^{\phi}_{13}$ \Subref{FIGnnVarFitDeltaPhiApp3}
    and $\Delta^{\phi}_{23}$ \Subref{FIGnnVarFitDeltaPhiApp4}, which are defined in
    \autoref{chapDoublePartonScattering}, \autoref{eqMassMeasurementPhasespace},
    for four-jet events in the 2010 data, for the signal and background input samples of the NN and
    for the sum of the two NN inputs (denoted by ``NN fit''), where the signal and background samples
    are each normalized according to the fit to the fraction of DPS events, as explained in \autoref{chapTrainingOutputOfNN}.
    }
  \end{center}
  \end{figure} 
  \begin{figure}[htp]
  \begin{center}
  \subfloat[]{\label{FIGnnVarFitDeltaEtaApp1}\includegraphics[width=.52\textwidth]{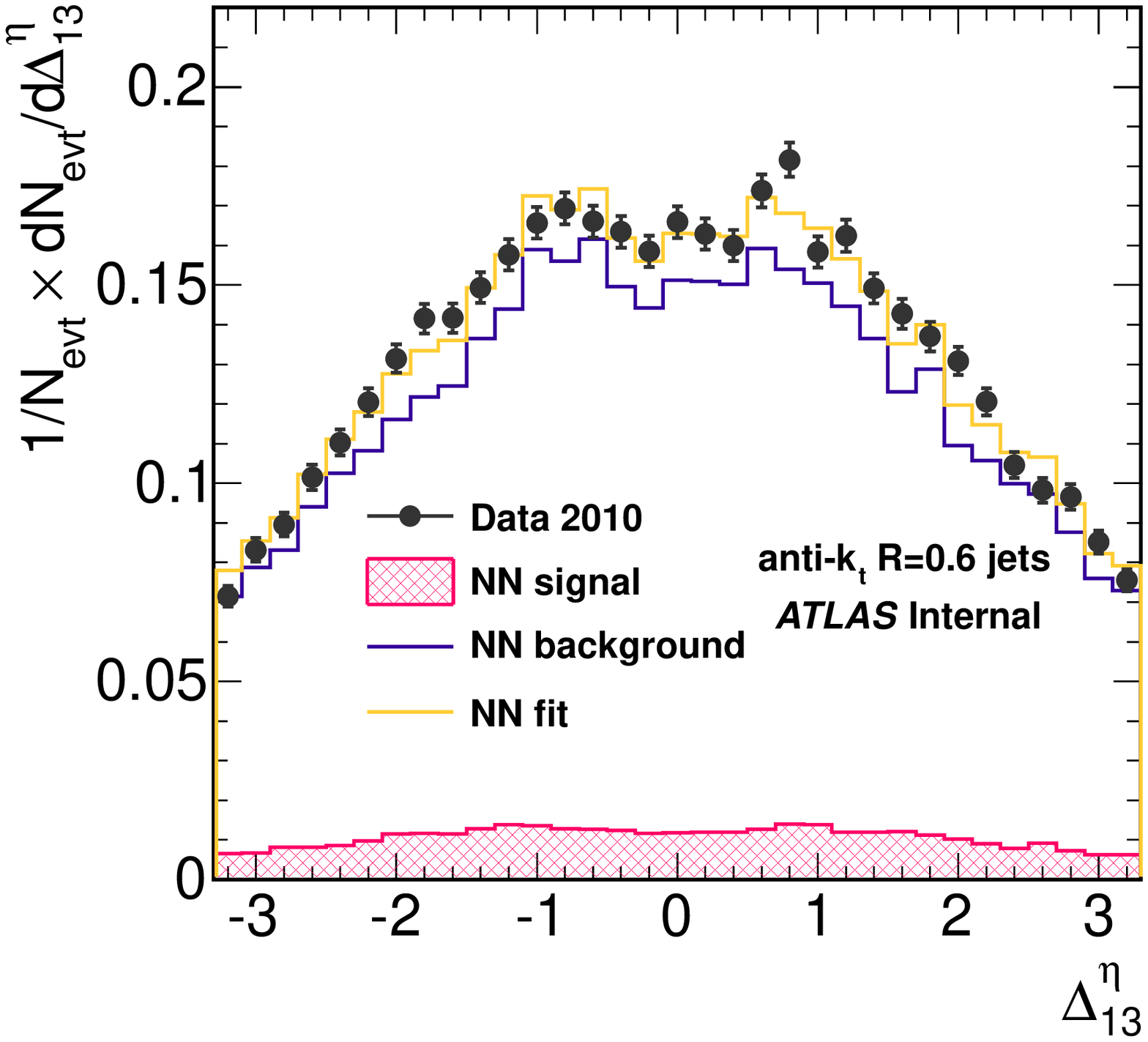}}
  \subfloat[]{\label{FIGnnVarFitDeltaEtaApp2}\includegraphics[width=.52\textwidth]{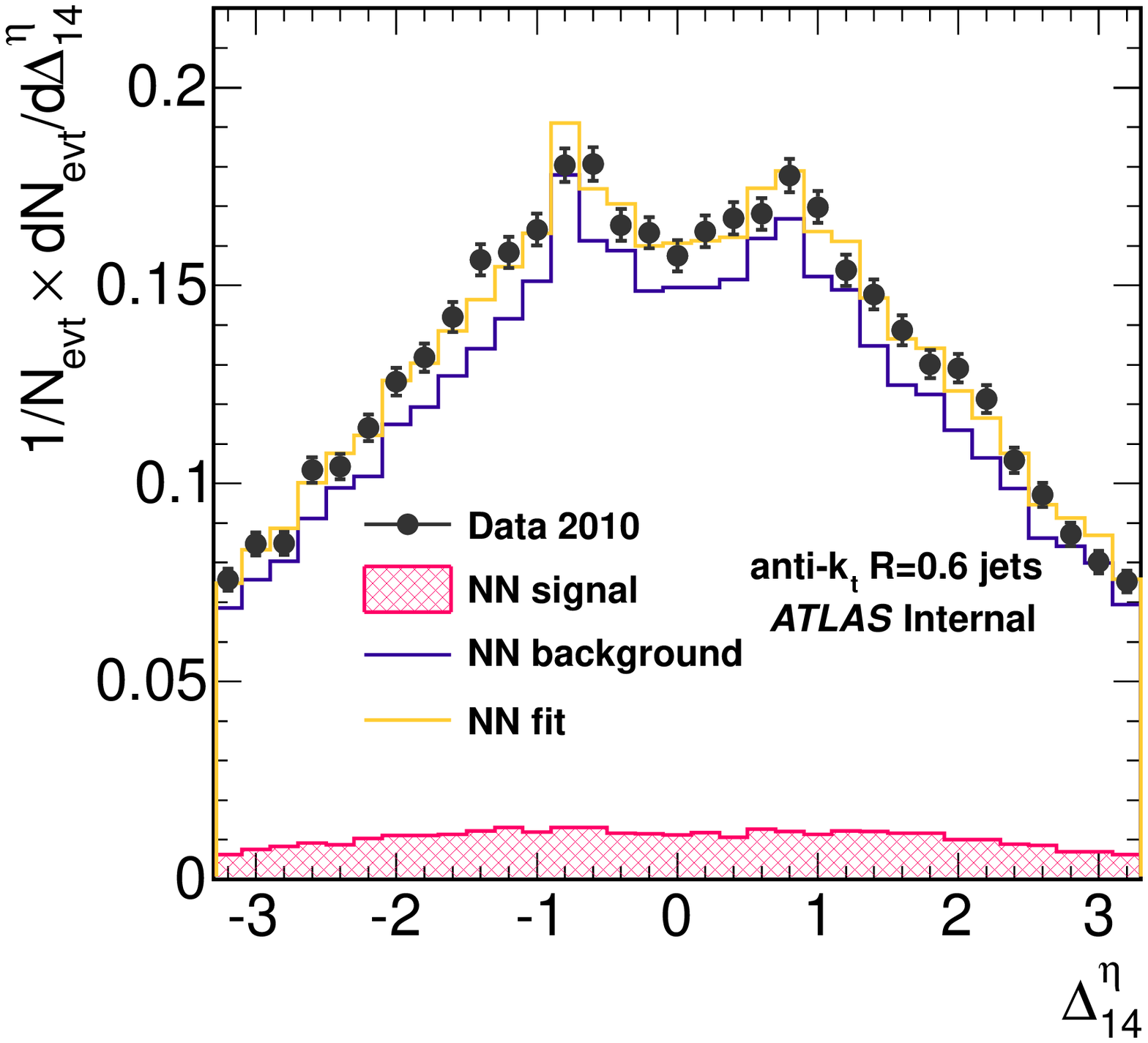}} \\
  \subfloat[]{\label{FIGnnVarFitDeltaEtaApp3}\includegraphics[width=.52\textwidth]{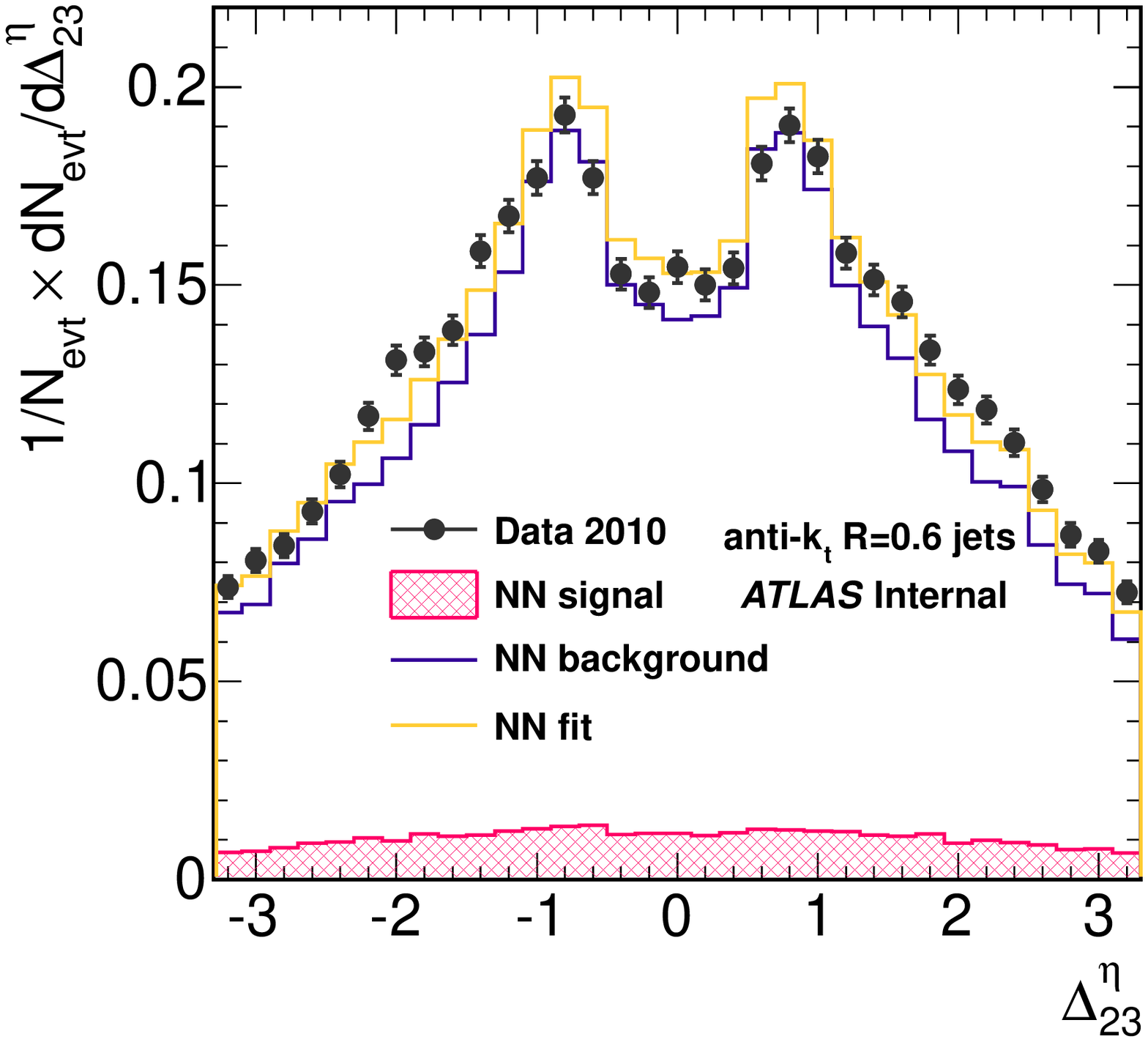}}
  \subfloat[]{\label{FIGnnVarFitDeltaEtaApp4}\includegraphics[width=.52\textwidth]{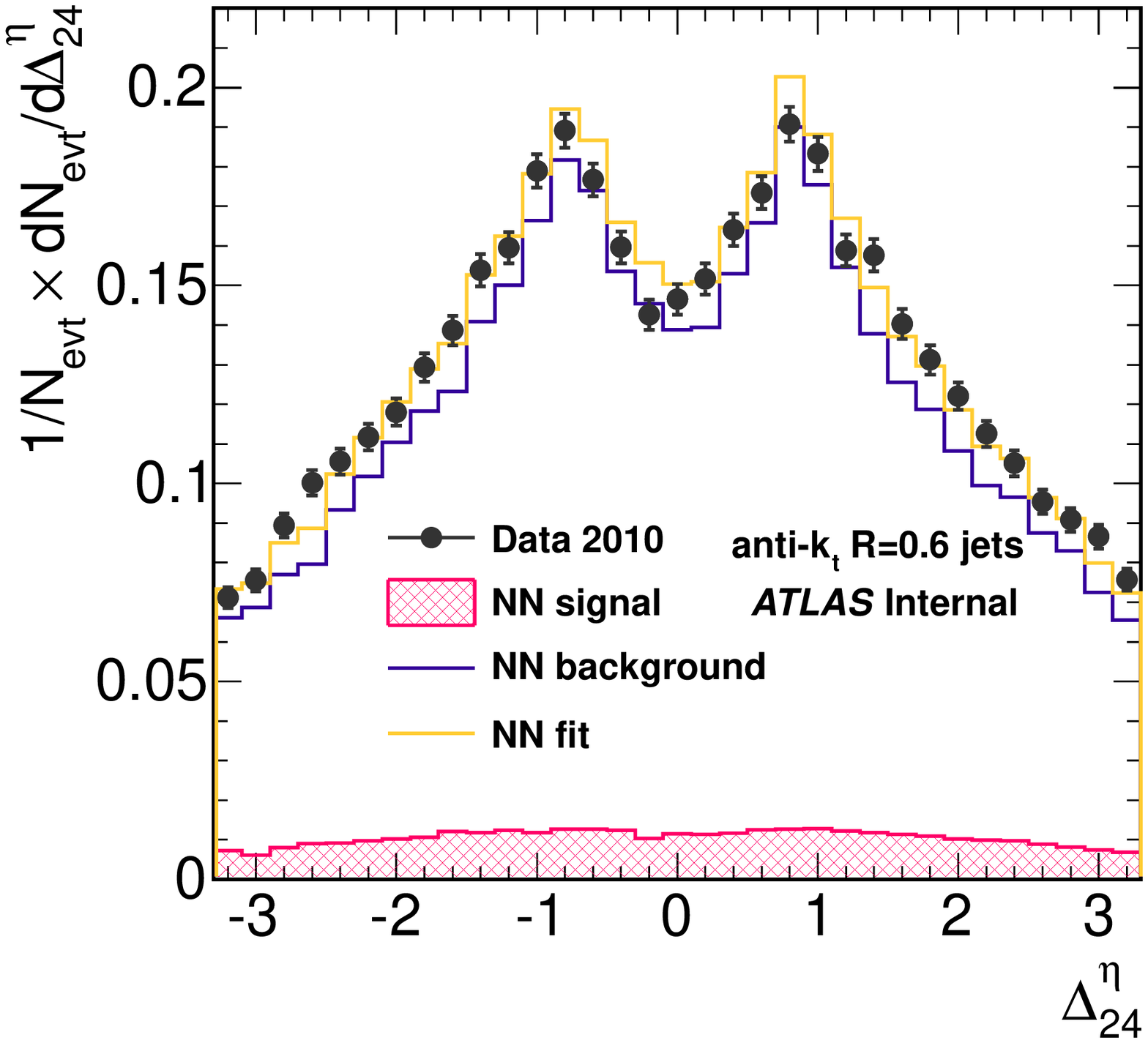}}
    \caption{\label{FIGnnVarFitDeltaEtaApp}Differential distributions of the variables, $\Delta^{\eta}_{13}$ \Subref{FIGnnVarFitDeltaEtaApp1},
    $\Delta^{\eta}_{14}$ \Subref{FIGnnVarFitDeltaEtaApp2}, $\Delta^{\eta}_{23}$ \Subref{FIGnnVarFitDeltaEtaApp3}
    and $\Delta^{\eta}_{24}$ \Subref{FIGnnVarFitDeltaEtaApp4}, which are defined in 
    \autoref{chapDoublePartonScattering}, \autoref{eqMassMeasurementPhasespace},
    for four-jet events in the 2010 data, for the signal and background input samples of the NN and
    for the sum of the two NN inputs (denoted by ``NN fit''), where the signal and background samples
    are each normalized according to the fit to the fraction of DPS events, as explained in \autoref{chapTrainingOutputOfNN}.
    }
  \end{center}
  \end{figure} 

}{} 


 }{}

\bibliographystyle{packages/atlasnote} \bibliography{phd}

\end{document}